# Unravelling carbene moiety role in the N-heterocyclic carbene-phosphinidenes coordination chemistry: bonding and reactivity in gold complexes

Guillaume Thiam,[*a] Leonardo Belpassi,[*b] and Paola Belanzoni[*ab]

[a]*Department of Chemistry, Biology and Biotechnologies, University of Perugia, Via Elce di Sotto, 8 – 06123, Perugia, Italy*

[b]*CNR Institute of Chemical Science and Technologies "Giulio Natta" (CNR-SCITEC), Via Elce di Sotto, 8 – 06123, Perugia, Italy*

***Guillaume Thiam ORCID: orcid.org/0000-0001-6154-8457***

***Email: guillaume.thiam@unipg.it***

***Leonardo Belpassi ORCID: orcid.org/0000-0002-2888-4990***

***Email: leonardo.belpassi@cnr.it***

***Paola Belanzoni ORCID: orcid.org/0000-0002-1286-9294***

***Email: paola.belanzoni@unipg.it***

## Abstract

The coordination properties of the emerging class of N-heterocyclic carbene-phosphinidene (NHCP) ligands have been reported to be strongly affected by the carbene (NHC) moiety (*Chem* **2025**, *11*, 102649), but a clear understanding remains limited. In this work the bonding features and reactivity of 13 NHCP gold hydride complexes, [NHCPAuH], bearing different classes of NHCs (i.e. CAACs, unconjugated DACs and conjugated DACs) have been systematically explored and compared to carbene analogues, using a consistent computational protocol. The analyses reveal that, although the π-acceptor ability trend of NHCPs qualitatively parallels that of NHCs, the σ-donor ability trend reverses, with more σ-donating NHC moieties generating less σ-donating NHCP ligands. Concurrently, the nature of the NHC moiety impacts the coordination geometry at the P center, which ranges from nearly "planar" (CAACs), to pyramidal (unconjugated DACs) and $PH_3$-like (conjugated DACs), with the $C_{NHC}$-P-Au bond angle appearing as a structural descriptor to monitor and/or design NHCP ligands with desired σ-donor properties. The NHCPs stronger σ-donor and weaker π-acceptor abilities compared to NHCs and their unique structural flexibility and electronic adaptability have been showcased to be directly controlled by the carbene moiety via modulation of the HOMO lone pair energy and its atomic phosphorus 3p character. The reactivity of [NHCPAuH] complexes with $CO_2$, taken as a probe for potential NHCP applications in small molecule activation processes, demonstrates qualitatively similar mechanisms, with remarkably different activation barriers, which correlate with the NHCPs σ-donor ability, eventually modulated by second sphere effects, thus reflecting the high tunability of their bonding/reactivity properties. This work provides insights and perspectives on the design principle of NHCP ligands within the chemical space investigated, with a spotlight on the pivotal role of NHC moiety in modulating their electronic and steric properties, offering opportunities for burgeoning applications across diverse fields.

## Introduction

Since their first isolation and structural characterization by Arduengo in 1991, (1) N-heterocyclic carbenes (NHCs) have been considered as privileged ligands, because of their extreme versatility in modulating both electronic and steric properties. Transition-metal NHC complexes are heavily employed across the chemical industry for applications in several disparate fields ranging from catalysis to material sciences and metallopharmaceuticals (for a recent review, see 2). The structure of an NHC ligand can be suitably modified to significantly impact their coordination behavior. Key features affecting their coordination properties are the presence and location of heteroatoms (e.g. nitrogen), substituents at proximal positions, backbone structure (e.g. unsaturation) and substituents, and the size of the ring. (3) The NHCs display peculiar features, such as strong electron-donating ability, remarkable tunability of both electronic and steric properties, straightforward synthetic accessibility and high affinity towards metal centers as well as non-metallic species. Such features have led to impressive advances in the development of many different classes of carbenic species, including cyclic (alkyl)(amino) carbenes (CAACs), (4) diaminocarbenes with an unconjugated cyclic ring (unconjugated DACs) and imidazole-2-ylidenes (conjugated DACs). (5)

Over the years, the electronic (6) and steric features (7) of carbenes have been quantitatively evaluated using experimental parameters. Concerning the NHC's electronic properties, the Tolman electronic parameter (TEP) (8) is widely used to correlate the ligand donor-acceptor properties with the infrared stretching frequency of carbon monoxide in $(NHC)Ni(CO)_3$, cis-$(NHC)IrCl(CO)_2$ or cis-$(NHC)RhCl(CO)_2$ complexes; (9) to evaluate the extent of σ-donation,

Huynh et al. introduced a parameter based on the $^{13}C$ NMR chemical shift of trans-[$PdBr_2(^iPr_2$-bimy)(L)] complexes. (10) Both $^{31}P$ or $^{77}Se$ NMR spectroscopy of NHC-phosphinidene and NHC-selenium adducts were introduced respectively by Bertrand (11) and Ganter (12) to assess the extent of π-backdonation. It was found that reduced electron density at the probe atom, caused by backbonding, can induce a downfield shift of the corresponding resonance.

In 2018, a comprehensive overview by Slootweg et al., covering the synthesis and applications of N-heterocyclic carbene-phosphinidenes (NHCPs), ratified a novel class of NHC derivatives formed by linking the 2-position of the N-heterocycle with an exocyclic phosphorous fragment. (13) Since then, NHCPs have become very relevant ligands, widely applied in both main-group and transition metal chemistry. Several examples of NHCP-stabilized transition metal complexes and their catalytic applications have been reported, with M = Mn, Fe, Co, Rh, Ir, Cu, Ag, Au. (14-20)

The bonding between carbene and exocyclic phosphorus atom in NHCP can be described by three resonance structures, as illustrated in Scheme 1.

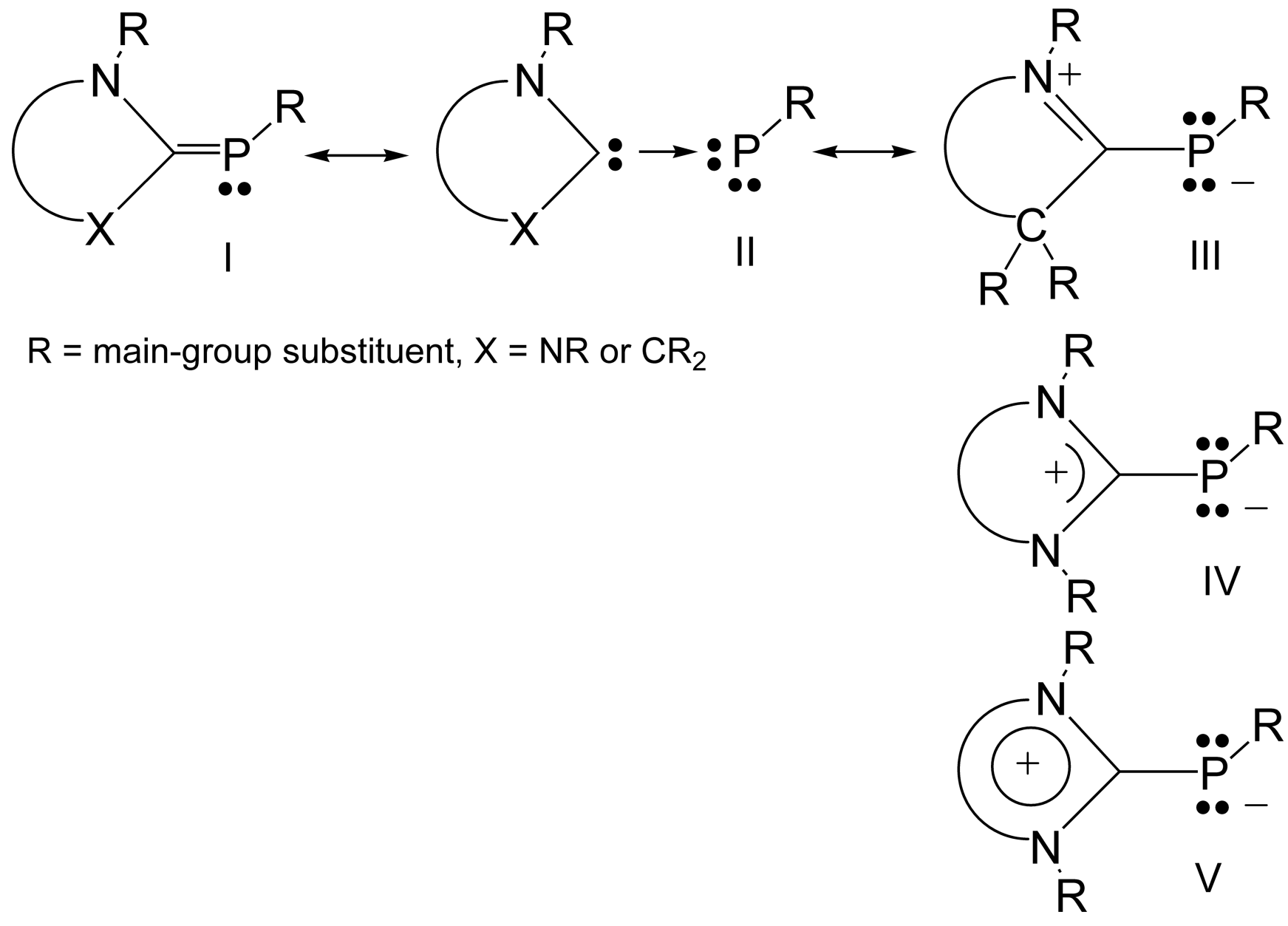

**Scheme 1**: Canonical forms of neutral NHCP: I) phosphorus atom as a 2σ electron donor; II) donor-acceptor carbene-phosphinidene complex; III)-V) phosphorus atom as a 2σ and 2π/2σ electron donor; zwitterionic resonance structure for CAACs (III), unconjugated DACs (IV) and conjugated DACs (V).

Canonical form I corresponds to a typical phosphaalkene showing a polar ($C^{\delta-} – P^{\delta+}$) formal C=P double bond with a lone pair at the phosphorus atom. Conversely, canonical forms II and III-V represent a C-P single bond with a dative C → P donor-acceptor interaction (II) and a zwitterionic character (III-V), where the stronger C to P σ-donation leads to a reversed-polarized C-P bond ($C^{\delta+} – P^{\delta-}$). The exact description of a given NHCPR adduct depends on the phosphorus substituent and the nature of the carbene. The $^{31}$P-NMR spectroscopic data show high electron density on the phosphorus atom and the degree of polarization of the carbon-phosphorus bond has been suggested to serve as an indicator of the π-accepting properties of the corresponding NHC. (11,21) However, it is worth reminding here that a systematic computational study by some of us (3) demonstrated that $^{31}$P-NMR data cannot be applied as a general tool for quantifying the π-accepting properties of carbenes, giving a qualitative correlation, which becomes effectively quantitative only when carbenes of similar structure are compared in a limited range of π-backdonation values. (3) Resonance structures I and II-V (Scheme 1) each possess one or two lone pairs of electrons on phosphorus available for metal coordination, forming mono- or bidentate complexes or mono- or dinuclear metal complexes.

More recently, a perspective by Inoue and Zhu demonstrated the impact of structurally and electronically distinct carbene moieties on the design principles of NHCP ligands, suggesting, among the potential future directions for the development of NHCPs, their applications in small molecule activation. (22) Yet, this highly attractive research field remains significantly underexplored to date. In ref. (22), the impact of the carbene moiety in NHCP has been described with reference to the three different categories of NHCs, i.e. CAACs, unconjugated DACs and conjugated DACs. Based on $^{31}$P and $^{77}$Se NMR spectroscopy of their corresponding carbene adducts, (11, 12) a progressive increase in π-acidity from conjugated DACs to CAACs has been presented, influencing the electronic properties and bonding of NHCP. The most pronounced π-acidity of CAACs is proposed to allow for a bonding in CAAC phosphinidenes best described by canonical structures I and II (Scheme 1). In contrast, the weakest π-acceptor ability of conjugated DACs is suggested to make the conjugated DAC phosphinidenes to preferentially adopt resonance structures II and V. For unconjugated DACs, due to their moderate π-acidity, the corresponding unconjugated DAC phosphinidenes may exhibit potentially resonance structures I, II and IV, depending on the type of substituents.

Although the carbene moiety influence on the NHCP π-backdonation properties has been widely discussed in previous works (11, 21) and in ref. 22, its effect on their σ-donor properties remains mostly unexplored, which is a fundamental topic to get a comprehensive view of the NHCP coordination properties.

Inspired by the paper by Inoue and Zhu, (22) we decided to systematically investigate the coordination chemistry of NHCPs, focusing on the impact of carbene moiety on the bonding nature of NHCPs with an Au-H metal fragment, in the corresponding gold hydride [(NHCP)AuH] complexes. Reactivity of these complexes is investigated by selecting $CO_2$ as a probe for small molecules activation. We chose Au–H for our computational modelling because NHC ligands are well known to stabilize gold–hydride, as demonstrated by Tsui's pioneering synthesis of the versatile [(IPr)AuH] synthon (IPr = 1,3-bis(2,6-diisopropylphenyl)imidazol-2-ylidene) (23). This complex exhibits reactivity towards dimethyl acetylenedicarboxylate, ethyl diazoacetate, and $O_2$, though not $CO_2$. On the other hand, mono-, bi- and tetranuclear coinage metal complexes of NHCP, [(IPrPPh)MCl] (M = Cu, Ag, Au), [(IPrPPh)$(MCl)_2$] (M = Cu, Au), and [$(IPrPPh)_2M_4Cl_2$]$X_2$ (M = Cu, Au; X = $BAr^F$, $SbF_6$), were also reported, and the cationic tetra- and dinuclear gold complexes were used as catalysts for enyne cyclization and carbene transfer reactions. (24) Similar [(IMesPPh)$(MCl)_2$] (IMes = 1,3-bis(2,4,6-trimethyl-phenyl)imidazolin-2-ylidene; M = Cu, Ag, Au) chloride complexes have been characterized. (25) However, to the best of our knowledge, no gold hydride [(NHCP)AuH] complexes have been reported so far, which could be extremely valuable for $CO_2$ activation, and therefore all the investigated complexes are hypothetical computational models. Indeed, a recently reported NHC-stabilized diphosphene Au(I) hydride [(NHCPP)AuH] (NHC = $IMe_4$ = 1,3,4,5-tetramethylimidazol-2-ylidene) complex showed facile and reversible reaction with carbon dioxide yielding the corresponding Au formate (Scheme 2). (26,27) The mechanism of this reactivity has been computationally elucidated by some of us in a previous work, where the ligand electronic effect in the reaction between a number of $[LAu(I)H]^{0/-}$ hydride species and $CO_2$ has been unveiled. (28)

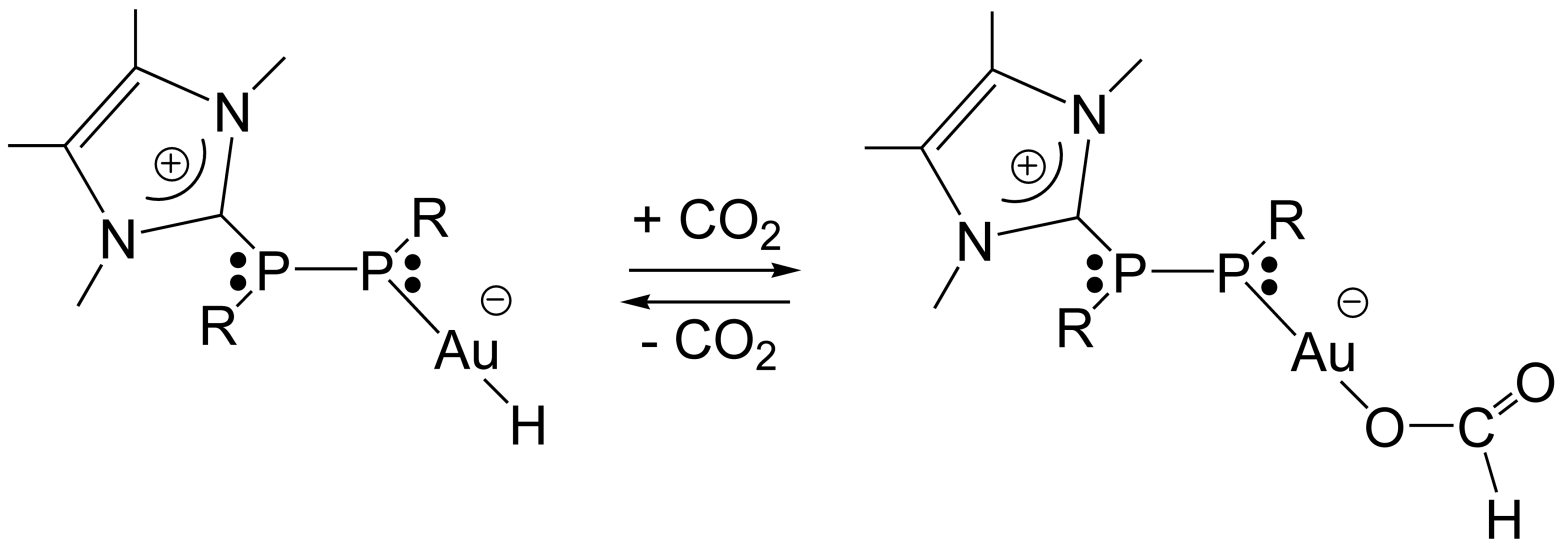

**Scheme 2**: Reversible reaction of NHCPP-AuH with $CO_2$ (R = 2,6-$Mes_2$-$C_6H_3$)

A direct correlation between the ligand σ-donor ability, which affects the electron density at the hydride H reactive site, and the kinetic activation barriers of this reaction has been found. However, the main obstacle detected for this reactivity is thermodynamic, rather than kinetic, as the product was found generally to be inherently unstable. The ability of the carbene moiety to stabilize a positive charge has been proven to be crucial for the $CO_2$ activation, which is enhanced when the AuH moiety is bonded to an anionic or zwitterionic ligand. From a ligand design perspective, based on our previous work, (28) strong σ-donor ligands, preferentially featuring a zwitterionic structure, emerged as promising and experimentally accessible potential candidates for the stoichiometric or catalytic $CO_2$ activation.

In this framework, understanding of the NHCP-AuH bond/reactivity relationship is compelling to advance our knowledge on the properties of this new type of ligands and on their possible tuning through carbene and phosphorus moiety modifications, opening up new avenues for organocatalysis and beyond, broadening their applicability across diverse domains and also driving significant innovations.

In the present work, we systematically investigate the features of the NHCP-AuH bond using the series of carbene moieties of NHCP listed in ref. (22) in close relationship with the NHC-AuH bond in carbene analogues to unveil the impact of the carbene on the PR group. To directly compare the NHCP bonding properties towards AuH, a common R = Ph bonded to P has been selected. An extensive set of 26 different gold hydride complexes of the type [LAuH] with L=NHC, NHCP have been investigated in our study, which are all sketched in Scheme 3.

**CAACs**

a)

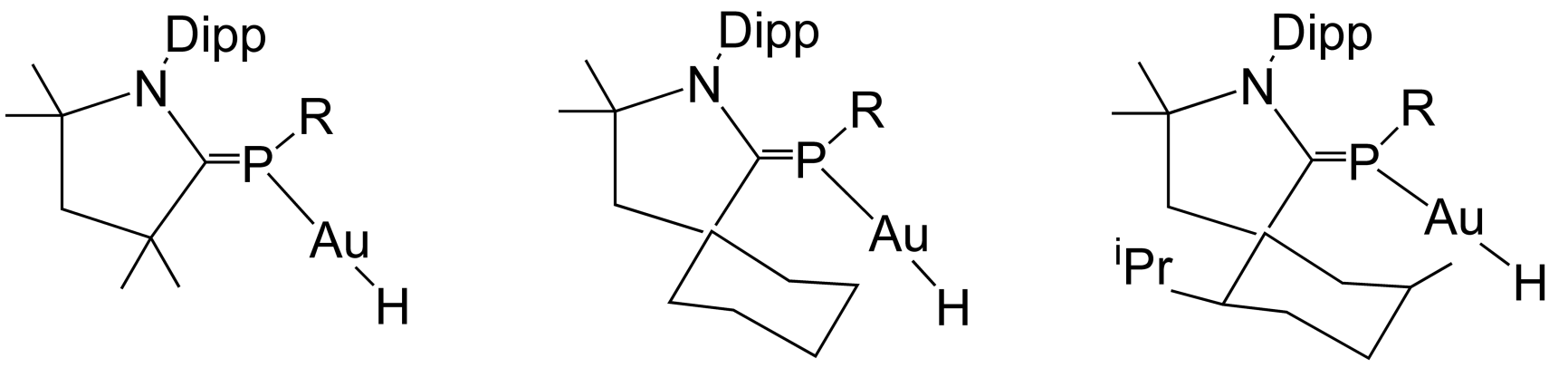

**1-P** **2-P** **3-P**

b)

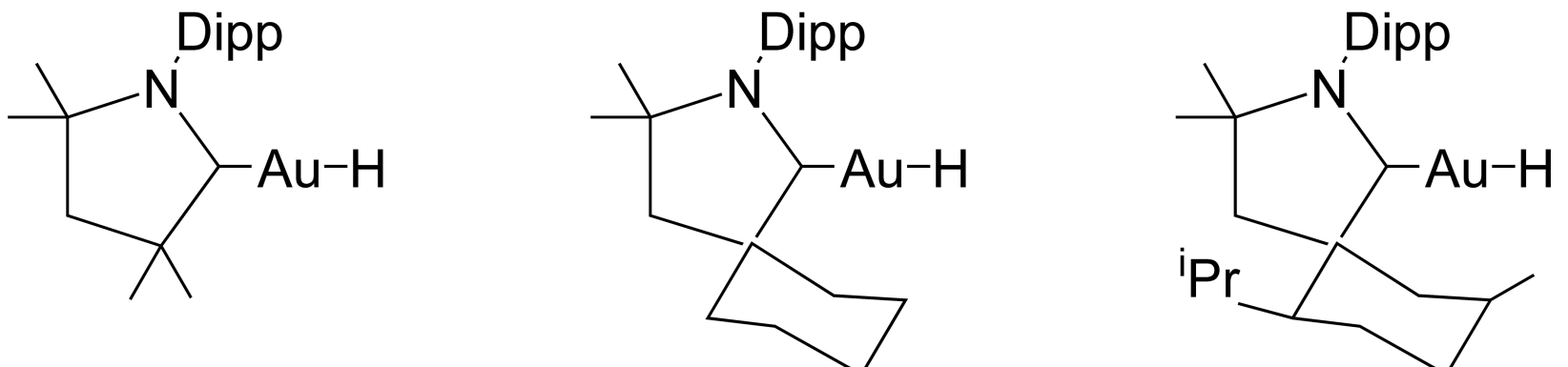

**1** **2** **3**

**Unconjugated DACs**

a)

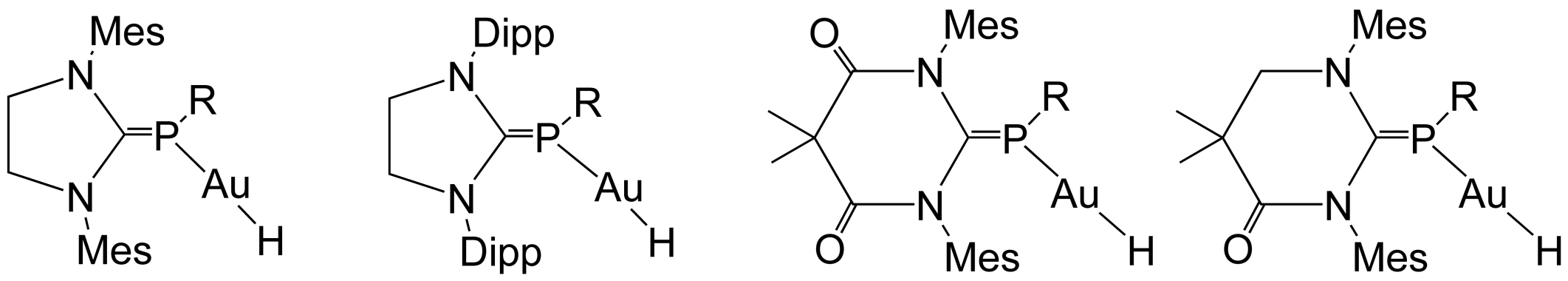

**4-P** **5-P** **6-P** **7-P**

b)

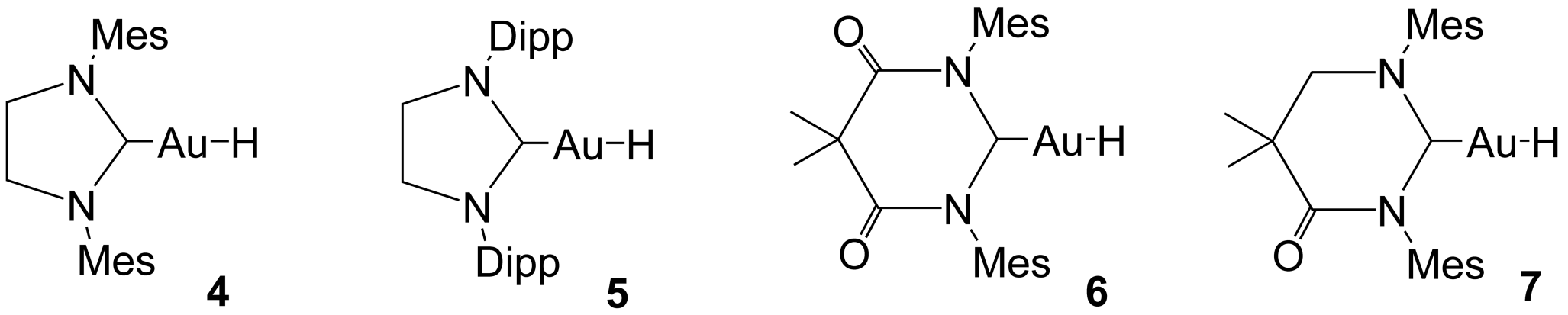

**4** **5** **6** **7**

**Conjugated DACs**

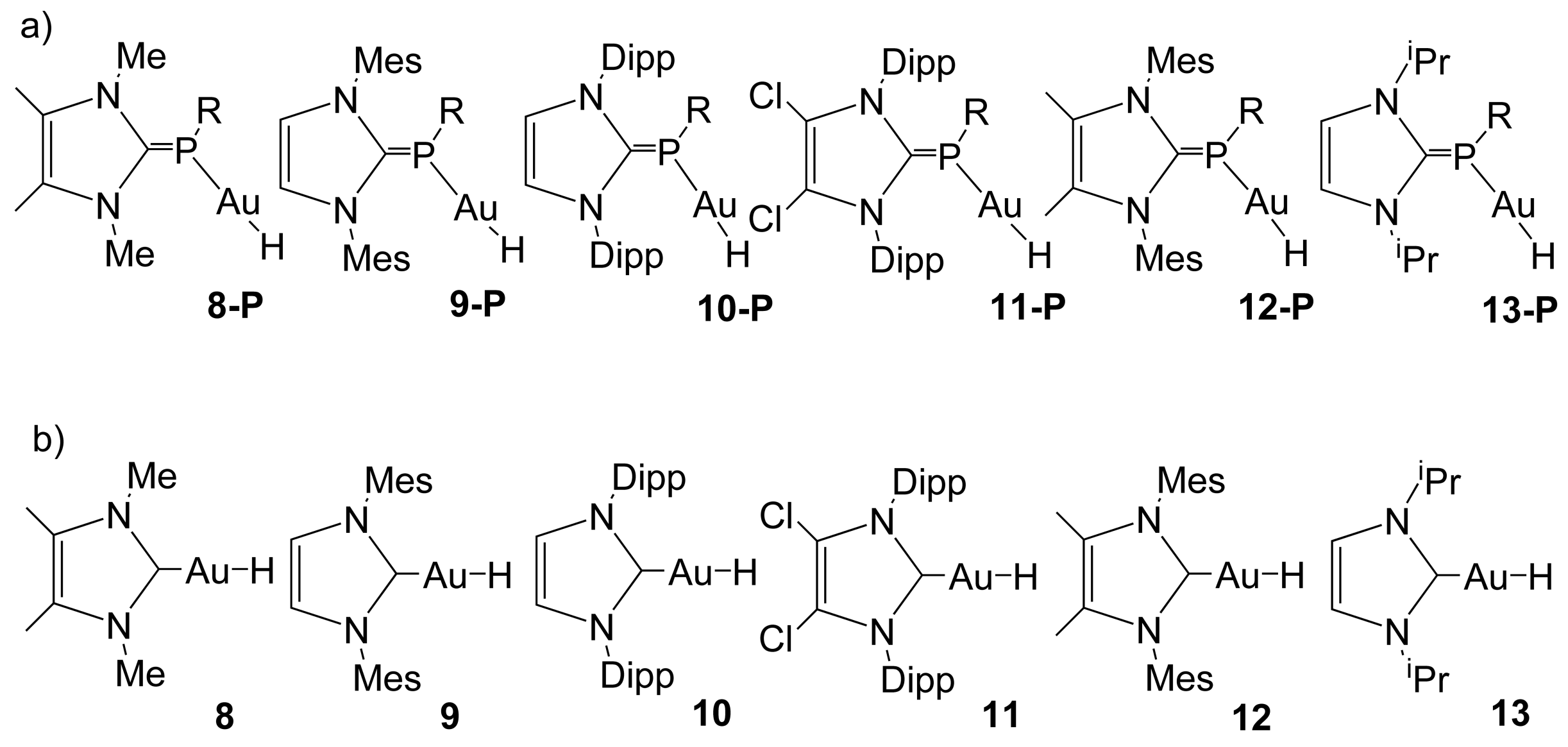

**Scheme 3**: Gold hydride complexes in this study: panel a) [NHCP-AuH]; panel b) [NHC-AuH] analogues. Me = methyl, Mes = 2,4,6-$Me_3C_6H_2$, $^{i}$Pr = isopropyl, Dipp = 2,6-diisopropylphenyl, R = phenyl. "DAC" notation, strictly representing diamidocarbenes **6**, **6-P**, is extended here to diaminocarbenes to maintain the same classification and abbreviations used in ref. (22).

This computational study aims at answering three fundamental questions: 1) if and to which extent the bonding properties of NHC towards a transition metal fragment can be extended to NHCP (Scheme 4); 2) how the nature of the NHC moiety affects the electronic and geometrical structures of NHCPs; 3) what is the NHCP-AuH bond nature/reactivity relationship, selecting $CO_2$ reduction to explore transition metal NHCP complexes applicability in small molecule activation processes. In particular, the insertion of $CO_2$ into metal hydrides and the reverse decarboxylation of metal formates are important elementary steps in catalytic cycles for both $CO_2$ hydrogenation to formic acid or methanol and formic acid dehydrogenation. (29,30)

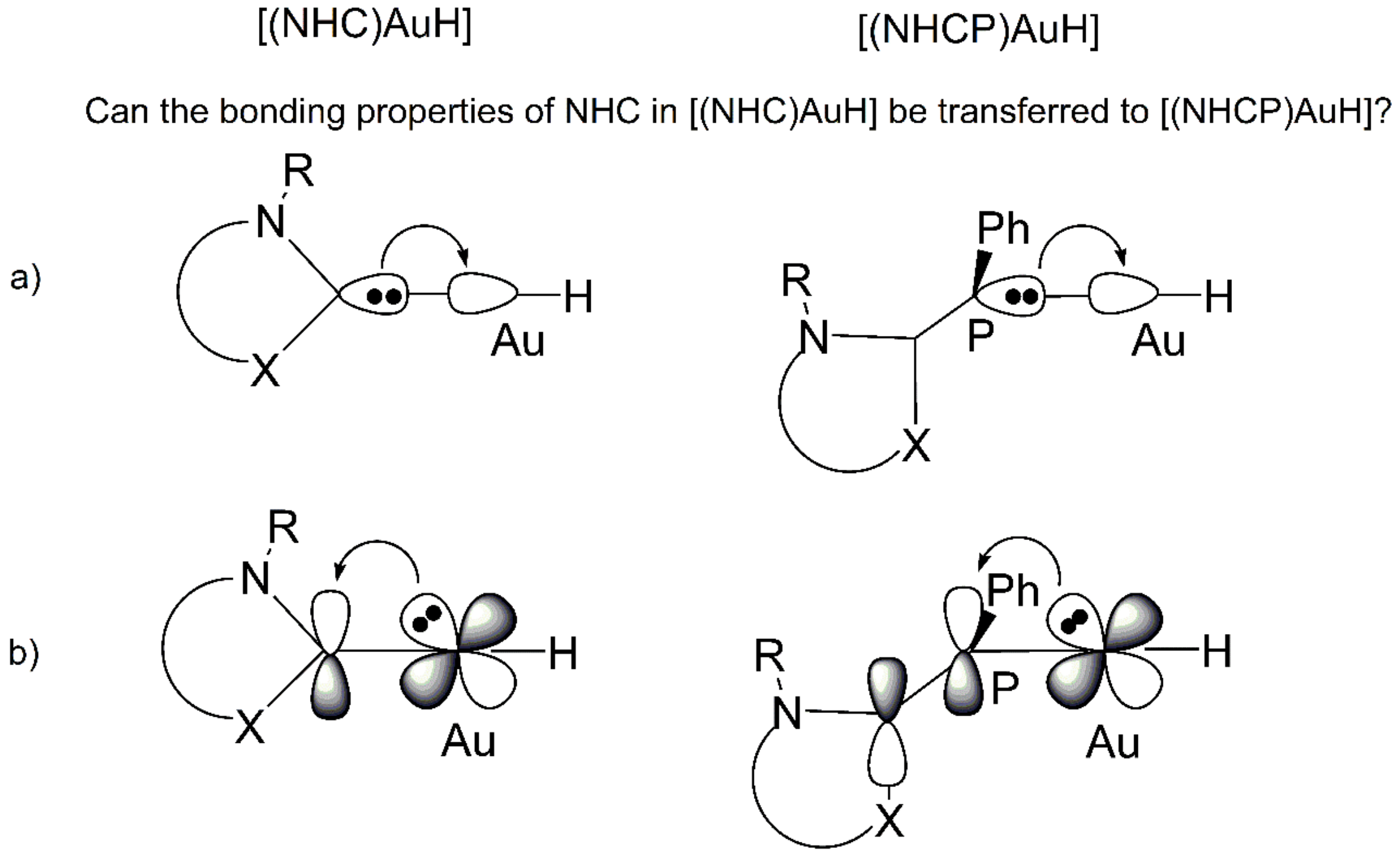

**Scheme 4**: Schematic representation of the relationship between a) σ-donation and b) π-backdonation bond components in [(NHC)AuH] (left) and [(NHCP)AuH] (right) complexes.

To properly describe the nature of the NHC-AuH and NHCP-AuH bonds, theoretical descriptors have been used, as exploited in computational tools such as Energy Decomposition Analysis (EDA), (31) Natural Orbitals for Chemical Valence (NOCV) (32,33) and Charge Displacement (CD) analysis, (34, 35) which are combined in a consistent computational protocol. (36)

Unlike the NHC ligands, which utilize a carbon $sp^2$ lone pair, NHCPs form a σ bond with Au–H through a highly flexible phosphorus 3p lone pair. This fundamental difference enhances bond tunability, allowing various NHC moieties to directly govern both the geometric and σ-donating properties of the NHCP ligand by modulating the energy and contribution of the phosphorus 3p lone pair.

In particular, the NHCP ligands are found to be stronger σ-donor than NHCs, with a σ-donor trend opposite to that of NHCs, i.e. σ–donation increases from CAACs to conjugated DACs, and much less π-acceptor compared to NHCs, although showing the same trend, i.e. π-backdonation increases from conjugated DACs to CAACs. The coordination geometry around P accordingly shifts from “nearly planar” to $PH_3$-like (bond angles: $C_{NHC}$-P-Au ~ 120° with CAACs, $C_{NHC}$-P-Au ~ 109° with unconjugated DACs and $C_{NHC}$-P-Au ~ 90° for conjugated DACs). As a result of the relevant tunability of their electronic and geometric structures, the NHCP complexes under study show remarkable differences when reacting with $CO_2$, with kinetic activation barriers for hydrogenation closely correlating with the NHCP σ-donor ability, eventually modulated by second sphere effects triggered by the flexibility imparted by the PPh unit.

## Results and discussion

### Comparative analysis of the NHCP-AuH bond

#### Geometries

Selected geometrical parameters of the optimized structures of complexes **1-P-13-P** in Scheme 3 a (depicted in Figure S1 in the Supporting Information) are collected in Table 1.

| | Au-H | Au-P | P-$C_{NHC}$ | P-$C_{Ph}$ | $C_{NHC}$-P-Au | $C_{Ph}$-P-Au | Au-P-$C_{Ph}$-$C_{NHC}$ |
|---|---|---|---|---|---|---|---|
| **1-P** | 1.622 | 2.338 | 1.747 | 1.831 | 120.4 | 111.1 | 133.4 |
| **2-P** | 1.622 | 2.339 | 1.750 | 1.833 | 119.3 | 111.3 | 132.1 |
| **3-P** | 1.622 | 2.336 | 1.751 | 1.830 | 121.5 | 111.4 | 136.8 |
| **4-P** | 1.628 | 2.365 | 1.802 | 1.835 | 104.9 | 109.5 | 112.7 |
| **5-P** | 1.627 | 2.347 | 1.792 | 1.835 | 112.8 | 109.7 | 123.2 |
| **6-P** | 1.619 | 2.342 | 1.764 | 1.820 | 113.9 | 112.5 | 126.6 |
| **7-P** | 1.624 | 2.359 | 1.797 | 1.832 | 109.1 | 110.5 | 117.9 |
| **8-P** | 1.631 | 2.375 | 1.818 | 1.840 | 98.7 | 108.3 | 115.2 |
| **9-P** | 1.630 | 2.369 | 1.808 | 1.836 | 103.3 | 108.9 | 109.9 |
| **10-P** | 1.629 | 2.362 | 1.804 | 1.831 | 104.6 | 108.4 | 110.6 |
| **11-P** | 1.629 | 2.384 | 1.809 | 1.846 | 95.7 | 105.8 | 100.2 |
| **12-P** | 1.633 | 2.375 | 1.813 | 1.845 | 101.1 | 108.4 | 106.2 |
| **13-P** | 1.631 | 2.367 | 1.820 | 1.838 | 97.8 | 110.7 | 102.3 |

**Table 1:** Selected structural data for the complexes **1-P-13-P**. Distances in Å, angles in degrees.

Along the series, the Au-H (1.62-1.63 Å), Au-P (2.34-2.38 Å), and P-$C_{Ph}$ (1.82-1.85 Å) bond distances as well as the $C_{Ph}$-P-Au (105.8-112.5°) bond angles do not change significantly. On the other hand, geometrical parameters involving carbene moiety show a clear trend: P-$C_{NHC}$ bond length slightly increases from CAACs (**1-P-3-P**, 1.75 Å) to unconjugated DACs (**4-P-7-P** 1.76-1.80 Å) to conjugated DACs (**8-P-13-P**, 1.80-1.82 Å), $C_{NHC}$-P-Au bond angle significantly decreases in the same order (**1-P-3-P**, 121.5-119.3°), (**4-P-7-P**, 113.9-104.9°), and (**8-P-13-P**, 104.6-95.7°) as well as the dihedral angle at the phosphorus atom, Au-P-$C_{Ph}$-$C_{NHC}$, (**1-P-3-P**, 136.8-132.1°), (**4-P-7-P**, 126.6-112.7°), and (**8-P-13-P**, 115.2-100.2°), suggesting that the nature of the carbene does have an impact on the NHCP-AuH bond description. Specifically, the $C_{NHC}$-P-Au bond angles in conjugated DACs are considerably compressed, which is characteristic of trigonal pyramidal phosphorous compounds such as phosphines ($PR_3$, $PH_3$ or $PF_3$ angles ranging from 90° to 104°, due to lone-pair on P), indicating a minimum hybridization at P and prominent use of pure p-orbitals in bonding. In unconjugated DACs, typical $sp^3$ hybridization angles (e.g., $NH_3$, 107.8°) are preferred and, in CAACs, angles close to 120° suggest a near "trigonal planar" coordination around a $sp^2$-like hybridized P.
A Mayer bond-order (MBO) analysis (37) for the Au-P and P-$C_{NHC}$ bonds across the series has been performed to evaluate the bond nature (see Table S1 and Figure S2 in the Supporting Information). As expected, a linear inverse correlation between Au-P bond distance and MBO is observed ($r^2$ = 0.709), with bond lengths between 2.336 Å and 2.384 Å and MBO values around 0.666-0.730, thus reflecting single bond behavior. The P-$C_{NHC}$ bond distance and MBO displays a stronger linear inverse correlation across the series ($r^2$ = 0.972), with the P-$C_{NHC}$ bond exhibiting two distinct structural regimes. Complexes **1-P-3-P** and **6-P** fall into a high bond-order regime (MBO = 1.206-1.241; distance = 1.747-1.764 Å), reflecting partial double bond character and strong electronic communication. Conversely, complexes **4-P-7-P** and **8-P-13-P** demonstrate single bond behavior, with distances extending up to 1.820 Å and MBO values dropping around 1.00. In terms of resonance structures (Scheme 1), CAACs –based NHCP coordination to AuH preserve a formal description of the PPh bonding to carbene as in I, unconjugated DACs-based as in II and conjugated DACs-based as in V, with consequences on the NHCP-AuH bond nature and reactivity.

## Bond nature assessment

By relying on the use of EDA, (31) and CD (34, 35) analysis, combined with NOCV (32, 33) approach (see Methodological Section in the Supporting Information), the features of the bond between the carbene-phosphinidenes and the gold hydride fragments are investigated within a quantitative and comparative framework, to unravel how such features can be modulated by the carbene-type moiety. This computational protocol has been successfully used by some of us in previous works. (38-42)

A comparative EDA is first carried out using neutral NHCP and AuH fragments as the most suitable fragmentation scheme for the description of the dative NHCP-AuH bond (see Table S2 in the Supporting Information for the results).

EDA scheme allows one to decompose the NHCP-AuH interaction energy ($\Delta E_{int}$) into contributions associated with the Pauli ($\Delta E_{Pauli}$), electrostatic ($\Delta E_{elstat}$) and orbital interactions ($\Delta E_{oi}$), the latter accounting for electron pair bonding, charge transfer and polarization. The interaction energy $\Delta E_{int}$ values fall within the range -75.8/-64.9 kcal/mol, and a clear trend cannot be found. The major contribution to the fragment interaction is given by the $\Delta E_{oi}$ term (-79.2/-64.3), since the steric term ($\Delta E_{Pauli}$ + $\Delta E_{elstat}$) is generally small (+9.3/-1.1 kcal/mol).

Coupling EDA results with CD and NOCV approaches allows us to quantitatively assess both the NHCP-AuH bond nature and polarity in all the complexes under study. Results of the CD-NOCV analysis are reported in Table 2. The CD-NOCV curves associated with the first four NOCV deformation densities ($\Delta\rho_1$, $\Delta\rho_2$, $\Delta\rho_3$ and $\Delta\rho_4$) and the total $\Delta\rho$', together with the corresponding isosurfaces, are depicted in Figure 1, using complex **8-P** (Scheme 3 a) as an illustration. The CD-NOCV curves for all the NHCP in Scheme 3 a are reported in the Supporting Information (Figures S3-S15).

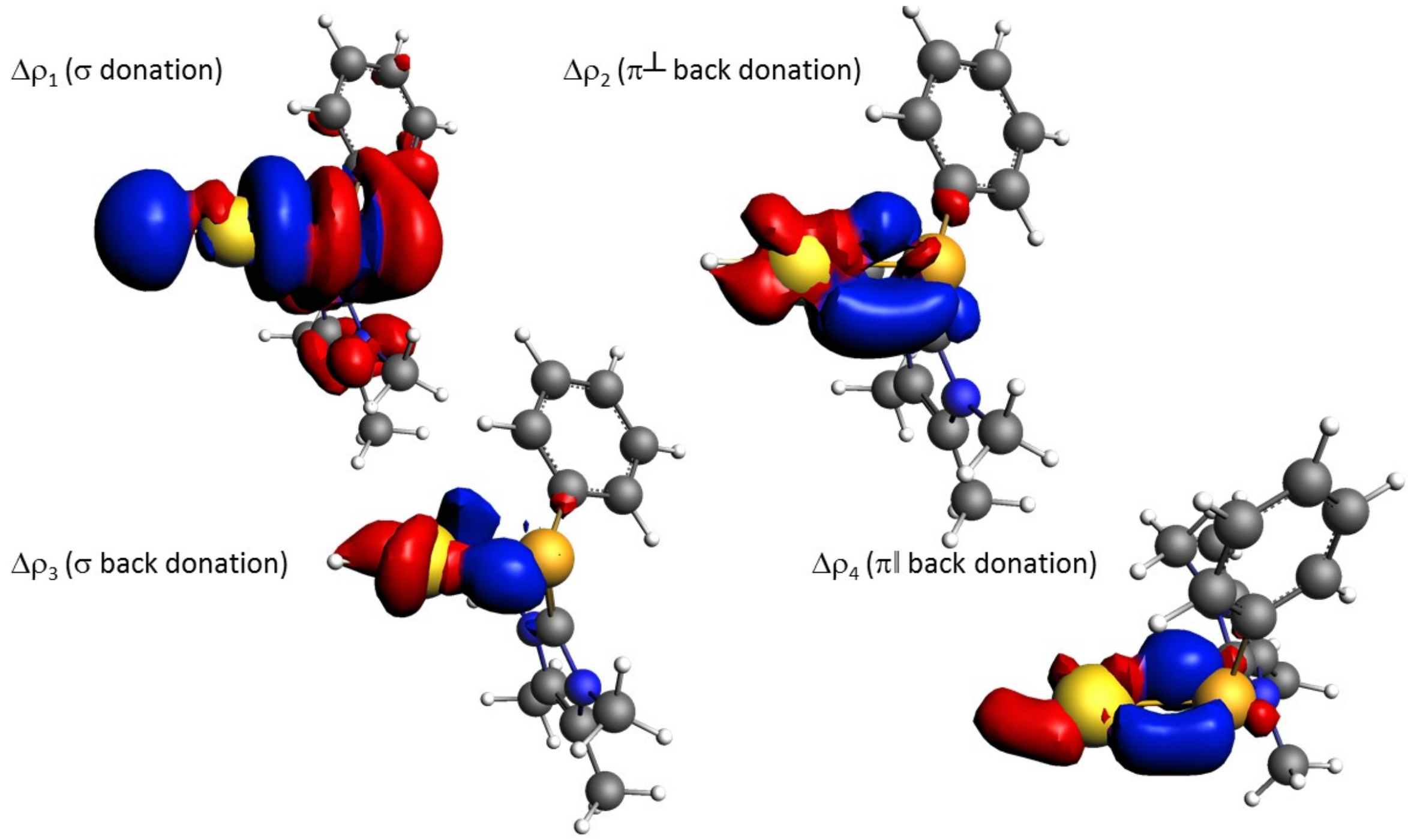
Δρ1 (σ donation)
Δρ2 (π⊥ back donation)
Δρ3 (σ back donation)
Δρ4 (π∥ back donation)

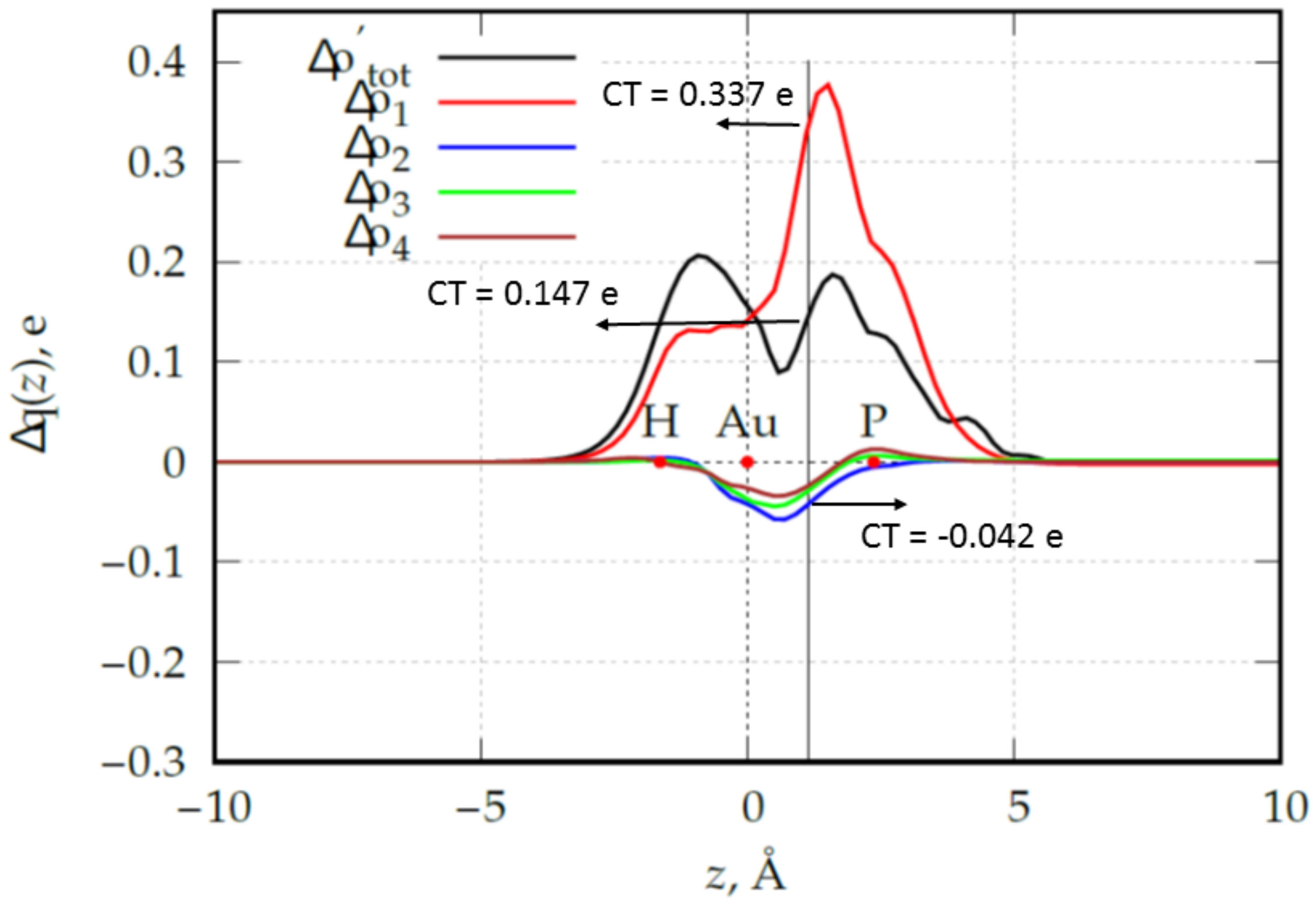
Δρ'tot
Δρ1
Δρ2
Δρ3
Δρ4
CT = 0.337 e
CT = 0.147 e
CT = -0.042 e
H
Au
P
0.4
0.3
0.2
0.1
0
−0.1
−0.2
−0.3
−10
−5
0
5
10
Δq(z), e
z, Å

**Figure 1**. The NHCP-AuH bond in complex **8-P**. Top: isodensity surfaces associated with the $\Delta\rho_1$, $\Delta\rho_2$, $\Delta\rho_3$ and $\Delta\rho_4$ NOCV components of the total $\Delta\rho'$ deformation density (isodensity value ± 5 me/$a_0^3$); blue regions correspond to electron density accumulation areas, red regions correspond to electron density depletion areas. Bottom: CD curves associated with the total $\Delta\rho'$ and its components. Red dots indicate the position of the nuclei along the z axis. The grey vertical line marks the isodensity boundary between the NHCP and AuH fragments. Positive (negative) values of the curve indicate right-to-left (left-to-right) charge transfer.

Figure 1 top shows that predominantly four NOCV Δρ' components characterize the bonding between NHCP and gold fragment AuH. The first contribution, $\Delta\rho_1$, captures the main σ-component of the NHCP-AuH bond and it represents the electron density donation from the lone pair of phosphorus atom and the charge accumulation in the P-Au bonding region as well as on the hydrogen atom. In addition, charge depletion appears on the carbon and nitrogen atoms of the NHC moiety and carbon atoms of Ph, suggesting a participation of both NHC and Ph groups. The second and fourth components, $\Delta\rho_2$ and $\Delta\rho_4$, clearly indicate the π-backbonding character of the NHCP-AuH bond, i.e. the electron charge transfer from the metal into the phosphinidene ligand. It is clear from Figure 1 top that π-backdonation is realized by the electron transfer from a filled d orbital of the Au atom into an empty orbital of NHCP mostly exhibiting phosphorus 3p character, occurring on perpendicular planes ($\pi^{\perp}$, $\pi^{\parallel}$). To avoid any confusion, in the absence of a symmetry axis, we clarify that the $\pi^{\perp}$ notation refers to the π-backdonation occurring perpendicularly with respect to the NHC ring plane, and the $\pi^{\parallel}$ to the π-backdonation taking place in a direction parallel to the NHC ring plane. Notably, in the $\Delta\rho_2$ component, i.e. the $\pi^{\perp}$-backdonation towards NHCP, charge accumulation at the carbon atom of the NHC moiety is shown, indicating phosphorus 3p orbital involvement in a π* interaction with the NHC carbon atom 2p orbital (perpendicular to the NHC plane, see Scheme 4), causing delocalization of the electron charge by resonance. The third component, $\Delta\rho_3$, shows a charge depletion on the gold region, along the Au-H bond, and accumulation on the Au-P bond localized close to P, depicting a σ-backdonation contribution to the NHCP-AuH bond. The orbital nature of the σ-backdonation can be envisaged as an interaction between the occupied $d_{z2}$ orbital at gold and the partially emptied (due to the main σ-donation) lone pair orbital of the P atom.

In Figure 1 bottom, the corresponding CD curves are shown. The red line refers to σ-donation, with a charge transfer ($CT_{\sigma\ don}$) of 0.337 e (see Table 2). The $\pi^{\perp}$-backdonation component, as well as, the σ-backdonation and the $\pi\|$-backdonation components, have only small CT values ($CT_{\pi\perp\ back}$ -0.042 e, $CT_{\sigma\ back}$ -0.028 e and $CT_{\pi\|\ back}$ -0.023 e, respectively), with the minus sign indicating the flux of charge from the left to the right. The net CD curve (black line) shows a CT ($CT_{net}$) of 0.147 e from NHCP to AuH. Note that both $CT_{net}$ and net CD contain contributions from all the NOCV Δρ' components and $CT_{net}$ quantifies a $NHCP^{\delta+}$ - $AuH^{\delta-}$ bond polarity.

The total (black) and σ-donation (red) curves are positive throughout the entire molecule region, whereas those representing the backdonation contributions are negative. Remarkably, the total CD (black line) shows a maximum value between Au and P, closer to P, and a slightly higher maximum between Au and H, with a minimum near Au, along the Au-P bond, extending to the outer region of P, i.e. towards Ph and NHC moieties.

The NOCV results for all the **1-P-13-P** complexes show that the NHCP-AuH bond can be described within the same picture (σ-donation, σ-backdonation, π∥-backdonation and π⊥ -backdonation). The CTs data are reported in Table 2 (extended data can be found in Tables S3 and S4 in the Supporting Information).

| NHCP | $CT_{\sigma\ don}$ | $CT_{\pi\perp\ back}$ | $CT_{\pi\parallel\ back}$ | $CT_{\pi\ total\ back}$ | $CT_{\sigma\ back}$ | $CT_{net}$ |
|---|---|---|---|---|---|---|
| **1-P** | 0.300 (-33.1) | -0.053 (-7.6) | -0.044 (-6.4) | -0.097 (-14.0) | -0.021 (-5.5) | 0.083 (-72.7) |
| **2-P** | 0.300 (-33.1) | -0.054 (-7.7) | -0.043 (-6.3) | -0.097 (-14.0) | -0.021 (-5.5) | 0.077 (-74.2) |
| **3-P** | 0.300 (-33.0) | -0.054 (-7.6) | -0.042 (-6.3) | -0.096 (-13.9) | -0.021 (-5.7) | 0.077 (-77.5) |
| **4-P** | 0.325 (-35.7) | -0.051 (-6.9) | -0.024 (-4.8) | -0.075 (-11.7) | -0.025 (-5.0) | 0.079 (-72.3) |
| **5-P** | 0.317 (-34.9) | -0.056 (-7.4) | -0.023 (-5.5) | -0.079 (-12.9) | -0.030 (-5.5) | 0.090 (-77.6) |
| **6-P** | 0.287 (-32.7) | -0.060 (-9.4) | -0.041 (-5.9) | -0.101 (-15.3) | -0.012 (-4.8) | 0.039 (-74.2) |
| **7-P** | 0.311 (-34.7) | -0.053 (-7.8) | -0.020 (-4.7) | -0.073 (-12.5) | -0.027 (-4.9) | 0.061 (-75.3) |
| **8-P** | 0.337 (-37.2) | -0.042 (-5.6) | -0.023 (-4.5) | -0.065 (-10.1) | -0.028 (-4.9) | 0.147 (-64.3) |
| **9-P** | 0.333 (-36.5) | -0.049 (-6.7) | -0.025 (-4.9) | -0.074 (-11.6) | -0.023 (-4.6) | 0.078 (-72.4) |
| **10-P** | 0.337 (-36.9) | -0.051 (-7.7) | -0.025 (-4.9) | -0.076 (-12.6) | -0.024 (-4.7) | 0.079 (-78.3) |
| **11-P** | 0.343 (-37.1) | -0.041 (-6.8) | -0.023 (-4.5) | -0.064 (-11.3) | -0.013 (-4.3) | 0.121 (-79.2) |
| **12-P** | 0.350 (-38.1) | -0.042 (-6.4) | -0.023 (-4.9) | -0.065 (-11.3) | -0.014 (-4.7) | 0.105 (-77.6) |
| **13-P** | 0.339 (-37.3) | -0.044 (-6.0) | -0.026 (-4.3) | -0.070 (-10.3) | -0.026 (-5.2) | 0.142 (-66.7) |

**Table 2**: Charge transfer (CT) values (in electrons) for the NHCP-AuH bond components (first four NOCV deformation densities and total) in the **1-P-13-P** complexes listed in Scheme 3,

panel a. Values in parenthesis are the orbital interaction energy contributions associated with the NOCV partitioning ($\Delta\varepsilon_1$, $\Delta\varepsilon_2$, $\Delta\varepsilon_3$, and $\Delta\varepsilon_4$, in kcal/mol).

Notably, the two π-backdonation contributions to the bond ($CT_{\pi\perp\,back}$ and $CT_{\pi\parallel\,back}$) are small and very similar for all the complexes (in the range -0.060/-0.041 e for $CT_{\pi\perp\,back}$ and -0.044/-0.020 e for $CT_{\pi\parallel\,back}$), with $CT_{\pi\perp\,back}$ values generally slightly larger than the $CT_{\pi\parallel\,back}$ ones. Since the total CT values ($CT_{net}$) in Table 2 contain all the contributions to the NHCP-AuH bond and $CT_{\sigma\,back}$ is small (in the range -0.012/-0.030 e) and does not show a clear trend along the series, in the following we will focus on the $CT_{\sigma\,don}$ and $CT_{\pi\,total\,back}$ values for a comparative description of the NHCP-AuH bond features. As a general trend, we note that the σ-donation bond component increases from CAACs (**1-P** – **3-P**) to unconjugated DACs (**4-P** – **7-P**) to conjugated DACs (**8-P** – **13-P**), spanning a range from 0.300 to 0.350 e, whereas the total π-backdonation decreases in the same order from CAACs to conjugated DACs (from -0.097 to -0.064 e). Within the entire series, complex **6-P** with two oxygen atoms on the backbone shows peculiar bond characteristic, having the lowest $CT_{\sigma\,don}$ value (0.287 e) and the largest $CT_{\pi\,total\,back}$ value (-0.101 e), which can be rationalized in terms of the high ability of oxygen to accept π charge through the mesomeric effect (3). Notably, a strong correlation ($r^2$ = 0.94) is observed between the $CT_{\sigma\,don}$ values and the Au-H bond lengths across the **1-P**-**13-P** series, suggesting a possible trans effect with a relevant role in the Au-H reactivity (Figure S16 in the Supporting Information). The $CT_{net}$ values vary substantially along the series (from 0.039 to 0.147 e), suggesting in all complexes a $NHCP^{\delta+}$ - $AuH^{\delta-}$ bond polarity, higher for conjugated DACs-based **8-P**, **11-P**, **12-P** and **13-P**. The orbital energy contributions associated to each bond component (Table 2) also show that the NHCP-AuH bond is dominated by σ-donation (with energies in a range from -32.7 to -38.1 kcal/mol), with much smaller and comparable π-backdonation (both $\pi\perp_{back}$: -5.6/-7.8 kcal/mol, slightly larger for **6-P**: -9.4 kcal/mol, and $\pi\parallel_{back}$: -4.5/-6.4 kcal/mol) and σ-backdonation ($\sigma_{back}$: -4.3/-5.7 kcal/mol) contributions. We note that, though the orbital interaction energy contributions associated to individual σ-donation/backdonation components show a clear trend, the total $\Delta E_{oi}$ as well as $\Delta E_{int}$ show no obvious systematic trends. This discrepancy is only apparent since, within the NOCV framework, $\Delta E_{oi}$ is partitioned into contributions associated with all the different NOCV deformation densities involved in the NHCP-AuH bond formation.

Remarkably, within each series (**1-P-3-P** CAACs, **4-P-7-P** unconjugated DACs and **8-P-13-P** conjugated DACs), both $CT_{\sigma\ don}$ and $CT_{\pi\ total\ back}$ values do not vary substantially, thus suggesting a minor tunability of carbene-phosphinidenes – AuH bond upon NHC moiety scaffold changes. On the other hand, along the entire **1-P-13-P** series, the results of the CD-NOCV analysis clearly point out that different classes of carbenes, i.e. CAACs, unconjugated DACs and conjugated DACs, appear to be relevant for tuning the features of the NHCP-AuH bond, generating NHCPs with distinct electronic structures.

**NHC-AuH vs. NHCP-AuH bond comparison**

Based on the results illustrated in the previous section, the main features of the bond between carbene-phosphinidenes and gold hydride fragments are significantly affected by the different class of the carbene moiety at the phosphinidene site, highlighting substantial decrease of σ-donation and increase of π-backdonation from conjugated DACs to unconjugated DACs to CAACs. To investigate the effect of the carbene moiety nature, we quantitatively assess the NHC-AuH/NHCP-AuH ligand relationship as coordination ligands.

By relying on the same EDA-CD-NOCV-based protocol used in the previous section, we analyze the features of the Au-C bond in the corresponding **1-13** carbene-AuH complexes [NHCAuH] (Scheme 3, panel b), which represent the analogues of the [NHCPAuH] complexes object of this study.

Results of EDA approach, using closed-shell gold hydride AuH and carbene NHC fragments, are collected in Table S5 in the Supporting Information. The interaction energy $\Delta E_{int}$ values for these complexes are included in the range -92.8/-76.0 kcal/mol, and do not show a clear trend along the series. Similar to NHCPs, the largest contribution is given by the $\Delta E_{oi}$ term (-84.0/-63.5 kcal/mol).

The CD-NOCV results similarly depict the NHC-AuH bond in complexes **1-13** as a "classical" coordination metal-ligand bond. The CD-NOCV curves associated with the first four NOCV deformation densities ($\Delta\rho_1$, $\Delta\rho_2$, $\Delta\rho_3$ and $\Delta\rho_4$) and the total $\Delta\rho'$, together with the corresponding isosurfaces, are depicted in Figure 2, using complex **8** (Scheme 3 b) for a direct comparison with complex **8-P** (Figure 1). The CD-NOCV curves for all the NHC in Scheme 3 b are reported in the Supporting Information (Figures S17-S29).

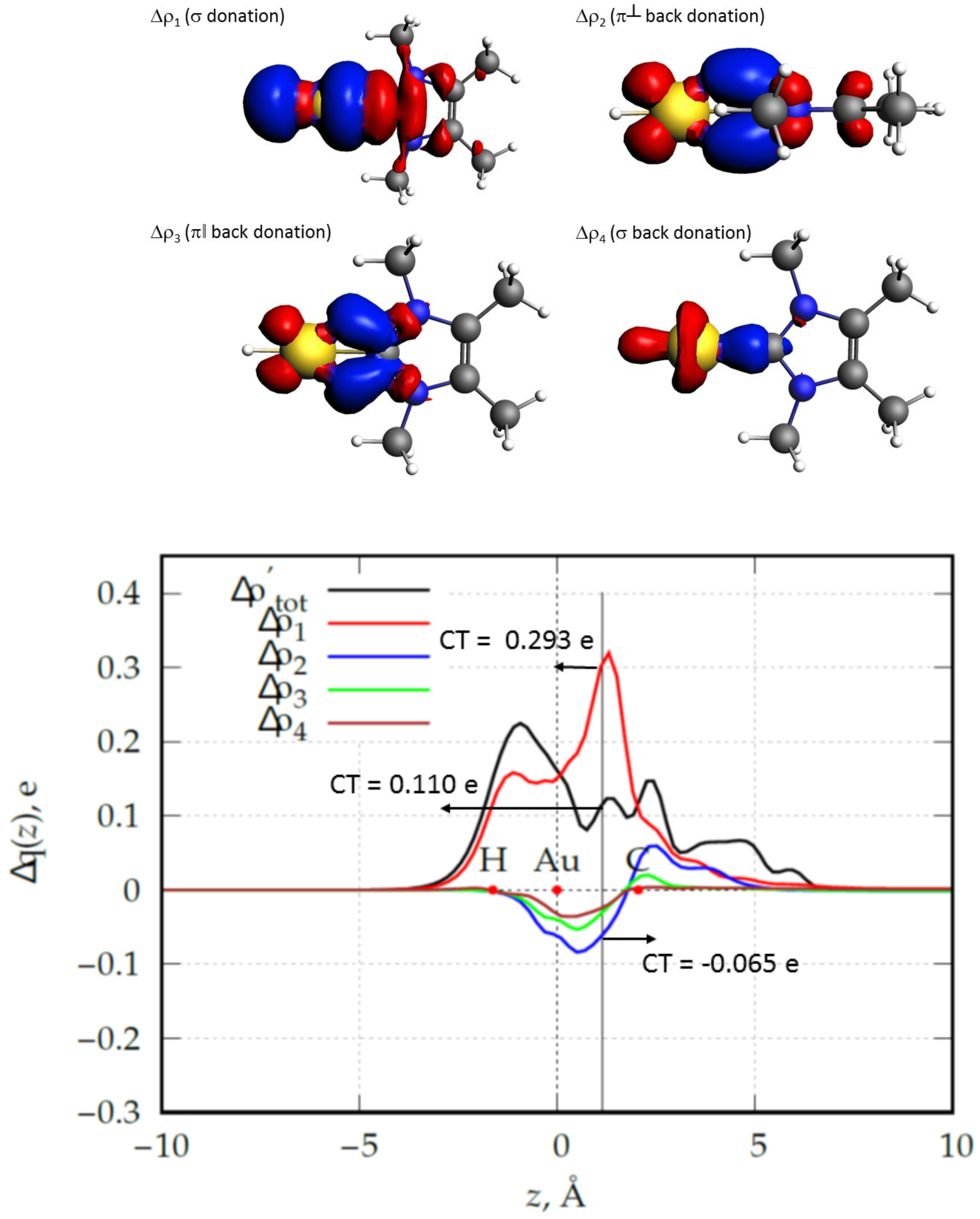

Δρ1 (σ donation)
Δρ2 (π⊥ back donation)
Δρ3 (π∥ back donation)
Δρ4 (σ back donation)
Δρ'tot
Δρ1
Δρ2
Δρ3
Δρ4
CT = 0.293 e
CT = 0.110 e
CT = -0.065 e
H
Au
C
Δq(z), e
z, Å
0.4
0.3
0.2
0.1
0
−0.1
−0.2
−0.3
−10
−5
0
5
10

**Figure 2**. The NHC-AuH bond in complex **8**. Top: isodensity surfaces associated with the $\Delta\rho_1$, $\Delta\rho_2$, $\Delta\rho_3$ and $\Delta\rho_4$ NOCV components of the total $\Delta\rho'$ deformation density (isodensity value ± 5 me/$a_0^3$); blue regions correspond to electron density accumulation areas, red regions correspond to electron density depletion areas. Bottom: CD curves associated with the total $\Delta\rho'$ and its components. Red dots indicate the position of the nuclei along the z axis. The grey vertical line marks the isodensity boundary between the NHC and AuH fragments. Positive (negative) values of the curve indicate right-to-left (left-to-right) charge transfer.

Consistent with the results reported in the literature for the Au(I)-NHC bond, (3, 43) the main NOCV component is, in all cases, a $C_{NHC}$-to-Au charge transfer of σ symmetry that represents a σ-donation ($\Delta\rho_1$) (Figure 2 top). Additionally, a gold-to-carbon π-backdonation component perpendicular to the Au-N-C-N plane can be recognized ($\Delta\rho_2$) as well as a gold-to-carbon π-backdonation component on the Au-N-C-N plane ($\Delta\rho_3$). Finally, an additional component (namely a σ-backdonation) is found ($\Delta\rho_4$). The corresponding CT and associated $\Delta E_{oi}$ values are shown in Table 3, where results for all **1-13** complexes are collected (extended data can be found in Tables S6 and S7 in the Supporting Information).

| NHC | $CT_{\sigma\ don}$ | $CT_{\pi\perp\ back}$ | $CT_{\pi\parallel\ back}$ | $CT_{\pi\ total\ back}$ | $CT_{\sigma\ back}$ | $CT_{net}$ |
|---|---|---|---|---|---|---|
| **1** | 0.312<br>(-37.6) | -0.113<br>(-11.4) | -0.027<br>(-5.8) | -0.140<br>(-17.2) | -0.017<br>(-4.3) | 0.042<br>(-78.9) |
| **2** | 0.312<br>(-37.7) | -0.111<br>(-11.3) | -0.027<br>(-5.8) | -0.138<br>(-17.1) | -0.018<br>(-4.4) | 0.046<br>(-80.6) |
| **3** | 0.311<br>(-37.4) | -0.106<br>(-11.2) | -0.026<br>(-5.7) | -0.132<br>(-16.9) | -0.017<br>(-4.8) | 0.047<br>(-84.0) |
| **4** | 0.300<br>(-36.0) | -0.087<br>(-9.7) | -0.032<br>(-5.8) | -0.119<br>(-15.5) | -0.030<br>(-4.4) | 0.101<br>(-70.8) |
| **5** | 0.298<br>(-35.7) | -0.087<br>(-10.0) | -0.027<br>(-5.9) | -0.114<br>(-15.9) | -0.016<br>(-4.4) | 0.029<br>(-79.7) |
| **6** | 0.293<br>(-34.4) | -0.138<br>(-13.4) | -0.029<br>(-5.5) | -0.167<br>(-18.9) | -0.035<br>(-5.1) | 0.016<br>(-76.3) |
| **7** | 0.297<br>(-35.1) | -0.106<br>(-10.9) | -0.033<br>(-5.2) | -0.139<br>(-16.1) | -0.033<br>(-4.9) | 0.016<br>(-74.7) |
| **8** | 0.293<br>(-35.8) | -0.065<br>(-8.2) | -0.034<br>(-5.5) | -0.099<br>(-13.7) | -0.026<br>(-4.8) | 0.110<br>(-63.5) |
| **9** | 0.294<br>(-35.5) | -0.080<br>(-9.3) | -0.030<br>(-5.8) | -0.110<br>(-15.1) | -0.034<br>(-4.9) | 0.116<br>(-69.5) |
| **10** | 0.292<br>(-35.3) | -0.075<br>(-9.2) | -0.027<br>(-5.9) | -0.102<br>(-15.1) | -0.033<br>(-4.9) | 0.020<br>(-78.7) |
| **11** | 0.294<br>(-35.0) | -0.079<br>(-9.6) | -0.028<br>(-6.1) | -0.107<br>(-15.7) | -0.036<br>(-5.2) | 0.019<br>(-78.5) |
| **12** | 0.296<br>(-36.0) | -0.076<br>(-9.2) | -0.031<br>(-5.8) | -0.107<br>(-15.0) | -0.034<br>(-4.8) | 0.129<br>(-71.3) |
| **13** | 0.293<br>(-35.7) | -0.066<br>(-8.1) | -0.028<br>(-5.5) | -0.094<br>(-13.6) | -0.033<br>(-5.0) | 0.114<br>(-65.3) |

**Table 3**: Charge transfer (CT) values (in electrons) for the NHC-AuH bond components (first four NOCV deformation densities and total) in the **1-13** complexes listed in Scheme 3, panel b. Values in parenthesis are the orbital interaction energy contributions associated with the NOCV partitioning ($\Delta\varepsilon_1$, $\Delta\varepsilon_2$, $\Delta\varepsilon_3$, and $\Delta\varepsilon_4$, in kcal/mol).

Overall, we note from Table 3 that the σ-donation bond component decreases from CAACs (**1** – **3**) to unconjugated DACs (**4** – **7**) to conjugated DACs (**8** – **13**), spanning a range from 0.312 to 0.292 e, as well as the total π-backdonation, decreasing in the same order from CAACs to conjugated DACs (from -0.140 to -0.094 e). Similar to **6-P**, complex **6**, with two oxygen atoms on the backbone, has the lowest $CT_{\sigma\ don}$ value (0.293 e) and the largest $CT_{\pi\ total\ back}$ value (-0.167 e). Contrary to NHCPs, one π-backdonation component ($CT_{\pi\perp\ back}$) is significantly larger than the other ($CT_{\pi\parallel\ back}$) for all the complexes. The $CT_{net}$ values also vary substantially along the series (from 0.016 to 0.129 e), suggesting in all complexes a $NHC^{\delta+}$ - $AuH^{\delta-}$ bond polarity, higher for conjugated DACs complexes **8**, **9**, **12** and **13**. Similar to NHCPs, orbital energy contributions associated to each bond component (Table 3) show that the NHC-AuH bond is mainly described by σ-donation (with associated energies in a narrow range from -34.4 to -37.7 kcal/mol). For the backdonation components, the major contribution is given by $\pi\perp_{back}$ (energies range from -8.1 to -11.4 kcal/mol, larger for **6**: -13.4 kcal/mol), with smaller and comparable $\pi_{\parallel\ back}$ (-4.8/-6.1 kcal/mol) and $\sigma_{back}$ (-4.4/-5.5 kcal/mol) contributions.

From the above analysis, complexes **6-P** and **6** emerge as behaving distinctly from the other members placed in the same unconjugated DACs group, i.e. **4-P**, **5-P** and **7-P** and **4**, **5** and **7**, respectively, both structurally and electronically, consistent with the literature reporting diamidocarbenes as significantly different from diaminocarbenes in terms of their electronic structure, stability and chemical reactivity (44,45). For this reason, in all correlation analysis presented hereafter, complexes **6** and **6-P** will be not included. For the sake of completeness, plots comprising these systems are reported in the Supporting Information (Plots including **6** and **6-P** section, Figures S42-S48).

Though the NHCP/NHC-AuH bond can be described in terms of analogous components, differences emerge between NHCP-AuH and NHC-AuH bonding properties when comparing Tables 2 and 3: 1) NHCP ligands are more σ-donor than NHCs ($CT_{\sigma\ don}$: 0.300-0.350e for NHCPs vs. 0.292-0.312e for NHCs) and, remarkably, the σ –donor trend is opposite to that in NHCs ,i.e. σ–donation increases from CAACs to conjugated DACs in the NHCP series, whereas σ–donation decreases from CAACs to conjugated DACs in the NHCs one, with the NHCPs σ-donor ability towards Au-H being modulated more efficiently compared to that of NHCs (larger $CT_{\sigma\ don}$ variation range); 2) NHCPs are much less π-acceptor and tunable compared to NHCs ($CT_{\pi\ total\ back}$: -0.097/-0.064 e for NHCPs vs. -0.140/-0.094 e for NHCs or $CT_{\pi\perp\,back}$: -0.056/-0.041 e for NHCPs vs. -0.113/-0.065 e for NHCs), although the two ligands show the same trend, i.e. π-backdonation increases from conjugated DACs to CAACs, consistent with the NHCs π-acceptor properties being transferable to the PPh moiety. (3) These results suggest that, although the π-acceptor ability trend of NHCs qualitatively parallels that of NHCPs, the bonding properties of the NHCs towards the AuH metal fragment cannot be transferred to the NHCPs, i.e. the PPh moiety does change the coordination chemistry of this new class of ligands. This study clearly demonstrates that PPh coordination to NHC strongly depends on the nature of the carbene, with more σ-donating NHC moieties generating less σ-donating NHCP ligands.

Based on $\Delta\rho_1$ isosurfaces, in combination with canonical forms of NHCPs (Scheme 1), the σ-donor properties of NHCPs can be rationalized in terms of the nature of the phosphorus orbital involved in the σ bonding. In a simple qualitative picture, the σ-donation is expected to heavily depend on the orbital hybridization at the phosphorus center, with σ-donor ability decreasing in the order: pure or un-hybridized p orbitals (structure V) > $sp^3$ (structure II) > $sp^2$ (structure I). As previously observed from the geometrical analysis of the **1-P**-**13-P** complexes, CAACs-based NHCP are mainly described by canonical form I, where orbital hybridization on P is essentially $sp^2$-like, unconjugated DACs-based NHCP by canonical form II, where hybridization on P is $sp^3$-like, and conjugated DACs-based NHCP by canonical form V, with prominent use of pure p-type orbitals of P in bonding, which would be consistent with the σ-donor ability order.

On a quantitative ground, the peculiarity of the two distinct lone pairs of NHCPs with respect to the nature of the carbene moiety is strikingly evident by inspection of the isodensity pictures, energies and composition of the corresponding HOMO and HOMO – 1 molecular orbitals (MOs) of the NHCP fragment in **1-P**, **4-P** and **8-P** as CAACs-based, unconjugated DACs-based and conjugated DACs-based representative, respectively, depicted in Figure 3.

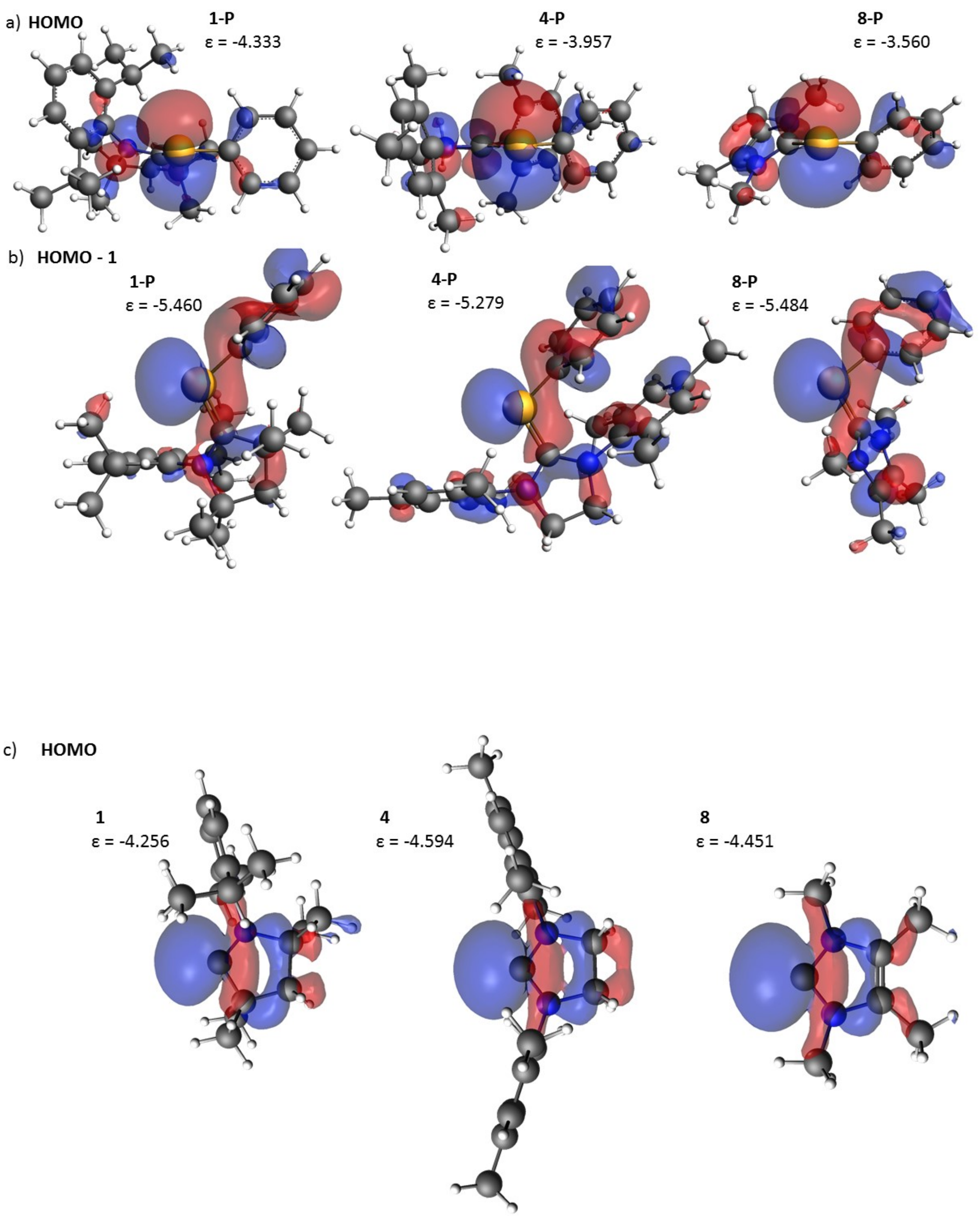

**Figure 3.** Isosurfaces of the: a) HOMO (front view) and b) HOMO-1 (side view) of **1-P**, **4-P** and **8-P** NHCP fragments; c) HOMO (side view) of **1**, **4** and **8**. Isovalue for all surfaces is ± 25 me/$a_0^3$. Orbital energy (in eV) is also reported.

For the three NHCP fragments, the HOMO lone pair is mainly centered at the P site, perpendicular to the $C_{NHC}$-Au-$C_{Ph}$ plane, although it also contains contributions from the NHC and Ph ligands. Notably, the HOMO energy increases from **1-P** to **4-P** by 8.7 kcal/mol and from **4-P** to **8-P** by 9.2 kcal/mol, consistent with the increasing P $3p_z$ character from **1-P** (36%) to **4-P** (49%) to **8-P** (55%) (Table S8 in the Supporting Information). The HOMO-1 lone pair, although centered at the P site, on the $C_{NHC}$-Au-$C_{Ph}$ plane, is much less diffuse and delocalizes over the NHC and Ph ligand scaffolds to a greater extent, highlighting why the structure of the NHC moiety (and R substituent) represents a control factor on their electronic and geometrical properties. In addition, the HOMO - 1 energy is lower than the corresponding HOMO's one by 26.0 kcal/mol for **1-P**, 30.5 kcal/mol for **4-P** and 44.4 kcal/mol for **8-P**, due to the P 3s orbital contribution (14% for **1-P**, 10% for **4-P** and 13% for **8-P**) (Table S8 in the Supporting Information), making it less available for bonding with AuH and weakening its ability to act as a σ-donor. However, this second lone pair on P can furnish an additional coordination opportunity, accounting for the ability of phosphorus to serve as a four-electron donor *via* the use of the two lone pairs. (25)

This comparative analysis clarifies that, despite being isolobal, CAACs-based, unconjugated DACs-based and conjugated DACs-based NHCPs display a different behavior as coordination ligands, both in terms of geometry and NHCP-AuH bonding properties. For the three NHCP fragments, the HOMO involved in the NHCP-AuH bond very closely resembles a "pure" P 3p orbital, sitting perpendicular to the plane of the $C_{NHC}$-P-$C_{Ph}$ bonding framework, maintaining its distinct lobe shape and higher energy relative to the strongly stabilized HOMO-1 orbital. However, it contains different 3p percentages, explaining why the $C_{NHC}$-P-Au bond angle drops to nearly 98.7° in **8-P** (largest 3p character) compared to **4-P** (104.9°) (intermediate 3p character) and **1-P** (120.4°) (lowest 3p character) (Figure 4).

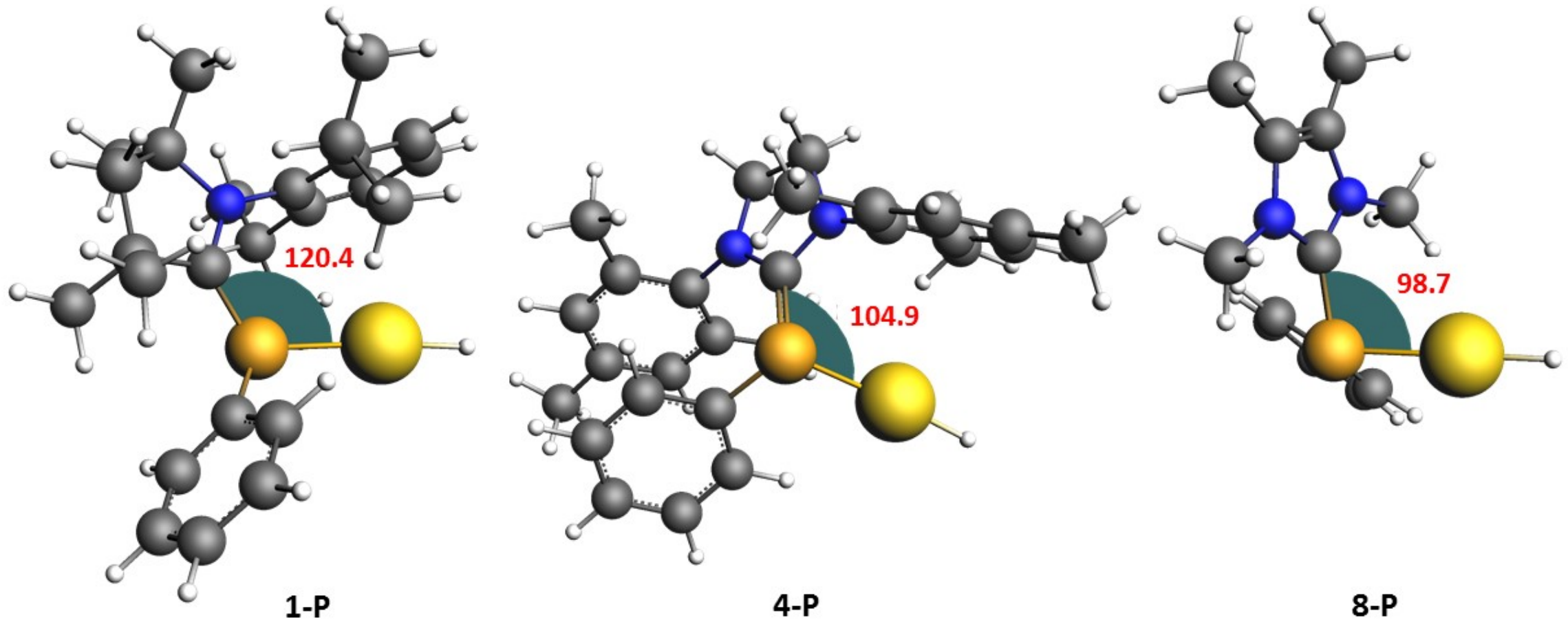

**Figure 4.** Optimized structures of **1-P**, **4-P** and **8-P**. The $C_{NHC}$-P-Au bond angle (in degrees) is reported for comparison.

Simultaneously, the different P 3p percentages in the HOMO composition are consistent with the largest σ-donor ability in AuH bonding for **8-P**, a decreased σ-donor ability for **4-P** and the lowest σ-donor ability for **1-P**. The different nature of the HOMO can be conceived as an effect of the NHC bonding to P: more σ-donor NHCs (i.e. CAACs) force the phosphorus 3s and 3p orbitals to mix, thus imparting an $sp^2$ –like hybridization at P, whereas less σ-donor NHCs (i.e. conjugated DACs) make hybridization at P unfavorable, thus tuning the energy, the composition and the σ-donor character of the HOMO lone pair in NHCPs, which is decreased by stronger σ-donor NHC moieties.

In Figure 3 panel c) isosurfaces of the HOMO of **1**, **4** and **8** NHC fragment analogues involved in the bonding with AuH are also shown for comparison, where similarity with NHCP fragment HOMO-1 is strikingly evident. The NHC fragment lone pair HOMO, centered at the C site, delocalizes over the ligand scaffold to a large extent, lying on the plane of the $N_{NHC}$-C-N(C)$_{NHC}$ $sp^2$ bonding framework. The $C_{NHC}$ $2p_z$ contribution to the HOMO lone pair involved in the σ-donation slightly decreases from 52.2 (**1**) to 50.5 (**4**) and 49.7 (**8**) (Table S8 in the Supporting Information), rationalizing the less tunable σ-donor ability along the NHC series as well as the σ-donation trend.

The π-backdonation ability order of NHCPs along the series, instead, parallels that of NHCs, consistent with the $^{31}P$ NMR chemical shift correlating with the π-acceptor properties of the NHCs. This result can be rationalized based on $\Delta\rho_2$ and the π-backdonation bond component as sketched in Scheme 4 b. The π-backdonation ($\pi^{\perp}$) from a filled d orbital of the Au atom into an empty 3p orbital of P involved in a π* antibonding interaction with the empty 2p orbital of $C_{NHC}$ perpendicular to the NHC plane, allows for resonant delocalization of π electron density from the Au center to NHC moiety, via the P atom, which increases on increasing the π-acceptor ability of NHC. Moreover, the high energy of the involved empty π* P 3p – $C_{NHC}$ 2p MO is consistent with both the lower π-backacceptor ability of NHCPs and its trend (LUMO energy **1-P**: -1.579 eV, **4-P**: -1.436 eV, **8-P**: -1.172 eV) (Figure S30 in the Supporting Information).

In summary, from this analysis NHCPs emerge as a class of ligands possessing highly distinct coordination properties: a strong σ-donor ability and a unique structural flexibility and electronic adaptability, directly controlled by the carbene moiety, through modulation of the HOMO lone pair energy and P 3p contribution. In particular, introducing strong σ-donor CAACs moieties, NHCPs with a reduced σ-donor ability and a nearly "planar" coordination around P are generated, while inserting weaker σ-donor conjugated DACs moieties, NHCPs with an increased σ-donor ability and a $PH_3$-like coordination around P are formed, with respect to NHCPs σ-donor ability with a $NH_3$-like coordination around P constructed by inserting medium strength σ-donor unconjugated DACs moieties.

To quantitatively demonstrate the $C_{NHC}$–P–Au angle as a structural descriptor for monitoring or designing the σ-donor properties of NHCP ligands, correlation between the $C_{NHC}$–P–Au angle vs. the NHCP fragment HOMO P 3p percentage contribution plot for the **1-P-13-P** series is shown in Figure 5 panel a), which exhibits a strong linear correlation, with $r^2 = 0.93$, fully consistent with the qualitative interpretation of phosphorus "hybridization". Similarly, the $C_{NHC}$–P–Au angle vs. $CT_{\sigma\ don}$ plot shown in Figure 5 panel b) displays a satisfactory linear trend with $r^2 = 0.88$.

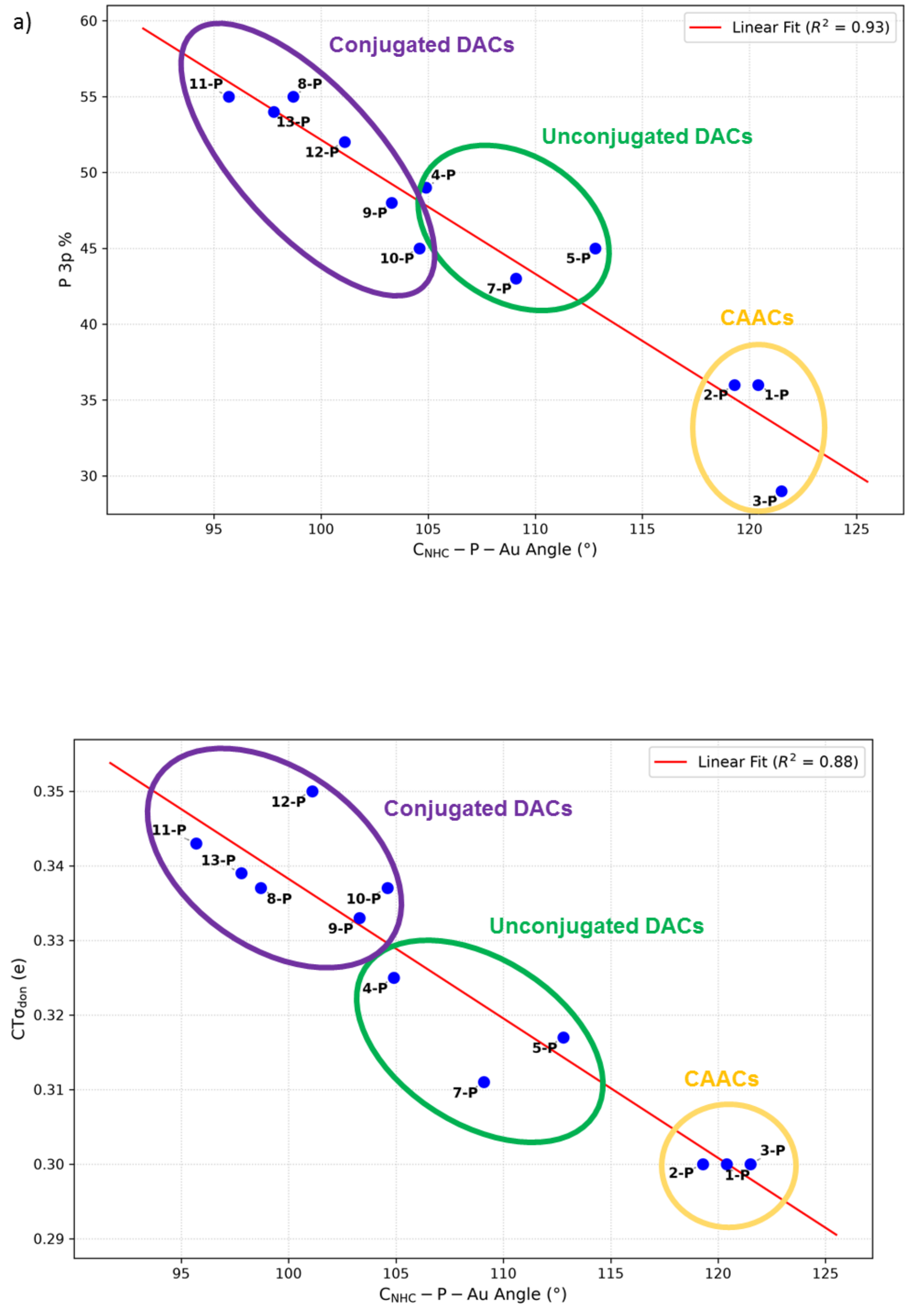

**Figure 5**: Correlation plots between the $C_{NHC}$-P-Au bond angles and panel a) the NHCP fragment HOMO P 3p percentage contribution; panel b) the $CT_{\sigma\ don}$ values for the **1-P-13-P** series. CAACs-based complexes are circled in yellow, unconjugated DACs-based complexes in green and conjugated DACs-based complexes in purple.

As proposed for the **1-P**, **4-P** and **8-P** representative, the NHCP σ-donor ability also demonstrates a reasonable correlation with both the NHCP HOMO fragment energy (Figure S31 in the Supporting Information, with $r^2 = 0.82$) and P 3p percentage contribution (Figure S32 in the Supporting Information, with $r^2 = 0.81$) across the **1-P-13-P** series. Remarkably, in all the NHCP fragments, the two available lone pairs have the same energy ordering and nature, i.e. they represent the HOMO (mainly a P 3p orbital perpendicular to the $C_{NHC}$-Au-$C_{Ph}$ plane) and the HOMO-1 (mainly localized at the P site, on the $C_{NHC}$-Au-$C_{Ph}$ plane), showing that in the complete **1-P-13-P** series the same nature lone pair is involved in the σ-donation towards the metal fragment (Table S9 in the Supporting Information). This analysis establishes that the $C_{NHC}$–P–Au bond angle is genuinely predictive for the investigated AuH complexes, although it should not automatically be assumed to be transferable to unrelated metal fragments. For the reader's convenience, a compact Table (Table 4) is provided, collecting the key structural, orbital and CD-NOCV descriptors for the complete NHCP **1-P-13-P** series.

| | CT $_{\sigma\ don}$ | NHCP fragment HOMO ε | NHCP fragment HOMO P 3p % | $C_{NHC}$-P-Au angle |
|---|---|---|---|---|
| **1-P** | 0.300 | -4.333 | 36 | 120.4 |
| **2-P** | 0.300 | -4.331 | 36 | 119.3 |
| **3-P** | 0.300 | -4.326 | 29 | 121.5 |
| **4-P** | 0.325 | -3.957 | 49 | 104.9 |
| **5-P** | 0.317 | -4.007 | 45 | 112.8 |
| **6-P** | 0.287 | -5.104 | 33 | 113.9 |
| **7-P** | 0.311 | -4.425 | 43 | 109.1 |
| **8-P** | 0.337 | -3.560 | 55 | 98.7 |
| **9-P** | 0.333 | -3.726 | 48 | 103.3 |
| **10-P** | 0.337 | -3.830 | 45 | 104.6 |
| **11-P** | 0.343 | -3.945 | 55 | 95.7 |
| **12-P** | 0.350 | -3.500 | 52 | 101.1 |
| **13-P** | 0.339 | -3.628 | 54 | 97.8 |

**Table 4**: Charge transfer values for the NHCP-AuH σ-donor bond component, CT $_{\sigma\ don}$ (in electrons), NHCP fragment HOMO energy, ε (in eV) and P 3p percentage contribution, P 3p %, and $C_{NHC}$-P-Au angle (in °) for the complete NHCP **1-P-13-P** series.

The reversal of the σ-donor trend, where more strongly σ-donating NHC moieties produce less σ-donating NHCP ligands, is summarized in Figure S33 in the Supporting Information, where $CT_{\sigma\ don}$ for the series of [(NHCP)AuH] is shown to inversely correlates with $CT_{\sigma\ don}$ for the series of [(NHC)AuH] ($r^2 = 0.77$).

We emphasize that the σ-donor ability trend of the NHCP ligands is obtained from a specific ligand-AuH interaction and therefore reflects both the electronic properties of the ligand and characteristics of the AuH acceptor fragment. To get an estimate of the metal fragment effect, we have performed optimization calculations of all the isolated NHCP ligands, thus relaxing their geometries with respect to those they acquire in the corresponding AuH complexes. Comparison of the phosphorus 3p % character of the P-centered lone pair (HOMO) in the NHCP fragment (i.e. in the complex) and in the relaxed NHCP (i.e. without the metal fragment) is reported in Table S10 in the Supporting Information. Analogously, in all the 13 NHCPs the HOMO represents the “p-type” lone pair and the HOMO-1 the isolobal lone pair with that of the corresponding NHC, both located at the P center. The phosphorus 3p character of the P-centered lone pair HOMO in the relaxed NHCPs is much larger than that in the fragment and it is almost unchanged along the series, in the range 53-59 % (except for the particular **6-P** case, 44%). Inspection of all relaxed HOMOs shows that the P lone-pair is involved in π interactions with carbon and/or nitrogen atoms of the NHC moiety ring to different extent, meaning that the AuH fragment coordination drives *simultaneously* both the energy and P 3p % contribution to the P-centered lone pair HOMO and the change in the geometry at the P center (i.e. the $C_{NHC}$–P–Au angle) and, consequently, the σ-donor ability of the NHCP ligand. This analysis demonstrates that it is not possible to disentangle electronic and geometric effects and to assess to what extent the observed σ-donor inversion is an intrinsic electronic effect of the NHC moiety or is mediated by structural relaxation at phosphorus. Furthermore, this investigation confirms that the bonding picture obtained with the AuH cannot be assumed to be applicable to other metals, although one should expect this trend to be transferable to analogous metal fragments featuring similar σ-acceptor properties. This interesting topic deserves a dedicated study. Finally, the proper schematic representation of the σ-donation bond components in [(NHC)AuH] and [(NHCP)AuH] complexes is depicted in Scheme 5, highlighting the different nature of the lone pairs involved in the interaction with the AuH metal fragment.

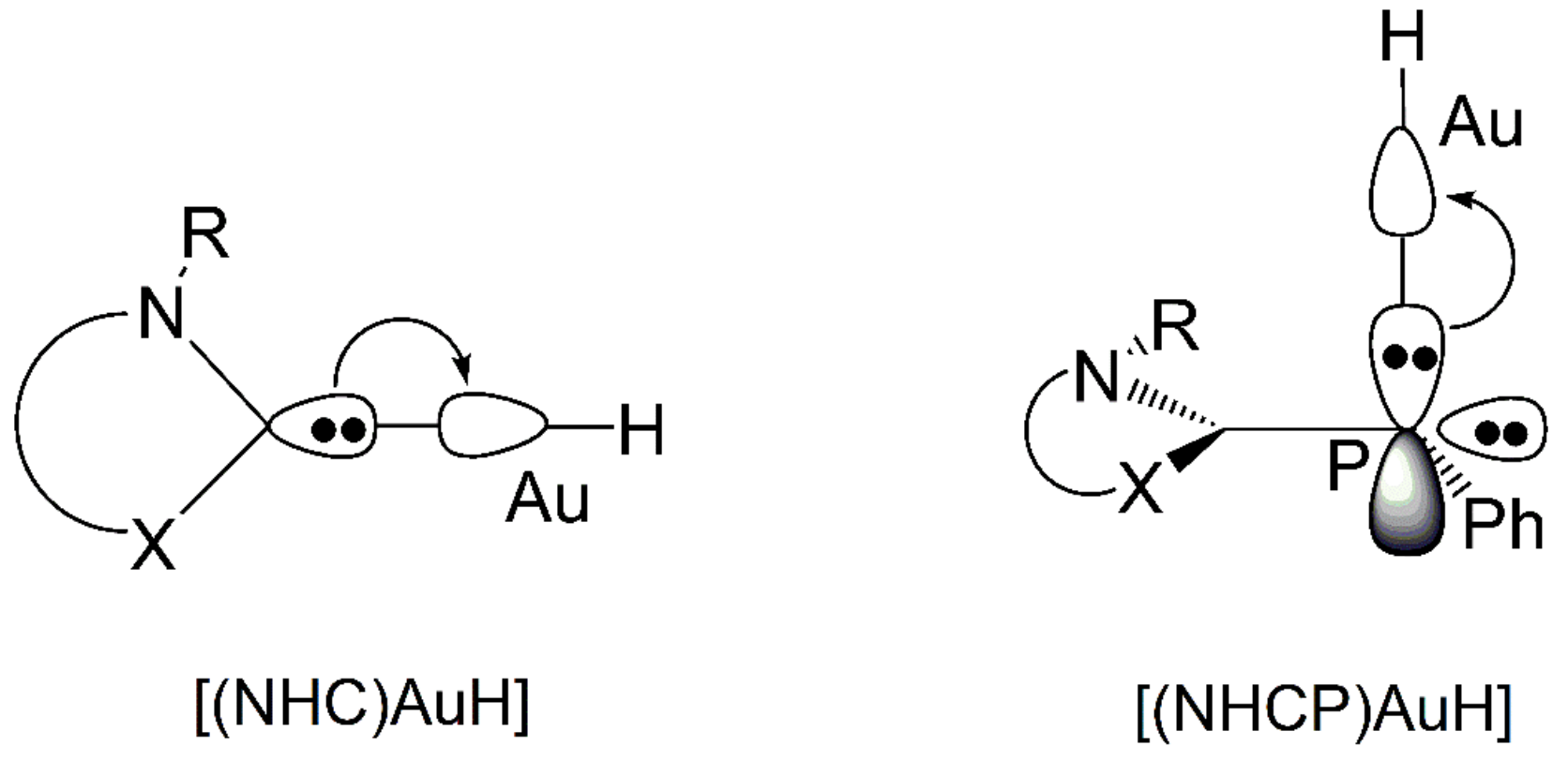

Scheme 5: Schematic representation of the relationship between σ-donation bond components in [(NHC)AuH] (left) and [(NHCP)AuH] (right) complexes.

In the following section we will discuss the impact of such different coordination behavior in the reaction of [NHCPAuH] and [NHCAuH] complexes with $CO_2$.
Beyond these electronic effects, the steric bulk of the NHC moiety and the choice of the substituent R on phosphorus provide further handles for tuning the coordination and reactivity of NHCPs.

**Bonding nature/reactivity relationship**

The free energy profiles for the $CO_2$ insertion into the Au-H bond in NHCAuH **1**, NHCAuH **8**, NHCPAuH **1-P**, and NHCPAuH **8-P** complexes, selected as lower and upper limits of the σ-donation and π-backdonation series for NHC and NHCP moieties, and in NHCPPAuH (**8-PP**) as the experimental active complex in $CO_2$ reduction, have been calculated. They are compared in Figure 6. Optimized structures of all transition states (TS) and product complexes (PC) along the reaction profiles for **1**, **1-P**, **8** and **8-P** are reported in the Supporting Information (Figure S34).

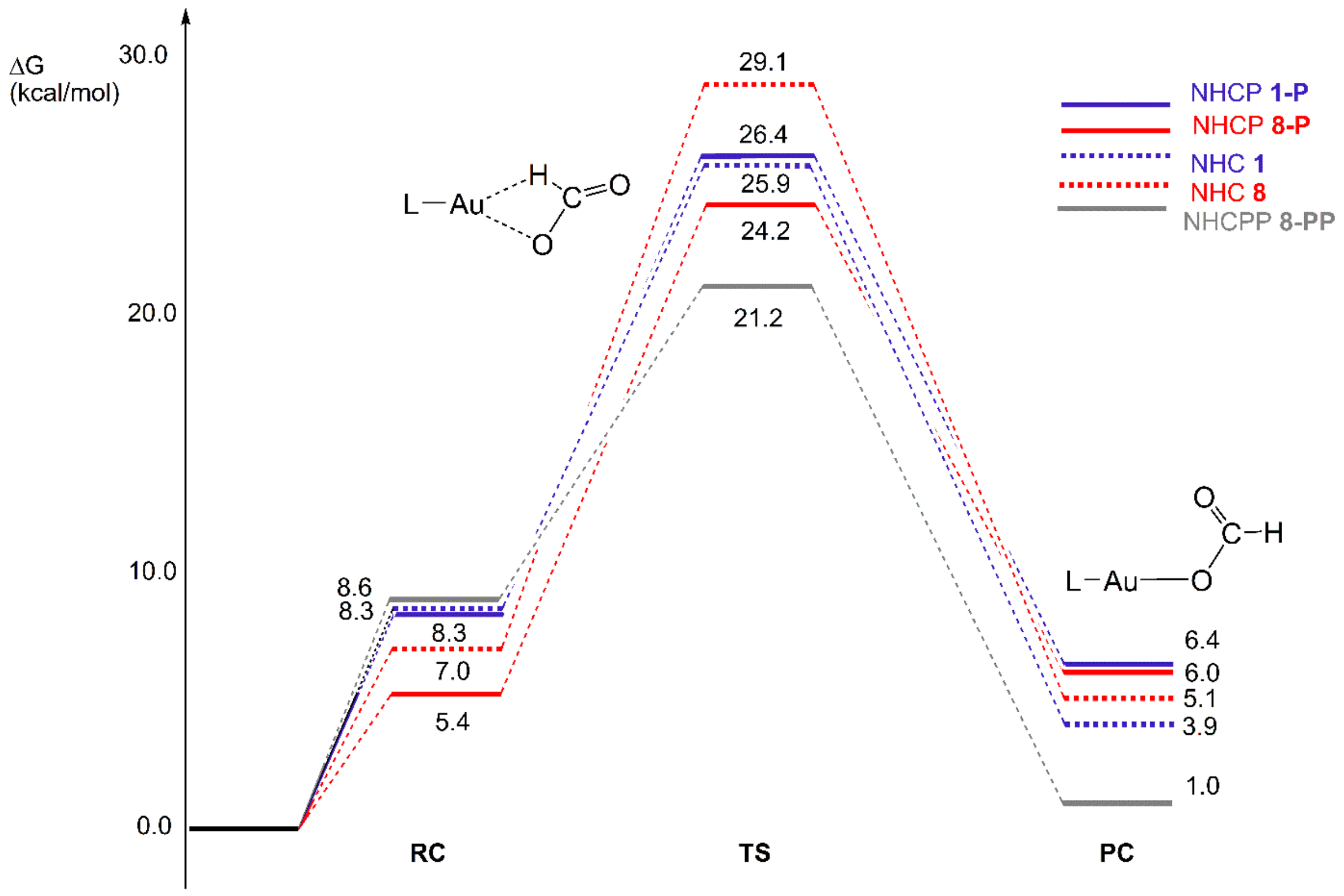

**Figure 6.** Free energy reaction profile for the $CO_2$ insertion into the Au-H bond in the NHCAuH **1** (dashed blue lines), NHCPAuH **1-P** (blue lines), NHCAuH **8** (dashed red lines), NHCPAuH **8-P** (red lines) and NHCPPAuH **8-PP** (grey lines) complexes. Δ*G* values refer to the energy of the separated reactants taken as zero. Sketched structures of transition states and products are shown. RC = reactant complex, TS = transition state, PC = product complex.

The reaction profiles in Figure 6 are qualitatively similar. The process occurs *via* a four-center concerted transition state, where the hydrogen atom binds to the carbon atom of $CO_2$ with simultaneous Au-H bond breaking and bond forming between one oxygen atom of $CO_2$ and the gold center, directly leading to the formate product. The activation energy barriers $\Delta G^{\ddagger}$, calculated relative to the separated [LAuH] + $CO_2$ reactants, span a range from 29.1 to 21.2 kcal/mol and the ΔGs for the product formation show an endergonic reaction, with values from 6.4 to 1.0 kcal/mol. We clarify that in Figure 6 only the most stable $\kappa^1$-O-formate product (PC) is shown. The slightly less stable gold-formate $\eta^1$-O-species, with the carboxylate H atom pointing towards the metal center, directly connected to the transition state, represents a local minimum and these two rotational conformers are calculated to be both thermodynamically accessible in solution, due to the facile rotation of the formate around the Au-O(OCHO) bond (see Supporting Information for a more detailed discussion, Figures S35-S36). As expected, the free energy profile for **8-PP** is consistent with the experimental feasibility ($\Delta G^{\ddagger}$ = 21.2 kcal/mol) and reversibility ($\Delta G$ = 1.0 kcal/mol) of this reaction. Structures of transition state TS and PC for **8-PP** complex are depicted in Figure S37 in the Supporting Information. At the TSs, $CO_2$ displays a significant bending (OCO angle range: 136.8°-132.5°) and relatively close C-H (1.268-1.202 Å) and OCO-Au (2.590-2.512 Å) distances. The Au-H bond elongation is sizable, in the range 1.978-1.832 Å. The comparison of the TS structures reveals no substantial differences (TS for **8-PP** is slightly more reactant-like). Apart from complex **8** ($\Delta G^{\ddagger}$ = 29.1 kcal/mol), $CO_2$ activation with **1**, **1-P** and **8-P** is calculated to be kinetically feasible. Very interestingly, the activation energy barriers closely correlate with the σ-donation component of the NHC/NHCP-AuH bond, i.e. with $CT_{\sigma\ don}$ values in Tables 2 and 3. Within the NHCP-AuH complex series, **8-P**, having a larger σ-donation component ($CT_{\sigma\ don}$ = 0.337 e), exhibits a lower barrier ($\Delta G^{\ddagger}$ = 24.2 kcal/mol) than that of **1-P** ($\Delta G^{\ddagger}$ = 26.4 kcal/mol), where the σ-donation component is lower ($CT_{\sigma\ don}$ = 0.300 e). The same correlation, resulting in an opposite reactivity trend, can be found within the NHC-AuH complex series, with **1** showing a lower barrier than **8** (for **1**: $CT_{\sigma\ don}$ = 0.312 e, $\Delta G^{\ddagger}$ = 25.9 kcal/mol; for **8**: $CT_{\sigma\ don}$ = 0.293 e, $\Delta G^{\ddagger}$ = 29.1 kcal/mol). As a proof of concept, the activation free energy barrier has been calculated for complex **10**, which shows a low σ-donation NHC-AuH bond component ($CT_{\sigma\ don}$ = 0.292 e), and which has been experimentally reported to fail in reacting with $CO_2$. The $\Delta G^{\ddagger}$ amounts to 29.0 kcal/mol, fully consistent with the value calculated for **8** and with experiment. On the other hand, the activation free energy barrier for the corresponding phosphinidene complex **10-P** ($CT_{\sigma\ don}$ = 0.337 e) is computed to be 24.7 kcal/mol, essentially the same as that calculated for **8-P**. For the experimentally active **8-PP** complex, a $CT_{\sigma\ don}$ has been calculated by analyzing the NHCPP-AuH bond, i.e. using NHCPP and AuH fragments, amounting to 0.344 e. The significantly lower calculated activation barrier (21.2 for **8-PP** vs. 24.2 kcal/mol for **8-P**) highlights the limitations of the $CT_{\sigma\ don}$ as a reactivity descriptor suggesting a beneficial effect of the additional P center in the ligand which stems from the added coordination flexibility, allowing for the incipient $HCO_2^-$ anion in the transition state structure of **8-PP** to be stabilized by second sphere C-H…O hydrogen bonding with the two methyl groups of the NHC ring (Figure S37 in the Supporting Information). The same second sphere hydrogen bonding is responsible for the stabilization of the corresponding formate product PC, leading to an almost thermoneutral reaction ($\Delta G$ = 1.0 kcal/mol, Figure 6 and Figure S37).

Although the differences in $CO_2$ activation barriers can be attributed primarily to variations in NHCP σ-donor ability, changing the carbene moiety may simultaneously also alter the steric environment around the reactive Au–H unit. To clarify whether steric effects could contribute to the activation barrier differences, the percent buried volume (%$V_{bur}$) (46, 47) has been employed as a simple steric descriptor for the selected representative systems **1**, **1-P**, **8**, **8-P** and **8-PP** complexes. Results are reported and discussed in the Supporting Information (Table S11), suggesting that steric effects could contribute to the activation barrier differences for **1-13** NHC complex series, particularly within the same class, with a less relevant impact on the activation barrier differences for **1-P-13-P** NHCP series, due to the peculiar PPh moiety spatial orientation.

To strengthen and generalize the proposed correlation between NHCP σ-donor ability and $CO_2$ activation barrier supported by the selected **1**, **1-P**, **8**, **8-P**, **8-PP**, **10** and **10-P** complexes, we have calculated the reaction paths (both free energy barriers and thermodynamics) for the complete **1P-13P** and **1-13** series. Results are reported in Table S12 in the Supporting Information and in Figure 7.

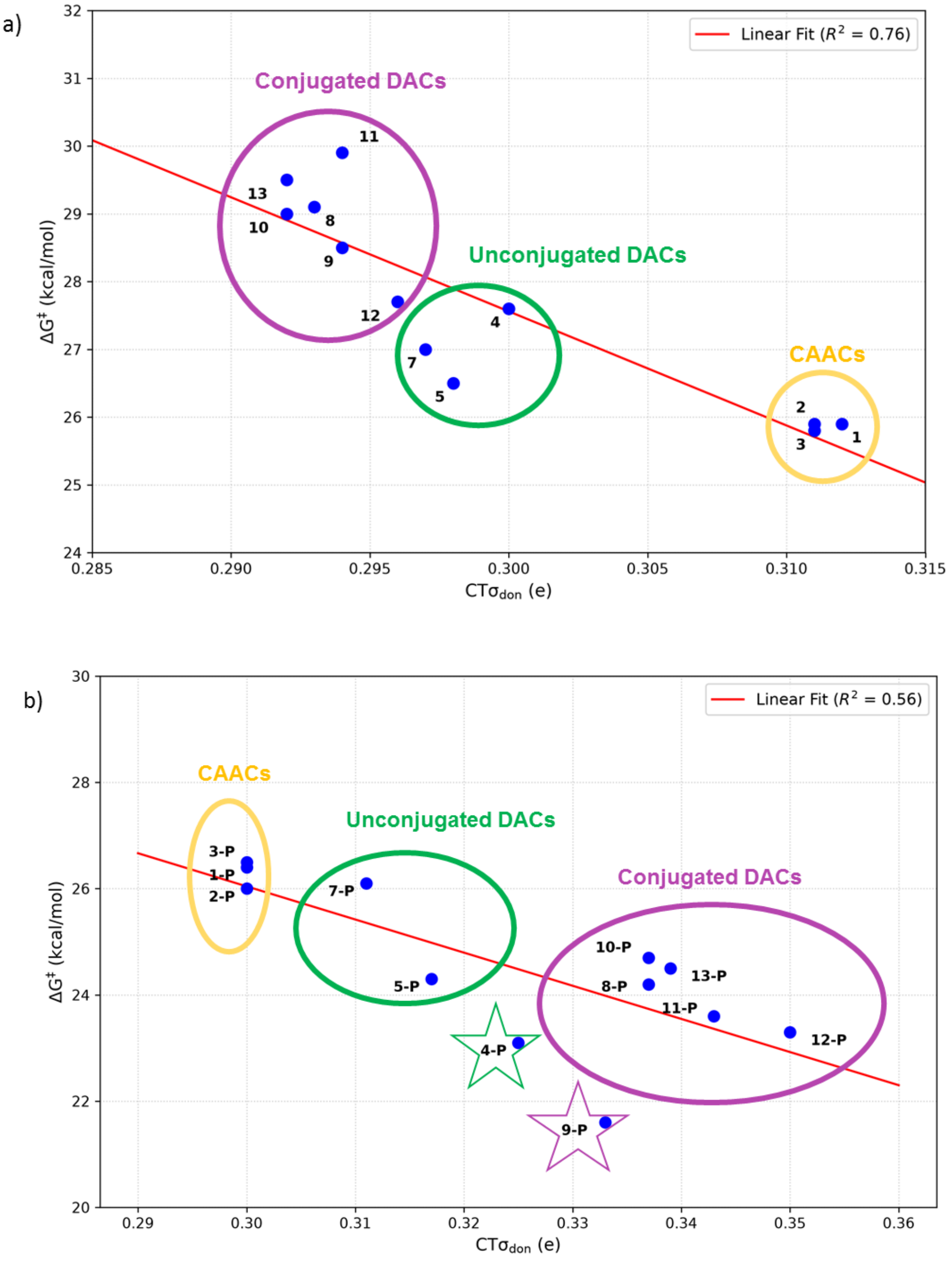

**Figure 7**: Correlation plots between $CT_{\sigma\ don}$ and activation energy barrier $\Delta G^{\ddagger}$ for the panel a) NHC **1-13** series and panel b) NHCP **1-P-13-P** series. CAACs-based complexes are circled in yellow, unconjugated DACs-based complexes in green and conjugated DACs-based complexes in purple.

The imaginary frequencies and corresponding normal modes for all the transition states are reported in the Supporting Information. All of them contain one imaginary frequency in the range 240.8*i* and 195.9*i* $cm^{-1}$ for the **1-P-13-P** NHCP series and 242.2*i* and 173.6*i* $cm^{-1}$ for the **1-13** NHC series, which is associated to a complex vibrational motion involving the hydrogen atom and the carbon and both oxygen atoms of $CO_2$. A concerted transition state describing the H-$CO_2$ interaction is clearly suggested in all complexes. The relatively small imaginary frequencies indicate a flat potential energy surface and are consistent with the imaginary frequency calculated in ref. (48) for the transition state for the $CO_2$ insertion into the Au-Al bond in a gold-aluminyl complex (147.1*i* $cm^{-1}$).
For the NHC **1-13** series, a linear fit between $CT_{\sigma\ don}$ and $\Delta G^{\ddagger}$ values has been found, with an $r^2$ coefficient of 0.76, which suggests that the NHC σ-donor ability mainly accounts for the barrier. Interestingly, although a quantitative trend for $CT_{\sigma\ don}$ and energy barrier $\Delta G^{\ddagger}$ can be identified between the three carbene classes, minor variations of the barrier within each class are visible (Figure 7 panel a), indicating that other factors than σ-donor ability (such as steric effects) can also contribute to it. For instance, for **8**, **9**, **10**, **11**, **12** and **13** conjugated DAC-based complexes, covering a $CT_{\sigma\ don}$ of 0.292-0.296 e, the energy barrier $\Delta G^{\ddagger}$ span only a small range from 28.5 (for **9**) to 29.9 (for **11**) kcal/mol. For the NHCP series, instead, the linear fit between $CT_{\sigma\ don}$ and $\Delta G^{\ddagger}$ values shows an $r^2$ coefficient of 0.56 (Figure 7 panel b), which strongly suggests that the NHCP σ-donor ability is not the only descriptor accounting for the barrier. Similar to NHC complexes, a more quantitative trend for $CT_{\sigma\ don}$ and energy barrier $\Delta G^{\ddagger}$ can be identified between the three carbene classes, with more significant variations within each class. In particular, although in the conjugated DACs-based series the **8-P**, **10-P**, **11-P**, **12-P** and **13-P** complexes cover a $CT_{\sigma\ don}$ of 0.350-0.337 e, spanning a small $\Delta G^{\ddagger}$ range from 23.6 to 24.7 kcal/mol, the **9-P** is an eye-catching outlier, showing a lower barrier ($\Delta G^{\ddagger}$ = 21.6 kcal/mol) than that expected from the smaller $CT_{\sigma\ don}$ of 0.333 e. In the unconjugated DACs-based series (**4-P**, **5-P**, and **7-P**) the results are even more scattered. This different behavior for the NHCP reactivity with respect to that of NHC is consistent with the additional degree of freedom triggered by the PPh unit in the NHC moiety, making the coordination structurally more flexible. Indeed, inspection of the transition state structures for the most outliers **4-P** and **9-P** complexes in Figure 7 panel b) reveals that the incipient formate $HCO_2^-$ interacts via anion…π interactions arising between the electron-rich oxygens of $HCO_2^-$ and the face of the aromatic Mes ring of the NHC moiety as well as that π-π stacking interactions occurs between the second aromatic Mes ring of the NHC moiety and the Ph ring bonded to P, thus stabilizing the transition state and lowering the expected higher barrier (Figure S38 in the Supporting Information). Notably, this analysis suggests that the NHCP reactivity can be further modulated through second sphere effects, which is a relevant aspect in ligand design. We finally emphasize that all the investigated $CO_2$ insertion reactions are calculated to be endergonic, with $\Delta G$ in the range 3.6-6.4 kcal/mol for the NHC **1-13** and 4.2-6.7 kcal/mol for the NHCP **1-P-13-P** series (Table S12 in the Supporting Information), confirming that the thermodynamic stabilization of the formate product may be a major limitation.

This reactivity can be well rationalized in terms of molecular orbitals (MOs) of $CO_2$ and **1-13**, **1-P-13P** and **8-PP** complexes. The main driving force is the charge transfer from the complex HOMO, also including NHC/NHCP-AuH contributions, towards the empty π* LUMO of $CO_2$, thus suggesting a role of the ligand in the reactivity of these complexes. This picture is strongly reminiscent of the interaction scheme depicted for $CO_2$ insertion reaction into the Au−Al bond in [$^t$Bu$_3$PAuAl(NON)] complex, (48) with the Au−H bond (analogous to the Au−Al bond) acting as the actual nucleophile. The HOMO percentage contribution of H 1s varies across the three classes of carbenes from 3-4% in **1-P-3-P**, to 6-7% in **4-P-7-P** to 7-12% in **8-P-13-P** series, and amounts to 12% in **8-PP** (Table S13 in the Supporting Information), consistent with the σ-donor ability (and its trans influence, see Figure S16 in the Supporting Information) and the energy barrier trends. The largest hydride character in conjugated DACs-based NHCP series suggests a stronger $Au^{\delta+}$-$H^{\delta-}$ polarization compared to unconjugated DACs and CAACs, which makes the conjugated DACs-based NHCP complexes more reactive for the carbon dioxide insertion. This analysis confirms that the Au–H polarization has a role in the energy barrier, with the enhanced basicity of the hydride making $CO_2$ reduction process kinetically more favorable. We emphasize that this conclusion about reactivity is derived for AuH complexes. In this framework, in a previous work by some of us (49), we have demonstrated that the reactivity of NHC copper-, silver- and gold-hydride complexes with carbon dioxide is significantly affected by the M-H bond polarity, which decreases from Cu, to Ag and to Au, with the most relevant $Cu^{\delta+}$-$H^{\delta-}$ bond polarization resulting in a more kinetically favorable $CO_2$ reduction.

The major issue for this $CO_2$ activation reactivity is the thermodynamic stabilization of the PC product. Remarkably, all reactions for both **1-13** and **1-P-13-P** series are endergonic (Table

S12), with similar ΔG values. Based on the high structural flexibility of the NHCP ligands, product stabilization might be similarly attained via second sphere interactions, which represents an interesting prospect for ligand design. We underline that the **1-13/1-P-13-P** reactivity with $CO_2$ has been here investigated with the sole aim of elucidating the bonding/reactivity relationship without attempting to study the catalytic $CO_2$ conversion. However, this study also provides useful information concerning one elementary step, namely $CO_2$ insertion into an Au–H bond and formation of an Au-formate species, within a possible catalytic cycle. In this view, these systems may certainly represent promising candidates for future catalytic studies. In particular, an almost thermoneutral or slightly endergonic formation of formate is a favorable opportunity in driving the reversible hydride transfer between formate and carbon dioxide for the NHCP-AuH complexes.

In summary, this investigation suggests that the rate of $CO_2$ insertion into the Au-H bond can be significantly affected by modifying the NHC/NHCP ligand, demonstrating a correlation between the bonding and reactivity: increasing the electron density on the metal fragment by incorporating a stronger σ-donor NHC/NHCP, i.e. a CAAC for NHCs and a conjugated DAC-based for NHCPs, would result in faster $CO_2$ insertion. For NHCP, due to the increased structural flexibility of the ligand imparted by the PPh group bonded to the NHC moiety, the reactivity can be additionally controlled by second sphere effects.

## Conclusions

Over the past two decades, N-heterocyclic carbene-phosphinidenes (NHCPs) ligands have been proven to hold great potential in coordination chemistry, bond activation, catalysis, and polymer science and increased attention is currently directed to their development based on the fine-tuning of their electronic structure through introduction of different classes of carbene moieties. Having control over the possibility of modulating the features of the NHCP bonding and reactivity by selecting a proper NHC moiety on a quantitative ground is a challenge, both experimentally and computationally. In this work, we used a consistent computational protocol which combines EDA, NOCV and CD analyses to quantitatively assess the NHCP bonding/reactivity relationship in transition metal chemistry. For this purpose, we chose the AuH fragment as a probe for interrogating the coordination bonding properties as well as $CO_2$ as a probe for the reactivity in small molecules activation processes. We investigated the NHCPR-AuH (R = Ph) bond for a set of 13 NHC moieties, categorized into three types (CAACs, unconjugated DACs and conjugated DACs), and the corresponding NHC-AuH bond for the carbene analogues, to understand if the σ-donation and π-acceptor NHC bond component trends could be transferred to [(NHCPPh)AuH] complexes. We clearly demonstrated that NHCPs are different from NHCs when used as coordination ligands to gold. Indeed, while the $C_{NHC}$ $sp^2$ lone pair is involved in the σ bonding with AuH in all the NHC complexes, a P 3p lone pair, directed perpendicular to the $C_{NHC}$-P-$C_{Ph}$ bonding plane, is used to form a σ bond with AuH in the NHCP complexes, which results both electronically and sterically highly tunable through introduction of a distinct class of NHC moiety. We demonstrated that, within the series studied here, the NHCP ligands are stronger σ-donor than NHCs, with a σ-donor trend opposite to that of NHCs directly bonded to the metal center, i.e. σ–donation increases from CAACs to conjugated DACs, and much weaker π-acceptor compared to NHCs, although showing the same trend, i.e. π-backdonation increases from conjugated DACs to CAACs, consistent with the use of the experimental $^{31}P$ NMR chemical shift to measure the NHC π-acceptor ability in their corresponding [(NHCPPh)] adducts. Although clear trends are identified between the three carbene classes, variations of σ-donor/π-acceptor properties within each class are rather small, meaning that the NHCP scaffold-dependent tunability within each carbene family can be described as minor. The coordination geometry at the P center is accordingly very flexible, shifting from "nearly planar" (in CAACs-based NHCP) to $PH_3$-like (in conjugated DACs-based NHCP), driven by the lone pair P 3p percentage and energy under NHC moiety control, with the $C_{NHC}$-P-Au bond angle that can be regarded as a structural descriptor to monitor and/or design NHCP ligands with desired σ-donor properties. A second lone pair on P, available for additional coordination, bears strict isolobal analogies with the $C_{NHC}$ lone pair in NHCs, and as such can be more directly modulated by both the NHC and R substituent structures. Since a NHC-stabilized diphosphene [(NHCPP)AuH] complex has been experimentally reported (26,27) to easily and reversibly react with $CO_2$, yielding the corresponding Au formate complex, we explored the mechanism and the activation energy barriers for a comparative analysis of the reactivity of all the NHC and NHCP complexes. The NHCP complexes under study showed remarkable quantitative differences when reacting with $CO_2$, with kinetic activation barriers closely correlating with the tunable NHCP σ-donor ability. In addition to the NHCP σ-donor ability, the higher structural flexibility imparted by the PPh group allows for the NHCP activation barriers modulation through second sphere effects. All investigated $CO_2$ insertion reactions were calculated to be endergonic, meaning that the kinetic tuning of $CO_2$ insertion (i.e. lowering the barrier by maximizing the ligand σ-donation) does not necessarily improve product thermodynamic stabilization of the resulting formate complex, which may be a major limitation in the proposed ligand-design principles. However, introduction of suitable second sphere coordination effects may play a relevant role to overcome it. We emphasize that conclusions regarding the NHCP coordination chemistry and reactivity studied here are drawn within the chemical space actually investigated, and the effects of changing the metal fragment or the substituent attached to phosphorus need to be systematically explored.

This work is devoted to understanding the NHCP highly tunable steric and electronic peculiar features as coordination ligands, as well as exploring their potential applications in small molecule activation and catalysis, fitting in the currently expanding literature concerning the emerging class of NHCPs.

**Computational details**

Geometry optimizations and frequency calculations of NHC- and NHCP- gold hydride complexes (**1-13**, **1-P-13-P**) as well as minima and transition states for reaction paths for all 26 complexes plus **8-PP** (minima with zero imaginary frequencies and transition states with one imaginary frequency) have been carried out using the Amsterdam Density Functional (ADF) code(50) in combination with the related Quantum-regions Interconnected by Local Description (QUILD) program.(51) The BP86 (52-54) GGA exchange-correlation (XC) functional, the TZ2P basis set with a small frozen core approximation for all atoms, the ZORA Hamiltonian (55-57) for treating scalar relativistic effects and the Grimme's D3-BJ dispersion correction were used.(58,59) Solvent effects were modeled employing the Conductor-like Screening Model (COSMO) with the default parameters for toluene as implemented in the ADF code.(60) The standard state used for solution-phase (COSMO) free energies calculations are T = 298.15 K, P = 1 atm. A 1 atm to 1M correction has not been applied since it is a systematic correction along the reaction paths for all the systems. The harmonic oscillator model, with cutting of the low-lying frequencies below a threshold of 100 $cm^{-1}$, was applied. Intrinsic Reaction Coordinate (IRC) calculations (61,62) have been performed for **8-P** and **8** complexes, as NHCP and NHC representative, demonstrating that the reported transition states connect the reactant and formate product local minima. Results are provided in the SI (Figures S39 and S40). Reaction energy barriers have been validated through single-point calculations on corresponding stationary points using a hybrid (B3LYP) (63) and a range-separated hybrid (ωB97x) (64) functional. Results are reported in the Supporting Information (Table S14 and Figure S41) and show that the reactivity trends for **1**, **1-P**, **8**, **8-P**, **10**, **10-P**, and **8-PP** are not significantly dependent on the chosen computational method. The same computational setup has also been used for the EDA (31) and CD-NOCV (32-35) calculations. This protocol has been used successfully in Ref. (28) to study the ligand effect on the reactivity of monomeric gold hydrides for $CO_2$ reduction as well as in Ref. (65) for a consistent assignment of the redox character in main-group bond activation reactivity, revealing a central role of ligand electronics. The CD method has been also validated against experimental values in previous works by Belpassi et al. demonstrating that NMR spectroscopy can separately probe σ-donation (via $^{13}C$ chemical shifts) (66) and π back- donation (via C–N bond rotational barriers, $\Delta H_r^{\ddagger}$) in nitrogen acyclic carbene gold(I) complexes. (67)

For details and description of the methods used in this work, see the Methodology section in the Supporting Information. Additional bonding studies using independent bond descriptors, i.e. Mayer bond order (MBO), (37) Gophinatan-Jug (G-J) bond order (68) and Nalewajski-Mrozek (N-M) bond order (69) analysis, as well as charge analysis using Mulliken, (70) Hirshfeld, (71) Voronoi Deformation Density (VDD) (72) and QTAIM (73,74) methods for representative **1-P**, **4-P** and **8-P** complexes have been reported in the Supporting Information (Tables S15-S17), which confirm the trends obtained with the EDA/CD-NOCV decomposition scheme.

**Supporting Information**

The Supporting Information is available free of charge at…….

All the relevant data (including methodology description, mechanistic, CD-NOCV and EDA data discussed in the manuscript and xyz geometries) are provided within the Supporting Information.

**Conflicts of interest**

The authors declare no competing financial interests.

**Acknowledgements**

This work was supported by the European Union – NextGenerationEU under the Italian Ministry of University and Research (MUR) National Innovation Ecosystem grant ECS00000041 – VITALITY – CUP B43C22000470005 and – CUP J97G22000170005 and by the European Union – NextGenerationEU, AdP POR H2 project L. A. 1.1.35 CUP B93C22000630006. P.B. and L.B. also thank Università degli Studi di Perugia, MUR and CNR for support within the project Vitality. G.T. acknowledges funding from the European Research Council (ERC) under the European Union's Horizon Europe Research and Innovation Program (Grant ERC-StG-2021-101040197-QED-SPIN).

**References**

1. Arduengo III, A.J.; Harlow, R.L.; Kline, M. A stable crystalline carbene. *J. Am. Chem. Soc.* **1991**, *113* (1), 361-363, DOI: 10.1021/ja00001a054.
2. Bellotti, P.; Koy, M.; Hopkinson, M.N.; Glorius, F. Recent advances in the chemistry and applications of N-heterocyclic carbenes. *Nat. Rev. Chem.* **2021**, *5*, 711-725, DOI: 10.1038/s41570-021-00321-1.
3. Gaggioli, C.A.; Bistoni, G.; Ciancaleoni, G.; Tarantelli, F.; Belpassi, L.; Belanzoni, P. Modulating the bonding properties of N-heterocyclic carbenes (NHCs): a systematic charge-displacement analysis. *Chem. Eur. J.* **2017**, *23*, 7558-7569, DOI: 10.1002/chem.201700638.
4. Soleilhavoup, M.; Bertrand, G. Cyclic (alkyl)(amino)carbenes (CAACs): stable carbenes on the rise. *Acc. Chem. Res.* **2015**, *48*, 256-266, DOI: 10.1021/ar5003494.
5. Moerdyk, J.P.; Schilter, D.; Bielawski, C.W. N,N-diamidocarbenes: isolable divalent carbons with bona fide carbene reactivity. *Acc. Chem. Res.* **2016**, *49*, 1458-1468, DOI: 10.1021/acs.accounts.6b00080.
6. Huynh, H.V. Electronic properties of N-heterocyclic carbenes and their experimental determination. *Chem. Rev.* **2018**, *118*, 9457-9492, DOI: 10.1021/acs.chemrev.8b00067.
7. Gómez-Suárez, A.; Nelson, D.J.; Nolan, S.P. Quantifying and understanding the steric properties of N-heterocyclic carbenes. *Chem. Commun.* **2017**, *53*, 2650-2660, DOI: 10.1039/c7cc00255f.
8. Tolman, C.A. Electron donor-acceptor properties of phosphorous ligands. Substituent additivity. *J. Am. Chem. Soc.* **1970**, *92*, 2953-2956, DOI: 10.1021/ja00713a006.
9. Chianese, A.R.; Li, X.; Janzen, M.C.; Faller, J.W., Crabtree, R.H. Rhodium and iridium complexes of N-heterocyclic carbenes via transmetalation: structure and dynamics. *Organometallics* **2003**, *22*, 1663-1667, DOI: 10.1021/om021029+.

10. Huynh, H.V.; Han, Y.; Jothibasu, R.; Yang, J.A. $^{13}$C NMR spectroscopic determination of ligand donor strengths using N-heterocyclic carbene complexes of palladium (II). *Organometallics* **2009**, *28*, 5395-5404, DOI: 10.1021/om900667d.
11. Back, O.; Henry-Ellinger, M.; Martin, C.D.; Martin, D.; Bertrand, G. $^{31}$P NMR chemical shifts of carbene-phosphinidene adducts as an indicator of the π-accepting properties of carbenes. *Angew. Chem. Int. Ed.* **2013**, *52*, 2939-2943, DOI: 10.1002/anie.201209109.
12. Liske, A.; Verlinden, K.; Buhl, H.; Schaper, K.; Ganter, C. Determining the π-acceptor properties of N-heterocyclic carbenes by measuring the $^{77}$Se NMR chemical shifts of their selenium adducts. *Organometallics* **2013**, *32*, 5269-5272, DOI: 10.1021/om400858y.
13. Krachko, T.; Slootweg, J.C. N-heterocyclic carbene-phosphinidenes adducts: synthesis, properties, and applications. *Eur. J. Inorg. Chem.* **2018**, 2734-2754, DOI: 10.1002/ejic.201800459.
14. Doddi, A.; Bockfeld, D.; Bannenberg, T.; Tamm, M. N-heterocyclic carbene analogues of nucleophilic phosphinidene transition metal complexes. *Chem. Eur. J.* **2020**, *26*, 14878-14887, DOI: 10.1002/chem.202003099.
15. Rodriguez Villanueva, J.E.; Wiebe, M.A.; Lavoie, G.G. Coordination and reactivity studies of titanium complexes of monoanionic inversely polarized phosphaalkene-ethenolate ligands. *Organometallics* **2020**, *39*, 3260-3267, DOI: 10.1021/acs.organomet.0c00476.
16. Bhattacharjee, J.; Bockfeld, D.; Tamm, M. N-heterocyclic carbene-phosphinidenide complexes as hydroboration catalysts. *J. Org. Chem.* **2022**, *87*, 1098-1109, DOI: 10.1021/acs.joc.1c02377.
17. Peters, M.; Bockfeld, D.; Tamm, M. Cationic iridium(I) NHC-phosphinidene complexes and their application in hydrogen isotope exchange reactions. *Eur. J. Inorg. Chem.* **2022**, e202200148, DOI: 10.1002/ejic.202200148.
18. Doddi, A.; Peters, M.; Bockfeld, D.; Tamm, M. Copper and silver complexes of N-heterocyclic carbene-parent phosphinidenes adducts. *Z. Anorg. Allg. Chem.* **2023**, *649*, e202200364, DOI: 10.1002/ zaa.202200364.
19. Hadlington, T.J.; Kostenko, A.; Driess, M. Synthesis and coordination ability of a donor-stabilised bis-phosphinidene. *Chem. Eur. J.* **2021**, *27*, 2476-2482, DOI: 10.1002/chem.202004300.
20. Zhong, M.; Yuan, M. Recent advances in the use of N-heterocyclic carbene adducts of N, P, C elements as supporting ligands in organometallic chemistry. *RSC Adv.* **2025**, *15*, 15052-15085, DOI: 10.1039/D5RA02549D.
21. Dutta, S.; Maity, B.; Thirumalai, D.; Koley, D. Computational investigation of carbene-phosphinidenes: correlation between $^{31}$P chemical shifts and bonding features to estimate the π-backdonation of carbenes. *Inorg. Chem.* **2018**, *57*, 3993-4008, DOI: 10.1021/acs.inorgchem.8b00174.
22. Zhu, H.; Inoue, S. N-heterocyclic carbene-phospinidenes in main group chemistry. *Chem* **2025**, *11*, 102649, DOI: 10.1016/j.chempr.2025.102649.
23. Tsui, E.Y.; Müller, P.; Sadighi, J.P. Reactions of a stable monomeric gold(I) hydride complex. *Angew. Chem. Int. Ed.* **2008**, *47*, 8937-8940, DOI: 10.1002/anie.200803842.
24. Doddi, A.; Bockfeld, D.; Nasr, A.; Bannenberg, T.; Jones, P.G.; Tamm, M. N-heterocyclic carbene-phosphinidene complexes of the coinage metals. *Chem. Eur. J.* **2015**, *21*, 16178-16189, DOI: 10.1002/chem.201502208.

25. Adiraju, V.A.K.; Yousufuddin, M.; Dias, H.V.R. Copper(I), silver(I) and gold(I) complexes of N-heterocyclic carbene-phosphinidene. *Dalton Trans.* **2015**, *44*, 4449-4454, DOI: 10.1039/c4dt03285c.
26. Dhara, D.; Das, S.; Pati, S.K.; Scheschkewitz, D.; Chandrasekhar, V.; Jana, A. NHC-coordinated diphosphene-stabilized gold(I) hydride and its reversible conversion to gold(I) formate with $CO_2$. *Angew. Chem. Int. Ed.* **2019**, *58*, 15367-15371, DOI: 10.1002/anie.201909798.
27. Dhara, D.; Scheschkewitz, D.; Chandrasekhar, V.; Yildiz, C.B.; Jana, A. Reactivity of NHC/diphosphene-coordinated Au(I)-hydride. *Chem. Commun.* **2021**, *57*, 809-812, DOI: 10.1039/d0cc05461e.
28. Rossi, E.; Sorbelli, D.; Belanzoni, P.; Belpassi, L.; Ciancaleoni, G. Monomeric gold hydrides for carbon dioxide reduction: ligand effect on the reactivity. *Chem. Eur. J.* **2024**, *30*, e202303512, DOI: 10.1002/chem.202303512.
29. Hazari, N.; Heimann, J.E. Carbon dioxide insertion into Group 9 and 10 metal-element σ bonds. *Inorg. Chem*. **2017**, *56*, 13655–13678, DOI:10.1021/acs.inorgchem.7b02315.
30. Cokoja, M.; Bruckmeier, C.; Rieger, B.; Herrmann, W.A.; Kühn, F.E. Transformation of carbon dioxide with homogeneous transition-metal catalysts: a molecular solution to a global challenge? *Angew. Chem. Int. Ed.* **2011**, *50*, 8510–8537, DOI: 10.1002/anie.201102010.
31. Zhao, L.; von Hopffgarten, M.; Andrada, D.M.; Frenking, G. Energy decomposition analysis. *Wiley Interdiscip. Rev.: Comput. Mol. Sci.*, **2018**, *8*, e1345, DOI: 10.1002/wcms.1345.
32. Mitoraj, M.; Michalak, A. Natural orbitals for chemical valence as descriptors of chemical bonding in transition metal complexes. *J. Mol. Model.*, **2007**, *13*, 347–355, DOI: 10.1007/s00894-006-0149-4.
33. Michalak, A.; Mitoraj, M.; Ziegler, T. Bond orbitals from chemical valence theory. *J. Phys. Chem. A*, **2008**, *112*, 1933–1939, DOI: 10.1021/jp075460u.
34. Belpassi, L.; Infante, I.; Tarantelli, F.; Visscher, L. The chemical bond between Au(I) and the noble gases. Comparative study of NgAuF and $NgAu^+$ (Ng = Ar, Kr, Xe) by density functional and coupled cluster methods. *J. Am. Chem. Soc.*, **2008**, *130*, 1048–1060, DOI: 10.1021/ja0772647.
35. Bistoni, G.; Rampino, S.; Tarantelli, F.; Belpassi, L. Charge-displacement analysis via natural orbitals for chemical valence: Charge transfer effects in coordination chemistry. *J. Chem. Phys*., **2015**, *142*, 084112, DOI: 10.1063/1.4908537.
36. Sorbelli, D.; Belpassi, L.; Belanzoni, P. Cooperative small molecule activation by apolar and weakly polar bonds through the lens of a suitable computational protocol. *Chem. Commun*., **2024**, *60*, 1222-1238, DOI: 10.1039/d3cc05614g.
37. Bridgeman, A. J.; Cavigliasso, G.; Ireland, L.R.; Rothery, J. The Mayer bond order as a tool in inorganic chemistry. *J. Chem. Soc., Dalton Trans.,* **2001**, *30*, 2095-2108, DOI: 10.1039/b102094n.
38. Sorbelli, D.; Nunes dos Santos Comprido, L.; Knizia, G.; Hashmi, A.S.K.; Belpassi, L.; Belanzoni, P.; Klein, J.E.M.N. Cationic gold(I) diarylallenylidene complexes: bonding features and ligand effects. *ChemPhysChem* **2019**, *20*, 1671-1679, DOI: 10.1002/cphc.201900411.
39. Sorbelli, D.; Belanzoni, P.; Belpassi, L. Tuning the gold(I)-carbon σ bond in gold-alkynyl complexes through structural modifications of the NHC ancillary ligand: effect on spectroscopic observables and reactivity. *Eur. J. Inorg. Chem.* **2021**, *2021*, 2401-2416, DOI: 10.1002/ejic.202100260.

40. Sorbelli, D.; Belpassi, L.; Belanzoni, P. What singles out aluminyl anions? A comparative computational study of the carbon dioxide insertion reaction in gold–aluminyl, −gallyl, and −indyl complexes. *Inorg. Chem.* **2022**, *61*, 1704–1716, DOI: 10.1021/acs.inorgchem.1c03579.
41. Sorbelli, D; Belpassi, L.; Belanzoni, P. Unraveling differences in aluminyl and carbene coordination chemistry: bonding in gold complexes and reactivity with carbon dioxide. *Chem. Sci.* **2022**, *13*, 4623–4634, DOI: 10.1039/d2sc00630h.
42. Sorbelli, D.; Rossi, E.; Havenith, R.W.A.; Klein, J.E.M.N.; Belpassi, L.; Belanzoni, P. Gold-aluminyl and gold-diarylboryl complexes: bonding and reactivity with carbon dioxide. *Inorg. Chem.* **2022**, *61*, 7327–7337, DOI: 10.1021/acs.inorgchem.2c00174.
43. Marchione, D.; Belpassi, L.; Bistoni, G.; Macchioni, A.; Tarantelli, F.; Zuccaccia, D. The chemical bond in gold(I) complexes with N-heterocyclic carbenes. *Organometallics* **2014**, *33*, 4200–4208, DOI: 10.1021/om5003667.
44. Blake, G.A.; Moerdyk, J.P.; Bielawski, C.W. Tuning the electronic properties of carbenes: a systematic comparison of neighboring amino versus amido groups. *Organometallics* **2012**, *31*, 3373-3378, DOI: 10.1021/om3001586.
45. Makhloufi, A.; Frank, W.; Ganter, C. Diamino- and mixed amino-amido-N-heterocyclic carbenes based on triazine backbones. *Organometallics* **2012**, *31*, 2001-2008, DOI: 10.1021/om201275z.
46. Poater, A.; Cosenza, B.; Correa, A.; Giudice, S.; Ragone, F.; Scarano, V.; Cavallo, L. SambVca: a web application for the calculation of the buried volume of N-heterocyclic carbene ligands. *Eur. J. Inorg. Chem.* **2009**, 1759-1766, DOI: 10.1002/ejic.200801160.
47. Software Platform Sambvca 2.1 (Steric Map Analysis of Buried Volume), reference manual: https://www.aocdweb.com/OMtools/sambvca2.1/help/help.html
48. Sorbelli, D.; Belpassi, L.; Belanzoni, P. Reactivity of a gold-aluminyl complex with carbon dioxide: a nucleophilic gold? *J. Am. Chem. Soc.* **2021**, *143*, 14433–14437, DOI:10.1021/jacs.1c06728.
49. Sorbelli, D.; Belpassi, L.; Belanzoni, P. Coinage metal effect on the reduction of carbon dioxide with monomeric metal-hydride complexes. *Eur. J. Inorg. Chem*. **2024**, *27*, e202400280, DOI: 10.1002/ejic.202400280.
50. te Velde, G.; Bickelhaupt, F. M.; Baerends, E. J.; Fonseca Guerra, C.; van Gisbergen, S. J. A.; Snijders, J. G.; Ziegler, T. Chemistry with ADF. *J. Comput. Chem.* **2001**, *22*, 931–967, DOI:10.1002/jcc.1056.
51. Swart, M.; Bickelhaupt, F. M. QUILD: QUantum-Regions Interconnected by Local Descriptions. *J. Comput. Chem.* **2008**, *29*, 724–734, DOI:10.1002/jcc.20834.57.
52. Becke, A.D. Density-functional exchange-energy approximation with correct asymptotic behavior. *Phys. Rev. A* **1988**, *38*, 3098, DOI:10.1103/PhysRevA.38.3098.
53. Perdew, J.P.; Wang, Y. Density-functional approximation for the correlation energy of the inhomogeneous electron gas. *Phys. Rev. B* **1986**, *33*, 8822, DOI: 10.1103/PhysRevB.33.8822
54. Perdew, J.P. Erratum: Density-functional approximation for the correlation energy of the inhomogeneous electron gas. *Phys. Rev. B* **1986**, *34*, 7406, DOI: 10.1103/PhysRevB.34.7406
55. Van Lenthe, E.; Baerends, E. J.; Snijders, J. G. Relativistic regular two-component Hamiltonians. *J. Chem. Phys.* **1993**, *99*, 4597–4610, DOI:10.1063/1.466059.
56. van Lenthe, E.; Baerends, E. J.; Snijders, J. G. Relativistic total energy using regular approximations. *J. Chem. Phys.* **1994**, *101*, 9783–9792, DOI: 10.1063/1.467943.60.

57. van Lenthe, E.; Ehlers, A.; Baerends, E. J. Geometry optimizations in the Zero Order Regular Approximation for relativistic effects. *J. Chem. Phys.* **1999**, *110*, 8943–8953, DOI: 10.1063/1.478813.61.
58. Grimme, S.; Antony, J.; Ehrlich, S.; Krieg, H. A consistent and accurate Ab Initio parametrization of density functional dispersion correction (DFT-D) for the 94 elements H-Pu. *J. Chem. Phys.* **2010**, *132*, 154104, DOI:10.1063/1.3382344.62.
59. Grimme, S.; Ehrlich, S.; Goerigk, L. Effect of the damping function in dispersion corrected density functional theory. *J. Comput. Chem.* **2011**, *32*, 1456–1465, DOI: 10.1002/jcc.21759.
60. Pye, C. C.; Ziegler, T. An implementation of the conductor-like screening model of solvation within the Amsterdam Density Functional package. *Theor. Chem. Acc.* **1999**, *101*, 396–408, DOI: 10.1007/s002140050457.
61. Deng, L.; Ziegler, T.; Fan, L. A combined density functional and intrinsic reaction coordinate study on the ground state energy surface of $H_2CO$. *J. Chem. Phys.* **1993**, *99*, 3823-3835, DOI: 10.1063/1.466129.
62. Deng, L.; Ziegler, T. The determination of intrinsic reaction coordinates by density functional theory. *Int. J. Quantum Chem.* **1994**, *52*, 731-765, DOI: 10.1002/qua.560520406.
63. Stephens, P.J.; Devlin, F.J.; Chabalowski, C.F.; Frisch, M.J. Ab initio calculation of vibrational absorption and circular dichroism spectra using density functional force fields. *J. Phys. Chem.* **1994**, *98*, 11623-11627, DOI: 10.1021/j100096a001.
64. Chai, J.- D.; Head-Gordon, M. Systematic optimization of long-range corrected hybrid density functionals. *J. Chem. Phys.* **2008**, *128*, 084106, DOI: 10.1063/1.2834918.
65. Belanzoni, P.; Comas-Vilà, G.; Sorbelli, D. Diagnosing the redox character of bond activation by main group centers: reductive addition vs redox-neutral insertion as a case study. *Inorg. Chem.* **2026**, *65*, 9711-9718 DOI: 10.1021/acs.inorgchem.6c01103.
66. Marchione, D.; Izquierdo, M.A.; Bistoni, G.; Havenith, R.W.A.; Macchioni, A.; Zuccaccia, D.; Tarantelli, F.; Belpassi, L. $^{13}C$ NMR spectroscopy of N-Heterocyclic carbenes can selectively probe σ donation in gold(I) complexes. *Chem. Eur. J.* **2017**, *23*, 2722-2728 DOI: 10.1002/chem.201605502.
67. Ciancaleoni, G.; Biasiolo, L.; Bistoni, G.; Macchioni, A.; Tarantelli, F.; Zuccaccia, D.; Belpassi, L. Selectively measuring π back-donation in gold(I) complexes by NMR spectroscopy. *Chem. Eur. J.* **2015**, *21*, 2467-2473 DOI: 10.1002/chem.201406049.
68. Gopinathan, M.S.; Jug, K. Valency I. A quantum chemical definition and properties. *Theoret. Chim. Acta* **1983**, *63*, 497-509, DOI: 10.1007/BF02394809.
69. Michalak, A.; DeKock, R.L.; Ziegler, T. Bond multiplicity in transition-metal complexes: applications of two-electron valence indices. *J. Phys. Chem. A* **2008**, *112*, 7256-7263, DOI: 10.1021/jp800139g.
70. Mulliken, R.S. Electronic population analysis on LCAO-MO molecular wave functions. I *J. Chem. Phys.* **1955**, *23*, 1833-1840, DOI: 10.1063/1.1740588.
71. Hirshfeld, F.L. Bonded-atom fragments for describing molecular charge densities. *Theoret. Chim. Acta* **1977**, *44*, 129-138, DOI: 10.1007/BF00549096.
72. Fonseca Guerra, C.; Handgraaf, J.-W.; Baerends, E.J.; Bickelhaupt, F.M. Voronoi deformation density (VDD) charges: assessment of the Mulliken, Bader, Hirshfeld, Weinhold, and VDD methods for charge analysis. *J. Comput. Chem.* **2004**, *25*, 189-210, DOI: 10.1002/jcc.10351.

73. Rodríguez, J.I.; Bader, R.F.W.; Ayers, P.W.; Michel, C.; Götz, A.W.; Bo, C. A high performance grid-based algorithm for computing QTAIM properties. *Chem. Phys. Lett.* **2009**, *472*, 149-152, DOI: 10.1016/j.cplett.2009.02.081.
74. Rodríguez, J.I.; Köster, A.M.; Ayers, P.W.; Santos-Valle, A.; Vela, A.; Merino, G. An efficient grid-based scheme to compute QTAIM atomic properties without explicit calculation of zero-flux surfaces. *J. Comput. Chem.* **2009**, *30*, 1082-1092, DOI: 10.1002/jcc.21134.

Graphical Abstract

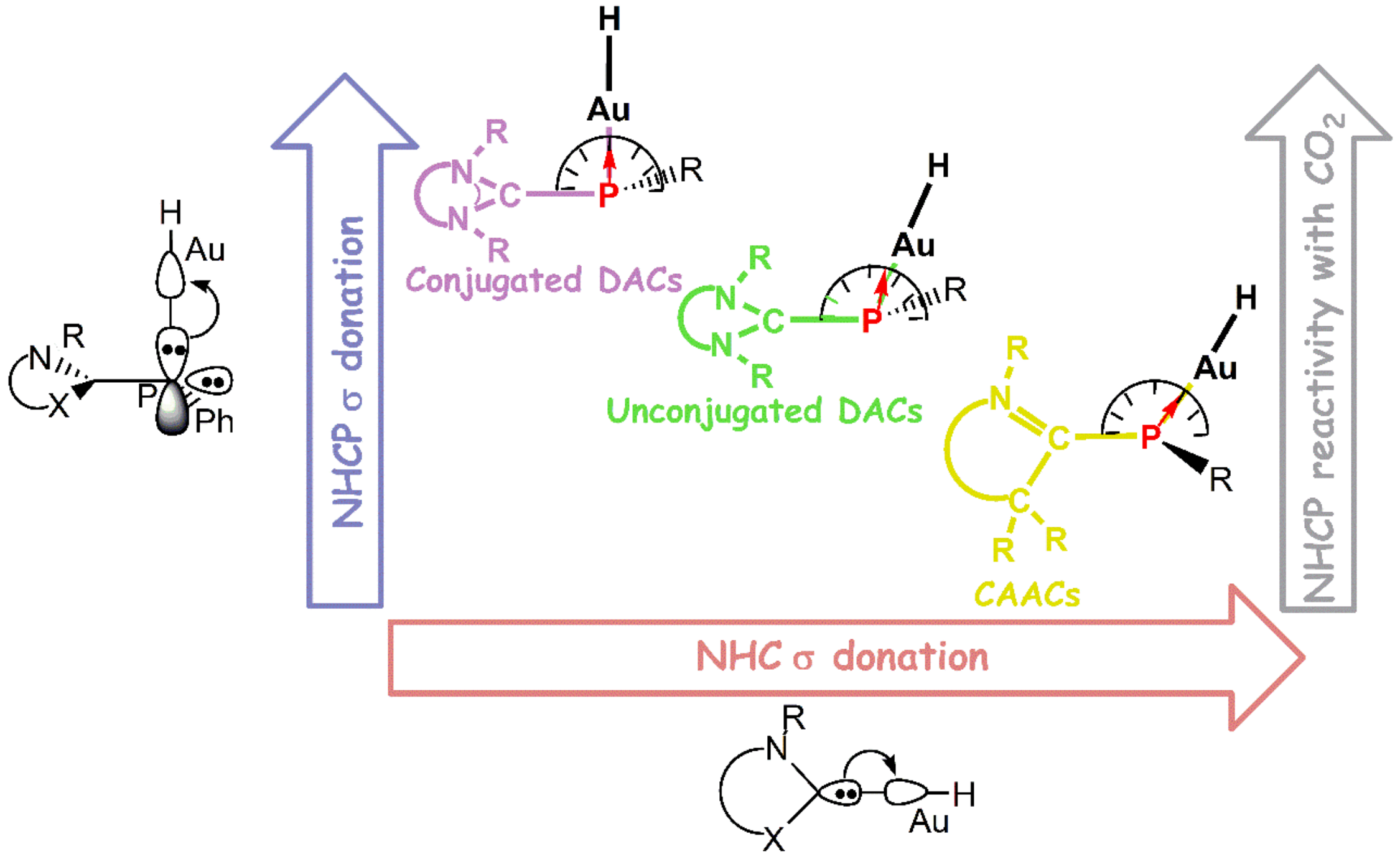

# Supporting Information

## Unravelling carbene moiety role in the N-heterocyclic carbene-phosphinidenes coordination chemistry: bonding and reactivity in gold complexes

Guillaume Thiam,*[a] Leonardo Belpassi,*[b] and Paola Belanzoni*[ab]

[a]*Department of Chemistry, Biology and Biotechnologies, University of Perugia, Via Elce di Sotto, 8 – 06123, Perugia, Italy. Paola Belanzoni email: paola.belanzoni@unipg.it; Guillaume Thiam email:guillaume.thiam @unipg.it.*

[b]*CNR Institute of Chemical Science and Technologies "Giulio Natta" (CNR-SCITEC), Via Elce di Sotto, 8 – 06123, Perugia, Italy. Leonardo Belpassi email: leonardo.belpassi@cnr.it*

## Table of Contents

## Methodology

- **Energy Decomposition Analysis (EDA)**

In this work the Energy Decomposition Analysis (EDA)[1–3] has been applied to get insights into the interaction between NHCP and Au-H fragments in **1-P-13-P** complexes and between NHC and AuH fragments in **1-13** analogue complexes. With this approach, the interaction energy between the fragments can be decomposed in different contributions as follows:

$$\Delta E_{\mathrm{int}} = \Delta E_{Pauli} + \Delta E_{elstat} + \Delta E_{oi} + \Delta E_{disp}$$

[1]

where $\Delta E_{\mathrm{Pauli}}$ represents the Pauli repulsion interaction between occupied orbitals on the two fragments, $\Delta E_{\mathrm{elstat}}$ is the quasiclassical electrostatic interaction between the unperturbed charge distribution of the fragments at their final positions, $\Delta E_{\mathrm{disp}}$ takes into account the dispersion contribution and $\Delta E_{\mathrm{oi}}$ is the orbital interaction, which arises from the orbital relaxation and the orbital mixing between the fragments, and accounts for electron pair bonding, charge transfer, and polarization. The sum $\Delta E_{Pauli} + \Delta E_{elstat}$ is commonly defined as $\Delta E_{\mathrm{steric}}$ roughly accounting for ionic character of the bond.

- **Natural Orbitals for Chemical Valence (NOCV) and Charge Displacement (CD) analysis**

The Natural Orbitals for Chemical Valence (NOCV)[4,5] is a suitable approach for the description of chemical bonding and is based on the rearrangement of the electron density occurring when a chemical bond is formed. In general, arrangement can be expressed as electron density difference between the formed adduct (AB) and sum of the densities of the two non-interacting fragments (A and B) frozen in the geometries they have in the adduct.

This deformation density can be brought into diagonal contributions in terms of NOCVs. In the NOCV scheme, the charge rearrangement taking place upon bond formation is obtained from the occupied orbitals of the two fragments suitably orthogonalized to each other and renormalized (*promolecule*). The resulting electron density rearrangement ($\Delta\rho'$) can be expressed in terms of NOCV pairs which are defined as the eigenfunctions of the so-called "valence operator"[6–8] as follows:

$$\Delta\rho' = \sum_k \nu_k (|\phi_{+k}|^2 - |\phi_{-k}|^2) = \sum_k \Delta\rho'_k \qquad [2]$$

where $\phi_{+k}$ and $\phi_{-k}$ are the NOCV pairs orbitals and $\nu_{\pm k}$ are the corresponding eigenvalues. When the adduct is formed from the promolecule, a fraction $\nu_k$ of electrons is transferred from the $\phi_{-k}$ to the $\phi_{-k}$ orbital, which are envisaged as donor and acceptor orbitals, respectively. For the sake of interpretation, a population analysis can also be performed in order to single out, for $\phi_{-k}$ and $\phi_{+k}$ orbitals, which molecular orbitals (MOs) of the two constituting fragments contribute to the interaction (with a resulting associated coefficient accounting for the magnitude of the contribution).

The NOCV scheme can be coupled with the framework of the Charge Displacement (CD) analysis. The CD analysis allows to quantify the amount of electronic charge that is transferred between the two fragments upon the formation of the A-B bond. The Charge Displacement function ($\Delta$q) is defined as the partial progressive integration on a suitable z-axis of the deformation density $\Delta\rho$:[9]

$$\Delta q(z) = \int_{-\infty}^{z} dz' \int_{-\infty}^{+\infty} \int_{-\infty}^{+\infty} \Delta\rho(x,y,z') dx dy \qquad [3]$$

The CD function, $\Delta$q(z), quantifies at each point of the bond axis the exact amount of electron charge that, upon formation of the bond, is transferred from the right to the left across a plane perpendicular to the bond axis through z.

The CD and NOCV frameworks can be coupled in the CD-NOCV scheme.[10] In the latter, the density rearrangement due to the bond formation between two fragments, ($\Delta\rho'$), can be partitioned in different NOCV deformation densities ($\Delta\rho_k$) and therefore one is able to quantify the charge transfer (CT) associated to each different component. It must be noted that only few of the NOCV pairs contributes to the chemical bond. Therefore, when the CD-NOCV analysis is carried out, usually only the first $\Delta\rho_k$ components are investigated in order to understand which significant chemical contribution to the bond they represent.

In equation [3], the integration axis is usually conveniently chosen as the bond axis between the two fragments constituting the adduct and usually we choose to evaluate the charge transfer between A and B by taking the CD value at the "isodensity boundary", i.e. the z-point where equally valued isodensity surfaces of the isolated fragments become tangent.[9]

Ultimately, this approach can also be expressed in the CD-NOCV framework. By combining Equations [2] and [3], we can use to this approach for calculating the charge transfer associated to each NOCV deformation density as follows:

$$CT_k = \int_A \Delta\, \rho_k(r)dr = -\int_B \Delta\, \rho_k(r)dr \qquad [7]$$

In the framework of NOCV, the orbital interaction term $\Delta E_{oi}$ from EDA analysis can be also further decomposed within the Extended Transition State ETS-NOCV[11] scheme into NOCV pairwise orbital terms ($\Delta E_{oi} = \sum_k \Delta\, E_{oi}^k$) which associates an energy contribution ($E_{oi}^k$) to each NOCV deformation density ($\Delta\rho_k$).

## Optimized structures of [NHCPAuH] complexes

### CAACs

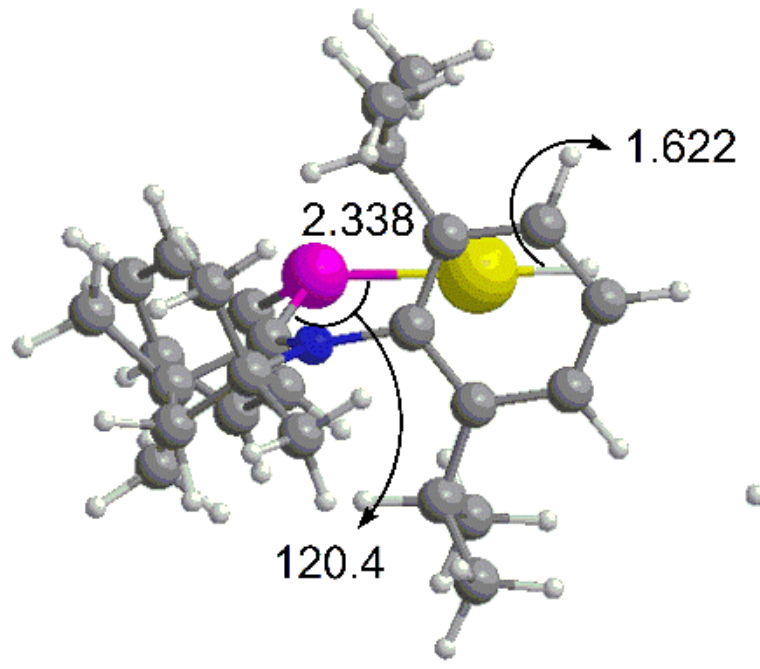

**1-P**

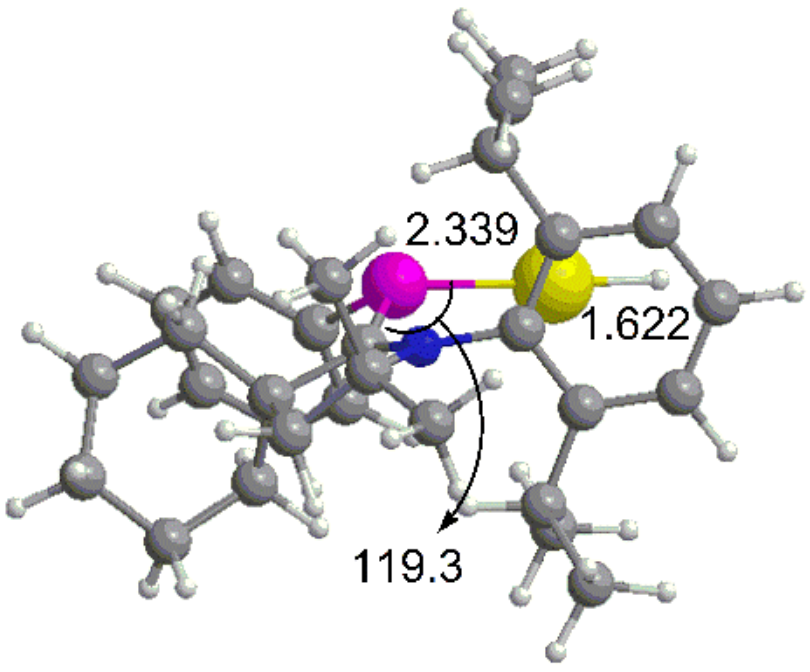

**2-P**

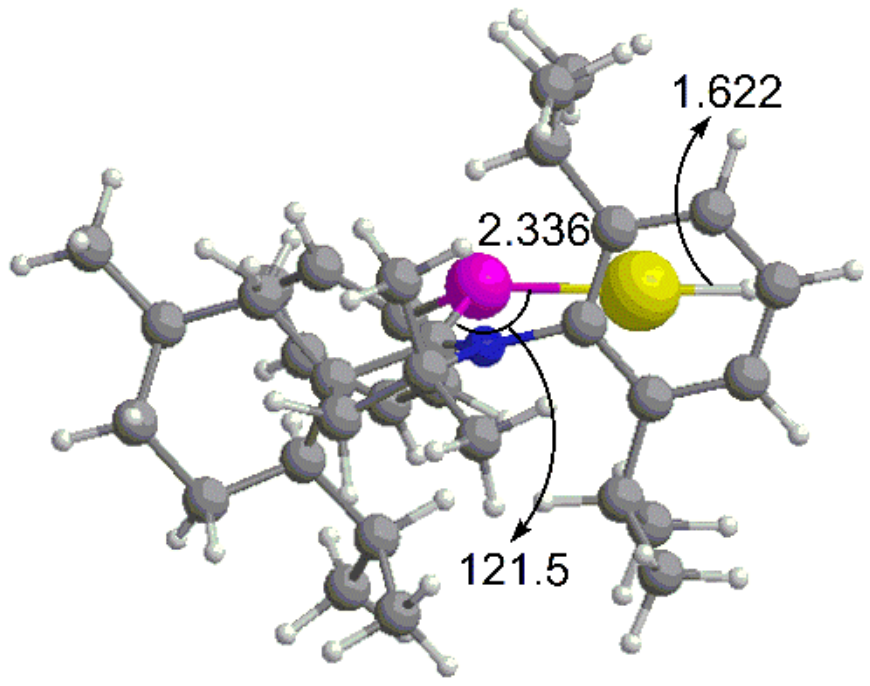

**3-P**

### Unconjugated DACs

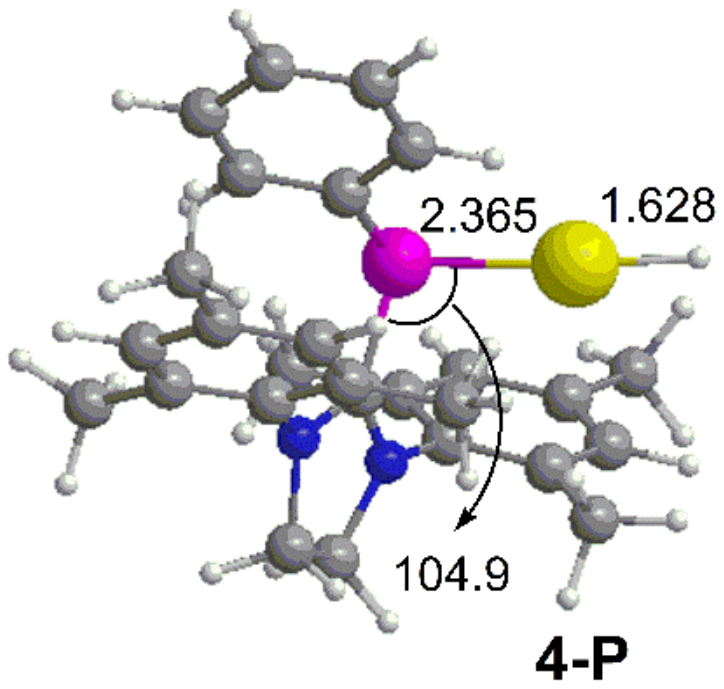

**4-P**

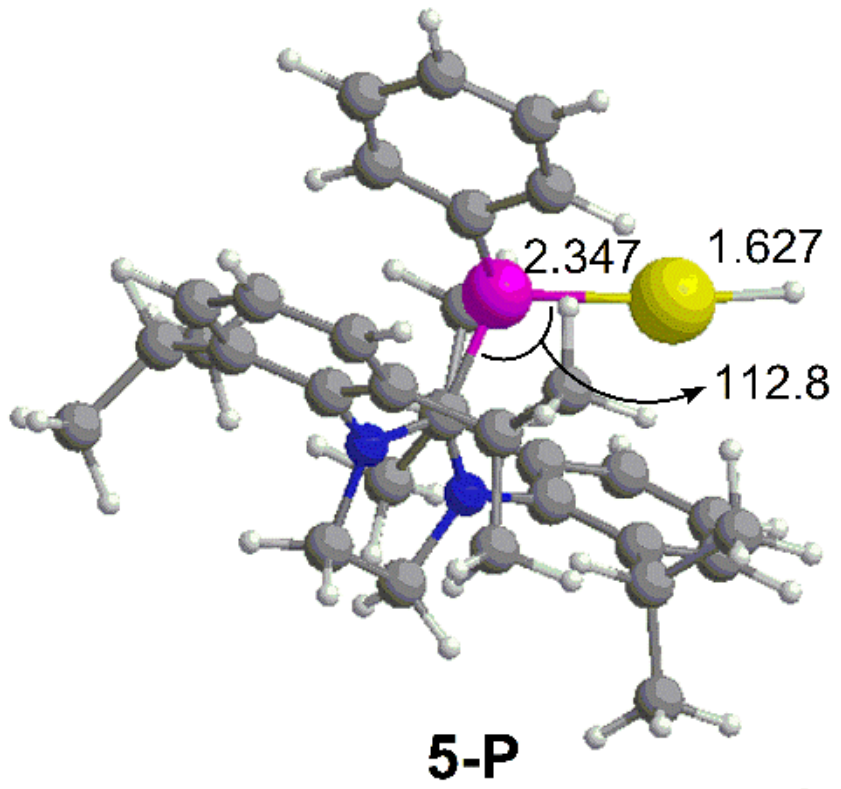

**5-P**

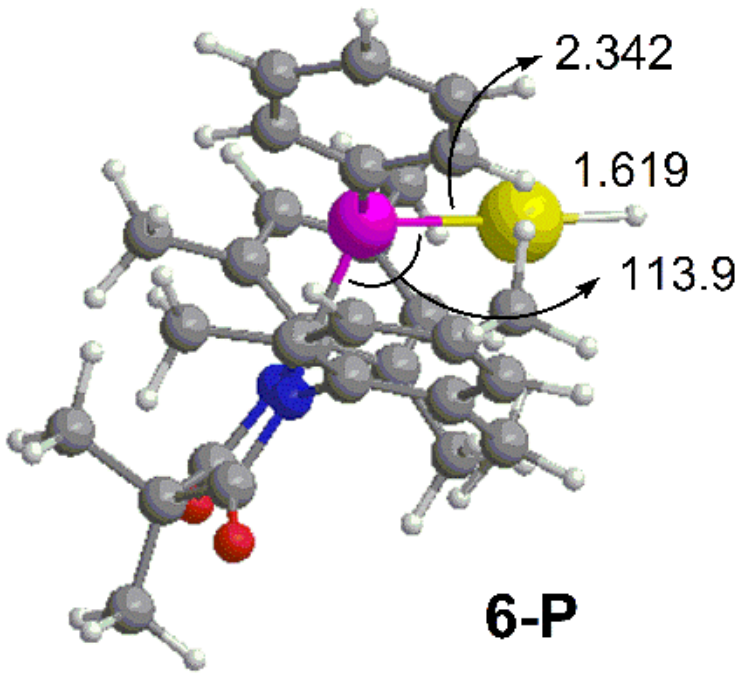

**6-P**

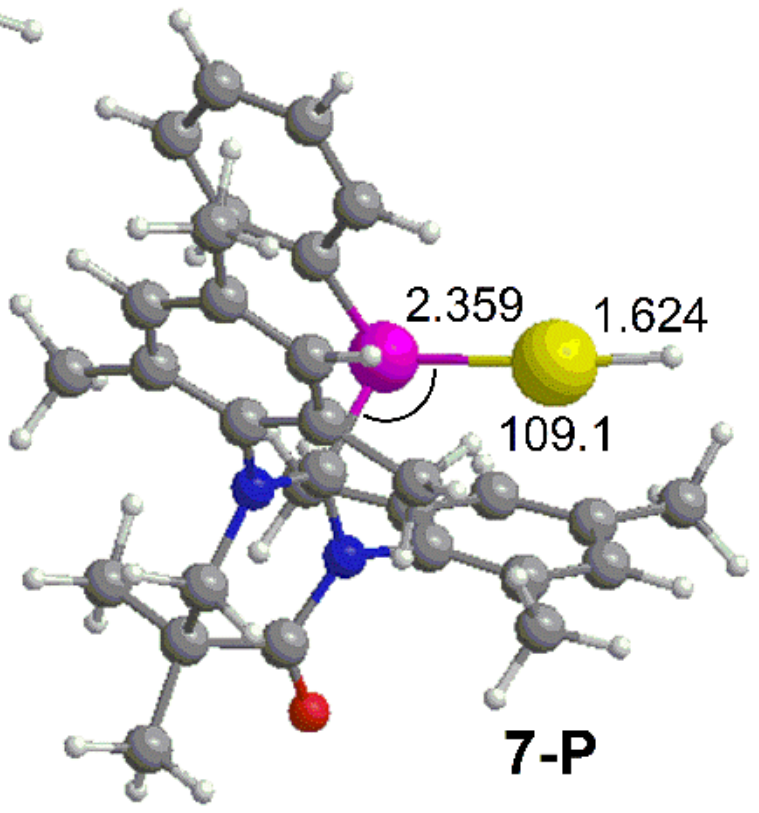

**7-P**

**Conjugated DACs**

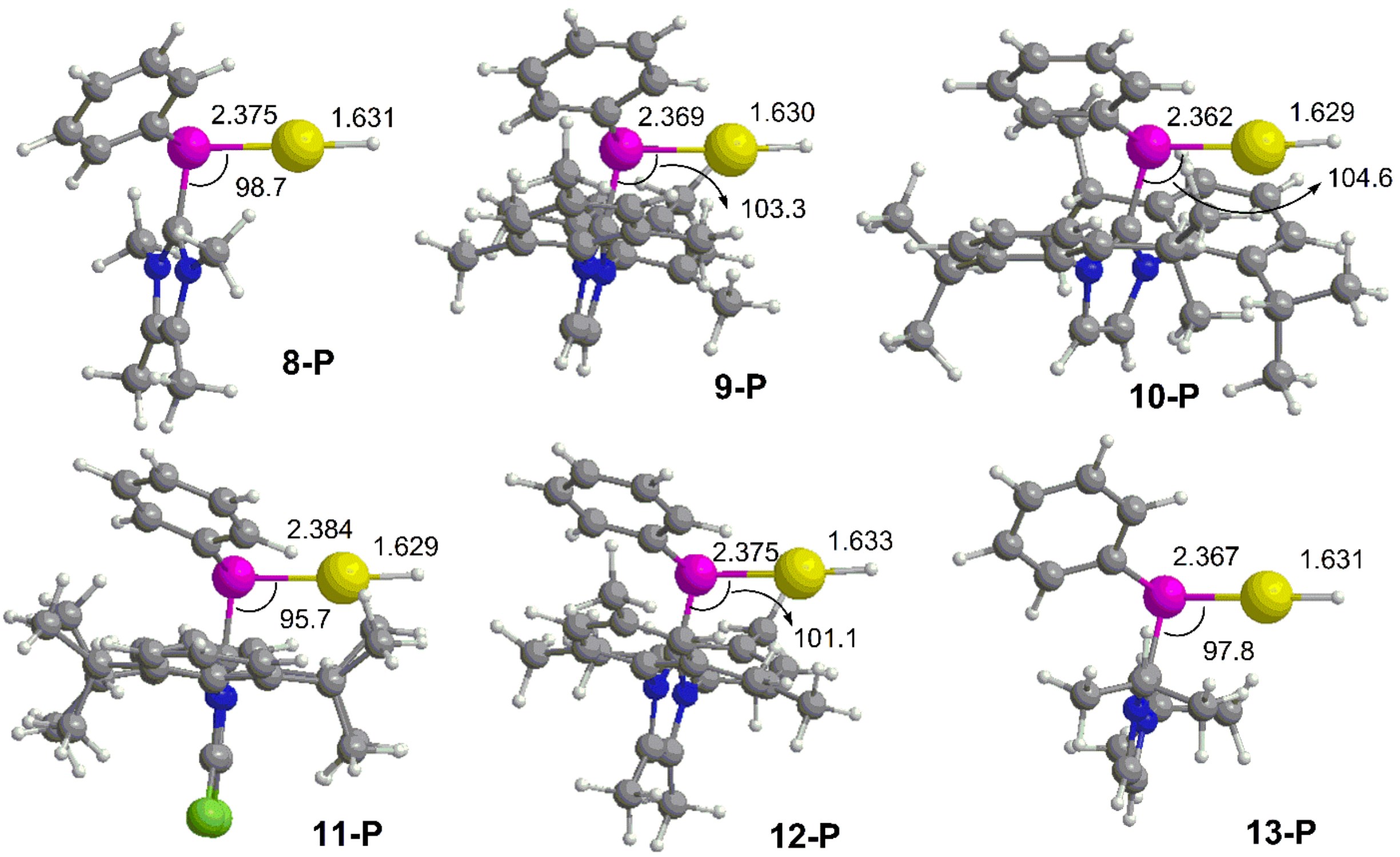

**Figure S1.** Optimized structures of all the **1-P-13-P** [NHCPAuH] complexes. Main geometrical parameters (Au-H and P-Au bond lengths and $C_{NHC}$-P-Au bond angles) are reported (bond in Å, angles in degrees).

## Structural & Mayer bond order (MBO) analysis

| Complex | Au-P Bond Order (Distance, Å) | P-$C_{NHC}$ Bond Order (Distance, Å) |
|---|---|---|
| 1-P | 0.7212 (2.3378) | 1.2358 (1.7473) |
| 2-P | 0.7252 (2.3395) | 1.2413 (1.7502) |
| 3-P | 0.7288 (2.3364) | 1.2326 (1.7511) |
| 4-P | 0.7039 (2.3649) | 1.0649 (1.8016) |
| 5-P | 0.7297 (2.3468) | 1.0850 (1.7917) |
| 6-P | 0.7140 (2.3416) | 1.2058 (1.7637) |
| 7-P | 0.7013 (2.3588) | 1.1150 (1.7972) |
| 8-P | 0.6946 (2.3751) | 0.9922 (1.8182) |
| 9-P | 0.7046 (2.3695) | 1.0265 (1.8077) |
| 10-P | 0.7116 (2.3616) | 1.0298 (1.8044) |
| 11-P | 0.6664 (2.3839) | 1.0032 (1.8092) |
| 12-P | 0.7102 (2.3745) | 1.0122 (1.8128) |
| 13-P | 0.7028 (2.3675) | 0.9877 (1.8199) |

**Table S1**. Mayer bond order analysis (MBO)[12] and interatomic bond distances (Å, shown in parentheses) for Au–P and P–$C_{NHC}$ interactions across complexes **1-P** to **13-P**.

## Regression Analysis

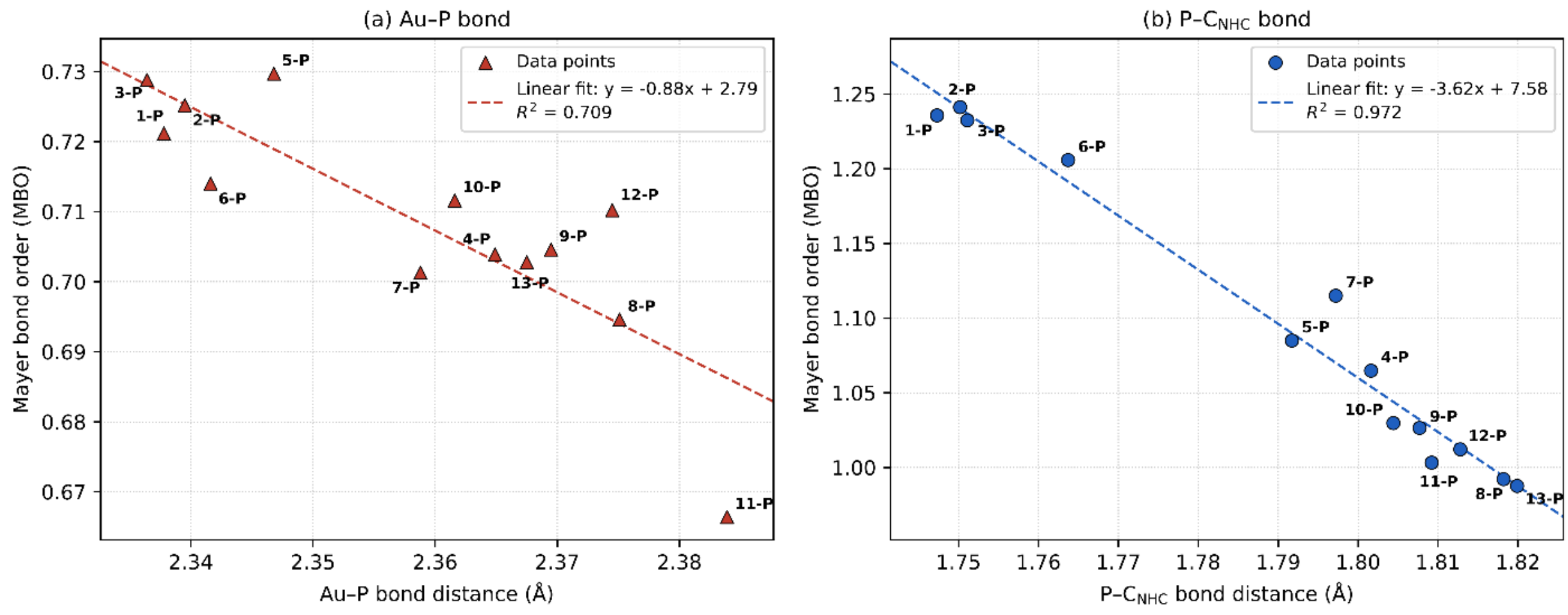

**Figure S2.** Structural correlation and linear regression analysis for complexes **1-P** to **13-P**. (a) Mayer bond order versus Au–P bond distance (Å) fitted with y = -0.88x + 2.79 ($R^2$ = 0.709). (b) Mayer bond order versus P–$C_{NHC}$ bond distance (Å) fitted with y = -3.62x + 7.58 ($R^2$ = 0.972).

| | CAACs | | | Unconjugated DACs | | | | Conjugated DACs | | | | | |
|---|---|---|---|---|---|---|---|---|---|---|---|---|---|
| | 2-P | 3-P | 1-P | 6-P | 7-P | 5-P | 4-P | 10-P | 11-P | 9-P | 12-P | 13-P | 8-P |
| $\Delta E_{Pauli}$ | 126.26 | 129.05 | 125.76 | 116.88 | 123.35 | 137.36 | 130.55 | 134.91 | 130.67 | 133.21 | 140.45 | 131.96 | 130.46 |
| $\Delta E_{elstat}$ | -121.38 | -124.17 | -120.73 | -107.57 | -117.34 | -134.9 | -128.83 | -132.35 | -127.24 | -131.83 | -138.58 | -133.04 | -131.25 |
| $\Delta E_{oi}$ | -74.17 | -77.51 | -72.67 | -74.16 | -75.34 | -77.55 | -72.34 | -78.25 | -79.24 | -72.43 | -77.63 | -66.7 | -64.34 |
| $\Delta E_{\mathrm{int}}$ | -69.29 | -72.63 | -67.64 | -64.86 | -69.33 | -75.09 | -70.62 | -75.7 | -75.81 | -71.04 | -75.75 | -67.79 | -65.13 |

**Table S2.** Energy Decomposition Analysis of the interaction between NHCP and AuH fragments in **1-P-13P** complexes. The overall interaction energy ($\Delta E_{int}$), the Pauli repulsion ($\Delta E_{Pauli}$), electrostatic ($\Delta E_{elstat}$), and orbital interaction ($\Delta E_{oi}$) contributions are reported in kcal/mol.

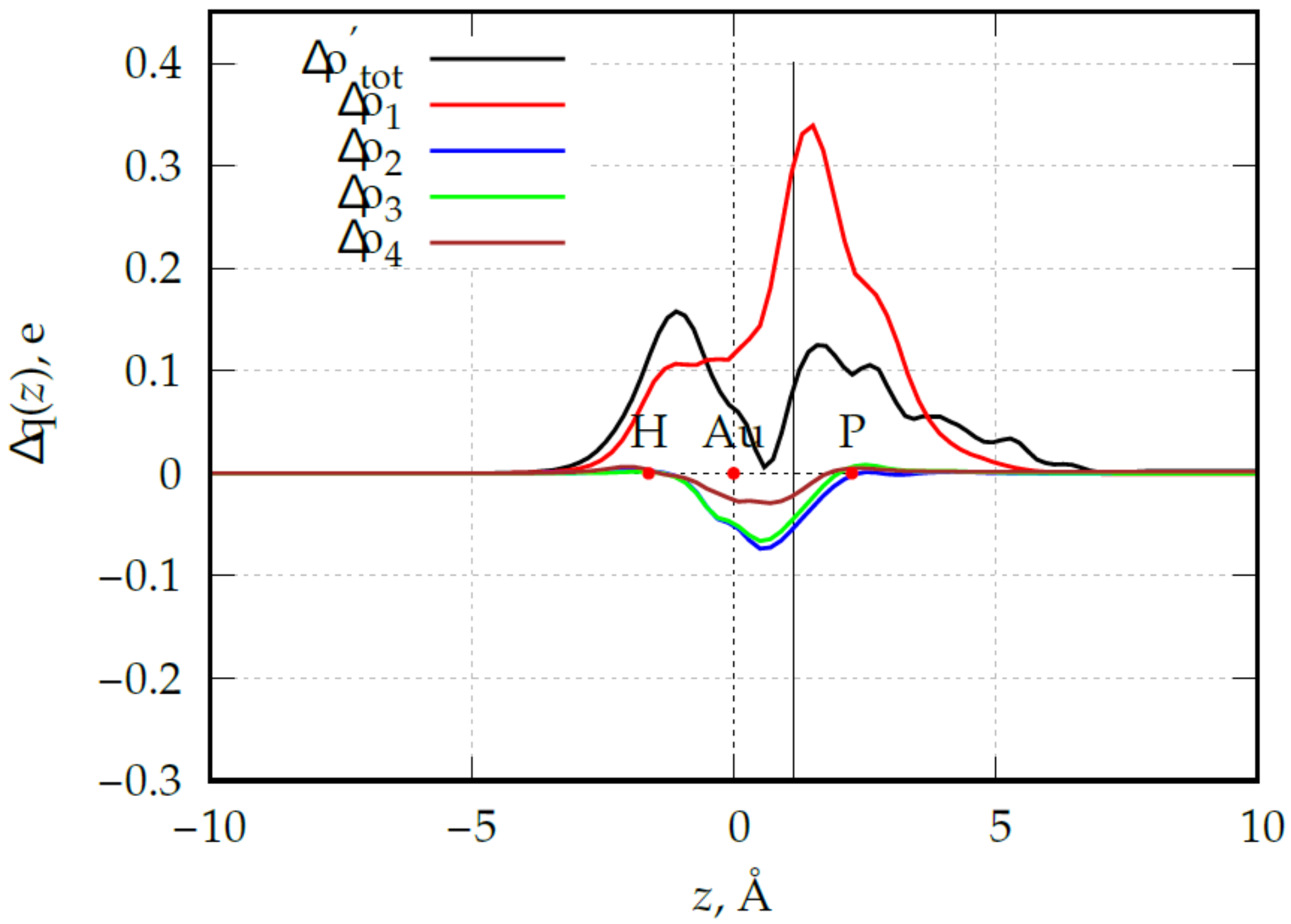

**Figure S3.** The NHCP-AuH bond in complex **1-P**. CD curves associated with the total Δρ' and its components. Red dots indicate the position of the nuclei along the z axis. The grey vertical line marks the isodensity boundary between the NHCP and AuH fragments. Positive (negative) values of the curve indicate right-to-left (left-to-right) charge transfer.

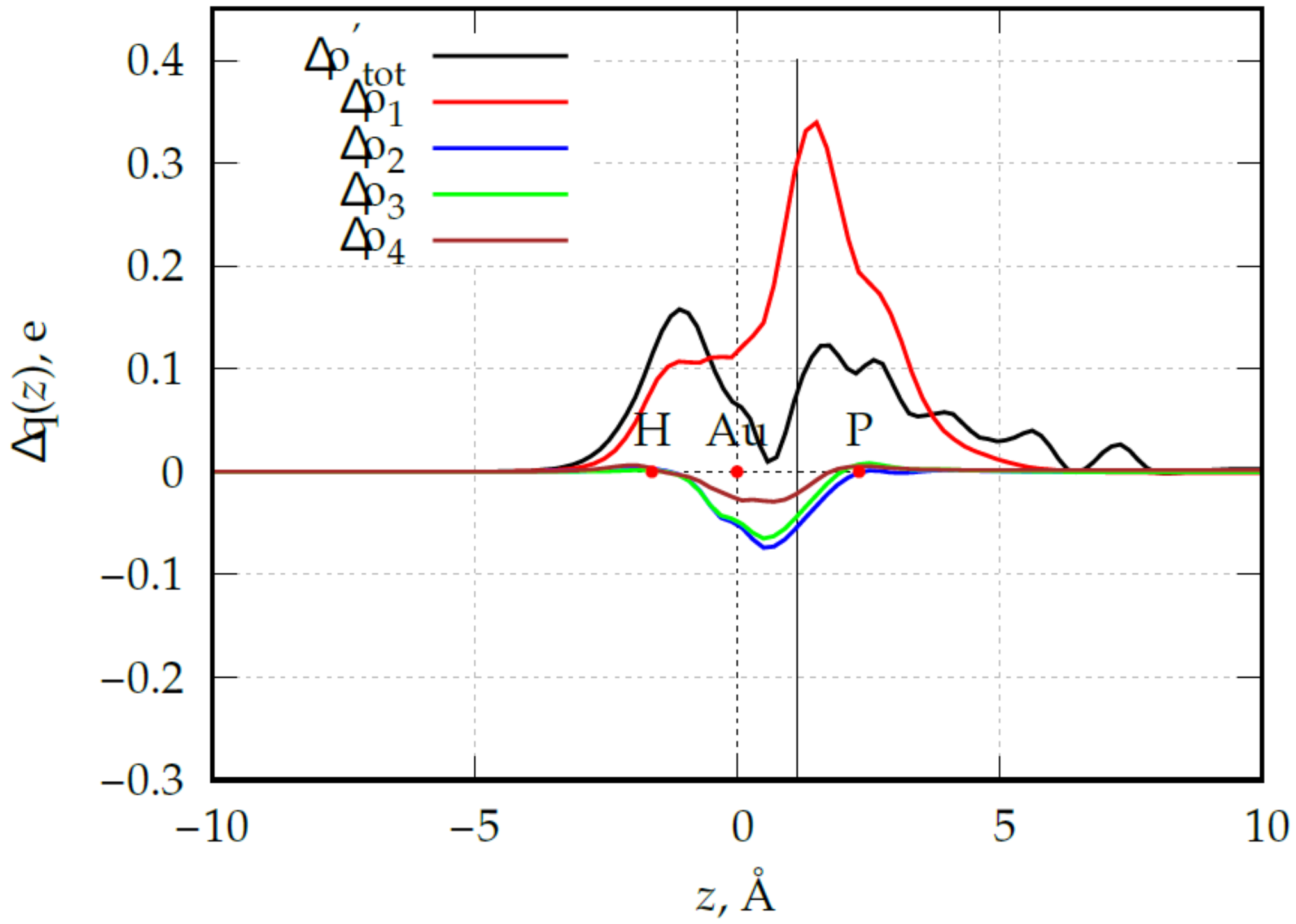

**Figure S4.** The NHCP-AuH bond in complex **2-P**. CD curves associated with the total Δρ' and its components. Red dots indicate the position of the nuclei along the z axis. The grey

vertical line marks the isodensity boundary between the NHCP and AuH fragments. Positive (negative) values of the curve indicate right-to-left (left-to-right) charge transfer.

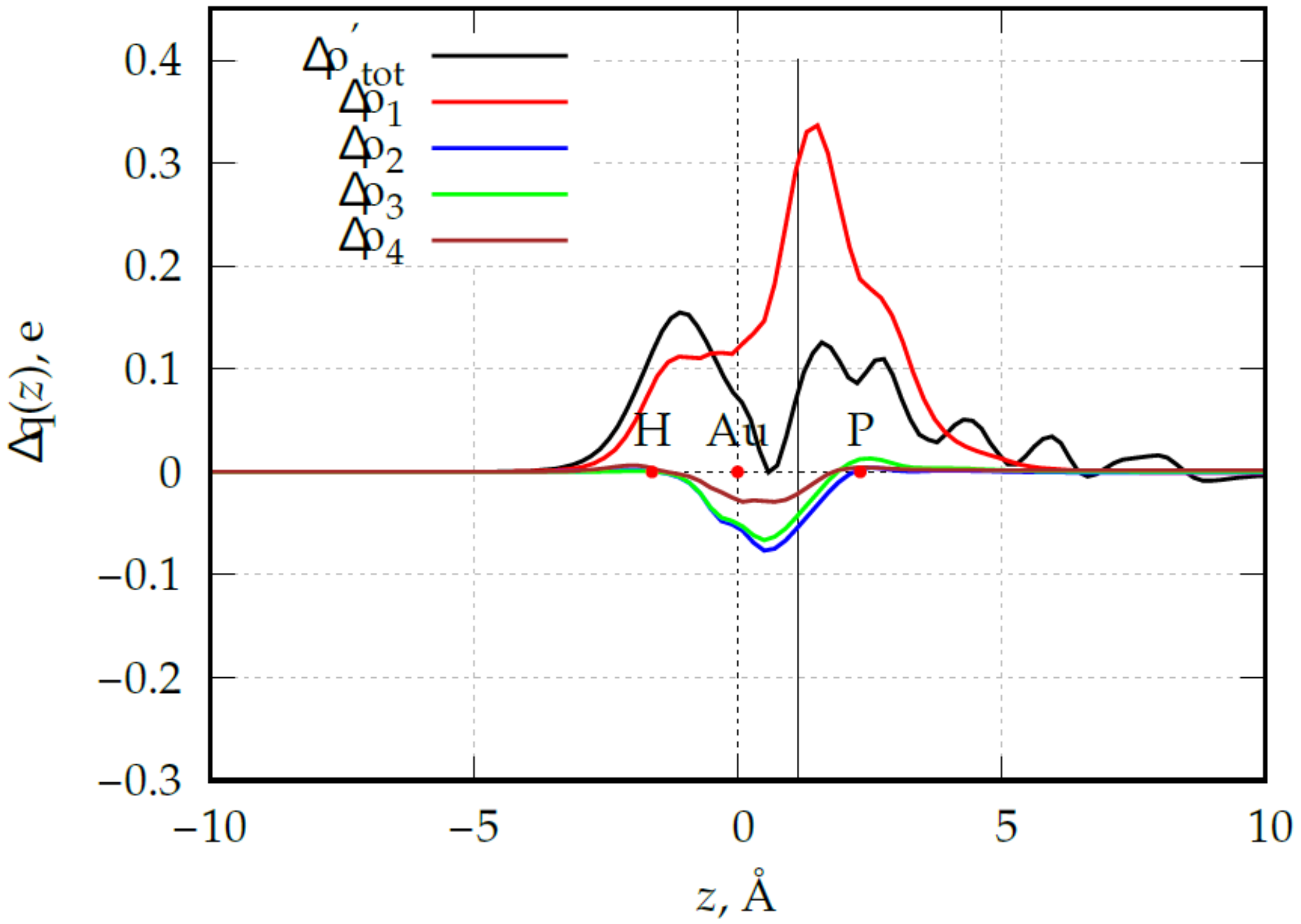

**Figure S5.** The NHCP-AuH bond in complex 3-P. CD curves associated with the total Δρ' and its components. Red dots indicate the position of the nuclei along the z axis. The grey vertical line marks the isodensity boundary between the NHCP and AuH fragments. Positive (negative) values of the curve indicate right-to-left (left-to-right) charge transfer.

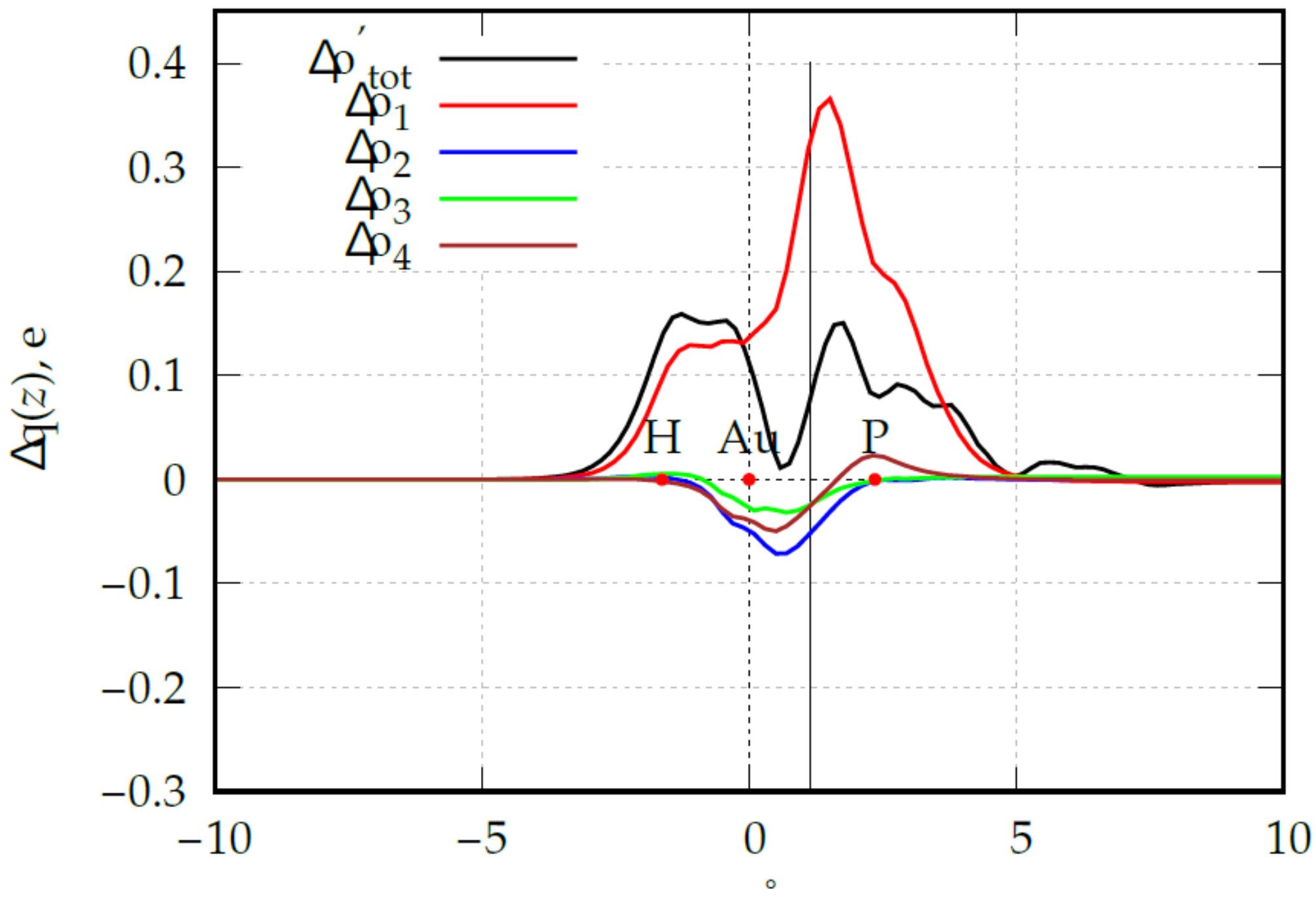

**Figure S6.** The NHCP-AuH bond in complex 4-P. CD curves associated with the total Δρ' and its components. Red dots indicate the position of the nuclei along the z axis. The grey vertical line marks the isodensity boundary between the NHCP and AuH fragments. Positive (negative) values of the curve indicate right-to-left (left-to-right) charge transfer.

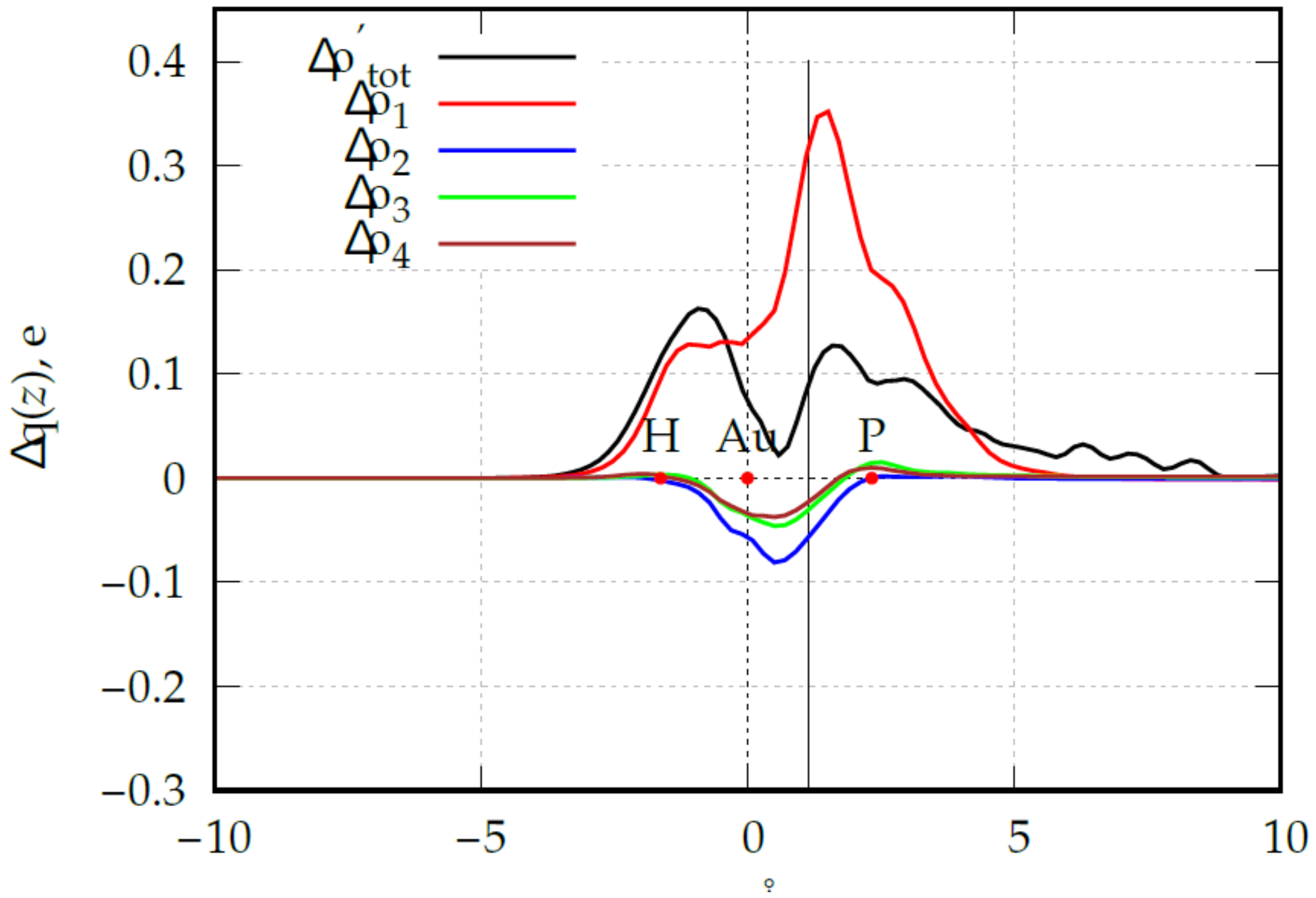

**Figure S7.** The NHCP-AuH bond in complex 5-P. CD curves associated with the total Δρ' and its components. Red dots indicate the position of the nuclei along the z axis. The grey vertical line marks the isodensity boundary between the NHCP and AuH fragments. Positive (negative) values of the curve indicate right-to-left (left-to-right) charge transfer.

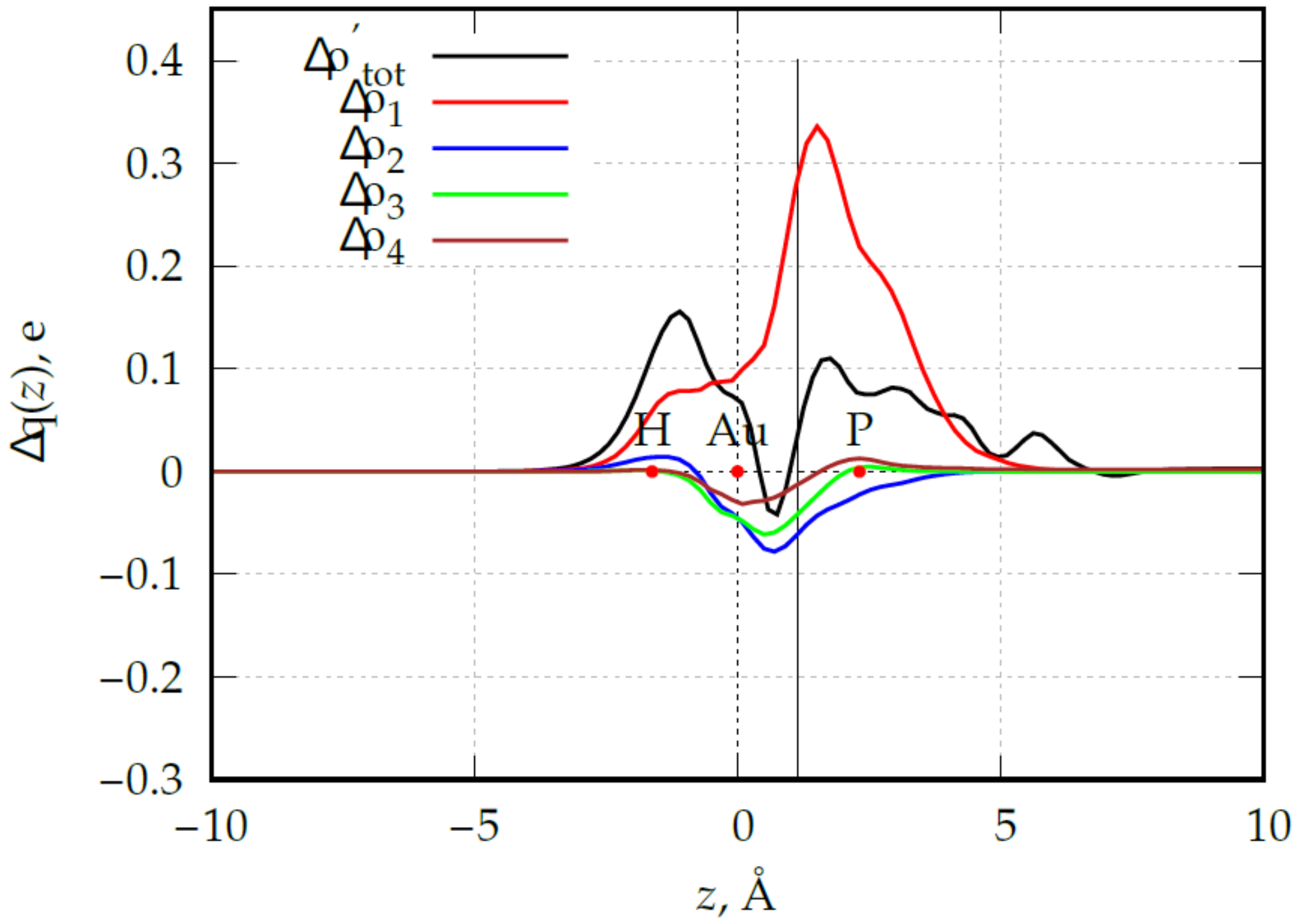

**Figure S8.** The NHCP-AuH bond in complex 6-P. CD curves associated with the total Δρ' and its components. Red dots indicate the position of the nuclei along the z axis. The grey vertical line marks the isodensity boundary between the NHCP and AuH fragments. Positive (negative) values of the curve indicate right-to-left (left-to-right) charge transfer.

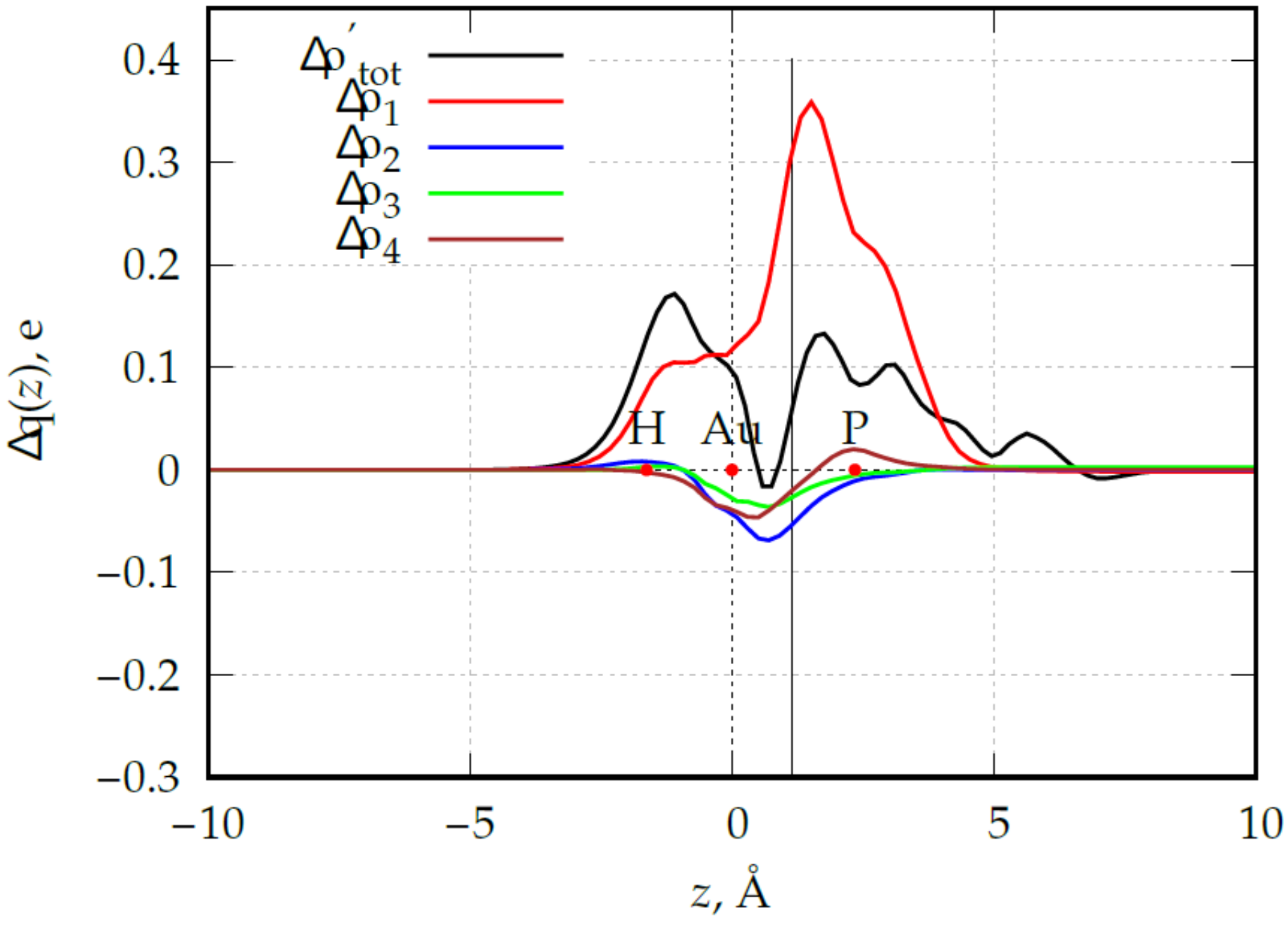

**Figure S9.** The NHCP-AuH bond in complex 7-P. CD curves associated with the total Δρ' and its components. Red dots indicate the position of the nuclei along the z axis. The grey vertical line marks the isodensity boundary between the NHCP and AuH fragments. Positive (negative) values of the curve indicate right-to-left (left-to-right) charge transfer.

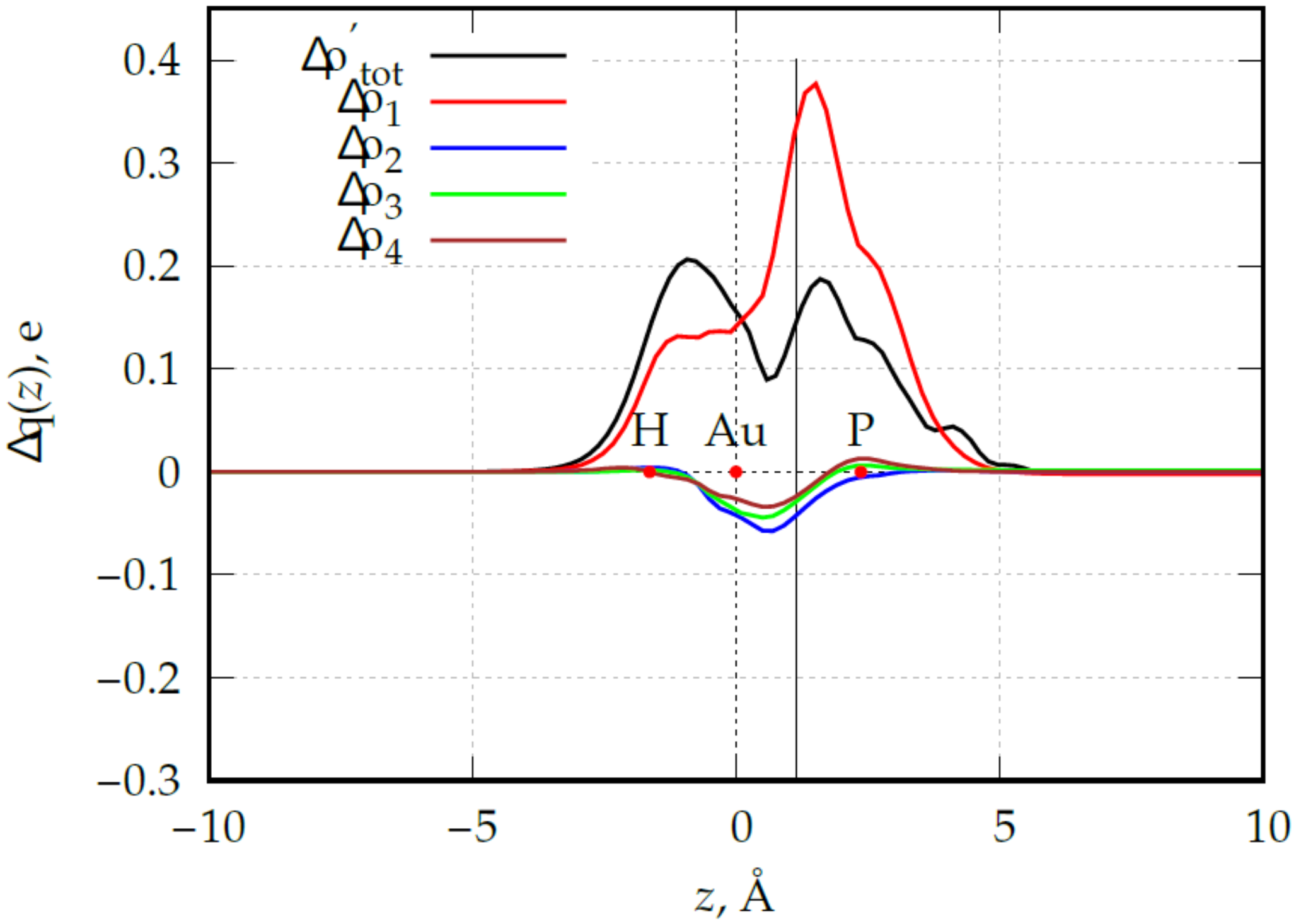

**Figure S10.** The NHCP-AuH bond in complex 8-P. CD curves associated with the total Δρ' and its components. Red dots indicate the position of the nuclei along the z axis. The grey vertical line marks the isodensity boundary between the NHCP and AuH fragments. Positive (negative) values of the curve indicate right-to-left (left-to-right) charge transfer.

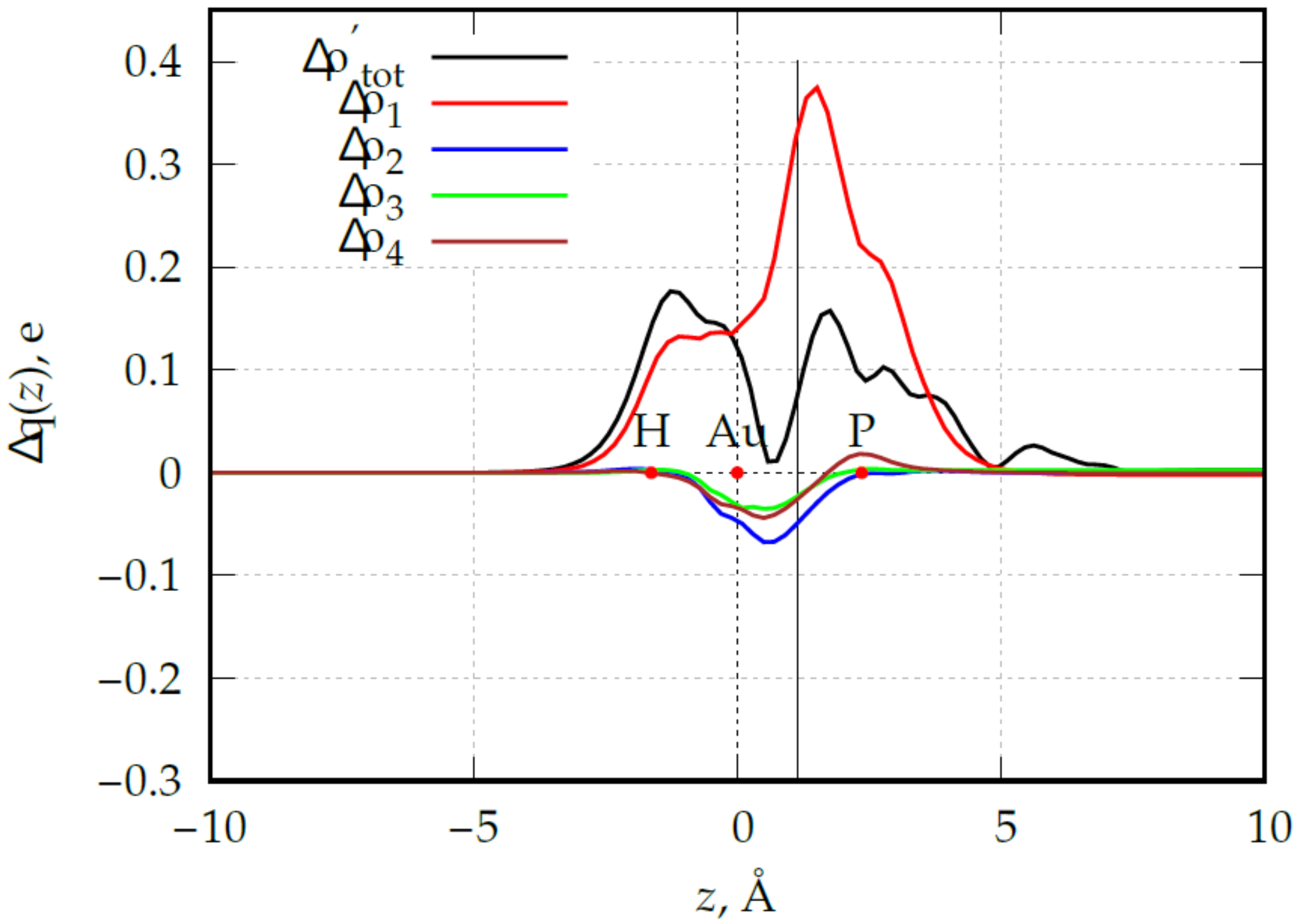

**Figure S11.** The NHCP-AuH bond in complex 9-P. CD curves associated with the total Δρ' and its components. Red dots indicate the position of the nuclei along the z axis. The grey vertical line marks the isodensity boundary between the NHCP and AuH fragments. Positive (negative) values of the curve indicate right-to-left (left-to-right) charge transfer.

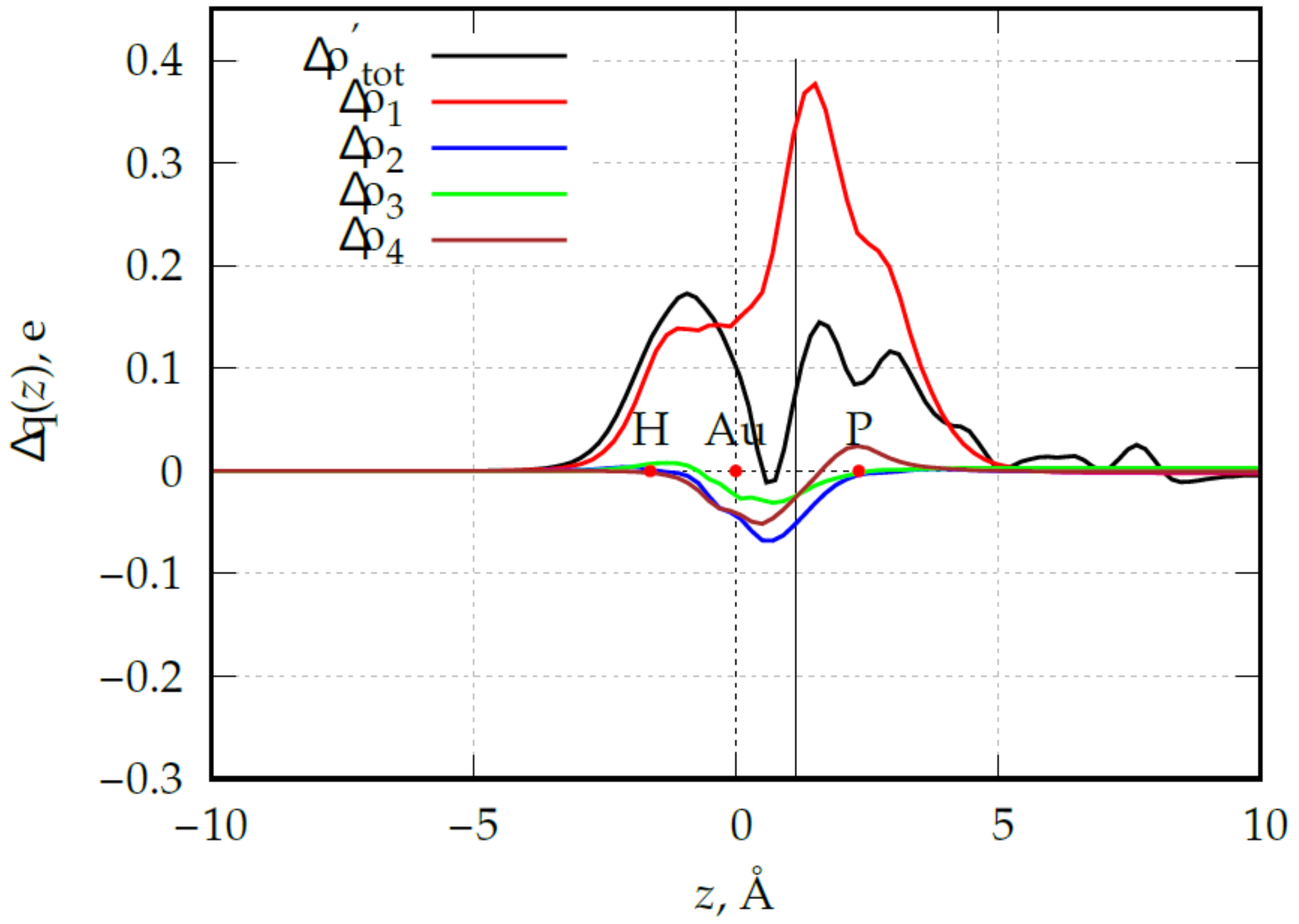

**Figure S12.** The NHCP-AuH bond in complex 10-P. CD curves associated with the total Δρ' and its components. Red dots indicate the position of the nuclei along the z axis. The grey vertical line marks the isodensity boundary between the NHCP and AuH fragments. Positive (negative) values of the curve indicate right-to-left (left-to-right) charge transfer.

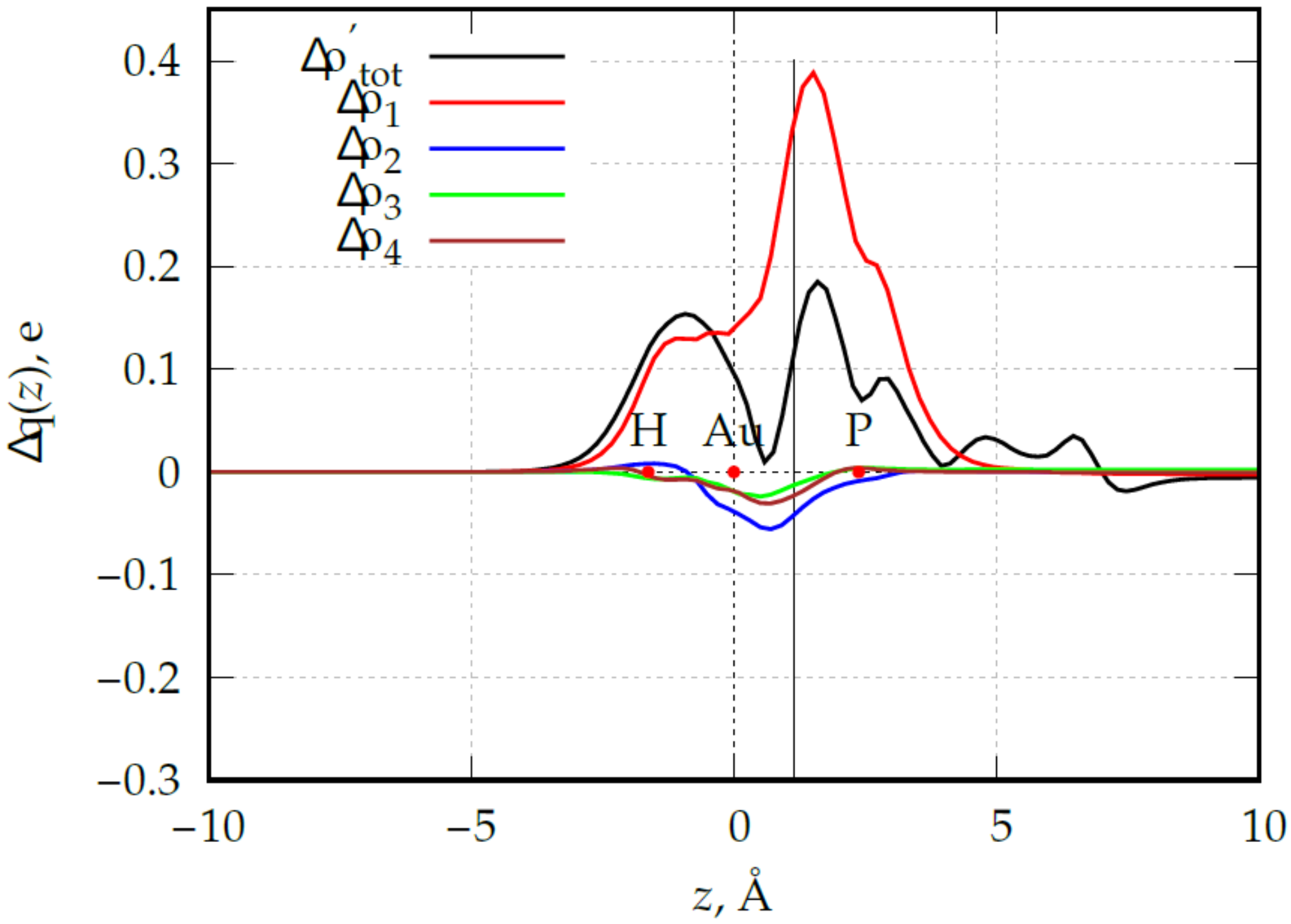

**Figure S13.** The NHCP-AuH bond in complex 11-P. CD curves associated with the total Δρ' and its components. Red dots indicate the position of the nuclei along the z axis. The grey vertical line marks the isodensity boundary between the NHCP and AuH fragments. Positive (negative) values of the curve indicate right-to-left (left-to-right) charge transfer.

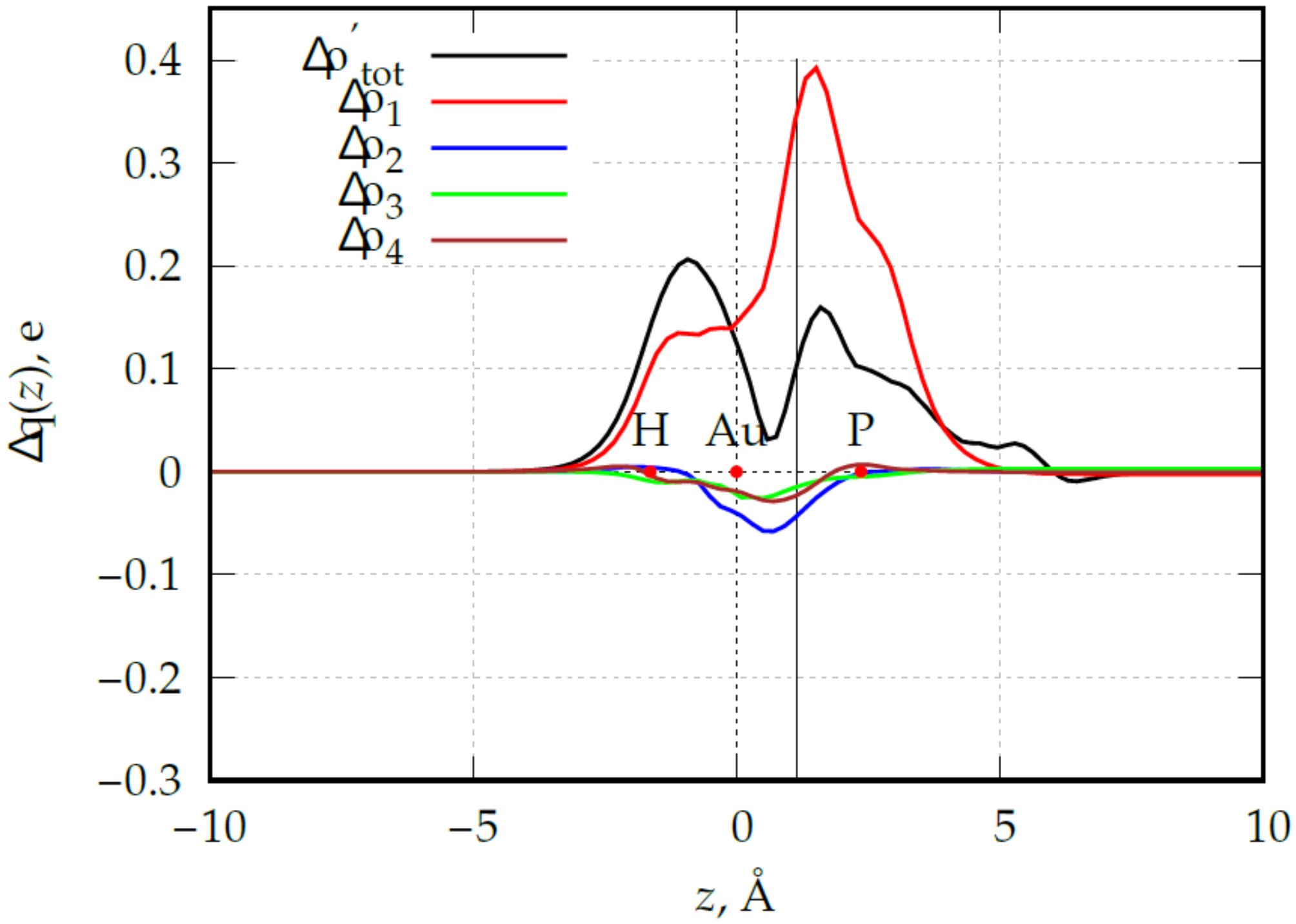

**Figure S14.** The NHCP-AuH bond in complex 12-P. CD curves associated with the total Δρ' and its components. Red dots indicate the position of the nuclei along the z axis. The grey vertical line marks the isodensity boundary between the NHCP and AuH fragments. Positive (negative) values of the curve indicate right-to-left (left-to-right) charge transfer.

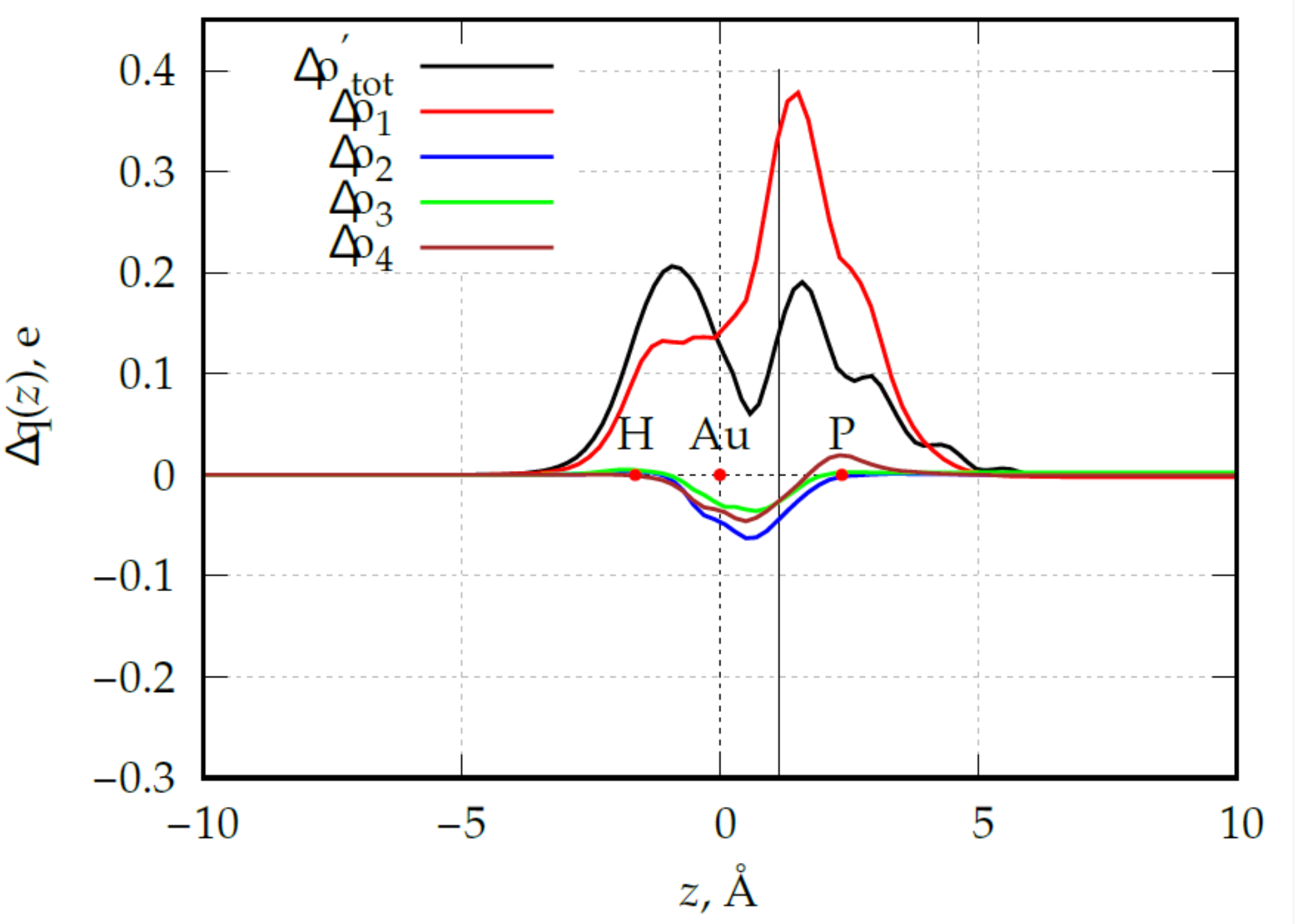

**Figure S15.** The NHCP-AuH bond in complex 13-P. CD curves associated with the total Δρ' and its components. Red dots indicate the position of the nuclei along the z axis. The grey vertical line marks the isodensity boundary between the NHCP and AuH fragments. Positive (negative) values of the curve indicate right-to-left (left-to-right) charge transfer.

| | CAACs | | | Unconjugated DACs | | | | Conjugated DACs | | | | | |
|---|---|---|---|---|---|---|---|---|---|---|---|---|---|
| NOCV Energies (kcal/mol) | 2-P | 3-P | 1-P | 6-P | 7-P | 5-P | 4-P | 10-P | 11-P | 9-P | 12-P | 13-P | 8-P |
| Δε1 | -33.09492 | -32.96762 | -33.0665 | -32.73126 | -34.74681 | -34.85444 | -35.68723 | -36.92736 | -37.06512 | -36.49338 | -38.10722 | -37.26861 | -37.23901 |
| Δε 2 | -7.66175 | -7.62923 | -7.61038 | -9.36475 | -7.77234 | -7.39371 | -6.91126 | -7.70532 | -6.80092 | -6.69023 | -6.36042 | -6.02236 | -5.59301 |
| Δε 3 | -6.26104 | -6.33147 | -6.37028 | -5.93782 | -4.67225 | -5.4972 | -4.97407 | -4.68881 | -4.48199 | -4.99037 | -4.94093 | -5.16877 | -4.92767 |
| Δε 4 | -5.50056 | -5.6567 | -5.51737 | -4.78901 | -4.91884 | -5.53114 | -4.75342 | -4.92806 | -4.32597 | -4.56852 | -4.72043 | -4.33432 | -4.49302 |
| Δε 5 | -2.15096 | -2.36153 | -2.17251 | -2.59199 | -3.76674 | -2.64253 | -2.82338 | -3.26198 | -4.04815 | -2.97621 | -4.05559 | -2.27279 | -2.18693 |
| Δε 6 | -1.24076 | -1.37682 | -1.25847 | -1.16922 | -1.36453 | -1.35783 | -1.13845 | -1.22927 | -1.61988 | -1.20578 | -1.89038 | -1.1164 | -1.21081 |

**Table S3.** Orbital interaction energy contributions (Δε) associated to the first 6 NOCV components for **1-P-13P** complexes.

| | CAACs | | | Unconjugated DACs | | | | Conjugated DACs | | | | | |
|---|---|---|---|---|---|---|---|---|---|---|---|---|---|
| NOCV | 2-P | 3-P | 1-P | 6-P | 7-P | 5-P | 4-P | 10-P | 11-P | 9-P | 12-P | 13-P | 8-P |
| **Isovalue** | 2.168164 | 2.166016 | 2.168164 | 2.19082 | 2.189844 | 2.170703 | 2.18125 | 2.182227 | 2.201953 | 2.182227 | 2.186133 | 2.182227 | 2.186914 |
| 1 | 0.300127 | 0.299859 | 0.29963 | 0.287022 | 0.311486 | 0.316644 | 0.325102 | 0.336987 | 0.343126 | 0.333418 | 0.349659 | 0.338703 | 0.3373 |
| 2 | -0.053826 | -0.053761 | -0.053474 | -0.060349 | -0.053093 | -0.056479 | -0.051445 | -0.050992 | -0.041301 | -0.048609 | -0.042458 | -0.043709 | -0.041851 |
| 3 | -0.043353 | -0.042234 | -0.044291 | -0.040963 | -0.026554 | -0.030402 | -0.023925 | -0.02407 | -0.012506 | -0.022615 | -0.014402 | -0.026445 | -0.02817 |
| 4 | -0.021257 | -0.021387 | -0.021421 | -0.01248 | -0.019813 | -0.02292 | -0.024961 | -0.02536 | -0.022512 | -0.025256 | -0.022537 | -0.025816 | -0.023284 |
| 5 | -0.011864 | -0.009134 | -0.011408 | -0.016506 | -0.019314 | -0.006666 | -0.012845 | -0.012266 | -0.021492 | -0.012601 | -0.021193 | -0.008027 | -0.013832 |
| 6 | 0.011789 | 0.011857 | 0.011815 | 0.005363 | 0.004928 | 0.007098 | 0.006765 | 0.007178 | 0.005762 | 0.006091 | -0.007011 | 0.0093 | 0.007721 |

**Table S4.** Charge transfer values (CT) calculated at the isodensity boundary associated to the first 6 NOCV components for **1-P-13P** complexes. Values are in electrons (e).

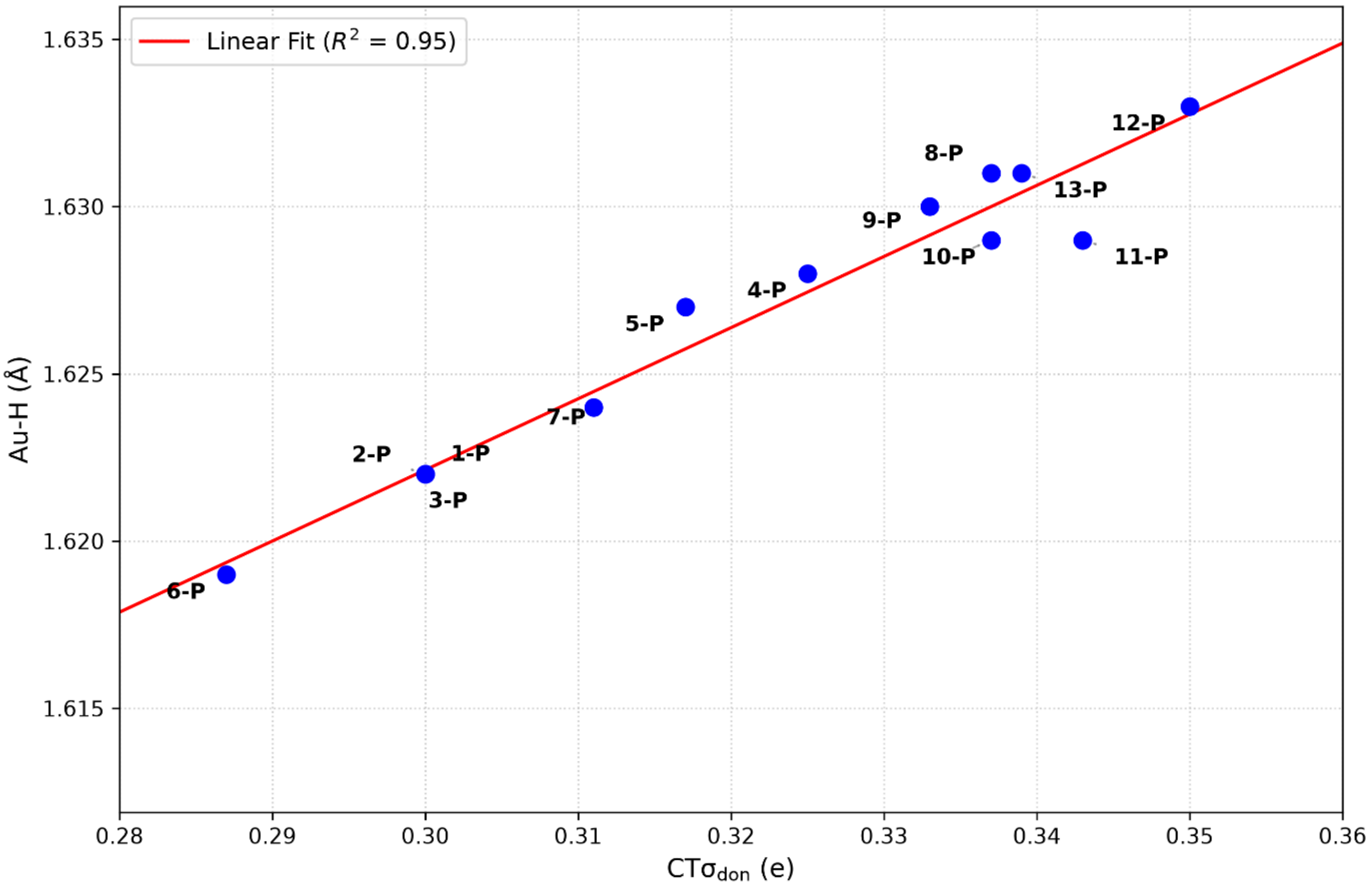

Linear Fit ($R^2$ = 0.95)
1.635
1.630
1.625
1.620
1.615
Au-H (Å)
0.28
0.29
0.30
0.31
0.32
0.33
0.34
0.35
0.36
$CT\sigma_{don}$ (e)
1-P
2-P
3-P
4-P
5-P
6-P
7-P
8-P
9-P
10-P
11-P
12-P
13-P

**Figure S16.** Correlation plot between CT $_{\sigma\ don}$ and Au-H bond distance for the complete **1-P-13-P** series.

| | CAACs | | | Unconjugated DACs | | | | Conjugated DACs | | | | | |
|---|---|---|---|---|---|---|---|---|---|---|---|---|---|
| | 1 | 2 | 3 | 5 | 4 | 6 | 7 | 11 | 9 | 10 | 12 | 8 | 13 |
| $\Delta E_{Pauli}$ | 188.43 | 189.75 | 189.99 | 175 | 169.44 | 171.67 | 171.84 | 167.17 | 168.16 | 174.42 | 170.66 | 172.72 | 171.6 |
| $\Delta E_{elstat}$ | -197.43 | -198.84 | -198.8 | -182.65 | -179 | -175.34 | -179.07 | -173.19 | -178.34 | -183.02 | -181.56 | -185.29 | -183.26 |
| $\Delta E_{oi}$ | -78.87 | -80.6 | -83.98 | -79.74 | -70.75 | -76.34 | -74.7 | -78.47 | -69.53 | -78.7 | -71.26 | -63.45 | -65.31 |
| $\Delta E_{\mathrm{int}}$ | -87.87 | -89.7 | -92.79 | -87.39 | -80.31 | -80.01 | -81.92 | -84.5 | -79.7 | -87.31 | -82.16 | -76.02 | -76.98 |

**Table S5.** Energy Decomposition Analysis of the interaction between NHC and AuH fragments in **1-13** complexes. The overall interaction energy ($\Delta E_{\mathrm{int}}$), the Pauli repulsion ($\Delta E_{Pauli}$), electrostatic ($\Delta E_{elstat}$), and orbital interaction ($\Delta E_{oi}$) contributions are reported in kcal/mol.

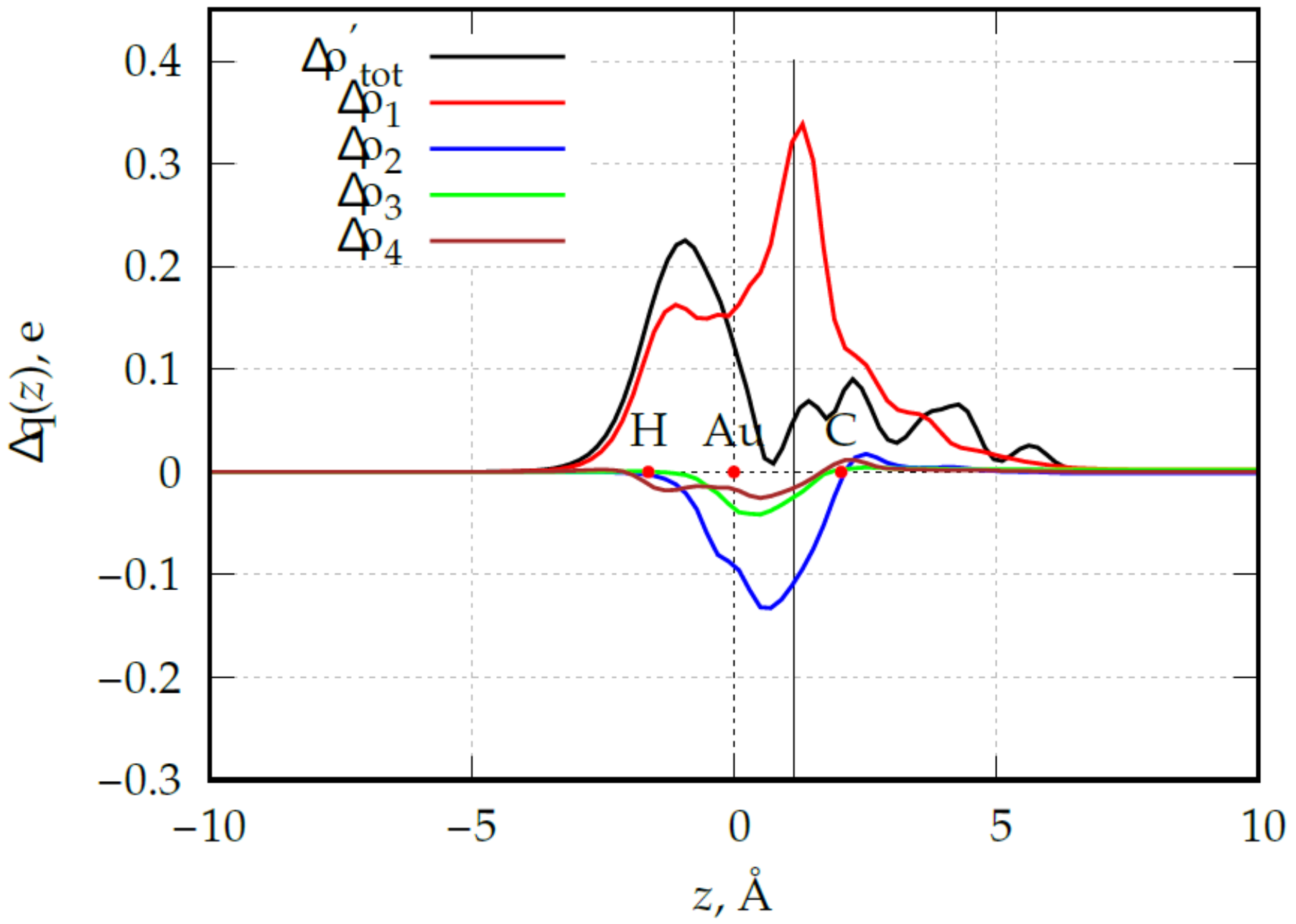

**Figure S17.** The NHC-AuH bond in complex **1**. CD curves associated with the total Δρ' and its components. Red dots indicate the position of the nuclei along the z axis. The grey vertical line marks the isodensity boundary between the NHC and AuH fragments. Positive (negative) values of the curve indicate right-to-left (left-to-right) charge transfer.

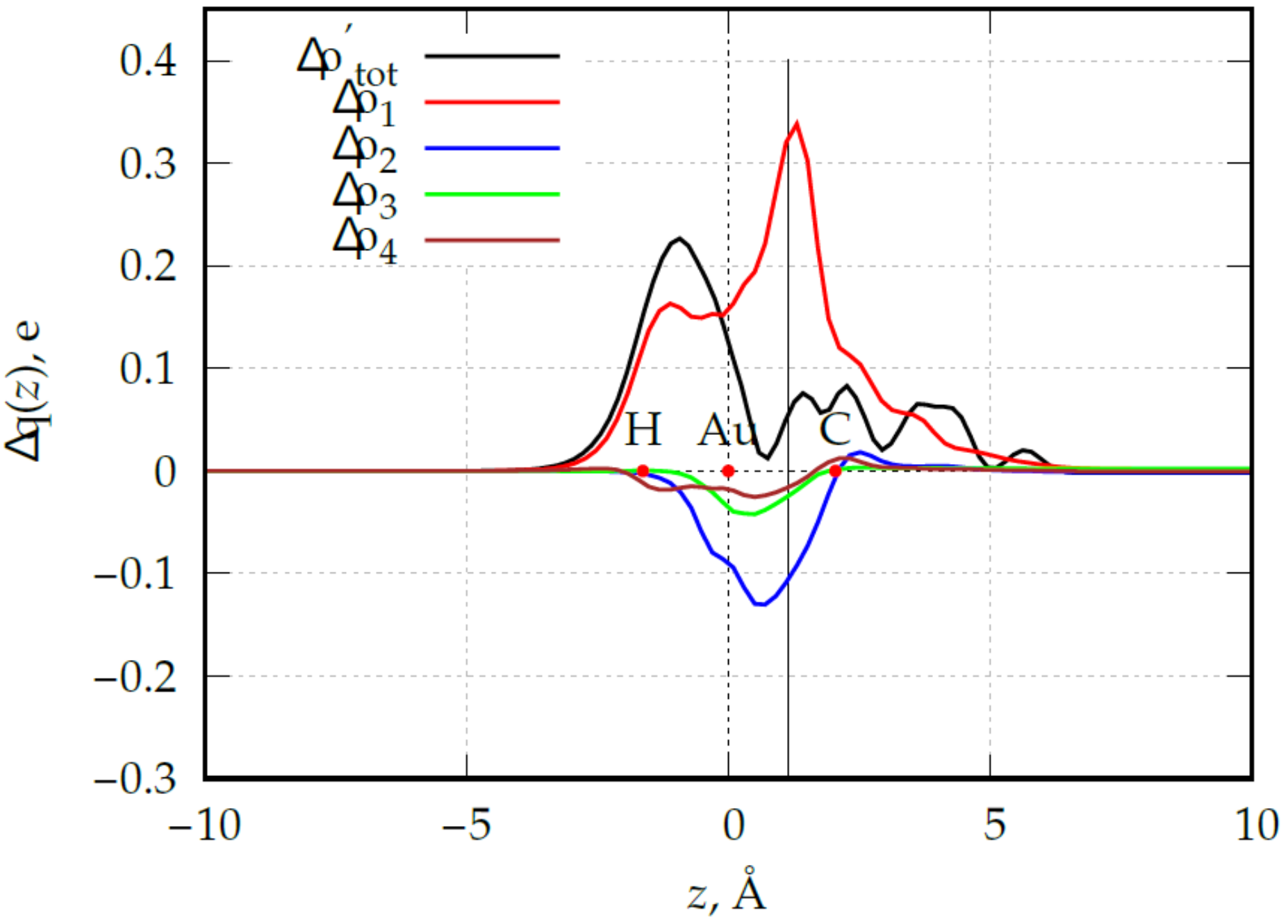

**Figure S18.** The NHC-AuH bond in complex **2**. CD curves associated with the total Δρ' and its components. Red dots indicate the position of the nuclei along the z axis. The grey vertical line marks the isodensity boundary between the NHC and AuH fragments. Positive (negative) values of the curve indicate right-to-left (left-to-right) charge transfer.

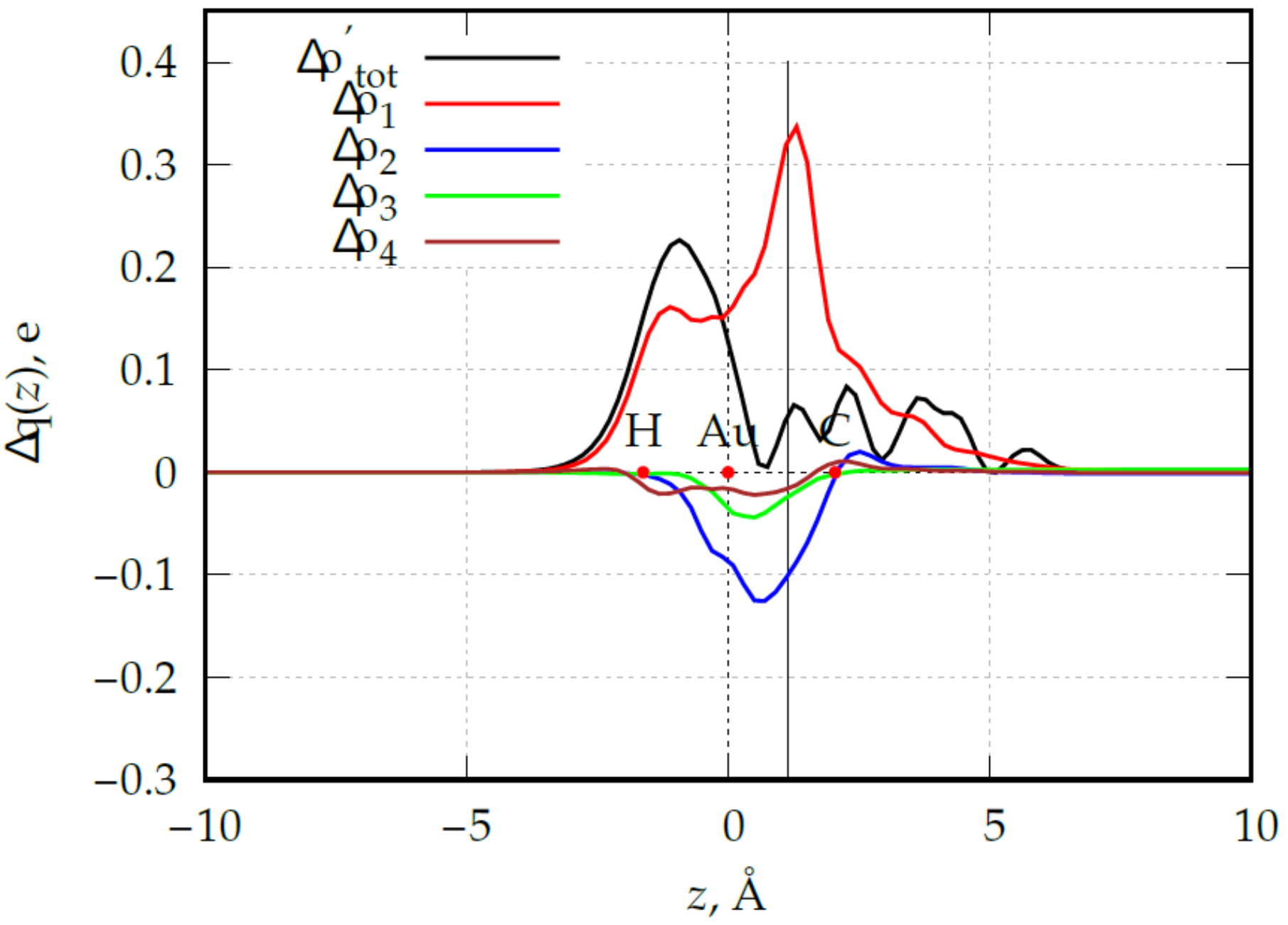

**Figure S19.** The NHC-AuH bond in complex **3**. CD curves associated with the total Δρ' and its components. Red dots indicate the position of the nuclei along the z axis. The grey vertical line marks the isodensity boundary between the NHC and AuH fragments. Positive (negative) values of the curve indicate right-to-left (left-to-right) charge transfer.

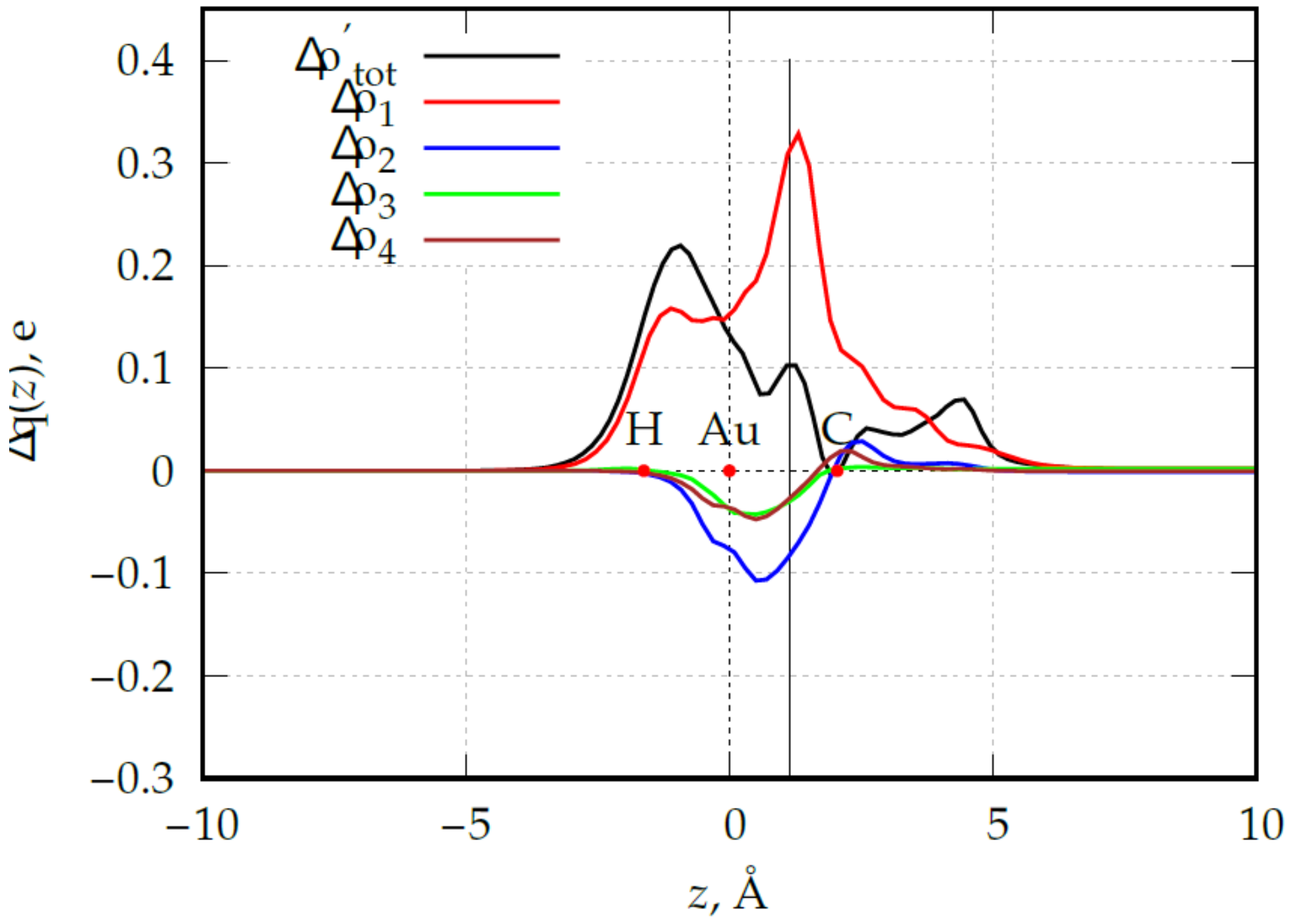

**Figure S20.** The NHC-AuH bond in complex **4**. CD curves associated with the total Δρ' and its components. Red dots indicate the position of the nuclei along the z axis. The grey vertical line marks the isodensity boundary between the NHC and AuH fragments. Positive (negative) values of the curve indicate right-to-left (left-to-right) charge transfer.

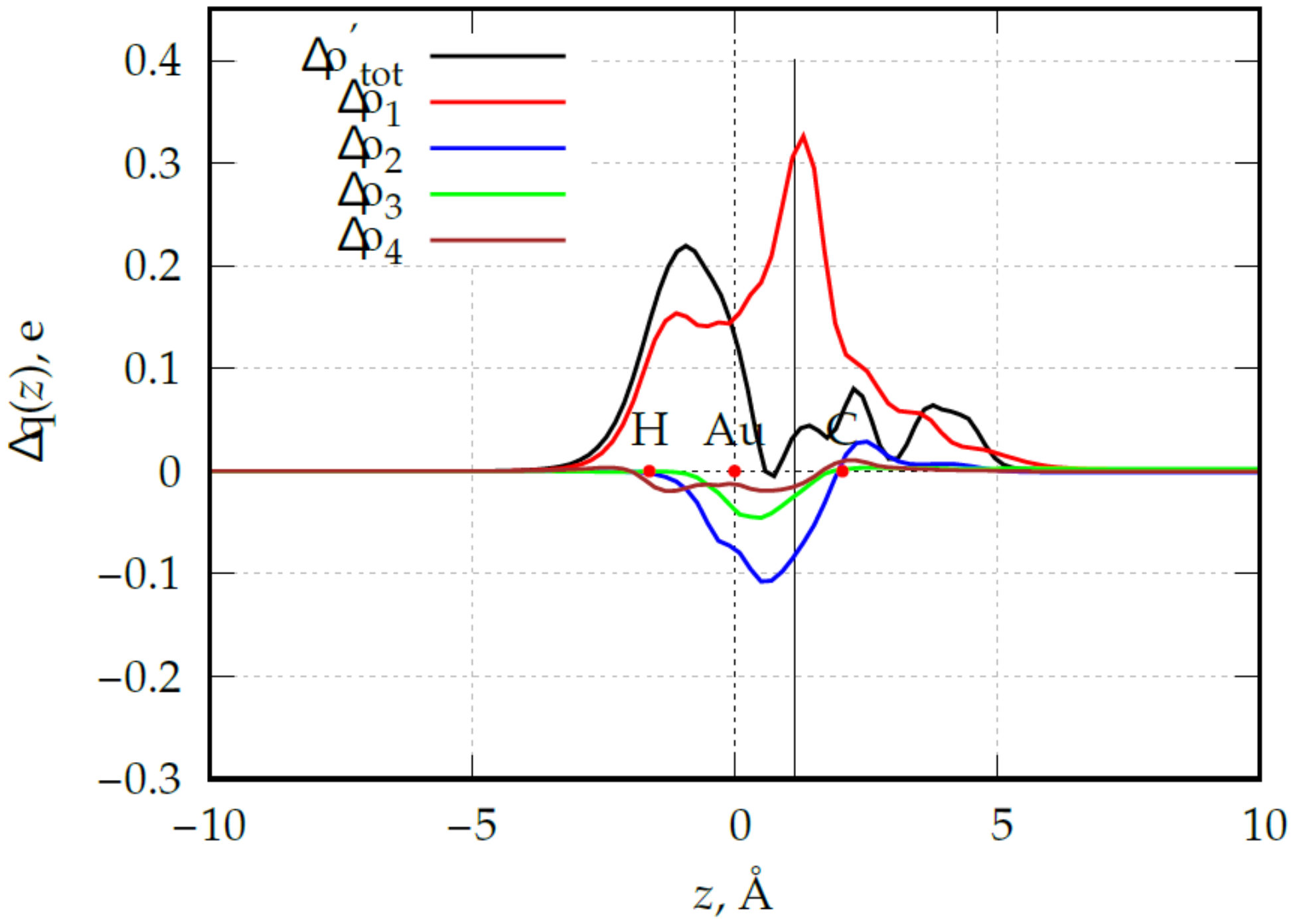

**Figure S21.** The NHC-AuH bond in complex **5**. CD curves associated with the total Δρ' and its components. Red dots indicate the position of the nuclei along the z axis. The grey vertical line marks the isodensity boundary between the NHC and AuH fragments. Positive (negative) values of the curve indicate right-to-left (left-to-right) charge transfer.

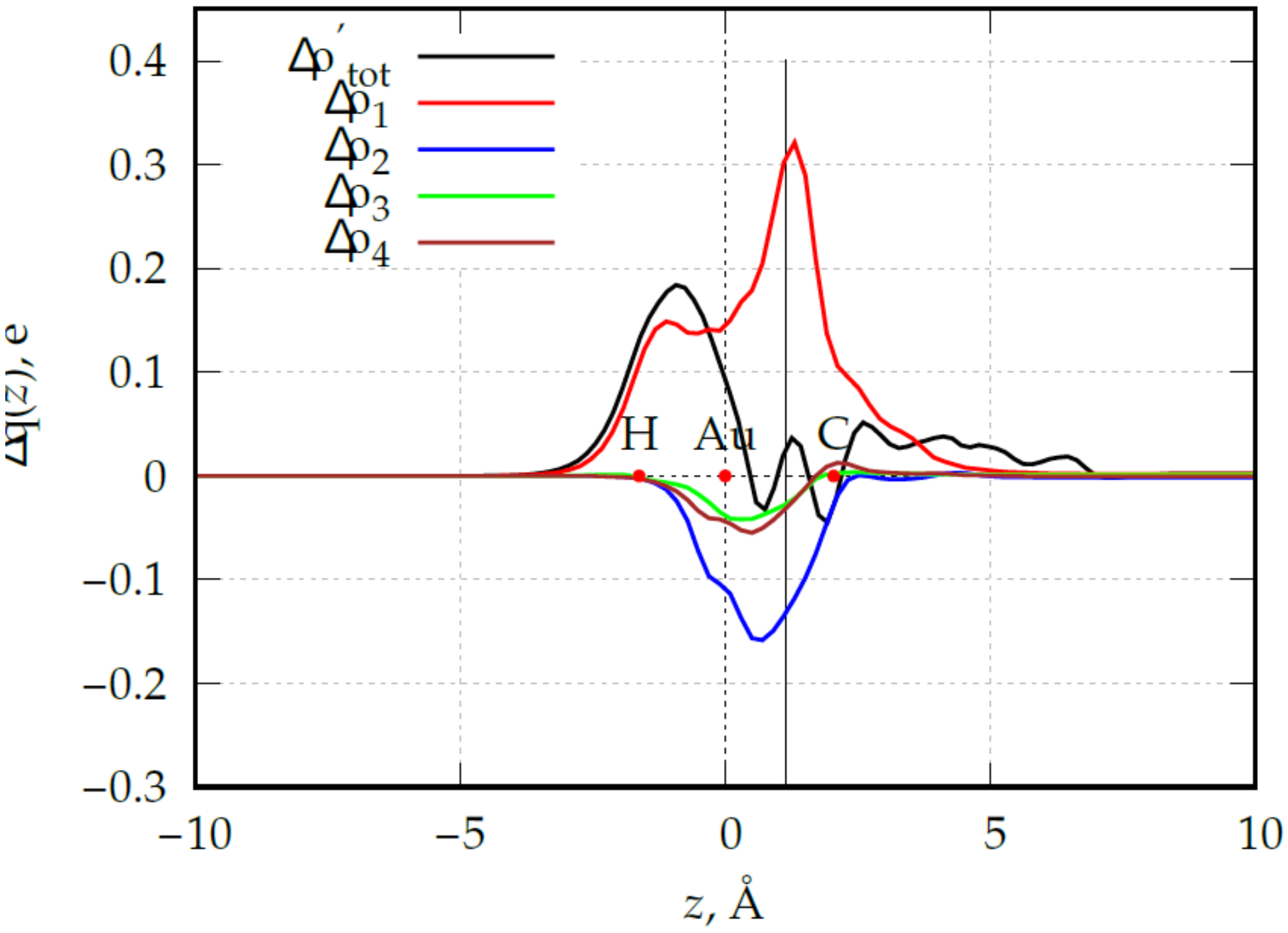

**Figure S22.** The NHC-AuH bond in complex **6**. CD curves associated with the total Δρ' and its components. Red dots indicate the position of the nuclei along the z axis. The grey vertical line marks the isodensity boundary between the NHC and AuH fragments. Positive (negative) values of the curve indicate right-to-left (left-to-right) charge transfer.

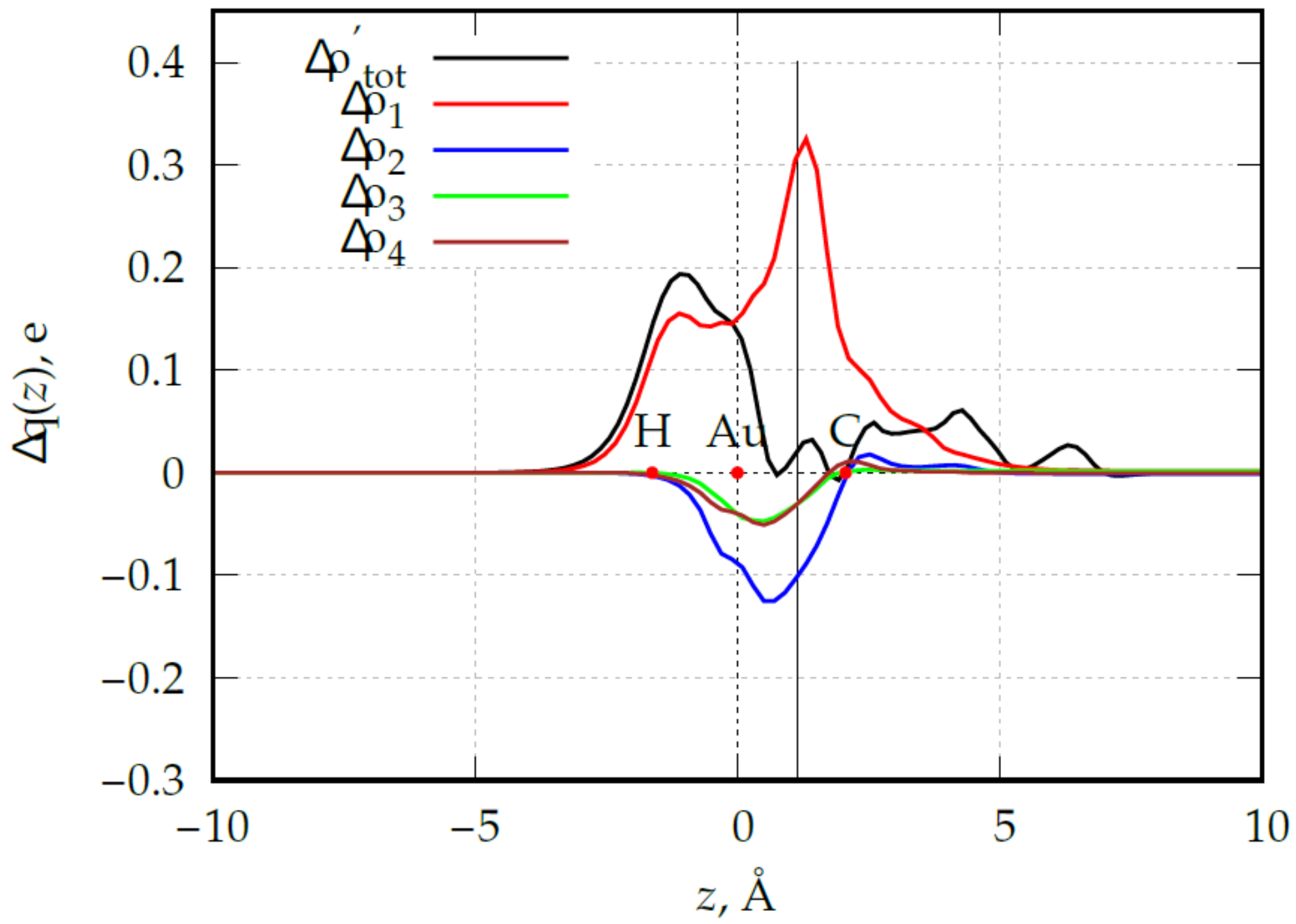

**Figure S23.** The NHC-AuH bond in complex **7**. CD curves associated with the total Δρ' and its components. Red dots indicate the position of the nuclei along the z axis. The grey vertical line marks the isodensity boundary between the NHC and AuH fragments. Positive (negative) values of the curve indicate right-to-left (left-to-right) charge transfer.

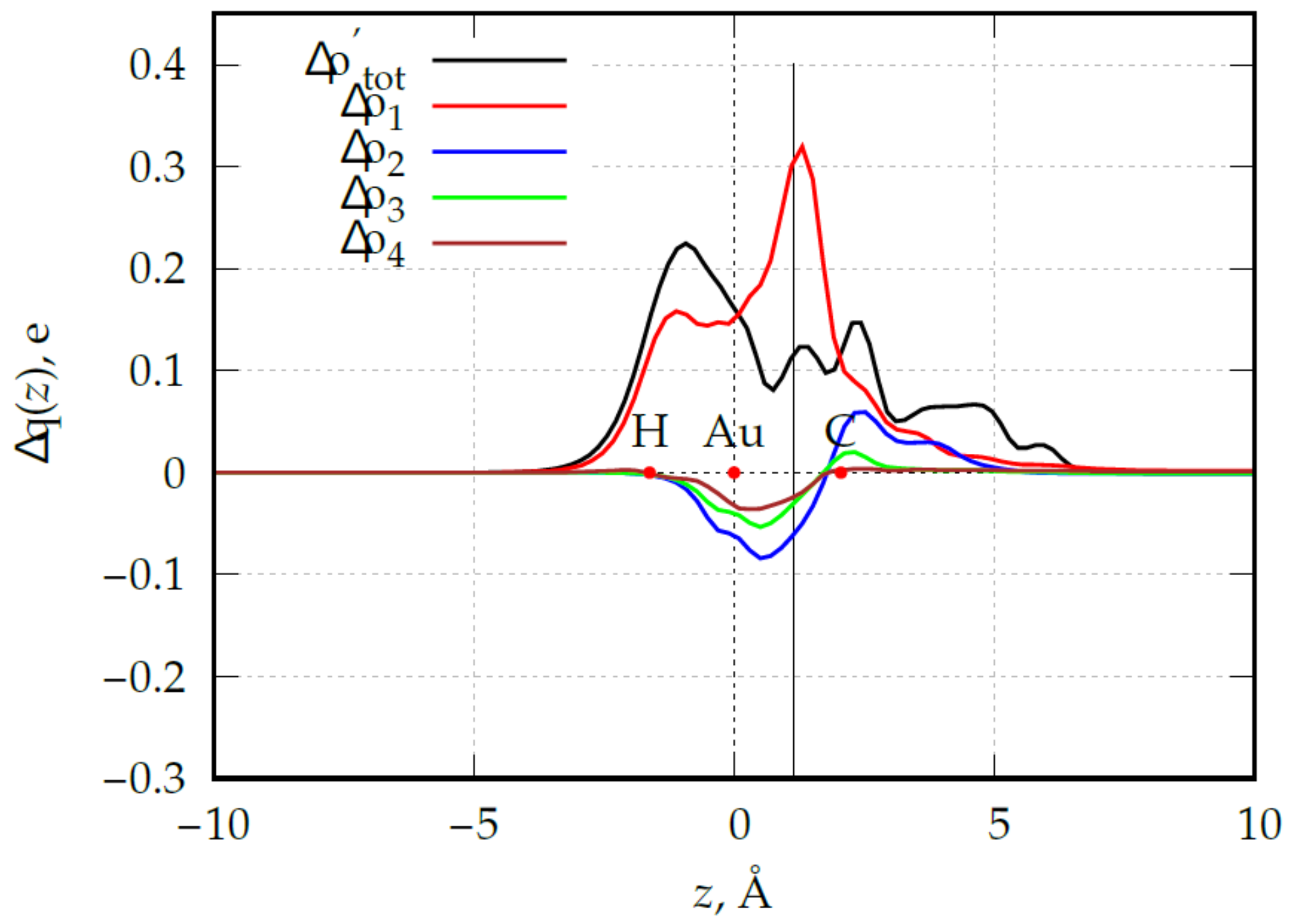

**Figure S24.** The NHC-AuH bond in complex **8**. CD curves associated with the total Δρ' and its components. Red dots indicate the position of the nuclei along the z axis. The grey vertical line marks the isodensity boundary between the NHC and AuH fragments. Positive (negative) values of the curve indicate right-to-left (left-to-right) charge transfer.

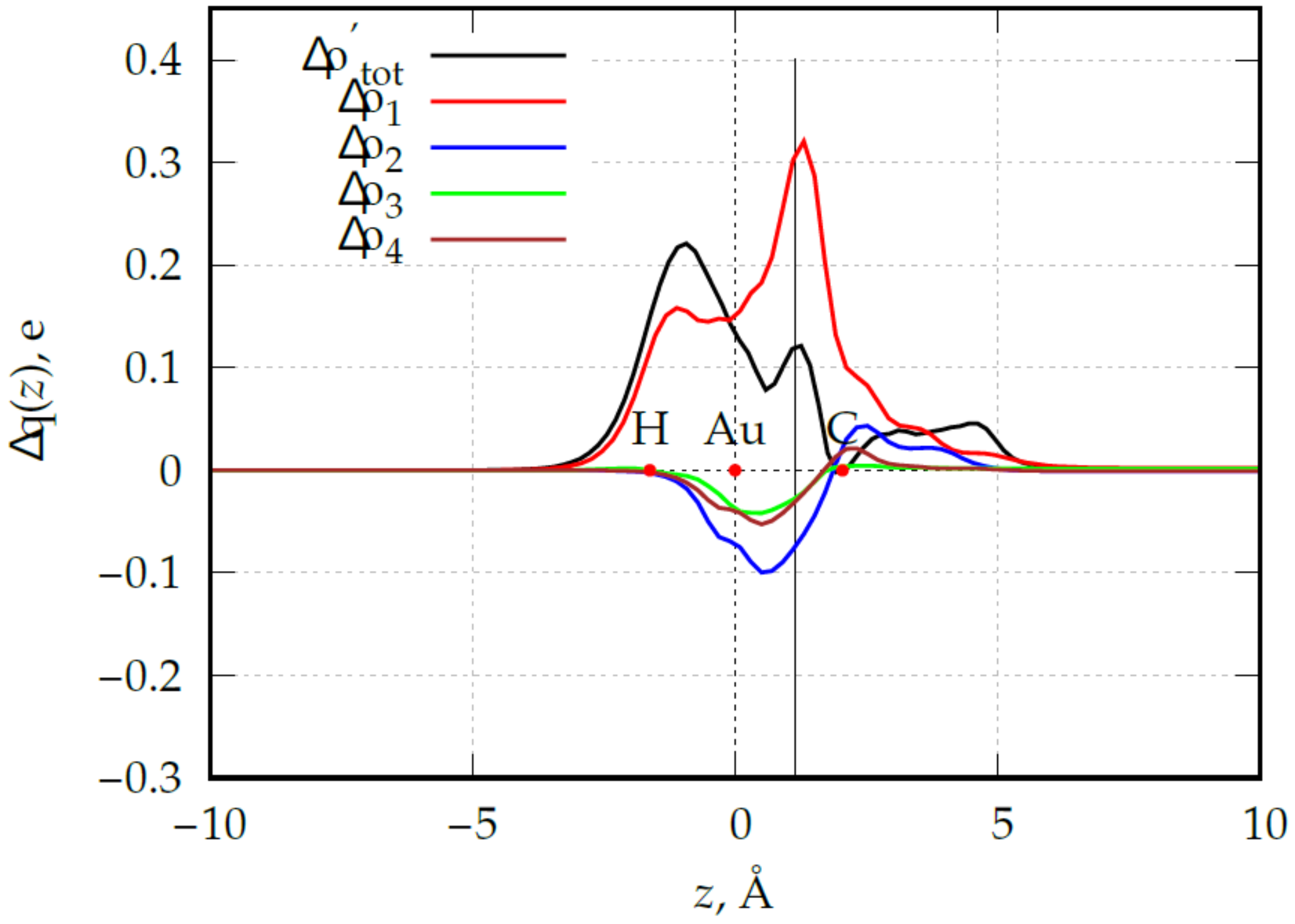

**Figure S25.** The NHC-AuH bond in complex **9**. CD curves associated with the total Δρ' and its components. Red dots indicate the position of the nuclei along the z axis. The grey vertical line marks the isodensity boundary between the NHC and AuH fragments. Positive (negative) values of the curve indicate right-to-left (left-to-right) charge transfer.

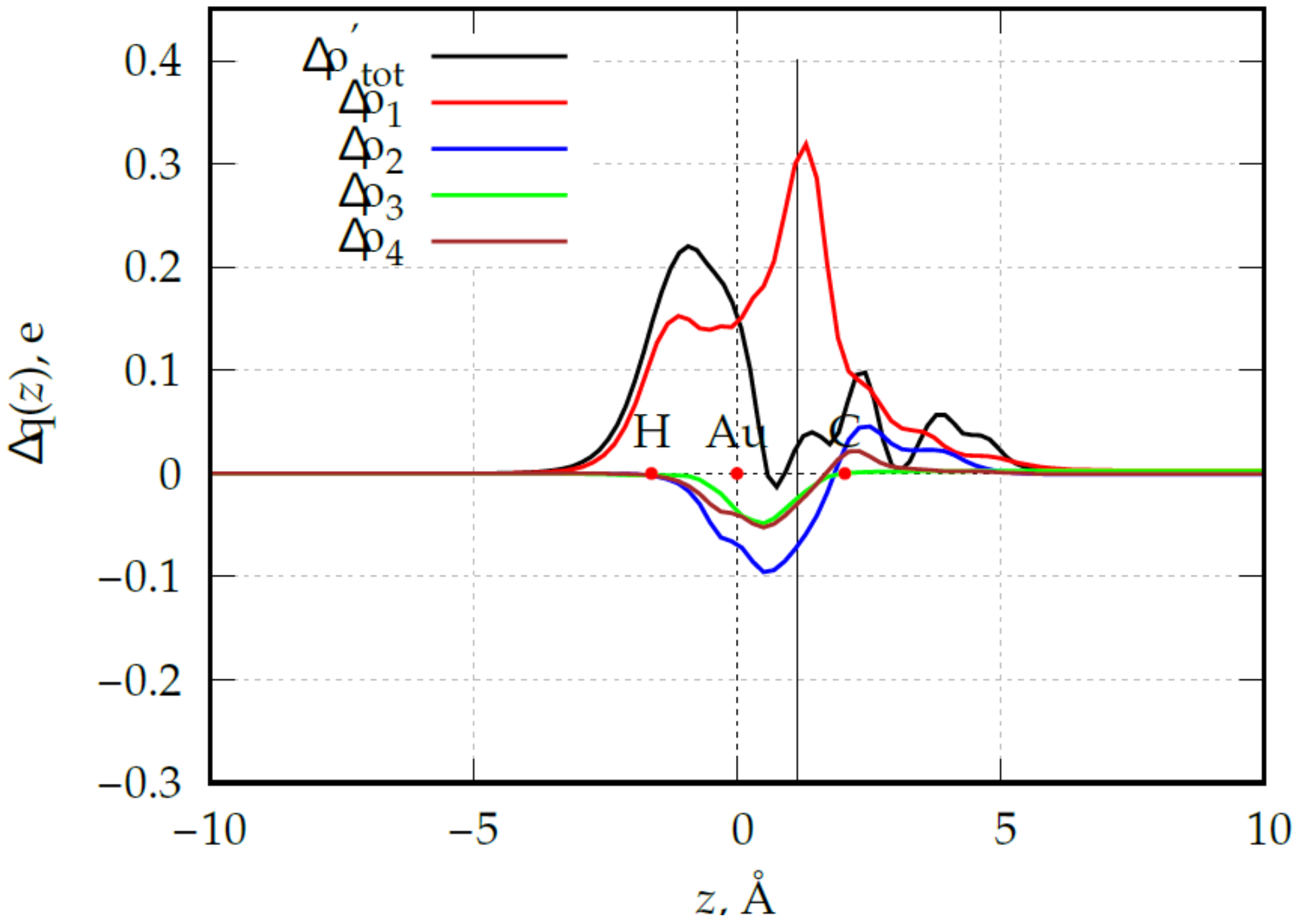

**Figure S26.** The NHC-AuH bond in complex **10**. CD curves associated with the total Δρ' and its components. Red dots indicate the position of the nuclei along the z axis. The grey vertical line marks the isodensity boundary between the NHC and AuH fragments. Positive (negative) values of the curve indicate right-to-left (left-to-right) charge transfer.

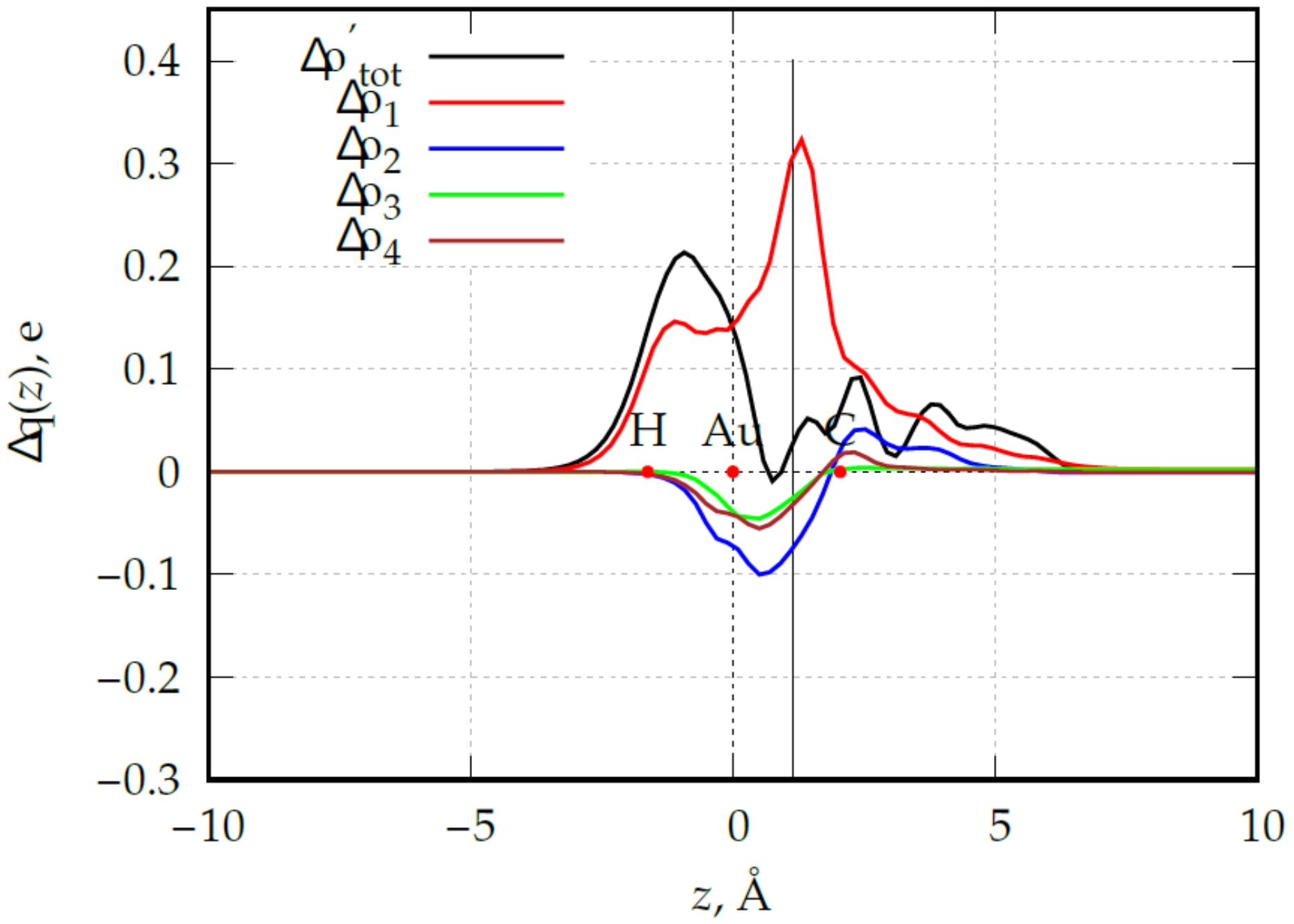

**Figure S27.** The NHC-AuH bond in complex **11**. CD curves associated with the total Δρ' and its components. Red dots indicate the position of the nuclei along the z axis. The grey vertical

line marks the isodensity boundary between the NHC and AuH fragments. Positive (negative) values of the curve indicate right-to-left (left-to-right) charge transfer.

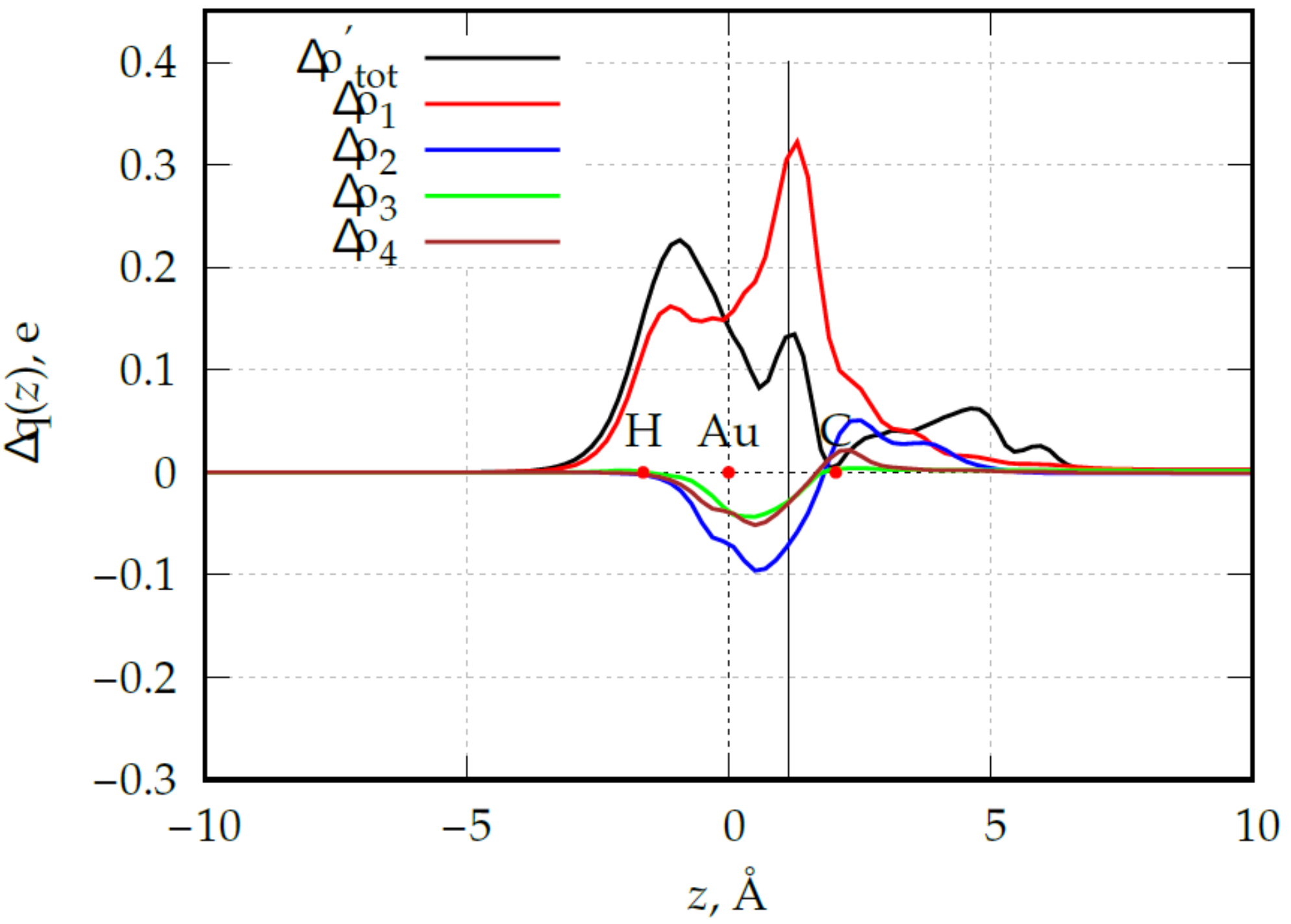

**Figure S28.** The NHC-AuH bond in complex **12**. CD curves associated with the total Δρ' and its components. Red dots indicate the position of the nuclei along the z axis. The grey vertical line marks the isodensity boundary between the NHC and AuH fragments. Positive (negative) values of the curve indicate right-to-left (left-to-right) charge transfer.

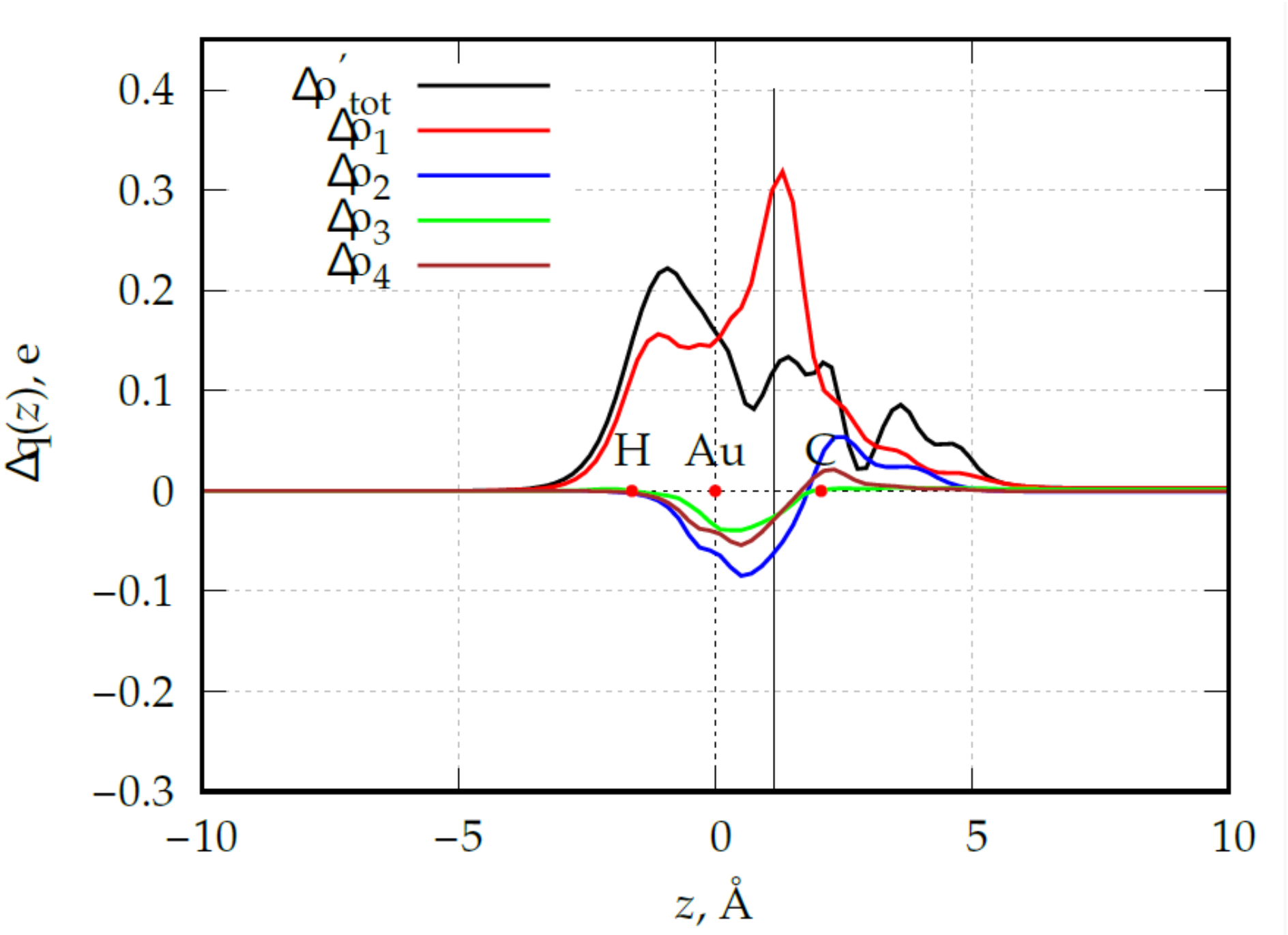

**Figure S29.** The NHC-AuH bond in complex **13**. CD curves associated with the total Δρ' and its components. Red dots indicate the position of the nuclei along the z axis. The grey vertical line marks the isodensity boundary between the NHC and AuH fragments. Positive (negative) values of the curve indicate right-to-left (left-to-right) charge transfer.

| | | |
|---|---|---|
| CAACs | Unconjugated DACs | Conjugated DACs |

| NOCV Energies (kcal/mol) | 1 | 2 | 3 | 6 | 7 | 5 | 4 | 11 | 9 | 10 | 12 | 8 | 13 |
|---|---|---|---|---|---|---|---|---|---|---|---|---|---|
| Δε1 | -37.61817 | -37.68241 | -37.42544 | -34.40861 | -35.12988 | -35.66593 | -36.02741 | -34.99125 | -35.54891 | -35.34264 | -35.9727 | -35.78937 | -35.68323 |
| Δε 2 | -11.38195 | -11.29003 | -11.15282 | -13.40885 | -10.90734 | -10.02902 | -9.6569 | -9.63922 | -9.3041 | -9.23069 | -9.18235 | -8.17545 | -8.07949 |
| Δε 3 | -5.82326 | -5.75445 | -5.70396 | -5.5166 | -5.16928 | -5.88425 | -5.83646 | -6.10838 | -5.82581 | -5.85408 | -5.75365 | -4.84982 | -5.53198 |
| Δε 4 | -4.27576 | -4.43963 | -4.78142 | -5.07859 | -4.87834 | -4.37721 | -4.41982 | -5.15734 | -4.88781 | -4.90238 | -4.78912 | -5.45318 | -4.95874 |
| Δε 5 | -3.74461 | -3.73788 | -3.91114 | -2.18998 | -2.70566 | -4.33473 | -2.25028 | -4.04247 | -2.11869 | -4.51508 | -2.23698 | -2.49654 | -2.8011 |
| Δε 6 | -1.20259 | -1.22229 | -1.38284 | -1.17593 | -1.03795 | -1.39419 | -0.59567 | -1.35904 | -0.53685 | -1.4851 | -0.58611 | -0.77694 | -0.90769 |

**Table S6.** Orbital interaction energy contributions (Δε) associated to the first 6 NOCV components for **1-13** complexes.

| | CAACs | | | Unconjugated DACs | | | | Conjugated DACs | | | | | |
|---|---|---|---|---|---|---|---|---|---|---|---|---|---|
| NOCV | 1 | 2 | 3 | 6 | 7 | 5 | 4 | 11 | 9 | 10 | 12 | 8 | 13 |
| **Isovalue** | 2.024414 | 2.024414 | 2.027148 | 2.025195 | 2.031055 | 2.030273 | 2.029102 | 2.030078 | 2.027148 | 2.029102 | 2.027148 | 2.033008 | 2.033984 |
| 1 | 0.31178 | 0.311551 | 0.310595 | 0.292958 | 0.297472 | 0.298085 | 0.300376 | 0.29381 | 0.293973 | 0.292082 | 0.296466 | 0.293408 | 0.292539 |
| 2 | -0.113464 | -0.111236 | -0.106081 | -0.137827 | -0.1063 | -0.087333 | -0.087333 | -0.079081 | -0.079778 | -0.075327 | -0.075525 | -0.065179 | -0.066121 |
| 3 | -0.027118 | -0.026899 | -0.026346 | -0.029274 | -0.032933 | -0.027089 | -0.032039 | -0.028279 | -0.029748 | -0.027214 | -0.030632 | -0.033755 | -0.027663 |
| 4 | -0.017073 | -0.017724 | -0.01711 | -0.034918 | -0.033492 | -0.016256 | -0.030371 | -0.035779 | -0.034279 | -0.033336 | -0.033619 | -0.025685 | -0.032506 |
| 5 | -0.011247 | -0.010775 | -0.013587 | -0.003562 | -0.004271 | -0.019549 | -0.003366 | -0.008532 | -0.003242 | -0.007357 | -0.002372 | -0.007752 | -0.007311 |
| 6 | 0.000107 | 0.00072 | 0.006181 | 0.012869 | 0.006175 | 0.001371 | -0.003008 | 0.003175 | -0.004105 | 0.001959 | -0.003698 | 0.001074 | 0.004467 |

**Table S7.** Charge transfer values (CT) calculated at the isodensity boundary associated to the first 6 NOCV components for **1-13** complexes. Values are in electrons (e).

| | HOMO (P %) | HOMO-1 (P %) |
|---|---|---|
| **1-P** | 35.8 $3p_z$ + 14.1 $3p_y$ + 5.9 $3p_x$ | 14.1 3s + 12.2 $3p_x$ + 9.2 $3p_z$ + 9.0 $3p_y$ |
| **4-P** | 49.3 $3p_z$ + 9.9 $3p_x$ + 3.4 $3p_y$ | 9.6 3s + 9.0 $3p_x$ +6.7 $3p_z$ + 1.4 $3p_y$ |
| **8-P** | 55.4 $3p_z$ + 8.1 $3p_y$ | 19.5 $3p_y$ + 13.1 3s + 4.2 $3p_z$ + 3.3 $3p_x$ |
| | HOMO (C%) | |
| **1** | 52.2 $2p_z$ +21.5 2s | |
| **4** | 50.5 $2p_z$ + 28.1 2s | |
| **8** | 49.7 $2p_z$ + 31.5 2s | |

**Table S8**: Percentage contribution of the P atom to the HOMO and HOMO -1 MOs of **1-P**, **4-P** and **8-P**; percentage contribution of $C_{NHC}$ to the HOMO of **1**, **4** and **8**.

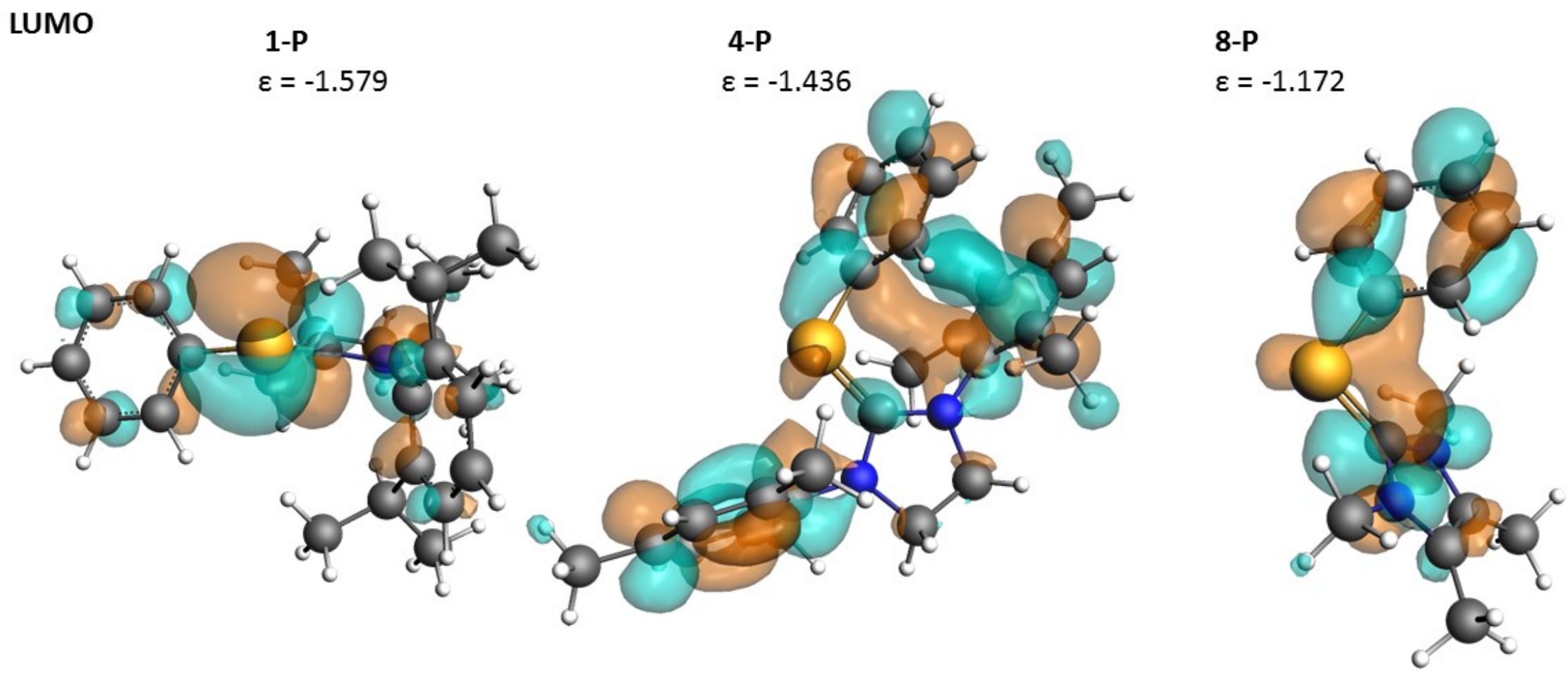

**Figure S30.** Isosurfaces of the LUMO of **1-P**, **4-P** and **8-P** NHCP fragments. Isovalue for all surfaces is ± 25 me/$a_0^3$. Orbital energy (in eV) is also reported.

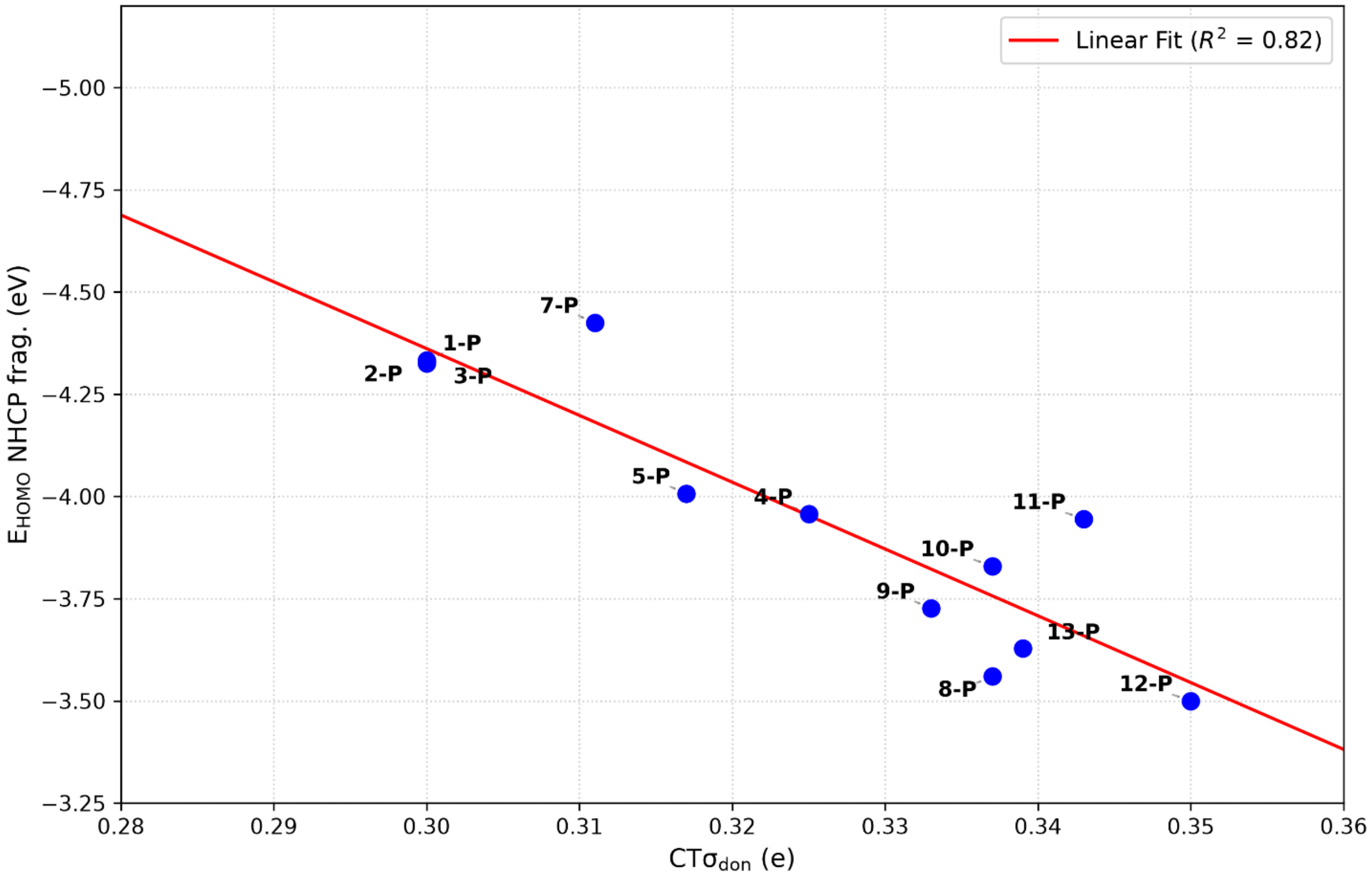

**Figure S31.** Correlation plot between CT $_{\sigma\ don}$ and NHCP HOMO fragment energy for the complete **1-P-13-P** series.

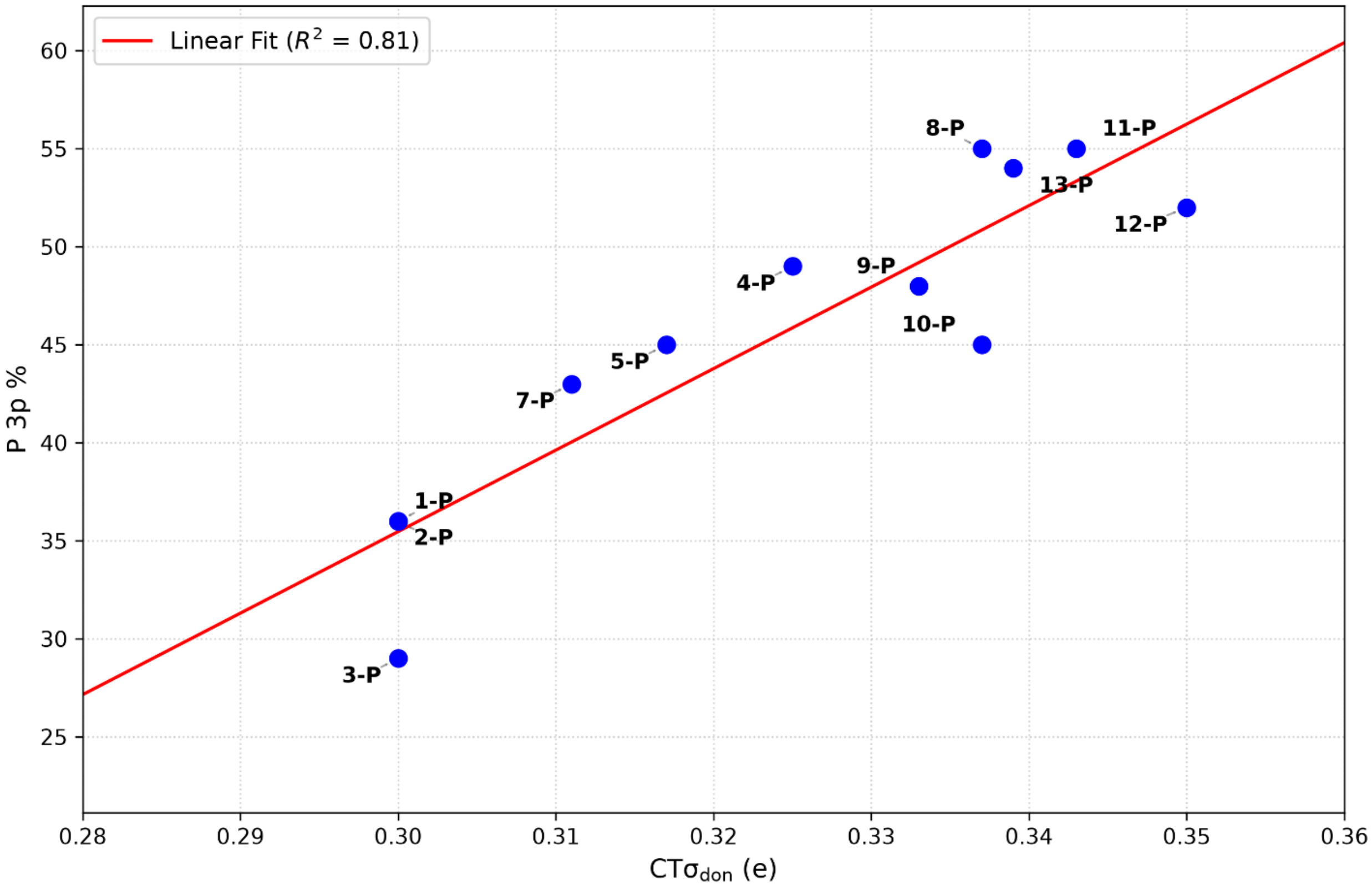

**Figure S32.** Correlation plot between CT $_{\sigma\ don}$ and NHCP HOMO fragment P 3p percentage contribution for the complete **1-P-13-P** series.

| | CT $_{\sigma\ don}$ | NHCP fragment HOMO ε | NHCP fragment HOMO P 3p % | NHCP fragment HOMO -1 ε | NHCP fragment HOMO -1 P 3s % | $C_{NHC}$-P-Au angle |
|---|---|---|---|---|---|---|
| **1-P** | 0.300 | -4.333 | 36 | -5.460 | 14 | 120.4 |
| **2-P** | 0.300 | -4.331 | 36 | -5.439 | 14 | 119.3 |
| **3-P** | 0.300 | -4.326 | 29 | -5.393 | 15 | 121.5 |
| **4-P** | 0.325 | -3.957 | 49 | -5.279 | 10 | 104.9 |
| **5-P** | 0.317 | -4.007 | 45 | -5.338 | 10 | 112.8 |
| **6-P** | 0.287 | -5.104 | 33 | -5.713 | 10 | 113.9 |
| **7-P** | 0.311 | -4.425 | 43 | -5.433 | 12 | 109.1 |
| **8-P** | 0.337 | -3.560 | 55 | -5.484 | 13 | 98.7 |
| **9-P** | 0.333 | -3.726 | 48 | -5.382 | 11 | 103.3 |
| **10-P** | 0.337 | -3.830 | 45 | -5.453 | 11 | 104.6 |
| **11-P** | 0.343 | -3.945 | 55 | -5.643 | 12 | 95.7 |
| **12-P** | 0.350 | -3.500 | 52 | -5.297 | 13 | 101.1 |
| **13-P** | 0.339 | -3.628 | 54 | -5.590 | 14 | 97.8 |

**Table S9**: CT $_{\sigma\ don}$ (e), energy (eV) and percentage contribution of the P 3p atomic orbital to the HOMO, energy (eV) and percentage contribution of the P 3s atomic contribution to the HOMO -1 and $C_{NHC}$-P-Au angle (°) of the complete **1-P-13-P** series.

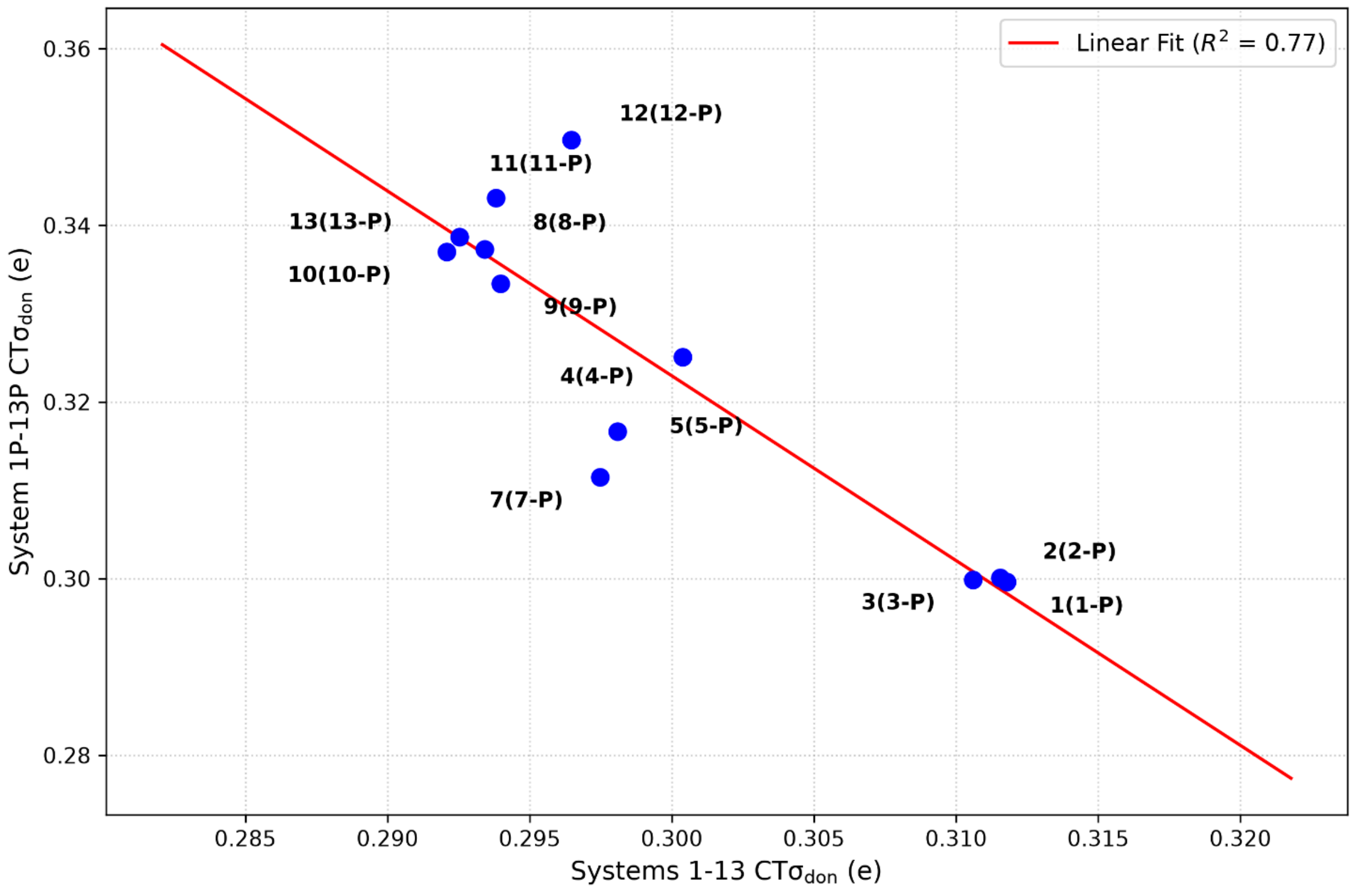

**Figure S33.** Correlation plot between CT $_{\sigma\ don}$ for the complete NHCP **1-P-13-P** series and CT $_{\sigma\ don}$ for the complete NHC **1-13** series.

| | NHCP fragment HOMO ε | NHCP fragment HOMO P 3p % | NHCP relaxed HOMO ε | NHCP relaxed HOMO P 3p % |
|---|---|---|---|---|
| **1-P** | -4.333 | 36 | -4.468 | 53 |
| **2-P** | -4.331 | 36 | -4.463 | 55 |
| **3-P** | -4.326 | 29 | -4.427 | 54 |
| **4-P** | -3.957 | 49 | -4.250 | 58 |
| **5-P** | -4.007 | 45 | -4.289 | 57 |
| **6-P** | -5.104 | 33 | -5.323 | 44 |
| **7-P** | -4.425 | 43 | -4.665 | 56 |
| **8-P** | -3.560 | 55 | -3.860 | 59 |
| **9-P** | -3.726 | 48 | -3.957 | 56 |
| **10-P** | -3.830 | 45 | -3.994 | 55 |
| **11-P** | -3.945 | 55 | -4.162 | 56 |
| **12-P** | -3.500 | 52 | -3.774 | 56 |
| **13-P** | -3.628 | 54 | -3.885 | 59 |

**Table S10**: Energies (eV) and phosphorus 3p orbital percentage contribution (P 3p %) to the HOMO of the NHCP ligands in the geometry within the corresponding complexes **1-P-13-P** (NHCP fragment) and in their optimized geometry (NHCP relaxed).

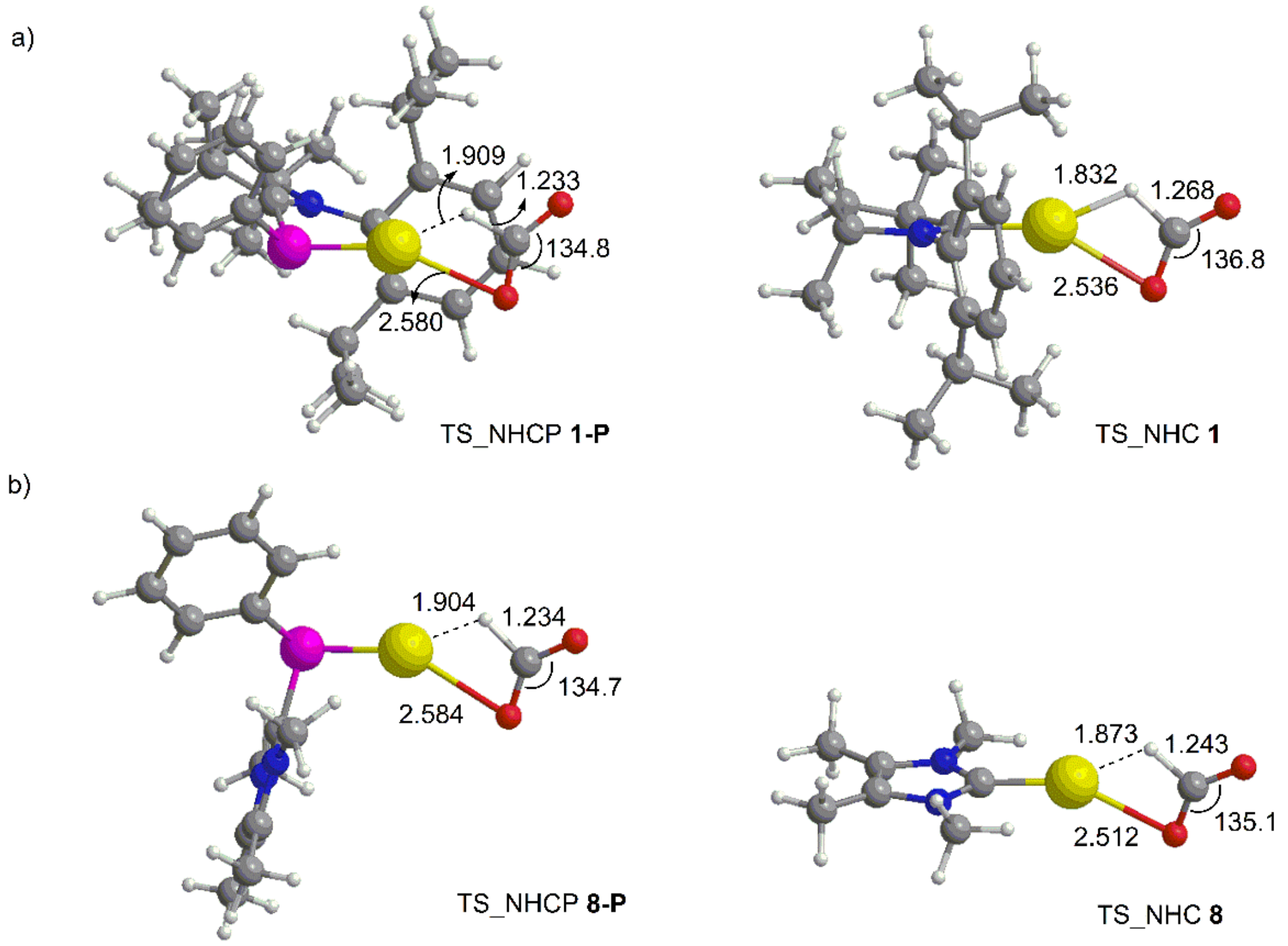

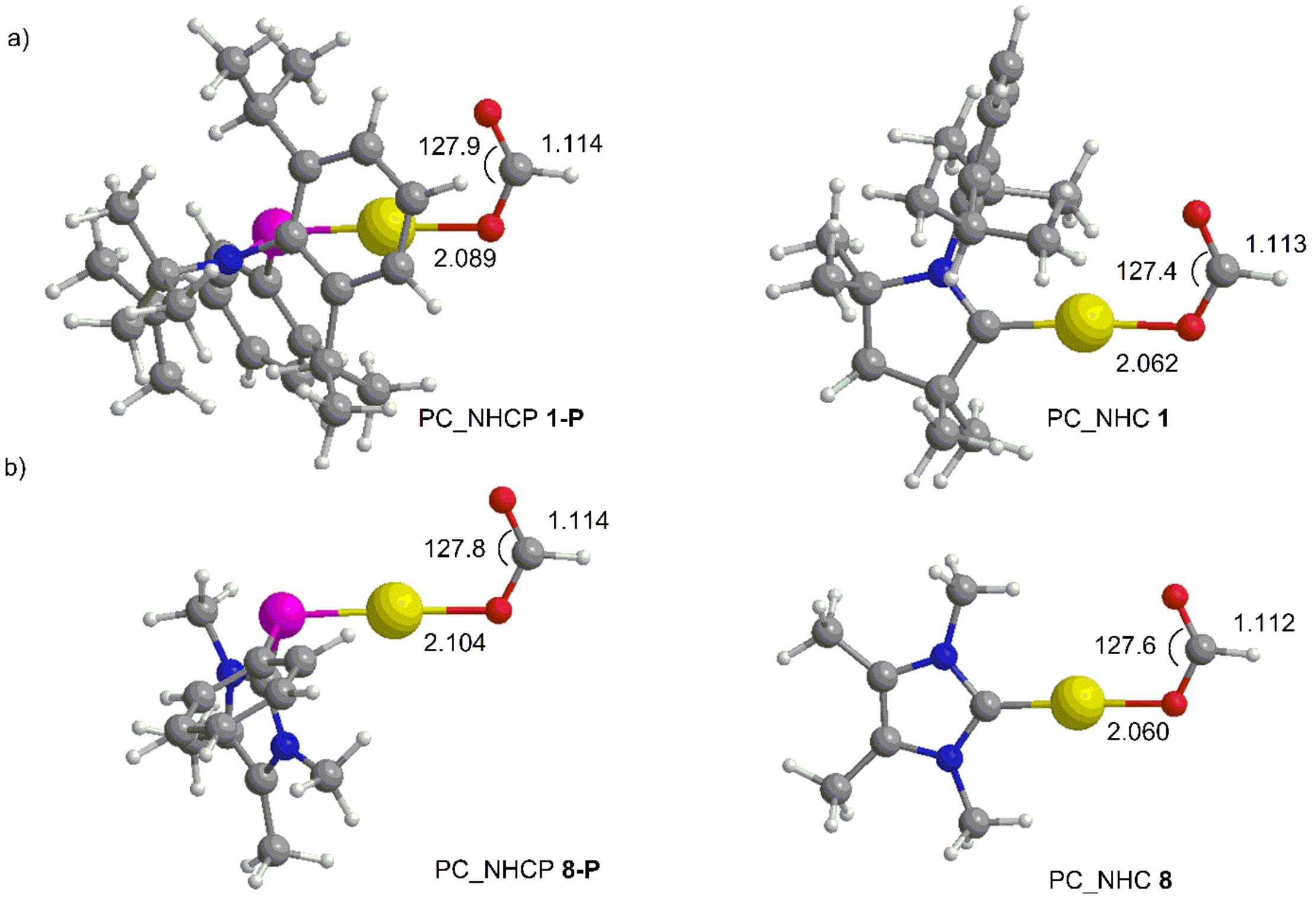

**Figure S34.** Optimized structures of the transition states (TS_NHC, TS_NHCP) and product complexes (PC_NHC, PC_NHCP) for **1**, **1-P**, **8**, and **8-P** reaction profiles for $CO_2$ insertion into the Au-H bond. Distances (in Å) and angles (in degrees) are also shown.

**Formate product – rotational conformers**

In Figure 6 in the main text, the reported transition states do not connect directly the formate product minima (PC). More precisely, the $CO_2$ insertion leads first to the formation of a gold-formate $\eta^1$-O-species (OCHO dihedral angle = 180°), with carboxylate H atom pointing towards the metal center, which represents a local minimum. Subsequently, a relatively facile formate rotation around the Au-O(OCHO) bond, resulting in the H atom pointing outside, allows for a lower energy minimum to be reached, a $\kappa^1$-O-formate complex (OCHO dihedral angle = 0°), which is the final product (PC) of this reaction step. These two species represent rotational conformers existing in a dynamic equilibrium. For **8-P** and **8** complexes, as NHCP and NHC representative, the electronic energy barrier for the rotation of the formate group around the Au-O(OCHO) bond has been estimated through single point calculations where the OCHO dihedral angle is changed from 180° to 0° in steps of 30°. Results are reported in Figures S34 and S35, which depict the potential energy surface (PES) scan for the OCHO rotation step for **8-P** and **8**, respectively.

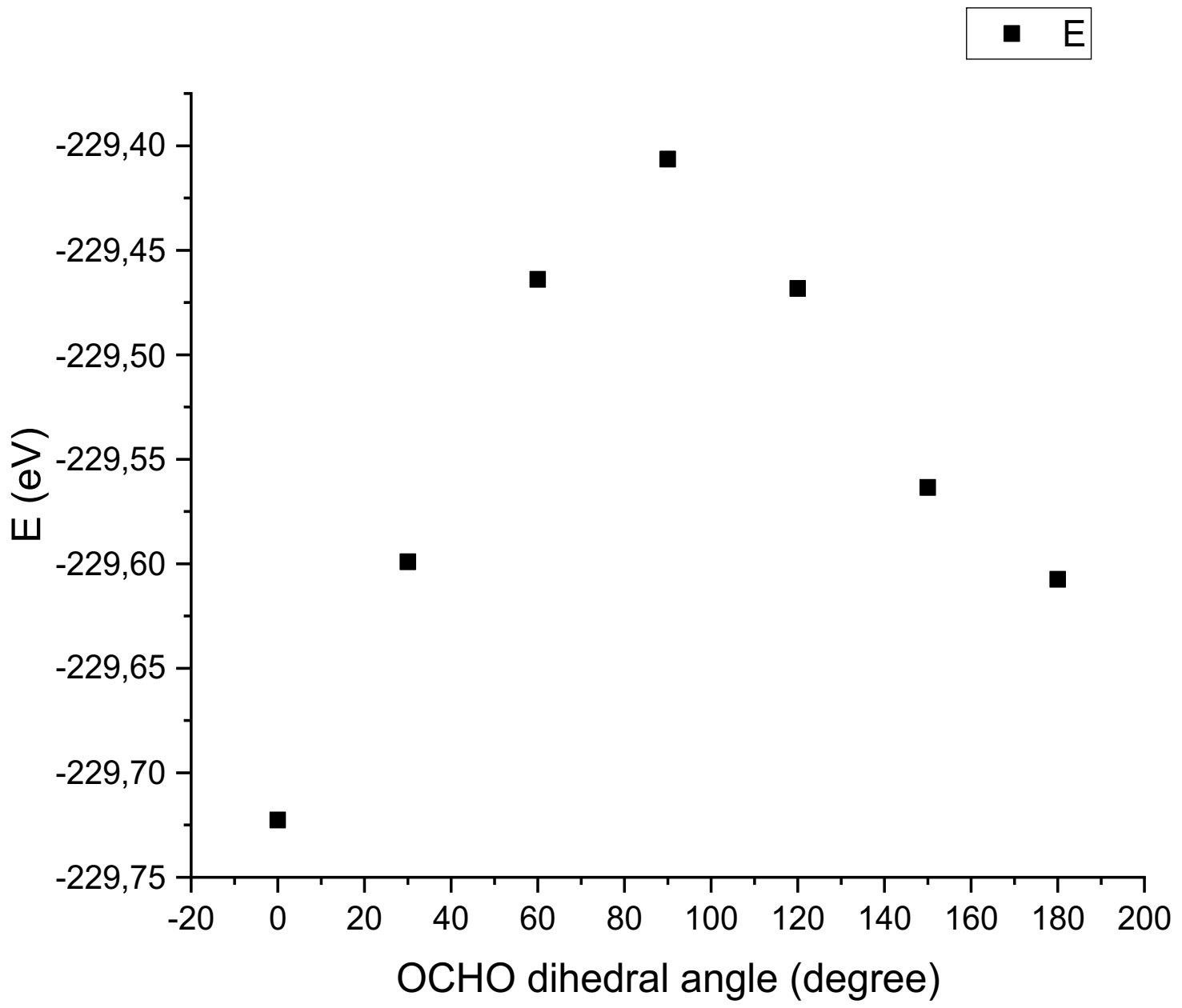

**Figure S35.** PES scan using the OCHO dihedral angle (in degrees) as reaction coordinate for formate rotation around the Au-O(OCHO) bond for complex **8-P**. The point at 180° represents the optimized local minimum and the point at 0° represents the optimized PC species.

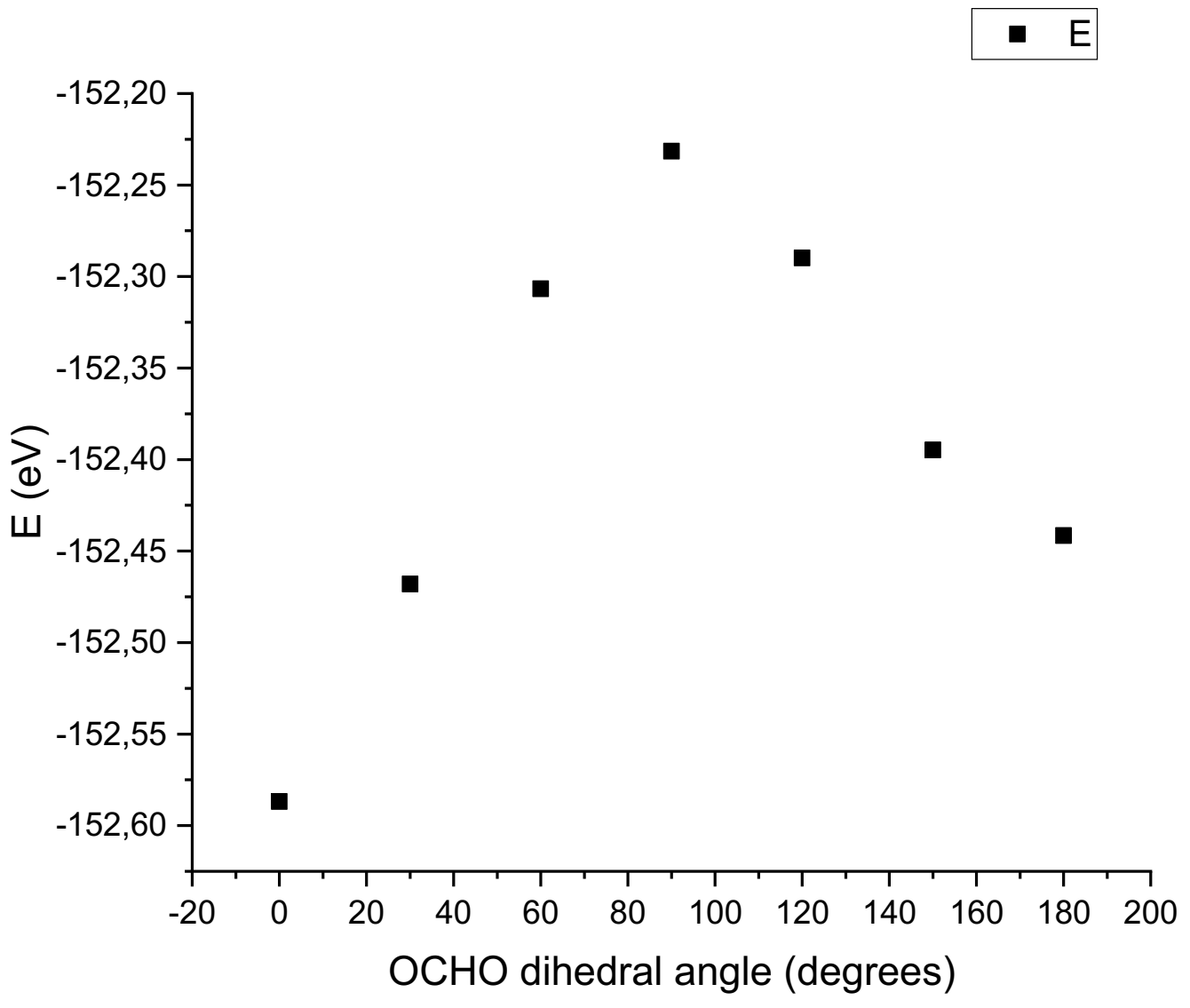

**Figure S36.** PES scan using the OCHO dihedral angle (in degrees) as reaction coordinate for formate rotation around the Au-O(OCHO) bond for complex **8**. The point at 180° represents the optimized local minimum and the point at 0° represents the optimized PC species.

Starting from the 180° conformer, the maximum is calculated at OCHO = 90°, lying 4.6 and 4.8 kcal/mol above it for **8-P** and **8**, respectively, and the final PC product (0° conformer) is 2.7 and 3.4 kcal/mol more stable than the 180° conformer for **8-P** and **8**, respectively. In terms of ΔG, PC is more stable than this 180° conformer by 2.5 kcal/mol for **8-P** and 2.9 kcal/mol for **8**. This test calculation demonstrates a facile rotation of the formate around the Au-O(OCHO) bond, involving a significantly lower barrier than that for the $CO_2$ insertion step. Based on that, we have refrained from calculating the barrier for the formate rotation step, since these two rotational conformers are both thermodynamically accessible in solution. To compute the thermodynamics of the $CO_2$ insertion step for all the 26 complexes, the slightly more stable PC structure has been considered.

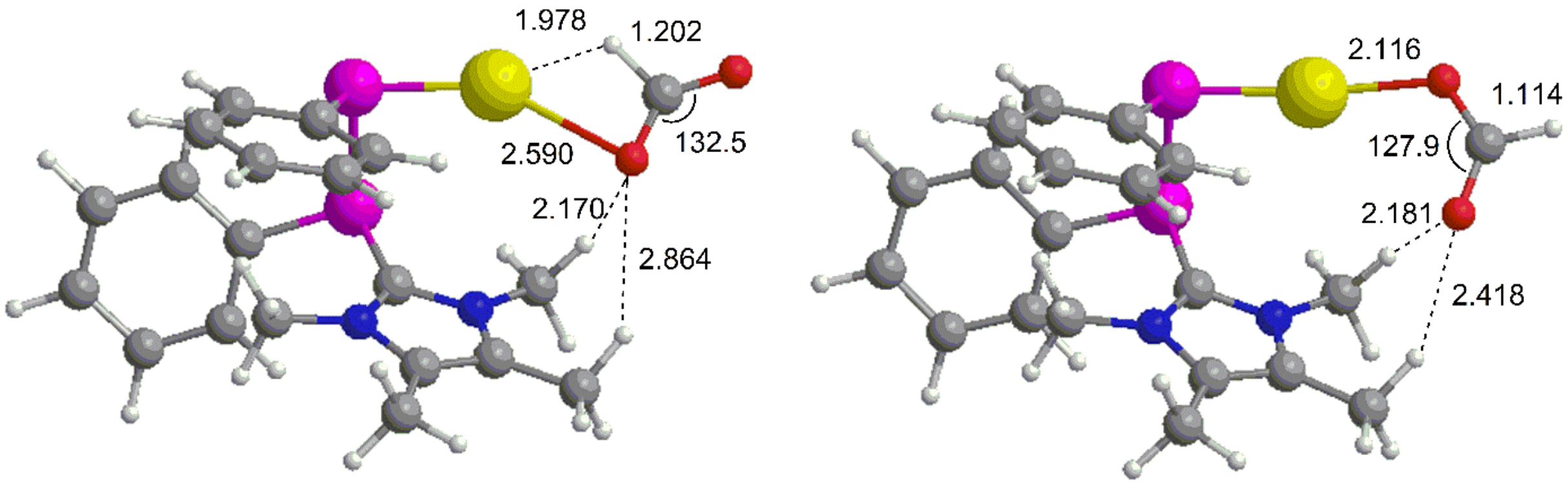

TS_NHCPP **8-PP** PC_NHCPP **8-PP**

**Figure S37.** Optimized structures of transition state (TS_NHCPP) and product complex (PC_NHCPP) for **8-PP** complex involved in the $CO_2$ insertion reaction into the Au-H bond. Distances (in Å) and angles (in degrees) are also shown.

## %Buried Volume Analysis

Quantitative Evaluation of Ligand Steric Hindrance using Sambvca 2.1

| | |
|---|---|
| **Software Platform:** | Sambvca 2.1 (Steric Map Analysis of Buried Volume) |
| **Reference Manual:** | https://www.aocdweb.com/OMtools/sambvca2.1/help/help.html |

The Sambvca 2.1 web tool evaluates the steric hindrance and active-site topology of ligands in metal coordination complexes. By defining a spherical domain around the catalytic center, Au, in the present case the software computes two primary descriptors:

**Percent Buried Volume (%V_bur):** The fraction of the coordination sphere occupied by the ligand framework, quantifying steric bulk.

Based on the official Sambvca 2.1 technical guidelines, the steric evaluation relies on a systematic protocol:

**A. Structural Alignment & Axis Definition**

- **Coordinate Input:** Supported formats include .xyz
- **Coordination Center:** Defined at the active metal center (M) placed at origin (0, 0, 0).
- **Z-Axis Alignment:** Oriented along the coordination line. The primary reaction domain is set to the z > 0 half-sphere.
- **Atom Deletion:** Non-analyzed ligands and spectator groups are deleted prior to integration.

**B. Calculation Parameters**

| | |
|---|---|
| **Sphere Radius (R)** | 3.5 Å |
| **Mesh Spacing (s)** | 0.10 Å |
| **Hydrogen Atoms** | Included |

> *Percent Buried Volume Calculation:*
> *V_Sphere = ∑ v_xyz = V_Free + V_Bur*
> *%V_bur = 100 × (V_Bur / V_Sphere)*

Total $\%V_{bur}$ for our complexes:

| | **1** | **1-P** | **8** | **8-P** | **8-PP** |
|---|---|---|---|---|---|
| $\%V_{bur}$ | 41.2% | 45.0% | 25.4% | 29.0% | 34.5% |

**Table S11**: Steric parameter buried volume (%$V_{bur}$) calculated for the NHC/NHCP/NHCPP in **1**, **1-P**, **8**, **8-P** and **8-PP** complexes.

Steric parameter buried volume (%$V_{bur}$) has been calculated for the NHC/NHCP/NHCPP in **1**, **1-P**, **8**, **8-P** and **8-PP** complexes as selected representative systems reported in Figure 6 in the main text, by using SambVca 2.1 software developed by Cavallo and co-workers.[13-15] Results for the studied ligands are reported in Table S11 and span the range from 25.4 to 45.0%. Introduction of a PPh group to NHC leads to an increase in the value of %$V_{bur}$ by 3.8 % (from **1** to **1-P**) and 3.6% (from **8** to **8-P**), and introduction of an additional PPh group by further 5.5% (from **8-P** to **8-PP**). Since %$V_{bur}$ gives a measure of the average space occupied by the NHC ligand in the first coordination sphere of the metal center, the steric bulkiness of NHC ligands increase on adding a PPh group, which is consistent with the reactivity order **8-PP** > **8-P** > **8**, i.e. a high %$V_{bur}$ lowers the activation barrier (in reaction where the metal center sheds a ligand). However, the higher %$V_{bur}$ for **1-P** with respect to that of **1** (45.0% vs 41.2%) is not consistent with the higher barrier calculated for **1-P** ($\Delta G^{\ddagger}$ = 26.4 vs 25. 9 kcal/mol for 1). In addition, although the higher %$V_{bur}$ for **1** compared to **8** (41.2% vs 25.4%) is consistent with the reactivity order **1** > **8** ($\Delta G^{\ddagger}$ = 25.9 vs 29. 1 kcal/mol for **1** and **8**, respectively), the higher %$V_{bur}$ for **1-P** compared to **8-P** (45.0% vs 29.0%) is not consistent with the reactivity order **8-P** > **1-P** ($\Delta G^{\ddagger}$ = 24.2 vs 26. 4 kcal/mol for **8-P** and **1-P**, respectively). Based on this analysis, we expect that steric effects could contribute to the activation barrier differences for **1-13** NHC complex series, particularly within the same class, with a less relevant impact on the activation barrier differences for **1-P-13-P** NHCP series, where the PPh moiety coordination orientation allows the reactive Au-H unit to be easily accessed by the $CO_2$.

| | CT $_{\sigma\ don}$ | ΔG‡ | ΔG | | CT $_{\sigma\ don}$ | ΔG‡ | ΔG |
|---|---|---|---|---|---|---|---|
| **1-P** | 0.300 | 26.4 | 6.4 | **1** | 0.312 | 25.9 | 3.9 |
| **2-P** | 0.300 | 26.0 | 6.2 | **2** | 0.312 | 25.9 | 4.4 |
| **3-P** | 0.300 | 26.5 | 6.7 | **3** | 0.311 | 25.8 | 4.2 |
| **4-P** | 0.325 | 23.1 | 5.2 | **4** | 0.300 | 27.6 | 5.3 |
| **5-P** | 0.317 | 24.3 | 5.7 | **5** | 0.298 | 26.5 | 3.6 |
| **6-P** | 0.287 | 29.3 | 8.6 | **6** | 0.293 | 26.9 | 7.0 |
| **7-P** | 0.311 | 26.1 | 6.4 | **7** | 0.297 | 27.0 | 6.1 |
| **8-P** | 0.337 | 24.2 | 6.0 | **8** | 0.293 | 29.1 | 5.1 |
| **9-P** | 0.333 | 21.6 | 4.7 | **9** | 0.294 | 28.5 | 6.1 |
| **10-P** | 0.337 | 24.7 | 5.0 | **10** | 0.292 | 29.0 | 5.9 |
| **11-P** | 0.343 | 23.6 | 5.3 | **11** | 0.294 | 29.9 | 6.4 |
| **12-P** | 0.350 | 23.3 | 5.1 | **12** | 0.296 | 27.7 | 5.0 |
| **13-P** | 0.339 | 24.5 | 4.2 | **13** | 0.293 | 29.5 | 5.7 |

**Table S12**: CT $_{\sigma\ don}$ values (e), activation free energy barriers ΔG‡ (kcal/mol) and thermodynamics ΔG (kcal/mol) of the $CO_2$ insertion into the Au-H bond reaction for the complete series of NHCP **1-P-13-P** and NHC **1-13** complexes. All ΔG values are calculated for the more stable rotational conformer PC.

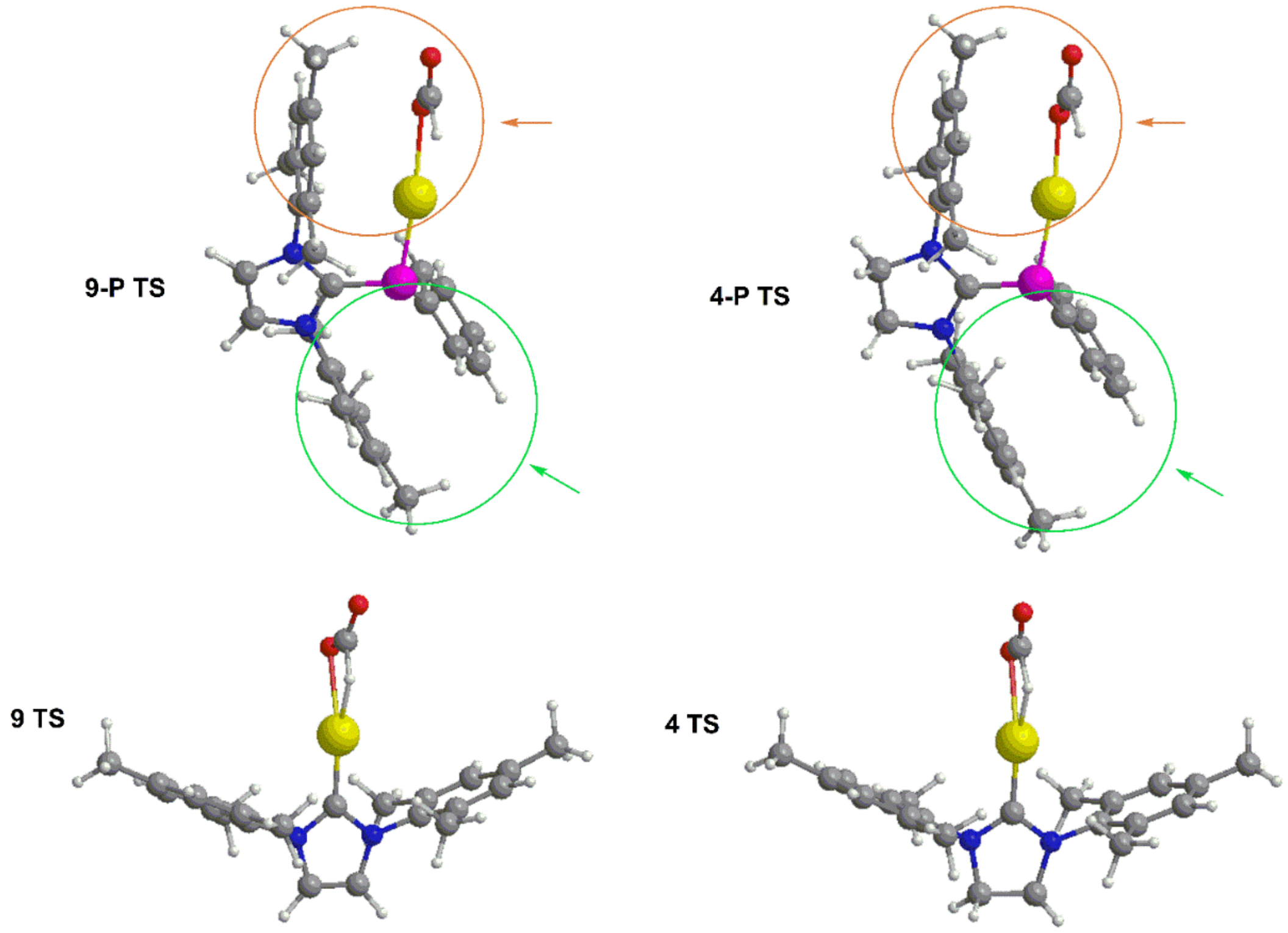

**Figure S38.** Optimized structures of transition states for **4-P** (4-P TS), **9-P** (9-P TS), **4** (4 TS) and **9** (9 TS) involved in the $CO_2$ insertion reaction into the Au-H bond. Distances (in Å) and angles (in degrees) are also shown.

Comparison of the transition state structures for **4** and **9** and for **4-P** and **9-P** highlights the incipient formate $HCO_2^-$ interaction via anion…π interactions arising between the electron-rich oxygens of $HCO_2^-$ and the face of the aromatic Mes ring of the NHC moiety (orange circle) as well as that π-π stacking interactions occurring between the second aromatic Mes ring of the NHC moiety and the Ph ring bonded to P (green circle), introduced by the PPh moiety in **4-P** TS and **9-P** TS.

| | HOMO ε | HOMO character % | HOMO-1 ε | HOMO-1 character % |
|---|---|---|---|---|
| 1-P | -5.080 | 42 3p P (18 2p N + 4 5d Au + 4 1s H) | -5.531 | 45 6s Au + 14 1s H |
| 2-P | -5.061 | 43 3p P (17 2p N + + 3 Au 5d + 4 1s H) | -5.521 | 44 6s Au + 14 1s H |
| 3-P | -4.999 | 44 3p P (17 2p N + 4 Au 5d + 3 1s H) | -5.523 | 43 6s Au + 16 1s H |
| 4-P | -4.930 | 36 3p P(+13 + 2 Au 5d + 7 1s H) | -5.505 | 39 6s Au + 17 1s H |
| 5-P | -4.942 | 31 3p P(+20 + 3 Au 5d + 6 1s H) | -5.387 | 40 6s Au + 16 1s H |
| 6-P | -5.547 | 28 3p P (7 2p C + 4 (+4) Au 5d + 6 1s H) | -5.845 | 27 6s Au + 11 1s H |
| 7-P | -5.169 | 34 3p P (+9 +3 au 5d + 6 1s H) | -5.587 | 33 6s Au + 15 1s H |
| 8-P | -4.760 | 41 3p P (+ 6 + 2 Au 5d + 10 1s H) | -5.634 | 44 6s Au + 19 1s H |
| 9-P | -4.750 | 31 3p P (+16 + 1 Au 5d + 8 1s H) | -5.538 | 39 6s Au + 20 1s H |
| 10-P | -4.769 | 24 3p P (+23 + 1 Au 5d + 7 1s H) | -5.537 | 39 6s Au + 20 1s H |
| 11-P | -4.998 | 24 3p P (+17 + 1 Au 5d + 12 1s H) | -5.583 | 40 6s Au + 16 1s H |
| 12-P | -4.685 | 30 3p P (12 + 9 + 1 Au 5d + 9 1s H) | -5.465 | 34 6s Au + 19 1s H |
| 13-P | -4.803 | 36 3p P (+12 + 1 Au 5d + 10 1s H) | -5.666 | 44 6s Au + 19 1s H |
| 8-PP | -4.869 | 43 3p P1 (+13 + 5 Au 6p + 2 Au 5d +12 1s H) | -5.440 | 20 3p P2 + 14 1s H + 13 6s Au |

**Table S13**: HOMO and HOMO-1 energies ε (in eV) and P/Au percentage character for the complete **1-P-13-P** series and **8-PP** complex.

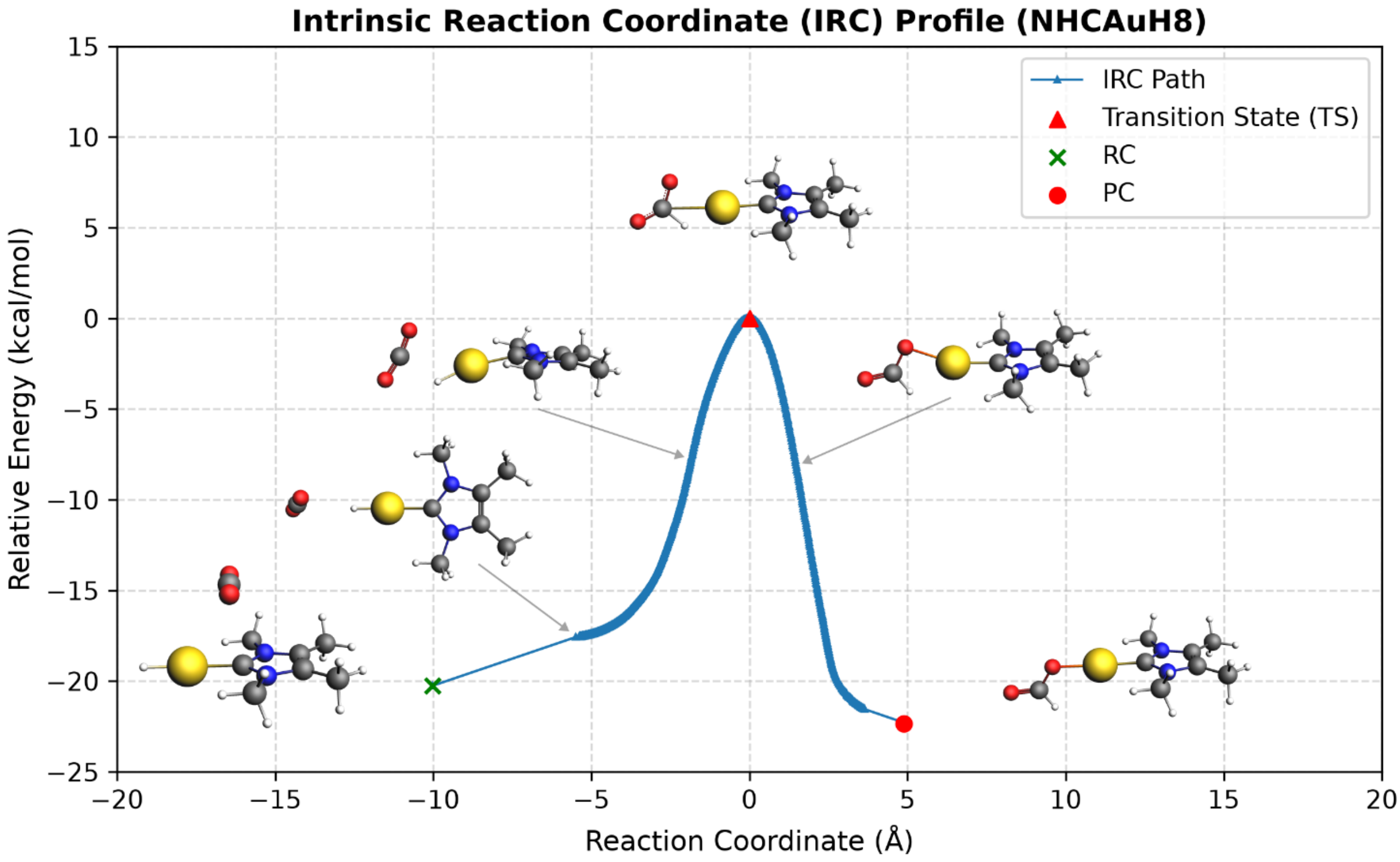

**Figure S39.** Intrinsic Reaction Coordinate (IRC) profile for complex **8**. Structures of reactant complex RC, transition state TS and product complex PC (local minimum rotational conformer) are shown together with three additional sample structures along the path.

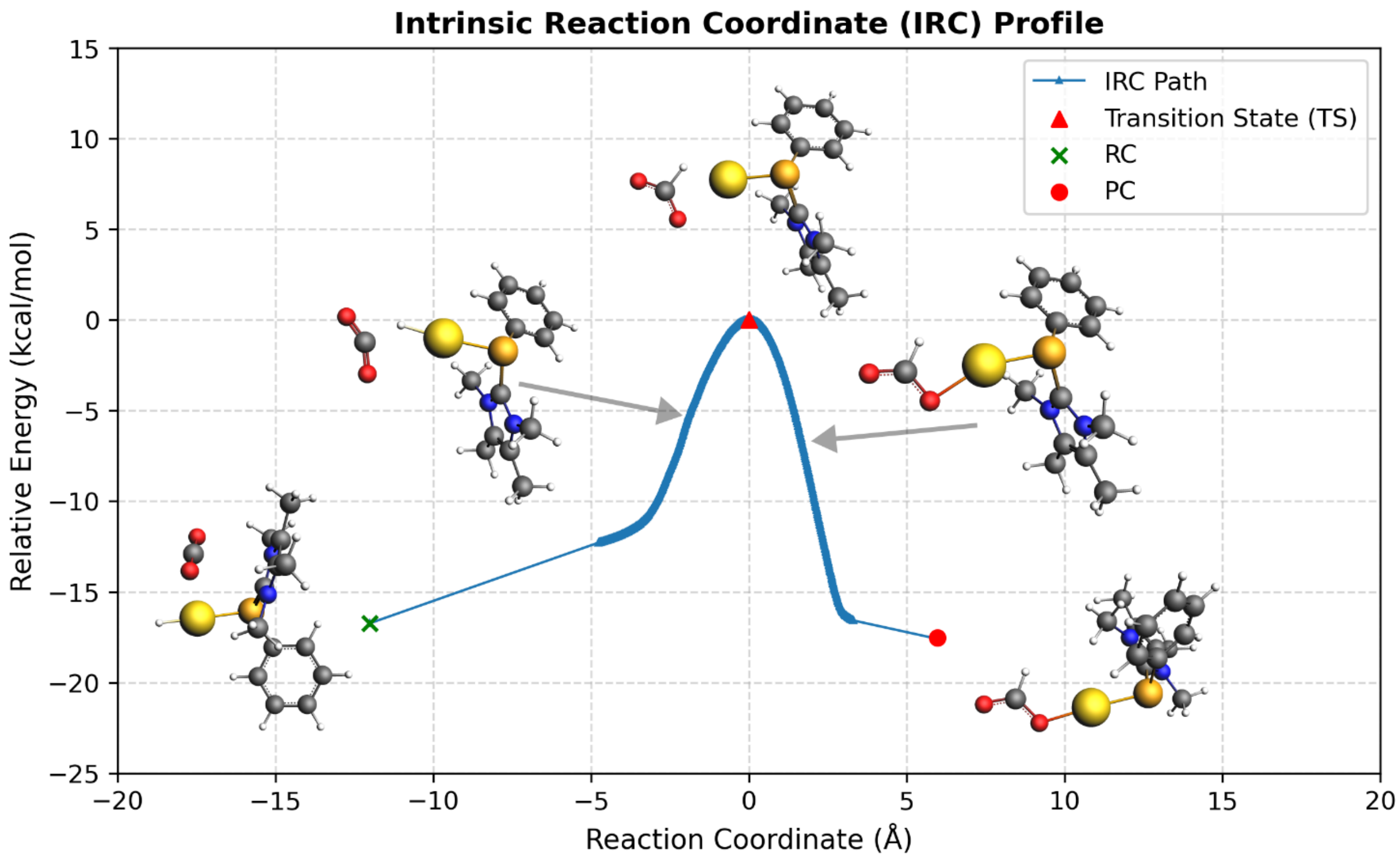

**Figure S40.** Intrinsic Reaction Coordinate (IRC) profile for complex **8-P**. Structures of reactant complex RC, transition state TS and product complex PC (local minimum rotational conformer) are shown together with three additional sample structures along the path.

**Benchmark study on the reaction barriers**

| | ΔE‡ BP86-D3 all electrons, gas phase | ΔE‡ B3LYP-D3 all electrons, gas phase | ΔE‡ ωB97x-D3 all electrons, gas phase |
|---|---|---|---|
| **1** | 16.2 | 21.1 | 21.1 |
| **1-P** | 18.3 | 21.7 | 20.0 |
| **8** | 21.5 | 25.6 | 24.8 |
| **8-P** | 17.1 | 20.4 | 20.1 |
| **10** | 19.9 | 23.7 | 23.2 |
| **10-P** | 16.6 | 19.2 | 17.4 |
| **8-PP** | 11.1 | 13.0 | 12.4 |

**Table S14**. Electronic energy barriers for $CO_2$ insertion into the Au-H bond calculated for **1**, **1-P**, **8**, **8-P**, **10**, **10-P** and **8-PP** representative complexes using different exchange-correlation functionals; first column: BP86-D3/all electrons/gas phase; second column: B3LYP-D3/all electrons/gas phase; third column: ωB97x-D3/all electrons/gas phase.

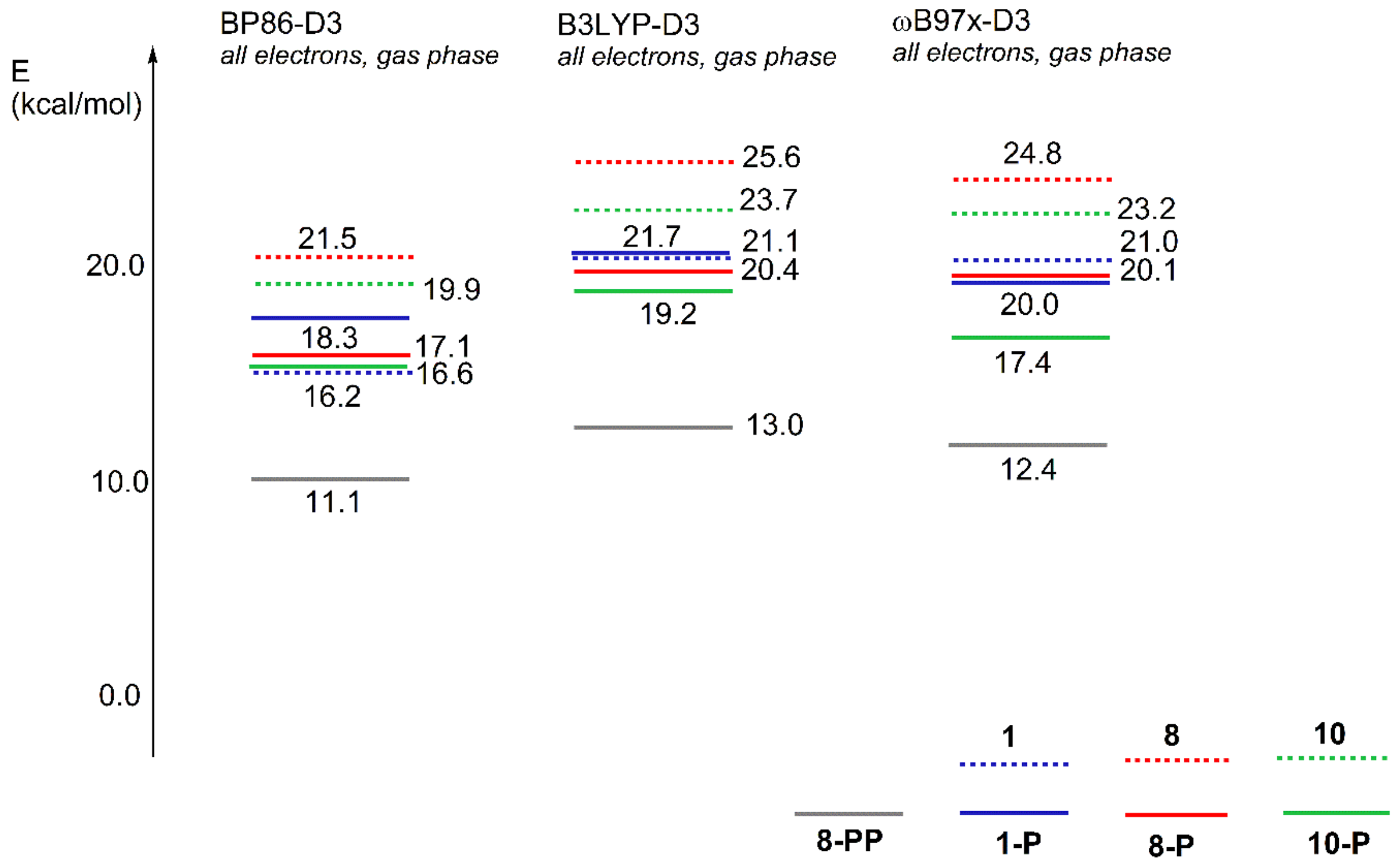

**Figure S41**. Electronic energy barriers order for $CO_2$ insertion into the Au-H bond calculated for **1**, **1-P**, **8**, **8-P**, **10**, **10-P** and **8-PP** representative complexes at different level of theory; left: BP86-D3/all electrons/gas phase; middle: B3LYP-D3/all electrons/gas phase; right: ωB97x-D3/all electrons/gas phase. Data taken from Table S14.

**Bonding analysis: Mayer bond order (MBO), Gophinatan-Jug (G-J) bond order, Nalewajski-Mrozek (N-M) bond order analysis, and charge analysis**

| | Au-P (Å) | MBO | G-J | N-M |
|---|---|---|---|---|
| **1-P** | 2.338 | 0.72 | 0.37 | 0.38 |
| **4-P** | 2.365 | 0.71 | 0.38 | 0.40 |
| **8-P** | 2.375 | 0.69 | 0.38 | 0.42 |

**Table S15**. Au-P bond distance (Å), Mayer bond order (MBO), Gophinatan-Jug bond order (G-J) and Nalewajski-Mrozek bond order (N-M) values for **1-P**, **4-P** and **8-P** using atomic P and Au fragments.

| | Au-P (Å) | Mayer | G-J | N-M |
|---|---|---|---|---|
| **1-P** | 2.338 | 1.22 | 0.33 | 0.26 |
| **4-P** | 2.365 | 1.26 | 0.34 | 0.27 |
| **8-P** | 2.375 | 1.12 | 0.34 | 0.26 |

**Table S16**. Au-P bond distance (Å), Mayer bond order (MBO), Gophinatan-Jug bond order (G-J) and Nalewajski-Mrozek bond order (N-M) values for **1-P**, **4-P** and **8-P** using NHCP and AuH fragments.

| | **1-P** | **4-P** | **8-P** |
|---|---|---|---|
| Mulliken | -0.380 | -0.362 | -0.377 |
| Hirshfeld | -0.141 | -0.160 | -0.164 |
| VDD | -0.132 | -0.145 | -0.156 |
| QTAIM | -0.321 | -0.361 | -0.402 |

**Table S17**. Atomic charge (e) on the AuH fragment calculated using Mulliken, Hirshfeld, Voronoi Deformation Density (VDD) and QTAIM methods for **1-P**, **4-P** and **8-P**.

A correlation between Au-P bond distance and Mayer, G-J and N-M bond order values is observed from Table S15, with bond lengths restricted between 2.338 Å (**1-P**), 2.365 Å (**4-P**) and 2.371 Å (**8-P**) and MBO maintained around 0.72-0.69, G-J around 0.37-0.38 and N-M around 0.38-0.42, all results indicating a single bond. However, whereas MBO values decreases in the order 1-P > 4-P > 8-P, a reverse order is found with N-M bond order, consistent with EDA/NOCV results. The same analysis based on the AuH-NHCP fragments (Table S16) shows a not clear trend for all the Mayer, G-J and N-M bond order values. These different trends are however expected since different bond order descriptors have different reference state, basis set sensivity and performance in describing bonding of different nature (covalent, ionic, dative, ecc..).

Results of charge analysis on the AuH fragment for **1-P**, **4-P** and **8-P**, using Mulliken, Hirshfeld, Voronoi Deformation Density (VDD) and QTAIM methods are collected in Table S17. A different trend can be observed only according to Mulliken, where the negative charge on AuH decreases in the order **1-P** > **8-P** > **4-P**, whereas according to Hirshfeld, VDD and QTAIM the negative charge on AuH decreases in the order **8-P** > **4-P** > **1-P**, consistent with EDA/CD-NOCV results, i.e. σ-donor strength decreasing in the order conjugated DACs- > unconjugated DACs- > CAACs-based NHCP.

**Plots including 6 and 6-P complexes**

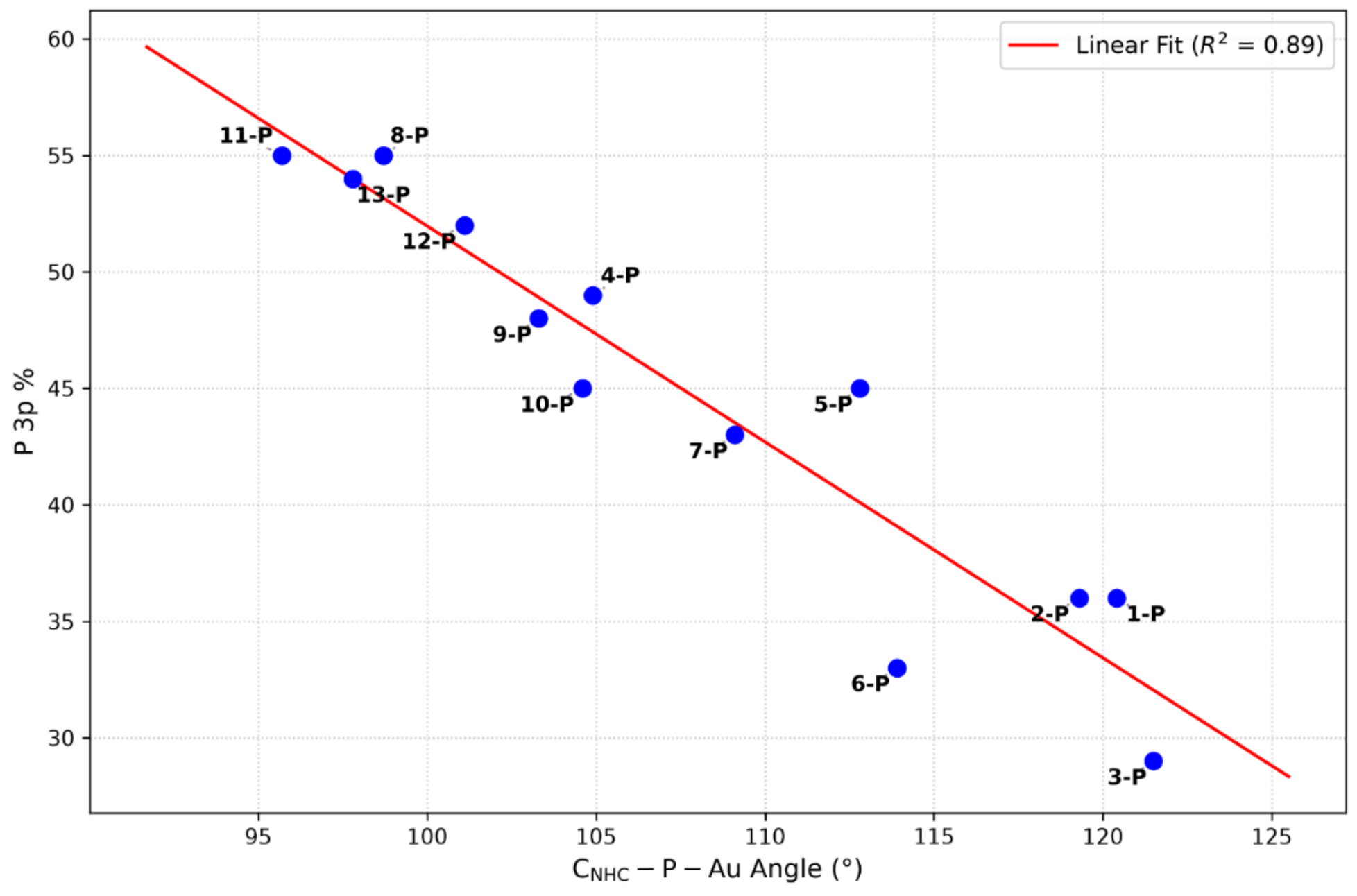

**Figure S42:** Correlation plots between the $C_{NHC}$-P-Au bond angles and the NHCP fragment HOMO P 3p percentage contribution for the complete **1-P-13-P** series.

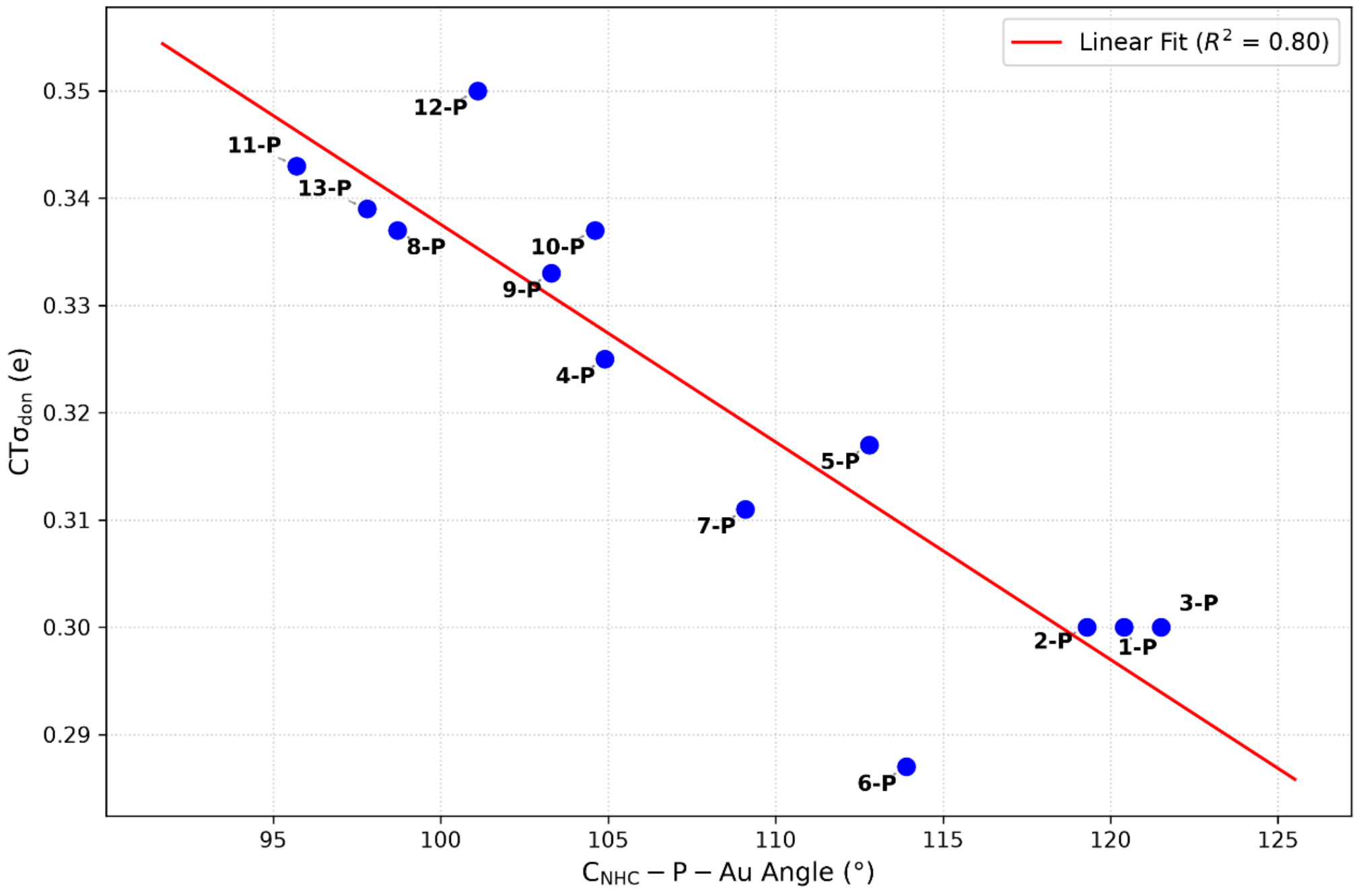

**Figure S43:** Correlation plots between the $C_{NHC}$-P-Au bond angles and the $CT_{\sigma\ don}$ values for the complete **1-P-13-P** series.

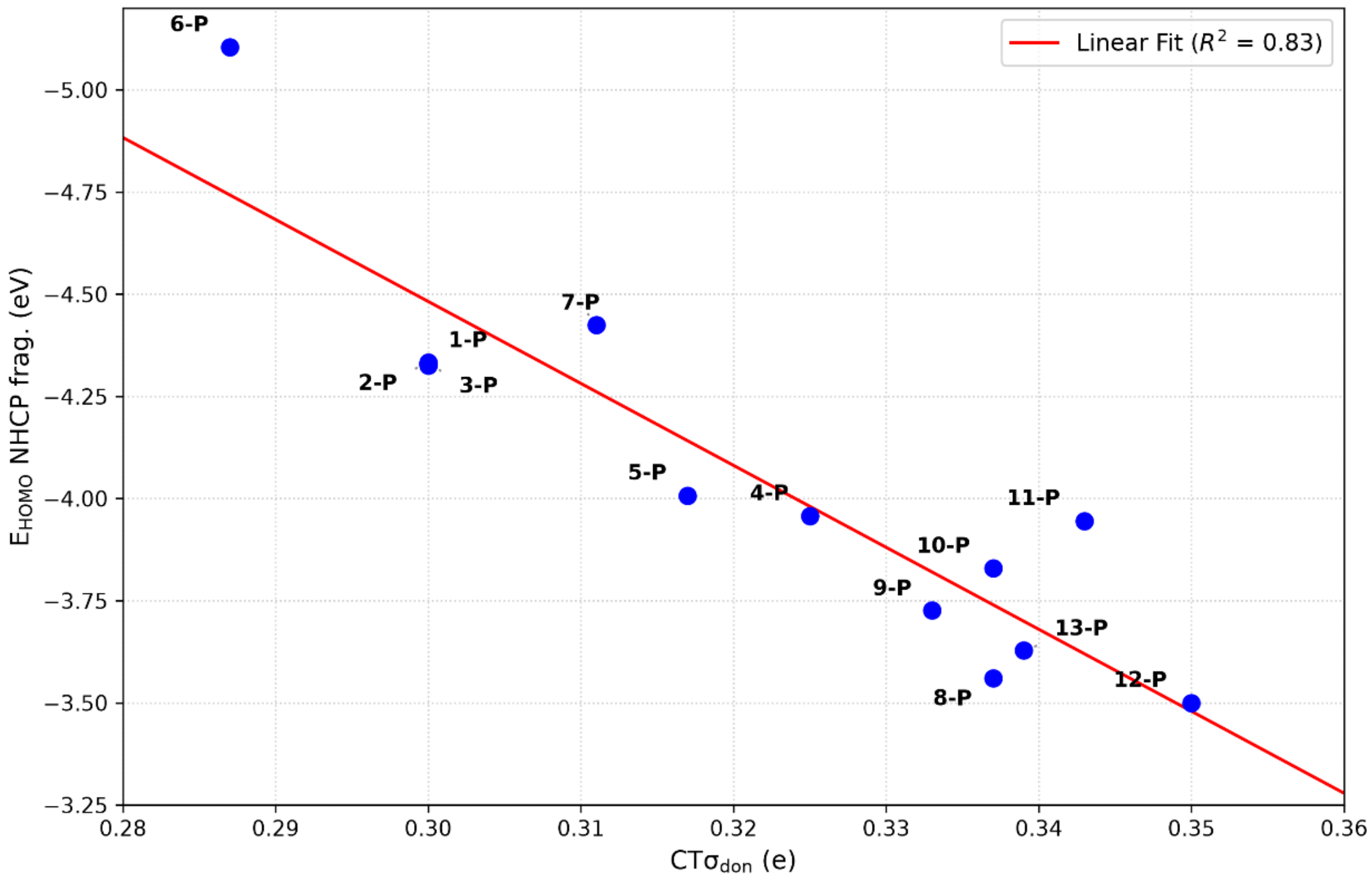

**Figure S44.** Correlation plot between CT $_{\sigma\ don}$ and NHCP HOMO fragment energy for the complete **1-P-13-P** series.

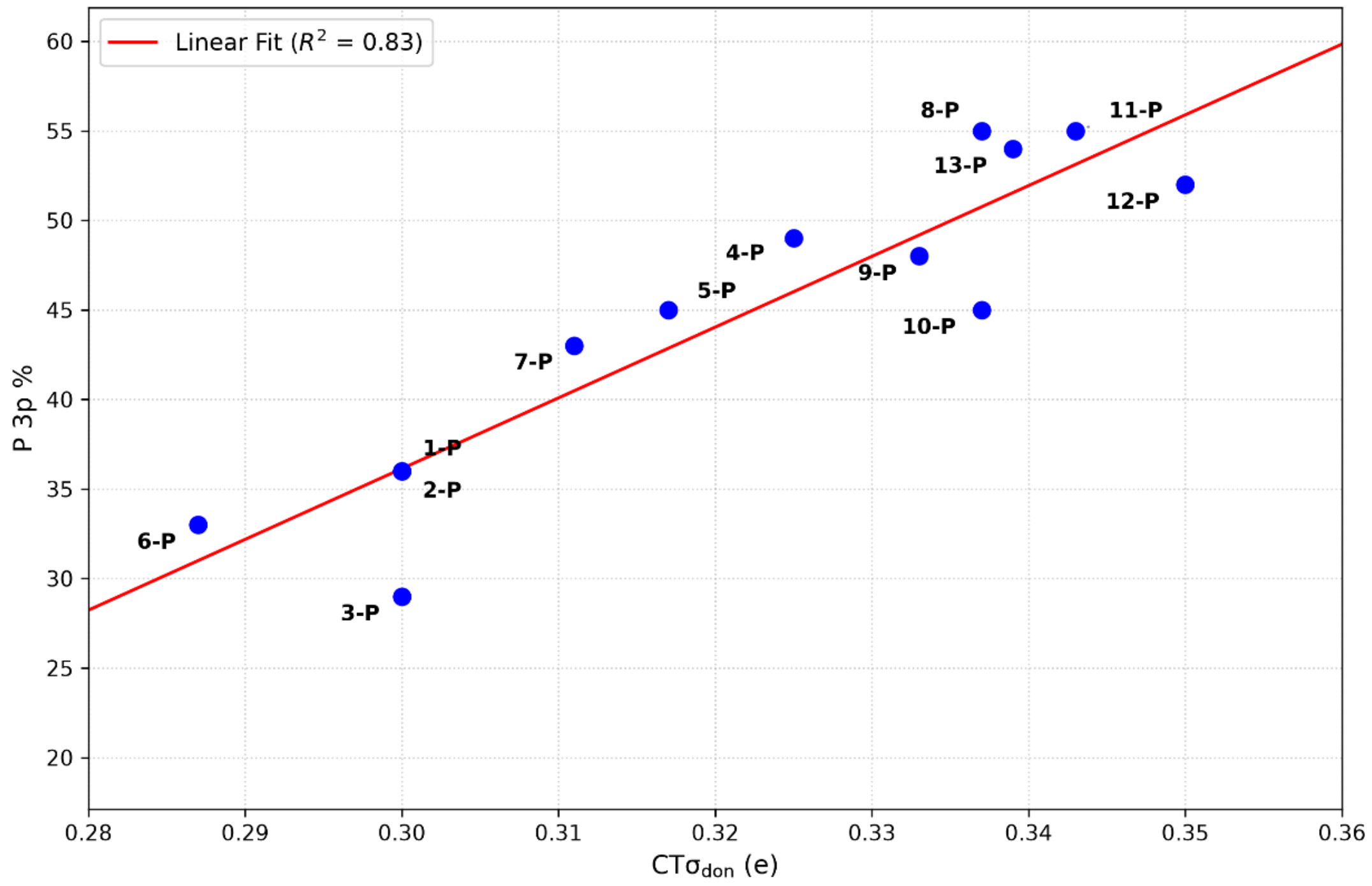

**Figure S45.** Correlation plot between CT $_{\sigma\ don}$ and NHCP HOMO fragment P 3p percentage contribution for the complete **1-P-13-P** series.

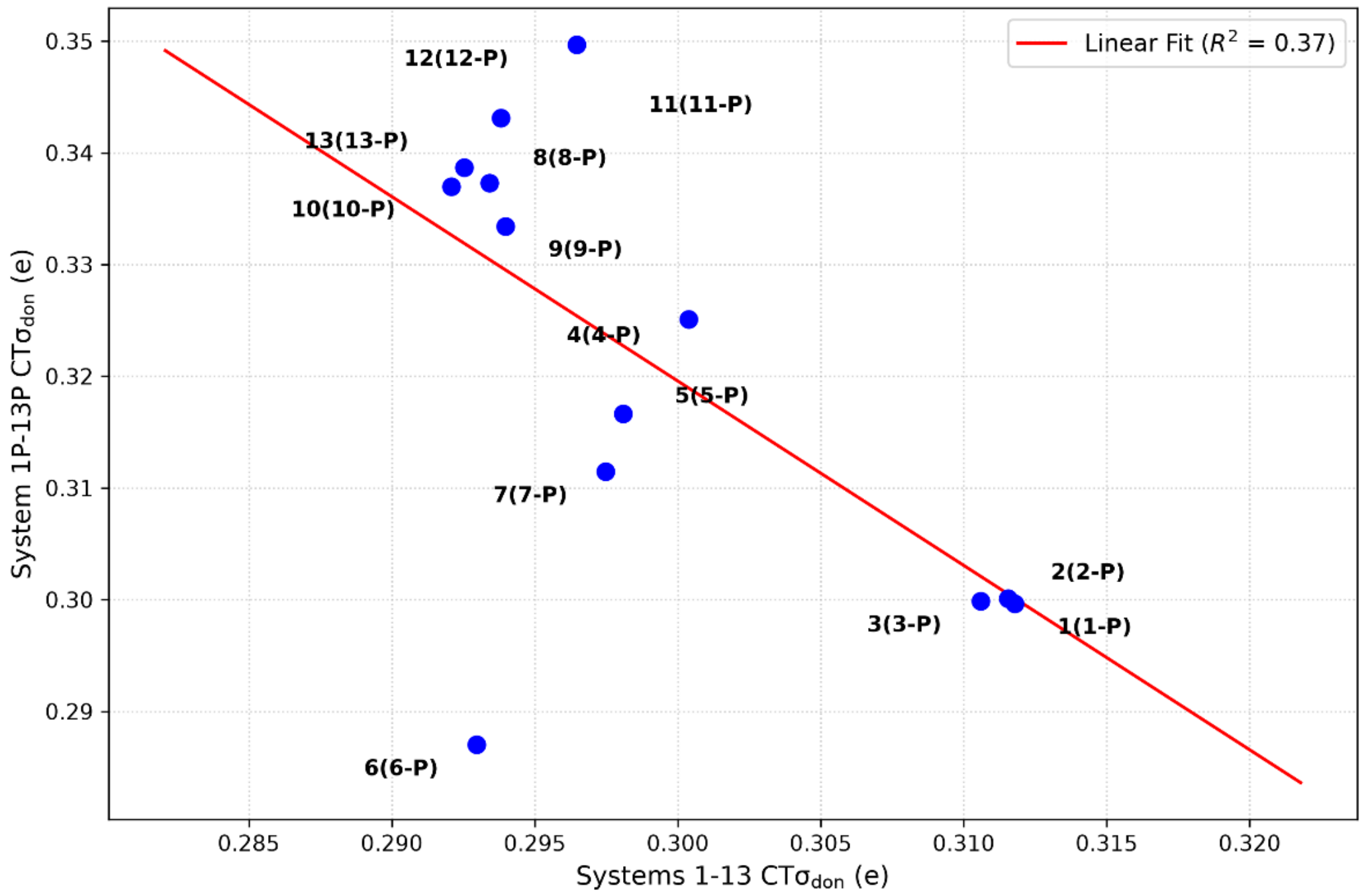

**Figure S46.** Correlation plot between CT $_{\sigma\ \mathrm{don}}$ for the complete NHCP **1-P-13-P** series and CT $_{\sigma\ \mathrm{don}}$ for the complete NHC **1-13** series.

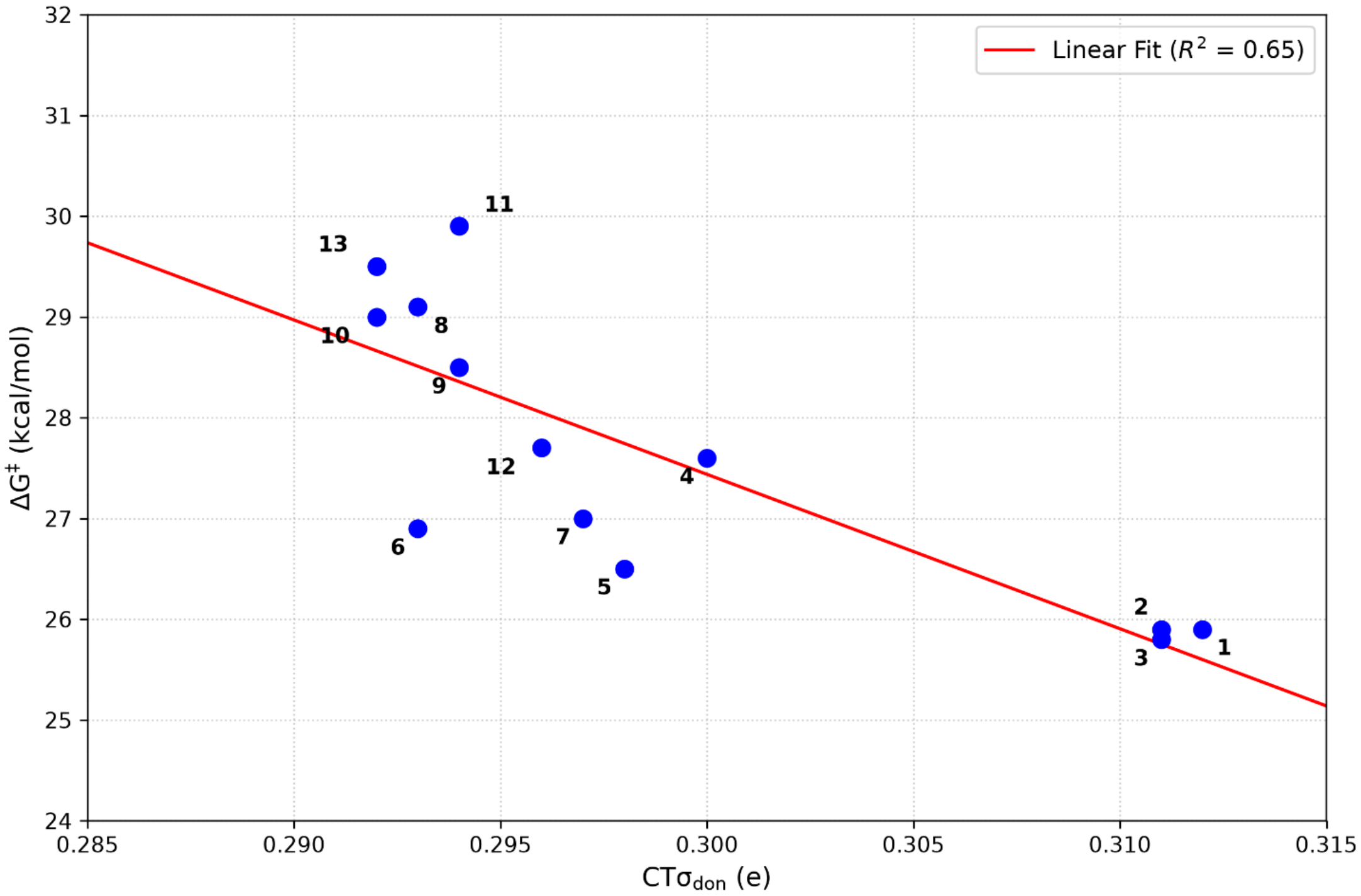

**Figure S47.** Correlation plot between CT $_{\sigma\ \mathrm{don}}$ and activation energy barrier $\Delta G^{\ddagger}$ for the complete NHC **1-13** series.

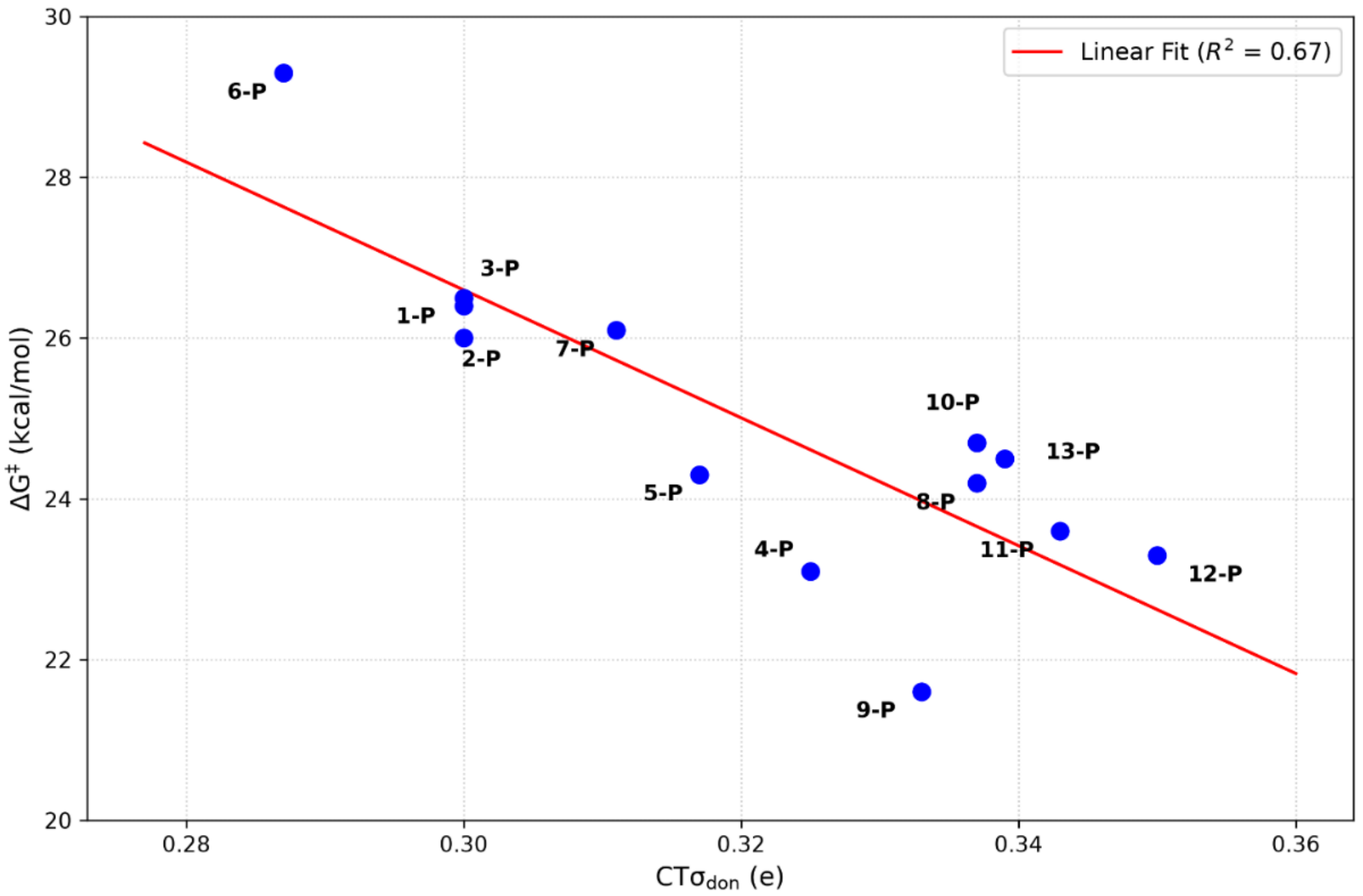

**Figure S48.** Correlation plot between CT $_{\sigma\ don}$ and activation energy barrier ΔG‡ for the complete NHCP **1-P-13-P** series.

## References

1 K. Morokuma, *J. Chem. Phys.*, 1971, **55**, 1236–1244.
2 T. Ziegler and A. Rauk, *Theor. Chim. Acta*, 1977, **46**, 1–10.
3 L. Zhao, M. von Hopffgarten, D. M. Andrada and G. Frenking, *WIREs Comput. Mol. Sci.*, 2018, **8**, e1345.
4 M. Mitoraj and A. Michalak, *J. Mol. Model.*, 2007, **13**, 347–355.
5 A. Michalak, M. Mitoraj and T. Ziegler, *J. Phys. Chem. A*, 2008, **112**, 1933–1939.
6 R. F. Nalewajski and J. Mrozek, *Int. J. Quantum Chem.*, 1994, **51**, 187–200.
7 R. F. Nalewajski, J. Mrozek and A. Michalak, *Int. J. Quantum Chem.*, 1997, **61**, 589–601.
8 T. Lu and F. Chen, *J. Phys. Chem. A*, 2013, **117**, 3100–3108.
9 L. Belpassi, I. Infante, F. Tarantelli and L. Visscher, *J. Am. Chem. Soc.*, 2008, **130**, 1048–1060.
10 G. Bistoni, S. Rampino, F. Tarantelli and L. Belpassi, *J. Chem. Phys.*, 2015, **142**, 084112.
11 E. Ramos-Cordoba, V. Postils and P. Salvador, *J. Chem. Theory Comput.*, 2015, **11**, 1501–1508.
12 I. Mayer, *Chem. Phys. Lett.*, 1983, **97**, 270-274.
13 https://www.aocdweb.com/OMtools/sambvca2.1/help/help.html
14 L. Falivene, L., Z. Cao, A. Petta, L. Serra, A. Poater, R. Oliva, V. Scarano, L. Cavallo, *Nat. Chem.*, 2019, **11**, 872–879. DOI: 10.1038/s41557-019-0319-5.
15 A. Poater, F. Ragone, S. Giudice, C. Costabile, R. Dorta, S.P. Nolan, L. Cavallo, *Organometallics,* 2008, **27**, 2679–2681. DOI: 10.1021/om8001119

**xyz geometries**

1-P ligand

64

| | | | |
|---|---|---|---|
| C | -2.303560 | 0.825824 | 0.951470 |
| C | -1.817324 | -0.278246 | 0.212900 |
| C | -2.223820 | -1.596394 | 0.522053 |
| C | -3.202565 | -1.775373 | 1.506379 |
| C | -3.742769 | -0.692581 | 2.191447 |
| C | -3.279676 | 0.592204 | 1.926338 |
| N | -0.862138 | -0.064958 | -0.837138 |
| C | 0.482744 | -0.001956 | -0.589974 |
| C | 1.190701 | 0.361517 | -1.901005 |
| C | 0.102104 | 0.052360 | -2.958555 |
| C | -1.263276 | 0.158420 | -2.264745 |
| P | 1.067304 | -0.297424 | 1.019320 |
| C | 1.562613 | 1.858006 | -1.892426 |
| C | 2.442857 | -0.470966 | -2.203495 |
| C | -1.930698 | 1.526432 | -2.457977 |
| C | -2.235187 | -0.916167 | -2.754231 |
| C | -1.587726 | -2.824917 | -0.106169 |
| C | -0.763957 | -3.595029 | 0.941101 |
| C | -1.754747 | 2.233651 | 0.792493 |
| C | -2.851290 | 3.261225 | 0.475616 |
| C | 2.889715 | -0.022773 | 0.909599 |
| C | 3.768896 | -1.120294 | 0.895251 |
| C | 5.151201 | -0.935411 | 0.923121 |
| C | 5.686086 | 0.353841 | 0.982369 |
| C | 4.827225 | 1.453984 | 1.034084 |
| C | 3.444530 | 1.264809 | 1.012714 |
| C | -2.624546 | -3.755728 | -0.753542 |
| C | -0.983152 | 2.652561 | 2.056700 |
| H | -3.671407 | 1.432682 | 2.498860 |
| H | -3.534003 | -2.784578 | 1.750001 |
| H | -0.888884 | -2.480271 | -0.878169 |
| H | 2.780004 | 2.126522 | 1.078520 |
| H | 5.234894 | 2.463048 | 1.100110 |
| H | 6.765859 | 0.499392 | 1.001858 |
| H | 5.814424 | -1.800334 | 0.896495 |
| H | 3.357892 | -2.129021 | 0.841944 |
| H | 0.174894 | 0.721493 | -3.824377 |
| H | 1.914030 | 2.157639 | -2.889752 |
| H | 0.709946 | 2.491462 | -1.622740 |
| H | 2.364727 | 2.046335 | -1.171817 |
| H | 2.241614 | -1.541463 | -2.071047 |
| H | 2.741935 | -0.300475 | -3.247409 |
| H | 3.282473 | -0.196989 | -1.559466 |
| H | -2.185749 | 1.658240 | -3.516995 |
| H | -2.859496 | 1.582344 | -1.878429 |
| H | -1.281676 | 2.354360 | -2.157933 |
| H | -1.784873 | -1.912446 | -2.714545 |
| H | -3.154866 | -0.917573 | -2.155658 |
| H | -2.509503 | -0.705739 | -3.795521 |
| H | 0.240403 | -0.976940 | -3.316696 |
| H | -4.507521 | -0.852167 | 2.951443 |
| H | -2.120280 | -4.584501 | -1.268716 |
| H | -3.286797 | -4.193565 | 0.005525 |
| H | -3.252269 | -3.228203 | -1.481100 |
| H | -0.253535 | -4.446239 | 0.469410 |
| H | -0.008028 | -2.940116 | 1.390718 |
| H | -1.410631 | -3.986042 | 1.739005 |
| H | -0.526613 | 3.641159 | 1.910878 |
| H | -1.654563 | 2.712664 | 2.924558 |
| H | -0.188410 | 1.929217 | 2.275918 |
| H | -1.037168 | 2.219533 | -0.035296 |
| H | -2.404460 | 4.246942 | 0.288466 |
| H | -3.438020 | 2.977928 | -0.406097 |
| H | -3.546321 | 3.367599 | 1.319486 |

2-P ligand

71

| | | | |
|---|---|---|---|
| C | 3.138200 | -2.564205 | 1.039488 |
| C | 2.235218 | -1.320331 | 1.012875 |
| C | 1.192417 | -1.391556 | -0.121112 |
| C | 1.925472 | -1.470882 | -1.503949 |
| C | 3.411248 | -1.827315 | -1.375688 |
| C | 3.609573 | -2.972983 | -0.379577 |
| C | 0.258872 | -2.606496 | 0.124788 |
| C | -1.149846 | -2.178899 | -0.306878 |
| N | -1.059204 | -0.703580 | -0.062339 |
| C | 0.217062 | -0.211098 | -0.073238 |
| C | -2.194278 | 0.134425 | 0.193612 |
| C | -2.878455 | 0.764865 | -0.866837 |
| C | -3.999714 | 1.546142 | -0.560684 |
| C | -4.417015 | 1.725275 | 0.753304 |
| C | -3.698772 | 1.140765 | 1.792718 |
| C | -2.576660 | 0.346558 | 1.540042 |
| C | -2.418752 | 0.678038 | -2.310982 |
| C | -1.995915 | 2.060504 | -2.836805 |
| C | -1.779949 | -0.200882 | 2.712770 |
| C | -2.655327 | -0.984211 | 3.702170 |
| P | 0.463056 | 1.510587 | -0.030927 |
| C | 2.299488 | 1.732595 | -0.017018 |
| C | 2.954127 | 2.064339 | 1.183637 |
| C | 4.318288 | 2.352881 | 1.205126 |
| C | 5.057891 | 2.334059 | 0.019541 |
| C | 4.417239 | 2.043681 | -1.186367 |
| C | 3.050643 | 1.758443 | -1.204014 |
| C | -1.413985 | -2.486961 | -1.788400 |
| C | -2.247943 | -2.824193 | 0.534564 |
| C | -1.036402 | 0.936277 | 3.435877 |
| C | -3.499819 | 0.069211 | -3.217740 |
| H | -4.545093 | 2.033640 | -1.368608 |
| H | -4.011077 | 1.311388 | 2.822690 |
| H | -1.019440 | -0.881015 | 2.309935 |
| H | 2.556264 | 1.552920 | -2.153388 |
| H | 4.981920 | 2.043651 | -2.118960 |
| H | 6.124252 | 2.557485 | 0.034368 |
| H | 4.806499 | 2.592804 | 2.150012 |
| H | 2.383348 | 2.075692 | 2.112751 |
| H | 0.594116 | -3.504497 | -0.405907 |
| H | -1.345641 | -3.570315 | -1.952216 |
| H | -2.418796 | -2.165170 | -2.081400 |
| H | -0.686872 | -1.992847 | -2.441639 |
| H | -2.057823 | -2.703600 | 1.605373 |
| H | -3.229105 | -2.391629 | 0.300999 |
| H | -2.288417 | -3.897644 | 0.312624 |
| H | 0.241066 | -2.837024 | 1.199185 |
| H | -5.292986 | 2.336514 | 0.970513 |
| H | -2.030329 | -1.442618 | 4.480233 |
| H | -3.375260 | -0.323500 | 4.203687 |
| H | -3.224742 | -1.778673 | 3.203698 |
| H | -0.407935 | 0.529373 | 4.239975 |
| H | -0.397120 | 1.482259 | 2.731374 |
| H | -1.747988 | 1.644059 | 3.883764 |
| H | -1.612265 | 1.972703 | -3.862612 |
| H | -2.846321 | 2.756142 | -2.852059 |
| H | -1.206323 | 2.483818 | -2.204603 |
| H | -1.531625 | 0.035431 | -2.337663 |
| H | -3.108684 | -0.082538 | -4.232695 |
| H | -3.857382 | -0.896449 | -2.838345 |
| H | -4.369582 | 0.736323 | -3.291048 |
| H | 1.810334 | -0.517842 | -2.031355 |
| H | 1.429989 | -2.233506 | -2.120484 |
| H | 3.805781 | -2.099493 | -2.364647 |
| H | 3.978623 | -0.944328 | -1.046514 |
| H | 4.661432 | -3.287639 | -0.356570 |
| H | 3.034661 | -3.844650 | -0.729472 |
| H | 2.615535 | -3.405198 | 1.516448 |
| H | 4.005529 | -2.341708 | 1.676395 |
| H | 2.851193 | -0.427949 | 0.877564 |
| H | 1.715573 | -1.197249 | 1.973424 |

3-P ligand

83

| | | | |
|---|---|---|---|
| C | -3.118718 | 0.311964 | -2.311578 |
| C | -2.187595 | 0.621789 | -1.124107 |
| C | -0.998425 | -0.380896 | -1.121540 |
| C | -1.565860 | -1.825033 | -0.880412 |
| C | -3.055432 | -1.983390 | -1.215704 |
| C | -3.376395 | -1.201963 | -2.492763 |
| C | -0.197864 | -0.342425 | -2.451886 |
| C | 1.285630 | -0.548079 | -2.101424 |
| N | 1.295026 | -0.170200 | -0.651143 |
| C | 0.065603 | -0.090831 | -0.062952 |
| C | 2.504686 | 0.026040 | 0.097816 |
| C | 3.095460 | -1.056832 | 0.788948 |
| C | 4.302548 | -0.830188 | 1.460684 |
| C | 4.893438 | 0.428581 | 1.483820 |
| C | 4.260491 | 1.499581 | 0.861071 |
| C | 3.054823 | 1.326493 | 0.173927 |
| C | 2.445741 | -2.426006 | 0.903879 |
| C | 2.028047 | -2.705921 | 2.358641 |
| C | 2.348162 | 2.544849 | -0.392561 |
| C | 3.233806 | 3.346565 | -1.357338 |
| P | -0.032326 | 0.271004 | 1.640160 |
| C | -1.835770 | 0.241743 | 2.041788 |
| C | -2.480097 | 1.447783 | 2.371517 |
| C | -3.808715 | 1.465069 | 2.797125 |
| C | -4.522944 | 0.271438 | 2.916304 |
| C | -3.886280 | -0.940557 | 2.637703 |
| C | -2.554750 | -0.954002 | 2.224344 |
| C | 1.737360 | -1.999965 | -2.302436 |
| C | 2.202690 | 0.356620 | -2.924600 |
| C | -1.856118 | 2.128582 | -0.973781 |
| C | -3.144634 | 2.916462 | -0.693786 |
| C | -3.426676 | -3.460971 | -1.326670 |
| C | 1.832832 | 3.448537 | 0.740760 |
| C | 3.358243 | -3.552886 | 0.395780 |
| C | -1.097890 | 2.806343 | -2.122693 |
| H | 4.778269 | -1.655324 | 1.990414 |
| H | 4.701740 | 2.494174 | 0.920595 |
| H | 1.468853 | 2.187458 | -0.939103 |
| H | -2.060914 | -1.907948 | 2.045295 |
| H | -4.425455 | -1.880947 | 2.754375 |
| H | -5.563008 | 0.283423 | 3.241473 |
| H | -4.289167 | 2.415065 | 3.032191 |
| H | -1.932384 | 2.385705 | 2.275690 |
| H | -0.544488 | -1.113501 | -3.148055 |
| H | 1.673898 | -2.254126 | -3.368123 |
| H | 2.779296 | -2.126270 | -1.987048 |
| H | 1.114955 | -2.705684 | -1.743122 |
| H | 1.883895 | 1.402490 | -2.880300 |
| H | 3.241021 | 0.288763 | -2.575500 |
| H | 2.175468 | 0.036788 | -3.973739 |
| H | -0.322752 | 0.621593 | -2.949245 |
| H | 5.836067 | 0.580685 | 2.009712 |
| H | 2.654021 | 4.158675 | -1.816503 |
| H | 4.080739 | 3.803857 | -0.827827 |
| H | 3.640058 | 2.717217 | -2.157911 |
| H | 1.254729 | 4.284237 | 0.323035 |
| H | 1.184533 | 2.879048 | 1.417722 |
| H | 2.667084 | 3.867222 | 1.320932 |
| H | 1.501171 | -3.668080 | 2.420241 |
| H | 2.905895 | -2.756715 | 3.017687 |
| H | 1.358603 | -1.917169 | 2.722247 |
| H | 1.530731 | -2.412966 | 0.301409 |
| H | 2.819997 | -4.510157 | 0.406237 |
| H | 3.712146 | -3.370383 | -0.625612 |
| H | 4.241321 | -3.662037 | 1.039831 |
| H | -1.386278 | -2.124279 | 0.156963 |
| H | -0.995216 | -2.526962 | -1.505950 |
| H | -3.643428 | -1.549502 | -0.390815 |
| H | -4.419725 | -1.376987 | -2.791417 |
| H | -2.754566 | -1.603849 | -3.308958 |
| H | -2.710661 | 0.734808 | -3.240398 |
| H | -4.072985 | 0.824994 | -2.135823 |
| H | -2.754709 | 0.390253 | -0.212984 |
| H | -1.220822 | 2.204776 | -0.077731 |
| H | -2.905995 | 3.934663 | -0.357990 |
| H | -3.765222 | 3.005286 | -1.596663 |
| H | -3.745293 | 2.434039 | 0.087817 |
| H | -1.038475 | 3.887282 | -1.934907 |
| H | -0.068973 | 2.445363 | -2.223212 |
| H | -1.604455 | 2.670551 | -3.088739 |
| H | -4.503875 | -3.587873 | -1.501640 |
| H | -2.890297 | -3.935513 | -2.162384 |
| H | -3.165999 | -4.006679 | -0.408962 |

4-P ligand

61

| | | | |
|---|---|---|---|
| C | 3.650016 | -0.264109 | 1.207228 |
| C | 3.033630 | -0.466862 | -0.040204 |
| C | 3.636455 | -0.042857 | -1.235034 |
| C | 4.887634 | 0.578684 | -1.158158 |
| C | 5.531968 | 0.789232 | 0.063949 |
| C | 4.896476 | 0.362902 | 1.236929 |
| N | 1.767231 | -1.122963 | -0.064153 |
| C | 0.561958 | -0.453723 | -0.081240 |
| N | -0.424319 | -1.401580 | -0.241615 |
| C | 0.137026 | -2.766552 | -0.186759 |
| C | 1.622917 | -2.528829 | -0.446243 |
| P | 0.580550 | 1.300134 | 0.092728 |
| C | -1.815796 | -1.235010 | 0.010365 |
| C | -2.270100 | -0.851704 | 1.285982 |
| C | -3.639364 | -0.640039 | 1.460103 |
| C | -4.556924 | -0.837522 | 0.424893 |
| C | -4.079105 | -1.291208 | -0.807281 |
| C | -2.716440 | -1.488382 | -1.038495 |
| C | -1.318338 | -0.657125 | 2.429574 |
| C | -6.014016 | -0.518052 | 0.620810 |
| C | -2.225474 | -1.900629 | -2.399324 |
| C | 2.940518 | -0.221128 | -2.556988 |
| C | 6.867700 | 1.482819 | 0.128021 |
| C | 2.956436 | -0.691674 | 2.471171 |
| C | -1.185969 | 1.796806 | -0.039879 |
| C | -1.975257 | 1.568630 | -1.181186 |
| C | -3.282694 | 2.042179 | -1.258044 |
| C | -3.823688 | 2.792142 | -0.210295 |
| C | -3.041338 | 3.065784 | 0.912663 |
| C | -1.738863 | 2.573056 | 0.995810 |
| H | -4.781503 | -1.454386 | -1.626469 |
| H | -3.997185 | -0.307450 | 2.435969 |
| H | 5.381720 | 0.528801 | 2.200509 |
| H | 5.364607 | 0.917161 | -2.079664 |
| H | -2.962813 | -1.649727 | -3.170461 |
| H | -1.276368 | -1.405634 | -2.640678 |
| H | -2.048222 | -2.985519 | -2.456488 |
| H | -1.861731 | -0.600834 | 3.379330 |
| H | -0.589102 | -1.475736 | 2.491993 |
| H | -0.734782 | 0.268241 | 2.300803 |
| H | -6.339694 | -0.732947 | 1.646421 |
| H | -6.192790 | 0.552884 | 0.438582 |
| H | -6.648751 | -1.083104 | -0.072850 |
| H | 3.550960 | -0.424177 | 3.351883 |
| H | 1.973776 | -0.205419 | 2.548422 |
| H | 2.782567 | -1.776571 | 2.488236 |
| H | 3.450648 | 0.347783 | -3.342207 |
| H | 2.920409 | -1.275985 | -2.867876 |
| H | 1.900651 | 0.127568 | -2.491063 |
| H | 7.296939 | 1.616833 | -0.871881 |
| H | 6.770420 | 2.476426 | 0.589190 |
| H | 7.582691 | 0.913136 | 0.737004 |
| H | -1.138804 | 2.777702 | 1.883871 |
| H | -3.449711 | 3.656477 | 1.733105 |
| H | -4.844943 | 3.167234 | -0.274088 |
| H | -3.882628 | 1.827794 | -2.142579 |
| H | -1.555686 | 1.004570 | -2.013485 |
| H | 2.272474 | -3.166680 | 0.164770 |
| H | 1.880868 | -2.680175 | -1.507579 |
| H | -0.045053 | -3.202498 | 0.808780 |
| H | -0.331549 | -3.409733 | -0.939484 |

5-P ligand

79

79

| | | | |
|---|---|---|---|
| C | 1.781039 | -1.709454 | -2.403690 |
| C | 1.421021 | -0.426301 | -1.951572 |
| C | 2.393696 | 0.586292 | -2.025718 |
| C | 3.688166 | 0.312501 | -2.460800 |
| C | 4.035103 | -0.974922 | -2.877993 |
| C | 3.070432 | -1.982473 | -2.862725 |
| P | -0.358521 | -0.163813 | -1.533955 |
| C | -0.410996 | 0.247661 | 0.176385 |
| N | -1.649025 | 0.459489 | 0.755621 |
| C | -1.566367 | 0.548278 | 2.215755 |
| C | -0.085249 | 0.857999 | 2.430076 |
| N | 0.523780 | 0.423328 | 1.161750 |
| C | -2.874984 | 0.114035 | 0.110604 |
| C | -3.293117 | -1.233029 | 0.083435 |
| C | -4.512349 | -1.528162 | -0.533616 |
| C | -5.284303 | -0.522956 | -1.113433 |
| C | -4.848136 | 0.798830 | -1.085863 |
| C | -3.636986 | 1.142145 | -0.474100 |
| C | -2.459838 | -2.343944 | 0.696835 |
| C | -3.171002 | -2.970632 | 1.906724 |
| C | -3.152009 | 2.579227 | -0.474927 |
| C | -4.164134 | 3.522330 | 0.190968 |
| C | 1.933899 | 0.255569 | 1.056612 |
| C | 2.447486 | -1.063951 | 1.019814 |
| C | 3.831501 | -1.229506 | 0.961693 |
| C | 4.685969 | -0.128586 | 0.962473 |
| C | 4.160159 | 1.154456 | 0.988118 |
| C | 2.775941 | 1.381671 | 1.016257 |
| C | 1.515497 | -2.266501 | 1.034107 |
| C | 0.937410 | -2.525494 | 2.436575 |
| C | 2.345148 | 2.847316 | 0.960068 |
| C | 2.476672 | 3.512600 | 2.343559 |
| C | 2.165943 | -3.550330 | 0.511340 |
| C | 0.995376 | 3.190963 | 0.311789 |
| C | -2.807113 | 3.036046 | -1.901313 |
| C | -2.083077 | -3.409202 | -0.342613 |
| H | 4.830295 | 2.014340 | 0.945737 |
| H | 4.251567 | -2.231333 | 0.907971 |
| H | -4.860147 | -2.560809 | -0.568002 |
| H | -5.454292 | 1.576195 | -1.551136 |
| H | 3.110684 | 3.317175 | 0.322369 |
| H | 0.678020 | -2.028476 | 0.360312 |
| H | -1.520616 | -1.900917 | 1.048363 |
| H | -2.223084 | 2.603932 | 0.109685 |
| H | 1.035868 | -2.505563 | -2.379118 |
| H | 3.323013 | -2.988481 | -3.199200 |
| H | 5.047369 | -1.186959 | -3.221206 |
| H | 4.429907 | 1.111120 | -2.479183 |
| H | 2.135492 | 1.601034 | -1.732825 |
| H | -1.865335 | -0.402675 | 2.687152 |
| H | -2.222639 | 1.341797 | 2.592201 |
| H | 0.354649 | 0.311006 | 3.271574 |
| H | 0.081811 | 1.931291 | 2.596668 |
| H | -6.230091 | -0.773704 | -1.593760 |
| H | 5.764732 | -0.276114 | 0.916713 |
| H | -2.401188 | 4.056895 | -1.888970 |
| H | -2.057135 | 2.367356 | -2.342777 |
| H | -3.698924 | 3.032269 | -2.544045 |
| H | -3.764667 | 4.544805 | 0.230946 |
| H | -5.108836 | 3.554851 | -0.369292 |
| H | -4.392758 | 3.201361 | 1.215902 |
| H | -1.422282 | -4.161455 | 0.109569 |
| H | -2.971636 | -3.928854 | -0.726222 |
| H | -1.556063 | -2.941809 | -1.183825 |
| H | -2.531068 | -3.729131 | 2.377505 |
| H | -3.419699 | -2.213240 | 2.662251 |
| H | -4.107342 | -3.458943 | 1.603541 |
| H | 2.275166 | 4.589949 | 2.270501 |
| H | 1.765795 | 3.083237 | 3.061842 |
| H | 3.485607 | 3.377024 | 2.752693 |
| H | 0.993450 | 4.256660 | 0.047740 |
| H | 0.799555 | 2.615121 | -0.599890 |
| H | 0.146423 | 3.030183 | 0.985378 |
| H | 1.398878 | -4.326493 | 0.392218 |
| H | 2.644915 | -3.390068 | -0.460882 |
| H | 2.919566 | -3.938936 | 1.210984 |
| H | 0.262103 | -3.391442 | 2.412431 |
| H | 1.743505 | -2.741923 | 3.151315 |
| H | 0.368696 | -1.671573 | 2.817124 |

6-P ligand

68

| | | | |
|---|---|---|---|
| C | 2.098243 | 1.981453 | 0.203994 |
| N | 1.743096 | 0.654354 | -0.008499 |
| C | 0.427479 | 0.153032 | 0.052284 |
| N | -0.580074 | 1.117436 | -0.066358 |
| C | -0.342753 | 2.495704 | 0.036672 |
| C | 0.978960 | 2.901858 | 0.663733 |
| C | 2.838770 | -0.289354 | -0.130802 |
| C | 3.300044 | -0.629986 | -1.408911 |
| C | 4.371821 | -1.520643 | -1.499052 |
| C | 4.984982 | -2.054481 | -0.360375 |
| C | 4.512566 | -1.663442 | 0.896314 |
| C | 3.446072 | -0.772093 | 1.036783 |
| C | 2.679603 | -0.038916 | -2.641295 |
| C | 2.978517 | -0.332103 | 2.394854 |
| C | 6.122706 | -3.032323 | -0.489566 |
| P | 0.292863 | -1.579811 | 0.149253 |
| C | -1.454838 | -1.903979 | 0.620807 |
| C | -1.837506 | -1.756446 | 1.965021 |
| C | -3.100094 | -2.162158 | 2.394141 |
| C | -3.992026 | -2.742994 | 1.491149 |
| C | -3.607988 | -2.930145 | 0.161435 |
| C | -2.345971 | -2.526044 | -0.268424 |
| C | -1.918193 | 0.749490 | -0.480578 |
| C | -2.149527 | 0.366508 | -1.806653 |
| C | -3.457443 | 0.030674 | -2.170620 |
| C | -4.513344 | 0.080664 | -1.259681 |
| C | -4.246375 | 0.515660 | 0.042315 |
| C | -2.961620 | 0.869498 | 0.450243 |
| O | -1.191016 | 3.305686 | -0.306595 |
| C | -1.042558 | 0.282917 | -2.816054 |
| C | -5.900034 | -0.352256 | -1.650533 |
| C | -2.718917 | 1.385853 | 1.838289 |
| C | 0.837271 | 2.724673 | 2.202348 |
| C | 1.309912 | 4.355782 | 0.325183 |
| O | 3.261045 | 2.350426 | 0.096506 |
| H | -5.060230 | 0.574990 | 0.766194 |
| H | -3.649876 | -0.281143 | -3.198403 |
| H | 4.745866 | -1.792107 | -2.487525 |
| H | 4.992129 | -2.053692 | 1.795090 |
| H | -3.539309 | 1.101616 | 2.505627 |
| H | -1.784536 | 0.996826 | 2.257890 |
| H | -2.645915 | 2.482971 | 1.829739 |
| H | -1.445005 | 0.305426 | -3.834713 |
| H | -0.328266 | 1.107887 | -2.707300 |
| H | -0.479278 | -0.654719 | -2.693486 |
| H | -6.016882 | -0.393308 | -2.739983 |
| H | -6.112853 | -1.356099 | -1.252223 |
| H | -6.661697 | 0.325319 | -1.243471 |
| H | 3.210896 | -0.369512 | -3.540051 |
| H | 1.628428 | -0.338466 | -2.734772 |
| H | 2.705904 | 1.058621 | -2.608777 |
| H | 3.474029 | -0.910298 | 3.181865 |
| H | 3.204451 | 0.730576 | 2.560363 |
| H | 1.893052 | -0.466022 | 2.499354 |
| H | 6.732966 | -3.058220 | 0.421166 |
| H | 5.742678 | -4.050005 | -0.663873 |
| H | 6.771887 | -2.777468 | -1.336824 |
| H | -2.054846 | -2.674333 | -1.307780 |
| H | -4.296166 | -3.392460 | -0.545809 |
| H | -4.980213 | -3.058630 | 1.824437 |
| H | -3.386868 | -2.025358 | 3.436686 |
| H | -1.137413 | -1.318814 | 2.676875 |
| H | 1.396840 | 4.497931 | -0.758082 |
| H | 0.517137 | 5.012070 | 0.696718 |
| H | 2.262235 | 4.630303 | 0.788642 |
| H | 0.030707 | 3.370353 | 2.568259 |
| H | 0.610441 | 1.687435 | 2.473383 |
| H | 1.775166 | 3.019434 | 2.686788 |

7-P ligand

69

| | | | |
|---|---|---|---|
| C | 2.076469 | -2.057809 | -0.224308 |
| N | 1.710294 | -0.740082 | 0.045413 |
| C | 0.387193 | -0.228627 | -0.005854 |
| N | -0.607296 | -1.166455 | 0.051226 |
| C | -0.291324 | -2.591428 | 0.171976 |
| C | 0.941305 | -2.984233 | -0.632178 |
| C | 2.802976 | 0.207806 | 0.161471 |
| C | 3.259096 | 0.564259 | 1.436957 |
| C | 4.325087 | 1.463086 | 1.523712 |
| C | 4.935757 | 1.994231 | 0.382767 |
| C | 4.467684 | 1.590760 | -0.871720 |
| C | 3.408000 | 0.690721 | -1.007731 |
| C | 2.626700 | -0.003722 | 2.674873 |
| C | 2.938037 | 0.247975 | -2.363775 |
| C | 6.062309 | 2.986510 | 0.505776 |
| P | 0.273860 | 1.522828 | -0.107019 |
| C | -1.478877 | 1.812287 | -0.607134 |
| C | -1.848353 | 1.639481 | -1.952252 |
| C | -3.123047 | 1.988491 | -2.396405 |
| C | -4.046417 | 2.540746 | -1.506863 |
| C | -3.680686 | 2.753845 | -0.175928 |
| C | -2.406644 | 2.402694 | 0.266707 |
| C | -1.948432 | -0.845953 | 0.460306 |
| C | -2.200808 | -0.453584 | 1.785301 |
| C | -3.523806 | -0.195721 | 2.156557 |
| C | -4.583326 | -0.323939 | 1.257061 |
| C | -4.299164 | -0.755483 | -0.041827 |
| C | -2.997009 | -1.033476 | -0.457337 |
| C | -1.093140 | -0.259769 | 2.778149 |
| C | -5.989357 | 0.033922 | 1.655776 |
| C | -2.733327 | -1.512590 | -1.855841 |
| C | 0.709322 | -2.796186 | -2.144939 |
| C | 1.308013 | -4.437326 | -0.323456 |
| O | 3.252855 | -2.403887 | -0.188963 |
| H | -5.112186 | -0.863422 | -0.761371 |
| H | -3.725616 | 0.123795 | 3.180465 |
| H | 4.693463 | 1.746910 | 2.511028 |
| H | 4.943077 | 1.982083 | -1.772466 |
| H | -3.558909 | -1.235155 | -2.520045 |
| H | -1.806839 | -1.085551 | -2.252175 |
| H | -2.630309 | -2.608068 | -1.890749 |
| H | -1.483674 | -0.255757 | 3.801983 |
| H | -0.326634 | -1.040369 | 2.701124 |
| H | -0.583228 | 0.699403 | 2.594385 |
| H | -6.108405 | 0.041776 | 2.745856 |
| H | -6.247202 | 1.037685 | 1.284698 |
| H | -6.719427 | -0.666544 | 1.230241 |
| H | 3.190127 | 0.289273 | 3.567432 |
| H | 1.596704 | 0.358721 | 2.785603 |
| H | 2.586758 | -1.100234 | 2.634398 |
| H | 3.426285 | 0.829932 | -3.152884 |
| H | 3.168177 | -0.813806 | -2.529756 |
| H | 1.851017 | 0.378295 | -2.457903 |
| H | 6.674265 | 3.011387 | -0.404014 |
| H | 5.670577 | 4.001622 | 0.669658 |
| H | 6.713193 | 2.747494 | 1.356473 |
| H | -2.131480 | 2.567523 | 1.308223 |
| H | -4.393267 | 3.192842 | 0.522443 |
| H | -5.043820 | 2.813097 | -1.851114 |
| H | -3.394619 | 1.832173 | -3.440597 |
| H | -1.123164 | 1.225947 | -2.653747 |
| H | 1.489389 | -4.581736 | 0.749135 |
| H | 0.490191 | -5.100998 | -0.632872 |
| H | 2.217143 | -4.723740 | -0.861878 |
| H | -0.105691 | -3.451481 | -2.477799 |
| H | 0.442993 | -1.761745 | -2.389324 |
| H | 1.616908 | -3.060954 | -2.700118 |
| H | -1.165169 | -3.157190 | -0.174611 |
| H | -0.141209 | -2.846566 | 1.236938 |

8-P ligand

33

| | | | |
|---|---|---|---|
| N | 2.039413 | 0.957531 | -0.205679 |
| C | 0.826344 | 0.498258 | 0.252365 |
| N | 1.014760 | -0.852292 | 0.441261 |
| C | 2.316331 | -1.222020 | 0.094084 |
| C | 2.964269 | -0.082323 | -0.294934 |
| P | -0.560421 | 1.551362 | 0.651038 |
| C | 0.045746 | -1.721676 | 1.080122 |
| C | 2.797514 | -2.625425 | 0.178307 |
| C | 4.361078 | 0.131975 | -0.754796 |
| C | 2.297978 | 2.343390 | -0.548661 |
| C | -1.961638 | 0.516950 | 0.094844 |
| C | -1.921790 | -0.367640 | -1.005391 |
| C | -3.058033 | -1.060347 | -1.419896 |
| C | -4.282111 | -0.874783 | -0.769967 |
| C | -4.348301 | 0.016019 | 0.306495 |
| C | -3.208565 | 0.692975 | 0.735365 |
| H | -3.274562 | 1.369380 | 1.589569 |
| H | -5.295088 | 0.176561 | 0.823954 |
| H | -5.170919 | -1.411493 | -1.099940 |
| H | -2.990898 | -1.740838 | -2.269896 |
| H | -0.986191 | -0.499625 | -1.550972 |
| H | 3.823888 | -2.693221 | -0.197005 |
| H | 2.795351 | -3.000020 | 1.212545 |
| H | 2.172968 | -3.302846 | -0.420994 |
| H | 4.917425 | -0.810360 | -0.721780 |
| H | 4.400345 | 0.507822 | -1.787629 |
| H | 4.887843 | 0.857111 | -0.118264 |
| H | 2.990087 | 2.392840 | -1.394881 |
| H | 1.335959 | 2.801694 | -0.814633 |
| H | 2.722618 | 2.891657 | 0.302348 |
| H | 0.571458 | -2.521132 | 1.609237 |
| H | -0.533777 | -1.117381 | 1.788966 |
| H | -0.651690 | -2.152576 | 0.350767 |

9-P ligand

59

| | | | |
|---|---|---|---|
| C | -1.935375 | -1.606707 | 1.048062 |
| C | -1.193694 | -1.789523 | -0.134046 |
| C | -1.799919 | -2.510385 | -1.180999 |
| C | -3.101360 | -2.999101 | -1.063572 |
| C | -3.834541 | -2.771399 | 0.102100 |
| C | -3.243240 | -2.071252 | 1.157778 |
| P | 0.569238 | -1.303313 | -0.314006 |
| C | 0.562616 | 0.435318 | 0.027642 |
| N | -0.394722 | 1.435490 | 0.127833 |
| C | 0.217507 | 2.676236 | 0.334622 |
| C | 1.555472 | 2.473311 | 0.351818 |
| N | 1.766605 | 1.111927 | 0.171327 |
| C | -1.809759 | 1.282762 | -0.001115 |
| C | -2.349293 | 0.941176 | -1.250807 |
| C | -3.727181 | 0.728662 | -1.325523 |
| C | -4.560210 | 0.879930 | -0.214008 |
| C | -3.989571 | 1.279330 | 0.998266 |
| C | -2.613728 | 1.475505 | 1.132168 |
| C | -1.472280 | 0.798232 | -2.458503 |
| C | -2.010158 | 1.810846 | 2.469121 |
| C | -6.027667 | 0.562763 | -0.308691 |
| C | 3.047579 | 0.475324 | 0.107808 |
| C | 3.604193 | -0.038709 | 1.285576 |
| C | 4.857469 | -0.653387 | 1.198131 |
| C | 5.538428 | -0.763752 | -0.018030 |
| C | 4.939530 | -0.245525 | -1.173574 |
| C | 3.690497 | 0.376624 | -1.135410 |
| C | 2.850199 | 0.041472 | 2.582829 |
| C | 6.882412 | -1.439794 | -0.094689 |
| C | 3.027782 | 0.899574 | -2.379544 |
| H | -0.369421 | 3.578923 | 0.428682 |
| H | 2.381119 | 3.161249 | 0.470504 |
| H | -4.625664 | 1.398438 | 1.876788 |
| H | -4.159043 | 0.430709 | -2.281951 |
| H | 5.454354 | -0.334886 | -2.131653 |
| H | 5.308283 | -1.061943 | 2.103786 |
| H | -2.735617 | 1.638887 | 3.271904 |
| H | -1.119561 | 1.198919 | 2.665177 |
| H | -1.689876 | 2.861014 | 2.525354 |
| H | -2.075010 | 0.669213 | -3.363961 |
| H | -0.827075 | 1.678346 | -2.585546 |
| H | -0.801133 | -0.070240 | -2.351544 |
| H | -6.418626 | 0.763089 | -1.313873 |
| H | -6.195647 | -0.504588 | -0.098082 |
| H | -6.614290 | 1.139955 | 0.416670 |
| H | 3.655516 | 0.721429 | -3.259342 |
| H | 2.059966 | 0.399980 | -2.532162 |
| H | 2.827911 | 1.977632 | -2.308247 |
| H | 3.428094 | -0.404735 | 3.399300 |
| H | 2.614671 | 1.080740 | 2.849765 |
| H | 1.894628 | -0.496716 | 2.493166 |
| H | 7.241935 | -1.728282 | 0.899938 |
| H | 6.829898 | -2.347063 | -0.713355 |
| H | 7.631914 | -0.779979 | -0.553091 |
| H | -1.242589 | -2.673062 | -2.104817 |
| H | -3.548780 | -3.549148 | -1.892074 |
| H | -4.855146 | -3.142798 | 0.192213 |
| H | -3.802915 | -1.892678 | 2.076169 |
| H | -1.474752 | -1.090482 | 1.889853 |

10-P ligand

77

| | | | |
|---|---|---|---|
| C | 3.570197 | -0.913164 | -0.050121 |
| C | 2.927264 | 0.332090 | -0.163481 |
| C | 3.463842 | 1.520208 | 0.358297 |
| C | 4.718262 | 1.445162 | 0.973850 |
| C | 5.392366 | 0.230237 | 1.075733 |
| C | 4.821458 | -0.937888 | 0.574671 |
| N | 1.655222 | 0.377460 | -0.819142 |
| C | 0.453702 | 0.145459 | -0.158986 |
| N | -0.487658 | 0.126333 | -1.182624 |
| C | 0.137421 | 0.350877 | -2.417268 |
| C | 1.462318 | 0.493533 | -2.188179 |
| P | 0.474619 | -0.139006 | 1.587349 |
| C | -1.889085 | -0.161285 | -1.088902 |
| C | -2.298550 | -1.461006 | -0.722066 |
| C | -3.672216 | -1.694896 | -0.623346 |
| C | -4.598545 | -0.697744 | -0.922788 |
| C | -4.167004 | 0.561579 | -1.319970 |
| C | -2.802642 | 0.863220 | -1.394391 |
| C | -1.285053 | -2.566747 | -0.484066 |
| C | -0.654576 | -3.027441 | -1.811716 |
| C | -2.359270 | 2.265487 | -1.779609 |
| C | -2.593457 | 2.518892 | -3.279175 |
| C | 2.716106 | 2.836355 | 0.269052 |
| C | 3.441599 | 3.822166 | -0.660166 |
| C | 2.939475 | -2.187629 | -0.579171 |
| C | 3.729084 | -2.739788 | -1.775720 |
| C | -1.303953 | -0.179677 | 2.027944 |
| C | -2.196292 | 0.871359 | 1.748917 |
| C | -3.526908 | 0.813810 | 2.150714 |
| C | -3.992558 | -0.278935 | 2.888291 |
| C | -3.109974 | -1.308145 | 3.219702 |
| C | -1.783119 | -1.260622 | 2.790766 |
| C | -1.853318 | -3.772374 | 0.268055 |
| C | -3.057525 | 3.351264 | -0.945836 |
| C | 2.480647 | 3.436135 | 1.662972 |
| C | 2.778771 | -3.235235 | 0.531221 |
| H | -0.432135 | 0.364034 | -3.334806 |
| H | 2.290090 | 0.662132 | -2.862459 |
| H | -4.899444 | 1.335254 | -1.547081 |
| H | -4.026281 | -2.675155 | -0.312551 |
| H | 5.352544 | -1.884204 | 0.674790 |
| H | 5.169874 | 2.348061 | 1.384031 |
| H | -1.281124 | 2.343962 | -1.586543 |
| H | -0.482440 | -2.145451 | 0.143611 |
| H | 1.931440 | -1.941196 | -0.933308 |
| H | 1.730122 | 2.625784 | -0.165168 |
| H | -1.106051 | -2.081865 | 3.029335 |
| H | -3.459260 | -2.161139 | 3.802469 |
| H | -5.033840 | -0.323263 | 3.206165 |
| H | -4.205158 | 1.627043 | 1.892275 |
| H | -1.829760 | 1.738410 | 1.201682 |
| H | -5.664827 | -0.907580 | -0.842620 |
| H | -1.037297 | -4.460035 | 0.524707 |
| H | -2.347729 | -3.467372 | 1.197871 |
| H | -2.575285 | -4.332457 | -0.343320 |
| H | 0.104523 | -3.798094 | -1.620120 |
| H | -1.420266 | -3.459828 | -2.471151 |
| H | -0.169544 | -2.203789 | -2.348299 |
| H | -2.653738 | 4.339117 | -1.203911 |
| H | -4.138101 | 3.373288 | -1.138531 |
| H | -2.910991 | 3.193319 | 0.128941 |
| H | -2.225112 | 3.513424 | -3.564661 |
| H | -2.085164 | 1.773024 | -3.902835 |
| H | -3.665809 | 2.471507 | -3.513484 |
| H | 6.368588 | 0.191724 | 1.558823 |
| H | 2.262798 | -4.122854 | 0.140744 |
| H | 3.751529 | -3.558545 | 0.926364 |
| H | 2.183484 | -2.822463 | 1.355568 |
| H | 3.239846 | -3.638349 | -2.175007 |
| H | 3.795356 | -1.997574 | -2.582216 |
| H | 4.752548 | -3.011783 | -1.482600 |
| H | 1.880924 | 4.352921 | 1.584934 |
| H | 1.941511 | 2.720189 | 2.296255 |
| H | 3.428548 | 3.694931 | 2.154374 |
| H | 2.873143 | 4.757974 | -0.745537 |
| H | 4.440129 | 4.066645 | -0.271969 |
| H | 3.565679 | 3.401020 | -1.666594 |

11-P ligand

77

| | | | |
|---|---|---|---|
| C | 3.374603 | -0.013253 | -1.368715 |
| C | 2.735765 | -0.514253 | -0.223098 |
| C | 3.173109 | -1.658456 | 0.465380 |
| C | 4.314623 | -2.302306 | -0.022760 |
| C | 4.976120 | -1.824831 | -1.152499 |
| C | 4.511132 | -0.693144 | -1.819151 |
| N | 1.567813 | 0.159933 | 0.255164 |
| C | 0.273387 | -0.167328 | -0.145356 |
| N | -0.540237 | 0.656523 | 0.622563 |
| C | 0.246462 | 1.451306 | 1.479559 |
| C | 1.546131 | 1.136253 | 1.246461 |
| P | 0.059827 | -1.444520 | -1.346748 |
| C | -1.965237 | 0.758952 | 0.542875 |
| C | -2.508217 | 1.893491 | -0.087955 |
| C | -3.902066 | 1.965360 | -0.174816 |
| C | -4.701862 | 0.948641 | 0.339589 |
| C | -4.132290 | -0.153106 | 0.972680 |
| C | -2.745412 | -0.269403 | 1.102437 |
| CL | -0.424140 | 2.524412 | 2.615968 |
| CL | 2.966811 | 1.728321 | 1.964332 |
| C | -1.613115 | 2.983102 | -0.665241 |
| C | -2.307690 | 4.345651 | -0.753322 |
| C | -2.096653 | -1.446900 | 1.814631 |
| C | -1.483096 | -1.029614 | 3.164390 |
| C | 2.444003 | -2.186928 | 1.686509 |
| C | 3.341567 | -2.138527 | 2.932509 |
| C | 2.861464 | 1.215914 | -2.093170 |
| C | 3.876851 | 2.366483 | -2.013911 |
| C | -1.758231 | -1.599967 | -1.609965 |
| C | -2.338885 | -2.855353 | -1.350894 |
| C | -3.676939 | -3.110255 | -1.655638 |
| C | -4.469356 | -2.111201 | -2.221490 |
| C | -3.902127 | -0.866664 | -2.507655 |
| C | -2.560850 | -0.623329 | -2.225865 |
| C | -1.021313 | 2.593554 | -2.031501 |
| C | -3.049961 | -2.624628 | 2.031634 |
| C | 1.895688 | -3.599831 | 1.435260 |
| C | 2.481920 | 0.889102 | -3.544691 |
| H | -4.776547 | -0.936499 | 1.364266 |
| H | -4.368972 | 2.821079 | -0.658245 |
| H | 5.036356 | -0.334218 | -2.703789 |
| H | 4.687352 | -3.191329 | 0.485187 |
| H | -1.276549 | -1.799676 | 1.167870 |
| H | -0.763465 | 3.107482 | 0.018939 |
| H | 1.946034 | 1.543432 | -1.583911 |
| H | 1.581267 | -1.534165 | 1.872550 |
| H | -2.127799 | 0.337958 | -2.483020 |
| H | -4.507508 | -0.081108 | -2.959785 |
| H | -5.517922 | -2.302731 | -2.447577 |
| H | -4.102894 | -4.090418 | -1.439447 |
| H | -1.730552 | -3.635261 | -0.891308 |
| H | -5.785537 | 1.015191 | 0.245414 |
| H | -0.326083 | 3.372385 | -2.372247 |
| H | -0.472839 | 1.644146 | -1.987404 |
| H | -1.814890 | 2.497769 | -2.785299 |
| H | -1.577466 | 5.114404 | -1.036544 |
| H | -3.099303 | 4.350114 | -1.515236 |
| H | -2.753628 | 4.633434 | 0.207347 |
| H | -2.487627 | -3.483958 | 2.418378 |
| H | -3.828394 | -2.382144 | 2.769022 |
| H | -3.536819 | -2.926905 | 1.097903 |
| H | -1.034755 | -1.904130 | 3.654696 |
| H | -0.700364 | -0.272679 | 3.054318 |
| H | -2.256705 | -0.623526 | 3.830314 |
| H | 5.862924 | -2.342101 | -1.518357 |
| H | 2.052827 | 1.774452 | -4.032755 |
| H | 3.359606 | 0.575000 | -4.126184 |
| H | 1.738187 | 0.082557 | -3.569697 |
| H | 3.473741 | 3.265622 | -2.498742 |
| H | 4.114013 | 2.612748 | -0.970980 |
| H | 4.814472 | 2.100756 | -2.521053 |
| H | 1.322998 | -3.944290 | 2.306749 |
| H | 1.234292 | -3.598245 | 0.559735 |
| H | 2.709274 | -4.317067 | 1.259823 |
| H | 2.782648 | -2.470449 | 3.817507 |
| H | 4.213248 | -2.797080 | 2.817601 |
| H | 3.707385 | -1.120446 | 3.116730 |

12-P ligand

65

| | | | |
|---|---|---|---|
| C | -3.567477 | -0.388603 | -1.187219 |
| C | -2.964345 | -0.159691 | 0.058487 |
| C | -3.562425 | 0.634784 | 1.044964 |
| C | -4.811460 | 1.196493 | 0.763340 |
| C | -5.453327 | 0.982773 | -0.460721 |
| C | -4.816697 | 0.188877 | -1.423148 |
| N | -1.696359 | -0.759494 | 0.332027 |
| C | -0.486014 | -0.143894 | 0.082633 |
| N | 0.454460 | -1.064581 | 0.502531 |
| C | -0.172746 | -2.231114 | 0.991326 |
| C | -1.514990 | -2.037226 | 0.881828 |
| P | -0.393735 | 1.438601 | -0.706411 |
| C | 1.848980 | -1.006968 | 0.186306 |
| C | 2.246696 | -1.322671 | -1.121297 |
| C | 3.606592 | -1.243608 | -1.428781 |
| C | 4.553324 | -0.856386 | -0.475886 |
| C | 4.118987 | -0.577618 | 0.822927 |
| C | 2.771119 | -0.653938 | 1.180203 |
| C | 0.614900 | -3.389984 | 1.482174 |
| C | -2.672027 | -2.892948 | 1.248348 |
| C | 1.229789 | -1.680007 | -2.166551 |
| C | 6.000475 | -0.684434 | -0.850303 |
| C | 2.328091 | -0.353768 | 2.584142 |
| C | -2.850377 | 0.893970 | 2.342170 |
| C | -6.790192 | 1.611996 | -0.755618 |
| C | -2.863757 | -1.206889 | -2.234250 |
| C | 1.343330 | 1.945094 | -0.311117 |
| C | 1.667595 | 2.462547 | 0.956275 |
| C | 2.943409 | 2.953644 | 1.231935 |
| C | 3.922114 | 2.961496 | 0.235549 |
| C | 3.610537 | 2.478802 | -1.036948 |
| C | 2.335710 | 1.980906 | -1.306230 |
| H | 4.845323 | -0.267661 | 1.575772 |
| H | 3.930344 | -1.469027 | -2.446299 |
| H | -5.301558 | 0.020834 | -2.386338 |
| H | -5.291636 | 1.820570 | 1.518767 |
| H | 3.019621 | 0.347438 | 3.064154 |
| H | 1.322543 | 0.080188 | 2.599743 |
| H | 2.303434 | -1.268964 | 3.194735 |
| H | 1.715767 | -2.045739 | -3.077691 |
| H | 0.530743 | -2.447486 | -1.806790 |
| H | 0.624660 | -0.793476 | -2.417324 |
| H | 6.245774 | -1.227296 | -1.771025 |
| H | 6.223381 | 0.379942 | -1.020960 |
| H | 6.668212 | -1.032088 | -0.051602 |
| H | -3.468057 | -1.278933 | -3.144933 |
| H | -1.898436 | -0.747372 | -2.489626 |
| H | -2.649290 | -2.224566 | -1.879367 |
| H | -3.455404 | 1.523040 | 3.004070 |
| H | -2.613260 | -0.040738 | 2.869053 |
| H | -1.895268 | 1.403307 | 2.144201 |
| H | -7.245990 | 2.026743 | 0.151126 |
| H | -6.683667 | 2.430326 | -1.482620 |
| H | -7.486194 | 0.882009 | -1.190070 |
| H | 2.103663 | 1.593478 | -2.298297 |
| H | 4.366495 | 2.482016 | -1.822879 |
| H | 4.919100 | 3.347184 | 0.447639 |
| H | 3.172483 | 3.339040 | 2.226031 |
| H | 0.900338 | 2.478126 | 1.731125 |
| H | -0.049346 | -4.235883 | 1.688072 |
| H | 1.358897 | -3.708223 | 0.738731 |
| H | 1.159908 | -3.153007 | 2.407037 |
| H | -2.323391 | -3.860750 | 1.623365 |
| H | -3.289670 | -2.425795 | 2.028881 |
| H | -3.327502 | -3.072991 | 0.384711 |

13-P ligand

39

| | | | |
|---|---|---|---|
| N | 2.226122 | 0.064195 | -0.244572 |
| C | 0.913938 | -0.081158 | 0.157252 |
| N | 0.595453 | -1.386981 | -0.150658 |
| C | 1.687702 | -2.024466 | -0.717822 |
| C | 2.707201 | -1.127085 | -0.763073 |
| P | 0.002588 | 1.187135 | 1.032332 |
| C | -0.609178 | -2.092942 | 0.327284 |
| C | -0.246363 | -2.957667 | 1.534783 |
| C | 2.983167 | 1.317824 | -0.102101 |
| C | 3.934573 | 1.524622 | -1.275691 |
| C | -1.661945 | 1.001830 | 0.294730 |
| C | -1.908867 | 0.625376 | -1.044256 |
| C | -3.201898 | 0.588132 | -1.559162 |
| C | -4.294747 | 0.955207 | -0.766339 |
| C | -4.069027 | 1.355804 | 0.554019 |
| C | -2.777806 | 1.371308 | 1.078061 |
| C | -1.262594 | -2.876091 | -0.806291 |
| C | 3.690210 | 1.358661 | 1.252783 |
| H | 1.650311 | -3.054618 | -1.043774 |
| H | 3.715185 | -1.232650 | -1.137509 |
| H | -2.621736 | 1.666712 | 2.117225 |
| H | -4.908590 | 1.647595 | 1.186237 |
| H | -5.305161 | 0.934088 | -1.173207 |
| H | -3.358387 | 0.285632 | -2.595569 |
| H | -1.068519 | 0.365734 | -1.688980 |
| H | -1.289656 | -1.300182 | 0.651313 |
| H | -1.150816 | -3.426073 | 1.941817 |
| H | 0.458063 | -3.753321 | 1.256915 |
| H | 0.208794 | -2.341785 | 2.320035 |
| H | -2.194534 | -3.328160 | -0.446713 |
| H | -1.503584 | -2.215466 | -1.646690 |
| H | -0.614856 | -3.687028 | -1.165950 |
| H | 2.199965 | 2.094522 | -0.112727 |
| H | 4.194818 | 2.323992 | 1.383670 |
| H | 2.967179 | 1.234906 | 2.068168 |
| H | 4.442635 | 0.561049 | 1.323981 |
| H | 4.371689 | 2.527718 | -1.207372 |
| H | 4.763208 | 0.803645 | -1.263960 |
| H | 3.408258 | 1.442601 | -2.234309 |

1-P

66

| | | | |
|---|---|---|---|
| C | -3.384618 | 0.002564 | -0.867493 |
| C | -2.847016 | -0.388580 | 0.369576 |
| C | -3.718515 | -0.808670 | 1.384328 |
| C | -5.092983 | -0.887105 | 1.152412 |
| C | -5.613096 | -0.530374 | -0.092528 |
| C | -4.756333 | -0.076718 | -1.099604 |
| P | -1.060506 | -0.178899 | 0.712876 |
| AU | -0.313370 | 1.928705 | 0.030792 |
| C | -0.254426 | -1.625138 | 0.154605 |
| C | -0.845819 | -3.027812 | -0.038241 |
| C | 0.392760 | -3.852127 | -0.478875 |
| C | 1.636336 | -3.082547 | -0.019701 |
| N | 1.085488 | -1.675205 | 0.004711 |
| C | -1.469273 | -3.570611 | 1.259377 |
| C | 1.926348 | -0.517728 | -0.165577 |
| C | 2.120075 | -0.043782 | -1.484962 |
| C | 2.970942 | 1.048178 | -1.671669 |
| C | 3.602164 | 1.662354 | -0.595002 |
| C | 3.364742 | 1.208937 | 0.696174 |
| C | 2.516461 | 0.121892 | 0.941831 |
| C | 1.386680 | -0.624023 | -2.683176 |
| C | 2.231039 | -0.271896 | 2.380037 |
| C | 2.090267 | -3.505301 | 1.383379 |
| H | 3.828296 | 1.720543 | 1.538908 |
| H | 3.127790 | 1.434913 | -2.677653 |
| C | 1.563418 | 0.882972 | 3.146432 |
| C | 3.508529 | -0.713349 | 3.112937 |
| C | 2.331385 | -0.960516 | -3.846867 |
| H | 0.897093 | -1.551984 | -2.363052 |
| C | 0.280072 | 0.335901 | -3.152832 |
| H | -3.316565 | -1.068628 | 2.363591 |
| H | -5.758430 | -1.219846 | 1.948702 |
| H | -6.685795 | -0.590255 | -0.274030 |
| H | -5.159763 | 0.220001 | -2.067536 |
| H | -2.717937 | 0.366147 | -1.648447 |
| H | 0.078554 | 3.437692 | -0.417237 |
| C | -1.897140 | -3.058886 | -1.162345 |
| H | 0.367361 | -4.869179 | -0.071622 |
| H | -1.792891 | -4.608133 | 1.100660 |
| H | -0.760179 | -3.553745 | 2.094253 |
| H | -2.345017 | -2.978700 | 1.543494 |
| H | -1.530987 | -2.540976 | -2.057866 |
| H | -2.097132 | -4.105897 | -1.428169 |
| H | -2.838452 | -2.595237 | -0.856652 |
| C | 2.809084 | -3.216299 | -0.984422 |
| H | 2.354143 | -4.570336 | 1.368882 |
| H | 2.975050 | -2.942279 | 1.694853 |
| H | 1.303478 | -3.356198 | 2.130319 |
| H | 2.514037 | -2.997358 | -2.014482 |
| H | 3.632773 | -2.548971 | -0.703194 |
| H | 3.179162 | -4.248204 | -0.947544 |
| H | 0.404672 | -3.928609 | -1.574410 |
| H | 4.259596 | 2.514902 | -0.761925 |
| H | 1.777439 | -1.468084 | -4.647576 |
| H | 2.771103 | -0.050670 | -4.276162 |
| H | 3.156190 | -1.611040 | -3.530195 |
| H | -0.311993 | -0.123338 | -3.956183 |
| H | -0.384848 | 0.600377 | -2.319840 |
| H | 0.714011 | 1.270269 | -3.533347 |
| H | 1.280419 | 0.553005 | 4.155060 |
| H | 2.248985 | 1.735057 | 3.247341 |
| H | 0.664166 | 1.232563 | 2.625264 |
| H | 1.523868 | -1.110432 | 2.365053 |
| H | 3.263769 | -1.095094 | 4.113021 |
| H | 4.050570 | -1.495700 | 2.567597 |
| H | 4.194680 | 0.135198 | 3.237250 |

2-P

73

| | | | |
|---|---|---|---|
| C | -2.417368 | -1.273571 | -0.950886 |
| C | -1.520896 | -1.461124 | 0.117901 |
| C | -1.918808 | -1.275123 | 1.462879 |
| C | -3.253065 | -0.944504 | 1.711649 |
| C | -4.164510 | -0.796451 | 0.670773 |
| C | -3.745302 | -0.951181 | -0.644437 |
| N | -0.151931 | -1.841055 | -0.116560 |
| C | 0.873632 | -0.974370 | -0.241296 |
| C | 2.199209 | -1.715211 | -0.052277 |
| C | 1.725529 | -3.139796 | 0.344988 |
| C | 0.292699 | -3.283796 | -0.178123 |
| P | 0.613807 | 0.678394 | -0.755280 |
| AU | -1.266067 | 1.834819 | 0.020467 |
| C | 3.077885 | -1.709246 | -1.347083 |
| C | 3.003406 | -1.102695 | 1.117606 |
| C | 0.266327 | -3.781638 | -1.629782 |
| C | -0.578667 | -4.185525 | 0.686967 |
| C | -0.939991 | -1.352911 | 2.622944 |
| C | -0.586559 | 0.056504 | 3.127180 |
| C | -1.997514 | -1.357748 | -2.406823 |
| C | -2.737014 | -2.489032 | -3.140027 |
| C | 2.155961 | 1.606621 | -0.407141 |
| C | 2.298331 | 2.302462 | 0.805496 |
| C | 3.434303 | 3.070535 | 1.051721 |
| C | 4.436715 | 3.171432 | 0.082707 |
| C | 4.286917 | 2.516372 | -1.140428 |
| C | 3.148987 | 1.745865 | -1.386773 |
| C | -1.457988 | -2.223044 | 3.777034 |
| C | -2.227469 | -0.020786 | -3.131440 |
| H | -4.454988 | -0.798251 | -1.456600 |
| H | -3.580693 | -0.786039 | 2.738089 |
| H | -0.012565 | -1.805397 | 2.251247 |
| H | 3.028944 | 1.250027 | -2.349373 |
| H | 5.054135 | 2.609265 | -1.908880 |
| H | 5.324606 | 3.772930 | 0.275580 |
| H | 3.537464 | 3.594926 | 2.001469 |
| H | 1.512319 | 2.230024 | 1.556401 |
| H | -2.501150 | 2.749357 | 0.540043 |
| H | 2.383863 | -3.920635 | -0.049678 |
| H | 0.741238 | -4.769611 | -1.680844 |
| H | -0.760661 | -3.881608 | -1.993526 |
| H | 0.806976 | -3.102655 | -2.298515 |
| H | -0.540935 | -3.897386 | 1.741838 |
| H | -1.623798 | -4.164007 | 0.354570 |
| H | -0.214563 | -5.216595 | 0.600899 |
| H | 1.712960 | -3.231020 | 1.439723 |
| H | -5.200039 | -0.535454 | 0.886185 |
| H | -0.679082 | -2.335612 | 4.542785 |
| H | -2.331427 | -1.766027 | 4.260730 |
| H | -1.752425 | -3.223296 | 3.434158 |
| H | 0.182871 | 0.003222 | 3.909525 |
| H | -0.220322 | 0.683271 | 2.304182 |
| H | -1.473476 | 0.551202 | 3.545487 |
| H | -1.831176 | -0.072932 | -4.154407 |
| H | -3.298240 | 0.214918 | -3.193852 |
| H | -1.730245 | 0.801685 | -2.602485 |
| H | -0.920352 | -1.564733 | -2.432867 |
| H | -2.344055 | -2.609033 | -4.158441 |
| H | -2.643211 | -3.449523 | -2.617520 |
| H | -3.808973 | -2.262368 | -3.218299 |
| C | 4.573030 | -1.532095 | -1.053827 |
| H | 2.728246 | -0.918972 | -2.020959 |
| H | 2.925003 | -2.659800 | -1.875933 |
| H | 5.151130 | -1.806004 | -1.946944 |
| H | 4.790173 | -0.472278 | -0.854632 |
| C | 4.999881 | -2.378697 | 0.148164 |
| H | 6.086212 | -2.318116 | 0.295324 |
| C | 4.258222 | -1.932852 | 1.432758 |
| H | 4.776936 | -3.435145 | -0.067243 |
| H | 3.993549 | -2.811408 | 2.036973 |
| H | 4.920299 | -1.318611 | 2.058566 |
| H | 3.296647 | -0.082967 | 0.856076 |
| H | 2.357790 | -1.023742 | 2.003360 |

3-P

85

| | | | |
|---|---|---|---|
| C | -2.360194 | 1.612473 | 1.150621 |
| C | -1.581203 | 1.629708 | -0.026122 |
| C | -2.169192 | 1.436547 | -1.299180 |
| C | -3.557891 | 1.303523 | -1.373613 |
| C | -4.346822 | 1.342249 | -0.229789 |
| C | -3.747871 | 1.480341 | 1.015487 |
| N | -0.151480 | 1.819001 | 0.033601 |
| C | 0.737446 | 0.820832 | 0.212419 |
| C | 2.160964 | 1.365651 | 0.096180 |
| C | 1.932388 | 2.829532 | -0.364660 |
| C | 0.479694 | 3.191413 | -0.025243 |
| P | 0.224388 | -0.762764 | 0.755777 |
| AU | -1.802631 | -1.695528 | 0.062938 |
| C | 2.853270 | 1.329105 | 1.504030 |
| C | 3.049022 | 0.567559 | -0.906084 |
| C | 0.364779 | 3.893290 | 1.332152 |
| C | -0.170760 | 4.069893 | -1.089448 |
| C | -1.347916 | 1.302943 | -2.567943 |
| C | -1.362125 | -0.144035 | -3.087730 |
| C | -1.777574 | 1.659440 | 2.553464 |
| C | -2.300579 | 2.862650 | 3.355587 |
| C | 1.609638 | -1.940063 | 0.543409 |
| C | 1.568519 | -2.845343 | -0.529346 |
| C | 2.530364 | -3.847432 | -0.651341 |
| C | 3.546812 | -3.963433 | 0.298501 |
| C | 3.578920 | -3.089659 | 1.387767 |
| C | 2.608380 | -2.096902 | 1.517821 |
| C | -1.807974 | 2.266728 | -3.671542 |
| C | -2.089425 | 0.361795 | 3.320987 |
| H | -4.366083 | 1.464793 | 1.912225 |
| H | -4.025081 | 1.147691 | -2.345120 |
| H | -0.310645 | 1.540115 | -2.313346 |
| H | 2.620762 | -1.444717 | 2.389279 |
| H | 4.352842 | -3.191482 | 2.148254 |
| H | 4.301425 | -4.743218 | 0.200469 |
| H | 2.487004 | -4.537387 | -1.493580 |
| H | 0.771228 | -2.758978 | -1.266965 |
| H | -3.165049 | -2.469537 | -0.355819 |
| H | 2.630609 | 3.516015 | 0.124580 |
| H | 0.918760 | 4.839585 | 1.292206 |
| H | -0.679676 | 4.123841 | 1.562447 |
| H | 0.777297 | 3.286037 | 2.144437 |
| H | -0.035318 | 3.659135 | -2.093984 |
| H | -1.243959 | 4.195688 | -0.899508 |
| H | 0.297913 | 5.060988 | -1.062102 |
| H | 2.087637 | 2.921670 | -1.441761 |
| H | -5.428248 | 1.234249 | -0.307114 |
| H | -1.117014 | 2.220636 | -4.523978 |
| H | -2.806064 | 1.995623 | -4.040757 |
| H | -1.853339 | 3.304048 | -3.318623 |
| H | -0.687122 | -0.245088 | -3.948349 |
| H | -1.049145 | -0.844808 | -2.301302 |
| H | -2.372357 | -0.435905 | -3.404692 |
| H | -1.583625 | 0.371346 | 4.295864 |
| H | -3.168628 | 0.262734 | 3.498993 |
| H | -1.754225 | -0.518655 | 2.761680 |
| H | -0.687228 | 1.737255 | 2.466353 |
| H | -1.782106 | 2.930016 | 4.321221 |
| H | -2.163194 | 3.811967 | 2.825030 |
| H | -3.373843 | 2.752115 | 3.560681 |
| C | 4.363135 | 1.051965 | 1.461837 |
| H | 2.356556 | 0.587364 | 2.139500 |
| H | 2.692303 | 2.301939 | 1.990527 |
| C | 5.014634 | 1.436363 | 2.789202 |
| H | 4.513461 | -0.029777 | 1.311949 |
| C | 4.986685 | 1.783022 | 0.270700 |
| H | 6.078814 | 1.662078 | 0.281737 |
| C | 4.406245 | 1.277534 | -1.068787 |
| H | 4.798441 | 2.862234 | 0.389074 |
| H | 4.318828 | 2.114864 | -1.774135 |
| H | 5.104920 | 0.564633 | -1.525834 |
| H | 3.250429 | -0.386167 | -0.402661 |
| C | 2.407559 | 0.145257 | -2.253970 |
| C | 2.199541 | 1.237930 | -3.311859 |
| C | 3.240073 | -0.981359 | -2.885175 |
| H | 1.419925 | -0.275553 | -2.006201 |
| H | 2.698860 | -1.432689 | -3.727344 |
| H | 4.194291 | -0.599116 | -3.273874 |
| H | 3.457961 | -1.773834 | -2.159803 |
| H | 1.805978 | 0.784143 | -4.231281 |
| H | 1.486680 | 2.011794 | -3.010653 |
| H | 3.145026 | 1.732397 | -3.573017 |
| H | 6.082598 | 1.179746 | 2.794578 |
| H | 4.925597 | 2.518615 | 2.966902 |
| H | 4.537548 | 0.918653 | 3.633124 |

4-P

63

| | | | |
|---|---|---|---|
| C | -2.561333 | 0.109622 | -2.110515 |
| C | -1.638888 | 0.777010 | -1.289510 |
| C | -1.963301 | 2.061494 | -0.822583 |
| C | -3.189825 | 2.644709 | -1.137887 |
| C | -4.113623 | 1.954366 | -1.923960 |
| C | -3.794792 | 0.683491 | -2.409052 |
| P | 0.046543 | 0.100344 | -1.028574 |
| AU | 1.324294 | 1.633460 | 0.240140 |
| C | -0.154864 | -1.334203 | 0.042585 |
| N | -1.256565 | -1.970582 | 0.515790 |
| C | -0.885260 | -3.133728 | 1.355069 |
| C | 0.601275 | -3.320354 | 1.042612 |
| N | 0.945550 | -2.045646 | 0.396262 |
| C | -2.571659 | -1.430132 | 0.636199 |
| C | -2.796099 | -0.313283 | 1.460311 |
| C | -4.090974 | 0.205887 | 1.512718 |
| C | -5.148899 | -0.369928 | 0.803591 |
| C | -4.898081 | -1.518300 | 0.047971 |
| C | -3.616623 | -2.061998 | -0.055359 |
| C | -1.685986 | 0.320366 | 2.246617 |
| C | -3.358270 | -3.256097 | -0.933030 |
| C | -6.510919 | 0.267671 | 0.810488 |
| C | 2.286227 | -1.724546 | 0.022184 |
| C | 2.753527 | -2.051097 | -1.259978 |
| C | 4.065332 | -1.694457 | -1.586703 |
| C | 4.901784 | -1.043081 | -0.675148 |
| C | 4.408961 | -0.764034 | 0.604520 |
| C | 3.105708 | -1.095671 | 0.976557 |
| C | 1.869769 | -2.748442 | -2.257420 |
| C | 6.296932 | -0.628939 | -1.062702 |
| C | 2.589797 | -0.797963 | 2.357382 |
| H | -5.713363 | -1.980387 | -0.510867 |
| H | -4.274217 | 1.092402 | 2.121683 |
| H | 5.048956 | -0.257907 | 1.328379 |
| H | 4.441212 | -1.931104 | -2.583317 |
| H | -4.145993 | -3.359351 | -1.687932 |
| H | -2.391063 | -3.170334 | -1.443995 |
| H | -3.337940 | -4.190377 | -0.351814 |
| H | -2.087621 | 1.022509 | 2.984971 |
| H | -1.083487 | -0.429956 | 2.776399 |
| H | -0.988553 | 0.872460 | 1.594343 |
| H | -6.729130 | 0.742604 | 1.775206 |
| H | -6.561266 | 1.052900 | 0.040290 |
| H | -7.300247 | -0.462041 | 0.592806 |
| H | 3.270724 | -0.122591 | 2.885568 |
| H | 1.604133 | -0.316930 | 2.310795 |
| H | 2.490977 | -1.716908 | 2.954771 |
| H | 2.435829 | -3.014349 | -3.156902 |
| H | 1.435926 | -3.668623 | -1.842849 |
| H | 1.035034 | -2.095162 | -2.550854 |
| H | 6.611526 | -1.106946 | -1.997822 |
| H | 6.349515 | 0.460087 | -1.207158 |
| H | 7.021873 | -0.886694 | -0.279340 |
| H | -1.240412 | 2.596235 | -0.204535 |
| H | -3.426133 | 3.640130 | -0.761844 |
| H | -5.074046 | 2.407964 | -2.167516 |
| H | -4.507271 | 0.140469 | -3.029638 |
| H | -2.308635 | -0.870650 | -2.514894 |
| H | 2.182728 | 2.756123 | 1.048009 |
| H | 1.212568 | -3.477049 | 1.938105 |
| H | 0.781833 | -4.153229 | 0.346501 |
| H | -1.066803 | -2.885477 | 2.410645 |
| H | -1.492015 | -4.006377 | 1.092249 |

5-P

81

| | | | |
|---|---|---|---|
| C | -2.646598 | -0.895401 | -1.538020 |
| C | -2.007407 | -1.534483 | -0.455032 |
| C | -2.701070 | -1.929779 | 0.701832 |
| C | -4.076916 | -1.682390 | 0.751466 |
| C | -4.734052 | -1.072493 | -0.313220 |
| C | -4.024738 | -0.682641 | -1.446152 |
| N | -0.611330 | -1.822918 | -0.566047 |
| C | 0.412208 | -1.027281 | -0.157893 |
| N | 1.575383 | -1.543162 | -0.625779 |
| C | 1.373753 | -2.828214 | -1.325820 |
| C | -0.152852 | -2.967896 | -1.367488 |
| P | 0.099688 | 0.256025 | 1.052728 |
| AU | -1.775552 | 1.565121 | 0.526198 |
| C | 2.864068 | -0.932407 | -0.594458 |
| C | 3.090904 | 0.187001 | -1.432650 |
| C | 4.365994 | 0.752950 | -1.430324 |
| C | 5.388105 | 0.218410 | -0.647979 |
| C | 5.137071 | -0.873009 | 0.169892 |
| C | 3.867420 | -1.466118 | 0.232799 |
| C | 1.986749 | 0.757243 | -2.312956 |
| C | 3.715768 | -2.611537 | 1.232882 |
| C | -1.994257 | -2.605885 | 1.861136 |
| C | -1.879290 | -0.444072 | -2.767465 |
| C | 1.591892 | 1.320888 | 1.143883 |
| C | 2.708243 | 0.961779 | 1.914991 |
| C | 3.776968 | 1.840256 | 2.073266 |
| C | 3.731189 | 3.112172 | 1.498439 |
| C | 2.605527 | 3.500876 | 0.771778 |
| C | 1.545303 | 2.612078 | 0.594841 |
| H | 5.928497 | -1.263055 | 0.811183 |
| H | 4.566082 | 1.628218 | -2.044030 |
| H | -4.545365 | -0.184904 | -2.263628 |
| H | -4.640920 | -1.969802 | 1.638193 |
| H | 4.381516 | -2.322109 | 2.061328 |
| C | 2.337466 | -2.844984 | 1.869618 |
| C | 4.292351 | -3.921966 | 0.662644 |
| C | 2.239053 | 2.201937 | -2.754880 |
| C | 1.753693 | -0.113729 | -3.560660 |
| H | 1.059312 | 0.759307 | -1.720862 |
| C | -2.019047 | 1.065315 | -3.008195 |
| H | -0.815241 | -0.638199 | -2.583999 |
| C | -2.301139 | -1.241469 | -4.012032 |
| C | -2.259383 | -1.881298 | 3.187865 |
| C | -2.381458 | -4.091604 | 1.943150 |
| H | -0.914901 | -2.546480 | 1.669411 |
| H | 0.661400 | 2.911934 | 0.031205 |
| H | 2.551461 | 4.499012 | 0.337195 |
| H | 4.564587 | 3.801852 | 1.628962 |
| H | 4.645354 | 1.533476 | 2.655485 |
| H | 2.741698 | -0.014140 | 2.394023 |
| H | -3.047991 | 2.555599 | 0.310600 |
| H | -0.558508 | -2.901045 | -2.385789 |
| H | -0.501959 | -3.906181 | -0.919039 |
| H | 1.834564 | -2.793089 | -2.319078 |
| H | 1.844189 | -3.643104 | -0.763269 |
| H | -5.805975 | -0.886031 | -0.253867 |
| H | 6.377189 | 0.675221 | -0.659905 |
| H | -1.667277 | -2.336941 | 3.992685 |
| H | -1.980554 | -0.823580 | 3.108379 |
| H | -3.318005 | -1.940509 | 3.474736 |
| H | -1.831208 | -4.588997 | 2.753372 |
| H | -3.456115 | -4.205156 | 2.142073 |
| H | -2.159001 | -4.614472 | 1.003290 |
| H | -1.387177 | 1.375387 | -3.851712 |
| H | -3.056215 | 1.339006 | -3.242658 |
| H | -1.720537 | 1.625072 | -2.110028 |
| H | -1.688006 | -0.954788 | -4.876951 |
| H | -2.194967 | -2.323805 | -3.857183 |
| H | -3.352382 | -1.044186 | -4.262427 |
| H | 4.280274 | -4.708874 | 1.428876 |
| H | 3.708046 | -4.277977 | -0.196102 |
| H | 5.327056 | -3.783950 | 0.325220 |
| H | 2.466466 | -3.446576 | 2.778781 |
| H | 1.835910 | -1.911422 | 2.152195 |
| H | 1.654860 | -3.401744 | 1.217561 |
| H | 1.336121 | 2.597461 | -3.236561 |
| H | 2.484263 | 2.847072 | -1.904665 |
| H | 3.056835 | 2.263819 | -3.486356 |
| H | 0.959684 | 0.321956 | -4.181098 |
| H | 2.669366 | -0.166858 | -4.165350 |
| H | 1.456317 | -1.136670 | -3.311015 |

6-P

70

| | | | |
|---|---|---|---|
| C | -2.973125 | 0.745752 | 0.871593 |
| C | -2.284038 | 1.304432 | -0.213804 |
| C | -2.807063 | 1.317760 | -1.513239 |
| C | -4.028618 | 0.673456 | -1.724954 |
| C | -4.727101 | 0.056387 | -0.682558 |
| C | -4.194996 | 0.122570 | 0.609407 |
| N | -0.988597 | 1.915226 | 0.023951 |
| C | 0.149867 | 1.131534 | -0.137528 |
| N | 1.351578 | 1.681931 | 0.272673 |
| C | 1.491282 | 3.040507 | 0.634834 |
| P | -0.098802 | -0.449162 | -0.879231 |
| AU | -1.260015 | -1.961510 | 0.480010 |
| C | 2.486118 | 0.829678 | 0.572549 |
| C | 2.432772 | -0.015750 | 1.687613 |
| C | 3.542496 | -0.831206 | 1.934548 |
| C | 4.675687 | -0.803825 | 1.120839 |
| C | 4.705408 | 0.096587 | 0.051046 |
| C | 3.628626 | 0.933116 | -0.235589 |
| C | 1.238103 | -0.082065 | 2.592197 |
| C | 5.829701 | -1.736850 | 1.363190 |
| C | 3.714433 | 1.925016 | -1.359340 |
| C | -2.095547 | 2.016271 | -2.637398 |
| C | -6.016897 | -0.674681 | -0.942411 |
| C | -2.442037 | 0.856676 | 2.272095 |
| C | 1.521509 | -1.088517 | -1.407703 |
| C | 2.078623 | -0.598890 | -2.599818 |
| C | 3.232174 | -1.175994 | -3.127015 |
| C | 3.824179 | -2.263713 | -2.482518 |
| C | 3.252884 | -2.779752 | -1.316665 |
| C | 2.102109 | -2.204258 | -0.785404 |
| H | 5.589014 | 0.144144 | -0.586528 |
| H | 3.512355 | -1.506389 | 2.790734 |
| H | -4.741015 | -0.326201 | 1.439706 |
| H | -4.445454 | 0.659135 | -2.732789 |
| H | 4.489272 | 1.631490 | -2.075239 |
| H | 2.765317 | 2.012994 | -1.900405 |
| H | 3.966273 | 2.921799 | -0.969614 |
| H | 1.516784 | -0.493060 | 3.568342 |
| H | 0.790013 | 0.906498 | 2.750770 |
| H | 0.458608 | -0.736264 | 2.163445 |
| H | 5.790670 | -2.172611 | 2.368410 |
| H | 5.809066 | -2.564418 | 0.637750 |
| H | 6.792419 | -1.224075 | 1.241308 |
| H | -3.073429 | 0.295957 | 2.968574 |
| H | -1.423385 | 0.459607 | 2.346495 |
| H | -2.416711 | 1.907489 | 2.593450 |
| H | -2.609749 | 1.840798 | -3.588026 |
| H | -2.061606 | 3.101071 | -2.465041 |
| H | -1.061897 | 1.657228 | -2.737362 |
| H | -6.473196 | -0.359237 | -1.888144 |
| H | -5.837874 | -1.758458 | -1.002370 |
| H | -6.739283 | -0.508868 | -0.133079 |
| H | 1.647777 | -2.608090 | 0.118469 |
| H | 3.705667 | -3.636999 | -0.819429 |
| H | 4.723794 | -2.718519 | -2.895821 |
| H | 3.663555 | -0.782527 | -4.046818 |
| H | 1.599851 | 0.232846 | -3.116924 |
| H | -1.876011 | -3.122340 | 1.425884 |
| C | 0.380670 | 3.969993 | 0.181151 |
| C | -0.975890 | 3.291072 | 0.277707 |
| C | 0.628678 | 4.280358 | -1.323380 |
| C | 0.393307 | 5.262195 | 0.999974 |
| H | 0.236653 | 5.054611 | 2.064617 |
| H | 1.358302 | 5.764220 | 0.883450 |
| H | -0.407740 | 5.921050 | 0.652296 |
| H | 1.602646 | 4.770098 | -1.434661 |
| H | 0.616760 | 3.371827 | -1.935951 |
| H | -0.153913 | 4.958300 | -1.682314 |
| O | 2.481947 | 3.422320 | 1.230591 |
| O | -2.015981 | 3.897720 | 0.476202 |

7-P

71

| | | | |
|---|---|---|---|
| C | 3.016371 | -0.526582 | 0.880281 |
| C | 2.369536 | -1.207502 | -0.159893 |
| C | 2.892653 | -1.267749 | -1.457970 |
| C | 4.058325 | -0.543222 | -1.720315 |
| C | 4.704565 | 0.199180 | -0.727106 |
| C | 4.182303 | 0.176025 | 0.570188 |
| N | 1.122512 | -1.890629 | 0.133060 |
| C | -0.072784 | -1.182749 | -0.049335 |
| N | -1.201937 | -1.767007 | 0.396240 |
| C | -1.166430 | -3.088796 | 1.036772 |
| P | 0.090494 | 0.352441 | -0.969332 |
| AU | 1.006742 | 2.004585 | 0.443055 |
| C | -2.415418 | -1.027721 | 0.644789 |
| C | -2.456471 | -0.117281 | 1.712715 |
| C | -3.660099 | 0.556060 | 1.947364 |
| C | -4.796035 | 0.337034 | 1.167069 |
| C | -4.725954 | -0.614163 | 0.144357 |
| C | -3.553602 | -1.317960 | -0.126731 |
| C | -1.260157 | 0.167952 | 2.569454 |
| C | -6.056752 | 1.126502 | 1.390458 |
| C | -3.524493 | -2.350711 | -1.217092 |
| C | 2.238418 | -2.093511 | -2.528461 |
| C | 5.926209 | 1.018719 | -1.046368 |
| C | 2.495324 | -0.583523 | 2.288176 |
| C | -1.609692 | 0.812007 | -1.474546 |
| C | -2.137048 | 0.216366 | -2.631328 |
| C | -3.362396 | 0.632333 | -3.149984 |
| C | -4.062824 | 1.670857 | -2.534015 |
| C | -3.526969 | 2.297350 | -1.405906 |
| C | -2.306896 | 1.876564 | -0.882689 |
| H | -5.604546 | -0.803842 | -0.473613 |
| H | -3.700997 | 1.278021 | 2.764278 |
| H | 4.689538 | 0.724166 | 1.364668 |
| H | 4.470264 | -0.561108 | -2.730290 |
| H | -4.324165 | -2.166234 | -1.942079 |
| H | -2.568587 | -2.345959 | -1.750100 |
| H | -3.670898 | -3.363481 | -0.811339 |
| H | -1.562164 | 0.619699 | 3.520367 |
| H | -0.678466 | -0.736616 | 2.785997 |
| H | -0.576477 | 0.874376 | 2.059731 |
| H | -6.067239 | 1.595601 | 2.381281 |
| H | -6.141934 | 1.927406 | 0.640312 |
| H | -6.948915 | 0.494765 | 1.292864 |
| H | 3.128835 | 0.008534 | 2.956469 |
| H | 1.476681 | -0.181989 | 2.349238 |
| H | 2.473642 | -1.619514 | 2.654754 |
| H | 2.733827 | -1.938580 | -3.492776 |
| H | 2.293191 | -3.163830 | -2.284813 |
| H | 1.179295 | -1.823506 | -2.640083 |
| H | 6.407687 | 0.677871 | -1.970778 |
| H | 5.653748 | 2.075969 | -1.182021 |
| H | 6.661172 | 0.974481 | -0.232516 |
| H | -1.883657 | 2.365837 | -0.005860 |
| H | -4.063852 | 3.117898 | -0.930418 |
| H | -5.018205 | 2.001541 | -2.940543 |
| H | -3.765024 | 0.152506 | -4.041874 |
| H | -1.574833 | -0.573990 | -3.129876 |
| H | 1.559302 | 3.237394 | 1.344064 |
| C | -0.117299 | -4.008841 | 0.423853 |
| C | 1.208465 | -3.264441 | 0.408805 |
| C | -0.460252 | -4.362451 | -1.037817 |
| C | -0.001585 | -5.279410 | 1.269283 |
| H | 0.258188 | -5.042726 | 2.308461 |
| H | -0.957087 | -5.819427 | 1.261180 |
| H | 0.776019 | -5.935163 | 0.865180 |
| H | -1.416979 | -4.898585 | -1.071682 |
| H | -0.541793 | -3.469355 | -1.667768 |
| H | 0.316904 | -5.010476 | -1.459054 |
| O | 2.296694 | -3.801799 | 0.545785 |
| H | -2.165028 | -3.530882 | 0.941347 |
| H | -0.971077 | -2.949779 | 2.113927 |

8-P

35

| | | | |
|---|---|---|---|
| C | -0.547970 | -3.009937 | 0.700829 |
| C | 0.439194 | -2.013876 | 0.599126 |
| C | 1.710348 | -2.379895 | 0.126400 |
| C | 1.982157 | -3.698016 | -0.240542 |
| C | 0.986472 | -4.674372 | -0.157131 |
| C | -0.282158 | -4.322600 | 0.313251 |
| P | 0.127184 | -0.302487 | 1.197341 |
| AU | 1.411485 | 1.196856 | -0.123087 |
| C | -1.493637 | -0.034360 | 0.418304 |
| N | -1.901497 | -0.300844 | -0.855119 |
| C | -3.220506 | 0.106676 | -1.041328 |
| C | -3.635132 | 0.656274 | 0.144902 |
| N | -2.564757 | 0.547361 | 1.028442 |
| C | -1.034410 | -0.840661 | -1.892959 |
| C | -2.572722 | 1.018355 | 2.406150 |
| C | -3.938022 | -0.056199 | -2.332316 |
| C | -4.930102 | 1.271965 | 0.534196 |
| H | 2.481685 | -1.614681 | 0.030012 |
| H | 2.974667 | -3.960608 | -0.607442 |
| H | 1.196772 | -5.701301 | -0.454211 |
| H | -1.065716 | -5.077225 | 0.388311 |
| H | -1.536237 | -2.752643 | 1.085678 |
| H | 2.297493 | 2.251829 | -0.995689 |
| H | -4.974037 | 0.281919 | -2.229396 |
| H | -3.470853 | 0.534149 | -3.133241 |
| H | -3.954866 | -1.106009 | -2.655034 |
| H | -5.613204 | 1.283883 | -0.320737 |
| H | -5.417906 | 0.715570 | 1.347027 |
| H | -4.795546 | 2.308489 | 0.873150 |
| H | -3.512717 | 0.732206 | 2.887973 |
| H | -1.725577 | 0.551599 | 2.921289 |
| H | -2.457826 | 2.107956 | 2.440141 |
| H | -1.427311 | -0.551784 | -2.870119 |
| H | -0.032576 | -0.411551 | -1.750360 |
| H | -0.970720 | -1.932078 | -1.818793 |

9-P

61

| | | | |
|---|---|---|---|
| C | 2.803169 | 0.199562 | 1.461762 |
| C | 2.495337 | 1.405449 | 0.815715 |
| C | 3.461748 | 2.188009 | 0.170401 |
| C | 4.773115 | 1.709899 | 0.149670 |
| C | 5.117009 | 0.485747 | 0.730148 |
| C | 4.124977 | -0.244645 | 1.390816 |
| N | 1.135907 | 1.862216 | 0.815880 |
| C | 0.080531 | 1.298992 | 0.144700 |
| N | -1.006815 | 2.060466 | 0.499909 |
| C | -0.635169 | 3.067958 | 1.378655 |
| C | 0.701657 | 2.952347 | 1.572098 |
| P | -0.112939 | -0.053652 | -1.038762 |
| C | -2.351879 | 1.787640 | 0.075984 |
| C | -2.766078 | 2.225209 | -1.187928 |
| C | -4.068020 | 1.904315 | -1.584784 |
| C | -4.933246 | 1.181174 | -0.758688 |
| C | -4.481997 | 0.787840 | 0.507070 |
| C | -3.191851 | 1.079911 | 0.950461 |
| C | -1.839412 | 2.992144 | -2.088777 |
| C | -6.319274 | 0.815554 | -1.219988 |
| C | -2.708764 | 0.641852 | 2.304844 |
| C | 3.094525 | 3.467221 | -0.531441 |
| C | 6.509943 | -0.065701 | 0.599431 |
| C | 1.755935 | -0.585883 | 2.192445 |
| H | 1.391185 | 3.519261 | 2.181211 |
| H | -1.360610 | 3.762328 | 1.779028 |
| H | 5.536372 | 2.290861 | -0.369761 |
| H | 4.381571 | -1.194551 | 1.861054 |
| H | -5.146858 | 0.226210 | 1.164269 |
| H | -4.411545 | 2.228431 | -2.568185 |
| H | 3.903878 | 3.784666 | -1.197647 |
| H | 2.180637 | 3.346602 | -1.128495 |
| H | 2.904788 | 4.285148 | 0.177951 |
| H | 2.211682 | -1.397174 | 2.769441 |
| H | 1.183700 | 0.053398 | 2.878503 |
| H | 1.021796 | -1.031084 | 1.497893 |
| H | 6.813497 | -0.611712 | 1.501352 |
| H | 6.553790 | -0.774017 | -0.242012 |
| H | 7.242359 | 0.727185 | 0.406235 |
| H | -3.479734 | 0.057935 | 2.817418 |
| H | -1.816618 | 0.004515 | 2.203062 |
| H | -2.439632 | 1.497680 | 2.939081 |
| H | -2.376889 | 3.376020 | -2.962404 |
| H | -1.375932 | 3.838610 | -1.564378 |
| H | -1.026972 | 2.337468 | -2.438855 |
| H | -6.579012 | 1.328544 | -2.153224 |
| H | -6.394491 | -0.267095 | -1.397024 |
| H | -7.071431 | 1.071912 | -0.461982 |
| C | 1.600987 | -0.618301 | -1.378266 |
| C | 1.980542 | -1.938216 | -1.082303 |
| C | 3.228871 | -2.424936 | -1.468441 |
| C | 4.124010 | -1.598922 | -2.149762 |
| C | 3.753515 | -0.288072 | -2.460254 |
| C | 2.496366 | 0.189073 | -2.097889 |
| H | 1.283889 | -2.577203 | -0.537567 |
| H | 3.505774 | -3.451134 | -1.227067 |
| H | 5.102655 | -1.975968 | -2.446010 |
| H | 4.443790 | 0.362800 | -2.996365 |
| H | 2.203738 | 1.202301 | -2.374072 |
| AU | -1.235086 | -1.720851 | 0.216523 |
| H | -2.009857 | -2.916392 | 1.007806 |

10-P

79

| | | | |
|---|---|---|---|
| C | -2.530021 | 0.178988 | -2.007759 |
| C | -1.483439 | 0.963970 | -1.501009 |
| C | -1.654679 | 2.355770 | -1.445585 |
| C | -2.855478 | 2.939919 | -1.847711 |
| C | -3.907143 | 2.142295 | -2.300751 |
| C | -3.739477 | 0.757276 | -2.380969 |
| P | 0.136250 | 0.206546 | -1.108205 |
| AU | 1.525646 | 1.868375 | -0.167481 |
| C | -0.239356 | -0.901612 | 0.265441 |
| N | -1.353845 | -1.205450 | 1.008621 |
| C | -1.074205 | -2.263009 | 1.879067 |
| C | 0.218143 | -2.616235 | 1.686515 |
| N | 0.726169 | -1.777152 | 0.707421 |
| C | -2.638894 | -0.557870 | 0.999228 |
| C | -2.731602 | 0.778212 | 1.440441 |
| C | -4.000934 | 1.362432 | 1.422605 |
| C | -5.125284 | 0.641374 | 1.026207 |
| C | -5.002985 | -0.682341 | 0.623485 |
| C | -3.752869 | -1.309204 | 0.585282 |
| C | -1.515174 | 1.540642 | 1.933592 |
| C | -3.647147 | -2.746237 | 0.098181 |
| C | 2.072713 | -1.793684 | 0.202566 |
| C | 2.305280 | -2.418443 | -1.034436 |
| C | 3.613274 | -2.392094 | -1.527909 |
| C | 4.633210 | -1.769269 | -0.812588 |
| C | 4.371684 | -1.169155 | 0.416173 |
| C | 3.080798 | -1.163772 | 0.955392 |
| C | 1.202918 | -3.153381 | -1.775522 |
| C | 2.803613 | -0.569157 | 2.324979 |
| H | -1.823481 | -2.636319 | 2.560967 |
| H | 0.825885 | -3.378636 | 2.150479 |
| H | -5.887398 | -1.233773 | 0.307555 |
| H | -4.113091 | 2.399272 | 1.729787 |
| H | 5.180901 | -0.688135 | 0.962074 |
| H | 3.835917 | -2.864402 | -2.483319 |
| C | -4.348023 | -2.962715 | -1.252096 |
| H | -2.584158 | -2.983051 | -0.043548 |
| C | -4.216861 | -3.720872 | 1.144041 |
| C | -1.730084 | 3.051945 | 2.028283 |
| C | -1.011810 | 0.994938 | 3.282143 |
| H | -0.704007 | 1.392789 | 1.201076 |
| C | 3.598276 | 0.710720 | 2.605010 |
| H | 1.736358 | -0.307848 | 2.362693 |
| C | 3.086985 | -1.614605 | 3.420972 |
| C | 1.211490 | -2.867725 | -3.280443 |
| C | 1.288360 | -4.660525 | -1.479286 |
| H | 0.241583 | -2.791584 | -1.389859 |
| H | -0.836607 | 2.971587 | -1.069366 |
| H | -2.974492 | 4.022069 | -1.792973 |
| H | -4.851995 | 2.596671 | -2.597220 |
| H | -4.553318 | 0.128445 | -2.741247 |
| H | -2.387447 | -0.897534 | -2.099972 |
| H | 2.461173 | 3.090415 | 0.367803 |
| H | -6.103620 | 1.120972 | 1.028348 |
| H | -0.768840 | 3.537993 | 2.236081 |
| H | -2.117993 | 3.461108 | 1.088056 |
| H | -2.424245 | 3.318707 | 2.837883 |
| H | -0.098571 | 1.528277 | 3.577042 |
| H | -1.767033 | 1.144215 | 4.066459 |
| H | -0.775619 | -0.074350 | 3.239297 |
| H | -4.187762 | -3.993460 | -1.594198 |
| H | -5.431331 | -2.804574 | -1.171950 |
| H | -3.965489 | -2.284248 | -2.022501 |
| H | -4.085286 | -4.759787 | 0.813820 |
| H | -3.730539 | -3.607267 | 2.120640 |
| H | -5.291157 | -3.541771 | 1.288150 |
| H | 5.645071 | -1.752705 | -1.216909 |
| H | 3.278708 | 1.140074 | 3.563547 |
| H | 4.676160 | 0.512215 | 2.677401 |
| H | 3.425551 | 1.457200 | 1.819679 |
| H | 2.829971 | -1.210153 | 4.408941 |
| H | 2.517676 | -2.539572 | 3.271711 |
| H | 4.153804 | -1.877833 | 3.424905 |
| H | 0.349662 | -3.351702 | -3.758703 |
| H | 1.146881 | -1.787706 | -3.462284 |
| H | 2.117709 | -3.252961 | -3.766438 |
| H | 0.468689 | -5.197088 | -1.975949 |
| H | 2.239039 | -5.075244 | -1.841686 |
| H | 1.224680 | -4.857023 | -0.400879 |

11-P

79

| | | | |
|---|---|---|---|
| C | -2.052982 | 0.267507 | 2.996588 |
| C | -1.723976 | -0.502002 | 1.870860 |
| C | -2.727442 | -1.282645 | 1.276555 |
| C | -4.033192 | -1.252299 | 1.757971 |
| C | -4.352945 | -0.466124 | 2.867915 |
| C | -3.357681 | 0.285317 | 3.493145 |
| P | 0.057282 | -0.617984 | 1.401410 |
| AU | 0.255579 | -2.597775 | 0.088317 |
| C | 0.226271 | 0.533166 | 0.015946 |
| N | -0.644309 | 1.108626 | -0.871858 |
| C | 0.063067 | 1.900630 | -1.791116 |
| C | 1.382216 | 1.815989 | -1.461921 |
| N | 1.477365 | 0.972057 | -0.355104 |
| C | -2.081322 | 1.001586 | -0.893777 |
| C | -2.659422 | -0.005061 | -1.684003 |
| C | -4.054938 | -0.004337 | -1.783831 |
| C | -4.819678 | 0.940062 | -1.107106 |
| C | -4.213214 | 1.886691 | -0.284939 |
| C | -2.822269 | 1.943408 | -0.155281 |
| C | -1.820691 | -1.085254 | -2.346449 |
| C | -2.154584 | 2.935014 | 0.790182 |
| C | 2.650770 | 0.884529 | 0.478887 |
| C | 2.818135 | 1.893744 | 1.443724 |
| C | 3.972167 | 1.842376 | 2.229865 |
| C | 4.910995 | 0.829439 | 2.051063 |
| C | 4.709888 | -0.157918 | 1.090244 |
| C | 3.568949 | -0.157112 | 0.279098 |
| C | 1.789104 | 2.987871 | 1.661563 |
| C | 3.356218 | -1.230030 | -0.771157 |
| CL | -0.683336 | 2.782826 | -3.032256 |
| CL | 2.718942 | 2.606293 | -2.141357 |
| H | -4.833270 | 2.582012 | 0.275893 |
| H | -4.548451 | -0.768220 | -2.380996 |
| H | 5.448109 | -0.948644 | 0.971017 |
| H | 4.135280 | 2.602976 | 2.992565 |
| C | -3.094481 | 3.417800 | 1.900415 |
| H | -1.327066 | 2.394804 | 1.275348 |
| C | -1.561787 | 4.162983 | 0.075202 |
| C | -2.553494 | -2.429119 | -2.433671 |
| C | -1.311834 | -0.681686 | -3.739438 |
| H | -0.941090 | -1.256364 | -1.699109 |
| C | 3.741414 | -2.623828 | -0.261711 |
| H | 2.277970 | -1.263254 | -0.999651 |
| C | 4.112866 | -0.893078 | -2.066891 |
| C | 1.147375 | 2.857792 | 3.050819 |
| C | 2.388434 | 4.383413 | 1.439467 |
| H | 0.986728 | 2.860419 | 0.924358 |
| H | -2.465341 | -1.918558 | 0.433325 |
| H | -4.802736 | -1.850609 | 1.271154 |
| H | -5.372815 | -0.449686 | 3.251499 |
| H | -3.594737 | 0.885151 | 4.371819 |
| H | -1.277660 | 0.851379 | 3.492500 |
| H | 0.365617 | -3.998235 | -0.736742 |
| H | -5.905583 | 0.920367 | -1.195726 |
| H | -1.839302 | -3.211043 | -2.717910 |
| H | -2.995049 | -2.717253 | -1.471835 |
| H | -3.353438 | -2.410253 | -3.186549 |
| H | -0.733610 | -1.507866 | -4.173307 |
| H | -2.153832 | -0.461179 | -4.410236 |
| H | -0.665039 | 0.200497 | -3.709172 |
| H | -2.515385 | 3.960019 | 2.658593 |
| H | -3.851763 | 4.110883 | 1.509079 |
| H | -3.605541 | 2.582765 | 2.390265 |
| H | -1.173631 | 4.869296 | 0.820959 |
| H | -0.738474 | 3.909376 | -0.597681 |
| H | -2.334046 | 4.677425 | -0.511702 |
| H | 5.806106 | 0.805413 | 2.672087 |
| H | 3.435894 | -3.381058 | -0.993852 |
| H | 4.825512 | -2.717355 | -0.110635 |
| H | 3.232569 | -2.850980 | 0.683312 |
| H | 3.926008 | -1.666064 | -2.823768 |
| H | 3.803456 | 0.074343 | -2.479142 |
| H | 5.195278 | -0.850176 | -1.880641 |
| H | 0.334089 | 3.586666 | 3.166709 |
| H | 0.739058 | 1.848631 | 3.188308 |
| H | 1.882847 | 3.040080 | 3.845717 |
| H | 1.612433 | 5.151622 | 1.554045 |
| H | 3.182268 | 4.595134 | 2.168069 |
| H | 2.817188 | 4.473858 | 0.433556 |

12-P

67

| | | | |
|---|---|---|---|
| C | 1.788995 | -0.418834 | 2.827635 |
| C | 1.465076 | 0.380139 | 1.719140 |
| C | 2.431194 | 1.276706 | 1.235902 |
| C | 3.693698 | 1.345178 | 1.821528 |
| C | 4.014800 | 0.519032 | 2.900845 |
| C | 3.056986 | -0.362694 | 3.405410 |
| P | -0.270150 | 0.345324 | 1.092135 |
| AU | -0.558374 | 2.158304 | -0.414023 |
| C | -0.238903 | -1.058283 | -0.054629 |
| N | 0.750431 | -1.727655 | -0.713075 |
| C | 0.215907 | -2.765591 | -1.494661 |
| C | -1.136848 | -2.735115 | -1.313757 |
| N | -1.394200 | -1.686594 | -0.423357 |
| C | 2.141176 | -1.374973 | -0.721144 |
| C | 2.575022 | -0.398756 | -1.626524 |
| C | 3.932019 | -0.062320 | -1.607595 |
| C | 4.829022 | -0.669870 | -0.724832 |
| C | 4.354477 | -1.664178 | 0.137098 |
| C | 3.011741 | -2.040868 | 0.151916 |
| C | 1.611287 | 0.282661 | -2.551316 |
| C | 2.516608 | -3.121450 | 1.071371 |
| C | 6.268023 | -0.235677 | -0.662462 |
| C | -2.694001 | -1.349873 | 0.082782 |
| C | -3.099491 | -1.921276 | 1.296121 |
| C | -4.372296 | -1.597284 | 1.772530 |
| C | -5.221718 | -0.736918 | 1.068874 |
| C | -4.777698 | -0.198439 | -0.144672 |
| C | -3.513385 | -0.490027 | -0.662554 |
| C | -2.173986 | -2.818810 | 2.067597 |
| C | -6.579814 | -0.377887 | 1.611736 |
| C | -3.031962 | 0.125238 | -1.943909 |
| C | 1.080445 | -3.635539 | -2.331291 |
| C | -2.229352 | -3.580405 | -1.857123 |
| H | 5.043244 | -2.146358 | 0.831799 |
| H | 4.290657 | 0.708033 | -2.291736 |
| H | -5.431225 | 0.471814 | -0.704972 |
| H | -4.706199 | -2.027968 | 2.717587 |
| H | 3.230404 | -3.292569 | 1.883947 |
| H | 1.548619 | -2.859121 | 1.513448 |
| H | 2.383207 | -4.071808 | 0.533292 |
| H | 2.141124 | 0.887754 | -3.294458 |
| H | 0.967640 | -0.436433 | -3.075870 |
| H | 0.935624 | 0.952562 | -1.983023 |
| H | 6.565555 | 0.305770 | -1.568155 |
| H | 6.423822 | 0.436085 | 0.195284 |
| H | 6.941411 | -1.092080 | -0.529665 |
| H | -3.850943 | 0.627404 | -2.469627 |
| H | -2.250591 | 0.879553 | -1.727722 |
| H | -2.580617 | -0.614428 | -2.617700 |
| H | -2.682857 | -3.254884 | 2.933625 |
| H | -1.782010 | -3.634747 | 1.445406 |
| H | -1.308839 | -2.240331 | 2.426604 |
| H | -6.873793 | -1.046917 | 2.428989 |
| H | -6.580612 | 0.650187 | 2.002936 |
| H | -7.347949 | -0.426981 | 0.828790 |
| H | 2.181172 | 1.914065 | 0.388155 |
| H | 4.433472 | 2.041682 | 1.427134 |
| H | 5.004721 | 0.569544 | 3.353589 |
| H | 3.293868 | -0.999856 | 4.257617 |
| H | 1.032453 | -1.086259 | 3.242280 |
| H | -0.785087 | 3.456754 | -1.377479 |
| H | 0.464617 | -4.330310 | -2.911163 |
| H | 1.683554 | -3.040766 | -3.030574 |
| H | 1.777592 | -4.223920 | -1.719164 |
| H | -1.819134 | -4.347533 | -2.521277 |
| H | -2.783645 | -4.081413 | -1.051665 |
| H | -2.954244 | -2.982231 | -2.425997 |

13-P

41

| | | | |
|---|---|---|---|
| C | -2.425137 | 1.731614 | 1.181679 |
| C | -1.884131 | 0.441964 | 1.032374 |
| C | -2.767037 | -0.644611 | 0.912277 |
| C | -4.147557 | -0.444059 | 0.924503 |
| C | -4.673013 | 0.844289 | 1.045931 |
| C | -3.803920 | 1.932061 | 1.172873 |
| P | -0.072614 | 0.156610 | 1.155582 |
| AU | 0.564877 | -1.732767 | -0.120643 |
| C | 0.523681 | 1.440844 | 0.012227 |
| N | 0.060503 | 1.848964 | -1.205514 |
| C | 0.895100 | 2.821839 | -1.725198 |
| C | 1.900659 | 3.015283 | -0.827652 |
| N | 1.654507 | 2.172985 | 0.241642 |
| C | -1.034513 | 1.202276 | -1.964452 |
| C | 2.526007 | 1.978312 | 1.420500 |
| H | 0.709347 | 3.292130 | -2.680641 |
| H | 2.748763 | 3.683574 | -0.857697 |
| H | -2.358824 | -1.648260 | 0.786969 |
| H | -4.816552 | -1.298988 | 0.825632 |
| H | -5.751219 | 1.000553 | 1.045203 |
| H | -4.203416 | 2.941241 | 1.276454 |
| H | -1.756953 | 2.585552 | 1.302155 |
| H | 1.047886 | -3.041701 | -0.964790 |
| C | -2.055527 | 2.239469 | -2.417050 |
| C | -0.439702 | 0.374674 | -3.100884 |
| H | -1.506307 | 0.524133 | -1.248314 |
| H | -1.243791 | -0.157656 | -3.622976 |
| H | 0.085950 | 1.007356 | -3.828300 |
| H | 0.258236 | -0.370355 | -2.695506 |
| H | -2.892938 | 1.729731 | -2.907659 |
| H | -2.447646 | 2.798333 | -1.559636 |
| H | -1.624634 | 2.946924 | -3.138098 |
| C | 3.521600 | 0.853083 | 1.141361 |
| C | 3.191376 | 3.285442 | 1.830933 |
| H | 1.832527 | 1.648071 | 2.208662 |
| H | 4.114626 | 0.650794 | 2.041516 |
| H | 2.996194 | -0.068968 | 0.858510 |
| H | 4.205122 | 1.131081 | 0.327969 |
| H | 3.713100 | 3.132165 | 2.782575 |
| H | 3.938690 | 3.612993 | 1.095924 |
| H | 2.453466 | 4.085085 | 1.968351 |

1

54

| | | | |
|---|---|---|---|
| C | 2.045646 | 0.888682 | 1.029629 |
| C | 1.613910 | 0.104889 | -0.058511 |
| C | 1.956352 | 0.413487 | -1.389151 |
| C | 2.841366 | 1.477445 | -1.599476 |
| C | 3.339958 | 2.222406 | -0.535965 |
| C | 2.923937 | 1.943927 | 0.762329 |
| N | 0.762502 | -1.034243 | 0.209552 |
| C | 1.351113 | -2.421966 | 0.470114 |
| C | 0.091187 | -3.172991 | 0.940577 |
| C | -1.126989 | -2.319350 | 0.504171 |
| C | -0.544415 | -0.950500 | 0.222099 |
| C | 1.941211 | -2.990567 | -0.821586 |
| C | 2.442084 | -2.365841 | 1.535921 |
| AU | -1.620375 | 0.759714 | -0.117288 |
| C | -1.771426 | -2.824352 | -0.802907 |
| C | -2.192998 | -2.230598 | 1.604076 |
| C | 1.307660 | -0.258744 | -2.587178 |
| C | 2.324278 | -0.746936 | -3.627336 |
| C | 1.497051 | 0.711284 | 2.434978 |
| C | 0.474398 | 1.823092 | 2.727991 |
| C | 0.288447 | 0.703432 | -3.222344 |
| C | 2.594759 | 0.670684 | 3.506015 |
| H | -2.504355 | 2.103887 | -0.373997 |
| H | 3.123396 | 1.740173 | -2.618235 |
| H | 3.271049 | 2.569298 | 1.583717 |
| H | 0.951422 | -0.238758 | 2.472062 |
| H | 0.055730 | -4.187461 | 0.527341 |
| H | -2.214816 | -3.814609 | -0.633785 |
| H | -1.036475 | -2.908969 | -1.613059 |
| H | -2.559397 | -2.134952 | -1.130443 |
| H | -1.772050 | -1.812837 | 2.527648 |
| H | -2.586636 | -3.232307 | 1.824292 |
| H | -3.022196 | -1.586609 | 1.285707 |
| H | 2.360675 | -3.981830 | -0.610702 |
| H | 2.751113 | -2.353797 | -1.195964 |
| H | 1.186108 | -3.100958 | -1.606779 |
| H | 2.053632 | -2.011831 | 2.495846 |
| H | 3.270502 | -1.717381 | 1.225188 |
| H | 2.839680 | -3.377454 | 1.683787 |
| H | 0.104992 | -3.260019 | 2.034651 |
| H | 4.029887 | 3.044894 | -0.722417 |
| H | 2.152684 | 0.466083 | 4.489996 |
| H | 3.122203 | 1.631065 | 3.575885 |
| H | 3.339624 | -0.106786 | 3.294214 |
| H | 0.031134 | 1.682555 | 3.723071 |
| H | -0.333175 | 1.810384 | 1.979972 |
| H | 0.952804 | 2.811341 | 2.698139 |
| H | -0.227600 | 0.214666 | -4.059756 |
| H | 0.785577 | 1.606649 | -3.601273 |
| H | -0.464532 | 1.010270 | -2.480117 |
| H | 0.743982 | -1.126561 | -2.228587 |
| H | 1.807658 | -1.278870 | -4.436991 |
| H | 3.064161 | -1.427571 | -3.187568 |
| H | 2.868260 | 0.092839 | -4.079496 |

2

61

| | | | |
|---|---|---|---|
| C | -4.398970 | -1.465666 | -0.118031 |
| C | -3.030321 | -0.982817 | -0.628223 |
| C | -2.016642 | -0.833378 | 0.519798 |
| C | -2.518328 | 0.274802 | 1.509170 |
| C | -4.034549 | 0.510771 | 1.422990 |
| C | -4.785294 | -0.808495 | 1.228753 |
| C | -1.709824 | -2.190517 | 1.201521 |
| C | -0.199490 | -2.197082 | 1.497652 |
| N | 0.268079 | -1.119852 | 0.520101 |
| C | -0.676898 | -0.383008 | -0.011446 |
| C | 1.670288 | -0.924358 | 0.225592 |
| C | 2.417894 | -0.012279 | 0.994561 |
| C | 3.787673 | 0.091469 | 0.724886 |
| C | 4.382625 | -0.656746 | -0.285212 |
| C | 3.605436 | -1.495365 | -1.078422 |
| C | 2.233199 | -1.640026 | -0.849253 |
| C | 1.783340 | 0.939474 | 1.993305 |
| C | 1.745775 | 2.357330 | 1.396935 |
| C | 1.390976 | -2.441926 | -1.826719 |
| C | 1.965826 | -3.835045 | -2.111528 |
| AU | -0.381427 | 1.164193 | -1.323010 |
| C | 0.131817 | -1.781022 | 2.932728 |
| C | 0.461723 | -3.540269 | 1.202808 |
| C | 1.211428 | -1.636367 | -3.125689 |
| C | 2.487349 | 0.934465 | 3.356391 |
| H | 4.392142 | 0.788856 | 1.303487 |
| H | 4.067757 | -2.034640 | -1.904124 |
| H | 0.391921 | -2.569791 | -1.393442 |
| H | -2.294641 | -2.350987 | 2.113635 |
| H | -0.229048 | -2.558282 | 3.617318 |
| H | 1.214790 | -1.684267 | 3.072080 |
| H | -0.344686 | -0.834493 | 3.207445 |
| H | 0.213740 | -3.904392 | 0.201086 |
| H | 1.552731 | -3.476907 | 1.298626 |
| H | 0.099272 | -4.276785 | 1.930589 |
| H | -1.943931 | -3.008093 | 0.506432 |
| H | 5.451234 | -0.561726 | -0.476252 |
| H | 1.282045 | -4.399037 | -2.759442 |
| H | 2.931810 | -3.772510 | -2.629489 |
| H | 2.116661 | -4.407370 | -1.187291 |
| H | 0.562758 | -2.179167 | -3.826339 |
| H | 0.754518 | -0.657789 | -2.912518 |
| H | 2.180114 | -1.462174 | -3.613551 |
| H | 1.243223 | 3.048513 | 2.086945 |
| H | 2.761572 | 2.731820 | 1.210892 |
| H | 1.198221 | 2.358259 | 0.442021 |
| H | 0.743602 | 0.631539 | 2.146126 |
| H | 1.949877 | 1.583609 | 4.060003 |
| H | 2.535327 | -0.073324 | 3.787935 |
| H | 3.514797 | 1.313745 | 3.276753 |
| H | -1.986907 | 1.214049 | 1.299016 |
| H | -2.246600 | -0.022943 | 2.530977 |
| H | -4.371352 | 1.024052 | 2.333824 |
| H | -4.259182 | 1.186694 | 0.584526 |
| H | -5.870396 | -0.648050 | 1.277099 |
| H | -4.536564 | -1.483486 | 2.062390 |
| H | -4.396356 | -2.559306 | -0.011188 |
| H | -5.154060 | -1.233533 | -0.881747 |
| H | -3.122854 | -0.006544 | -1.124275 |
| H | -2.633828 | -1.674744 | -1.384528 |
| H | -0.182397 | 2.396271 | -2.369889 |

3

73

| | | | |
|---|---|---|---|
| C | 4.137432 | -0.971381 | -0.554568 |
| C | 2.764423 | -0.835812 | 0.125553 |
| C | 1.758906 | -0.153380 | -0.841938 |
| C | 2.232334 | 1.324932 | -1.039070 |
| C | 3.764085 | 1.507943 | -0.964630 |
| C | 4.479184 | 0.242151 | -1.445099 |
| C | 1.540779 | -0.878522 | -2.197906 |
| C | 0.022263 | -0.858583 | -2.470807 |
| N | -0.509748 | -0.464211 | -1.094820 |
| C | 0.387730 | -0.084352 | -0.219788 |
| C | -1.925144 | -0.494532 | -0.794807 |
| C | -2.703999 | 0.658677 | -1.009228 |
| C | -4.081036 | 0.562713 | -0.775838 |
| C | -4.655336 | -0.619379 | -0.320044 |
| C | -3.850466 | -1.719672 | -0.041433 |
| C | -2.469824 | -1.679391 | -0.260911 |
| C | -2.099980 | 2.011852 | -1.343147 |
| C | -2.133289 | 2.909541 | -0.093871 |
| C | -1.605008 | -2.840539 | 0.196084 |
| C | -2.096954 | -4.198142 | -0.320158 |
| AU | -0.013369 | 0.541712 | 1.694258 |
| C | -0.374141 | 0.198709 | -3.504065 |
| C | -0.528529 | -2.212873 | -2.909543 |
| C | 2.272714 | -2.120124 | 0.841206 |
| C | 3.143257 | -2.393341 | 2.074464 |
| C | 4.193509 | 2.749647 | -1.745084 |
| C | -1.503343 | -2.833278 | 1.730918 |
| C | -2.785583 | 2.698146 | -2.532163 |
| C | 2.164420 | -3.389939 | -0.009357 |
| H | -4.709096 | 1.438831 | -0.932702 |
| H | -4.297719 | -2.622442 | 0.372527 |
| H | -0.591449 | -2.680735 | -0.186550 |
| H | 2.078977 | -0.382718 | -3.012151 |
| H | 0.017010 | -0.102713 | -4.483510 |
| H | -1.464266 | 0.281189 | -3.585478 |
| H | 0.038791 | 1.183231 | -3.259820 |
| H | -0.237426 | -3.014488 | -2.223839 |
| H | -1.622716 | -2.192512 | -2.985247 |
| H | -0.123376 | -2.451924 | -3.900395 |
| H | 1.900808 | -1.909955 | -2.154598 |
| H | -5.730112 | -0.672664 | -0.148576 |
| H | -1.392049 | -4.988299 | -0.029609 |
| H | -3.075556 | -4.457728 | 0.104853 |
| H | -2.192174 | -4.206089 | -1.413158 |
| H | -0.823624 | -3.625591 | 2.072307 |
| H | -1.119355 | -1.865457 | 2.088075 |
| H | -2.487287 | -3.000154 | 2.189763 |
| H | -1.652515 | 3.874306 | -0.304446 |
| H | -3.167027 | 3.099898 | 0.225217 |
| H | -1.598769 | 2.427907 | 0.739323 |
| H | -1.045310 | 1.862067 | -1.597518 |
| H | -2.270028 | 3.636928 | -2.773371 |
| H | -2.780591 | 2.064323 | -3.428018 |
| H | -3.830350 | 2.945596 | -2.302363 |
| H | 1.756240 | 1.958518 | -0.276474 |
| H | 1.867429 | 1.677157 | -2.015345 |
| H | 4.037042 | 1.661679 | 0.091924 |
| H | 5.565641 | 0.407217 | -1.458951 |
| H | 4.188483 | 0.046673 | -2.489028 |
| H | 4.177511 | -1.890906 | -1.154911 |
| H | 4.901615 | -1.081530 | 0.226569 |
| H | 2.867645 | -0.104390 | 0.944012 |
| H | 1.263912 | -1.882210 | 1.218056 |
| H | 2.729241 | -3.222113 | 2.663979 |
| H | 4.169219 | -2.668674 | 1.791829 |
| H | 3.192200 | -1.508583 | 2.723295 |
| H | 1.849046 | -4.231022 | 0.622646 |
| H | 1.427487 | -3.302169 | -0.815127 |
| H | 3.126910 | -3.665193 | -0.462517 |
| H | 5.269050 | 2.939728 | -1.627200 |
| H | 3.987420 | 2.622201 | -2.818473 |
| H | 3.652041 | 3.642605 | -1.403581 |
| H | -0.294995 | 1.037064 | 3.219014 |

4

51

| | | | |
|---|---|---|---|
| C | -3.038847 | 1.042643 | 1.231964 |
| C | -2.316458 | 1.047547 | 0.028329 |
| C | -2.889567 | 0.631052 | -1.181838 |
| C | -4.228285 | 0.227175 | -1.169195 |
| C | -4.985349 | 0.224072 | 0.005982 |
| C | -4.372742 | 0.631769 | 1.197095 |
| N | -0.957665 | 1.489865 | 0.044610 |
| C | 0.103212 | 0.669695 | 0.068369 |
| N | 1.217472 | 1.412068 | -0.017146 |
| C | 0.943745 | 2.868919 | -0.018841 |
| C | -0.588453 | 2.912213 | -0.144603 |
| AU | 0.022972 | -1.377899 | 0.221585 |
| C | 2.544525 | 0.890605 | 0.071270 |
| C | 3.064870 | 0.539694 | 1.327447 |
| C | 4.375345 | 0.058440 | 1.383689 |
| C | 5.160583 | -0.078657 | 0.233906 |
| C | 4.603415 | 0.271742 | -1.000193 |
| C | 3.296236 | 0.754587 | -1.105585 |
| C | 2.218355 | 0.632890 | 2.566515 |
| C | 6.559606 | -0.630654 | 0.319259 |
| C | 2.694813 | 1.085878 | -2.443992 |
| C | -2.069880 | 0.569441 | -2.440690 |
| C | -6.418228 | -0.241027 | 0.002442 |
| C | -2.370742 | 1.423874 | 2.524420 |
| H | 5.197267 | 0.156376 | -1.908439 |
| H | 4.792606 | -0.218916 | 2.353204 |
| H | -4.944499 | 0.617906 | 2.126521 |
| H | -4.687899 | -0.101018 | -2.102822 |
| H | 3.340585 | 0.735837 | -3.256646 |
| H | 1.708376 | 0.616877 | -2.554087 |
| H | 2.554899 | 2.169486 | -2.571807 |
| H | 2.822886 | 0.471544 | 3.465730 |
| H | 1.719840 | 1.607488 | 2.652091 |
| H | 1.425057 | -0.130266 | 2.538803 |
| H | 7.041789 | -0.352762 | 1.264789 |
| H | 6.545908 | -1.729738 | 0.268763 |
| H | 7.182816 | -0.271125 | -0.508550 |
| H | -3.056531 | 1.296397 | 3.368959 |
| H | -1.483934 | 0.798848 | 2.697718 |
| H | -2.031294 | 2.469477 | 2.518221 |
| H | -2.697296 | 0.320939 | -3.303487 |
| H | -1.558813 | 1.519411 | -2.647136 |
| H | -1.289669 | -0.200841 | -2.345748 |
| H | -6.817984 | -0.295154 | -1.017069 |
| H | -6.502896 | -1.242739 | 0.448804 |
| H | -7.057288 | 0.430842 | 0.590498 |
| H | -1.064028 | 3.539069 | 0.618104 |
| H | -0.924228 | 3.255474 | -1.133036 |
| H | 1.305536 | 3.314294 | 0.918250 |
| H | 1.456476 | 3.357881 | -0.854695 |
| H | -0.043236 | -2.994562 | 0.348105 |

5
69

| | | | |
|---|---|---|---|
| C | -3.314171 | -1.021403 | 0.537011 |
| C | -2.542888 | 0.147288 | 0.692160 |
| C | -3.029164 | 1.416522 | 0.316456 |
| C | -4.319658 | 1.503375 | -0.214625 |
| C | -5.103284 | 0.365375 | -0.372158 |
| C | -4.598689 | -0.877869 | -0.004782 |
| N | -1.227878 | 0.076472 | 1.255330 |
| C | -0.097603 | 0.108976 | 0.530685 |
| N | 0.942782 | 0.201490 | 1.371024 |
| C | 0.545052 | 0.262151 | 2.795263 |
| C | -0.991582 | 0.250862 | 2.706688 |
| AU | 0.023674 | 0.014774 | -1.519871 |
| C | 2.317253 | 0.257240 | 0.976151 |
| C | 2.894459 | 1.511451 | 0.710286 |
| C | 4.244700 | 1.544467 | 0.345423 |
| C | 4.989324 | 0.371830 | 0.253585 |
| C | 4.394617 | -0.859450 | 0.518437 |
| C | 3.047534 | -0.941672 | 0.885628 |
| C | 2.077637 | 2.788856 | 0.741642 |
| C | 1.826836 | 3.279617 | -0.693201 |
| C | 2.382772 | -2.288303 | 1.093118 |
| C | 2.090206 | -2.936553 | -0.270475 |
| C | -2.188022 | 2.674790 | 0.436915 |
| C | -1.878731 | 3.246968 | -0.954407 |
| C | -2.896914 | -2.433635 | 0.931499 |
| C | -1.436471 | -2.826528 | 0.680454 |
| C | 2.725138 | 3.876953 | 1.608436 |
| C | 3.195230 | -3.221197 | 1.999947 |
| C | -2.855550 | 3.719177 | 1.344212 |
| C | -3.313634 | -2.739961 | 2.381650 |
| H | -5.208216 | -1.771378 | -0.145609 |
| H | -4.709993 | 2.474275 | -0.518755 |
| H | 4.716471 | 2.500292 | 0.119448 |
| H | 4.983232 | -1.771526 | 0.427037 |
| H | -3.507163 | -3.086220 | 0.288188 |
| H | -1.227648 | 2.402944 | 0.892851 |
| H | 1.098777 | 2.550387 | 1.179510 |
| H | 1.414861 | -2.110168 | 1.579622 |
| H | 0.943644 | 1.173277 | 3.258373 |
| H | 0.948712 | -0.603997 | 3.336175 |
| H | -1.446735 | 1.190768 | 3.047084 |
| H | -1.442455 | -0.572872 | 3.270962 |
| H | 6.039254 | 0.416746 | -0.035241 |
| H | -6.104838 | 0.444415 | -0.794216 |
| H | 1.549440 | -3.883331 | -0.140618 |
| H | 1.479505 | -2.266478 | -0.893816 |
| H | 3.025917 | -3.141997 | -0.808418 |
| H | 2.636221 | -4.148242 | 2.184101 |
| H | 4.152301 | -3.500185 | 1.539248 |
| H | 3.410514 | -2.749900 | 2.968025 |
| H | 1.204214 | 4.183877 | -0.689508 |
| H | 2.775393 | 3.516742 | -1.193968 |
| H | 1.314096 | 2.503936 | -1.280391 |
| H | 2.073271 | 4.759224 | 1.658194 |
| H | 2.901762 | 3.520202 | 2.631673 |
| H | 3.689022 | 4.200332 | 1.193159 |
| H | -3.112409 | -3.792508 | 2.622662 |
| H | -2.753649 | -2.120274 | 3.094768 |
| H | -4.382497 | -2.546472 | 2.537278 |
| H | -1.347059 | -3.920554 | 0.699369 |
| H | -1.078990 | -2.464873 | -0.291791 |
| H | -0.764851 | -2.430214 | 1.450338 |
| H | -1.242514 | 4.137522 | -0.871074 |
| H | -1.353486 | 2.499688 | -1.566050 |
| H | -2.801286 | 3.533613 | -1.476866 |
| H | -2.202818 | 4.594491 | 1.461116 |
| H | -3.807432 | 4.065541 | 0.919535 |
| H | -3.063571 | 3.308176 | 2.341096 |
| H | 0.117003 | -0.070761 | -3.136891 |

6

58

| | | | |
|---|---|---|---|
| C | 1.070108 | 2.629265 | -0.328316 |
| N | 1.028988 | 1.214551 | -0.228707 |
| C | -0.100019 | 0.457456 | -0.198375 |
| N | -1.293287 | 1.103559 | -0.279084 |
| C | -1.463475 | 2.507842 | -0.388933 |
| C | -0.240981 | 3.350856 | -0.075340 |
| C | 2.303631 | 0.514606 | -0.233412 |
| C | 2.853806 | 0.125971 | -1.459709 |
| C | 4.082546 | -0.536016 | -1.432300 |
| C | 4.746831 | -0.807053 | -0.230678 |
| C | 4.156405 | -0.396715 | 0.968408 |
| C | 2.929638 | 0.272044 | 0.992551 |
| C | 2.132909 | 0.388981 | -2.751305 |
| C | 2.289260 | 0.689515 | 2.286220 |
| C | 6.056249 | -1.550321 | -0.231520 |
| AU | -0.005183 | -1.586197 | -0.022967 |
| C | -2.494907 | 0.286826 | -0.332800 |
| C | -2.945870 | -0.166506 | -1.575650 |
| C | -4.107226 | -0.943732 | -1.594201 |
| C | -4.799435 | -1.261408 | -0.421804 |
| C | -4.308676 | -0.780611 | 0.797900 |
| C | -3.154688 | 0.001329 | 0.867581 |
| O | -2.551728 | 2.980291 | -0.644765 |
| C | -2.191755 | 0.147040 | -2.836571 |
| C | -6.034334 | -2.121964 | -0.462699 |
| C | -2.619319 | 0.495846 | 2.181718 |
| C | -0.290092 | 3.605366 | 1.464457 |
| C | -0.286241 | 4.678960 | -0.837249 |
| O | 2.119610 | 3.204195 | -0.530866 |
| H | -4.835843 | -1.021569 | 1.722140 |
| H | -4.478147 | -1.310200 | -2.552218 |
| H | 4.530647 | -0.849274 | -2.376326 |
| H | 4.660414 | -0.602694 | 1.913611 |
| H | -3.249970 | 0.156409 | 3.009716 |
| H | -1.599523 | 0.120977 | 2.352836 |
| H | -2.578988 | 1.593518 | 2.214475 |
| H | -2.723507 | -0.237442 | -3.712927 |
| H | -2.052771 | 1.228700 | -2.967429 |
| H | -1.193760 | -0.315238 | -2.814655 |
| H | -6.451767 | -2.173521 | -1.475150 |
| H | -5.802551 | -3.148784 | -0.143670 |
| H | -6.809133 | -1.738329 | 0.213639 |
| H | 2.725990 | 0.043635 | -3.604294 |
| H | 1.168841 | -0.139993 | -2.771081 |
| H | 1.925959 | 1.459003 | -2.888631 |
| H | 2.914255 | 0.403670 | 3.138385 |
| H | 2.134416 | 1.776682 | 2.328425 |
| H | 1.306964 | 0.209364 | 2.406696 |
| H | 6.614056 | -1.377582 | 0.696594 |
| H | 5.886323 | -2.633585 | -0.321194 |
| H | 6.685397 | -1.247728 | -1.078134 |
| H | -0.246919 | 4.511438 | -1.919869 |
| H | -1.215240 | 5.205228 | -0.599834 |
| H | 0.569460 | 5.296175 | -0.548830 |
| H | -1.217549 | 4.136806 | 1.705095 |
| H | -0.256559 | 2.669387 | 2.033748 |
| H | 0.567110 | 4.226483 | 1.746897 |
| H | 0.068234 | -3.194356 | 0.158676 |

7

59

| | | | |
|---|---|---|---|
| C | -1.068432 | 2.667595 | 0.348763 |
| N | -0.991558 | 1.260156 | 0.297879 |
| C | 0.167399 | 0.518817 | 0.178308 |
| N | 1.314613 | 1.169676 | 0.324121 |
| C | 1.397042 | 2.596347 | 0.669910 |
| C | 0.240495 | 3.398193 | 0.073617 |
| C | -2.243555 | 0.525615 | 0.312269 |
| C | -2.754499 | 0.100888 | 1.543742 |
| C | -3.960012 | -0.604048 | 1.539176 |
| C | -4.643629 | -0.884471 | 0.350785 |
| C | -4.097272 | -0.436314 | -0.855707 |
| C | -2.895463 | 0.275952 | -0.899867 |
| C | -2.017943 | 0.384542 | 2.822592 |
| C | -2.305806 | 0.733183 | -2.203696 |
| C | -5.925697 | -1.674718 | 0.372883 |
| AU | 0.096339 | -1.507055 | -0.205786 |
| C | 2.552725 | 0.422491 | 0.357931 |
| C | 2.947755 | -0.170055 | 1.565496 |
| C | 4.176169 | -0.837267 | 1.592282 |
| C | 4.992244 | -0.927461 | 0.461451 |
| C | 4.551017 | -0.335932 | -0.727376 |
| C | 3.333771 | 0.343584 | -0.804061 |
| C | 2.055706 | -0.157465 | 2.775599 |
| C | 6.298201 | -1.675605 | 0.506184 |
| C | 2.853433 | 0.926360 | -2.102403 |
| C | 0.361729 | 3.506804 | -1.461268 |
| C | 0.212803 | 4.790879 | 0.707571 |
| O | -2.138112 | 3.228367 | 0.512477 |
| H | 5.166492 | -0.410688 | -1.625242 |
| H | 4.497840 | -1.304516 | 2.524189 |
| H | -4.373943 | -0.943808 | 2.489869 |
| H | -4.616991 | -0.647452 | -1.791511 |
| H | 3.559019 | 0.705148 | -2.910010 |
| H | 1.874281 | 0.507069 | -2.373632 |
| H | 2.735607 | 2.016610 | -2.044410 |
| H | 2.599089 | -0.498045 | 3.663242 |
| H | 1.643154 | 0.837280 | 2.986026 |
| H | 1.203063 | -0.834808 | 2.614418 |
| H | 6.653906 | -1.797760 | 1.536061 |
| H | 6.184127 | -2.679613 | 0.071365 |
| H | 7.074398 | -1.157386 | -0.071336 |
| H | -2.569267 | -0.007780 | 3.683637 |
| H | -1.023063 | -0.082081 | 2.811701 |
| H | -1.872733 | 1.463704 | 2.970047 |
| H | -2.949293 | 0.447544 | -3.042413 |
| H | -2.176778 | 1.823941 | -2.227215 |
| H | -1.316019 | 0.278439 | -2.359617 |
| H | -6.507670 | -1.517607 | -0.543280 |
| H | -5.716727 | -2.751873 | 0.453425 |
| H | -6.549366 | -1.398628 | 1.232762 |
| H | 0.085740 | 4.728498 | 1.795659 |
| H | 1.153779 | 5.314082 | 0.494813 |
| H | -0.615614 | 5.379636 | 0.301226 |
| H | 1.287227 | 4.038293 | -1.718283 |
| H | 0.381106 | 2.522139 | -1.940579 |
| H | -0.485414 | 4.073325 | -1.865110 |
| H | 2.356938 | 2.979523 | 0.303323 |
| H | 1.402247 | 2.689911 | 1.768139 |
| H | 0.047225 | -3.095644 | -0.526081 |

8

23

| | | | |
|---|---|---|---|
| C | 2.869443 | 0.683565 | 0.009705 |
| C | 2.868532 | -0.685914 | 0.009591 |
| N | 1.528669 | -1.073428 | 0.002818 |
| C | 0.691870 | 0.000339 | -0.001579 |
| N | 1.530114 | 1.072940 | 0.002730 |
| C | 3.987990 | -1.663196 | 0.012267 |
| C | 1.069576 | -2.454464 | -0.000245 |
| AU | -1.371753 | 0.001260 | -0.005106 |
| C | 1.073209 | 2.454665 | -0.001434 |
| C | 3.990089 | 1.659422 | 0.014084 |
| H | -2.993390 | 0.001916 | -0.005962 |
| H | 4.949902 | 1.132379 | 0.020147 |
| H | 3.957922 | 2.310523 | 0.899304 |
| H | 3.967386 | 2.307098 | -0.873948 |
| H | 4.948431 | -1.137362 | 0.021682 |
| H | 3.965848 | -2.307747 | -0.878085 |
| H | 3.953634 | -2.317349 | 0.895111 |
| H | 1.445647 | -2.981047 | -0.885183 |
| H | -0.024612 | -2.436751 | -0.022731 |
| H | 1.409078 | -2.974000 | 0.903649 |
| H | 1.417060 | 2.975453 | 0.900042 |
| H | -0.021052 | 2.438980 | -0.019770 |
| H | 1.446717 | 2.978833 | -0.888922 |

9
49

| | | | |
|---|---|---|---|
| C | -3.022296 | -0.978523 | -1.269390 |
| C | -2.302977 | -1.132803 | -0.076291 |
| C | -2.860302 | -0.875420 | 1.181613 |
| C | -4.198860 | -0.472443 | 1.225115 |
| C | -4.960531 | -0.320444 | 0.063166 |
| C | -4.355700 | -0.575743 | -1.173812 |
| N | -0.928681 | -1.545185 | -0.149999 |
| C | 0.115878 | -0.670467 | -0.113992 |
| N | 1.216492 | -1.469425 | -0.199578 |
| C | 0.868561 | -2.813115 | -0.286383 |
| C | -0.490699 | -2.861157 | -0.255785 |
| AU | 0.051585 | 1.371074 | 0.048305 |
| C | 2.559546 | -0.960654 | -0.165311 |
| C | 3.148448 | -0.526913 | -1.359698 |
| C | 4.454450 | -0.034634 | -1.293761 |
| C | 5.153995 | 0.038047 | -0.083924 |
| C | 4.518679 | -0.392434 | 1.085628 |
| C | 3.213943 | -0.892419 | 1.070881 |
| C | 2.379994 | -0.549095 | -2.651316 |
| C | 6.547517 | 0.607623 | -0.036961 |
| C | 2.513294 | -1.303631 | 2.336280 |
| C | -2.033870 | -0.983900 | 2.432427 |
| C | -6.393964 | 0.136496 | 0.132231 |
| C | -2.360738 | -1.200336 | -2.601362 |
| H | 1.611929 | -3.595179 | -0.360681 |
| H | -1.179458 | -3.693930 | -0.295749 |
| H | 5.047098 | -0.328985 | 2.037915 |
| H | 4.933881 | 0.307112 | -2.212238 |
| H | -4.932961 | -0.447564 | -2.090819 |
| H | -4.654079 | -0.266122 | 2.194774 |
| H | 3.161641 | -1.152200 | 3.205641 |
| H | 1.598441 | -0.710713 | 2.477250 |
| H | 2.210719 | -2.359433 | 2.310654 |
| H | 3.026148 | -0.281151 | -3.493934 |
| H | 1.940158 | -1.535991 | -2.847344 |
| H | 1.549009 | 0.170521 | -2.608633 |
| H | 7.110646 | 0.360921 | -0.945694 |
| H | 6.513673 | 1.704655 | 0.039799 |
| H | 7.103971 | 0.233225 | 0.830678 |
| H | -3.059174 | -0.997320 | -3.419978 |
| H | -1.491092 | -0.537615 | -2.713689 |
| H | -1.994456 | -2.230487 | -2.708788 |
| H | -2.662368 | -0.887460 | 3.324119 |
| H | -1.495801 | -1.939600 | 2.483950 |
| H | -1.276238 | -0.186060 | 2.453817 |
| H | -6.771013 | 0.114841 | 1.161444 |
| H | -6.492361 | 1.166323 | -0.240753 |
| H | -7.043403 | -0.495863 | -0.487920 |
| H | 0.001547 | 2.986279 | 0.180133 |

10

67

| | | | |
|---|---|---|---|
| C | 3.036897 | 0.799383 | 1.173645 |
| C | 2.248195 | -0.362527 | 1.197060 |
| C | 2.730123 | -1.628957 | 0.834125 |
| C | 4.072758 | -1.716827 | 0.448402 |
| C | 4.885256 | -0.586696 | 0.425118 |
| C | 4.371823 | 0.658908 | 0.780751 |
| N | 0.869602 | -0.239289 | 1.588256 |
| C | -0.146399 | -0.040846 | 0.701677 |
| N | -1.255104 | 0.039065 | 1.490528 |
| C | -0.939057 | -0.100946 | 2.837280 |
| C | 0.408185 | -0.280400 | 2.898837 |
| AU | -0.027744 | 0.107570 | -1.343616 |
| C | -2.579976 | 0.261304 | 0.976476 |
| C | -2.973508 | 1.577253 | 0.692978 |
| C | -4.267876 | 1.764327 | 0.192593 |
| C | -5.119213 | 0.682634 | -0.010679 |
| C | -4.692398 | -0.614658 | 0.269568 |
| C | -3.407554 | -0.855047 | 0.766339 |
| C | -2.041655 | 2.760781 | 0.876598 |
| C | -2.629617 | 3.800620 | 1.841552 |
| C | -2.890459 | -2.266944 | 0.977252 |
| C | -3.942803 | -3.216733 | 1.560295 |
| C | 1.831667 | -2.850961 | 0.794587 |
| C | 2.412800 | -4.021200 | 1.599944 |
| C | 2.445970 | 2.164605 | 1.475139 |
| C | 3.353652 | 3.024723 | 2.363289 |
| C | -1.690471 | 3.379384 | -0.485375 |
| C | -2.323311 | -2.807636 | -0.347999 |
| C | 1.551231 | -3.248343 | -0.663700 |
| C | 2.103878 | 2.880841 | 0.156864 |
| H | -1.694145 | -0.060521 | 3.610333 |
| H | 1.075057 | -0.430128 | 3.736664 |
| H | -5.364404 | -1.451970 | 0.089299 |
| H | -4.609275 | 2.771339 | -0.044558 |
| H | 5.015314 | 1.536450 | 0.741729 |
| H | 4.483208 | -2.681623 | 0.152598 |
| H | -2.060220 | -2.217833 | 1.695258 |
| H | -1.105977 | 2.391865 | 1.316130 |
| H | 1.504520 | 2.011813 | 2.020229 |
| H | 0.870501 | -2.579756 | 1.250789 |
| H | -6.123315 | 0.849144 | -0.400180 |
| H | -0.985308 | 4.211032 | -0.357914 |
| H | -1.225386 | 2.626415 | -1.138211 |
| H | -2.588009 | 3.764515 | -0.987969 |
| H | -1.916759 | 4.621871 | 1.993810 |
| H | -3.558470 | 4.233094 | 1.445860 |
| H | -2.854600 | 3.353376 | 2.818642 |
| H | -1.875434 | -3.799258 | -0.199138 |
| H | -3.121604 | -2.896992 | -1.097613 |
| H | -1.555211 | -2.133829 | -0.753812 |
| H | -3.484886 | -4.189543 | 1.781441 |
| H | -4.372913 | -2.819414 | 2.488864 |
| H | -4.763789 | -3.396036 | 0.853378 |
| H | 5.927193 | -0.674904 | 0.118127 |
| H | 1.614749 | 3.843352 | 0.357378 |
| H | 3.015432 | 3.070808 | -0.426355 |
| H | 1.429790 | 2.266727 | -0.458359 |
| H | 2.845943 | 3.963753 | 2.619685 |
| H | 3.609215 | 2.505462 | 3.296174 |
| H | 4.289793 | 3.287032 | 1.852651 |
| H | 0.858937 | -4.099502 | -0.703775 |
| H | 1.101891 | -2.407138 | -1.212157 |
| H | 2.479774 | -3.534348 | -1.176547 |
| H | 1.711234 | -4.865828 | 1.598564 |
| H | 3.358942 | -4.374683 | 1.168620 |
| H | 2.602888 | -3.732295 | 2.641879 |
| H | 0.066773 | 0.224371 | -2.956267 |

11

67

| | | | |
|---|---|---|---|
| C | -3.052609 | 0.348359 | -0.927061 |
| C | -2.352176 | -0.655626 | -0.244065 |
| C | -2.904887 | -1.424338 | 0.791934 |
| C | -4.231327 | -1.157599 | 1.143893 |
| C | -4.958784 | -0.167453 | 0.485916 |
| C | -4.376143 | 0.578647 | -0.534875 |
| N | -0.982094 | -0.900554 | -0.605608 |
| C | 0.070825 | -0.211338 | -0.080359 |
| N | 1.169595 | -0.773927 | -0.660009 |
| C | 0.813044 | -1.804111 | -1.528731 |
| C | -0.552247 | -1.882110 | -1.496796 |
| AU | 0.018071 | 1.309464 | 1.294747 |
| C | 2.518442 | -0.362318 | -0.379930 |
| C | 3.122043 | 0.558916 | -1.250658 |
| C | 4.430211 | 0.951317 | -0.953856 |
| C | 5.090831 | 0.444948 | 0.164668 |
| C | 4.458591 | -0.462638 | 1.009473 |
| C | 3.150527 | -0.890770 | 0.753855 |
| CL | 1.947667 | -2.715447 | -2.401619 |
| CL | -1.615064 | -2.919378 | -2.318451 |
| C | 2.353643 | 1.160162 | -2.412761 |
| C | 3.210502 | 1.337347 | -3.671065 |
| C | 2.456355 | -1.861055 | 1.690497 |
| C | 3.223538 | -3.186827 | 1.797317 |
| C | -2.076009 | -2.443809 | 1.550595 |
| C | -2.822823 | -3.763347 | 1.779362 |
| C | -2.404278 | 1.191608 | -2.007975 |
| C | -3.212417 | 1.171111 | -3.313097 |
| C | 1.712652 | 2.489290 | -1.974350 |
| C | 2.234551 | -1.213302 | 3.065817 |
| C | -1.583176 | -1.829429 | 2.872276 |
| C | -2.182238 | 2.624467 | -1.497119 |
| H | 4.985163 | -0.842026 | 1.884332 |
| H | 4.936565 | 1.667663 | -1.598456 |
| H | -4.954566 | 1.356920 | -1.031078 |
| H | -4.698152 | -1.724157 | 1.948085 |
| H | 1.465421 | -2.086037 | 1.274709 |
| H | 1.537312 | 0.470861 | -2.670272 |
| H | -1.415996 | 0.764706 | -2.225248 |
| H | -1.188131 | -2.675868 | 0.946553 |
| H | 6.109218 | 0.766388 | 0.381320 |
| H | 1.092546 | 2.901884 | -2.781165 |
| H | 1.082596 | 2.349661 | -1.084240 |
| H | 2.489305 | 3.224779 | -1.723784 |
| H | 2.582795 | 1.676301 | -4.505220 |
| H | 3.994854 | 2.091681 | -3.524520 |
| H | 3.691051 | 0.394633 | -3.962444 |
| H | 1.692730 | -1.899928 | 3.729192 |
| H | 3.191217 | -0.960041 | 3.542169 |
| H | 1.645183 | -0.290647 | 2.963656 |
| H | 2.674696 | -3.891958 | 2.435324 |
| H | 3.357777 | -3.646925 | 0.809734 |
| H | 4.217465 | -3.037721 | 2.240138 |
| H | -5.990029 | 0.029059 | 0.777911 |
| H | -1.668825 | 3.228150 | -2.256859 |
| H | -3.139784 | 3.108636 | -1.261877 |
| H | -1.569070 | 2.617232 | -0.583932 |
| H | -2.687263 | 1.740227 | -4.091267 |
| H | -3.357136 | 0.144547 | -3.673811 |
| H | -4.202461 | 1.626823 | -3.178377 |
| H | -0.927593 | -2.532703 | 3.402527 |
| H | -1.024448 | -0.900441 | 2.686914 |
| H | -2.432343 | -1.588198 | 3.526009 |
| H | -2.156189 | -4.492901 | 2.257205 |
| H | -3.690733 | -3.629258 | 2.438287 |
| H | -3.175976 | -4.188527 | 0.831149 |
| H | -0.021861 | 2.507174 | 2.382444 |

12

55

| | | | |
|---|---|---|---|
| C | -3.008502 | 0.742563 | 1.199170 |
| C | -2.329412 | 0.900868 | -0.016358 |
| C | -2.933518 | 0.655832 | -1.254692 |
| C | -4.274845 | 0.260684 | -1.255755 |
| C | -4.996097 | 0.104791 | -0.068238 |
| C | -4.346587 | 0.346945 | 1.148538 |
| N | -0.957821 | 1.318646 | 0.009918 |
| C | 0.086184 | 0.447766 | 0.033495 |
| N | 1.181898 | 1.253318 | 0.042808 |
| C | 0.836866 | 2.611202 | 0.025991 |
| C | -0.530046 | 2.652974 | 0.004895 |
| AU | 0.022540 | -1.598935 | 0.053064 |
| C | 2.524585 | 0.750407 | 0.048637 |
| C | 3.137431 | 0.477564 | 1.276898 |
| C | 4.449010 | -0.005859 | 1.254460 |
| C | 5.132481 | -0.221805 | 0.053644 |
| C | 4.475733 | 0.053916 | -1.151554 |
| C | 3.166347 | 0.538797 | -1.178684 |
| C | 1.861283 | 3.686513 | 0.038648 |
| C | -1.487027 | 3.788239 | -0.024298 |
| C | 2.385759 | 0.658073 | 2.565844 |
| C | 6.533500 | -0.776140 | 0.052794 |
| C | 2.444904 | 0.793476 | -2.472966 |
| C | -2.145390 | 0.770038 | -2.529556 |
| C | -6.433522 | -0.345365 | -0.090740 |
| C | -2.295795 | 0.950479 | 2.506152 |
| H | 4.991810 | -0.121367 | -2.096803 |
| H | 4.945117 | -0.225850 | 2.200912 |
| H | -4.892575 | 0.216293 | 2.084195 |
| H | -4.764855 | 0.063452 | -2.210334 |
| H | 3.073958 | 0.526672 | -3.328722 |
| H | 1.520952 | 0.200165 | -2.519266 |
| H | 2.152073 | 1.847984 | -2.575411 |
| H | 3.040982 | 0.491394 | 3.427384 |
| H | 1.949667 | 1.662885 | 2.647292 |
| H | 1.550903 | -0.055986 | 2.619342 |
| H | 7.028904 | -0.611594 | 1.017078 |
| H | 6.520617 | -1.860227 | -0.133657 |
| H | 7.144127 | -0.317411 | -0.735362 |
| H | -2.983227 | 0.827970 | 3.349734 |
| H | -1.478821 | 0.222320 | 2.613177 |
| H | -1.841196 | 1.948561 | 2.570260 |
| H | -2.794964 | 0.649206 | -3.402935 |
| H | -1.630598 | 1.737091 | -2.607180 |
| H | -1.368072 | -0.007658 | -2.563401 |
| H | -6.862666 | -0.261952 | -1.096105 |
| H | -6.516385 | -1.396373 | 0.222865 |
| H | -7.047833 | 0.247945 | 0.599284 |
| H | 1.379537 | 4.669694 | 0.047675 |
| H | 2.508244 | 3.612122 | 0.923781 |
| H | 2.513644 | 3.631921 | -0.844161 |
| H | -0.946939 | 4.740667 | -0.017306 |
| H | -2.117874 | 3.757299 | -0.923713 |
| H | -2.160700 | 3.768487 | 0.843644 |
| H | -0.029058 | -3.220750 | 0.070772 |

13

29

| | | | |
|---|---|---|---|
| N | -0.247618 | -1.803688 | -0.159669 |
| C | 0.370820 | -0.595479 | -0.040293 |
| N | 1.698967 | -0.883104 | -0.138427 |
| C | 1.905556 | -2.241590 | -0.317233 |
| C | 0.674328 | -2.823941 | -0.330564 |
| AU | -0.506024 | 1.251069 | 0.227120 |
| C | 2.759038 | 0.141235 | -0.069834 |
| C | 3.499410 | 0.229391 | -1.402029 |
| C | -1.712402 | -1.973381 | -0.111785 |
| C | -2.108816 | -2.864759 | 1.062565 |
| C | 3.681890 | -0.123419 | 1.117023 |
| C | -2.234258 | -2.477914 | -1.454855 |
| H | 2.889643 | -2.678496 | -0.418225 |
| H | 0.389097 | -3.860862 | -0.444745 |
| H | 2.204449 | 1.075408 | 0.100413 |
| H | 4.229377 | 1.047005 | -1.366381 |
| H | 4.042404 | -0.700647 | -1.618661 |
| H | 2.798121 | 0.424542 | -2.222099 |
| H | 4.413575 | 0.688800 | 1.203436 |
| H | 3.108618 | -0.175864 | 2.050322 |
| H | 4.234982 | -1.064229 | 0.991553 |
| H | -2.083855 | -0.953381 | 0.062905 |
| H | -3.329329 | -2.530548 | -1.430105 |
| H | -1.936252 | -1.801441 | -2.264824 |
| H | -1.851860 | -3.483059 | -1.678893 |
| H | -3.201796 | -2.908237 | 1.139871 |
| H | -1.739214 | -3.890425 | 0.928488 |
| H | -1.709163 | -2.468358 | 2.003485 |
| H | -1.195449 | 2.703389 | 0.437667 |

CO2

3

| | | | |
|---|---|---|---|
| C | 0.000000 | 0.000000 | 0.000000 |
| O | 0.000000 | 0.000000 | 1.169408 |
| O | 0.000000 | 0.000000 | -1.169408 |

1-P RC

69

| | | | |
|---|---|---|---|
| C | -1.725782 | -1.639878 | -1.066892 |
| C | -0.968628 | -1.691329 | 0.119692 |
| C | -1.529770 | -1.351874 | 1.373378 |
| C | -2.890425 | -1.040191 | 1.419415 |
| C | -3.669506 | -1.047150 | 0.266887 |
| C | -3.085114 | -1.320496 | -0.963018 |
| N | 0.405469 | -2.124405 | 0.108117 |
| C | 1.472419 | -1.310080 | -0.008361 |
| C | 2.748351 | -2.076067 | 0.368919 |
| C | 2.184333 | -3.431716 | 0.867771 |
| C | 0.776776 | -3.580646 | 0.282004 |
| P | 1.345538 | 0.277554 | -0.731265 |
| AU | -0.524474 | 1.617061 | -0.327063 |
| C | 3.689658 | -2.248021 | -0.835921 |
| C | 3.513159 | -1.397325 | 1.519857 |
| C | 0.785714 | -4.291288 | -1.076445 |
| C | -0.180944 | -4.307109 | 1.219633 |
| C | -0.700616 | -1.246951 | 2.642559 |
| C | -0.454845 | 0.230093 | 2.996014 |
| C | -1.128744 | -1.852412 | -2.446387 |
| C | -1.805318 | -3.014580 | -3.192010 |
| C | 2.868938 | 1.186636 | -0.271979 |
| C | 2.935010 | 1.893997 | 0.939582 |
| C | 4.066247 | 2.640963 | 1.261418 |
| C | 5.140491 | 2.707003 | 0.369594 |
| C | 5.068761 | 2.039272 | -0.853936 |
| C | 3.935082 | 1.290485 | -1.176049 |
| C | -1.337725 | -1.979575 | 3.832728 |
| C | -1.229717 | -0.567504 | -3.286819 |
| H | -3.690331 | -1.273129 | -1.867180 |
| H | -3.344838 | -0.774496 | 2.372274 |
| H | 0.277704 | -1.703165 | 2.446637 |
| H | 3.873820 | 0.784234 | -2.139412 |
| H | 5.893677 | 2.105400 | -1.562922 |
| H | 6.024894 | 3.291267 | 0.622236 |
| H | 4.109615 | 3.176136 | 2.209729 |
| H | 2.094508 | 1.849058 | 1.630832 |
| H | -1.703493 | 2.703157 | -0.081797 |
| H | 2.834746 | -4.268228 | 0.588247 |
| H | 4.538741 | -2.881355 | -0.545759 |
| H | 3.185204 | -2.715721 | -1.688631 |
| H | 4.081130 | -1.278917 | -1.160647 |
| H | 2.833169 | -1.117149 | 2.334278 |
| H | 4.250561 | -2.108368 | 1.917306 |
| H | 4.043182 | -0.500841 | 1.188967 |
| H | 1.189464 | -5.303771 | -0.949341 |
| H | -0.228490 | -4.379982 | -1.476929 |
| H | 1.403822 | -3.763949 | -1.810478 |
| H | -0.174588 | -3.871846 | 2.222886 |
| H | -1.207215 | -4.290395 | 0.833237 |
| H | 0.135622 | -5.354190 | 1.300769 |
| H | 2.113279 | -3.411791 | 1.963616 |
| H | -4.732415 | -0.815003 | 0.324824 |
| H | -0.653271 | -1.968232 | 4.691122 |
| H | -2.268380 | -1.489916 | 4.147934 |
| H | -1.575269 | -3.023698 | 3.593797 |
| H | 0.218675 | 0.311962 | 3.859874 |
| H | -0.016234 | 0.766970 | 2.143859 |
| H | -1.400564 | 0.730515 | 3.243017 |
| H | -0.711447 | -0.700763 | -4.245835 |
| H | -2.278738 | -0.319521 | -3.496607 |
| H | -0.781984 | 0.284208 | -2.759924 |
| H | -0.063479 | -2.084150 | -2.321121 |
| H | -1.286528 | -3.214815 | -4.138754 |
| H | -1.814326 | -3.938890 | -2.601383 |
| H | -2.848562 | -2.766355 | -3.428976 |
| C | -4.172801 | 2.145191 | 0.540043 |
| O | -4.460918 | 2.130670 | -0.593331 |
| O | -3.934036 | 2.157625 | 1.685301 |

8-P RC

38

| | | | |
|---|---|---|---|
| C | -1.024582 | -3.266975 | -0.099952 |
| C | -1.590827 | -2.028224 | 0.243858 |
| C | -2.957729 | -1.823315 | -0.013646 |
| C | -3.725863 | -2.814024 | -0.623907 |
| C | -3.150715 | -4.042150 | -0.963182 |
| C | -1.798980 | -4.265616 | -0.690283 |
| P | -0.587134 | -0.772849 | 1.136407 |
| AU | 1.649597 | -0.912592 | 0.345687 |
| C | -1.102542 | 0.709995 | 0.221933 |
| N | -1.353265 | 1.921376 | 0.793808 |
| C | -1.639541 | 2.875280 | -0.179366 |
| C | -1.590496 | 2.228667 | -1.387911 |
| N | -1.248439 | 0.904569 | -1.121043 |
| C | -1.260458 | 2.193184 | 2.220743 |
| C | -0.952052 | -0.103476 | -2.128814 |
| C | -1.813501 | 2.739040 | -2.764333 |
| C | -1.916584 | 4.295992 | 0.153754 |
| H | 0.037709 | -3.434370 | 0.082969 |
| H | -1.339422 | -5.218958 | -0.951765 |
| H | -3.752535 | -4.818673 | -1.434325 |
| H | -4.781355 | -2.630603 | -0.826838 |
| H | -3.423244 | -0.876310 | 0.264374 |
| H | 3.198120 | -0.993863 | -0.159472 |
| H | -2.127815 | 3.786923 | -2.728286 |
| H | -0.896877 | 2.684078 | -3.368508 |
| H | -2.594537 | 2.168382 | -3.284305 |
| H | -2.080180 | 4.871619 | -0.762708 |
| H | -2.811681 | 4.401280 | 0.782622 |
| H | -1.074201 | 4.750975 | 0.693190 |
| H | -2.088081 | 2.839844 | 2.526832 |
| H | -1.314063 | 1.233315 | 2.746375 |
| H | -0.304677 | 2.676752 | 2.453382 |
| H | -0.702570 | 0.396387 | -3.066910 |
| H | -0.089458 | -0.686130 | -1.777740 |
| H | -1.805339 | -0.775284 | -2.272180 |
| C | 1.763528 | 2.309376 | -0.604090 |
| O | 1.762609 | 2.628042 | 0.522011 |
| O | 1.755996 | 2.036191 | -1.741283 |

1 RC

57

| | | | |
|---|---|---|---|
| C | 2.486556 | -1.744858 | 0.044566 |
| N | 1.421129 | -0.656887 | -0.111044 |
| C | 0.198801 | -1.074258 | -0.326704 |
| C | 0.191689 | -2.587465 | -0.375928 |
| C | 1.590890 | -2.993356 | 0.153476 |
| C | 1.770213 | 0.746505 | -0.059727 |
| C | 2.093578 | 1.419351 | -1.253727 |
| C | 2.499387 | 2.755463 | -1.154659 |
| C | 2.555625 | 3.403186 | 0.074829 |
| C | 2.158163 | 2.736292 | 1.229665 |
| C | 1.746643 | 1.400209 | 1.187761 |
| C | 1.891178 | 0.806214 | -2.628359 |
| C | 0.638261 | 1.416450 | -3.280475 |
| C | 1.163845 | 0.758202 | 2.434744 |
| C | 2.053936 | 0.924550 | 3.672285 |
| AU | -1.439729 | 0.127804 | -0.592173 |
| C | -0.032508 | -2.994537 | -1.847704 |
| C | -0.940889 | -3.151390 | 0.490759 |
| C | 3.400363 | -1.764141 | -1.182715 |
| C | 3.333446 | -1.509323 | 1.292670 |
| C | -0.244791 | 1.326239 | 2.677860 |
| C | 3.115116 | 0.957022 | -3.540177 |
| C | -3.576059 | 0.210336 | 2.204661 |
| O | -3.824215 | 1.352523 | 2.220127 |
| O | -3.339568 | -0.935556 | 2.219899 |
| H | -2.766442 | 1.049072 | -0.793216 |
| H | 2.754030 | 3.301865 | -2.061854 |
| H | 2.146612 | 3.267921 | 2.180139 |
| H | 1.043315 | -0.314576 | 2.244107 |
| H | 2.015756 | -3.837645 | -0.401392 |
| H | -0.051900 | -4.089662 | -1.925379 |
| H | 0.762506 | -2.616145 | -2.502034 |
| H | -0.988474 | -2.597322 | -2.210566 |
| H | -0.843749 | -2.819255 | 1.532058 |
| H | -0.913769 | -4.249392 | 0.470738 |
| H | -1.915962 | -2.813412 | 0.119242 |
| H | 4.160911 | -2.541807 | -1.042504 |
| H | 3.916486 | -0.804436 | -1.301868 |
| H | 2.851739 | -1.987460 | -2.103459 |
| H | 2.728405 | -1.534786 | 2.204368 |
| H | 3.860950 | -0.548841 | 1.241598 |
| H | 4.084507 | -2.305917 | 1.361916 |
| H | 1.509621 | -3.293512 | 1.206194 |
| H | 2.878009 | 4.442708 | 0.129355 |
| H | 1.619494 | 0.384141 | 4.523621 |
| H | 2.143094 | 1.979249 | 3.963664 |
| H | 3.065634 | 0.536992 | 3.497919 |
| H | -0.718238 | 0.825161 | 3.533124 |
| H | -0.873100 | 1.176794 | 1.786477 |
| H | -0.203597 | 2.403615 | 2.886492 |
| H | 0.445068 | 0.943849 | -4.252842 |
| H | 0.764805 | 2.496173 | -3.438903 |
| H | -0.242383 | 1.263742 | -2.637675 |
| H | 1.692938 | -0.263146 | -2.500059 |
| H | 2.940271 | 0.439663 | -4.492644 |
| H | 4.019495 | 0.536749 | -3.082405 |
| H | 3.314817 | 2.011883 | -3.770166 |

8 RC

26

| | | | |
|---|---|---|---|
| C | 2.643033 | -0.197260 | 0.507501 |
| N | 1.534068 | -0.880952 | 0.009830 |
| C | 0.698575 | -0.055571 | -0.679181 |
| N | 1.304544 | 1.161744 | -0.609794 |
| C | 2.496849 | 1.106366 | 0.111507 |
| C | 1.278853 | -2.298493 | 0.216590 |
| AU | -1.106349 | -0.528701 | -1.561182 |
| C | 0.754109 | 2.372540 | -1.201221 |
| C | 3.355979 | 2.298234 | 0.336161 |
| C | 3.707322 | -0.864936 | 1.301244 |
| C | -1.133877 | 0.446354 | 1.846942 |
| O | -0.981833 | -0.662004 | 2.188570 |
| O | -1.282144 | 1.563993 | 1.535711 |
| H | -2.530162 | -0.898874 | -2.240636 |
| H | 4.237978 | 2.022838 | 0.923821 |
| H | 2.819990 | 3.087396 | 0.882246 |
| H | 3.704775 | 2.730775 | -0.612329 |
| H | 4.471374 | -0.138033 | 1.596361 |
| H | 4.203214 | -1.660247 | 0.726667 |
| H | 3.304338 | -1.320550 | 2.216860 |
| H | 2.093915 | -2.898829 | -0.203711 |
| H | 0.341871 | -2.542180 | -0.293229 |
| H | 1.177195 | -2.515497 | 1.286040 |
| H | 0.550435 | 3.117733 | -0.423658 |
| H | -0.182501 | 2.099758 | -1.696828 |
| H | 1.451876 | 2.790594 | -1.935869 |

1-P TS

69

| | | | |
|---|---|---|---|
| C | -2.197148 | 1.058956 | 1.094098 |
| C | -1.492387 | 1.280656 | -0.105876 |
| C | -1.930734 | 0.741223 | -1.340328 |
| C | -3.112616 | -0.004523 | -1.343092 |
| C | -3.836882 | -0.212746 | -0.174863 |
| C | -3.379285 | 0.312385 | 1.025902 |
| N | -0.279518 | 2.065999 | -0.124699 |
| C | 0.954086 | 1.583112 | 0.054396 |
| C | 1.995539 | 2.640359 | -0.329535 |
| C | 1.111702 | 3.773077 | -0.914326 |
| C | -0.303976 | 3.567977 | -0.370593 |
| P | 1.274977 | 0.067230 | 0.921044 |
| AU | -0.176859 | -1.536496 | 0.294717 |
| C | 2.819892 | 3.104790 | 0.884548 |
| C | 2.946476 | 2.129855 | -1.428706 |
| C | -0.527017 | 4.304762 | 0.954867 |
| C | -1.394449 | 3.971628 | -1.354681 |
| C | -1.146114 | 0.880150 | -2.636081 |
| C | -0.467544 | -0.446872 | -3.013956 |
| C | -1.716276 | 1.533106 | 2.454441 |
| C | -2.723720 | 2.499756 | 3.099945 |
| C | 2.965371 | -0.463867 | 0.451000 |
| C | 3.216096 | -1.165425 | -0.738140 |
| C | 4.501066 | -1.615719 | -1.030138 |
| C | 5.546288 | -1.388385 | -0.129596 |
| C | 5.298359 | -0.721232 | 1.070377 |
| C | 4.010924 | -0.266491 | 1.362789 |
| C | -2.028449 | 1.372140 | -3.794042 |
| C | -1.458757 | 0.346272 | 3.399045 |
| H | -3.931653 | 0.112050 | 1.942610 |
| H | -3.454727 | -0.457275 | -2.271728 |
| H | -0.349111 | 1.617193 | -2.477851 |
| H | 3.812931 | 0.236903 | 2.309271 |
| H | 6.104336 | -0.561953 | 1.785955 |
| H | 6.549502 | -1.746856 | -0.357429 |
| H | 4.687851 | -2.151352 | -1.960445 |
| H | 2.397333 | -1.354858 | -1.430924 |
| H | -0.747844 | -3.004587 | -0.783642 |
| H | 1.506319 | 4.763750 | -0.663381 |
| H | 3.486756 | 3.920179 | 0.575423 |
| H | 2.185708 | 3.467636 | 1.700480 |
| H | 3.437183 | 2.286640 | 1.268938 |
| H | 2.390073 | 1.634658 | -2.234414 |
| H | 3.474763 | 2.992280 | -1.856868 |
| H | 3.689233 | 1.429933 | -1.038845 |
| H | -0.406004 | 5.381626 | 0.783828 |
| H | -1.539040 | 4.134642 | 1.332899 |
| H | 0.188762 | 3.996591 | 1.723873 |
| H | -1.230433 | 3.544085 | -2.347216 |
| H | -2.385474 | 3.668461 | -0.995994 |
| H | -1.388013 | 5.063952 | -1.450817 |
| H | 1.091425 | 3.685016 | -2.008656 |
| H | -4.742494 | -0.816220 | -0.196638 |
| H | -1.410335 | 1.574954 | -4.678394 |
| H | -2.768765 | 0.612528 | -4.075809 |
| H | -2.573515 | 2.287575 | -3.533195 |
| H | 0.097644 | -0.336332 | -3.949113 |
| H | 0.219143 | -0.772139 | -2.221446 |
| H | -1.212098 | -1.242750 | -3.149048 |
| H | -1.049322 | 0.706451 | 4.352127 |
| H | -2.389389 | -0.195585 | 3.611356 |
| H | -0.748502 | -0.363801 | 2.961162 |
| H | -0.763010 | 2.058453 | 2.316391 |
| H | -2.311153 | 2.920273 | 4.026390 |
| H | -2.991395 | 3.328301 | 2.432877 |
| H | -3.652621 | 1.972804 | 3.355299 |
| C | -1.937937 | -3.219404 | -0.541902 |
| O | -2.256985 | -3.019538 | 0.655134 |
| O | -2.537059 | -3.592039 | -1.546358 |

2-P TS

76

| | | | |
|---|---|---|---|
| C | -4.817805 | -0.369015 | -1.675710 |
| C | -3.386079 | -0.032290 | -1.226006 |
| C | -2.930555 | -0.950492 | -0.065995 |
| C | -3.849242 | -0.712326 | 1.177175 |
| C | -5.181338 | -0.042768 | 0.813636 |
| C | -5.759477 | -0.631889 | -0.475524 |
| C | -2.934642 | -2.425769 | -0.552728 |
| C | -1.664296 | -3.069744 | 0.004681 |
| N | -0.761315 | -1.846547 | 0.075725 |
| C | -1.452278 | -0.713650 | 0.234644 |
| C | 0.672058 | -1.949530 | -0.070897 |
| C | 1.509201 | -2.138952 | 1.045072 |
| C | 2.884335 | -2.262717 | 0.813635 |
| C | 3.412425 | -2.185560 | -0.467821 |
| C | 2.569575 | -1.979024 | -1.554431 |
| C | 1.187861 | -1.858682 | -1.386881 |
| C | 1.002058 | -2.161089 | 2.475837 |
| C | 1.628362 | -1.029946 | 3.308043 |
| C | 0.316011 | -1.572375 | -2.599672 |
| C | 0.609013 | -2.529530 | -3.764483 |
| P | -0.747336 | 0.737625 | 0.985289 |
| C | -1.813569 | 2.158997 | 0.527510 |
| C | -1.585760 | 2.892998 | -0.647352 |
| C | -2.370521 | 4.003393 | -0.947064 |
| C | -3.380220 | 4.406785 | -0.068180 |
| C | -3.588033 | 3.705689 | 1.120008 |
| C | -2.802101 | 2.591820 | 1.421608 |
| C | -1.880351 | -3.621817 | 1.419143 |
| C | -1.088454 | -4.159998 | -0.887265 |
| AU | 1.320749 | 1.086302 | 0.159391 |
| C | 0.470952 | -0.112273 | -3.056025 |
| C | 1.277245 | -3.519043 | 3.143149 |
| H | 3.554969 | -2.388687 | 1.661934 |
| H | 2.996738 | -1.880712 | -2.550436 |
| H | -0.732471 | -1.709964 | -2.307477 |
| H | -2.955641 | 2.058445 | 2.359262 |
| H | -4.357130 | 4.030077 | 1.820382 |
| H | -3.991439 | 5.277681 | -0.302325 |
| H | -2.191089 | 4.558943 | -1.867123 |
| H | -0.790385 | 2.582248 | -1.323800 |
| H | -3.834016 | -2.963002 | -0.235527 |
| H | -2.651008 | -4.401933 | 1.381126 |
| H | -0.964442 | -4.071579 | 1.812539 |
| H | -2.212133 | -2.841224 | 2.112683 |
| H | -0.976053 | -3.827002 | -1.922761 |
| H | -0.114031 | -4.501846 | -0.518211 |
| H | -1.774813 | -5.015282 | -0.877789 |
| H | -2.895254 | -2.451613 | -1.649906 |
| H | 4.488211 | -2.253050 | -0.619777 |
| H | -0.118532 | -2.373308 | -4.571547 |
| H | 1.608202 | -2.352532 | -4.182415 |
| H | 0.560307 | -3.580398 | -3.452171 |
| H | -0.170975 | 0.088751 | -3.924185 |
| H | 0.199815 | 0.579076 | -2.247893 |
| H | 1.511564 | 0.099679 | -3.335300 |
| H | 1.200025 | -1.024674 | 4.319097 |
| H | 2.714399 | -1.161462 | 3.398416 |
| H | 1.446677 | -0.052465 | 2.846471 |
| H | -0.082410 | -1.993324 | 2.454673 |
| H | 0.809069 | -3.561591 | 4.135301 |
| H | 0.899694 | -4.357923 | 2.544902 |
| H | 2.357014 | -3.671499 | 3.273875 |
| H | -3.314895 | -0.109399 | 1.920629 |
| H | -4.048114 | -1.682212 | 1.652124 |
| H | -5.885064 | -0.165184 | 1.647868 |
| H | -5.031083 | 1.039895 | 0.688507 |
| H | -6.755271 | -0.215803 | -0.675599 |
| H | -5.900826 | -1.715003 | -0.337677 |
| H | -4.814350 | -1.237079 | -2.348287 |
| H | -5.187989 | 0.477781 | -2.269456 |
| H | -3.342989 | 1.008630 | -0.897155 |
| H | -2.683247 | -0.124880 | -2.065511 |
| C | 3.683968 | 1.269456 | -0.877871 |
| O | 3.898464 | 1.047059 | 0.337760 |
| O | 4.329016 | 1.124129 | -1.911767 |
| H | 2.572873 | 1.753904 | -1.110950 |

3-P TS

88

| | | | |
|---|---|---|---|
| C | 4.800564 | -0.217277 | 1.269451 |
| C | 3.329293 | 0.143784 | 0.993535 |
| C | 2.714492 | -0.905979 | 0.016961 |
| C | 3.459200 | -0.792890 | -1.360388 |
| C | 4.859802 | -0.170046 | -1.270798 |
| C | 5.561542 | -0.657482 | -0.001383 |
| C | 2.802253 | -2.348122 | 0.580928 |
| C | 1.481778 | -3.049102 | 0.239214 |
| N | 0.558990 | -1.852996 | 0.043291 |
| C | 1.216560 | -0.714102 | -0.193295 |
| C | -0.881942 | -2.005401 | 0.025683 |
| C | -1.571910 | -2.238713 | -1.184648 |
| C | -2.959139 | -2.415277 | -1.118390 |
| C | -3.647255 | -2.341279 | 0.084806 |
| C | -2.955267 | -2.076689 | 1.259921 |
| C | -1.569189 | -1.900278 | 1.260938 |
| C | -0.920469 | -2.244476 | -2.558733 |
| C | -1.485291 | -1.118768 | -3.443503 |
| C | -0.881019 | -1.548596 | 2.568350 |
| C | -1.207238 | -2.554554 | 3.682617 |
| P | 0.447251 | 0.671795 | -1.007364 |
| C | 1.474053 | 2.164450 | -0.723558 |
| C | 1.147873 | 3.073747 | 0.293250 |
| C | 1.838873 | 4.278384 | 0.412544 |
| C | 2.863741 | 4.591013 | -0.482449 |
| C | 3.176802 | 3.705047 | -1.515562 |
| C | 2.475461 | 2.507110 | -1.645833 |
| C | 1.576437 | -3.844566 | -1.065746 |
| C | 0.994383 | -3.975048 | 1.346627 |
| C | 2.521575 | 0.484623 | 2.273320 |
| C | 3.029027 | 1.810330 | 2.860458 |
| C | 5.658935 | -0.474281 | -2.537044 |
| AU | -1.622363 | 1.029897 | -0.196136 |
| C | -1.232837 | -0.124455 | 3.026801 |
| C | -1.111092 | -3.592194 | -3.275927 |
| C | 2.504893 | -0.559029 | 3.398620 |
| H | -3.512847 | -2.583090 | -2.040800 |
| H | -3.504524 | -1.971205 | 2.193594 |
| H | 0.198486 | -1.568209 | 2.391956 |
| H | 2.698358 | 1.840656 | -2.477547 |
| H | 3.956705 | 3.955001 | -2.234143 |
| H | 3.403903 | 5.532276 | -0.387414 |
| H | 1.576043 | 4.973515 | 1.209176 |
| H | 0.341082 | 2.831790 | 0.984448 |
| H | 3.651131 | -2.890869 | 0.153707 |
| H | 2.319232 | -4.641380 | -0.937412 |
| H | 0.617804 | -4.312890 | -1.306822 |
| H | 1.885079 | -3.217753 | -1.908901 |
| H | 1.001641 | -3.486530 | 2.324555 |
| H | -0.017501 | -4.344157 | 1.140320 |
| H | 1.666811 | -4.839820 | 1.395783 |
| H | 2.939542 | -2.330532 | 1.663893 |
| H | -4.729643 | -2.454772 | 0.103144 |
| H | -0.593752 | -2.345173 | 4.568971 |
| H | -2.259883 | -2.479015 | 3.984300 |
| H | -1.022781 | -3.588885 | 3.369103 |
| H | -0.726550 | 0.105022 | 3.973767 |
| H | -0.921528 | 0.615049 | 2.275900 |
| H | -2.314953 | -0.010068 | 3.173848 |
| H | -0.951701 | -1.094160 | -4.402928 |
| H | -2.550819 | -1.281759 | -3.650768 |
| H | -1.380480 | -0.140568 | -2.963163 |
| H | 0.153410 | -2.059321 | -2.433206 |
| H | -0.543120 | -3.605630 | -4.215503 |
| H | -0.787174 | -4.443187 | -2.666140 |
| H | -2.169195 | -3.749063 | -3.523614 |
| H | 2.846806 | -0.224587 | -2.070238 |
| H | 3.550393 | -1.803880 | -1.783139 |
| H | 4.748135 | 0.924003 | -1.196388 |
| H | 6.594341 | -0.283758 | 0.026111 |
| H | 5.635950 | -1.755882 | -0.044754 |
| H | 4.865773 | -1.004349 | 2.032488 |
| H | 5.286255 | 0.667660 | 1.700779 |
| H | 3.334961 | 1.083169 | 0.427169 |
| H | 1.480429 | 0.645514 | 1.948181 |
| H | 2.339243 | 2.177328 | 3.631809 |
| H | 4.012478 | 1.680919 | 3.333032 |
| H | 3.118254 | 2.584584 | 2.089830 |
| H | 1.951772 | -0.158149 | 4.258311 |
| H | 2.018952 | -1.499240 | 3.121243 |
| H | 3.520242 | -0.792826 | 3.745629 |
| H | 6.636255 | 0.026114 | -2.516676 |
| H | 5.834356 | -1.555823 | -2.636393 |
| H | 5.125067 | -0.137942 | -3.436692 |
| C | -4.012244 | 1.212674 | 0.747354 |
| O | -4.176548 | 0.912984 | -0.461137 |
| O | -4.680044 | 1.071696 | 1.768307 |
| H | -2.947682 | 1.781025 | 0.977447 |

4-P TS

66

| | | | |
|---|---|---|---|
| C | -2.666532 | 1.064727 | 1.079074 |
| C | -1.750037 | 1.897497 | 0.408746 |
| C | -2.134037 | 2.716802 | -0.664584 |
| C | -3.473216 | 2.682808 | -1.063603 |
| C | -4.413929 | 1.867513 | -0.426533 |
| C | -3.991449 | 1.069536 | 0.643010 |
| N | -0.399106 | 1.940229 | 0.876765 |
| C | 0.674316 | 1.292714 | 0.386512 |
| N | 1.754938 | 1.559565 | 1.148164 |
| C | 1.413113 | 2.410078 | 2.310733 |
| C | -0.047995 | 2.792346 | 2.033325 |
| P | 0.708064 | 0.406479 | -1.213218 |
| AU | -1.349370 | -0.522836 | -1.346774 |
| C | 3.109708 | 1.217369 | 0.851793 |
| C | 3.736110 | 0.208642 | 1.603509 |
| C | 5.047571 | -0.131799 | 1.274140 |
| C | 5.738824 | 0.508458 | 0.240508 |
| C | 5.098037 | 1.539218 | -0.451488 |
| C | 3.785210 | 1.919872 | -0.159209 |
| C | 3.020272 | -0.493304 | 2.723222 |
| C | 7.128286 | 0.068792 | -0.134027 |
| C | 3.120235 | 3.031308 | -0.923400 |
| C | -1.125781 | 3.557413 | -1.397109 |
| C | -5.836290 | 1.793070 | -0.906981 |
| C | -2.225092 | 0.157693 | 2.193506 |
| C | 1.874796 | -0.969610 | -0.836989 |
| C | 3.060562 | -1.045961 | -1.580152 |
| C | 3.955647 | -2.094474 | -1.373786 |
| C | 3.673051 | -3.078940 | -0.426191 |
| C | 2.481504 | -3.022775 | 0.301417 |
| C | 1.580950 | -1.981353 | 0.089075 |
| H | 5.630750 | 2.062156 | -1.247119 |
| H | 5.534717 | -0.936294 | 1.826795 |
| H | -4.703635 | 0.398393 | 1.123187 |
| H | -3.785431 | 3.300026 | -1.907035 |
| H | 3.854935 | 3.582409 | -1.520065 |
| H | 2.352526 | 2.633662 | -1.604160 |
| H | 2.618233 | 3.741257 | -0.252550 |
| H | 3.464242 | -1.477283 | 2.906879 |
| H | 3.087755 | 0.082705 | 3.658422 |
| H | 1.958281 | -0.635993 | 2.493252 |
| H | 7.756555 | -0.076051 | 0.754459 |
| H | 7.094570 | -0.893524 | -0.666326 |
| H | 7.617112 | 0.797583 | -0.790941 |
| H | -2.938699 | -0.662999 | 2.317012 |
| H | -1.241098 | -0.278611 | 1.981119 |
| H | -2.154643 | 0.694026 | 3.151812 |
| H | -1.624201 | 4.261921 | -2.071326 |
| H | -0.487461 | 4.128418 | -0.709958 |
| H | -0.460440 | 2.922720 | -2.002392 |
| H | -6.052348 | 2.571485 | -1.647840 |
| H | -6.012902 | 0.810983 | -1.368965 |
| H | -6.543295 | 1.896743 | -0.073121 |
| H | 0.636174 | -1.954481 | 0.631544 |
| H | 2.248226 | -3.799008 | 1.029828 |
| H | 4.374035 | -3.896766 | -0.262259 |
| H | 4.877417 | -2.139986 | -1.952880 |
| H | 3.287578 | -0.270407 | -2.310582 |
| H | -0.714915 | 2.569872 | 2.872464 |
| H | -0.172389 | 3.849060 | 1.760869 |
| H | 1.532371 | 1.839785 | 3.240452 |
| H | 2.085278 | 3.274773 | 2.341464 |
| C | -3.793028 | -1.514215 | -1.412148 |
| O | -3.214809 | -2.109642 | -0.469507 |
| O | -4.968795 | -1.409897 | -1.758406 |
| H | -3.051541 | -0.907422 | -2.168022 |

5-P TS

84

| | | | |
|---|---|---|---|
| C | 1.193586 | -2.712064 | -0.454119 |
| C | 1.589341 | -1.530029 | -1.096740 |
| C | 2.783275 | -1.525597 | -1.833791 |
| C | 3.594861 | -2.656485 | -1.873057 |
| C | 3.206123 | -3.820670 | -1.207393 |
| C | 1.998353 | -3.849321 | -0.509162 |
| P | 0.455043 | -0.087755 | -1.212877 |
| AU | -1.638881 | -0.771827 | -0.734916 |
| C | 0.902291 | 1.067182 | 0.126011 |
| N | 0.043501 | 2.025931 | 0.540327 |
| C | 0.707487 | 3.032640 | 1.392506 |
| C | 2.186621 | 2.657737 | 1.270726 |
| N | 2.138677 | 1.343633 | 0.593215 |
| C | -1.387561 | 2.029461 | 0.448829 |
| C | -2.125764 | 1.482185 | 1.521430 |
| C | -3.519193 | 1.541640 | 1.445297 |
| C | -4.152283 | 2.119676 | 0.348037 |
| C | -3.401319 | 2.658232 | -0.692102 |
| C | -2.002535 | 2.627228 | -0.663758 |
| C | -1.454301 | 0.811956 | 2.706977 |
| C | -1.770007 | 1.550350 | 4.016912 |
| C | -1.195490 | 3.195925 | -1.814656 |
| C | -1.329229 | 4.726043 | -1.876660 |
| C | 3.283007 | 0.486741 | 0.574913 |
| C | 3.286235 | -0.623666 | 1.455177 |
| C | 4.421468 | -1.434245 | 1.472041 |
| C | 5.520863 | -1.148117 | 0.665162 |
| C | 5.488202 | -0.058168 | -0.191408 |
| C | 4.366658 | 0.780870 | -0.270573 |
| C | 2.100735 | -0.930018 | 2.361330 |
| C | 2.060449 | 0.012438 | 3.577918 |
| C | 4.452110 | 1.913964 | -1.291055 |
| C | 5.242777 | 3.108075 | -0.721804 |
| C | 2.065584 | -2.379358 | 2.857174 |
| C | 3.154585 | 2.376799 | -1.967062 |
| C | -1.588263 | 2.544019 | -3.148300 |
| C | -1.829446 | -0.673535 | 2.805604 |
| H | 6.336841 | 0.142172 | -0.846403 |
| H | 4.447034 | -2.307433 | 2.119330 |
| H | -4.119985 | 1.096166 | 2.236212 |
| H | -3.909383 | 3.096650 | -1.550314 |
| H | 5.074252 | 1.493884 | -2.096717 |
| H | 1.184049 | -0.767830 | 1.773636 |
| H | -0.369484 | 0.854870 | 2.550106 |
| H | -0.139127 | 2.955490 | -1.632441 |
| H | 0.243895 | -2.737168 | 0.079950 |
| H | 1.677058 | -4.761261 | -0.006646 |
| H | 3.836391 | -4.708478 | -1.245914 |
| H | 4.528384 | -2.631332 | -2.433795 |
| H | 3.080172 | -0.631203 | -2.377946 |
| H | 0.324708 | 2.954371 | 2.419265 |
| H | 0.491608 | 4.038738 | 1.017625 |
| H | 2.700192 | 2.568789 | 2.232704 |
| H | 2.739958 | 3.372774 | 0.650043 |
| H | -5.239850 | 2.131326 | 0.297916 |
| H | 6.396799 | -1.795410 | 0.691139 |
| H | -0.939914 | 2.908765 | -3.955676 |
| H | -1.489238 | 1.453135 | -3.089083 |
| H | -2.626785 | 2.779188 | -3.416531 |
| H | -0.705726 | 5.134497 | -2.683060 |
| H | -2.369992 | 5.019317 | -2.070759 |
| H | -1.021873 | 5.193021 | -0.931513 |
| H | -1.310415 | -1.142483 | 3.652356 |
| H | -2.908986 | -0.807924 | 2.948988 |
| H | -1.549673 | -1.208139 | 1.886611 |
| H | -1.221536 | 1.097281 | 4.853448 |
| H | -1.494964 | 2.612001 | 3.958012 |
| H | -2.842097 | 1.496193 | 4.247880 |
| H | 5.392927 | 3.869242 | -1.498976 |
| H | 4.712450 | 3.580496 | 0.115262 |
| H | 6.226388 | 2.790459 | -0.354294 |
| H | 3.408955 | 2.949760 | -2.867992 |
| H | 2.516353 | 1.538296 | -2.272448 |
| H | 2.551889 | 3.037061 | -1.332624 |
| H | 1.105300 | -2.571733 | 3.351320 |
| H | 2.175616 | -3.092385 | 2.033900 |
| H | 2.858584 | -2.571785 | 3.592855 |
| H | 1.211190 | -0.243805 | 4.224440 |
| H | 2.981150 | -0.089049 | 4.168103 |
| H | 1.957244 | 1.064447 | 3.296002 |
| C | -4.013121 | -1.621474 | -0.058298 |
| O | -3.325708 | -2.669568 | -0.105244 |
| O | -5.083191 | -1.318220 | 0.462180 |
| H | -3.540194 | -0.666812 | -0.672786 |

6-P TS

73

| | | | |
|---|---|---|---|
| C | -0.188528 | 3.557063 | -0.553638 |
| N | 0.195270 | 2.295167 | -0.064650 |
| C | -0.664278 | 1.227496 | 0.111133 |
| N | -1.911718 | 1.315497 | -0.458029 |
| C | -2.400647 | 2.511344 | -1.056148 |
| C | -1.685428 | 3.788326 | -0.662209 |
| C | 1.585735 | 2.135281 | 0.323685 |
| C | 2.495790 | 1.673986 | -0.638254 |
| C | 3.809046 | 1.445748 | -0.222694 |
| C | 4.216460 | 1.685662 | 1.094483 |
| C | 3.286751 | 2.204838 | 2.000902 |
| C | 1.960146 | 2.449368 | 1.635726 |
| C | 2.100131 | 1.488848 | -2.075099 |
| C | 0.982198 | 3.041398 | 2.610209 |
| C | 5.628587 | 1.387900 | 1.519740 |
| P | -0.050842 | -0.087873 | 1.174202 |
| C | -1.470888 | -1.119978 | 1.658443 |
| C | -2.255443 | -0.668703 | 2.732115 |
| C | -3.264596 | -1.476067 | 3.253458 |
| C | -3.476764 | -2.751600 | 2.727618 |
| C | -2.670940 | -3.222033 | 1.687183 |
| C | -1.667954 | -2.415753 | 1.158479 |
| C | -2.710913 | 0.127412 | -0.694000 |
| C | -2.311419 | -0.779414 | -1.683528 |
| C | -3.109487 | -1.911971 | -1.878544 |
| C | -4.277032 | -2.129780 | -1.146465 |
| C | -4.669905 | -1.161859 | -0.215398 |
| C | -3.911049 | -0.017583 | 0.018866 |
| O | -3.366370 | 2.475955 | -1.790314 |
| C | -1.088558 | -0.570517 | -2.525518 |
| C | -5.091466 | -3.379656 | -1.336461 |
| C | -4.390493 | 1.032899 | 0.978622 |
| C | -2.186436 | 4.174775 | 0.761667 |
| C | -1.989506 | 4.907008 | -1.658920 |
| O | 0.645991 | 4.414100 | -0.775967 |
| AU | 1.409230 | -1.277028 | -0.062395 |
| H | -5.591340 | -1.303654 | 0.350406 |
| H | -2.806555 | -2.637527 | -2.634460 |
| H | 4.525292 | 1.061389 | -0.948413 |
| H | 3.599281 | 2.428680 | 3.021520 |
| H | -5.122902 | 0.613320 | 1.675729 |
| H | -3.568866 | 1.458471 | 1.565747 |
| H | -4.871647 | 1.857659 | 0.433193 |
| H | -1.137426 | -1.177160 | -3.435505 |
| H | -0.968285 | 0.479412 | -2.820001 |
| H | -0.175564 | -0.866172 | -1.977286 |
| H | -4.792686 | -3.919870 | -2.241987 |
| H | -4.957205 | -4.057357 | -0.480037 |
| H | -6.163220 | -3.152113 | -1.403021 |
| H | 2.797463 | 0.815606 | -2.581327 |
| H | 1.094642 | 1.067424 | -2.173589 |
| H | 2.102396 | 2.459028 | -2.592844 |
| H | 1.448374 | 3.175625 | 3.591666 |
| H | 0.628539 | 4.022930 | 2.264957 |
| H | 0.104651 | 2.391454 | 2.736782 |
| H | 5.861631 | 1.847568 | 2.487155 |
| H | 5.781090 | 0.302955 | 1.614814 |
| H | 6.350861 | 1.751076 | 0.777193 |
| H | -1.033510 | -2.779026 | 0.351139 |
| H | -2.824438 | -4.223341 | 1.286735 |
| H | -4.259585 | -3.387167 | 3.139691 |
| H | -3.875908 | -1.114348 | 4.079316 |
| H | -2.067554 | 0.314690 | 3.164037 |
| H | -1.655298 | 4.637364 | -2.666957 |
| H | -3.067233 | 5.092599 | -1.688991 |
| H | -1.469856 | 5.819417 | -1.352141 |
| H | -3.269463 | 4.337440 | 0.730635 |
| H | -1.965224 | 3.396792 | 1.500932 |
| H | -1.695236 | 5.104374 | 1.069790 |
| C | 3.008816 | -2.808505 | -1.465004 |
| O | 3.308379 | -1.628375 | -1.746082 |
| O | 3.382571 | -3.921289 | -1.800921 |
| H | 2.078811 | -2.918975 | -0.618607 |

7-P TS

74

| | | | |
|---|---|---|---|
| C | -0.165631 | 3.605372 | 0.765738 |
| N | -0.409141 | 2.313833 | 0.253631 |
| C | 0.570600 | 1.348446 | 0.037369 |
| N | 1.751672 | 1.517230 | 0.642477 |
| C | 2.004443 | 2.696702 | 1.486428 |
| C | 1.300224 | 3.947736 | 0.972787 |
| C | -1.766668 | 2.023712 | -0.174666 |
| C | -2.623945 | 1.399378 | 0.743684 |
| C | -3.893394 | 1.031704 | 0.295379 |
| C | -4.313800 | 1.291219 | -1.014311 |
| C | -3.451795 | 1.984230 | -1.869896 |
| C | -2.170677 | 2.374536 | -1.468383 |
| C | -2.207297 | 1.171304 | 2.168106 |
| C | -1.271876 | 3.151454 | -2.387822 |
| C | -5.663112 | 0.822042 | -1.486732 |
| P | 0.129600 | 0.096138 | -1.226067 |
| C | 1.680552 | -0.781090 | -1.648202 |
| C | 2.457987 | -0.232053 | -2.680338 |
| C | 3.574807 | -0.912535 | -3.164376 |
| C | 3.908773 | -2.162653 | -2.641723 |
| C | 3.117644 | -2.733765 | -1.641410 |
| C | 2.007557 | -2.052043 | -1.151823 |
| C | 2.713744 | 0.453001 | 0.807874 |
| C | 2.433043 | -0.581405 | 1.714019 |
| C | 3.403939 | -1.574051 | 1.884810 |
| C | 4.617275 | -1.550325 | 1.196420 |
| C | 4.873315 | -0.476494 | 0.337415 |
| C | 3.945422 | 0.543851 | 0.137864 |
| C | 1.140564 | -0.660844 | 2.469141 |
| C | 5.615107 | -2.665724 | 1.340765 |
| C | 4.265460 | 1.691856 | -0.775987 |
| C | 1.858521 | 4.396557 | -0.393718 |
| C | 1.443428 | 5.067475 | 2.006142 |
| O | -1.087978 | 4.381817 | 0.937499 |
| AU | -1.250616 | -1.336739 | -0.160719 |
| H | 5.817071 | -0.438387 | -0.208013 |
| H | 3.194115 | -2.392300 | 2.574872 |
| H | -4.562120 | 0.519537 | 0.986853 |
| H | -3.781067 | 2.227579 | -2.880858 |
| H | 5.055360 | 1.411670 | -1.480591 |
| H | 3.389293 | 2.007265 | -1.351412 |
| H | 4.620599 | 2.566084 | -0.209791 |
| H | 1.228147 | -1.345327 | 3.318865 |
| H | 0.818715 | 0.316838 | 2.848939 |
| H | 0.327995 | -1.034916 | 1.819830 |
| H | 5.397373 | -3.289185 | 2.215610 |
| H | 5.591252 | -3.314039 | 0.451917 |
| H | 6.638528 | -2.279593 | 1.431735 |
| H | -2.986660 | 0.632498 | 2.713364 |
| H | -1.293791 | 0.569212 | 2.226679 |
| H | -2.015433 | 2.129281 | 2.671464 |
| H | -1.746104 | 3.289490 | -3.365118 |
| H | -1.051252 | 4.144492 | -1.972106 |
| H | -0.317737 | 2.628215 | -2.542390 |
| H | -5.979004 | 1.352910 | -2.392474 |
| H | -5.634464 | -0.252524 | -1.719551 |
| H | -6.427621 | 0.963925 | -0.712377 |
| H | 1.383562 | -2.498145 | -0.378425 |
| H | 3.365534 | -3.716635 | -1.242236 |
| H | 4.775464 | -2.700351 | -3.024638 |
| H | 4.175050 | -0.472439 | -3.960081 |
| H | 2.177738 | 0.729224 | -3.112459 |
| H | 1.026925 | 4.768558 | 2.975950 |
| H | 2.503842 | 5.315581 | 2.140744 |
| H | 0.912995 | 5.964199 | 1.670680 |
| H | 2.922749 | 4.643036 | -0.293899 |
| H | 1.753358 | 3.618294 | -1.158425 |
| H | 1.325622 | 5.289783 | -0.739102 |
| H | 3.088784 | 2.851200 | 1.519714 |
| H | 1.673812 | 2.460775 | 2.511640 |
| C | -2.961784 | -2.972337 | 0.964569 |
| O | -2.805178 | -2.018991 | 1.763590 |
| O | -3.734512 | -3.920140 | 0.888907 |
| H | -2.154593 | -2.991680 | 0.024517 |

8-P TS

38

| | | | |
|---|---|---|---|
| C | -0.272777 | 3.145769 | 0.092992 |
| C | 0.783748 | 2.344076 | 0.548174 |
| C | 2.082023 | 2.879503 | 0.565608 |
| C | 2.320907 | 4.174803 | 0.108060 |
| C | 1.262283 | 4.965022 | -0.348405 |
| C | -0.034703 | 4.447977 | -0.347069 |
| P | 0.493214 | 0.671350 | 1.264445 |
| AU | -1.384378 | -0.151516 | 0.294103 |
| C | 1.751749 | -0.248941 | 0.301599 |
| N | 2.587521 | -1.179983 | 0.835169 |
| C | 3.409899 | -1.719263 | -0.147534 |
| C | 3.079154 | -1.091161 | -1.322916 |
| N | 2.053316 | -0.199360 | -1.024363 |
| C | 2.584724 | -1.601993 | 2.230622 |
| C | 1.334455 | 0.607391 | -2.002283 |
| C | 3.630108 | -1.258913 | -2.691913 |
| C | 4.414245 | -2.773230 | 0.144382 |
| H | -1.281196 | 2.731628 | 0.068512 |
| H | -0.866518 | 5.055837 | -0.702540 |
| H | 1.448293 | 5.978525 | -0.701816 |
| H | 3.335959 | 4.571839 | 0.116289 |
| H | 2.913392 | 2.277131 | 0.935654 |
| H | -3.188355 | -0.228222 | -0.309385 |
| H | 4.470201 | -1.960116 | -2.672176 |
| H | 2.876940 | -1.658058 | -3.385418 |
| H | 3.993580 | -0.306073 | -3.099319 |
| H | 4.944520 | -3.050911 | -0.771664 |
| H | 5.159854 | -2.431760 | 0.875955 |
| H | 3.939488 | -3.678442 | 0.547291 |
| H | 3.610822 | -1.626997 | 2.610149 |
| H | 1.992223 | -0.880440 | 2.801502 |
| H | 2.132562 | -2.595887 | 2.322185 |
| H | 1.481968 | 0.175545 | -2.993908 |
| H | 0.267414 | 0.587658 | -1.743813 |
| H | 1.688535 | 1.643718 | -1.984650 |
| C | -3.467604 | -1.285431 | -0.882197 |
| O | -4.615348 | -1.255767 | -1.312129 |
| O | -2.539698 | -2.129636 | -0.900514 |

9-P TS
64

| | | | |
|---|---|---|---|
| C | 3.115302 | 0.840342 | 1.645490 |
| C | 1.888087 | 0.802045 | 0.970120 |
| C | 1.537007 | 1.868348 | 0.128758 |
| C | 2.418145 | 2.929228 | -0.067316 |
| C | 3.649193 | 2.949760 | 0.593872 |
| C | 3.991188 | 1.908788 | 1.457648 |
| P | 0.760732 | -0.619626 | 1.316386 |
| AU | -1.298220 | 0.279055 | 1.584556 |
| C | 0.667337 | -1.367830 | -0.349842 |
| N | 1.727494 | -1.685407 | -1.154225 |
| C | 1.303741 | -2.439153 | -2.239972 |
| C | -0.036883 | -2.609672 | -2.104882 |
| N | -0.419107 | -1.950473 | -0.942769 |
| C | 3.096791 | -1.313661 | -0.918749 |
| C | 3.628120 | -0.240852 | -1.649718 |
| C | 4.943375 | 0.132236 | -1.375922 |
| C | 5.715089 | -0.535446 | -0.418487 |
| C | 5.153850 | -1.622724 | 0.255946 |
| C | 3.839413 | -2.037979 | 0.021405 |
| C | 2.812556 | 0.482288 | -2.684106 |
| C | 3.244518 | -3.199139 | 0.767496 |
| C | 7.108769 | -0.064318 | -0.104544 |
| C | -1.791692 | -1.877291 | -0.511273 |
| C | -2.247809 | -2.772541 | 0.464178 |
| C | -3.598311 | -2.700819 | 0.818963 |
| C | -4.467776 | -1.773929 | 0.235354 |
| C | -3.961657 | -0.894888 | -0.729972 |
| C | -2.623634 | -0.924066 | -1.121805 |
| C | -1.308534 | -3.734107 | 1.135885 |
| C | -5.903851 | -1.674327 | 0.666642 |
| C | -2.075665 | 0.061208 | -2.115441 |
| H | 2.000089 | -2.782752 | -2.992179 |
| H | -0.760023 | -3.133221 | -2.714338 |
| H | 5.749608 | -2.162271 | 0.993276 |
| H | 5.369552 | 0.980300 | -1.912858 |
| H | -4.618221 | -0.141373 | -1.164772 |
| H | -3.975654 | -3.379921 | 1.584206 |
| H | 4.016973 | -3.732461 | 1.331241 |
| H | 2.477808 | -2.851231 | 1.476849 |
| H | 2.757285 | -3.910064 | 0.086938 |
| H | 3.293248 | 1.426327 | -2.959772 |
| H | 2.699197 | -0.120926 | -3.596422 |
| H | 1.805447 | 0.707838 | -2.313774 |
| H | 7.674225 | 0.150831 | -1.020458 |
| H | 7.074100 | 0.866969 | 0.480195 |
| H | 7.662415 | -0.807207 | 0.480981 |
| H | -2.825199 | 0.822459 | -2.351277 |
| H | -1.192945 | 0.571270 | -1.705610 |
| H | -1.769442 | -0.427100 | -3.051264 |
| H | -1.861885 | -4.446997 | 1.755866 |
| H | -0.709051 | -4.297251 | 0.408300 |
| H | -0.605485 | -3.193782 | 1.788419 |
| H | -6.196027 | -2.525337 | 1.292540 |
| H | -6.046070 | -0.750736 | 1.246578 |
| H | -6.576504 | -1.628776 | -0.199980 |
| H | 0.561425 | 1.868968 | -0.357034 |
| H | 2.138411 | 3.748917 | -0.728536 |
| H | 4.334698 | 3.783229 | 0.444010 |
| H | 4.944348 | 1.925276 | 1.985214 |
| H | 3.388379 | 0.020943 | 2.309347 |
| C | -3.717648 | 1.319922 | 1.802833 |
| O | -3.116243 | 2.031243 | 0.961337 |
| O | -4.894069 | 1.224155 | 2.150471 |
| H | -3.006842 | 0.551881 | 2.430707 |

10-P TS

82

| | | | |
|---|---|---|---|
| C | 2.909608 | -1.987370 | 0.583391 |
| C | 1.622222 | -2.419222 | 0.220388 |
| C | 1.376747 | -3.375138 | -0.781106 |
| C | 2.488874 | -3.913801 | -1.435379 |
| C | 3.778266 | -3.511860 | -1.099619 |
| C | 3.985262 | -2.561834 | -0.102366 |
| N | 0.475499 | -1.870159 | 0.900987 |
| C | -0.359483 | -0.929288 | 0.358320 |
| N | -1.405596 | -0.829391 | 1.233023 |
| C | -1.206816 | -1.692755 | 2.309043 |
| C | -0.037311 | -2.345339 | 2.096157 |
| P | 0.087655 | -0.168504 | -1.248065 |
| AU | 1.470084 | 1.402137 | -0.359999 |
| C | -2.635802 | -0.082349 | 1.096036 |
| C | -2.620880 | 1.310455 | 1.290491 |
| C | -3.845579 | 1.975431 | 1.177518 |
| C | -5.023672 | 1.280737 | 0.915511 |
| C | -5.004349 | -0.098369 | 0.745385 |
| C | -3.804944 | -0.814287 | 0.821400 |
| C | -1.343107 | 2.053444 | 1.629237 |
| C | -0.901726 | 1.790827 | 3.079992 |
| C | -3.807670 | -2.321399 | 0.612421 |
| C | -4.373534 | -3.044185 | 1.848060 |
| C | -0.023342 | -3.846094 | -1.134073 |
| C | -0.254895 | -5.276524 | -0.619609 |
| C | 3.134525 | -0.960475 | 1.678803 |
| C | 3.246971 | -1.630324 | 3.059638 |
| C | -1.453053 | 0.708789 | -1.721582 |
| C | -2.563085 | -0.054790 | -2.110351 |
| C | -3.724351 | 0.562865 | -2.565929 |
| C | -3.774471 | 1.953966 | -2.684161 |
| C | -2.657735 | 2.719068 | -2.346418 |
| C | -1.503941 | 2.102207 | -1.863518 |
| C | -1.412702 | 3.558229 | 1.363758 |
| C | -4.589547 | -2.746520 | -0.639415 |
| C | -0.306644 | -3.730585 | -2.637889 |
| C | 4.368563 | -0.086455 | 1.425625 |
| H | -1.919402 | -1.745180 | 3.118452 |
| H | 0.470917 | -3.106834 | 2.668627 |
| H | -5.931002 | -0.628002 | 0.529729 |
| H | -3.875304 | 3.055525 | 1.298956 |
| H | 4.999306 | -2.250958 | 0.139758 |
| H | 2.341647 | -4.656629 | -2.218055 |
| H | -2.768423 | -2.651070 | 0.474460 |
| H | -0.546915 | 1.656632 | 0.972666 |
| H | 2.253500 | -0.297535 | 1.686113 |
| H | -0.744247 | -3.195266 | -0.621995 |
| H | -0.638004 | 2.700346 | -1.578346 |
| H | -2.687112 | 3.803725 | -2.446299 |
| H | -4.680893 | 2.438499 | -3.045025 |
| H | -4.590563 | -0.040468 | -2.836179 |
| H | -2.511932 | -1.142030 | -2.046251 |
| H | 2.999265 | 2.127435 | 0.592263 |
| H | -5.965026 | 1.823708 | 0.836822 |
| H | -0.404663 | 3.984755 | 1.428966 |
| H | -1.801060 | 3.770994 | 0.360902 |
| H | -2.050343 | 4.070377 | 2.097995 |
| H | 0.044823 | 2.310275 | 3.277251 |
| H | -1.655224 | 2.168011 | 3.785183 |
| H | -0.748528 | 0.724354 | 3.282739 |
| H | -4.483186 | -3.827928 | -0.795038 |
| H | -5.660868 | -2.531826 | -0.535930 |
| H | -4.230259 | -2.233926 | -1.537945 |
| H | -4.324356 | -4.132561 | 1.711691 |
| H | -3.825663 | -2.788814 | 2.762556 |
| H | -5.424801 | -2.766177 | 2.002727 |
| H | 4.632548 | -3.939214 | -1.623918 |
| H | 4.369304 | 0.767300 | 2.114175 |
| H | 5.298770 | -0.647126 | 1.589370 |
| H | 4.379118 | 0.304702 | 0.400800 |
| H | 3.406792 | -0.869148 | 3.834204 |
| H | 2.346110 | -2.195363 | 3.323832 |
| H | 4.098858 | -2.323661 | 3.078008 |
| H | -1.349087 | -4.004027 | -2.847459 |
| H | -0.139132 | -2.702637 | -2.982903 |
| H | 0.337210 | -4.400963 | -3.222201 |
| H | -1.281157 | -5.600344 | -0.837995 |
| H | 0.435153 | -5.981452 | -1.102431 |
| H | -0.096702 | -5.338435 | 0.464993 |
| C | 2.760983 | 3.268774 | 0.918033 |
| O | 1.944764 | 3.863419 | 0.164812 |
| O | 3.382556 | 3.599459 | 1.924670 |

11-P TS

82

| | | | |
|---|---|---|---|
| C | 1.472352 | -3.179355 | -0.323346 |
| C | 0.105434 | -2.881713 | -0.436782 |
| C | -0.773418 | -3.556785 | -1.301494 |
| C | -0.236056 | -4.588201 | -2.076344 |
| C | 1.112748 | -4.922092 | -1.981164 |
| C | 1.954898 | -4.225529 | -1.118538 |
| N | -0.464772 | -1.845470 | 0.390467 |
| C | -0.682017 | -0.555157 | -0.013330 |
| N | -1.442638 | 0.018759 | 0.963004 |
| C | -1.700230 | -0.912573 | 1.977752 |
| C | -1.090839 | -2.079162 | 1.612400 |
| P | 0.053763 | 0.024054 | -1.592850 |
| AU | 1.993018 | 0.793534 | -0.695482 |
| C | -1.961595 | 1.363503 | 1.007596 |
| C | -1.191069 | 2.341232 | 1.656640 |
| C | -1.773895 | 3.605442 | 1.798307 |
| C | -3.045610 | 3.865302 | 1.298145 |
| C | -3.754931 | 2.882461 | 0.611168 |
| C | -3.229619 | 1.597645 | 0.444362 |
| CL | -2.637461 | -0.566506 | 3.345117 |
| CL | -1.108498 | -3.590189 | 2.373262 |
| C | 0.225806 | 2.062709 | 2.133001 |
| C | 0.279886 | 1.522522 | 3.571476 |
| C | -3.965835 | 0.537395 | -0.367681 |
| C | -4.675368 | -0.524305 | 0.492520 |
| C | -2.238600 | -3.181294 | -1.429085 |
| C | -3.157163 | -4.353962 | -1.058495 |
| C | 2.394655 | -2.424921 | 0.615596 |
| C | 2.599876 | -3.194031 | 1.931423 |
| C | -0.911473 | 1.541710 | -1.997346 |
| C | -1.883999 | 1.416647 | -3.000286 |
| C | -2.577918 | 2.537666 | -3.459592 |
| C | -2.308371 | 3.795275 | -2.920189 |
| C | -1.328558 | 3.932940 | -1.933542 |
| C | -0.625384 | 2.817980 | -1.489187 |
| C | 1.142609 | 3.285114 | 2.007554 |
| C | -4.983260 | 1.144138 | -1.340104 |
| C | -2.539086 | -2.646544 | -2.837934 |
| C | 3.746033 | -2.104445 | -0.035806 |
| H | -4.724984 | 3.125107 | 0.184344 |
| H | -1.214869 | 4.399387 | 2.289052 |
| H | 3.008937 | -4.491177 | -1.065942 |
| H | -0.880923 | -5.132651 | -2.765065 |
| H | -3.203311 | 0.022920 | -0.973552 |
| H | 0.649035 | 1.288815 | 1.469974 |
| H | 1.919107 | -1.461418 | 0.856165 |
| H | -2.451830 | -2.367998 | -0.723010 |
| H | 0.158690 | 2.926491 | -0.743294 |
| H | -1.104910 | 4.913384 | -1.515062 |
| H | -2.851347 | 4.670094 | -3.276241 |
| H | -3.326441 | 2.426269 | -4.243374 |
| H | -2.091954 | 0.437780 | -3.432243 |
| H | -3.480633 | 4.856962 | 1.418503 |
| H | 2.176803 | 2.972071 | 2.188468 |
| H | 1.095582 | 3.732372 | 1.006463 |
| H | 0.887820 | 4.062272 | 2.740578 |
| H | 1.325740 | 1.356316 | 3.858167 |
| H | -0.162160 | 2.244422 | 4.272028 |
| H | -0.255004 | 0.574317 | 3.683575 |
| H | -5.336609 | 0.367753 | -2.030279 |
| H | -5.861557 | 1.532473 | -0.806593 |
| H | -4.546557 | 1.955736 | -1.930514 |
| H | -5.247452 | -1.198435 | -0.158359 |
| H | -3.981889 | -1.139270 | 1.072387 |
| H | -5.375876 | -0.050313 | 1.192321 |
| H | 1.514080 | -5.729920 | -2.592173 |
| H | 4.312938 | -1.430268 | 0.615585 |
| H | 4.343281 | -3.011530 | -0.198789 |
| H | 3.608665 | -1.602309 | -1.003169 |
| H | 3.245099 | -2.615319 | 2.604443 |
| H | 1.650637 | -3.387355 | 2.444267 |
| H | 3.083219 | -4.162026 | 1.739004 |
| H | -3.572186 | -2.279051 | -2.896067 |
| H | -1.857059 | -1.824571 | -3.089333 |
| H | -2.415057 | -3.432790 | -3.594387 |
| H | -4.208917 | -4.043895 | -1.112707 |
| H | -3.018697 | -5.197969 | -1.746975 |
| H | -2.953866 | -4.709374 | -0.040483 |
| C | 4.230833 | 1.710615 | 0.390775 |
| O | 3.664076 | 1.017611 | 1.269970 |
| O | 5.275440 | 2.351161 | 0.341743 |
| H | 3.622124 | 1.784157 | -0.676947 |

12-P TS

70

| | | | |
|---|---|---|---|
| C | 3.250924 | -0.966639 | 0.763616 |
| C | 2.408813 | -1.903645 | 0.144336 |
| C | 2.813813 | -2.695809 | -0.937276 |
| C | 4.110244 | -2.507351 | -1.424243 |
| C | 4.981353 | -1.573833 | -0.852799 |
| C | 4.538065 | -0.821931 | 0.241393 |
| N | 1.077283 | -2.066000 | 0.662130 |
| C | -0.014890 | -1.414189 | 0.175646 |
| N | -1.040193 | -1.768608 | 0.995529 |
| C | -0.593073 | -2.645714 | 1.995229 |
| C | 0.746113 | -2.830710 | 1.782647 |
| P | 0.111490 | -0.422016 | -1.360343 |
| AU | 0.711376 | 1.595707 | -0.521028 |
| C | -2.363859 | -1.210741 | 0.973830 |
| C | -2.567511 | 0.017573 | 1.615925 |
| C | -3.859456 | 0.550450 | 1.584628 |
| C | -4.911025 | -0.107232 | 0.940809 |
| C | -4.662226 | -1.345121 | 0.338265 |
| C | -3.393692 | -1.925162 | 0.348745 |
| C | -1.505226 | -3.172641 | 3.040363 |
| C | 1.756581 | -3.654610 | 2.491637 |
| C | -1.437025 | 0.744945 | 2.283611 |
| C | -6.276554 | 0.517929 | 0.854649 |
| C | -3.143314 | -3.262108 | -0.289680 |
| C | 1.874408 | -3.686467 | -1.564575 |
| C | 6.364753 | -1.364852 | -1.408786 |
| C | 2.783758 | -0.131562 | 1.921238 |
| C | -1.650288 | -0.270681 | -1.877350 |
| C | -2.143025 | -1.262693 | -2.739337 |
| C | -3.436008 | -1.178264 | -3.255796 |
| C | -4.243332 | -0.085130 | -2.938477 |
| C | -3.750389 | 0.923883 | -2.107792 |
| C | -2.464653 | 0.833195 | -1.581763 |
| H | -5.473810 | -1.867243 | -0.169850 |
| H | -4.041352 | 1.511976 | 2.066357 |
| H | 5.210153 | -0.097537 | 0.703409 |
| H | 4.446881 | -3.107834 | -2.270325 |
| H | -3.929203 | -3.494410 | -1.015807 |
| H | -2.178782 | -3.286266 | -0.809329 |
| H | -3.131117 | -4.066171 | 0.461138 |
| H | -1.807088 | 1.589817 | 2.872933 |
| H | -0.855762 | 0.086334 | 2.942835 |
| H | -0.727927 | 1.141275 | 1.533953 |
| H | -6.404306 | 1.310630 | 1.600970 |
| H | -6.427461 | 0.966417 | -0.138842 |
| H | -7.068935 | -0.227928 | 0.996840 |
| H | 3.630761 | 0.363242 | 2.407738 |
| H | 2.100159 | 0.660890 | 1.578243 |
| H | 2.246886 | -0.724041 | 2.672926 |
| H | 2.395787 | -4.305136 | -2.302408 |
| H | 1.418010 | -4.347423 | -0.815559 |
| H | 1.054797 | -3.157748 | -2.074810 |
| H | 6.660785 | -2.188149 | -2.069169 |
| H | 6.408903 | -0.433544 | -1.992201 |
| H | 7.107173 | -1.277925 | -0.604897 |
| H | -2.079609 | 1.617832 | -0.932061 |
| H | -4.371797 | 1.784827 | -1.863613 |
| H | -5.250644 | -0.012702 | -3.347346 |
| H | -3.807539 | -1.960694 | -3.917041 |
| H | -1.503560 | -2.102807 | -3.013157 |
| H | -0.941097 | -3.770592 | 3.762878 |
| H | -2.001595 | -2.354924 | 3.579777 |
| H | -2.292101 | -3.806545 | 2.610302 |
| H | 1.289945 | -4.191361 | 3.323383 |
| H | 2.210276 | -4.391884 | 1.815585 |
| H | 2.570330 | -3.035587 | 2.892025 |
| C | 1.462128 | 3.930196 | 0.511227 |
| O | 1.736309 | 3.065225 | 1.377414 |
| O | 1.568292 | 5.151457 | 0.453362 |
| H | 0.971192 | 3.483512 | -0.523645 |

13-P TS

44

| | | | |
|---|---|---|---|
| N | 1.648874 | -2.428017 | 0.025316 |
| C | 1.389121 | -1.098136 | -0.124790 |
| N | 1.670534 | -0.805302 | -1.424236 |
| C | 2.104662 | -1.942284 | -2.077086 |
| C | 2.080068 | -2.962338 | -1.173532 |
| P | 0.748899 | -0.049116 | 1.242821 |
| AU | -1.313994 | 0.340718 | 0.395013 |
| C | 1.373765 | 0.480467 | -2.101889 |
| C | 0.215164 | 0.278175 | -3.075490 |
| C | 1.339804 | -3.196402 | 1.255668 |
| C | 2.346580 | -4.319511 | 1.464682 |
| C | 1.776055 | 1.452923 | 0.947260 |
| C | 3.174433 | 1.323875 | 0.920903 |
| C | 3.987128 | 2.448513 | 0.787659 |
| C | 3.416107 | 3.720957 | 0.701424 |
| C | 2.027730 | 3.859802 | 0.750282 |
| C | 1.211632 | 2.734847 | 0.869918 |
| C | 2.636517 | 1.048023 | -2.738222 |
| C | -0.108462 | -3.681275 | 1.192320 |
| H | 2.389821 | -1.936947 | -3.119652 |
| H | 2.339185 | -4.005046 | -1.285006 |
| H | 0.126378 | 2.839785 | 0.882761 |
| H | 1.574051 | 4.848308 | 0.683123 |
| H | 4.051329 | 4.599929 | 0.598898 |
| H | 5.070255 | 2.331608 | 0.757173 |
| H | 3.630317 | 0.335963 | 1.001681 |
| H | 1.046114 | 1.146833 | -1.298893 |
| H | -0.052286 | 1.242472 | -3.523378 |
| H | 0.486014 | -0.411837 | -3.885355 |
| H | -0.666812 | -0.112541 | -2.551597 |
| H | 2.408916 | 2.025613 | -3.178442 |
| H | 3.422876 | 1.182426 | -1.986570 |
| H | 3.014241 | 0.398581 | -3.539035 |
| H | 1.434685 | -2.458467 | 2.064363 |
| H | -0.371812 | -4.184153 | 2.130204 |
| H | -0.796604 | -2.838975 | 1.046596 |
| H | -0.245397 | -4.391089 | 0.365966 |
| H | 2.161871 | -4.780094 | 2.441729 |
| H | 2.244327 | -5.106849 | 0.706554 |
| H | 3.376438 | -3.942762 | 1.450879 |
| C | -3.672337 | 0.607310 | -0.759707 |
| O | -3.172458 | -0.462911 | -1.185601 |
| O | -4.698221 | 1.233380 | -1.001500 |
| H | -2.999890 | 1.182898 | 0.096456 |

1 TS

57

| | | | |
|---|---|---|---|
| C | 0.186169 | -1.228078 | 0.032065 |
| N | 1.422133 | -0.784838 | 0.048611 |
| C | 2.494031 | -1.870724 | 0.084175 |
| C | 1.627195 | -3.116396 | 0.359972 |
| C | 0.164522 | -2.740277 | 0.014495 |
| C | 1.752081 | 0.623118 | -0.016817 |
| C | 1.905887 | 1.339880 | 1.186891 |
| C | 2.289509 | 2.682229 | 1.097150 |
| C | 2.489854 | 3.293936 | -0.135825 |
| C | 2.258542 | 2.583589 | -1.308508 |
| C | 1.870278 | 1.239481 | -1.279263 |
| C | 1.553732 | 0.760754 | 2.546472 |
| C | 2.675854 | 0.938372 | 3.578113 |
| C | 1.475175 | 0.559096 | -2.578771 |
| C | 0.150747 | 1.152924 | -3.089883 |
| C | 3.227543 | -1.931063 | -1.256278 |
| C | 3.501972 | -1.598461 | 1.196518 |
| C | -0.248339 | -3.180851 | -1.406652 |
| C | -0.832626 | -3.298349 | 1.036890 |
| AU | -1.399529 | -0.081618 | 0.061753 |
| C | 2.564994 | 0.649586 | -3.655721 |
| C | 0.244084 | 1.391657 | 3.051614 |
| H | -2.775254 | 0.868274 | 0.810683 |
| H | 2.357345 | 3.086219 | -2.269568 |
| H | 2.411923 | 3.262231 | 2.010879 |
| H | 1.365334 | -0.312834 | 2.424771 |
| H | 1.976303 | -3.979697 | -0.217144 |
| H | -0.256764 | -4.277168 | -1.458497 |
| H | 0.443335 | -2.803892 | -2.169530 |
| H | -1.252658 | -2.810098 | -1.645921 |
| H | -0.605655 | -2.937845 | 2.048030 |
| H | -0.784979 | -4.395307 | 1.038796 |
| H | -1.856751 | -2.993821 | 0.786891 |
| H | 3.996068 | -2.711245 | -1.199527 |
| H | 3.726392 | -0.979030 | -1.470831 |
| H | 2.556619 | -2.174198 | -2.086086 |
| H | 3.024129 | -1.588293 | 2.180904 |
| H | 4.019310 | -0.644211 | 1.039421 |
| H | 4.252888 | -2.397941 | 1.190945 |
| H | 1.692556 | -3.376308 | 1.424042 |
| H | 2.794106 | 4.339119 | -0.182953 |
| H | 2.400531 | 0.448863 | 4.521391 |
| H | 2.849344 | 2.000389 | 3.796065 |
| H | 3.622442 | 0.508312 | 3.229273 |
| H | -0.050656 | 0.945364 | 4.010555 |
| H | -0.568239 | 1.227577 | 2.330385 |
| H | 0.359484 | 2.474265 | 3.195708 |
| H | -0.172369 | 0.632063 | -4.000944 |
| H | 0.265874 | 2.219036 | -3.326899 |
| H | -0.645083 | 1.056750 | -2.339229 |
| H | 1.291515 | -0.500441 | -2.369940 |
| H | 2.257961 | 0.089821 | -4.548707 |
| H | 3.521089 | 0.240605 | -3.306544 |
| H | 2.736742 | 1.689886 | -3.961748 |
| C | -3.524925 | 1.393457 | -0.066909 |
| O | -4.415538 | 2.008230 | 0.495467 |
| O | -3.179671 | 1.164157 | -1.245679 |

2 TS
64

| | | | |
|---|---|---|---|
| C | 2.573079 | 0.172528 | 1.129630 |
| C | 2.004168 | 0.551929 | -0.104002 |
| C | 2.532600 | 0.128502 | -1.338175 |
| C | 3.718097 | -0.616005 | -1.312156 |
| C | 4.335310 | -0.949414 | -0.111814 |
| C | 3.753827 | -0.575858 | 1.095596 |
| N | 0.810459 | 1.367444 | -0.071926 |
| C | 0.849356 | 2.891600 | -0.078777 |
| C | -0.573921 | 3.201810 | 0.424153 |
| C | -1.433687 | 1.945445 | 0.130223 |
| C | -0.397921 | 0.857611 | -0.001077 |
| C | -2.255956 | 2.029202 | -1.201821 |
| C | -3.750137 | 2.298634 | -0.963786 |
| C | -3.952845 | 3.301920 | 0.173432 |
| C | -3.427926 | 2.727980 | 1.510685 |
| C | -2.386032 | 1.617332 | 1.297536 |
| C | 1.095804 | 3.392531 | -1.503731 |
| C | 1.941263 | 3.422935 | 0.842464 |
| AU | -0.796379 | -1.054782 | -0.137269 |
| C | 1.821722 | 0.336199 | -2.663864 |
| C | 2.720892 | 0.997503 | -3.717577 |
| C | 1.902997 | 0.442254 | 2.466619 |
| C | 1.274877 | -0.856560 | 3.002619 |
| C | 1.276964 | -1.007183 | -3.180034 |
| C | 2.861620 | 1.053605 | 3.497675 |
| H | 4.149092 | -0.960275 | -2.251322 |
| H | 4.213487 | -0.887266 | 2.032474 |
| H | 1.083238 | 1.153019 | 2.304915 |
| H | -0.973927 | 4.104939 | -0.047898 |
| H | 1.070568 | 4.488978 | -1.500868 |
| H | 2.080924 | 3.077805 | -1.865574 |
| H | 0.332490 | 3.032655 | -2.201702 |
| H | 1.772657 | 3.130099 | 1.883102 |
| H | 2.931916 | 3.070879 | 0.529663 |
| H | 1.938352 | 4.518516 | 0.792669 |
| H | -0.541040 | 3.375737 | 1.507894 |
| H | 5.257532 | -1.529450 | -0.115756 |
| H | 2.314947 | 1.308135 | 4.415031 |
| H | 3.654687 | 0.346367 | 3.773335 |
| H | 3.341994 | 1.963839 | 3.117298 |
| H | 0.740467 | -0.662222 | 3.942081 |
| H | 0.565041 | -1.283323 | 2.281466 |
| H | 2.049306 | -1.610559 | 3.197483 |
| H | 0.721625 | -0.860007 | -4.115757 |
| H | 2.092486 | -1.717270 | -3.370835 |
| H | 0.597624 | -1.457695 | -2.443399 |
| H | 0.958194 | 0.987791 | -2.491385 |
| H | 2.148348 | 1.194790 | -4.633189 |
| H | 3.138529 | 1.947721 | -3.361483 |
| H | 3.561706 | 0.344573 | -3.986471 |
| H | -2.134581 | 1.091023 | -1.760912 |
| H | -1.829487 | 2.828004 | -1.822958 |
| H | -4.201090 | 2.664668 | -1.895789 |
| H | -4.264709 | 1.357501 | -0.719129 |
| H | -5.012177 | 3.573998 | 0.266251 |
| H | -3.416670 | 4.231081 | -0.074535 |
| H | -2.998928 | 3.534063 | 2.121515 |
| H | -4.258691 | 2.309165 | 2.094783 |
| H | -2.881990 | 0.662883 | 1.070604 |
| H | -1.804255 | 1.450917 | 2.214870 |
| C | -1.237650 | -3.603259 | -0.163062 |
| O | -0.991077 | -3.297634 | 1.023898 |
| O | -1.428063 | -4.630308 | -0.793357 |
| H | -1.327381 | -2.615016 | -0.948017 |

3 TS

76

| | | | |
|---|---|---|---|
| C | -2.306081 | -1.354036 | 1.309896 |
| C | -1.737446 | -1.461904 | 0.023854 |
| C | -2.517941 | -1.418392 | -1.148050 |
| C | -3.908833 | -1.371979 | -0.995905 |
| C | -4.498468 | -1.337509 | 0.262858 |
| C | -3.700436 | -1.305050 | 1.401781 |
| N | -0.298201 | -1.576914 | -0.075944 |
| C | 0.416046 | -2.922588 | -0.094182 |
| C | 1.885877 | -2.480311 | 0.078711 |
| C | 1.939798 | -0.956910 | -0.211744 |
| C | 0.495516 | -0.535459 | -0.157644 |
| C | 2.463682 | -0.646955 | -1.654715 |
| C | 3.962020 | -0.287849 | -1.713735 |
| C | 4.728753 | -1.038549 | -0.622440 |
| C | 4.242338 | -0.635834 | 0.786959 |
| C | 2.790627 | -0.125769 | 0.790162 |
| C | 2.188345 | 0.087621 | 2.204961 |
| C | 2.262032 | -1.093163 | 3.179644 |
| C | 4.517815 | -0.555507 | -3.111625 |
| C | 0.168714 | -3.616359 | -1.434644 |
| C | -0.075347 | -3.817951 | 1.038243 |
| AU | -0.123352 | 1.319548 | -0.293686 |
| C | -1.925790 | -1.290775 | -2.541152 |
| C | -2.464926 | -2.346143 | -3.516590 |
| C | -1.472553 | -1.172107 | 2.565863 |
| C | -1.568556 | 0.286839 | 3.046846 |
| C | -2.181171 | 0.125979 | -3.084846 |
| C | -1.860219 | -2.144961 | 3.686881 |
| C | 2.812310 | 1.332316 | 2.849632 |
| H | -4.538133 | -1.336948 | -1.884221 |
| H | -4.165564 | -1.219299 | 2.382795 |
| H | -0.423548 | -1.355478 | 2.309430 |
| H | 2.534843 | -3.041162 | -0.600992 |
| H | 0.700113 | -4.575653 | -1.437631 |
| H | -0.898055 | -3.819700 | -1.581026 |
| H | 0.535127 | -3.019090 | -2.276506 |
| H | 0.121694 | -3.376985 | 2.020003 |
| H | -1.149880 | -4.019210 | 0.949115 |
| H | 0.456424 | -4.775463 | 0.983358 |
| H | 2.221437 | -2.694205 | 1.096474 |
| H | -5.583403 | -1.302240 | 0.356270 |
| H | -1.176743 | -2.027802 | 4.538017 |
| H | -2.877256 | -1.949701 | 4.051271 |
| H | -1.818347 | -3.188820 | 3.352029 |
| H | -0.914421 | 0.443998 | 3.914764 |
| H | -1.270111 | 0.989129 | 2.256976 |
| H | -2.596755 | 0.532474 | 3.344098 |
| H | -1.709119 | 0.248221 | -4.068544 |
| H | -3.256820 | 0.319453 | -3.191467 |
| H | -1.762850 | 0.881900 | -2.406183 |
| H | -0.839640 | -1.414445 | -2.465787 |
| H | -1.960170 | -2.254541 | -4.487058 |
| H | -2.311204 | -3.366559 | -3.144816 |
| H | -3.540915 | -2.211868 | -3.689722 |
| H | 1.878359 | 0.180350 | -2.080840 |
| H | 2.272787 | -1.526968 | -2.286433 |
| H | 4.063523 | 0.790626 | -1.511341 |
| H | 5.806521 | -0.848151 | -0.719244 |
| H | 4.595476 | -2.120673 | -0.779461 |
| H | 4.346342 | -1.490052 | 1.469559 |
| H | 4.886360 | 0.157905 | 1.188243 |
| H | 2.792920 | 0.884539 | 0.348410 |
| H | 1.123043 | 0.322630 | 2.050559 |
| H | 2.319396 | 1.561198 | 3.803463 |
| H | 3.881949 | 1.185707 | 3.056037 |
| H | 2.705313 | 2.208453 | 2.196701 |
| H | 1.826338 | -0.799028 | 4.143769 |
| H | 1.706606 | -1.972518 | 2.834577 |
| H | 3.298032 | -1.403525 | 3.371347 |
| H | 5.562043 | -0.224711 | -3.192038 |
| H | 4.483351 | -1.630530 | -3.343030 |
| H | 3.935265 | -0.027853 | -3.879284 |
| C | -0.877693 | 3.796036 | -0.281795 |
| O | -0.852781 | 3.421451 | 0.910782 |
| O | -1.115286 | 4.832889 | -0.879832 |
| H | -0.589660 | 2.894486 | -1.118709 |

4 TS

54

| | | | |
|---|---|---|---|
| C | -3.222529 | 1.091542 | 1.157897 |
| C | -2.534133 | 1.298809 | -0.046209 |
| C | -3.074889 | 0.929250 | -1.287754 |
| C | -4.350093 | 0.359677 | -1.301762 |
| C | -5.073643 | 0.145069 | -0.123503 |
| C | -4.493693 | 0.514809 | 1.094753 |
| N | -1.225283 | 1.874928 | -0.000349 |
| C | -0.096341 | 1.153992 | 0.016906 |
| N | 0.966293 | 1.967467 | -0.043576 |
| C | 0.577191 | 3.394959 | -0.024828 |
| C | -0.958929 | 3.323004 | -0.152721 |
| C | 2.319724 | 1.503968 | -0.038193 |
| C | 2.948524 | 1.259076 | -1.269245 |
| C | 4.259856 | 0.780901 | -1.251227 |
| C | 4.937055 | 0.536921 | -0.050662 |
| C | 4.272833 | 0.785249 | 1.154427 |
| C | 2.960895 | 1.267142 | 1.186393 |
| C | 2.205698 | 1.452673 | -2.562455 |
| C | 2.238423 | 1.475891 | 2.488714 |
| C | 6.336411 | -0.020105 | -0.062983 |
| AU | -0.001949 | -0.798082 | 0.200619 |
| C | -2.278766 | 1.085743 | -2.554223 |
| C | -6.427056 | -0.513463 | -0.164893 |
| C | -2.580936 | 1.430890 | 2.474978 |
| H | -5.039827 | 0.340611 | 2.023146 |
| H | -4.783990 | 0.064302 | -2.258338 |
| H | 4.759715 | 0.579094 | -2.199856 |
| H | 4.782958 | 0.586227 | 2.098013 |
| H | -3.262072 | 1.218625 | 3.305881 |
| H | -1.665170 | 0.839669 | 2.618856 |
| H | -2.292066 | 2.489416 | 2.529937 |
| H | -2.873872 | 0.793732 | -3.425959 |
| H | -1.937302 | 2.118795 | -2.704243 |
| H | -1.382103 | 0.449811 | -2.524865 |
| H | -6.968621 | -0.256234 | -1.083597 |
| H | -6.323190 | -1.608558 | -0.141371 |
| H | -7.040971 | -0.222302 | 0.695911 |
| H | 2.861597 | 1.266429 | -3.419422 |
| H | 1.354899 | 0.759353 | -2.625364 |
| H | 1.799642 | 2.468967 | -2.655957 |
| H | 2.875245 | 1.197009 | 3.334900 |
| H | 1.931162 | 2.522280 | 2.624095 |
| H | 1.327090 | 0.861934 | 2.526196 |
| H | 6.825610 | 0.104295 | 0.910072 |
| H | 6.321616 | -1.095197 | -0.294930 |
| H | 6.953293 | 0.468323 | -0.828371 |
| H | 1.051118 | 3.932453 | -0.853623 |
| H | 0.898606 | 3.855831 | 0.918544 |
| H | -1.318323 | 3.667596 | -1.131665 |
| H | -1.478282 | 3.890583 | 0.627167 |
| C | 0.204732 | -3.364055 | 0.320977 |
| O | 0.158585 | -3.050924 | -0.891401 |
| O | 0.356306 | -4.396804 | 0.954681 |
| H | 0.057800 | -2.403158 | 1.116405 |

5 TS

72

| | | | |
|---|---|---|---|
| C | 2.980025 | 1.155738 | -0.816369 |
| C | 2.378548 | -0.068186 | -1.159210 |
| C | 3.045622 | -1.301661 | -1.044789 |
| C | 4.362761 | -1.286096 | -0.576915 |
| C | 4.985585 | -0.086364 | -0.238344 |
| C | 4.300559 | 1.119492 | -0.354172 |
| N | 1.027957 | -0.073473 | -1.634253 |
| C | -0.048411 | 0.054465 | -0.844563 |
| N | -1.150506 | 0.180365 | -1.604983 |
| C | -0.847235 | -0.099380 | -3.029206 |
| C | 0.677337 | 0.066632 | -3.064503 |
| C | -2.486514 | 0.125081 | -1.085188 |
| C | -3.208917 | 1.322661 | -0.911562 |
| C | -4.519574 | 1.217069 | -0.430840 |
| C | -5.086466 | -0.017702 | -0.126244 |
| C | -4.342503 | -1.181412 | -0.283913 |
| C | -3.028668 | -1.134530 | -0.762981 |
| C | -2.644634 | 2.722600 | -1.112716 |
| C | -1.973072 | 2.996591 | -2.465861 |
| C | -2.232652 | -2.422420 | -0.888362 |
| C | -2.850492 | -3.362621 | -1.934987 |
| AU | -0.017308 | -0.032621 | 1.117244 |
| C | 2.329839 | -2.609384 | -1.329437 |
| C | 3.189965 | -3.604185 | -2.117532 |
| C | 2.238347 | 2.476960 | -0.900150 |
| C | 2.926457 | 3.449835 | -1.869815 |
| C | 1.825790 | -3.221884 | -0.010705 |
| C | 2.069278 | 3.097405 | 0.495197 |
| C | -2.093469 | -3.116036 | 0.474920 |
| C | -1.718812 | 3.116258 | 0.050511 |
| H | -5.097300 | 2.129800 | -0.282650 |
| H | -4.783405 | -2.142978 | -0.023035 |
| H | 4.905170 | -2.223490 | -0.462921 |
| H | 4.794706 | 2.048724 | -0.072248 |
| H | -3.521275 | 3.386045 | -1.062028 |
| H | -1.216771 | -2.171595 | -1.220083 |
| H | 1.444448 | -2.378184 | -1.938110 |
| H | 1.231152 | 2.278833 | -1.290967 |
| H | 1.184032 | -0.696949 | -3.663377 |
| H | 0.986860 | 1.059589 | -3.423043 |
| H | -1.160614 | -1.124978 | -3.272783 |
| H | -1.375621 | 0.598348 | -3.683577 |
| H | 6.011786 | -0.093413 | 0.128464 |
| H | -6.108500 | -0.069795 | 0.248070 |
| H | 1.493483 | 4.029724 | 0.430868 |
| H | 1.539792 | 2.412430 | 1.171003 |
| H | 3.044292 | 3.329628 | 0.943805 |
| H | 2.343670 | 4.376379 | -1.955579 |
| H | 3.931553 | 3.716654 | -1.516478 |
| H | 3.028055 | 3.010850 | -2.871128 |
| H | 1.260288 | -4.143437 | -0.202909 |
| H | 2.669775 | -3.465306 | 0.648858 |
| H | 1.173337 | -2.516591 | 0.522879 |
| H | 2.599412 | -4.494650 | -2.369484 |
| H | 3.561017 | -3.160352 | -3.050445 |
| H | 4.056909 | -3.940507 | -1.533613 |
| H | -1.844109 | 4.079129 | -2.595841 |
| H | -0.976804 | 2.543561 | -2.523459 |
| H | -2.579476 | 2.627207 | -3.302920 |
| H | -1.427440 | 4.171553 | -0.036559 |
| H | -2.211053 | 2.970633 | 1.019345 |
| H | -0.805371 | 2.508965 | 0.052886 |
| H | -1.473195 | -4.016855 | 0.382929 |
| H | -1.617619 | -2.444275 | 1.201545 |
| H | -3.071424 | -3.415089 | 0.874121 |
| H | -2.233786 | -4.263727 | -2.050303 |
| H | -3.857891 | -3.679611 | -1.633010 |
| H | -2.934803 | -2.872823 | -2.914114 |
| C | -0.064160 | -0.039233 | 3.693343 |
| O | 0.040146 | 1.154000 | 3.325673 |
| O | -0.221897 | -0.620898 | 4.754830 |
| H | 0.016793 | -0.878881 | 2.761895 |

6 TS

61

| | | | |
|---|---|---|---|
| C | -2.960152 | -0.432192 | -1.141940 |
| C | -2.389702 | -0.786232 | 0.086676 |
| C | -2.942045 | -0.422306 | 1.318966 |
| C | -4.110246 | 0.344480 | 1.295645 |
| C | -4.711462 | 0.735739 | 0.095119 |
| C | -4.126256 | 0.334156 | -1.111301 |
| N | -1.152999 | -1.550146 | 0.067713 |
| C | 0.012014 | -0.845407 | 0.050987 |
| N | 1.188358 | -1.529152 | 0.059943 |
| C | 1.303100 | -2.940766 | 0.101335 |
| C | 0.034933 | -3.729181 | -0.174592 |
| C | -1.245080 | -2.962403 | 0.107641 |
| O | 2.386434 | -3.461148 | 0.271905 |
| C | 2.414716 | -0.749522 | 0.075305 |
| C | 2.968579 | -0.383317 | 1.307514 |
| C | 4.136859 | 0.380810 | 1.282472 |
| C | 4.738735 | 0.769160 | 0.079693 |
| C | 4.150168 | 0.370544 | -1.124510 |
| C | 2.981721 | -0.394981 | -1.153186 |
| AU | -0.020637 | 1.114078 | 0.173215 |
| O | -2.319205 | -3.500737 | 0.281641 |
| C | 0.033563 | -3.995892 | -1.712195 |
| C | 0.048027 | -5.053539 | 0.597949 |
| C | 2.313947 | -0.780112 | 2.600244 |
| C | 5.979390 | 1.621847 | 0.084643 |
| C | 2.343596 | -0.804787 | -2.450837 |
| C | -2.284857 | -0.816240 | 2.611345 |
| C | -5.947210 | 1.595397 | 0.093712 |
| C | -2.321955 | -0.839493 | -2.440357 |
| H | -4.584627 | 0.629940 | -2.055825 |
| H | -4.557474 | 0.647621 | 2.243029 |
| H | 4.587018 | 0.681576 | 2.229401 |
| H | 4.608756 | 0.663793 | -2.069610 |
| H | -2.903641 | -0.469342 | -3.290487 |
| H | -1.303762 | -0.432504 | -2.524853 |
| H | -2.249453 | -1.932404 | -2.528228 |
| H | -2.866512 | -0.457965 | 3.466698 |
| H | -2.186332 | -1.906930 | 2.695485 |
| H | -1.275698 | -0.384310 | 2.682558 |
| H | -6.454705 | 1.569078 | 1.064963 |
| H | -5.687320 | 2.642629 | -0.119566 |
| H | -6.655707 | 1.272171 | -0.679684 |
| H | 2.890511 | -0.412971 | 3.455282 |
| H | 1.299760 | -0.359801 | 2.668542 |
| H | 2.227913 | -1.871696 | 2.687602 |
| H | 2.919975 | -0.428298 | -3.301814 |
| H | 2.280558 | -1.898176 | -2.540731 |
| H | 1.321580 | -0.406950 | -2.531873 |
| H | 6.533000 | 1.526840 | -0.856979 |
| H | 5.717709 | 2.682927 | 0.210531 |
| H | 6.646817 | 1.349755 | 0.911766 |
| H | 0.049941 | -4.876190 | 1.679757 |
| H | -0.842193 | -5.634583 | 0.340813 |
| H | 0.945790 | -5.620313 | 0.335389 |
| H | -0.856465 | -4.581040 | -1.968419 |
| H | 0.024363 | -3.062317 | -2.286991 |
| H | 0.931992 | -4.566087 | -1.972719 |
| C | -0.187956 | 3.664826 | 0.142277 |
| O | -0.134407 | 3.290903 | -1.044371 |
| O | -0.340383 | 4.692651 | 0.768934 |
| H | -0.029116 | 2.697894 | 1.007175 |

7 Ts

62

| | | | |
|---|---|---|---|
| C | -3.120122 | 0.735011 | 0.983271 |
| C | -2.428213 | 0.978083 | -0.212682 |
| C | -2.928617 | 0.584459 | -1.462784 |
| C | -4.172324 | -0.052830 | -1.492446 |
| C | -4.900627 | -0.306219 | -0.326459 |
| C | -4.356541 | 0.091667 | 0.899322 |
| N | -1.152371 | 1.658453 | -0.156128 |
| C | -0.038028 | 0.937732 | -0.092787 |
| N | 1.172656 | 1.597208 | -0.121928 |
| C | 1.334433 | 2.997343 | -0.038013 |
| C | 0.068618 | 3.784513 | 0.282172 |
| C | -1.136488 | 3.111629 | -0.376975 |
| O | 2.441180 | 3.495881 | -0.131017 |
| C | 2.367487 | 0.771737 | -0.156749 |
| C | 2.878219 | 0.379250 | -1.399885 |
| C | 3.995017 | -0.459024 | -1.404763 |
| C | 4.592740 | -0.896799 | -0.217264 |
| C | 4.058565 | -0.462848 | 0.999848 |
| C | 2.942655 | 0.375783 | 1.056602 |
| AU | -0.058762 | -1.034469 | -0.089122 |
| C | -0.061024 | 3.786798 | 1.819951 |
| C | 0.191796 | 5.216533 | -0.246000 |
| C | 2.223473 | 0.821601 | -2.677854 |
| C | 5.764354 | -1.841922 | -0.250360 |
| C | 2.358995 | 0.808004 | 2.372054 |
| C | -2.129779 | 0.770116 | -2.723481 |
| C | -6.220944 | -1.027007 | -0.379746 |
| C | -2.536039 | 1.110531 | 2.315750 |
| H | -4.903289 | -0.111980 | 1.821181 |
| H | -4.575390 | -0.369391 | -2.455381 |
| H | 4.403492 | -0.784297 | -2.362567 |
| H | 4.514996 | -0.792420 | 1.934078 |
| H | -3.192450 | 0.785562 | 3.129404 |
| H | -1.554388 | 0.637777 | 2.459379 |
| H | -2.392627 | 2.195318 | 2.407309 |
| H | -2.712427 | 0.464564 | -3.598596 |
| H | -1.812835 | 1.810102 | -2.873201 |
| H | -1.219417 | 0.153116 | -2.687374 |
| H | -6.660695 | -0.983638 | -1.382952 |
| H | -6.091933 | -2.087423 | -0.117379 |
| H | -6.936752 | -0.600036 | 0.334141 |
| H | 2.774376 | 0.445647 | -3.546155 |
| H | 1.192827 | 0.442713 | -2.738571 |
| H | 2.176409 | 1.916860 | -2.746394 |
| H | 2.937874 | 0.397192 | 3.205486 |
| H | 2.350412 | 1.902089 | 2.468831 |
| H | 1.321548 | 0.457517 | 2.475200 |
| H | 6.345026 | -1.791823 | 0.678513 |
| H | 5.418123 | -2.879428 | -0.369031 |
| H | 6.432464 | -1.622277 | -1.092484 |
| H | 0.321990 | 5.227581 | -1.335266 |
| H | -0.714101 | 5.782904 | 0.004187 |
| H | 1.054257 | 5.716598 | 0.205830 |
| H | -0.957315 | 4.350586 | 2.109148 |
| H | -0.142298 | 2.772414 | 2.224964 |
| H | 0.813041 | 4.271159 | 2.270326 |
| H | -2.069489 | 3.521695 | 0.026374 |
| H | -1.130724 | 3.300430 | -1.462086 |
| C | 0.027550 | -3.598413 | 0.038547 |
| O | -0.059299 | -3.180490 | 1.213275 |
| O | 0.175009 | -4.674735 | -0.513986 |
| H | -0.066605 | -2.692202 | -0.849755 |

8 TS

26

| | | | |
|---|---|---|---|
| C | 3.349733 | -0.652427 | 0.117297 |
| N | 2.014164 | -1.054113 | 0.179072 |
| C | 1.182143 | 0.019322 | 0.145139 |
| N | 1.999450 | 1.101977 | 0.069253 |
| C | 3.340275 | 0.714815 | 0.046575 |
| C | 1.570375 | -2.437284 | 0.276398 |
| AU | -0.790295 | 0.009454 | 0.123790 |
| C | 1.535537 | 2.481115 | 0.020277 |
| C | 4.450073 | 1.697725 | -0.044919 |
| C | 4.473709 | -1.623308 | 0.131298 |
| C | -3.379390 | -0.027202 | 0.206161 |
| O | -4.477701 | -0.194294 | -0.303464 |
| O | -2.962732 | 0.178385 | 1.373408 |
| H | -2.489631 | -0.074520 | -0.660150 |
| H | 5.413245 | 1.177730 | -0.057679 |
| H | 4.450867 | 2.389116 | 0.809330 |
| H | 4.379973 | 2.301016 | -0.961073 |
| H | 5.429468 | -1.093246 | 0.069478 |
| H | 4.417743 | -2.318294 | -0.718297 |
| H | 4.479157 | -2.223365 | 1.052078 |
| H | 1.936798 | -3.014823 | -0.579509 |
| H | 0.476307 | -2.439745 | 0.274628 |
| H | 1.933148 | -2.888962 | 1.206660 |
| H | 1.961775 | 3.053097 | 0.851849 |
| H | 0.445038 | 2.471693 | 0.108034 |
| H | 1.819969 | 2.946901 | -0.930077 |

9 TS

52

| | | | |
|---|---|---|---|
| C | -3.053912 | -1.180372 | -1.195130 |
| C | -2.405554 | -1.473575 | 0.011114 |
| C | -2.986793 | -1.236769 | 1.264152 |
| C | -4.276633 | -0.702141 | 1.285638 |
| C | -4.965699 | -0.394369 | 0.107326 |
| C | -4.341707 | -0.642942 | -1.119676 |
| N | -1.052916 | -1.961432 | -0.032232 |
| C | 0.020261 | -1.125875 | -0.063800 |
| N | 1.108876 | -1.941393 | -0.059482 |
| C | 0.720476 | -3.277416 | -0.026199 |
| C | -0.640478 | -3.289921 | -0.011150 |
| C | 2.455834 | -1.437388 | -0.043333 |
| C | 3.075010 | -1.239006 | 1.199187 |
| C | 4.362546 | -0.700459 | 1.198942 |
| C | 5.016367 | -0.355796 | 0.009365 |
| C | 4.355338 | -0.565330 | -1.204193 |
| C | 3.065536 | -1.103286 | -1.257937 |
| C | 2.353435 | -1.552903 | 2.480818 |
| C | 2.339583 | -1.268111 | -2.563700 |
| C | 6.390593 | 0.258383 | 0.045921 |
| AU | -0.022322 | 0.828909 | -0.205380 |
| C | -2.223750 | -1.494318 | 2.534040 |
| C | -6.332192 | 0.234710 | 0.163551 |
| C | -2.366398 | -1.390015 | -2.515411 |
| H | -1.349008 | -4.106305 | 0.012949 |
| H | 1.444237 | -4.080710 | -0.018114 |
| H | -4.863824 | -0.399824 | -2.045958 |
| H | -4.747699 | -0.505653 | 2.249676 |
| H | 4.861435 | -0.531700 | 2.154298 |
| H | 4.849542 | -0.293224 | -2.137654 |
| H | -3.035189 | -1.138750 | -3.345054 |
| H | -1.473436 | -0.752118 | -2.587750 |
| H | -2.032019 | -2.428649 | -2.640430 |
| H | -2.867859 | -1.354915 | 3.408319 |
| H | -1.808992 | -2.510315 | 2.565719 |
| H | -1.377318 | -0.797152 | 2.619413 |
| H | -6.914823 | -0.149288 | 1.010124 |
| H | -6.247504 | 1.324170 | 0.292340 |
| H | -6.895219 | 0.054062 | -0.759833 |
| H | 2.987769 | -1.334273 | 3.345937 |
| H | 1.437103 | -0.951562 | 2.566007 |
| H | 2.051700 | -2.607834 | 2.531490 |
| H | 3.010178 | -1.076689 | -3.407803 |
| H | 1.918815 | -2.276158 | -2.672992 |
| H | 1.499707 | -0.559986 | -2.623970 |
| H | 6.837666 | 0.298863 | -0.953892 |
| H | 6.344689 | 1.285540 | 0.436296 |
| H | 7.062173 | -0.308133 | 0.704427 |
| C | -0.190534 | 3.399444 | -0.301423 |
| O | -0.137914 | 3.071673 | 0.908131 |
| O | -0.347434 | 4.442854 | -0.916828 |
| H | -0.045507 | 2.451407 | -1.105982 |

10 TS

70

| | | | |
|---|---|---|---|
| C | 3.013981 | -1.158054 | 1.043168 |
| C | 2.404808 | 0.097745 | 1.193894 |
| C | 3.032502 | 1.308232 | 0.860126 |
| C | 4.342970 | 1.234531 | 0.376095 |
| C | 4.982981 | 0.006512 | 0.228417 |
| C | 4.324656 | -1.176320 | 0.554238 |
| N | 1.044260 | 0.141981 | 1.661781 |
| C | -0.028140 | 0.068394 | 0.828379 |
| N | -1.114205 | 0.135693 | 1.644539 |
| C | -0.726548 | 0.250227 | 2.974942 |
| C | 0.634560 | 0.254102 | 2.985857 |
| C | -2.466620 | 0.083253 | 1.154584 |
| C | -3.068885 | -1.175945 | 1.003448 |
| C | -4.371339 | -1.202642 | 0.493650 |
| C | -5.028563 | -0.024512 | 0.148467 |
| C | -4.395138 | 1.206774 | 0.296622 |
| C | -3.092790 | 1.288828 | 0.801055 |
| C | -2.329211 | -2.465859 | 1.306651 |
| C | -3.122956 | -3.376819 | 2.253641 |
| C | -2.379208 | 2.625118 | 0.889717 |
| C | -3.186715 | 3.658125 | 1.687338 |
| AU | -0.014836 | -0.057591 | -1.129065 |
| C | 2.311701 | 2.640833 | 0.946307 |
| C | 3.109954 | 3.679369 | 1.745831 |
| C | 2.276613 | -2.453373 | 1.328050 |
| C | 3.053823 | -3.355753 | 2.296914 |
| C | 1.977112 | 3.150887 | -0.465444 |
| C | 1.956695 | -3.179970 | 0.010942 |
| C | -2.042289 | 3.138387 | -0.520164 |
| C | -1.969380 | -3.187696 | -0.002930 |
| H | -1.450786 | 0.315604 | 3.774880 |
| H | 1.345545 | 0.323688 | 3.797195 |
| H | -4.871299 | -2.159897 | 0.352035 |
| H | -4.914406 | 2.118213 | 0.003291 |
| H | 4.863684 | 2.149925 | 0.098110 |
| H | 4.830266 | -2.130808 | 0.413878 |
| H | -1.387866 | -2.208484 | 1.810247 |
| H | -1.429032 | 2.469366 | 1.417989 |
| H | 1.360594 | 2.480136 | 1.471482 |
| H | 1.320518 | -2.203189 | 1.806725 |
| H | -6.040870 | -0.067447 | -0.252345 |
| H | -1.483646 | 4.081703 | -0.462665 |
| H | -1.429143 | 2.405476 | -1.062312 |
| H | -2.957055 | 3.314782 | -1.102040 |
| H | -2.618552 | 4.593107 | 1.778042 |
| H | -4.137176 | 3.893619 | 1.190402 |
| H | -3.413994 | 3.292498 | 2.697027 |
| H | -1.389631 | -4.095951 | 0.207225 |
| H | -2.876161 | -3.479032 | -0.549782 |
| H | -1.372147 | -2.540970 | -0.659823 |
| H | -2.534987 | -4.271249 | 2.497804 |
| H | -3.370146 | -2.859988 | 3.190036 |
| H | -4.061869 | -3.712245 | 1.793436 |
| H | 6.001712 | -0.029846 | -0.156530 |
| H | 1.413380 | 4.091395 | -0.411184 |
| H | 2.893445 | 3.331166 | -1.043687 |
| H | 1.370032 | 2.414178 | -1.009354 |
| H | 2.534451 | 4.609923 | 1.835658 |
| H | 3.337895 | 3.315170 | 2.755878 |
| H | 4.059504 | 3.922789 | 1.251052 |
| H | 1.379991 | -4.093135 | 0.207733 |
| H | 1.370394 | -2.539519 | -0.661603 |
| H | 2.879423 | -3.464105 | -0.512498 |
| H | 2.468993 | -4.256321 | 2.525708 |
| H | 4.008423 | -3.681057 | 1.862361 |
| H | 3.270095 | -2.835931 | 3.239308 |
| C | 0.000735 | -0.351168 | -3.697544 |
| H | -0.013856 | 0.590686 | -2.876877 |
| O | 0.011828 | -1.493734 | -3.178148 |
| O | 0.000612 | 0.089459 | -4.836202 |

11 TS

70

| | | | |
|---|---|---|---|
| C | -2.987887 | 0.324842 | 1.284375 |
| C | -2.465593 | 0.727036 | 0.046356 |
| C | -3.164106 | 0.623726 | -1.167227 |
| C | -4.459516 | 0.101712 | -1.109742 |
| C | -5.017401 | -0.294549 | 0.104230 |
| C | -4.289760 | -0.188394 | 1.285929 |
| N | -1.121128 | 1.237556 | 0.012196 |
| C | -0.017842 | 0.442673 | -0.040329 |
| N | 1.046499 | 1.291624 | -0.035383 |
| C | 0.614811 | 2.617248 | 0.024687 |
| C | -0.752950 | 2.583189 | 0.059503 |
| C | 2.412489 | 0.839049 | -0.024003 |
| C | 3.024214 | 0.531205 | -1.247737 |
| C | 4.335792 | 0.046288 | -1.195506 |
| C | 4.988746 | -0.120236 | 0.022515 |
| C | 4.345963 | 0.188909 | 1.219292 |
| C | 3.035628 | 0.676014 | 1.223545 |
| C | 2.304798 | 0.673832 | -2.575288 |
| C | 2.099501 | -0.701033 | -3.230392 |
| C | 2.300203 | 0.935750 | 2.525021 |
| C | 1.860983 | -0.398806 | 3.152534 |
| AU | 0.038972 | -1.512238 | -0.153418 |
| CL | -1.882807 | 3.842882 | 0.149859 |
| CL | 1.683204 | 3.932299 | 0.048858 |
| C | -2.515407 | 0.978419 | -2.492482 |
| C | -1.998540 | -0.301143 | -3.174597 |
| C | -2.185934 | 0.399510 | 2.569874 |
| C | -1.939686 | -1.008209 | 3.134621 |
| C | -3.449722 | 1.768446 | -3.416592 |
| C | -2.862563 | 1.317976 | 3.598337 |
| C | 3.126172 | 1.778454 | 3.505079 |
| C | 3.043215 | 1.642980 | -3.510289 |
| H | -4.734282 | -0.515922 | 2.224844 |
| H | -5.036201 | -0.005489 | -2.026896 |
| H | 4.847331 | -0.213374 | -2.121403 |
| H | 4.864483 | 0.035721 | 2.164450 |
| H | -1.204364 | 0.833543 | 2.337157 |
| H | -1.645214 | 1.616913 | -2.286015 |
| H | 1.309489 | 1.095137 | -2.380366 |
| H | 1.389731 | 1.506284 | 2.295679 |
| H | -6.028313 | -0.700420 | 0.126819 |
| H | -1.482685 | -0.056445 | -4.112377 |
| H | -1.297146 | -0.838554 | -2.521389 |
| H | -2.831875 | -0.978632 | -3.404727 |
| H | -2.909794 | 2.072529 | -4.322476 |
| H | -4.309857 | 1.164603 | -3.734118 |
| H | -3.829387 | 2.670929 | -2.921129 |
| H | -1.323031 | -0.952690 | 4.040804 |
| H | -2.885799 | -1.499849 | 3.396928 |
| H | -1.418901 | -1.640993 | 2.403476 |
| H | -2.243938 | 1.395345 | 4.501946 |
| H | -3.009503 | 2.327049 | 3.191952 |
| H | -3.844207 | 0.925325 | 3.895363 |
| H | 6.007873 | -0.505297 | 0.041039 |
| H | 1.540272 | -0.598234 | -4.169213 |
| H | 3.061207 | -1.180774 | -3.455305 |
| H | 1.534614 | -1.366530 | -2.563246 |
| H | 2.481420 | 1.769947 | -4.444908 |
| H | 3.165721 | 2.628251 | -3.042600 |
| H | 4.040930 | 1.262808 | -3.767341 |
| H | 1.279538 | -0.219302 | 4.066335 |
| H | 1.242939 | -0.980742 | 2.455855 |
| H | 2.734764 | -1.009706 | 3.415632 |
| H | 2.530920 | 2.005796 | 4.398811 |
| H | 4.027598 | 1.245575 | 3.834770 |
| H | 3.437830 | 2.725956 | 3.047355 |
| C | 0.205994 | -4.082282 | -0.175488 |
| O | 0.188911 | -3.707176 | 1.021651 |
| O | 0.345312 | -5.144810 | -0.758056 |
| H | 0.033065 | -3.160119 | -1.009045 |

12 TS

58

| | | | |
|---|---|---|---|
| C | 3.063402 | -0.981083 | 1.201622 |
| C | 2.410103 | -1.230771 | -0.012200 |
| C | 3.004431 | -0.985980 | -1.256624 |
| C | 4.308103 | -0.483299 | -1.263320 |
| C | 5.000041 | -0.218700 | -0.076835 |
| C | 4.364276 | -0.476026 | 1.143290 |
| N | 1.056452 | -1.707792 | 0.016203 |
| C | -0.014957 | -0.875181 | 0.047209 |
| N | -1.098706 | -1.691519 | 0.027265 |
| C | -0.715097 | -3.039763 | -0.011708 |
| C | 0.653336 | -3.049793 | -0.018673 |
| C | -2.447169 | -1.200233 | 0.007388 |
| C | -3.051912 | -0.963961 | -1.235309 |
| C | -4.353839 | -0.460842 | -1.235746 |
| C | -5.037317 | -0.187376 | -0.044447 |
| C | -4.390780 | -0.431121 | 1.171008 |
| C | -3.088332 | -0.937459 | 1.223163 |
| C | -2.298754 | -1.201175 | -2.514845 |
| C | -2.378809 | -1.149888 | 2.531076 |
| C | -6.429035 | 0.387253 | -0.079909 |
| AU | 0.018743 | 1.078522 | 0.209517 |
| C | 1.638405 | -4.159873 | -0.047957 |
| C | -1.715772 | -4.136133 | -0.028869 |
| C | 2.241136 | -1.208948 | -2.532624 |
| C | 6.384737 | 0.371970 | -0.109914 |
| C | 2.363353 | -1.204129 | 2.512861 |
| H | 4.889454 | -0.267656 | 2.076501 |
| H | 4.788922 | -0.280854 | -2.221286 |
| H | -4.842076 | -0.265915 | -2.191685 |
| H | -4.908068 | -0.214027 | 2.106485 |
| H | 3.047432 | -1.042222 | 3.352283 |
| H | 1.517689 | -0.508524 | 2.616648 |
| H | 1.951695 | -2.219521 | 2.589319 |
| H | 2.884587 | -1.049638 | -3.404055 |
| H | 1.823485 | -2.223172 | -2.589607 |
| H | 1.394103 | -0.510872 | -2.598966 |
| H | 6.893159 | 0.147726 | -1.055214 |
| H | 6.337231 | 1.466744 | -0.011232 |
| H | 7.000472 | -0.003775 | 0.716597 |
| H | -2.944218 | -1.033576 | -3.383217 |
| H | -1.440589 | -0.517576 | -2.587017 |
| H | -1.898064 | -2.222335 | -2.572104 |
| H | -3.057801 | -0.985386 | 3.374131 |
| H | -1.963051 | -2.163318 | 2.610655 |
| H | -1.534788 | -0.450852 | 2.624863 |
| H | -6.904473 | 0.346238 | 0.906784 |
| H | -6.404803 | 1.439857 | -0.397819 |
| H | -7.064144 | -0.152148 | -0.794841 |
| H | 1.121927 | -5.124838 | -0.063717 |
| H | 2.283870 | -4.099935 | -0.935194 |
| H | 2.294413 | -4.134442 | 0.833090 |
| H | -1.212500 | -5.107864 | -0.053589 |
| H | -2.358609 | -4.103625 | 0.861657 |
| H | -2.373069 | -4.066140 | -0.906593 |
| C | 0.167612 | 3.654450 | 0.338178 |
| O | 0.112355 | 3.344230 | -0.876408 |
| O | 0.318071 | 4.693696 | 0.964254 |
| H | 0.036634 | 2.697681 | 1.127371 |

13 TS

32

| | | | |
|---|---|---|---|
| C | -3.382557 | -0.646099 | 0.197350 |
| N | -2.058599 | -1.052623 | 0.143474 |
| C | -1.234405 | 0.028199 | 0.097980 |
| N | -2.051710 | 1.114064 | 0.140231 |
| C | -3.378342 | 0.715656 | 0.195427 |
| C | -1.580620 | -2.451083 | 0.098970 |
| C | -2.155435 | -3.253484 | 1.261222 |
| AU | 0.726584 | 0.009696 | -0.096903 |
| C | -1.566996 | 2.509904 | 0.092288 |
| C | -1.864106 | 3.124589 | -1.273107 |
| C | -2.132276 | 3.315088 | 1.257552 |
| C | -1.870903 | -3.064463 | -1.268380 |
| H | -4.197134 | 1.420191 | 0.237291 |
| H | -4.205568 | -1.345580 | 0.241053 |
| H | -0.478679 | 2.412603 | 0.213145 |
| H | -1.681611 | 4.314330 | 1.260210 |
| H | -3.220017 | 3.439072 | 1.171551 |
| H | -1.907561 | 2.830305 | 2.214861 |
| H | -1.437784 | 4.133537 | -1.324022 |
| H | -1.421369 | 2.519526 | -2.073242 |
| H | -2.945862 | 3.200698 | -1.445834 |
| H | -0.492513 | -2.358797 | 0.226494 |
| H | -1.707185 | -4.253802 | 1.267156 |
| H | -1.934611 | -2.768637 | 2.219416 |
| H | -3.242957 | -3.374726 | 1.169081 |
| H | -1.446952 | -4.074529 | -1.316852 |
| H | -2.951655 | -3.137102 | -1.448414 |
| H | -1.420676 | -2.460389 | -2.065145 |
| C | 3.305825 | -0.098221 | -0.241167 |
| O | 2.990077 | -0.061966 | 0.974211 |
| O | 4.352514 | -0.209273 | -0.861514 |
| H | 2.347942 | 0.004890 | -1.029526 |

1-P INT

69

| | | | |
|---|---|---|---|
| C | -3.839420 | 1.478796 | 1.141643 |
| C | -2.784577 | 1.266377 | 0.243817 |
| C | -2.846139 | 1.846762 | -1.032583 |
| C | -3.964942 | 2.580873 | -1.418985 |
| C | -5.028253 | 2.759679 | -0.529356 |
| C | -4.959855 | 2.215927 | 0.753777 |
| P | -1.298365 | 0.351966 | 0.799892 |
| AU | 0.597976 | 1.429029 | 0.293340 |
| O | 2.360347 | 2.437903 | -0.115412 |
| C | 3.366885 | 2.269455 | 0.705947 |
| O | 4.494480 | 2.719008 | 0.515507 |
| C | -1.424683 | -1.280845 | 0.138425 |
| N | -0.370921 | -2.095464 | -0.001561 |
| C | -0.750647 | -3.565890 | -0.064995 |
| C | -2.207876 | -3.452535 | -0.523071 |
| C | -2.727272 | -2.064167 | -0.067828 |
| C | 0.999763 | -1.666987 | -0.165845 |
| C | 1.869707 | -1.566993 | 0.937922 |
| C | 3.198243 | -1.199147 | 0.691032 |
| C | 3.644701 | -0.915962 | -0.593237 |
| C | 2.760820 | -0.987916 | -1.664137 |
| C | 1.427533 | -1.362137 | -1.480337 |
| C | 1.428299 | -1.774063 | 2.376097 |
| C | 1.587773 | -0.485128 | 3.200502 |
| C | 0.485885 | -1.334926 | -2.673339 |
| C | 1.026619 | -2.130916 | -3.870647 |
| C | -3.552397 | -2.143320 | 1.229390 |
| C | -3.594772 | -1.463062 | -1.189742 |
| C | -0.633145 | -4.202777 | 1.323836 |
| C | 0.126249 | -4.331292 | -1.047664 |
| C | 0.195787 | 0.117832 | -3.087740 |
| C | 2.210871 | -2.916161 | 3.045345 |
| H | 3.887117 | -1.105350 | 1.529188 |
| H | 3.108076 | -0.729398 | -2.663196 |
| H | -0.467100 | -1.785707 | -2.369271 |
| H | -3.779951 | 1.069466 | 2.150043 |
| H | -5.775945 | 2.371187 | 1.458678 |
| H | -5.902023 | 3.335439 | -0.832745 |
| H | -4.006667 | 3.018209 | -2.416089 |
| H | -2.015725 | 1.715292 | -1.724011 |
| H | 3.142411 | 1.678730 | 1.623312 |
| H | -2.824698 | -4.264571 | -0.122793 |
| H | -4.426288 | -2.786911 | 1.063964 |
| H | -2.971470 | -2.554532 | 2.062112 |
| H | -3.908737 | -1.149820 | 1.520625 |
| H | -2.991769 | -1.235310 | -2.077758 |
| H | -4.355324 | -2.203299 | -1.472696 |
| H | -4.104193 | -0.549528 | -0.874005 |
| H | -1.041055 | -5.220276 | 1.280519 |
| H | 0.411492 | -4.272143 | 1.639492 |
| H | -1.190572 | -3.641789 | 2.081206 |
| H | 0.025147 | -3.946219 | -2.066322 |
| H | 1.182549 | -4.289759 | -0.756546 |
| H | -0.186115 | -5.382621 | -1.048666 |
| H | -2.245399 | -3.510433 | -1.619003 |
| H | 4.676400 | -0.608855 | -0.756449 |
| H | 0.275621 | -2.164170 | -4.670964 |
| H | 1.928240 | -1.660313 | -4.283592 |
| H | 1.285366 | -3.160982 | -3.596243 |
| H | -0.562021 | 0.149414 | -3.882076 |
| H | -0.158866 | 0.702217 | -2.229104 |
| H | 1.107337 | 0.604339 | -3.459282 |
| H | 1.201579 | -0.635839 | 4.217185 |
| H | 2.644418 | -0.197059 | 3.279265 |
| H | 1.042177 | 0.348610 | 2.740178 |
| H | 0.361666 | -2.029944 | 2.370048 |
| H | 1.799613 | -3.131200 | 4.040173 |
| H | 2.187271 | -3.839327 | 2.453137 |
| H | 3.265231 | -2.636583 | 3.172210 |

1-P PC

69

| | | | |
|---|---|---|---|
| C | 2.348191 | 0.176586 | -0.955954 |
| C | 1.749315 | 0.747629 | 0.184669 |
| C | 1.954656 | 0.216886 | 1.480332 |
| C | 2.812428 | -0.877992 | 1.608723 |
| C | 3.437454 | -1.437556 | 0.499732 |
| C | 3.198841 | -0.918723 | -0.765605 |
| N | 0.878209 | 1.894192 | 0.073808 |
| C | -0.448464 | 1.835160 | -0.106488 |
| C | -1.068658 | 3.218668 | 0.127583 |
| C | 0.142626 | 4.039783 | 0.642478 |
| C | 1.409486 | 3.312966 | 0.183857 |
| P | -1.250455 | 0.424003 | -0.808461 |
| AU | -0.399437 | -1.539312 | -0.126301 |
| C | -1.663003 | 3.809765 | -1.162942 |
| C | -2.151588 | 3.184232 | 1.222317 |
| C | 1.893509 | 3.800373 | -1.186526 |
| C | 2.556832 | 3.418714 | 1.181325 |
| C | 1.222545 | 0.730216 | 2.710208 |
| C | 0.113639 | -0.248367 | 3.131990 |
| C | 2.074613 | 0.646473 | -2.373990 |
| C | 3.360758 | 1.140814 | -3.057206 |
| C | -3.026715 | 0.564122 | -0.379805 |
| C | -3.519341 | 0.076977 | 0.840795 |
| C | -4.880756 | 0.143929 | 1.128214 |
| C | -5.769891 | 0.681964 | 0.193078 |
| C | -5.293818 | 1.133648 | -1.038454 |
| C | -3.929608 | 1.065543 | -1.327264 |
| C | 2.174179 | 0.997756 | 3.885902 |
| C | 1.433494 | -0.468535 | -3.217210 |
| H | 3.651963 | -1.394655 | -1.633297 |
| H | 2.972551 | -1.316258 | 2.592467 |
| H | 0.733960 | 1.676455 | 2.446909 |
| H | -3.562071 | 1.397785 | -2.298142 |
| H | -5.985184 | 1.531583 | -1.780760 |
| H | -6.834460 | 0.733081 | 0.418879 |
| H | -5.250792 | -0.227425 | 2.083469 |
| H | -2.827131 | -0.351287 | 1.564547 |
| H | 1.490248 | -4.949993 | 0.310239 |
| H | 0.111552 | 5.074020 | 0.281921 |
| H | -2.001853 | 4.835633 | -0.966599 |
| H | -0.932697 | 3.837196 | -1.978894 |
| H | -2.525050 | 3.223851 | -1.496219 |
| H | -1.807826 | 2.622029 | 2.099667 |
| H | -2.360847 | 4.215738 | 1.536334 |
| H | -3.083129 | 2.736477 | 0.867856 |
| H | 2.147345 | 4.865196 | -1.113557 |
| H | 2.791153 | 3.259742 | -1.500167 |
| H | 1.128058 | 3.681724 | -1.960232 |
| H | 2.241585 | 3.164052 | 2.196834 |
| H | 3.392821 | 2.769088 | 0.895700 |
| H | 2.917914 | 4.454420 | 1.187419 |
| H | 0.121094 | 4.065097 | 1.740154 |
| H | 4.089459 | -2.301707 | 0.618628 |
| H | 1.628028 | 1.467607 | 4.714443 |
| H | 2.607518 | 0.063080 | 4.264832 |
| H | 3.003225 | 1.656363 | 3.598505 |
| H | -0.445272 | 0.150535 | 3.989249 |
| H | -0.583091 | -0.427156 | 2.302729 |
| H | 0.539690 | -1.219971 | 3.415429 |
| H | 1.193168 | -0.088641 | -4.219280 |
| H | 2.113616 | -1.322546 | -3.325289 |
| H | 0.514409 | -0.839777 | -2.750957 |
| H | 1.360105 | 1.477694 | -2.325347 |
| H | 3.128814 | 1.584240 | -4.034611 |
| H | 3.890930 | 1.888989 | -2.454933 |
| H | 4.052597 | 0.304551 | -3.224115 |
| C | 1.099886 | -4.031501 | -0.184859 |
| O | 1.411016 | -3.763854 | -1.345737 |
| O | 0.322967 | -3.354930 | 0.612362 |

2-P PC

76

| | | | |
|---|---|---|---|
| C | -4.949890 | -0.624876 | -1.341110 |
| C | -3.499038 | -0.206778 | -1.049861 |
| C | -2.892302 | -1.048500 | 0.097623 |
| C | -3.700231 | -0.791890 | 1.413435 |
| C | -5.085221 | -0.185533 | 1.154862 |
| C | -5.763731 | -0.855237 | -0.043627 |
| C | -2.880534 | -2.543611 | -0.321603 |
| C | -1.549032 | -3.120660 | 0.163669 |
| N | -0.688838 | -1.871037 | 0.110096 |
| C | -1.401880 | -0.746466 | 0.255923 |
| C | 0.733518 | -1.932191 | -0.128319 |
| C | 1.646892 | -2.032971 | 0.938533 |
| C | 3.009173 | -2.109130 | 0.624980 |
| C | 3.449308 | -2.076429 | -0.691243 |
| C | 2.531239 | -1.953142 | -1.729285 |
| C | 1.159829 | -1.875682 | -1.476190 |
| C | 1.231007 | -2.009195 | 2.398246 |
| C | 1.856785 | -0.817752 | 3.141173 |
| C | 0.200229 | -1.649905 | -2.633849 |
| C | 0.438245 | -2.626465 | -3.794416 |
| P | -0.679559 | 0.759997 | 0.839600 |
| C | -1.838112 | 2.114081 | 0.407988 |
| C | -1.723566 | 2.796207 | -0.814047 |
| C | -2.576679 | 3.855766 | -1.113010 |
| C | -3.544301 | 4.259999 | -0.188542 |
| C | -3.638440 | 3.614389 | 1.044826 |
| C | -2.783306 | 2.552567 | 1.345392 |
| C | -1.641004 | -3.624202 | 1.609714 |
| C | -1.003394 | -4.227171 | -0.728550 |
| AU | 1.371597 | 1.148083 | 0.009006 |
| C | 0.280921 | -0.195711 | -3.128820 |
| C | 1.605931 | -3.323187 | 3.104346 |
| H | 3.737073 | -2.158047 | 1.432589 |
| H | 2.887986 | -1.892605 | -2.756241 |
| H | -0.821039 | -1.808836 | -2.266021 |
| H | -2.849474 | 2.061787 | 2.315657 |
| H | -4.373592 | 3.941604 | 1.779707 |
| H | -4.211380 | 5.088661 | -0.423636 |
| H | -2.485329 | 4.370184 | -2.069221 |
| H | -0.963639 | 2.484631 | -1.529717 |
| H | -3.734413 | -3.094347 | 0.085910 |
| H | -2.384748 | -4.429532 | 1.659101 |
| H | -0.683002 | -4.028097 | 1.949230 |
| H | -1.946651 | -2.828225 | 2.297390 |
| H | -0.977498 | -3.928804 | -1.780643 |
| H | 0.007022 | -4.522610 | -0.421117 |
| H | -1.656980 | -5.103288 | -0.638742 |
| H | -2.922661 | -2.618458 | -1.416652 |
| H | 4.515631 | -2.118747 | -0.908800 |
| H | -0.340834 | -2.501520 | -4.557840 |
| H | 1.406059 | -2.442159 | -4.278386 |
| H | 0.428108 | -3.670832 | -3.456995 |
| H | -0.449002 | -0.020971 | -3.930821 |
| H | 0.084157 | 0.505509 | -2.308108 |
| H | 1.284173 | 0.026428 | -3.516221 |
| H | 1.490087 | -0.787513 | 4.176058 |
| H | 2.950847 | -0.894945 | 3.163936 |
| H | 1.607624 | 0.129594 | 2.649699 |
| H | 0.141084 | -1.887106 | 2.437284 |
| H | 1.196694 | -3.341627 | 4.123149 |
| H | 1.234924 | -4.204256 | 2.565345 |
| H | 2.697415 | -3.420902 | 3.178568 |
| H | -3.120888 | -0.141615 | 2.079166 |
| H | -3.818147 | -1.747953 | 1.940593 |
| H | -5.699324 | -0.291903 | 2.059452 |
| H | -4.990022 | 0.894121 | 0.967200 |
| H | -6.787843 | -0.479316 | -0.165411 |
| H | -5.854160 | -1.933734 | 0.157930 |
| H | -4.969765 | -1.530557 | -1.962312 |
| H | -5.412831 | 0.169370 | -1.942401 |
| H | -3.472828 | 0.850034 | -0.774705 |
| H | -2.877193 | -0.311112 | -1.949953 |
| O | 3.221013 | 1.565386 | -0.866558 |
| C | 4.272948 | 1.359713 | -0.126090 |
| O | 4.311700 | 1.002289 | 1.051388 |
| H | 5.213119 | 1.551385 | -0.692590 |

3-P PC

88

| | | | |
|---|---|---|---|
| C | 4.852103 | -0.522810 | 1.012564 |
| C | 3.388750 | -0.071157 | 0.852427 |
| C | 2.654471 | -1.030483 | -0.134285 |
| C | 3.316528 | -0.882428 | -1.549728 |
| C | 4.755681 | -0.345246 | -1.521181 |
| C | 5.505113 | -0.936219 | -0.325034 |
| C | 2.701483 | -2.505675 | 0.344045 |
| C | 1.330594 | -3.123042 | 0.042295 |
| N | 0.460733 | -1.879302 | -0.022193 |
| C | 1.156093 | -0.752951 | -0.231903 |
| C | -0.981847 | -1.963419 | 0.061027 |
| C | -1.770174 | -2.098386 | -1.103048 |
| C | -3.156338 | -2.208297 | -0.941932 |
| C | -3.748907 | -2.171420 | 0.312403 |
| C | -2.959659 | -1.999133 | 1.442954 |
| C | -1.571017 | -1.883316 | 1.345936 |
| C | -1.219087 | -2.061009 | -2.519766 |
| C | -1.783482 | -0.862471 | -3.302938 |
| C | -0.771414 | -1.598681 | 2.604316 |
| C | -1.049303 | -2.618829 | 3.718015 |
| P | 0.391409 | 0.712567 | -0.869589 |
| C | 1.515745 | 2.134079 | -0.603086 |
| C | 1.282458 | 3.020960 | 0.459032 |
| C | 2.043134 | 4.181772 | 0.592405 |
| C | 3.046071 | 4.474236 | -0.333606 |
| C | 3.266119 | 3.613524 | -1.411212 |
| C | 2.494314 | 2.461621 | -1.554919 |
| C | 1.310557 | -3.845066 | -1.308171 |
| C | 0.864964 | -4.088860 | 1.126025 |
| C | 2.687358 | 0.241433 | 2.200291 |
| C | 3.295644 | 1.514101 | 2.808728 |
| C | 5.458925 | -0.628860 | -2.847888 |
| AU | -1.656518 | 1.114567 | -0.043344 |
| C | -1.029855 | -0.172696 | 3.117267 |
| C | -1.531405 | -3.357810 | -3.286937 |
| C | 2.699643 | -0.855279 | 3.274266 |
| H | -3.784832 | -2.290072 | -1.826784 |
| H | -3.430493 | -1.924526 | 2.421859 |
| H | 0.289404 | -1.655466 | 2.343123 |
| H | 2.645380 | 1.818093 | -2.419837 |
| H | 4.028161 | 3.849124 | -2.153596 |
| H | 3.642395 | 5.379761 | -0.226707 |
| H | 1.852950 | 4.858117 | 1.425131 |
| H | 0.494365 | 2.795593 | 1.176975 |
| H | 3.498342 | -3.062075 | -0.159287 |
| H | 2.019648 | -4.681318 | -1.272218 |
| H | 0.317183 | -4.253343 | -1.516169 |
| H | 1.598157 | -3.183796 | -2.132181 |
| H | 0.944062 | -3.651434 | 2.125024 |
| H | -0.171257 | -4.406554 | 0.958296 |
| H | 1.502710 | -4.980562 | 1.095326 |
| H | 2.897466 | -2.556482 | 1.417034 |
| H | -4.831667 | -2.242903 | 0.405595 |
| H | -0.366377 | -2.451001 | 4.561405 |
| H | -2.074234 | -2.517954 | 4.098380 |
| H | -0.922824 | -3.650888 | 3.369629 |
| H | -0.412841 | 0.030615 | 4.002745 |
| H | -0.794547 | 0.569939 | 2.342300 |
| H | -2.084910 | -0.038238 | 3.389545 |
| H | -1.319174 | -0.815068 | -4.297067 |
| H | -2.869002 | -0.955176 | -3.434202 |
| H | -1.593383 | 0.081275 | -2.781266 |
| H | -0.130755 | -1.936275 | -2.462144 |
| H | -1.030792 | -3.350694 | -4.264170 |
| H | -1.213513 | -4.255516 | -2.744128 |
| H | -2.611160 | -3.446641 | -3.466105 |
| H | 2.693944 | -0.242787 | -2.186144 |
| H | 3.321626 | -1.871601 | -2.029740 |
| H | 4.714820 | 0.747993 | -1.386921 |
| H | 6.557930 | -0.622248 | -0.347004 |
| H | 5.512393 | -2.033226 | -0.426934 |
| H | 4.920840 | -1.351326 | 1.730088 |
| H | 5.413580 | 0.310915 | 1.453796 |
| H | 3.411221 | 0.896715 | 0.336677 |
| H | 1.635573 | 0.463857 | 1.957616 |
| H | 2.685842 | 1.866708 | 3.651090 |
| H | 4.307854 | 1.323830 | 3.191963 |
| H | 3.354379 | 2.323467 | 2.071871 |
| H | 2.219231 | -0.477097 | 4.186433 |
| H | 2.159224 | -1.761813 | 2.985090 |
| H | 3.724509 | -1.144442 | 3.542910 |
| H | 6.463747 | -0.185919 | -2.864556 |
| H | 5.564097 | -1.712266 | -3.008314 |
| H | 4.893477 | -0.218445 | -3.696023 |
| O | -3.528164 | 1.529986 | 0.786128 |
| C | -4.554213 | 1.333423 | 0.007714 |
| O | -4.551140 | 0.987539 | -1.173884 |
| H | -5.513322 | 1.521578 | 0.542086 |

4-P PC

66

| | | | |
|---|---|---|---|
| C | -2.664124 | 1.585265 | 0.997469 |
| C | -1.779156 | 2.181024 | 0.079102 |
| C | -2.197052 | 2.608512 | -1.189519 |
| C | -3.525582 | 2.365465 | -1.551799 |
| C | -4.422546 | 1.732441 | -0.687048 |
| C | -3.981105 | 1.375498 | 0.591578 |
| N | -0.408159 | 2.320546 | 0.458419 |
| C | 0.591188 | 1.493669 | 0.080056 |
| N | 1.760437 | 1.946365 | 0.585632 |
| C | 1.545775 | 3.111368 | 1.479480 |
| C | 0.081683 | 3.481715 | 1.219975 |
| P | 0.228080 | 0.172177 | -1.127507 |
| AU | -1.321798 | -1.081511 | -0.049909 |
| C | 2.999011 | 1.240899 | 0.666193 |
| C | 3.077904 | 0.066331 | 1.433907 |
| C | 4.298674 | -0.609256 | 1.462143 |
| C | 5.425102 | -0.131075 | 0.785693 |
| C | 5.320927 | 1.072999 | 0.084360 |
| C | 4.115385 | 1.772749 | 0.004342 |
| C | 1.890821 | -0.455929 | 2.190188 |
| C | 6.701946 | -0.925256 | 0.772754 |
| C | 4.010997 | 3.029444 | -0.815948 |
| C | -1.245210 | 3.277923 | -2.143055 |
| C | -5.818597 | 1.390173 | -1.126017 |
| C | -2.200835 | 1.172668 | 2.368458 |
| C | 1.783048 | -0.784852 | -1.314272 |
| C | 2.839339 | -0.256387 | -2.072150 |
| C | 3.979713 | -1.014416 | -2.325703 |
| C | 4.065444 | -2.329274 | -1.861996 |
| C | 3.004299 | -2.878955 | -1.141330 |
| C | 1.871551 | -2.113119 | -0.869374 |
| H | 6.190911 | 1.459477 | -0.448779 |
| H | 4.368053 | -1.539966 | 2.026762 |
| H | -4.669524 | 0.877668 | 1.274416 |
| H | -3.861974 | 2.666121 | -2.545297 |
| H | 4.820898 | 3.078339 | -1.552376 |
| H | 3.051772 | 3.080467 | -1.346641 |
| H | 4.084449 | 3.930805 | -0.189003 |
| H | 2.186443 | -1.273879 | 2.855344 |
| H | 1.422554 | 0.329471 | 2.799320 |
| H | 1.110709 | -0.834940 | 1.511550 |
| H | 6.865309 | -1.440568 | 1.727521 |
| H | 6.656648 | -1.696723 | -0.011336 |
| H | 7.571014 | -0.289741 | 0.564259 |
| H | -2.898956 | 0.452973 | 2.807997 |
| H | -1.210625 | 0.701627 | 2.327243 |
| H | -2.134597 | 2.037714 | 3.045604 |
| H | -1.783357 | 3.668306 | -3.013682 |
| H | -0.712334 | 4.112949 | -1.668222 |
| H | -0.487725 | 2.563535 | -2.498146 |
| H | -6.062094 | 1.849142 | -2.091323 |
| H | -5.909363 | 0.298678 | -1.224851 |
| H | -6.562943 | 1.715570 | -0.386682 |
| H | 1.040097 | -2.537753 | -0.305745 |
| H | 3.058452 | -3.907465 | -0.785017 |
| H | 4.952411 | -2.926229 | -2.072766 |
| H | 4.800647 | -0.580724 | -2.895792 |
| H | 2.767381 | 0.759154 | -2.461693 |
| H | -0.499752 | 3.607510 | 2.139857 |
| H | -0.023198 | 4.392110 | 0.613098 |
| H | 1.727792 | 2.799844 | 2.517110 |
| H | 2.243229 | 3.917284 | 1.229514 |
| O | -2.705360 | -2.252038 | 1.008793 |
| C | -3.919636 | -2.318858 | 0.542693 |
| O | -4.366756 | -1.813837 | -0.487896 |
| H | -4.579530 | -2.921679 | 1.209167 |

5-P PC

84

| | | | |
|---|---|---|---|
| C | 1.338695 | -2.798944 | -0.347757 |
| C | 1.562075 | -1.600257 | -1.042042 |
| C | 2.668536 | -1.517686 | -1.901073 |
| C | 3.564176 | -2.578191 | -2.010489 |
| C | 3.346368 | -3.757017 | -1.294216 |
| C | 2.224012 | -3.869020 | -0.473014 |
| P | 0.288347 | -0.276220 | -1.039512 |
| AU | -1.629591 | -1.090317 | -0.176069 |
| C | 0.862509 | 1.047044 | 0.064207 |
| N | 0.013957 | 2.036026 | 0.428595 |
| C | 0.708380 | 3.170027 | 1.063031 |
| C | 2.184478 | 2.764895 | 0.968250 |
| N | 2.122549 | 1.400043 | 0.400606 |
| C | -1.414942 | 2.003005 | 0.359810 |
| C | -2.126978 | 1.615786 | 1.513954 |
| C | -3.522363 | 1.595485 | 1.436948 |
| C | -4.177873 | 1.932253 | 0.256356 |
| C | -3.451601 | 2.315372 | -0.866935 |
| C | -2.054823 | 2.366010 | -0.838772 |
| C | -1.422247 | 1.225175 | 2.801345 |
| C | -1.717345 | 2.241455 | 3.916448 |
| C | -1.271890 | 2.802997 | -2.062837 |
| C | -1.406341 | 4.320417 | -2.276444 |
| C | 3.277627 | 0.561731 | 0.401920 |
| C | 3.336962 | -0.488006 | 1.351503 |
| C | 4.485642 | -1.279607 | 1.370283 |
| C | 5.545374 | -1.029868 | 0.500278 |
| C | 5.456941 | 0.000282 | -0.424019 |
| C | 4.316554 | 0.812174 | -0.510357 |
| C | 2.192712 | -0.751978 | 2.321952 |
| C | 2.175082 | 0.272762 | 3.470564 |
| C | 4.325339 | 1.860898 | -1.621739 |
| C | 5.150427 | 3.094869 | -1.206992 |
| C | 2.204481 | -2.164739 | 2.914414 |
| C | 2.983232 | 2.275804 | -2.243236 |
| C | -1.685424 | 2.023302 | -3.317726 |
| C | -1.773793 | -0.200677 | 3.247114 |
| H | 6.271801 | 0.166625 | -1.129614 |
| H | 4.553306 | -2.109535 | 2.069574 |
| H | -4.101420 | 1.281500 | 2.304412 |
| H | -3.978121 | 2.564034 | -1.786976 |
| H | 4.891426 | 1.375819 | -2.432563 |
| H | 1.250994 | -0.648780 | 1.761443 |
| H | -0.342211 | 1.237080 | 2.608813 |
| H | -0.212367 | 2.581428 | -1.878138 |
| H | 0.456085 | -2.882214 | 0.286535 |
| H | 2.034538 | -4.792913 | 0.073291 |
| H | 4.042045 | -4.590275 | -1.388119 |
| H | 4.429140 | -2.487822 | -2.666583 |
| H | 2.829463 | -0.617602 | -2.490578 |
| H | 0.361616 | 3.289554 | 2.097811 |
| H | 0.488318 | 4.094802 | 0.516620 |
| H | 2.695309 | 2.744655 | 1.936711 |
| H | 2.749909 | 3.423762 | 0.299713 |
| H | -5.264302 | 1.878673 | 0.204726 |
| H | 6.432848 | -1.661209 | 0.530146 |
| H | -1.041783 | 2.300910 | -4.162941 |
| H | -1.592068 | 0.943810 | -3.149327 |
| H | -2.723917 | 2.238545 | -3.601954 |
| H | -0.797173 | 4.644466 | -3.131169 |
| H | -2.450958 | 4.593676 | -2.479385 |
| H | -1.081480 | 4.880224 | -1.389102 |
| H | -1.234896 | -0.452348 | 4.170538 |
| H | -2.848906 | -0.310870 | 3.437219 |
| H | -1.501051 | -0.926250 | 2.468783 |
| H | -1.145447 | 1.995920 | 4.821257 |
| H | -1.459085 | 3.264168 | 3.609947 |
| H | -2.784270 | 2.233713 | 4.177481 |
| H | 5.244470 | 3.791656 | -2.050499 |
| H | 4.678362 | 3.633791 | -0.374914 |
| H | 6.158062 | 2.805033 | -0.884270 |
| H | 3.177009 | 2.760490 | -3.208768 |
| H | 2.317370 | 1.423482 | -2.424403 |
| H | 2.434136 | 3.000666 | -1.631883 |
| H | 1.266372 | -2.339141 | 3.456049 |
| H | 2.296867 | -2.928758 | 2.135366 |
| H | 3.028023 | -2.296223 | 3.629916 |
| H | 1.347441 | 0.052969 | 4.157465 |
| H | 3.113575 | 0.225752 | 4.039509 |
| H | 2.048351 | 1.300559 | 3.117050 |
| O | -3.383316 | -1.911140 | 0.635769 |
| C | -4.470809 | -1.758677 | -0.062572 |
| O | -4.592017 | -1.235839 | -1.171154 |
| H | -5.359003 | -2.174730 | 0.467336 |

6-P PC
73

| | | | |
|---|---|---|---|
| C | -0.276284 | 3.576148 | -0.395029 |
| N | 0.144177 | 2.282247 | -0.046999 |
| C | -0.706576 | 1.203868 | 0.132617 |
| N | -2.000962 | 1.333717 | -0.321859 |
| C | -2.529852 | 2.569968 | -0.781106 |
| C | -1.776501 | 3.812635 | -0.348530 |
| C | 1.557361 | 2.099764 | 0.235009 |
| C | 2.405068 | 1.708884 | -0.812901 |
| C | 3.734985 | 1.433292 | -0.495548 |
| C | 4.219454 | 1.540101 | 0.813500 |
| C | 3.355232 | 2.002815 | 1.809025 |
| C | 2.016440 | 2.305379 | 1.541571 |
| C | 1.911886 | 1.622558 | -2.229034 |
| C | 1.113142 | 2.839143 | 2.617211 |
| C | 5.629913 | 1.132483 | 1.129821 |
| P | -0.038147 | -0.174490 | 1.052253 |
| C | -1.419791 | -1.246080 | 1.546708 |
| C | -2.165687 | -0.872144 | 2.676075 |
| C | -3.144694 | -1.722928 | 3.184793 |
| C | -3.366965 | -2.964141 | 2.586084 |
| C | -2.598822 | -3.358498 | 1.486979 |
| C | -1.622343 | -2.510399 | 0.973590 |
| C | -2.813293 | 0.164074 | -0.601138 |
| C | -2.456683 | -0.681962 | -1.658866 |
| C | -3.266150 | -1.798501 | -1.894012 |
| C | -4.401660 | -2.063887 | -1.128024 |
| C | -4.750284 | -1.158947 | -0.119932 |
| C | -3.980156 | -0.030566 | 0.153881 |
| O | -3.550360 | 2.594717 | -1.439468 |
| C | -1.255887 | -0.433209 | -2.521502 |
| C | -5.222643 | -3.303005 | -1.357089 |
| C | -4.415255 | 0.949295 | 1.204763 |
| C | -2.146833 | 4.081755 | 1.139937 |
| C | -2.162708 | 5.006656 | -1.221209 |
| O | 0.537796 | 4.451358 | -0.627757 |
| AU | 1.536251 | -1.222229 | -0.160119 |
| H | -5.644520 | -1.339222 | 0.477743 |
| H | -2.994774 | -2.475737 | -2.704640 |
| H | 4.404884 | 1.103056 | -1.289535 |
| H | 3.726536 | 2.126723 | 2.826914 |
| H | -5.093022 | 0.472545 | 1.920284 |
| H | -3.564584 | 1.357673 | 1.762165 |
| H | -4.945477 | 1.794408 | 0.742402 |
| H | -1.344072 | -0.973115 | -3.470018 |
| H | -1.128835 | 0.633166 | -2.745356 |
| H | -0.334769 | -0.779286 | -2.021652 |
| H | -4.981020 | -3.773703 | -2.316954 |
| H | -5.030378 | -4.040439 | -0.563101 |
| H | -6.297247 | -3.080528 | -1.337638 |
| H | 2.677410 | 1.183077 | -2.876251 |
| H | 1.010268 | 1.005197 | -2.306513 |
| H | 1.664459 | 2.622324 | -2.612722 |
| H | 1.634445 | 2.866718 | 3.579532 |
| H | 0.782721 | 3.860155 | 2.380608 |
| H | 0.218448 | 2.211837 | 2.734210 |
| H | 5.919158 | 1.432149 | 2.143853 |
| H | 5.716672 | 0.039100 | 1.049028 |
| H | 6.340631 | 1.571928 | 0.417429 |
| H | -1.015830 | -2.815867 | 0.121889 |
| H | -2.761700 | -4.333039 | 1.028428 |
| H | -4.129122 | -3.632407 | 2.985165 |
| H | -3.726842 | -1.420722 | 4.054576 |
| H | -1.971046 | 0.085932 | 3.158370 |
| H | -1.921779 | 4.818800 | -2.273377 |
| H | -3.237945 | 5.192110 | -1.140116 |
| H | -1.612712 | 5.892752 | -0.891046 |
| H | -3.227606 | 4.244089 | 1.218766 |
| H | -1.864622 | 3.246984 | 1.790971 |
| H | -1.625816 | 4.984400 | 1.478004 |
| O | 2.866895 | -2.261220 | -1.384722 |
| C | 4.077094 | -2.402720 | -0.917851 |
| O | 4.514439 | -2.025419 | 0.168560 |
| H | 4.732392 | -2.933805 | -1.643742 |

7-P PC

74

| | | | |
|---|---|---|---|
| C | -0.053260 | 3.630633 | 0.563101 |
| N | -0.340719 | 2.304691 | 0.186073 |
| C | 0.615697 | 1.309447 | -0.016456 |
| N | 1.841718 | 1.519465 | 0.485540 |
| C | 2.161835 | 2.764575 | 1.200431 |
| C | 1.425693 | 3.974535 | 0.637028 |
| C | -1.720034 | 2.006347 | -0.158352 |
| C | -2.549597 | 1.462425 | 0.834291 |
| C | -3.835990 | 1.075288 | 0.461381 |
| C | -4.297785 | 1.222554 | -0.851901 |
| C | -3.463855 | 1.840692 | -1.786902 |
| C | -2.168926 | 2.256135 | -1.460593 |
| C | -2.079015 | 1.323316 | 2.254610 |
| C | -1.299930 | 2.948391 | -2.471685 |
| C | -5.649877 | 0.693184 | -1.237792 |
| P | 0.088308 | -0.045616 | -1.113298 |
| C | 1.613168 | -0.943391 | -1.579763 |
| C | 2.345421 | -0.447292 | -2.670259 |
| C | 3.436326 | -1.154963 | -3.172613 |
| C | 3.788804 | -2.382422 | -2.609172 |
| C | 3.040468 | -2.903142 | -1.550272 |
| C | 1.956171 | -2.192893 | -1.041961 |
| C | 2.796975 | 0.458202 | 0.697421 |
| C | 2.543149 | -0.502306 | 1.689448 |
| C | 3.505140 | -1.497716 | 1.890857 |
| C | 4.688427 | -1.545289 | 1.153289 |
| C | 4.924510 | -0.538727 | 0.211751 |
| C | 4.002653 | 0.479315 | -0.024625 |
| C | 1.284406 | -0.503946 | 2.502612 |
| C | 5.673666 | -2.666498 | 1.334723 |
| C | 4.305285 | 1.553459 | -1.029823 |
| C | 1.889787 | 4.303560 | -0.797328 |
| C | 1.646023 | 5.174055 | 1.561868 |
| O | -0.956046 | 4.432123 | 0.731741 |
| AU | -1.300535 | -1.344791 | 0.105233 |
| H | 5.845956 | -0.554310 | -0.371746 |
| H | 3.312746 | -2.260940 | 2.646157 |
| H | -4.484651 | 0.618984 | 1.209070 |
| H | -3.821782 | 1.993782 | -2.805799 |
| H | 5.036081 | 1.196668 | -1.763106 |
| H | 3.405909 | 1.869599 | -1.567481 |
| H | 4.731534 | 2.444854 | -0.544658 |
| H | 1.402659 | -1.120030 | 3.399992 |
| H | 0.991549 | 0.506426 | 2.815047 |
| H | 0.442443 | -0.915415 | 1.915217 |
| H | 5.487735 | -3.218663 | 2.263261 |
| H | 5.596960 | -3.378466 | 0.499205 |
| H | 6.706561 | -2.294954 | 1.349110 |
| H | -2.839183 | 0.823078 | 2.863060 |
| H | -1.158313 | 0.731593 | 2.315995 |
| H | -1.873997 | 2.309100 | 2.695253 |
| H | -1.800066 | 2.991146 | -3.444874 |
| H | -1.077487 | 3.977554 | -2.156975 |
| H | -0.346240 | 2.418182 | -2.601852 |
| H | -5.911933 | 0.967532 | -2.266187 |
| H | -5.646710 | -0.403157 | -1.155535 |
| H | -6.432723 | 1.070336 | -0.565730 |
| H | 1.366164 | -2.597728 | -0.220524 |
| H | 3.303269 | -3.868391 | -1.118597 |
| H | 4.636125 | -2.941298 | -3.005199 |
| H | 4.002805 | -0.754252 | -4.012852 |
| H | 2.049448 | 0.495622 | -3.131439 |
| H | 1.297728 | 4.959555 | 2.579829 |
| H | 2.714202 | 5.421852 | 1.601602 |
| H | 1.096864 | 6.045506 | 1.191641 |
| H | 2.957861 | 4.553795 | -0.789956 |
| H | 1.735415 | 3.462512 | -1.482966 |
| H | 1.333083 | 5.165621 | -1.182142 |
| H | 3.246024 | 2.912975 | 1.139048 |
| H | 1.909977 | 2.621852 | 2.264739 |
| O | -2.531384 | -2.569833 | 1.278129 |
| C | -3.744430 | -2.762802 | 0.843072 |
| O | -4.248888 | -2.349602 | -0.201437 |
| H | -4.337376 | -3.384773 | 1.551738 |

8-P INT

38

| | | | |
|---|---|---|---|
| C | -0.368709 | 2.932819 | 0.123400 |
| C | -1.186593 | 1.993858 | -0.522224 |
| C | -2.528303 | 2.324876 | -0.773855 |
| C | -3.047771 | 3.549632 | -0.357304 |
| C | -2.226104 | 4.476563 | 0.290732 |
| C | -0.884374 | 4.166300 | 0.521936 |
| P | -0.518952 | 0.407794 | -1.174689 |
| AU | 1.398225 | -0.104069 | -0.083834 |
| C | -1.662360 | -0.752660 | -0.342626 |
| N | -2.146757 | -1.881056 | -0.927597 |
| C | -2.981222 | -2.567776 | -0.051342 |
| C | -3.026158 | -1.830277 | 1.104801 |
| N | -2.198185 | -0.726760 | 0.907666 |
| C | -1.773003 | -2.348387 | -2.256111 |
| C | -1.847795 | 0.248085 | 1.933146 |
| C | -3.769417 | -2.066472 | 2.369386 |
| C | -3.632942 | -3.851894 | -0.415057 |
| H | 0.671967 | 2.679653 | 0.328308 |
| H | -0.236258 | 4.883185 | 1.025827 |
| H | -2.630022 | 5.436699 | 0.609980 |
| H | -4.093917 | 3.787043 | -0.549983 |
| H | -3.171058 | 1.615268 | -1.297635 |
| H | 4.214311 | -0.295908 | -0.861100 |
| H | -4.400839 | -2.954934 | 2.269319 |
| H | -3.090117 | -2.232565 | 3.217031 |
| H | -4.419052 | -1.216786 | 2.619072 |
| H | -4.235791 | -4.217949 | 0.421616 |
| H | -4.295229 | -3.740309 | -1.284735 |
| H | -2.888848 | -4.623315 | -0.657487 |
| H | -2.657634 | -2.726177 | -2.776757 |
| H | -1.347420 | -1.501102 | -2.804195 |
| H | -1.020902 | -3.142335 | -2.180005 |
| H | -2.059327 | -0.182677 | 2.914013 |
| H | -0.776826 | 0.465900 | 1.847401 |
| H | -2.414776 | 1.175772 | 1.800691 |
| C | 4.295971 | -0.610867 | 0.205644 |
| O | 5.387656 | -0.911919 | 0.683380 |
| O | 3.171329 | -0.629040 | 0.876788 |

8-P PC

38

| | | | |
|---|---|---|---|
| N | -2.129501 | -0.625869 | 0.844464 |
| C | -1.603701 | -0.683939 | -0.409910 |
| N | -2.260988 | -1.697062 | -1.036202 |
| C | -3.190010 | -2.281344 | -0.181335 |
| C | -3.114066 | -1.596052 | 1.005649 |
| P | -0.300230 | 0.345556 | -1.182375 |
| AU | 1.453292 | -0.196438 | 0.158956 |
| O | 3.007566 | -0.759350 | 1.460627 |
| C | 4.207697 | -0.772573 | 0.954770 |
| C | -2.006503 | -2.134615 | -2.402992 |
| C | -4.044274 | -3.424558 | -0.593186 |
| C | -3.872804 | -1.770872 | 2.270950 |
| C | -1.667973 | 0.281322 | 1.886676 |
| C | -0.946153 | 1.988168 | -0.645993 |
| C | -0.097269 | 2.966786 | -0.107876 |
| C | -0.585714 | 4.235680 | 0.205275 |
| C | -1.931821 | 4.544773 | -0.001632 |
| C | -2.784848 | 3.578352 | -0.542203 |
| C | -2.292929 | 2.317255 | -0.875197 |
| H | 0.947830 | 2.719251 | 0.080766 |
| H | 0.087674 | 4.982515 | 0.625797 |
| H | -2.314597 | 5.532182 | 0.254412 |
| H | -3.836131 | 3.810902 | -0.713103 |
| H | -2.965054 | 1.577070 | -1.312951 |
| H | -4.630830 | -2.550640 | 2.147550 |
| H | -3.215752 | -2.070327 | 3.099382 |
| H | -4.384300 | -0.844513 | 2.564857 |
| H | -4.673624 | -3.744175 | 0.243039 |
| H | -4.705129 | -3.157100 | -1.429665 |
| H | -3.438404 | -4.285291 | -0.907915 |
| H | -2.956869 | -2.263325 | -2.930349 |
| H | -1.404551 | -1.363687 | -2.894827 |
| H | -1.453815 | -3.080757 | -2.402258 |
| H | -1.908479 | -0.146284 | 2.862106 |
| H | -0.579200 | 0.389214 | 1.781828 |
| H | -2.137195 | 1.265167 | 1.777807 |
| H | 4.969583 | -1.081678 | 1.706750 |
| O | 4.543860 | -0.495046 | -0.197833 |

9-P PC

64

| | | | |
|---|---|---|---|
| C | 2.670687 | -0.159770 | 2.214766 |
| C | 1.738383 | 0.578117 | 1.468016 |
| C | 2.014862 | 1.926614 | 1.193798 |
| C | 3.203085 | 2.511385 | 1.629925 |
| C | 4.137940 | 1.757108 | 2.340038 |
| C | 3.868442 | 0.417175 | 2.629038 |
| P | 0.100448 | -0.169180 | 1.097800 |
| AU | -1.120381 | 1.341997 | -0.083401 |
| C | 0.474309 | -1.410617 | -0.191953 |
| N | 1.581508 | -1.764758 | -0.910321 |
| C | 1.296678 | -2.862927 | -1.720446 |
| C | -0.002866 | -3.193024 | -1.506091 |
| N | -0.492335 | -2.305353 | -0.562132 |
| C | 2.865169 | -1.120085 | -0.889099 |
| C | 3.002931 | 0.129933 | -1.506988 |
| C | 4.251520 | 0.749475 | -1.423108 |
| C | 5.332141 | 0.149531 | -0.770441 |
| C | 5.158205 | -1.120857 | -0.213317 |
| C | 3.926773 | -1.776461 | -0.253614 |
| C | 1.854389 | 0.784187 | -2.214216 |
| C | 3.738842 | -3.112352 | 0.411417 |
| C | 6.636626 | 0.882023 | -0.621337 |
| C | -1.839534 | -2.262210 | -0.059429 |
| C | -2.108869 | -2.848141 | 1.184202 |
| C | -3.408214 | -2.721293 | 1.682806 |
| C | -4.401636 | -2.030731 | 0.980291 |
| C | -4.087848 | -1.479163 | -0.266664 |
| C | -2.810000 | -1.589774 | -0.816687 |
| C | -1.035213 | -3.553713 | 1.964153 |
| C | -5.782183 | -1.862897 | 1.556922 |
| C | -2.496459 | -1.017690 | -2.170726 |
| H | 2.047460 | -3.283996 | -2.374043 |
| H | -0.628197 | -3.964442 | -1.932967 |
| H | 5.992365 | -1.600524 | 0.300260 |
| H | 4.376250 | 1.734922 | -1.873217 |
| H | -4.842448 | -0.918634 | -0.816931 |
| H | -3.644334 | -3.163616 | 2.651631 |
| H | 4.582207 | -3.332649 | 1.074421 |
| H | 2.816002 | -3.136855 | 1.006173 |
| H | 3.666713 | -3.927893 | -0.322039 |
| H | 2.190344 | 1.665553 | -2.769899 |
| H | 1.366658 | 0.093465 | -2.915269 |
| H | 1.074537 | 1.107136 | -1.501379 |
| H | 6.842909 | 1.519267 | -1.490114 |
| H | 6.598689 | 1.536144 | 0.263329 |
| H | 7.475891 | 0.189454 | -0.485940 |
| H | -3.227109 | -0.243904 | -2.423608 |
| H | -1.499047 | -0.559481 | -2.192163 |
| H | -2.520412 | -1.800429 | -2.943921 |
| H | -1.464907 | -4.090893 | 2.816451 |
| H | -0.485067 | -4.272334 | 1.342112 |
| H | -0.306126 | -2.824781 | 2.349832 |
| H | -5.890842 | -2.405767 | 2.503087 |
| H | -5.994788 | -0.801146 | 1.745153 |
| H | -6.549646 | -2.225749 | 0.859766 |
| H | 1.286805 | 2.511750 | 0.631272 |
| H | 3.401654 | 3.559270 | 1.406081 |
| H | 5.068641 | 2.212954 | 2.676591 |
| H | 4.589434 | -0.177830 | 3.188942 |
| H | 2.453294 | -1.196284 | 2.474916 |
| O | -2.120932 | 2.883110 | -1.103203 |
| C | -3.363320 | 2.718977 | -1.450298 |
| O | -4.084647 | 1.737878 | -1.255748 |
| H | -3.758963 | 3.612887 | -1.985301 |

10-P PC

82

| | | | |
|---|---|---|---|
| C | -1.068105 | 2.509796 | -1.466943 |
| C | -1.166535 | 1.112866 | -1.545734 |
| C | -2.292930 | 0.545580 | -2.159144 |
| C | -3.329965 | 1.348668 | -2.623875 |
| C | -3.232668 | 2.739183 | -2.526114 |
| C | -2.094077 | 3.315263 | -1.961286 |
| P | 0.217801 | 0.017159 | -1.053160 |
| AU | 1.661232 | 1.291594 | 0.160022 |
| C | -0.499084 | -0.941798 | 0.326761 |
| N | 0.180868 | -2.038579 | 0.790345 |
| C | -0.505494 | -2.634792 | 1.836004 |
| C | -1.633772 | -1.908721 | 2.028772 |
| N | -1.631052 | -0.874656 | 1.092432 |
| C | 1.328353 | -2.610820 | 0.132350 |
| C | 2.608938 | -2.371627 | 0.656073 |
| C | 3.686967 | -2.968573 | -0.006826 |
| C | 3.486809 | -3.758248 | -1.136198 |
| C | 2.202899 | -3.970009 | -1.632172 |
| C | 1.090199 | -3.398523 | -1.007997 |
| C | 2.815142 | -1.527530 | 1.899817 |
| C | 4.093736 | -0.685333 | 1.835594 |
| C | -0.306903 | -3.653375 | -1.544656 |
| C | -0.427046 | -3.273226 | -3.026540 |
| C | -2.770551 | 0.001259 | 0.950008 |
| C | -2.664748 | 1.343075 | 1.358644 |
| C | -3.811614 | 2.132485 | 1.229993 |
| C | -5.005494 | 1.601483 | 0.747821 |
| C | -5.077733 | 0.267916 | 0.364617 |
| C | -3.955486 | -0.563044 | 0.444197 |
| C | -1.373042 | 1.906153 | 1.919000 |
| C | -1.324460 | 3.434461 | 1.929611 |
| C | -4.053288 | -2.009897 | -0.016243 |
| C | -4.696478 | -2.145210 | -1.404695 |
| C | -1.069575 | 1.359183 | 3.325212 |
| C | -4.830948 | -2.860564 | 1.003206 |
| C | -0.725380 | -5.111171 | -1.294573 |
| C | 2.812363 | -2.406018 | 3.163326 |
| H | -2.439203 | -2.011526 | 2.740498 |
| H | -0.130243 | -3.519205 | 2.329506 |
| H | -6.012831 | -0.132008 | -0.024478 |
| H | -3.767673 | 3.181249 | 1.512777 |
| H | 4.697680 | -2.804580 | 0.361504 |
| H | 2.061575 | -4.587827 | -2.518120 |
| H | -3.035105 | -2.416686 | -0.088419 |
| H | -0.551701 | 1.570846 | 1.259678 |
| H | 1.966747 | -0.827656 | 1.961960 |
| H | -1.007831 | -3.013363 | -0.992913 |
| H | -0.190944 | 2.960279 | -0.999240 |
| H | -2.007792 | 4.399346 | -1.889855 |
| H | -4.040364 | 3.370068 | -2.895585 |
| H | -4.213277 | 0.891659 | -3.069604 |
| H | -2.352051 | -0.538071 | -2.264073 |
| H | -5.885053 | 2.238976 | 0.663137 |
| H | -0.301869 | 3.764592 | 2.143702 |
| H | -1.602550 | 3.850959 | 0.954569 |
| H | -1.993573 | 3.856737 | 2.692911 |
| H | -0.112007 | 1.764863 | 3.676479 |
| H | -1.853056 | 1.660590 | 4.034427 |
| H | -0.994877 | 0.265854 | 3.342890 |
| H | -4.659239 | -3.192537 | -1.731626 |
| H | -5.751081 | -1.841310 | -1.390251 |
| H | -4.181386 | -1.533919 | -2.153175 |
| H | -4.860185 | -3.910899 | 0.684189 |
| H | -4.380631 | -2.819157 | 2.001934 |
| H | -5.865379 | -2.501092 | 1.089762 |
| H | 4.341867 | -4.209701 | -1.638793 |
| H | 4.110824 | 0.027783 | 2.667827 |
| H | 4.994403 | -1.309913 | 1.907135 |
| H | 4.139249 | -0.107032 | 0.905110 |
| H | 2.935642 | -1.782494 | 4.058579 |
| H | 1.879264 | -2.972296 | 3.269862 |
| H | 3.641334 | -3.126438 | 3.129670 |
| H | -1.464498 | -3.395591 | -3.365032 |
| H | -0.130453 | -2.227878 | -3.178740 |
| H | 0.207160 | -3.909622 | -3.658018 |
| H | -1.752517 | -5.279815 | -1.644739 |
| H | -0.064958 | -5.805823 | -1.830984 |
| H | -0.680019 | -5.359152 | -0.226118 |
| O | 3.058184 | 2.429820 | 1.254741 |
| C | 2.799514 | 3.702110 | 1.302499 |
| O | 1.857486 | 4.301207 | 0.775981 |
| H | 3.554363 | 4.256968 | 1.905094 |

11-P PC

82

| | | | |
|---|---|---|---|
| C | 0.985491 | -3.446224 | -0.414097 |
| C | -0.300605 | -2.902067 | -0.541166 |
| C | -1.269201 | -3.389803 | -1.436064 |
| C | -0.905419 | -4.474292 | -2.239177 |
| C | 0.362508 | -5.042043 | -2.139248 |
| C | 1.295427 | -4.533175 | -1.239826 |
| N | -0.702145 | -1.806774 | 0.310698 |
| C | -0.675594 | -0.482850 | -0.045647 |
| N | -1.437296 | 0.168658 | 0.882191 |
| C | -1.920316 | -0.743674 | 1.829544 |
| C | -1.459252 | -1.976492 | 1.468448 |
| P | 0.360096 | 0.045182 | -1.464630 |
| AU | 2.196910 | 0.451047 | -0.180584 |
| C | -1.772009 | 1.571829 | 0.896724 |
| C | -0.991235 | 2.434408 | 1.680337 |
| C | -1.392034 | 3.774643 | 1.733464 |
| C | -2.504392 | 4.216275 | 1.027132 |
| C | -3.236620 | 3.335623 | 0.233293 |
| C | -2.891731 | 1.984800 | 0.146840 |
| CL | -2.929707 | -0.311481 | 3.118908 |
| CL | -1.757171 | -3.490127 | 2.165463 |
| C | 0.235565 | 1.963212 | 2.443347 |
| C | -0.076603 | 1.761095 | 3.936018 |
| C | -3.667509 | 1.026003 | -0.748450 |
| C | -4.632993 | 0.115538 | 0.032655 |
| C | -2.646875 | -2.764982 | -1.564260 |
| C | -3.762926 | -3.785712 | -1.300729 |
| C | 1.997882 | -2.905004 | 0.575712 |
| C | 2.037169 | -3.762574 | 1.850857 |
| C | -0.333173 | 1.680139 | -1.962090 |
| C | -1.179053 | 1.697189 | -3.081143 |
| C | -1.638344 | 2.905015 | -3.608537 |
| C | -1.254101 | 4.111996 | -3.023836 |
| C | -0.393501 | 4.107028 | -1.923325 |
| C | 0.075724 | 2.902244 | -1.407790 |
| C | 1.425289 | 2.918941 | 2.288731 |
| C | -4.454308 | 1.743346 | -1.849938 |
| C | -2.807763 | -2.085203 | -2.932145 |
| C | 3.395990 | -2.771762 | -0.041629 |
| H | -4.081443 | 3.710485 | -0.339259 |
| H | -0.814139 | 4.481117 | 2.326362 |
| H | 2.284662 | -4.982804 | -1.178685 |
| H | -1.623666 | -4.877903 | -2.951851 |
| H | -2.923716 | 0.383860 | -1.244611 |
| H | 0.546069 | 0.996789 | 2.016293 |
| H | 1.676372 | -1.888903 | 0.857175 |
| H | -2.745276 | -1.980968 | -0.802909 |
| H | 0.770497 | 2.897760 | -0.571416 |
| H | -0.078147 | 5.046077 | -1.470156 |
| H | -1.612549 | 5.055870 | -3.433407 |
| H | -2.291795 | 2.901322 | -4.480685 |
| H | -1.470473 | 0.757281 | -3.550072 |
| H | -2.796166 | 5.264905 | 1.077363 |
| H | 2.299965 | 2.494229 | 2.795584 |
| H | 1.705317 | 3.067792 | 1.239805 |
| H | 1.220189 | 3.898575 | 2.739656 |
| H | 0.820315 | 1.406146 | 4.459905 |
| H | -0.384767 | 2.711599 | 4.393463 |
| H | -0.879580 | 1.034550 | 4.096439 |
| H | -4.834632 | 1.004839 | -2.566940 |
| H | -5.322318 | 2.275963 | -1.437755 |
| H | -3.830716 | 2.459706 | -2.393599 |
| H | -5.198944 | -0.509227 | -0.670936 |
| H | -4.123262 | -0.553421 | 0.730780 |
| H | -5.350190 | 0.718602 | 0.604896 |
| H | 0.628049 | -5.887552 | -2.773139 |
| H | 4.054500 | -2.229850 | 0.647757 |
| H | 3.845959 | -3.753270 | -0.242694 |
| H | 3.357790 | -2.206610 | -0.981430 |
| H | 2.749313 | -3.335485 | 2.568721 |
| H | 1.053480 | -3.818272 | 2.332114 |
| H | 2.357972 | -4.787411 | 1.617250 |
| H | -3.767457 | -1.554424 | -2.986028 |
| H | -1.997492 | -1.364788 | -3.101120 |
| H | -2.779929 | -2.822034 | -3.745812 |
| H | -4.743254 | -3.293447 | -1.343705 |
| H | -3.755943 | -4.587305 | -2.051152 |
| H | -3.651434 | -4.245555 | -0.310836 |
| O | 3.916675 | 0.620038 | 1.028858 |
| C | 4.560561 | 1.744488 | 0.933590 |
| O | 4.288813 | 2.709792 | 0.214951 |
| H | 5.448413 | 1.773445 | 1.605582 |

12-P PC

70

| | | | |
|---|---|---|---|
| C | 3.312078 | -1.178639 | 0.598860 |
| C | 2.356890 | -1.987382 | -0.033109 |
| C | 2.629085 | -2.718424 | -1.196544 |
| C | 3.909662 | -2.609033 | -1.744096 |
| C | 4.892195 | -1.804243 | -1.158126 |
| C | 4.577397 | -1.103707 | 0.012435 |
| N | 1.040053 | -2.083820 | 0.534750 |
| C | -0.020217 | -1.329013 | 0.133061 |
| N | -1.069574 | -1.761376 | 0.883516 |
| C | -0.667812 | -2.781216 | 1.758847 |
| C | 0.667667 | -2.980865 | 1.539856 |
| P | 0.186608 | -0.065499 | -1.178021 |
| AU | 0.654725 | 1.728505 | 0.145784 |
| C | -2.403195 | -1.229292 | 0.848863 |
| C | -2.700950 | -0.120685 | 1.649974 |
| C | -3.997670 | 0.396559 | 1.573071 |
| C | -4.967624 | -0.168791 | 0.741153 |
| C | -4.633875 | -1.303858 | -0.005980 |
| C | -3.356015 | -1.860465 | 0.037593 |
| C | -1.618813 | -3.416220 | 2.705115 |
| C | 1.639355 | -3.927142 | 2.143444 |
| C | -1.659004 | 0.502659 | 2.530607 |
| C | -6.332469 | 0.450239 | 0.611551 |
| C | -3.014274 | -3.091197 | -0.754391 |
| C | 1.562333 | -3.549083 | -1.851124 |
| C | 6.253948 | -1.670057 | -1.786051 |
| C | 2.972106 | -0.396275 | 1.835001 |
| C | -1.537473 | 0.126403 | -1.805373 |
| C | -1.953969 | -0.742675 | -2.826411 |
| C | -3.207696 | -0.598141 | -3.419729 |
| C | -4.049922 | 0.442956 | -3.024746 |
| C | -3.631615 | 1.334943 | -2.034934 |
| C | -2.386593 | 1.178638 | -1.429812 |
| H | -5.382329 | -1.757633 | -0.656484 |
| H | -4.248226 | 1.273312 | 2.171647 |
| H | 5.334512 | -0.473426 | 0.480578 |
| H | 4.142466 | -3.163730 | -2.654049 |
| H | -3.749077 | -3.253843 | -1.549496 |
| H | -2.022920 | -3.012406 | -1.214987 |
| H | -3.006442 | -3.985702 | -0.113765 |
| H | -2.095377 | 1.277964 | 3.168283 |
| H | -1.163393 | -0.240646 | 3.169719 |
| H | -0.866832 | 0.969912 | 1.915457 |
| H | -6.535448 | 1.152662 | 1.428260 |
| H | -6.408104 | 1.005684 | -0.335402 |
| H | -7.121071 | -0.313330 | 0.602411 |
| H | 3.855850 | 0.115342 | 2.229037 |
| H | 2.211533 | 0.369661 | 1.600841 |
| H | 2.553279 | -1.031569 | 2.626616 |
| H | 1.979242 | -4.151853 | -2.664515 |
| H | 1.068426 | -4.221175 | -1.136522 |
| H | 0.783633 | -2.894495 | -2.272013 |
| H | 6.441041 | -2.471025 | -2.510939 |
| H | 6.340335 | -0.710764 | -2.316992 |
| H | 7.047108 | -1.693445 | -1.027645 |
| H | -2.060641 | 1.867375 | -0.651243 |
| H | -4.280275 | 2.154828 | -1.727655 |
| H | -5.025499 | 0.564841 | -3.494604 |
| H | -3.520341 | -1.288671 | -4.203002 |
| H | -1.283758 | -1.534932 | -3.162960 |
| H | -1.089575 | -4.123256 | 3.351542 |
| H | -2.106515 | -2.664637 | 3.340098 |
| H | -2.412614 | -3.961722 | 2.177090 |
| H | 1.150419 | -4.528858 | 2.915817 |
| H | 2.054405 | -4.607396 | 1.387310 |
| H | 2.484212 | -3.396144 | 2.602037 |
| O | 1.042219 | 3.328898 | 1.465476 |
| C | 1.815146 | 4.280168 | 1.029057 |
| O | 2.358829 | 4.367615 | -0.073617 |
| H | 1.960089 | 5.077604 | 1.793949 |

13-P PC

44

| | | | |
|---|---|---|---|
| N | 1.850118 | -2.206195 | 0.197886 |
| C | 1.402178 | -0.928100 | 0.044376 |
| N | 1.679412 | -0.586963 | -1.243781 |
| C | 2.295863 | -1.641373 | -1.886611 |
| C | 2.391750 | -2.662628 | -0.988434 |
| P | 0.575131 | 0.019446 | 1.382924 |
| AU | -1.468124 | 0.220282 | 0.418651 |
| C | 1.282935 | 0.680821 | -1.903042 |
| C | 0.330313 | 0.384809 | -3.057371 |
| C | 1.651042 | -3.007951 | 1.428608 |
| C | 2.814658 | -3.965521 | 1.649153 |
| C | 1.518573 | 1.594437 | 1.203780 |
| C | 2.915142 | 1.564986 | 1.356103 |
| C | 3.661729 | 2.740901 | 1.306574 |
| C | 3.022292 | 3.970780 | 1.126867 |
| C | 1.632739 | 4.012880 | 0.997033 |
| C | 0.885701 | 2.834457 | 1.032080 |
| C | 2.521648 | 1.476397 | -2.300039 |
| C | 0.292641 | -3.703832 | 1.365289 |
| H | 2.609245 | -1.585153 | -2.919596 |
| H | 2.800527 | -3.656583 | -1.096843 |
| H | -0.197288 | 2.865637 | 0.908275 |
| H | 1.125922 | 4.967687 | 0.857378 |
| H | 3.604755 | 4.890812 | 1.091688 |
| H | 4.745482 | 2.698555 | 1.415607 |
| H | 3.421473 | 0.611017 | 1.513089 |
| H | 0.731831 | 1.226364 | -1.130425 |
| H | 0.000286 | 1.334638 | -3.495290 |
| H | 0.820236 | -0.197905 | -3.848791 |
| H | -0.557597 | -0.154153 | -2.702008 |
| H | 2.211150 | 2.437863 | -2.725348 |
| H | 3.154607 | 1.673258 | -1.427258 |
| H | 3.112698 | 0.945415 | -3.058499 |
| H | 1.627819 | -2.258832 | 2.232692 |
| H | 0.108754 | -4.242223 | 2.302773 |
| H | -0.511769 | -2.971142 | 1.223511 |
| H | 0.261024 | -4.423849 | 0.537028 |
| H | 2.698564 | -4.440285 | 2.630076 |
| H | 2.835604 | -4.765713 | 0.897625 |
| H | 3.776102 | -3.438242 | 1.634108 |
| O | -3.419252 | 0.382425 | -0.353153 |
| C | -3.629313 | -0.190437 | -1.500849 |
| O | -2.824664 | -0.827817 | -2.187359 |
| H | -4.678758 | -0.052432 | -1.846978 |

1 INT

57

| | | | |
|---|---|---|---|
| C | 0.720433 | -1.161555 | 0.098206 |
| N | 1.611536 | -0.198274 | 0.060900 |
| C | 3.064052 | -0.651767 | 0.128814 |
| C | 2.876228 | -2.131511 | 0.525225 |
| C | 1.404653 | -2.503044 | 0.208192 |
| C | 1.221749 | 1.194893 | -0.044334 |
| C | 0.972723 | 1.910445 | 1.145104 |
| C | 0.605259 | 3.254961 | 1.028087 |
| C | 0.467852 | 3.859838 | -0.216829 |
| C | 0.659080 | 3.114117 | -1.374409 |
| C | 1.024735 | 1.764283 | -1.318306 |
| C | 0.963592 | 1.262180 | 2.519773 |
| C | 1.839481 | 2.014139 | 3.531021 |
| C | 1.062356 | 0.961097 | -2.607502 |
| C | -0.360920 | 0.808799 | -3.172423 |
| C | 3.731187 | -0.478712 | -1.237196 |
| C | 3.835560 | 0.151839 | 1.170979 |
| C | 1.244367 | -3.227376 | -1.144099 |
| C | 0.765083 | -3.340740 | 1.323556 |
| AU | -1.192780 | -0.760029 | 0.000572 |
| C | 2.000015 | 1.573101 | -3.657252 |
| C | -0.482709 | 1.140501 | 3.031744 |
| H | -2.511649 | 1.722571 | -0.136187 |
| H | 0.493132 | 3.578604 | -2.345490 |
| H | 0.397927 | 3.829327 | 1.929944 |
| H | 1.354448 | 0.242300 | 2.422531 |
| H | 3.587326 | -2.777356 | -0.001690 |
| H | 1.735320 | -4.207736 | -1.091929 |
| H | 1.693894 | -2.659080 | -1.967431 |
| H | 0.182438 | -3.377549 | -1.374035 |
| H | 0.803966 | -2.813401 | 2.285155 |
| H | 1.303167 | -4.292256 | 1.427299 |
| H | -0.285762 | -3.554912 | 1.092745 |
| H | 4.773810 | -0.810240 | -1.162208 |
| H | 3.732263 | 0.573906 | -1.542086 |
| H | 3.240182 | -1.072674 | -2.014894 |
| H | 3.424000 | 0.013859 | 2.175461 |
| H | 3.833355 | 1.221545 | 0.928417 |
| H | 4.876874 | -0.192479 | 1.180209 |
| H | 3.056643 | -2.244926 | 1.601855 |
| H | 0.176953 | 4.907531 | -0.285124 |
| H | 1.859692 | 1.474469 | 4.486810 |
| H | 1.444347 | 3.019334 | 3.727779 |
| H | 2.870635 | 2.123580 | 3.173307 |
| H | -0.501904 | 0.634180 | 4.005840 |
| H | -1.096233 | 0.560542 | 2.327873 |
| H | -0.943638 | 2.130190 | 3.147896 |
| H | -0.346823 | 0.189774 | -4.079265 |
| H | -0.791424 | 1.785800 | -3.428344 |
| H | -1.020741 | 0.328340 | -2.436460 |
| H | 1.423168 | -0.046222 | -2.373518 |
| H | 2.051011 | 0.924441 | -4.541421 |
| H | 3.017682 | 1.706367 | -3.269596 |
| H | 1.636906 | 2.555250 | -3.987680 |
| C | -3.414695 | 1.065732 | -0.143348 |
| O | -4.538439 | 1.552271 | -0.198721 |
| O | -3.169244 | -0.225337 | -0.089668 |

1 PC

57

| | | | |
|---|---|---|---|
| C | 1.241933 | 1.706556 | -1.241780 |
| C | 1.351901 | 1.074022 | 0.011773 |
| C | 1.285429 | 1.788282 | 1.223868 |
| C | 1.215575 | 3.182938 | 1.151132 |
| C | 1.187248 | 3.839040 | -0.075248 |
| C | 1.176151 | 3.104281 | -1.255543 |
| N | 1.516543 | -0.363193 | 0.068469 |
| C | 0.507724 | -1.201174 | 0.073710 |
| C | 1.020138 | -2.627052 | 0.087644 |
| C | 2.530294 | -2.469144 | 0.392548 |
| C | 2.900897 | -1.006063 | 0.081807 |
| AU | -1.380762 | -0.677894 | 0.024113 |
| O | -3.404372 | -0.284579 | -0.006084 |
| C | -3.773091 | 0.957081 | -0.189682 |
| C | 3.568853 | -0.830796 | -1.283166 |
| C | 3.786058 | -0.380783 | 1.156130 |
| C | 0.751684 | -3.234296 | -1.305034 |
| C | 0.299610 | -3.456963 | 1.158011 |
| C | 1.147073 | 1.103720 | 2.572698 |
| C | -0.312584 | 1.215410 | 3.049064 |
| C | 1.049765 | 0.945469 | -2.542041 |
| C | 2.024329 | 1.379616 | -3.644946 |
| C | 2.116046 | 1.653131 | 3.627584 |
| C | -0.408737 | 1.101034 | -3.008867 |
| H | -4.881365 | 1.058159 | -0.183143 |
| H | 1.086523 | 3.621896 | -2.209564 |
| H | 1.158840 | 3.761995 | 2.071773 |
| H | 1.358847 | 0.035816 | 2.440326 |
| H | 3.144130 | -3.166584 | -0.188402 |
| H | 1.118712 | -4.268818 | -1.327009 |
| H | 1.255858 | -2.671821 | -2.100267 |
| H | -0.323936 | -3.236543 | -1.519653 |
| H | 0.418907 | -3.007746 | 2.152221 |
| H | 0.718443 | -4.471971 | 1.182983 |
| H | -0.773255 | -3.524155 | 0.939072 |
| H | 4.557076 | -1.305257 | -1.253941 |
| H | 3.709474 | 0.231177 | -1.514175 |
| H | 2.991231 | -1.294595 | -2.089132 |
| H | 3.369533 | -0.523331 | 2.157981 |
| H | 3.931571 | 0.691699 | 0.978363 |
| H | 4.768520 | -0.867421 | 1.124440 |
| H | 2.710792 | -2.672220 | 1.455910 |
| H | 1.135216 | 4.926750 | -0.110333 |
| H | 2.024249 | 1.079464 | 4.559344 |
| H | 1.895678 | 2.702132 | 3.864321 |
| H | 3.158424 | 1.597093 | 3.289649 |
| H | -0.450839 | 0.669972 | 3.992221 |
| H | -0.997570 | 0.797689 | 2.297619 |
| H | -0.589890 | 2.265511 | 3.210793 |
| H | -0.593907 | 0.477419 | -3.893672 |
| H | -0.625331 | 2.145564 | -3.268814 |
| H | -1.106514 | 0.805217 | -2.213358 |
| H | 1.212207 | -0.120074 | -2.343632 |
| H | 1.877642 | 0.762078 | -4.540954 |
| H | 3.070713 | 1.282610 | -3.329887 |
| H | 1.856701 | 2.425517 | -3.933853 |
| O | -3.047421 | 1.935255 | -0.355948 |

2 PC

64

| | | | |
|---|---|---|---|
| C | 1.973888 | -1.605973 | -1.164177 |
| C | 1.236975 | -1.557345 | 0.035160 |
| C | 1.857282 | -1.536959 | 1.299250 |
| C | 3.248144 | -1.686051 | 1.337485 |
| C | 3.992674 | -1.814924 | 0.170438 |
| C | 3.361368 | -1.746828 | -1.067298 |
| N | -0.206502 | -1.507300 | -0.048071 |
| C | -1.048847 | -2.777852 | -0.067513 |
| C | -2.420359 | -2.194016 | -0.456998 |
| C | -2.369059 | -0.682890 | -0.125930 |
| C | -0.889128 | -0.387131 | -0.080417 |
| C | -2.965247 | -0.322345 | 1.279418 |
| C | -4.362478 | 0.310804 | 1.189629 |
| C | -5.202107 | -0.359553 | 0.100051 |
| C | -4.566186 | -0.143585 | -1.293939 |
| C | -3.062596 | 0.167005 | -1.205529 |
| C | -1.038969 | -3.426267 | 1.318340 |
| C | -0.519038 | -3.776315 | -1.092737 |
| AU | -0.086055 | 1.401303 | -0.042056 |
| C | 1.110599 | -1.220781 | 2.583894 |
| C | 1.382553 | -2.232008 | 3.705270 |
| C | 1.339241 | -1.375121 | -2.524542 |
| C | 1.666191 | 0.052381 | -2.998471 |
| C | 1.465883 | 0.207420 | 3.036360 |
| C | 1.761982 | -2.414705 | -3.570253 |
| H | 3.757124 | -1.672918 | 2.300314 |
| H | 3.958167 | -1.781427 | -1.977627 |
| H | 0.249782 | -1.432496 | -2.410218 |
| H | -3.234358 | -2.713176 | 0.059744 |
| H | -1.660834 | -4.329253 | 1.290194 |
| H | -0.022412 | -3.723480 | 1.600630 |
| H | -1.437394 | -2.758575 | 2.089004 |
| H | -0.543900 | -3.366495 | -2.107168 |
| H | 0.506903 | -4.084107 | -0.857261 |
| H | -1.155789 | -4.669303 | -1.071367 |
| H | -2.572622 | -2.325053 | -1.536738 |
| H | 5.074808 | -1.930036 | 0.224530 |
| H | 1.224715 | -2.241912 | -4.512166 |
| H | 2.835958 | -2.348299 | -3.787627 |
| H | 1.550406 | -3.437673 | -3.234521 |
| H | 1.156447 | 0.267277 | -3.947201 |
| H | 1.346988 | 0.791508 | -2.249959 |
| H | 2.747266 | 0.174466 | -3.148083 |
| H | 0.872413 | 0.487847 | 3.916899 |
| H | 2.530017 | 0.281188 | 3.296840 |
| H | 1.266692 | 0.929374 | 2.232077 |
| H | 0.036123 | -1.232011 | 2.369121 |
| H | 0.775060 | -1.988870 | 4.586904 |
| H | 1.145153 | -3.258205 | 3.398112 |
| H | 2.435995 | -2.210054 | 4.014044 |
| H | -2.282883 | 0.366075 | 1.797696 |
| H | -3.009649 | -1.239612 | 1.881941 |
| H | -4.855135 | 0.232869 | 2.168101 |
| H | -4.270940 | 1.385504 | 0.972949 |
| H | -6.230592 | 0.024412 | 0.112922 |
| H | -5.272845 | -1.436199 | 0.319360 |
| H | -4.731294 | -1.030857 | -1.920714 |
| H | -5.060112 | 0.691979 | -1.808831 |
| H | -2.900568 | 1.225156 | -0.957108 |
| H | -2.575802 | 0.001229 | -2.176620 |
| O | 0.630588 | 3.335053 | -0.019885 |
| C | 1.929745 | 3.487258 | 0.006768 |
| O | 2.785335 | 2.604457 | 0.020852 |
| H | 2.214253 | 4.563029 | 0.016588 |

3 PC

76

| | | | |
|---|---|---|---|
| C | -2.189015 | -1.436332 | 1.363572 |
| C | -1.567077 | -1.551233 | 0.104797 |
| C | -2.302413 | -1.654590 | -1.091231 |
| C | -3.693881 | -1.760061 | -0.985794 |
| C | -4.328115 | -1.731804 | 0.251483 |
| C | -3.582104 | -1.544078 | 1.410500 |
| N | -0.120913 | -1.535692 | 0.050539 |
| C | 0.693270 | -2.820304 | 0.137276 |
| C | 2.125534 | -2.255816 | 0.255891 |
| C | 2.052629 | -0.756689 | -0.135399 |
| C | 0.579779 | -0.438530 | -0.102499 |
| C | 2.535714 | -0.508613 | -1.603575 |
| C | 4.010778 | -0.066845 | -1.711056 |
| C | 4.831662 | -0.672776 | -0.570145 |
| C | 4.331345 | -0.182695 | 0.805559 |
| C | 2.843951 | 0.209704 | 0.788773 |
| C | 2.236931 | 0.483417 | 2.190274 |
| C | 2.352789 | -0.635542 | 3.230867 |
| C | 4.574841 | -0.418865 | -3.086981 |
| C | 0.490146 | -3.644285 | -1.135781 |
| C | 0.277500 | -3.655138 | 1.345129 |
| AU | -0.180622 | 1.361198 | -0.300688 |
| C | -1.670924 | -1.512996 | -2.465346 |
| C | -2.072749 | -2.634673 | -3.431571 |
| C | -1.423325 | -1.064184 | 2.620644 |
| C | -1.686330 | 0.415108 | 2.953456 |
| C | -2.027695 | -0.133762 | -3.047975 |
| C | -1.747279 | -1.965791 | 3.817975 |
| C | 2.817506 | 1.784335 | 2.758984 |
| H | -4.290807 | -1.839162 | -1.893446 |
| H | -4.090882 | -1.454773 | 2.369268 |
| H | -0.352348 | -1.157789 | 2.408537 |
| H | 2.806564 | -2.809225 | -0.398225 |
| H | 1.084050 | -4.563008 | -1.058201 |
| H | -0.561554 | -3.928949 | -1.255229 |
| H | 0.814487 | -3.101276 | -2.029874 |
| H | 0.435834 | -3.118989 | 2.285857 |
| H | -0.776193 | -3.952329 | 1.280383 |
| H | 0.888272 | -4.565815 | 1.366893 |
| H | 2.491453 | -2.376439 | 1.278757 |
| H | -5.412798 | -1.817318 | 0.309688 |
| H | -1.118250 | -1.691650 | 4.675089 |
| H | -2.794256 | -1.855530 | 4.129662 |
| H | -1.574514 | -3.024618 | 3.588376 |
| H | -1.060950 | 0.733127 | 3.798590 |
| H | -1.467113 | 1.056104 | 2.088083 |
| H | -2.739351 | 0.572767 | 3.221848 |
| H | -1.521628 | 0.016848 | -4.010890 |
| H | -3.110663 | -0.043952 | -3.206646 |
| H | -1.718722 | 0.666753 | -2.361172 |
| H | -0.581786 | -1.538337 | -2.347558 |
| H | -1.545338 | -2.515769 | -4.387180 |
| H | -1.833127 | -3.626780 | -3.029150 |
| H | -3.149587 | -2.610137 | -3.643884 |
| H | 1.896792 | 0.255578 | -2.069562 |
| H | 2.387665 | -1.436376 | -2.175830 |
| H | 4.047243 | 1.029102 | -1.601046 |
| H | 5.894245 | -0.422970 | -0.696727 |
| H | 4.767537 | -1.770365 | -0.633664 |
| H | 4.513165 | -0.959193 | 1.561217 |
| H | 4.915157 | 0.693166 | 1.118858 |
| H | 2.761700 | 1.180606 | 0.272927 |
| H | 1.163140 | 0.669926 | 2.022443 |
| H | 2.306087 | 2.061159 | 3.690357 |
| H | 3.887866 | 1.682609 | 2.986972 |
| H | 2.696614 | 2.612434 | 2.047790 |
| H | 1.922900 | -0.296919 | 4.183018 |
| H | 1.813038 | -1.544428 | 2.943405 |
| H | 3.398450 | -0.911332 | 3.424548 |
| H | 5.596347 | -0.033196 | -3.206383 |
| H | 4.605962 | -1.509998 | -3.225089 |
| H | 3.956930 | 0.003616 | -3.891229 |
| O | -0.828026 | 3.302908 | -0.548627 |
| C | -2.084170 | 3.548960 | -0.282683 |
| O | -2.932434 | 2.755072 | 0.120344 |
| H | -2.338293 | 4.617448 | -0.463263 |

4 PC
54

| | | | |
|---|---|---|---|
| C | -3.134943 | -0.991662 | -1.322099 |
| C | -2.466594 | -1.334375 | -0.137258 |
| C | -3.066047 | -1.194577 | 1.123627 |
| C | -4.376019 | -0.711766 | 1.175131 |
| C | -5.080152 | -0.368590 | 0.015929 |
| C | -4.444479 | -0.515470 | -1.221617 |
| N | -1.125429 | -1.821538 | -0.225377 |
| C | -0.035989 | -1.057197 | -0.063762 |
| N | 1.060923 | -1.820467 | -0.172224 |
| C | 0.741508 | -3.224947 | -0.514237 |
| C | -0.794628 | -3.253181 | -0.398047 |
| C | 2.398684 | -1.315123 | -0.160474 |
| C | 3.152335 | -1.432361 | 1.017886 |
| C | 4.461955 | -0.949302 | 1.009963 |
| C | 5.019434 | -0.356044 | -0.129517 |
| C | 4.231469 | -0.244505 | -1.278212 |
| C | 2.915039 | -0.714293 | -1.318275 |
| C | 2.540745 | -2.014103 | 2.262795 |
| C | 2.062137 | -0.524548 | -2.541794 |
| C | 6.426503 | 0.181234 | -0.101403 |
| AU | -0.045637 | 0.882404 | 0.258405 |
| C | -2.298012 | -1.499671 | 2.379952 |
| C | -6.476652 | 0.188727 | 0.101604 |
| C | -2.438642 | -1.089615 | -2.651462 |
| H | -4.975895 | -0.239337 | -2.133669 |
| H | -4.855229 | -0.592400 | 2.148101 |
| H | 5.058800 | -1.028805 | 1.920149 |
| H | 4.647357 | 0.228333 | -2.169097 |
| H | -3.084247 | -0.722015 | -3.456066 |
| H | -1.515216 | -0.494298 | -2.649712 |
| H | -2.155620 | -2.124418 | -2.890710 |
| H | -2.951044 | -1.450656 | 3.257838 |
| H | -1.834617 | -2.494685 | 2.349164 |
| H | -1.484461 | -0.771093 | 2.514075 |
| H | -7.019233 | -0.220924 | 0.962529 |
| H | -6.448811 | 1.282170 | 0.221042 |
| H | -7.050129 | -0.027819 | -0.807938 |
| H | 3.254463 | -1.991517 | 3.093403 |
| H | 1.648115 | -1.444457 | 2.556789 |
| H | 2.223058 | -3.056188 | 2.117673 |
| H | 2.673624 | -0.228507 | -3.400866 |
| H | 1.508573 | -1.434849 | -2.806576 |
| H | 1.319282 | 0.268150 | -2.363270 |
| H | 6.806686 | 0.357312 | -1.114532 |
| H | 6.465888 | 1.137407 | 0.440966 |
| H | 7.107706 | -0.510959 | 0.410393 |
| H | 1.235226 | -3.912017 | 0.181763 |
| H | 1.094972 | -3.447183 | -1.530213 |
| H | -1.143537 | -3.823482 | 0.473316 |
| H | -1.283877 | -3.654562 | -1.292453 |
| O | -0.082505 | 2.893792 | 0.700713 |
| C | 0.245448 | 3.712509 | -0.260811 |
| O | 0.561148 | 3.424739 | -1.415823 |
| H | 0.212299 | 4.772907 | 0.074573 |

5 PC

72

| | | | |
|---|---|---|---|
| C | 3.028541 | 1.520879 | -0.229369 |
| C | 2.353048 | 0.617158 | -1.067334 |
| C | 2.959454 | -0.543922 | -1.579356 |
| C | 4.292699 | -0.785468 | -1.234201 |
| C | 4.988256 | 0.098312 | -0.411661 |
| C | 4.361592 | 1.236973 | 0.087600 |
| N | 0.990777 | 0.877832 | -1.419664 |
| C | -0.064618 | 0.585493 | -0.646563 |
| N | -1.179770 | 1.062808 | -1.228530 |
| C | -0.909937 | 1.548376 | -2.602941 |
| C | 0.617740 | 1.688829 | -2.597797 |
| C | -2.509288 | 0.718139 | -0.814072 |
| C | -3.263499 | 1.641206 | -0.063094 |
| C | -4.573756 | 1.278110 | 0.271053 |
| C | -5.110587 | 0.052330 | -0.114570 |
| C | -4.335362 | -0.849676 | -0.834357 |
| C | -3.020054 | -0.537709 | -1.196229 |
| C | -2.734142 | 2.955918 | 0.494543 |
| C | -2.039129 | 3.891203 | -0.504887 |
| C | -2.191810 | -1.558373 | -1.957256 |
| C | -2.790844 | -1.839682 | -3.344579 |
| AU | 0.020213 | -0.410083 | 1.047805 |
| C | 2.169547 | -1.545908 | -2.399752 |
| C | 2.931311 | -2.035869 | -3.637418 |
| C | 2.340046 | 2.736183 | 0.363313 |
| C | 3.071389 | 4.038789 | 0.008185 |
| C | 1.730967 | -2.717583 | -1.504855 |
| C | 2.178910 | 2.573337 | 1.883288 |
| C | -2.023444 | -2.849118 | -1.142021 |
| C | -1.844418 | 2.710116 | 1.725284 |
| H | -5.176861 | 1.973733 | 0.855236 |
| H | -4.752570 | -1.816232 | -1.115021 |
| H | 4.789775 | -1.682310 | -1.601351 |
| H | 4.912286 | 1.911085 | 0.743038 |
| H | -3.629372 | 3.490058 | 0.847519 |
| H | -1.185242 | -1.146885 | -2.106594 |
| H | 1.256222 | -1.044060 | -2.748969 |
| H | 1.329267 | 2.794891 | -0.062717 |
| H | 1.094140 | 1.296381 | -3.502273 |
| H | 0.948282 | 2.727237 | -2.448907 |
| H | -1.258930 | 0.798584 | -3.327336 |
| H | -1.431519 | 2.490349 | -2.791348 |
| H | 6.025702 | -0.108189 | -0.149874 |
| H | -6.134294 | -0.203274 | 0.157708 |
| H | 1.629515 | 3.425178 | 2.305019 |
| H | 1.623849 | 1.653349 | 2.114678 |
| H | 3.157389 | 2.514971 | 2.378565 |
| H | 2.520221 | 4.903347 | 0.401115 |
| H | 4.080186 | 4.060068 | 0.441625 |
| H | 3.170108 | 4.156808 | -1.079045 |
| H | 1.115413 | -3.426997 | -2.073588 |
| H | 2.607090 | -3.255843 | -1.118325 |
| H | 1.146047 | -2.366905 | -0.643122 |
| H | 2.288820 | -2.692600 | -4.238581 |
| H | 3.253670 | -1.196758 | -4.267874 |
| H | 3.823189 | -2.613483 | -3.360230 |
| H | -1.936176 | 4.888969 | -0.058542 |
| H | -1.030782 | 3.541203 | -0.753845 |
| H | -2.616847 | 3.996425 | -1.432257 |
| H | -1.553086 | 3.665866 | 2.181528 |
| H | -2.369510 | 2.110290 | 2.478704 |
| H | -0.931298 | 2.165160 | 1.451015 |
| H | -1.409717 | -3.568955 | -1.698993 |
| H | -1.527492 | -2.654545 | -0.182066 |
| H | -2.994876 | -3.320065 | -0.940493 |
| H | -2.146181 | -2.529681 | -3.904995 |
| H | -3.783883 | -2.301167 | -3.259265 |
| H | -2.901905 | -0.917259 | -3.930678 |
| O | 0.128633 | -1.347433 | 2.877522 |
| C | -0.000923 | -2.645400 | 2.889423 |
| O | -0.181432 | -3.389469 | 1.925094 |
| H | 0.071620 | -3.054308 | 3.921479 |

6 PC

61

| | | | |
|---|---|---|---|
| C | -3.092015 | -0.587332 | -1.096873 |
| C | -2.403983 | -0.732654 | 0.113123 |
| C | -2.860595 | -0.194173 | 1.320291 |
| C | -4.058215 | 0.525002 | 1.289081 |
| C | -4.779477 | 0.704585 | 0.105262 |
| C | -4.282066 | 0.141692 | -1.076137 |
| N | -1.162417 | -1.489655 | 0.110854 |
| C | 0.002125 | -0.798494 | -0.009031 |
| N | 1.171672 | -1.481853 | 0.108185 |
| C | 1.275440 | -2.885820 | 0.303835 |
| C | 0.012107 | -3.684003 | 0.038976 |
| C | -1.255970 | -2.894481 | 0.306842 |
| O | 2.342218 | -3.389137 | 0.583491 |
| C | 2.407631 | -0.715625 | 0.106840 |
| C | 2.866943 | -0.178622 | 1.315047 |
| C | 4.056721 | 0.550760 | 1.280846 |
| C | 4.769988 | 0.743008 | 0.092539 |
| C | 4.271260 | 0.181893 | -1.087545 |
| C | 3.086685 | -0.558650 | -1.105252 |
| AU | -0.003913 | 1.143302 | -0.304084 |
| O | -2.318498 | -3.404947 | 0.589593 |
| C | 0.011535 | -3.977314 | -1.495102 |
| C | 0.017323 | -4.993472 | 0.831661 |
| C | 2.090985 | -0.348303 | 2.589853 |
| C | 6.033511 | 1.561279 | 0.083564 |
| C | 2.543116 | -1.139726 | -2.379732 |
| C | -2.076038 | -0.348413 | 2.591771 |
| C | -6.051535 | 1.509308 | 0.092439 |
| C | -2.549606 | -1.169787 | -2.371233 |
| H | -4.832544 | 0.274202 | -2.008602 |
| H | -4.433025 | 0.957599 | 2.217347 |
| H | 4.432850 | 0.981844 | 2.209403 |
| H | 4.815336 | 0.324071 | -2.022211 |
| H | -3.203077 | -0.929728 | -3.216047 |
| H | -1.548603 | -0.769883 | -2.590076 |
| H | -2.463273 | -2.263599 | -2.310395 |
| H | -2.648094 | 0.024005 | 3.447557 |
| H | -1.812270 | -1.396452 | 2.785237 |
| H | -1.139927 | 0.227061 | 2.537591 |
| H | -6.480948 | 1.593233 | 1.097676 |
| H | -5.860315 | 2.528561 | -0.274516 |
| H | -6.801155 | 1.059211 | -0.570751 |
| H | 2.661968 | 0.029435 | 3.444013 |
| H | 1.146229 | 0.213344 | 2.542634 |
| H | 1.843752 | -1.400811 | 2.780922 |
| H | 3.191178 | -0.891387 | -3.226320 |
| H | 2.464838 | -2.234356 | -2.323000 |
| H | 1.538340 | -0.746079 | -2.592625 |
| H | 6.655939 | 1.326554 | -0.788019 |
| H | 5.797827 | 2.635041 | 0.045536 |
| H | 6.625142 | 1.389514 | 0.991690 |
| H | 0.017435 | -4.799902 | 1.910265 |
| H | -0.874361 | -5.575684 | 0.581967 |
| H | 0.912691 | -5.569385 | 0.580605 |
| H | -0.879888 | -4.564630 | -1.741872 |
| H | 0.008066 | -3.056182 | -2.088445 |
| H | 0.906209 | -4.558948 | -1.743499 |
| O | -0.010485 | 3.145952 | -0.754607 |
| C | -0.012136 | 3.948374 | 0.275944 |
| O | -0.010956 | 3.629857 | 1.464647 |
| H | -0.014951 | 5.015419 | -0.034959 |

7 PC

62

| | | | |
|---|---|---|---|
| C | -3.177855 | 0.629138 | 0.932511 |
| C | -2.424917 | 0.940164 | -0.208811 |
| C | -2.872827 | 0.651764 | -1.505932 |
| C | -4.128444 | 0.052764 | -1.640649 |
| C | -4.917734 | -0.264386 | -0.531560 |
| C | -4.423421 | 0.026922 | 0.744576 |
| N | -1.145797 | 1.598861 | -0.049044 |
| C | -0.036494 | 0.869474 | -0.032843 |
| N | 1.170548 | 1.536782 | -0.013468 |
| C | 1.330217 | 2.925832 | 0.190997 |
| C | 0.067711 | 3.672192 | 0.605196 |
| C | -1.135562 | 3.067347 | -0.117999 |
| O | 2.431653 | 3.438662 | 0.116813 |
| C | 2.374129 | 0.738193 | -0.165814 |
| C | 2.853067 | 0.493779 | -1.457928 |
| C | 4.002846 | -0.288216 | -1.585390 |
| C | 4.664364 | -0.810634 | -0.468339 |
| C | 4.153797 | -0.531275 | 0.803144 |
| C | 3.006194 | 0.245378 | 0.981126 |
| AU | -0.085526 | -1.104194 | -0.035202 |
| C | -0.054844 | 3.509390 | 2.135603 |
| C | 0.183833 | 5.152574 | 0.233066 |
| C | 2.135640 | 1.035274 | -2.661983 |
| C | 5.880966 | -1.682215 | -0.635313 |
| C | 2.445685 | 0.509806 | 2.349293 |
| C | -2.007426 | 0.900097 | -2.710176 |
| C | -6.251768 | -0.940845 | -0.700443 |
| C | -2.639739 | 0.881984 | 2.311675 |
| H | -5.017920 | -0.229875 | 1.622469 |
| H | -4.492258 | -0.180794 | -2.642127 |
| H | 4.390407 | -0.494256 | -2.584203 |
| H | 4.657574 | -0.932120 | 1.683699 |
| H | -3.351087 | 0.544010 | 3.072294 |
| H | -1.694722 | 0.340282 | 2.461563 |
| H | -2.437148 | 1.947247 | 2.484020 |
| H | -2.571073 | 0.734103 | -3.634147 |
| H | -1.600522 | 1.918886 | -2.737716 |
| H | -1.151527 | 0.208434 | -2.704862 |
| H | -6.643571 | -0.804477 | -1.715276 |
| H | -6.161834 | -2.022209 | -0.519981 |
| H | -6.989101 | -0.551986 | 0.013397 |
| H | 2.667358 | 0.771851 | -3.582349 |
| H | 1.118002 | 0.624428 | -2.727505 |
| H | 2.045375 | 2.129204 | -2.617926 |
| H | 3.084612 | 0.069656 | 3.121737 |
| H | 2.359150 | 1.586247 | 2.549320 |
| H | 1.442211 | 0.069406 | 2.447274 |
| H | 6.504511 | -1.675173 | 0.266824 |
| H | 5.586292 | -2.724965 | -0.825799 |
| H | 6.493425 | -1.355214 | -1.484940 |
| H | 0.314734 | 5.280721 | -0.848713 |
| H | -0.725186 | 5.684678 | 0.540856 |
| H | 1.043286 | 5.606666 | 0.736334 |
| H | -0.947484 | 4.042663 | 2.487338 |
| H | -0.138396 | 2.457708 | 2.430682 |
| H | 0.822882 | 3.940329 | 2.630876 |
| H | -2.069663 | 3.429515 | 0.327166 |
| H | -1.135313 | 3.367255 | -1.178176 |
| O | -0.138133 | -3.158442 | -0.138713 |
| C | -0.139065 | -3.803290 | 0.995630 |
| O | -0.119740 | -3.323060 | 2.129053 |
| H | -0.160867 | -4.904564 | 0.840450 |

8 INT

26

| | | | |
|---|---|---|---|
| C | 3.385799 | -0.468349 | -0.004279 |
| C | 3.243708 | 0.893750 | -0.002597 |
| N | 1.872456 | 1.145512 | -0.002159 |
| C | 1.159038 | -0.012581 | -0.002972 |
| N | 2.095283 | -0.998681 | -0.004304 |
| C | 4.252226 | 1.984652 | -0.001235 |
| C | 1.276389 | 2.474147 | -0.000761 |
| AU | -0.804245 | -0.176692 | -0.002663 |
| O | -2.838075 | -0.370667 | -0.002239 |
| C | -3.596646 | 0.700236 | -0.004376 |
| O | -4.823328 | 0.673413 | -0.004818 |
| C | 1.787493 | -2.422233 | -0.006037 |
| C | 4.600013 | -1.324078 | -0.006057 |
| H | -3.051918 | 1.673446 | -0.006034 |
| H | 5.500025 | -0.700764 | -0.005651 |
| H | 4.638034 | -1.970888 | -0.893897 |
| H | 4.638932 | -1.973334 | 0.879948 |
| H | 5.263178 | 1.564878 | -0.002402 |
| H | 4.153405 | 2.624985 | 0.886684 |
| H | 4.152588 | 2.627950 | -0.886897 |
| H | 1.579810 | 3.026753 | 0.895496 |
| H | 0.189266 | 2.349915 | -0.001579 |
| H | 1.580928 | 3.029097 | -0.895176 |
| H | 2.199524 | -2.899574 | -0.902347 |
| H | 0.698915 | -2.529467 | -0.004368 |
| H | 2.202586 | -2.902344 | 0.887365 |

8 PC

26

| | | | |
|---|---|---|---|
| C | 3.158671 | -0.946955 | 0.007596 |
| N | 1.782141 | -1.170767 | -0.000052 |
| C | 1.091837 | 0.001583 | 0.013483 |
| N | 2.049798 | 0.967837 | 0.028948 |
| C | 3.328660 | 0.411937 | 0.026193 |
| C | 1.158581 | -2.487616 | -0.019549 |
| AU | -0.867662 | 0.232222 | 0.014362 |
| C | 1.771022 | 2.396556 | 0.048043 |
| C | 4.559655 | 1.243674 | 0.043358 |
| C | 4.149833 | -2.054067 | -0.004600 |
| C | -3.671098 | -0.485464 | 0.018462 |
| O | -3.326915 | -1.668450 | 0.016363 |
| O | -2.899903 | 0.567301 | 0.018124 |
| H | -4.747785 | -0.205497 | 0.021015 |
| H | 5.448026 | 0.603828 | 0.039823 |
| H | 4.605861 | 1.879430 | 0.938889 |
| H | 4.615407 | 1.902848 | -0.834486 |
| H | 5.166999 | -1.648880 | 0.005589 |
| H | 4.045817 | -2.681170 | -0.901311 |
| H | 4.038519 | -2.707600 | 0.872063 |
| H | 1.454502 | -3.033591 | -0.922519 |
| H | 0.073500 | -2.345547 | -0.017418 |
| H | 1.454538 | -3.059694 | 0.867070 |
| H | 2.187846 | 2.854057 | 0.952578 |
| H | 0.684190 | 2.521743 | 0.042889 |
| H | 2.199652 | 2.880179 | -0.837170 |

9 PC

52

| | | | |
|---|---|---|---|
| C | 3.042230 | 0.885455 | -1.359915 |
| C | 2.447343 | 1.373547 | -0.189015 |
| C | 3.108271 | 1.401704 | 1.044845 |
| C | 4.425666 | 0.937225 | 1.081435 |
| C | 5.067022 | 0.452587 | -0.063781 |
| C | 4.361914 | 0.435450 | -1.272310 |
| N | 1.085125 | 1.828951 | -0.254834 |
| C | 0.014634 | 1.003755 | -0.088601 |
| N | -1.070485 | 1.804171 | -0.279020 |
| C | -0.684022 | 3.109933 | -0.559843 |
| C | 0.676239 | 3.125737 | -0.544261 |
| C | -2.426504 | 1.328457 | -0.231194 |
| C | -3.153234 | 1.492103 | 0.955958 |
| C | -4.468487 | 1.024762 | 0.977465 |
| C | -5.047657 | 0.405365 | -0.137255 |
| C | -4.277649 | 0.250201 | -1.293169 |
| C | -2.955693 | 0.701044 | -1.365592 |
| C | -2.519083 | 2.114524 | 2.169260 |
| C | -2.120600 | 0.476889 | -2.595129 |
| C | -6.461830 | -0.109433 | -0.075848 |
| AU | 0.040750 | -0.916384 | 0.309897 |
| C | 2.403887 | 1.872610 | 2.286530 |
| C | 6.473758 | -0.079945 | 0.009562 |
| C | 2.267943 | 0.806361 | -2.645936 |
| H | 1.386034 | 3.923634 | -0.713539 |
| H | -1.407794 | 3.891209 | -0.746561 |
| H | 4.846727 | 0.051942 | -2.171007 |
| H | 4.960407 | 0.947536 | 2.032195 |
| H | -5.052761 | 1.140933 | 1.891579 |
| H | -4.712236 | -0.240439 | -2.165002 |
| H | 2.914666 | 0.489912 | -3.470740 |
| H | 1.448630 | 0.078471 | -2.552980 |
| H | 1.813192 | 1.770404 | -2.910086 |
| H | 3.084535 | 1.869741 | 3.144196 |
| H | 2.000057 | 2.886945 | 2.166781 |
| H | 1.554520 | 1.213190 | 2.517129 |
| H | 7.039558 | 0.392366 | 0.821743 |
| H | 6.465005 | -1.163546 | 0.199668 |
| H | 7.011832 | 0.082610 | -0.932421 |
| H | -3.220712 | 2.125887 | 3.009910 |
| H | -1.625983 | 1.549269 | 2.471260 |
| H | -2.195668 | 3.146759 | 1.977331 |
| H | -2.748023 | 0.178935 | -3.441748 |
| H | -1.555340 | 1.374781 | -2.876274 |
| H | -1.389686 | -0.325483 | -2.412892 |
| H | -6.854306 | -0.318314 | -1.077870 |
| H | -6.509536 | -1.043101 | 0.503647 |
| H | -7.127521 | 0.612218 | 0.415154 |
| O | 0.104698 | -2.906821 | 0.827343 |
| C | -0.161008 | -3.770013 | -0.114879 |
| O | -0.446100 | -3.534322 | -1.289284 |
| H | -0.104044 | -4.816058 | 0.260726 |

10 PC

70

| | | | |
|---|---|---|---|
| C | 3.452420 | 0.179700 | 1.138732 |
| C | 2.705304 | -0.864471 | 0.568885 |
| C | 3.124953 | -1.595497 | -0.552357 |
| C | 4.366706 | -1.260229 | -1.104218 |
| C | 5.139939 | -0.240396 | -0.556733 |
| C | 4.686926 | 0.473736 | 0.551155 |
| N | 1.423573 | -1.174093 | 1.144126 |
| C | 0.251193 | -0.664390 | 0.674996 |
| N | -0.701869 | -1.202387 | 1.484643 |
| C | -0.135467 | -2.031294 | 2.445123 |
| C | 1.207653 | -2.016916 | 2.227161 |
| C | -2.106781 | -0.919710 | 1.353564 |
| C | -2.891549 | -1.794449 | 0.584149 |
| C | -4.251444 | -1.491259 | 0.463924 |
| C | -4.789808 | -0.364134 | 1.081781 |
| C | -3.979778 | 0.487752 | 1.826765 |
| C | -2.612290 | 0.229947 | 1.978191 |
| C | -2.278333 | -2.968160 | -0.158701 |
| C | -2.018499 | -2.575272 | -1.623853 |
| C | -1.729665 | 1.183913 | 2.761127 |
| C | -1.669605 | 2.556054 | 2.073280 |
| AU | -0.000153 | 0.567694 | -0.834916 |
| C | 2.268933 | -2.680620 | -1.178955 |
| C | 1.820994 | -2.266412 | -2.589212 |
| C | 2.904875 | 1.012580 | 2.284008 |
| C | 2.261175 | 2.297156 | 1.732235 |
| C | 2.991355 | -4.035839 | -1.185898 |
| C | 3.962688 | 1.331641 | 3.347938 |
| C | -2.190935 | 1.299186 | 4.222330 |
| C | -3.126497 | -4.242443 | -0.064770 |
| H | 2.022521 | -2.516804 | 2.732171 |
| H | -0.738176 | -2.544322 | 3.181432 |
| H | 5.296344 | 1.279517 | 0.956918 |
| H | 4.727225 | -1.800574 | -1.978780 |
| H | -4.895847 | -2.138277 | -0.129280 |
| H | -4.411382 | 1.373118 | 2.292297 |
| H | 2.112661 | 0.431253 | 2.775641 |
| H | 1.362006 | -2.792214 | -0.570363 |
| H | -1.304998 | -3.188203 | 0.301098 |
| H | -0.709908 | 0.776877 | 2.768684 |
| H | 6.102530 | 0.007929 | -1.003014 |
| H | 1.160488 | -3.029838 | -3.020644 |
| H | 1.273199 | -1.314535 | -2.554218 |
| H | 2.682594 | -2.141265 | -3.258605 |
| H | 2.337299 | -4.812649 | -1.603354 |
| H | 3.901787 | -3.998818 | -1.799241 |
| H | 3.280161 | -4.336603 | -0.170277 |
| H | 1.809652 | 2.880204 | 2.545760 |
| H | 3.015456 | 2.923291 | 1.236607 |
| H | 1.478609 | 2.067837 | 0.995661 |
| H | 3.495699 | 1.850635 | 4.195085 |
| H | 4.441408 | 0.418217 | 3.724277 |
| H | 4.747935 | 1.990031 | 2.953712 |
| H | -5.851383 | -0.143744 | 0.972869 |
| H | -1.524496 | -3.395753 | -2.161276 |
| H | -2.963408 | -2.345913 | -2.134937 |
| H | -1.377604 | -1.684226 | -1.682486 |
| H | -2.599064 | -5.078825 | -0.541791 |
| H | -3.328719 | -4.512705 | 0.979623 |
| H | -4.089734 | -4.126314 | -0.578888 |
| H | -0.995016 | 3.225344 | 2.623445 |
| H | -1.303912 | 2.475161 | 1.040933 |
| H | -2.662375 | 3.025531 | 2.045430 |
| H | -1.511980 | 1.954995 | 4.783195 |
| H | -3.200197 | 1.727690 | 4.287216 |
| H | -2.208102 | 0.317708 | 4.714238 |
| O | -0.227724 | 1.753995 | -2.499201 |
| C | -0.538449 | 3.008110 | -2.311245 |
| O | -0.710733 | 3.585694 | -1.238229 |
| H | -0.640094 | 3.551548 | -3.276639 |

11 PC

70

| | | | |
|---|---|---|---|
| C | -3.043352 | -0.768191 | -1.207834 |
| C | -2.440266 | -0.763719 | 0.057641 |
| C | -3.079293 | -0.322113 | 1.226439 |
| C | -4.396474 | 0.129041 | 1.095724 |
| C | -5.030113 | 0.135478 | -0.145128 |
| C | -4.361397 | -0.305202 | -1.284345 |
| N | -1.076964 | -1.210534 | 0.162637 |
| C | 0.000430 | -0.397337 | -0.020126 |
| N | 1.085963 | -1.194439 | 0.185850 |
| C | 0.695084 | -2.494850 | 0.499964 |
| C | -0.673053 | -2.505985 | 0.481901 |
| C | 2.442401 | -0.719425 | 0.121464 |
| C | 2.988774 | -0.130055 | 1.270351 |
| C | 4.298096 | 0.354056 | 1.170810 |
| C | 5.010317 | 0.248539 | -0.020485 |
| C | 4.432521 | -0.342211 | -1.143025 |
| C | 3.128287 | -0.843247 | -1.096612 |
| C | 2.206587 | 0.010218 | 2.562378 |
| C | 1.968456 | 1.490373 | 2.897144 |
| C | 2.458886 | -1.423878 | -2.328236 |
| C | 1.954839 | -0.285330 | -3.232040 |
| AU | -0.003481 | 1.517859 | -0.442399 |
| CL | -1.766071 | -3.767980 | 0.773740 |
| CL | 1.800015 | -3.738015 | 0.825758 |
| C | -2.356544 | -0.264049 | 2.558889 |
| C | -1.909271 | 1.179054 | 2.849454 |
| C | -2.299624 | -1.206968 | -2.454892 |
| C | -2.033038 | 0.001064 | -3.367190 |
| C | -3.197738 | -0.837196 | 3.706443 |
| C | -3.041347 | -2.329727 | -3.194107 |
| C | 3.368294 | -2.391890 | -3.094780 |
| C | 2.899094 | -0.733600 | 3.714994 |
| H | -4.866562 | -0.283158 | -2.249122 |
| H | -4.929549 | 0.489056 | 1.974185 |
| H | 4.760597 | 0.826949 | 2.036267 |
| H | 5.000707 | -0.406695 | -2.069625 |
| H | -1.323248 | -1.605098 | -2.147630 |
| H | -1.450771 | -0.881557 | 2.481810 |
| H | 1.221812 | -0.453964 | 2.418222 |
| H | 1.580365 | -1.995757 | -1.998308 |
| H | -6.055651 | 0.494863 | -0.225769 |
| H | -1.338280 | 1.221257 | 3.786118 |
| H | -1.277359 | 1.575850 | 2.043030 |
| H | -2.780551 | 1.840559 | 2.947795 |
| H | -2.607480 | -0.852109 | 4.631729 |
| H | -4.090980 | -0.227356 | 3.895497 |
| H | -3.523555 | -1.862304 | 3.488132 |
| H | -1.455587 | -0.304061 | -4.249536 |
| H | -2.974902 | 0.449343 | -3.710810 |
| H | -1.464513 | 0.772482 | -2.829742 |
| H | -2.455148 | -2.664220 | -4.060018 |
| H | -3.209315 | -3.192081 | -2.536245 |
| H | -4.017767 | -1.988495 | -3.563068 |
| H | 6.027075 | 0.636270 | -0.078065 |
| H | 1.379698 | 1.579677 | 3.819364 |
| H | 2.920036 | 2.016979 | 3.050803 |
| H | 1.420834 | 2.001154 | 2.093827 |
| H | 2.289671 | -0.670719 | 4.626061 |
| H | 3.047889 | -1.793496 | 3.470807 |
| H | 3.881146 | -0.294987 | 3.936974 |
| H | 1.416358 | -0.690762 | -4.098721 |
| H | 1.277189 | 0.381598 | -2.680786 |
| H | 2.796803 | 0.317582 | -3.598318 |
| H | 2.813695 | -2.849037 | -3.924402 |
| H | 4.237132 | -1.875933 | -3.524353 |
| H | 3.735382 | -3.193574 | -2.441439 |
| O | -0.004000 | 3.500486 | -0.977920 |
| C | 0.022836 | 4.369744 | -0.002740 |
| O | 0.046931 | 4.137251 | 1.205225 |
| H | 0.021121 | 5.413327 | -0.387082 |

12 PC

58

| | | | |
|---|---|---|---|
| C | -3.117538 | -1.127858 | -1.151012 |
| C | -2.426796 | -1.138107 | 0.067563 |
| C | -3.011923 | -0.730979 | 1.272755 |
| C | -4.349446 | -0.326897 | 1.238323 |
| C | -5.083860 | -0.314978 | 0.047996 |
| C | -4.451294 | -0.713742 | -1.135592 |
| N | -1.064488 | -1.587905 | 0.086811 |
| C | 0.006704 | -0.761760 | -0.040740 |
| N | 1.084791 | -1.579973 | 0.080355 |
| C | 0.699418 | -2.911508 | 0.285886 |
| C | -0.668146 | -2.916592 | 0.289811 |
| C | 2.443668 | -1.120428 | 0.051669 |
| C | 3.034633 | -0.709436 | 1.254462 |
| C | 4.367451 | -0.296243 | 1.211832 |
| C | 5.094686 | -0.278113 | 0.015188 |
| C | 4.457444 | -0.679285 | -1.163134 |
| C | 3.125080 | -1.103727 | -1.170274 |
| C | 2.239376 | -0.679129 | 2.529473 |
| C | 2.428049 | -1.492301 | -2.444264 |
| C | 6.525494 | 0.192880 | -0.000036 |
| AU | -0.003272 | 1.180177 | -0.316919 |
| C | -1.654986 | -4.011752 | 0.465854 |
| C | 1.694996 | -3.999633 | 0.456125 |
| C | -2.210262 | -0.693434 | 2.543574 |
| C | -6.516258 | 0.150500 | 0.031840 |
| C | -2.425073 | -1.513122 | -2.428515 |
| H | -5.008436 | -0.698181 | -2.073483 |
| H | -4.826392 | -0.008182 | 2.166141 |
| H | 4.847688 | 0.026068 | 2.136879 |
| H | 5.008228 | -0.658250 | -2.104509 |
| H | -3.122560 | -1.490596 | -3.272285 |
| H | -1.600661 | -0.817018 | -2.640299 |
| H | -1.984543 | -2.517820 | -2.368957 |
| H | -2.849726 | -0.462812 | 3.401967 |
| H | -1.699783 | -1.647030 | 2.734499 |
| H | -1.430238 | 0.079225 | 2.476994 |
| H | -6.972690 | 0.072692 | 1.025714 |
| H | -6.577089 | 1.203920 | -0.279619 |
| H | -7.118234 | -0.433825 | -0.675541 |
| H | 2.877779 | -0.424889 | 3.381934 |
| H | 1.439883 | 0.072973 | 2.460262 |
| H | 1.753904 | -1.643378 | 2.732348 |
| H | 3.121723 | -1.469046 | -3.291148 |
| H | 1.990497 | -2.498069 | -2.381346 |
| H | 1.600878 | -0.798671 | -2.653394 |
| H | 7.023055 | -0.069348 | -0.940991 |
| H | 6.575540 | 1.285702 | 0.113289 |
| H | 7.096850 | -0.244835 | 0.828998 |
| H | -1.139546 | -4.966697 | 0.610431 |
| H | -2.300151 | -3.832609 | 1.337126 |
| H | -2.311953 | -4.104430 | -0.409974 |
| H | 1.187059 | -4.958099 | 0.603895 |
| H | 2.347381 | -4.087810 | -0.423589 |
| H | 2.344023 | -3.815980 | 1.323583 |
| O | -0.014748 | 3.198878 | -0.728027 |
| C | -0.052546 | 4.017872 | 0.286600 |
| O | -0.079154 | 3.731561 | 1.483937 |
| H | -0.058757 | 5.079848 | -0.046179 |

13 PC

32

| | | | |
|---|---|---|---|
| C | -3.043936 | -1.420276 | 0.053061 |
| N | -1.659591 | -1.443954 | 0.045457 |
| C | -1.157067 | -0.178240 | 0.060671 |
| N | -2.248314 | 0.636849 | 0.073800 |
| C | -3.415209 | -0.110048 | 0.073367 |
| C | -0.821493 | -2.661740 | 0.026825 |
| C | -1.045715 | -3.476664 | 1.297606 |
| AU | 0.748454 | 0.325753 | 0.078215 |
| C | -2.172661 | 2.110660 | 0.131100 |
| C | -2.966858 | 2.734222 | -1.012909 |
| C | -2.605789 | 2.606970 | 1.509018 |
| C | -1.066537 | -3.454130 | -1.254212 |
| H | -4.397779 | 0.340790 | 0.084377 |
| H | -3.643508 | -2.319919 | 0.040613 |
| H | -1.103137 | 2.320894 | -0.008993 |
| H | -2.477001 | 3.694389 | 1.567804 |
| H | -3.662987 | 2.377747 | 1.698301 |
| H | -1.999147 | 2.141819 | 2.295139 |
| H | -2.816205 | 3.820093 | -1.011648 |
| H | -2.635681 | 2.338484 | -1.980450 |
| H | -4.043430 | 2.545495 | -0.905623 |
| H | 0.207140 | -2.272938 | 0.022398 |
| H | -0.359643 | -4.331670 | 1.310099 |
| H | -0.855575 | -2.866421 | 2.188576 |
| H | -2.072218 | -3.864452 | 1.348677 |
| H | -0.379733 | -4.307742 | -1.293484 |
| H | -2.093046 | -3.843131 | -1.294772 |
| H | -0.892570 | -2.828094 | -2.137486 |
| O | 2.720560 | 0.918196 | 0.100917 |
| C | 3.615360 | -0.032937 | 0.112506 |
| O | 3.417189 | -1.248530 | 0.110695 |
| H | 4.649200 | 0.377598 | 0.125196 |

8-PP

47

| | | | |
|---|---|---|---|
| C | 0.849887 | 1.184689 | -0.852102 |
| N | -0.092343 | 2.067051 | -1.288484 |
| C | -0.019866 | 3.253895 | -0.570959 |
| C | 1.026573 | 3.116937 | 0.306322 |
| N | 1.544754 | 1.840036 | 0.119613 |
| C | -1.100533 | 1.774430 | -2.298252 |
| C | 2.670976 | 1.324758 | 0.887137 |
| C | 1.575470 | 4.051698 | 1.320529 |
| C | -0.960340 | 4.381671 | -0.791748 |
| H | -1.980558 | 1.314532 | -1.816057 |
| H | -1.375534 | 2.699589 | -2.810329 |
| H | -0.684862 | 1.056018 | -3.010347 |
| H | 2.515646 | 1.558722 | 1.944063 |
| H | 2.720812 | 0.244867 | 0.768756 |
| H | 3.605668 | 1.773899 | 0.532249 |
| P | 0.861746 | -0.548399 | -1.421845 |
| P | -0.238879 | -1.615826 | 0.155781 |
| H | -0.746302 | 5.191243 | -0.087269 |
| H | -0.876470 | 4.789170 | -1.808886 |
| H | -2.001518 | 4.066387 | -0.640859 |
| H | 1.490676 | 3.632486 | 2.332998 |
| H | 2.635978 | 4.271779 | 1.137446 |
| H | 1.024665 | 4.996898 | 1.297988 |
| AU | -2.367860 | -0.660145 | -0.265314 |
| H | -3.875550 | -0.074416 | -0.515086 |
| C | 0.374033 | -0.840279 | 1.719494 |
| C | -0.184311 | 0.344212 | 2.233825 |
| C | 0.275145 | 0.892638 | 3.430437 |
| C | 1.304024 | 0.266708 | 4.140987 |
| C | 1.859689 | -0.917545 | 3.650787 |
| C | 1.393127 | -1.468430 | 2.456160 |
| H | -0.992039 | 0.824250 | 1.679377 |
| H | -0.177820 | 1.806503 | 3.815803 |
| H | 1.661324 | 0.692997 | 5.078024 |
| H | 2.653624 | -1.418429 | 4.204874 |
| H | 1.822025 | -2.399885 | 2.087391 |
| C | 2.656031 | -0.930862 | -1.302120 |
| C | 3.103047 | -2.120740 | -0.713546 |
| C | 4.459570 | -2.448656 | -0.727056 |
| C | 5.384831 | -1.595684 | -1.331474 |
| C | 4.943222 | -0.415172 | -1.937471 |
| C | 3.588678 | -0.090577 | -1.933964 |
| H | 2.377283 | -2.783572 | -0.241285 |
| H | 4.794393 | -3.374354 | -0.259429 |
| H | 6.443917 | -1.850738 | -1.337959 |
| H | 5.657355 | 0.251012 | -2.421576 |
| H | 3.251830 | 0.826944 | -2.419217 |

8-PP RC

50

| | | | |
|---|---|---|---|
| C | 3.799122 | 0.730524 | -2.118605 |
| C | 3.059362 | -0.350940 | -1.611148 |
| C | 3.734406 | -1.518775 | -1.234090 |
| C | 5.124099 | -1.595541 | -1.337116 |
| C | 5.855859 | -0.508350 | -1.817748 |
| C | 5.187166 | 0.655733 | -2.209945 |
| P | 1.221367 | -0.281417 | -1.629907 |
| P | 0.424443 | -1.814037 | -0.268036 |
| C | 0.901056 | -1.188067 | 1.406022 |
| C | 0.151949 | -0.205219 | 2.078108 |
| C | 0.525610 | 0.236730 | 3.346480 |
| C | 1.654969 | -0.299001 | 3.973286 |
| C | 2.398943 | -1.288486 | 3.326285 |
| C | 2.022021 | -1.731128 | 2.057069 |
| C | 0.941591 | 1.285858 | -0.740330 |
| N | -0.148503 | 2.064259 | -0.992696 |
| C | -0.193712 | 3.140680 | -0.115839 |
| C | 0.913707 | 3.036212 | 0.687978 |
| N | 1.594270 | 1.890713 | 0.291512 |
| C | -1.164079 | 1.748890 | -1.988806 |
| C | -1.268831 | 4.165254 | -0.140416 |
| C | 1.385720 | 3.888624 | 1.807645 |
| C | 2.818525 | 1.437262 | 0.938476 |
| AU | -1.873318 | -1.332921 | -0.559401 |
| H | -1.773739 | 0.897752 | -1.631767 |
| H | -1.790043 | 2.627956 | -2.153370 |
| H | -0.676260 | 1.452260 | -2.922161 |
| H | 2.685778 | 1.496138 | 2.022581 |
| H | 3.007886 | 0.402627 | 0.662521 |
| H | 3.664893 | 2.059015 | 0.624772 |
| H | -1.081382 | 4.918280 | 0.631147 |
| H | -1.316306 | 4.681096 | -1.109176 |
| H | -2.254639 | 3.725646 | 0.057233 |
| H | 1.406768 | 3.326292 | 2.751872 |
| H | 2.397851 | 4.274644 | 1.625857 |
| H | 0.715270 | 4.743148 | 1.939429 |
| H | -3.484668 | -1.120571 | -0.744265 |
| H | -0.734130 | 0.204167 | 1.593098 |
| H | -0.072585 | 0.995283 | 3.852220 |
| H | 1.945147 | 0.043445 | 4.966481 |
| H | 3.272785 | -1.721031 | 3.813922 |
| H | 2.601904 | -2.510035 | 1.561754 |
| H | 3.158406 | -2.364743 | -0.856922 |
| H | 5.637036 | -2.509072 | -1.036699 |
| H | 6.941027 | -0.568159 | -1.894632 |
| H | 5.750139 | 1.505111 | -2.597060 |
| H | 3.284247 | 1.637711 | -2.439561 |
| C | -3.861301 | 1.400169 | 0.524117 |
| O | -4.689716 | 1.471555 | -0.297223 |
| O | -3.036937 | 1.386619 | 1.355807 |

8-PP TS

50

| | | | |
|---|---|---|---|
| C | 2.018399 | 1.691570 | -2.204437 |
| C | 0.881643 | 1.100226 | -1.626574 |
| C | 0.211798 | 0.091407 | -2.341015 |
| C | 0.677340 | -0.325060 | -3.587311 |
| C | 1.822415 | 0.255371 | -4.140850 |
| C | 2.490792 | 1.266065 | -3.446283 |
| P | 0.320468 | 1.746850 | 0.010948 |
| AU | -1.797246 | 0.944180 | 0.228021 |
| C | -4.219028 | -0.181524 | 0.475650 |
| P | 1.142250 | 0.283120 | 1.492082 |
| C | 2.967664 | 0.460807 | 1.473142 |
| C | 3.772887 | -0.582165 | 1.961441 |
| C | 5.152699 | -0.419713 | 2.060022 |
| C | 5.746816 | 0.791597 | 1.693120 |
| C | 4.949353 | 1.839730 | 1.231545 |
| C | 3.567404 | 1.678137 | 1.124125 |
| C | 0.927482 | -1.336579 | 0.680518 |
| N | 1.555134 | -1.923544 | -0.377075 |
| C | 0.879713 | -3.077600 | -0.756858 |
| C | -0.197294 | -3.203479 | 0.086903 |
| N | -0.129613 | -2.141859 | 0.974391 |
| C | 2.777713 | -1.470911 | -1.028782 |
| C | -1.092703 | -1.936015 | 2.052195 |
| C | -1.302493 | -4.193357 | 0.125450 |
| C | 1.322785 | -3.917848 | -1.896909 |
| H | -2.071720 | -1.703454 | 1.615795 |
| H | -1.148067 | -2.846581 | 2.658034 |
| H | -0.749873 | -1.101127 | 2.668340 |
| H | 2.654393 | -1.555517 | -2.111707 |
| H | 2.957455 | -0.429176 | -0.773316 |
| H | 3.626611 | -2.079211 | -0.696105 |
| H | -1.199027 | -4.907934 | -0.696669 |
| H | -1.309463 | -4.758151 | 1.067950 |
| H | -2.272686 | -3.686899 | 0.026903 |
| H | 1.319952 | -3.347793 | -2.836573 |
| H | 2.339021 | -4.305893 | -1.744101 |
| H | 0.648783 | -4.771313 | -2.016578 |
| H | -0.682575 | -0.360101 | -1.909067 |
| H | 0.139001 | -1.100190 | -4.132799 |
| H | 2.183855 | -0.069304 | -5.116177 |
| H | 3.375765 | 1.733444 | -3.877621 |
| H | 2.533531 | 2.494160 | -1.676613 |
| H | 2.946428 | 2.500347 | 0.767781 |
| H | 5.403297 | 2.790101 | 0.951950 |
| H | 6.825669 | 0.918526 | 1.774499 |
| H | 5.767239 | -1.239302 | 2.431900 |
| H | 3.316803 | -1.527135 | 2.260566 |
| O | -3.422496 | -1.063700 | 0.042744 |
| O | -5.405644 | -0.218617 | 0.791310 |
| H | -3.737360 | 0.911635 | 0.611738 |

8-PP INT
50

| | | | |
|---|---|---|---|
| N | -1.425733 | 1.942758 | -0.019660 |
| C | -0.892750 | 1.172469 | 0.968910 |
| N | 0.135151 | 1.901503 | 1.488392 |
| C | 0.279605 | 3.100203 | 0.806429 |
| C | -0.718527 | 3.133778 | -0.138023 |
| P | -1.180632 | -0.566622 | 1.454662 |
| P | -0.304089 | -1.772514 | -0.208452 |
| AU | 1.724800 | -0.777868 | 0.095603 |
| O | 3.436745 | 0.409466 | 0.436481 |
| C | 3.621754 | 1.377702 | -0.418894 |
| O | 4.199310 | 2.436878 | -0.166216 |
| C | 1.022445 | 1.481473 | 2.569276 |
| C | 1.361491 | 4.066984 | 1.118337 |
| C | -1.058025 | 4.163827 | -1.151208 |
| C | -2.612547 | 1.640589 | -0.811101 |
| C | -3.007795 | -0.679099 | 1.346576 |
| C | -3.818393 | 0.328343 | 1.896393 |
| C | -5.204483 | 0.187942 | 1.914175 |
| C | -5.801886 | -0.968080 | 1.403101 |
| C | -5.001005 | -1.984789 | 0.880567 |
| C | -3.613138 | -1.844036 | 0.854408 |
| C | -0.933185 | -0.972581 | -1.753011 |
| C | -0.263846 | 0.078628 | -2.405126 |
| C | -0.761090 | 0.619686 | -3.589505 |
| C | -1.945792 | 0.126699 | -4.143023 |
| C | -2.618237 | -0.922443 | -3.511927 |
| C | -2.109919 | -1.475155 | -2.335955 |
| H | 2.019059 | 1.262036 | 2.165669 |
| H | 1.083920 | 2.283751 | 3.311426 |
| H | 0.607268 | 0.580167 | 3.027187 |
| H | -2.426605 | 1.914425 | -1.852332 |
| H | -2.818566 | 0.573673 | -0.758062 |
| H | -3.471961 | 2.198091 | -0.421050 |
| H | 1.275533 | 4.946514 | 0.472891 |
| H | 1.309454 | 4.406035 | 2.162083 |
| H | 2.352260 | 3.618791 | 0.952428 |
| H | -0.972518 | 3.762689 | -2.170932 |
| H | -2.082676 | 4.538443 | -1.023234 |
| H | -0.372221 | 5.011914 | -1.064104 |
| H | 0.657858 | 0.465286 | -1.968169 |
| H | -0.219515 | 1.426316 | -4.084157 |
| H | -2.334987 | 0.548856 | -5.068901 |
| H | -3.535140 | -1.321864 | -3.944753 |
| H | -2.629384 | -2.308876 | -1.864725 |
| H | -2.990033 | -2.642060 | 0.450613 |
| H | -5.457392 | -2.893430 | 0.488556 |
| H | -6.885631 | -1.076862 | 1.418122 |
| H | -5.821707 | 0.982021 | 2.333879 |
| H | -3.361837 | 1.229999 | 2.307396 |
| H | 3.217180 | 1.196939 | -1.443795 |

8-PP PC

50

| | | | |
|---|---|---|---|
| C | -3.797455 | -0.498690 | -1.956833 |
| C | -2.972374 | 0.519783 | -1.449256 |
| C | -3.554872 | 1.730174 | -1.050344 |
| C | -4.937422 | 1.906806 | -1.125682 |
| C | -5.753855 | 0.880814 | -1.604062 |
| C | -5.177458 | -0.322223 | -2.022940 |
| P | -1.149074 | 0.314554 | -1.501146 |
| P | -0.244393 | 1.707819 | -0.023326 |
| C | -0.822937 | 1.055235 | 1.606403 |
| C | -0.171846 | 0.014582 | 2.292412 |
| C | -0.636868 | -0.421352 | 3.532407 |
| C | -1.763897 | 0.171597 | 4.109638 |
| C | -2.414339 | 1.212963 | 3.444006 |
| C | -1.942235 | 1.655969 | 2.207928 |
| C | -0.958194 | -1.323529 | -0.723270 |
| N | 0.086463 | -2.136866 | -1.036587 |
| C | 0.134664 | -3.225026 | -0.179701 |
| C | -0.940373 | -3.103529 | 0.666420 |
| N | -1.596467 | -1.927799 | 0.318184 |
| C | 1.068851 | -1.897894 | -2.087713 |
| C | 1.220397 | -4.234395 | -0.249579 |
| C | -1.394575 | -3.963864 | 1.786995 |
| C | -2.802215 | -1.464659 | 0.992513 |
| AU | 1.888488 | 0.951628 | -0.255455 |
| H | 2.036161 | -1.674422 | -1.619423 |
| H | 1.147032 | -2.793369 | -2.713329 |
| H | 0.733044 | -1.050611 | -2.690435 |
| H | -2.673287 | -1.586767 | 2.071267 |
| H | -2.955663 | -0.410543 | 0.773105 |
| H | -3.670306 | -2.038657 | 0.648227 |
| H | 1.105813 | -4.968659 | 0.553681 |
| H | 1.209915 | -4.774102 | -1.207039 |
| H | 2.198333 | -3.744824 | -0.143974 |
| H | -1.380200 | -3.416606 | 2.740184 |
| H | -2.417147 | -4.332728 | 1.628707 |
| H | -0.733039 | -4.829934 | 1.884131 |
| O | 3.937630 | 0.457783 | -0.439573 |
| H | 0.707638 | -0.445381 | 1.839332 |
| H | -0.112948 | -1.222402 | 4.054386 |
| H | -2.125486 | -0.168175 | 5.079874 |
| H | -3.285759 | 1.689462 | 3.893223 |
| H | -2.444368 | 2.480997 | 1.702659 |
| H | -2.918327 | 2.533177 | -0.678492 |
| H | -5.377855 | 2.851130 | -0.806285 |
| H | -6.833206 | 1.018469 | -1.658117 |
| H | -5.806384 | -1.124177 | -2.409231 |
| H | -3.355932 | -1.437469 | -2.295006 |
| C | 4.302282 | -0.773740 | -0.285021 |
| O | 3.581892 | -1.762230 | -0.082569 |
| H | 5.406620 | -0.897754 | -0.356474 |

FREQUENCIES ($cm^{-1}$) AND CORRESPONDING NORMAL MODES OF TRANSITION STATES

| 1-P TS | | | | 2-P TS | | | | 3-P TS | | | |
| --- | --- | --- | --- | --- | --- | --- | --- | --- | --- | --- | --- |
| | -236.590 | | | | -234.063 | | | | -240.756 | | |
| | ------------------------ | | | | ------------------------ | | | | ------------------------ | | |
| C | 0.001 | 0.004 | -0.000 | C | 0.001 | 0.001 | 0.000 | C | -0.000 | 0.000 | -0.001 |
| C | -0.003 | 0.002 | 0.001 | C | 0.000 | 0.001 | -0.002 | C | -0.001 | 0.000 | -0.002 |
| C | -0.003 | 0.000 | -0.000 | C | -0.001 | 0.001 | -0.002 | C | 0.000 | 0.001 | -0.003 |
| C | 0.001 | -0.004 | -0.002 | C | -0.005 | -0.004 | 0.001 | C | 0.006 | -0.003 | 0.001 |
| C | 0.003 | -0.004 | -0.004 | C | -0.004 | -0.001 | 0.001 | C | 0.004 | -0.001 | 0.001 |
| C | 0.001 | 0.002 | -0.001 | C | -0.002 | 0.000 | 0.002 | C | 0.003 | 0.000 | 0.001 |
| N | 0.002 | -0.003 | 0.004 | C | 0.002 | 0.002 | -0.000 | C | -0.003 | 0.002 | -0.000 |
| C | -0.005 | -0.007 | 0.022 | C | 0.002 | 0.001 | -0.003 | C | -0.003 | 0.001 | -0.002 |
| C | -0.001 | -0.002 | -0.000 | N | 0.001 | -0.003 | -0.005 | N | -0.002 | -0.002 | -0.003 |
| C | 0.002 | 0.001 | -0.001 | C | -0.004 | -0.002 | -0.023 | C | 0.002 | -0.001 | -0.024 |
| C | 0.002 | 0.001 | 0.003 | C | -0.001 | 0.003 | -0.000 | C | -0.001 | 0.004 | 0.000 |
| P | -0.003 | -0.044 | 0.059 | C | 0.002 | 0.003 | 0.001 | C | -0.004 | 0.003 | 0.001 |
| AU | -0.035 | 0.005 | -0.049 | C | 0.002 | 0.001 | 0.001 | C | -0.003 | 0.001 | 0.002 |
| C | -0.005 | -0.007 | -0.005 | C | 0.000 | -0.007 | 0.004 | C | -0.001 | -0.006 | 0.005 |
| C | 0.002 | 0.001 | -0.000 | C | -0.001 | -0.005 | 0.002 | C | 0.001 | -0.004 | 0.004 |
| C | -0.002 | -0.003 | 0.001 | C | -0.002 | -0.000 | 0.000 | C | 0.002 | 0.003 | 0.002 |
| C | 0.001 | -0.001 | 0.002 | C | 0.002 | 0.002 | 0.001 | C | -0.004 | 0.002 | 0.001 |
| C | -0.001 | -0.004 | -0.002 | C | 0.003 | 0.001 | 0.001 | C | -0.007 | -0.000 | 0.000 |
| C | -0.005 | -0.008 | -0.011 | C | -0.002 | -0.003 | 0.001 | C | 0.003 | -0.004 | 0.004 |
| C | 0.001 | 0.003 | -0.000 | C | -0.001 | -0.004 | -0.001 | C | 0.001 | -0.009 | -0.001 |
| C | 0.001 | 0.002 | -0.001 | P | -0.017 | -0.033 | -0.064 | P | 0.028 | -0.060 | -0.109 |
| C | 0.013 | -0.001 | 0.026 | C | 0.011 | -0.005 | -0.026 | C | -0.014 | -0.007 | -0.026 |
| C | 0.018 | -0.010 | 0.017 | C | 0.011 | -0.017 | -0.015 | C | -0.011 | -0.018 | -0.012 |
| C | 0.019 | -0.011 | 0.006 | C | 0.012 | -0.018 | -0.005 | C | -0.009 | -0.019 | -0.001 |
| C | 0.012 | -0.004 | -0.002 | C | 0.008 | -0.008 | 0.002 | C | -0.006 | -0.009 | 0.005 |
| C | 0.004 | -0.000 | 0.002 | C | 0.003 | -0.001 | -0.001 | C | -0.002 | -0.002 | 0.000 |
| C | 0.002 | 0.003 | 0.013 | C | 0.003 | 0.002 | -0.013 | C | -0.004 | 0.001 | -0.013 |
| C | 0.002 | -0.005 | 0.001 | C | -0.003 | -0.002 | -0.001 | C | 0.002 | -0.000 | -0.001 |
| C | 0.004 | 0.002 | 0.000 | C | 0.001 | -0.001 | -0.003 | C | -0.000 | -0.001 | -0.002 |
| H | 0.002 | 0.003 | -0.002 | AU | -0.032 | 0.014 | 0.049 | C | -0.000 | -0.001 | -0.001 |
| H | 0.002 | -0.010 | -0.005 | C | -0.008 | -0.005 | 0.007 | C | -0.000 | -0.001 | -0.000 |
| H | 0.001 | -0.006 | 0.002 | C | 0.001 | 0.002 | 0.001 | C | 0.003 | 0.001 | -0.000 |
| H | -0.007 | 0.003 | 0.015 | H | 0.003 | 0.001 | 0.002 | AU | 0.136 | 0.086 | 0.216 |
| H | -0.001 | -0.001 | -0.004 | H | -0.003 | -0.010 | 0.004 | C | 0.011 | -0.008 | 0.021 |
| H | 0.013 | -0.006 | -0.011 | H | -0.001 | -0.005 | -0.002 | C | -0.003 | 0.001 | 0.002 |
| H | 0.026 | -0.017 | 0.001 | H | -0.005 | 0.007 | -0.015 | C | 0.001 | -0.001 | -0.001 |
| H | 0.021 | -0.014 | 0.019 | H | -0.002 | 0.000 | 0.005 | H | -0.002 | 0.001 | 0.001 |
| H | -0.037 | -0.489 | 0.454 | H | 0.009 | -0.010 | 0.012 | H | 0.001 | -0.004 | 0.002 |
| H | 0.003 | -0.000 | -0.004 | H | 0.015 | -0.026 | 0.000 | H | 0.000 | -0.002 | -0.000 |
| H | -0.005 | -0.007 | -0.007 | H | 0.011 | -0.021 | -0.017 | H | 0.000 | 0.002 | -0.005 |
| H | -0.006 | -0.008 | -0.004 | H | 0.002 | 0.001 | 0.003 | H | 0.001 | 0.000 | 0.002 |
| H | -0.005 | -0.008 | -0.004 | H | -0.005 | 0.000 | -0.001 | H | -0.002 | -0.003 | 0.004 |
| H | 0.004 | 0.002 | 0.002 | H | -0.004 | -0.005 | -0.002 | H | -0.003 | -0.008 | 0.002 |
| H | -0.001 | 0.004 | -0.003 | H | -0.001 | -0.001 | -0.001 | H | -0.003 | -0.006 | -0.004 |
| H | 0.003 | 0.003 | 0.001 | H | 0.002 | -0.000 | -0.003 | H | -0.001 | 0.000 | 0.001 |
| H | -0.005 | -0.003 | 0.001 | H | 0.000 | -0.002 | -0.003 | H | 0.001 | 0.000 | -0.001 |
| H | -0.001 | -0.006 | 0.001 | H | -0.000 | -0.000 | -0.002 | H | 0.001 | -0.001 | -0.000 |
| H | -0.000 | -0.000 | 0.001 | H | 0.005 | 0.003 | -0.000 | H | -0.000 | 0.000 | -0.000 |
| H | 0.002 | -0.001 | 0.002 | H | 0.001 | -0.017 | 0.005 | H | 0.000 | -0.001 | -0.000 |
| H | 0.002 | -0.002 | 0.002 | H | -0.003 | -0.007 | -0.002 | H | -0.000 | -0.000 | -0.000 |
| H | 0.001 | -0.001 | 0.002 | H | -0.003 | -0.002 | 0.002 | H | -0.000 | -0.000 | -0.001 |
| H | 0.004 | 0.003 | -0.001 | H | 0.003 | -0.004 | -0.002 | H | -0.001 | 0.001 | -0.000 |

| | | | |
|---|---|---|---|
| H | 0.008 | -0.011 | -0.005 |
| H | 0.002 | -0.009 | 0.002 |
| H | -0.001 | -0.003 | -0.002 |
| H | 0.005 | -0.003 | 0.003 |
| H | 0.012 | -0.017 | -0.020 |
| H | -0.018 | -0.010 | -0.023 |
| H | -0.011 | -0.000 | 0.004 |
| H | 0.003 | 0.003 | 0.000 |
| H | 0.004 | 0.001 | 0.000 |
| H | 0.004 | 0.004 | 0.003 |
| H | 0.001 | 0.003 | -0.001 |
| H | 0.000 | 0.001 | -0.001 |
| H | -0.001 | 0.002 | -0.002 |
| H | 0.001 | 0.001 | -0.001 |
| C | 0.253 | -0.092 | 0.290 |
| O | 0.444 | 0.254 | 0.215 |
| O | -0.252 | -0.079 | -0.023 |

4-P TS

-215.860

------------------------

| | | | |
|---|---|---|---|
| C | -0.006 | 0.006 | -0.002 |
| C | -0.008 | 0.003 | 0.001 |
| C | -0.006 | -0.000 | 0.002 |
| C | -0.006 | -0.002 | 0.001 |
| C | -0.004 | -0.002 | -0.001 |
| C | -0.004 | 0.002 | -0.003 |
| N | -0.008 | -0.004 | 0.007 |
| C | -0.017 | -0.010 | 0.016 |
| N | -0.012 | -0.020 | 0.014 |
| C | -0.003 | -0.016 | 0.011 |
| C | -0.001 | -0.008 | 0.003 |
| P | -0.046 | 0.007 | 0.089 |
| AU | 0.091 | -0.062 | -0.272 |
| C | -0.007 | -0.012 | 0.008 |
| C | -0.005 | -0.004 | 0.004 |
| C | -0.002 | 0.001 | 0.001 |
| H | 0.001 | -0.017 | 0.015 |
| H | -0.019 | -0.003 | 0.014 |
| H | -0.008 | 0.004 | -0.006 |
| H | 0.004 | 0.002 | 0.001 |
| H | 0.003 | 0.001 | 0.002 |
| H | 0.003 | 0.002 | -0.001 |
| H | 0.002 | 0.002 | 0.002 |
| H | 0.000 | 0.001 | 0.002 |
| H | 0.000 | 0.002 | 0.002 |
| H | 0.001 | 0.000 | 0.000 |
| H | -0.004 | -0.006 | -0.001 |
| H | -0.008 | -0.005 | 0.004 |
| H | -0.005 | -0.000 | 0.002 |
| H | -0.002 | -0.001 | 0.001 |
| H | -0.002 | 0.000 | 0.004 |
| H | -0.002 | 0.000 | 0.003 |
| H | 0.002 | 0.001 | 0.000 |
| H | 0.001 | 0.001 | 0.001 |
| H | -0.000 | 0.001 | -0.002 |
| H | 0.002 | 0.001 | -0.003 |
| C | 0.215 | -0.147 | -0.303 |
| O | 0.534 | 0.049 | -0.183 |
| O | -0.274 | 0.034 | 0.015 |
| H | -0.194 | -0.324 | -0.532 |

5-P TS

-206.194

------------------------

| | | | |
|---|---|---|---|
| C | -0.024 | 0.007 | -0.001 |
| C | -0.042 | 0.007 | -0.007 |
| C | -0.036 | 0.014 | -0.000 |
| C | -0.023 | 0.022 | 0.006 |
| C | -0.013 | 0.019 | 0.005 |
| C | -0.013 | 0.014 | 0.004 |
| P | -0.040 | 0.033 | -0.014 |
| AU | 0.024 | -0.060 | 0.009 |
| C | -0.008 | 0.006 | -0.006 |
| N | -0.006 | 0.004 | 0.002 |
| C | -0.003 | 0.000 | 0.003 |
| C | -0.004 | -0.003 | -0.000 |
| N | -0.006 | -0.002 | -0.003 |
| C | -0.004 | -0.004 | -0.003 |
| C | -0.003 | -0.007 | -0.003 |
| C | -0.004 | -0.007 | -0.002 |
| H | -0.000 | -0.004 | 0.002 |
| H | 0.002 | -0.005 | -0.001 |
| H | 0.001 | -0.001 | 0.002 |
| H | -0.002 | -0.002 | -0.002 |
| H | -0.002 | -0.011 | 0.011 |
| H | 0.009 | 0.000 | 0.013 |
| H | 0.003 | 0.002 | -0.000 |
| H | -0.002 | 0.000 | -0.000 |
| H | -0.002 | -0.001 | 0.000 |
| H | -0.002 | 0.000 | -0.001 |
| H | -0.001 | 0.001 | 0.000 |
| H | -0.001 | 0.001 | 0.000 |
| H | -0.000 | 0.001 | 0.001 |
| H | -0.001 | 0.000 | 0.001 |
| H | 0.002 | -0.002 | -0.000 |
| H | 0.003 | -0.001 | 0.001 |
| H | 0.001 | -0.000 | 0.000 |
| H | 0.001 | -0.000 | 0.001 |
| H | 0.001 | 0.000 | 0.000 |
| H | -0.000 | 0.000 | -0.000 |
| H | -0.000 | 0.000 | 0.000 |
| H | -0.000 | 0.000 | -0.000 |
| H | -0.000 | -0.000 | -0.000 |
| H | 0.000 | -0.001 | 0.000 |
| H | 0.000 | -0.000 | -0.000 |
| H | -0.000 | 0.000 | 0.000 |
| H | 0.001 | -0.000 | 0.000 |
| H | 0.000 | -0.000 | 0.000 |
| H | 0.001 | -0.000 | -0.001 |
| H | 0.001 | 0.000 | -0.001 |
| H | 0.001 | 0.000 | -0.001 |
| H | 0.000 | 0.001 | 0.000 |
| C | -0.195 | -0.235 | -0.331 |
| O | -0.676 | 0.015 | -0.269 |
| O | 0.337 | 0.081 | 0.020 |
| H | 0.094 | -0.145 | -0.153 |

6-P TS

-230.375

------------------------

| | | | |
|---|---|---|---|
| C | 0.002 | -0.004 | 0.005 |
| N | -0.001 | -0.014 | 0.006 |
| C | -0.018 | -0.026 | -0.028 |
| N | -0.007 | -0.012 | 0.012 |
| C | -0.002 | -0.002 | 0.009 |
| C | -0.000 | -0.003 | 0.004 |
| C | -0.004 | -0.005 | 0.001 |
| C | -0.006 | 0.005 | -0.002 |
| C | -0.007 | 0.011 | -0.002 |
| C | -0.003 | 0.004 | -0.000 |
| C | -0.001 | -0.001 | 0.002 |
| C | -0.001 | -0.004 | 0.002 |
| C | -0.007 | -0.000 | -0.001 |
| C | 0.004 | -0.009 | 0.000 |
| C | -0.002 | 0.002 | 0.001 |
| P | -0.009 | -0.068 | -0.009 |

| | | | | | | | | | | | |
|---|---|---|---|---|---|---|---|---|---|---|---|
| C | -0.000 | 0.000 | 0.001 | C | -0.004 | -0.003 | -0.001 | C | 0.012 | -0.009 | -0.002 |
| C | 0.000 | -0.001 | 0.001 | C | -0.004 | -0.002 | -0.002 | C | 0.007 | -0.002 | -0.003 |
| C | -0.002 | -0.005 | 0.004 | C | -0.004 | -0.003 | -0.002 | C | 0.006 | -0.001 | 0.001 |
| C | -0.003 | -0.001 | 0.002 | C | -0.002 | -0.003 | -0.001 | C | 0.008 | -0.002 | 0.004 |
| C | 0.000 | 0.001 | -0.000 | C | -0.002 | 0.001 | -0.003 | C | 0.011 | -0.008 | 0.004 |
| C | -0.000 | -0.003 | 0.003 | C | -0.003 | -0.001 | -0.000 | C | 0.012 | -0.010 | 0.004 |
| C | -0.005 | -0.001 | 0.001 | C | -0.002 | -0.001 | 0.001 | C | -0.002 | -0.006 | 0.003 |
| C | -0.003 | -0.008 | -0.002 | C | -0.005 | -0.002 | -0.003 | C | -0.001 | -0.003 | -0.001 |
| C | -0.002 | 0.005 | -0.000 | C | -0.003 | -0.001 | -0.002 | C | 0.001 | -0.001 | -0.001 |
| C | -0.035 | -0.008 | 0.022 | C | -0.002 | 0.001 | -0.000 | C | 0.001 | -0.001 | -0.001 |
| C | -0.023 | -0.000 | 0.011 | C | -0.002 | 0.001 | 0.000 | C | 0.001 | -0.000 | -0.001 |
| C | -0.013 | 0.012 | 0.002 | C | -0.002 | 0.001 | 0.000 | C | -0.000 | -0.002 | 0.001 |
| C | -0.011 | 0.017 | 0.000 | C | -0.003 | -0.000 | -0.001 | O | -0.006 | -0.009 | 0.007 |
| C | -0.014 | 0.017 | 0.005 | C | -0.002 | -0.001 | -0.001 | C | 0.003 | -0.004 | -0.006 |
| C | -0.023 | 0.006 | 0.013 | C | -0.001 | -0.001 | -0.001 | C | 0.001 | -0.000 | -0.001 |
| H | 0.001 | 0.001 | 0.000 | C | -0.001 | 0.000 | -0.000 | C | -0.000 | -0.001 | 0.000 |
| H | -0.000 | 0.002 | -0.000 | C | -0.001 | -0.001 | 0.001 | C | -0.000 | -0.002 | 0.005 |
| H | -0.001 | -0.002 | 0.000 | C | -0.002 | -0.001 | -0.001 | C | 0.000 | -0.003 | 0.003 |
| H | -0.002 | -0.001 | 0.000 | C | -0.001 | 0.003 | -0.000 | O | 0.002 | -0.008 | 0.006 |
| H | 0.000 | -0.001 | 0.001 | C | -0.002 | 0.000 | -0.001 | AU | -0.033 | 0.042 | -0.030 |
| H | -0.000 | -0.000 | 0.001 | C | -0.002 | -0.003 | 0.004 | H | 0.001 | 0.001 | -0.002 |
| H | 0.000 | -0.001 | 0.001 | H | -0.002 | 0.002 | 0.001 | H | 0.003 | -0.000 | -0.003 |
| H | -0.001 | -0.000 | 0.000 | H | -0.002 | 0.002 | 0.001 | H | -0.009 | 0.013 | -0.002 |
| H | 0.000 | 0.000 | 0.001 | H | -0.004 | -0.008 | -0.003 | H | 0.000 | -0.003 | 0.003 |
| H | -0.001 | 0.000 | -0.000 | H | -0.005 | -0.001 | -0.000 | H | -0.001 | -0.002 | 0.001 |
| H | 0.000 | 0.001 | 0.000 | H | -0.002 | 0.000 | -0.000 | H | 0.000 | 0.000 | 0.000 |
| H | 0.000 | 0.000 | -0.001 | H | -0.003 | -0.001 | -0.000 | H | 0.003 | -0.002 | 0.000 |
| H | -0.000 | -0.000 | 0.000 | H | -0.002 | -0.002 | -0.003 | H | 0.010 | 0.002 | -0.009 |
| H | -0.001 | 0.003 | -0.001 | H | -0.003 | -0.002 | 0.000 | H | -0.001 | -0.003 | -0.003 |
| H | -0.002 | -0.000 | 0.001 | H | -0.023 | -0.003 | -0.001 | H | -0.000 | -0.008 | -0.014 |
| H | 0.002 | 0.001 | 0.000 | H | -0.007 | 0.014 | 0.009 | H | 0.001 | -0.001 | -0.000 |
| H | -0.001 | -0.001 | 0.001 | H | -0.007 | 0.024 | 0.008 | H | 0.001 | 0.000 | -0.000 |
| H | -0.002 | 0.000 | 0.000 | H | -0.023 | 0.031 | 0.009 | H | 0.001 | 0.000 | -0.001 |
| H | -0.000 | 0.000 | -0.000 | H | -0.040 | 0.013 | -0.004 | H | 0.005 | -0.016 | 0.007 |
| H | -0.001 | -0.007 | 0.001 | H | -0.002 | -0.001 | 0.003 | H | -0.003 | 0.013 | -0.002 |
| H | -0.001 | -0.003 | -0.007 | H | -0.000 | 0.002 | 0.006 | H | -0.024 | -0.007 | -0.010 |
| H | -0.001 | 0.001 | 0.001 | H | -0.003 | -0.006 | -0.001 | H | 0.006 | -0.006 | 0.001 |
| H | -0.007 | 0.001 | 0.004 | H | -0.004 | -0.003 | 0.001 | H | 0.010 | -0.010 | 0.001 |
| H | -0.003 | 0.008 | -0.000 | H | -0.004 | -0.001 | 0.001 | H | 0.001 | -0.014 | 0.001 |
| H | -0.002 | 0.007 | -0.002 | H | -0.001 | 0.002 | 0.002 | H | -0.001 | -0.000 | 0.002 |
| H | -0.003 | 0.005 | -0.000 | H | -0.003 | 0.002 | -0.000 | H | 0.001 | 0.002 | -0.002 |
| H | -0.006 | -0.002 | 0.004 | H | 0.001 | 0.001 | -0.004 | H | -0.003 | 0.002 | 0.001 |
| H | 0.001 | -0.003 | 0.000 | H | -0.002 | -0.000 | -0.001 | H | 0.008 | -0.013 | 0.007 |
| H | -0.000 | -0.002 | 0.000 | H | -0.001 | -0.001 | 0.002 | H | 0.012 | -0.009 | 0.010 |
| H | 0.000 | -0.005 | 0.004 | H | -0.002 | -0.000 | 0.000 | H | 0.006 | -0.001 | 0.009 |
| H | -0.000 | -0.006 | 0.004 | H | -0.002 | -0.002 | 0.001 | H | 0.003 | -0.001 | 0.003 |
| C | -0.092 | 0.058 | 0.417 | H | -0.007 | 0.003 | 0.011 | H | 0.003 | -0.003 | -0.005 |
| O | 0.302 | 0.288 | 0.598 | H | -0.002 | -0.004 | -0.002 | H | 0.000 | -0.003 | 0.003 |
| O | -0.231 | -0.092 | -0.272 | H | 0.004 | -0.010 | 0.010 | H | 0.000 | -0.003 | 0.004 |
| H | -0.161 | -0.029 | 0.122 | H | 0.000 | 0.004 | -0.003 | H | 0.000 | -0.003 | 0.003 |
| | | | | H | -0.003 | 0.002 | -0.006 | H | -0.000 | -0.002 | 0.003 |
| | | | | H | -0.001 | -0.000 | -0.001 | H | -0.000 | -0.001 | 0.006 |
| | | | | H | 0.000 | -0.000 | 0.001 | H | -0.001 | -0.002 | 0.002 |
| | | | | H | -0.001 | -0.001 | 0.001 | C | 0.279 | -0.239 | 0.237 |
| | | | | H | -0.001 | -0.001 | 0.001 | O | 0.438 | -0.021 | 0.386 |

| | | | |
|---|---|---|---|
| H | 0.000 | 0.003 | 0.000 |
| H | -0.003 | 0.004 | -0.000 |
| H | 0.000 | 0.003 | 0.000 |
| H | -0.001 | -0.001 | -0.001 |
| H | -0.001 | -0.001 | -0.001 |
| H | -0.001 | -0.000 | -0.001 |
| H | -0.000 | -0.002 | -0.000 |
| H | -0.001 | -0.001 | -0.002 |
| H | -0.002 | -0.001 | -0.000 |
| C | -0.123 | 0.375 | -0.092 |
| O | 0.231 | 0.543 | -0.101 |
| O | -0.136 | -0.205 | 0.104 |
| H | -0.552 | 0.286 | -0.007 |
| O | -0.226 | -0.034 | -0.203 |
| H | 0.005 | -0.598 | -0.012 |

7-P TS

-218.906

------------------------

| | | | |
|---|---|---|---|
| C | 0.001 | 0.007 | 0.006 |
| N | -0.004 | 0.013 | 0.005 |
| C | -0.020 | 0.028 | -0.012 |
| N | -0.011 | 0.017 | 0.012 |
| C | -0.002 | 0.011 | 0.017 |
| C | -0.000 | 0.007 | 0.008 |
| C | -0.006 | 0.003 | -0.000 |
| C | -0.008 | -0.004 | -0.002 |
| C | -0.008 | -0.007 | -0.002 |
| C | -0.004 | -0.002 | -0.000 |
| C | -0.003 | 0.002 | 0.001 |
| C | -0.003 | 0.003 | 0.001 |
| C | -0.005 | 0.001 | -0.001 |
| C | 0.001 | 0.007 | 0.000 |
| C | -0.003 | -0.001 | -0.000 |
| P | -0.017 | 0.052 | 0.021 |
| C | 0.002 | 0.012 | 0.007 |
| C | 0.004 | 0.004 | 0.001 |
| C | 0.006 | 0.002 | 0.000 |
| C | 0.007 | 0.002 | 0.001 |
| C | 0.009 | 0.005 | 0.001 |
| C | 0.005 | 0.010 | 0.005 |
| C | -0.005 | 0.010 | 0.004 |
| C | -0.000 | 0.004 | -0.002 |
| C | 0.002 | 0.001 | -0.003 |
| C | 0.001 | 0.001 | -0.001 |
| C | -0.001 | 0.001 | -0.000 |
| C | -0.003 | 0.004 | 0.002 |
| C | 0.004 | 0.000 | -0.009 |
| C | 0.002 | -0.001 | -0.001 |
| C | -0.003 | 0.003 | 0.001 |
| C | -0.002 | 0.002 | 0.007 |
| C | 0.002 | 0.009 | 0.004 |
| O | 0.002 | 0.008 | 0.006 |
| AU | -0.009 | -0.040 | -0.050 |
| AU | -0.009 | -0.040 | -0.050 |
| H | -0.000 | -0.002 | -0.001 |
| H | 0.005 | -0.002 | -0.006 |
| H | -0.008 | -0.005 | -0.001 |

8-P TS

-195.866

------------------------

| | | | |
|---|---|---|---|
| C | -0.002 | 0.017 | -0.010 |
| C | 0.003 | 0.018 | -0.002 |
| C | 0.001 | 0.009 | 0.001 |
| C | -0.003 | 0.006 | -0.001 |
| C | -0.007 | 0.008 | -0.004 |
| C | -0.006 | 0.015 | -0.008 |
| P | 0.049 | 0.040 | -0.007 |
| AU | -0.024 | -0.069 | -0.011 |
| C | 0.039 | 0.009 | -0.005 |
| N | 0.028 | 0.017 | 0.002 |
| C | 0.015 | 0.020 | 0.009 |
| C | 0.013 | 0.014 | 0.005 |
| N | 0.021 | 0.008 | -0.001 |
| C | 0.023 | 0.015 | 0.002 |
| C | 0.011 | 0.006 | -0.008 |
| C | 0.006 | 0.009 | 0.009 |
| C | 0.011 | 0.021 | 0.011 |
| H | -0.003 | 0.013 | -0.017 |
| H | -0.010 | 0.017 | -0.014 |
| H | -0.012 | 0.006 | -0.006 |
| H | -0.004 | 0.003 | -0.001 |
| H | 0.006 | 0.006 | 0.004 |
| H | 0.502 | 0.333 | -0.103 |
| H | 0.004 | 0.011 | 0.016 |
| H | 0.003 | 0.003 | 0.008 |
| H | 0.006 | 0.007 | 0.006 |
| H | 0.010 | 0.024 | 0.010 |
| H | 0.013 | 0.023 | 0.012 |
| H | 0.010 | 0.019 | 0.009 |
| H | 0.020 | 0.038 | 0.003 |
| H | 0.043 | 0.002 | 0.003 |
| H | 0.001 | 0.000 | -0.001 |
| H | 0.002 | 0.005 | -0.007 |
| H | 0.013 | 0.006 | -0.018 |
| H | 0.008 | 0.005 | -0.004 |
| C | 0.110 | 0.389 | 0.072 |
| O | 0.137 | -0.218 | -0.124 |
| O | -0.178 | 0.520 | 0.220 |

9-P TS

-211.333

------------------------

| | | | |
|---|---|---|---|
| C | -0.023 | -0.002 | 0.012 |
| C | -0.034 | 0.006 | 0.021 |
| C | -0.022 | -0.006 | 0.014 |
| C | -0.015 | -0.016 | 0.007 |
| C | -0.012 | -0.017 | 0.003 |
| C | -0.014 | -0.013 | 0.005 |
| P | -0.027 | -0.007 | 0.052 |
| AU | 0.022 | 0.017 | -0.062 |
| C | -0.016 | 0.007 | 0.016 |
| N | -0.009 | 0.015 | 0.011 |
| C | -0.003 | 0.016 | 0.006 |
| C | -0.001 | 0.009 | 0.001 |
| N | -0.007 | 0.003 | 0.006 |
| C | -0.006 | 0.010 | 0.007 |
| C | -0.004 | 0.004 | 0.004 |
| C | -0.002 | -0.001 | 0.001 |
| C | -0.000 | -0.000 | 0.001 |
| C | 0.000 | 0.000 | 0.002 |
| C | -0.002 | 0.004 | 0.004 |
| C | -0.003 | 0.001 | 0.003 |
| C | 0.000 | 0.002 | 0.003 |
| C | 0.000 | -0.001 | -0.000 |
| C | -0.007 | -0.002 | 0.001 |
| C | -0.005 | 0.001 | 0.002 |
| C | -0.005 | 0.002 | 0.001 |
| C | -0.004 | 0.001 | -0.001 |
| C | -0.004 | -0.003 | -0.002 |
| C | -0.005 | -0.006 | -0.002 |
| C | -0.004 | 0.000 | 0.001 |
| C | -0.003 | 0.007 | -0.000 |
| C | -0.002 | -0.004 | 0.000 |
| H | 0.002 | 0.022 | 0.005 |
| H | 0.003 | 0.008 | -0.004 |
| H | 0.003 | -0.002 | 0.000 |
| H | -0.001 | -0.004 | -0.000 |
| H | -0.003 | 0.002 | 0.001 |
| H | -0.006 | 0.004 | 0.001 |
| H | 0.001 | 0.001 | 0.003 |
| H | -0.002 | 0.003 | 0.002 |
| H | -0.004 | 0.003 | 0.002 |
| H | -0.003 | 0.001 | 0.001 |
| H | -0.001 | 0.004 | 0.001 |
| H | 0.011 | -0.006 | -0.012 |
| H | 0.006 | -0.001 | -0.006 |
| H | 0.001 | 0.004 | -0.018 |
| H | 0.002 | -0.001 | -0.001 |
| H | 0.002 | -0.000 | -0.001 |
| H | 0.001 | -0.001 | -0.000 |
| H | -0.006 | 0.002 | -0.000 |
| H | -0.005 | 0.003 | 0.005 |
| H | -0.004 | 0.004 | -0.007 |
| H | 0.003 | 0.006 | 0.001 |
| H | 0.005 | 0.007 | 0.001 |
| H | 0.000 | 0.011 | -0.001 |
| H | -0.002 | 0.002 | 0.001 |
| H | -0.003 | -0.000 | -0.004 |
| H | -0.004 | -0.002 | 0.001 |
| H | 0.002 | 0.011 | 0.008 |
| H | 0.011 | 0.005 | 0.002 |
| H | 0.008 | -0.000 | 0.002 |
| H | 0.006 | 0.001 | -0.000 |
| H | 0.001 | 0.005 | -0.000 |
| H | 0.002 | 0.012 | 0.005 |
| H | 0.002 | 0.011 | 0.004 |
| H | 0.003 | 0.008 | 0.002 |
| H | -0.002 | 0.002 | 0.006 |
| H | -0.002 | -0.000 | 0.010 |
| H | -0.002 | 0.001 | 0.004 |
| H | -0.001 | 0.016 | 0.022 |
| H | 0.003 | 0.015 | 0.017 |
| C | 0.119 | 0.245 | 0.311 |
| O | 0.277 | -0.002 | 0.511 |
| O | -0.187 | 0.042 | -0.226 |
| H | -0.134 | 0.594 | 0.127 |

10-P TS

-209.877

------------------------

| | | | |
|---|---|---|---|
| C | 0.001 | 0.002 | -0.001 |
| C | 0.001 | 0.007 | 0.007 |
| C | 0.001 | 0.003 | 0.002 |
| C | 0.001 | 0.001 | -0.001 |
| C | 0.001 | 0.000 | -0.002 |
| C | 0.001 | -0.000 | -0.003 |
| N | -0.000 | 0.012 | 0.012 |
| C | -0.003 | 0.012 | 0.008 |
| N | -0.001 | 0.004 | 0.004 |
| C | 0.001 | 0.002 | 0.008 |
| C | 0.004 | 0.007 | 0.012 |
| P | -0.039 | 0.025 | 0.023 |
| AU | 0.033 | -0.051 | -0.026 |
| C | -0.001 | 0.002 | 0.001 |
| C | -0.000 | 0.002 | 0.002 |
| C | -0.000 | 0.002 | 0.001 |

11-P TS

-203.271

------------------------

| | | | |
|---|---|---|---|
| C | -0.005 | 0.001 | -0.001 |
| C | -0.006 | -0.006 | -0.006 |
| C | -0.003 | -0.003 | -0.001 |
| C | -0.001 | 0.000 | 0.001 |
| C | -0.001 | 0.002 | 0.000 |
| C | -0.003 | 0.004 | -0.000 |
| N | -0.008 | -0.013 | -0.010 |
| C | -0.025 | -0.023 | -0.003 |
| N | -0.012 | -0.011 | -0.009 |
| C | -0.002 | -0.007 | -0.014 |
| C | -0.003 | -0.006 | -0.015 |
| P | -0.032 | -0.042 | -0.030 |
| AU | 0.004 | 0.032 | 0.061 |
| C | -0.004 | -0.007 | -0.004 |
| C | -0.000 | -0.004 | 0.002 |
| C | 0.002 | -0.002 | 0.003 |
| H | -0.000 | 0.000 | 0.004 |
| H | 0.002 | 0.002 | 0.002 |
| H | -0.003 | 0.001 | 0.003 |
| H | -0.000 | 0.001 | 0.004 |
| H | -0.004 | -0.000 | 0.001 |
| H | -0.000 | -0.001 | -0.001 |
| H | 0.001 | -0.001 | -0.000 |
| H | 0.001 | -0.001 | -0.001 |
| H | -0.002 | -0.007 | -0.002 |
| H | -0.005 | -0.001 | 0.004 |
| H | 0.003 | -0.002 | 0.002 |
| H | -0.003 | 0.002 | 0.002 |
| H | -0.006 | -0.001 | 0.001 |
| H | -0.001 | -0.001 | 0.001 |
| H | -0.003 | 0.015 | 0.003 |
| H | 0.001 | 0.009 | -0.012 |
| H | -0.004 | 0.002 | 0.003 |
| H | -0.023 | -0.003 | 0.014 |
| H | -0.012 | -0.025 | 0.004 |
| H | -0.008 | -0.023 | -0.001 |
| H | -0.010 | -0.019 | 0.002 |
| H | -0.022 | 0.004 | 0.014 |
| C | -0.093 | -0.078 | 0.376 |
| O | 0.225 | -0.250 | 0.471 |
| O | -0.175 | 0.088 | -0.214 |
| H | -0.509 | 0.064 | 0.377 |

12-P TS

-197.895

------------------------

| | | | |
|---|---|---|---|
| C | 0.006 | -0.003 | -0.002 |
| C | 0.003 | -0.012 | 0.004 |
| C | 0.000 | -0.008 | 0.002 |
| C | 0.002 | -0.002 | 0.000 |
| C | 0.003 | -0.000 | -0.000 |
| C | 0.006 | 0.001 | -0.002 |
| N | 0.000 | -0.022 | 0.008 |
| C | 0.006 | -0.038 | -0.001 |
| N | 0.004 | -0.021 | 0.009 |
| C | -0.002 | -0.011 | 0.019 |
| C | -0.001 | -0.012 | 0.019 |
| P | -0.001 | -0.055 | 0.024 |
| AU | 0.021 | 0.033 | -0.056 |
| C | -0.000 | -0.011 | 0.004 |
| C | -0.003 | -0.006 | -0.002 |
| C | -0.005 | -0.002 | -0.003 |

| | | | | | | | | | | | |
|---|---|---|---|---|---|---|---|---|---|---|---|
| C | -0.000 | 0.001 | -0.000 | C | 0.001 | -0.000 | 0.001 | C | -0.003 | -0.001 | -0.000 |
| C | -0.001 | 0.001 | -0.000 | C | -0.001 | 0.001 | 0.001 | C | -0.000 | -0.001 | -0.000 |
| C | -0.001 | 0.001 | 0.000 | C | -0.003 | -0.002 | -0.002 | C | 0.001 | -0.005 | 0.002 |
| C | 0.001 | 0.001 | 0.002 | CL | -0.004 | -0.002 | -0.011 | C | 0.001 | -0.007 | 0.017 |
| C | 0.000 | -0.001 | 0.002 | CL | 0.002 | -0.007 | -0.010 | C | -0.004 | -0.008 | 0.018 |
| C | -0.001 | 0.001 | 0.000 | C | 0.002 | -0.004 | 0.004 | C | -0.007 | -0.001 | -0.010 |
| C | 0.000 | 0.001 | -0.000 | C | 0.004 | -0.004 | 0.005 | C | -0.003 | 0.001 | 0.000 |
| C | 0.001 | 0.002 | 0.002 | C | -0.003 | -0.001 | -0.001 | C | 0.003 | -0.004 | 0.001 |
| C | 0.001 | 0.002 | 0.001 | C | -0.001 | -0.001 | -0.001 | C | -0.002 | -0.010 | 0.001 |
| C | 0.001 | -0.002 | -0.005 | C | -0.002 | -0.003 | 0.000 | C | 0.003 | 0.001 | 0.000 |
| C | 0.010 | -0.008 | -0.001 | C | -0.002 | -0.003 | -0.001 | C | 0.009 | 0.007 | -0.009 |
| C | -0.038 | -0.001 | 0.011 | C | -0.007 | 0.004 | 0.000 | C | -0.007 | -0.009 | 0.010 |
| C | -0.038 | 0.006 | 0.013 | C | -0.003 | 0.005 | 0.000 | C | -0.006 | -0.001 | 0.002 |
| C | -0.031 | 0.014 | 0.011 | C | -0.002 | -0.010 | -0.011 | C | -0.007 | 0.002 | 0.001 |
| C | -0.018 | 0.014 | 0.007 | C | 0.004 | -0.004 | -0.003 | C | -0.008 | 0.003 | 0.002 |
| C | -0.015 | 0.007 | 0.005 | C | 0.007 | -0.002 | -0.001 | C | -0.011 | -0.001 | 0.004 |
| C | -0.022 | -0.001 | 0.004 | C | 0.007 | -0.002 | -0.001 | C | -0.009 | -0.006 | 0.008 |
| C | -0.003 | 0.002 | -0.001 | C | 0.007 | -0.005 | -0.002 | H | 0.000 | 0.001 | -0.001 |
| C | -0.002 | 0.000 | -0.001 | C | 0.002 | -0.008 | -0.008 | H | -0.008 | 0.001 | -0.006 |
| C | -0.000 | 0.001 | 0.001 | C | -0.000 | -0.003 | 0.001 | H | 0.009 | 0.005 | -0.004 |
| C | -0.011 | 0.017 | 0.002 | C | -0.002 | -0.001 | -0.001 | H | 0.002 | -0.000 | -0.001 |
| H | 0.001 | -0.003 | 0.007 | C | -0.001 | -0.005 | 0.001 | H | 0.005 | -0.005 | 0.003 |
| H | 0.009 | 0.007 | 0.015 | C | -0.008 | 0.007 | -0.001 | H | 0.004 | -0.002 | -0.002 |
| H | -0.001 | 0.001 | -0.000 | H | -0.000 | 0.003 | 0.002 | H | -0.001 | -0.005 | 0.000 |
| H | -0.001 | 0.002 | 0.001 | H | 0.004 | -0.001 | 0.006 | H | -0.010 | 0.001 | -0.010 |
| H | 0.001 | -0.001 | -0.006 | H | -0.003 | 0.007 | 0.001 | H | -0.013 | 0.002 | -0.012 |
| H | 0.001 | -0.001 | -0.002 | H | 0.001 | 0.001 | 0.003 | H | -0.001 | -0.001 | -0.018 |
| H | -0.001 | 0.001 | 0.001 | H | -0.002 | -0.001 | -0.002 | H | -0.004 | 0.001 | -0.000 |
| H | 0.002 | 0.002 | 0.003 | H | 0.002 | -0.003 | 0.007 | H | -0.003 | 0.001 | -0.000 |
| H | -0.002 | -0.007 | -0.013 | H | -0.009 | 0.002 | -0.002 | H | -0.003 | 0.002 | 0.001 |
| H | 0.000 | 0.002 | 0.002 | H | -0.002 | -0.003 | 0.001 | H | 0.009 | -0.006 | 0.001 |
| H | -0.018 | -0.011 | -0.003 | H | -0.002 | -0.009 | -0.011 | H | -0.012 | 0.024 | -0.014 |
| H | -0.010 | 0.007 | 0.003 | H | 0.009 | -0.004 | -0.002 | H | 0.031 | 0.020 | -0.015 |
| H | -0.014 | 0.020 | 0.007 | H | 0.009 | -0.001 | -0.000 | H | -0.004 | -0.009 | 0.002 |
| H | -0.035 | 0.023 | 0.014 | H | 0.008 | -0.002 | 0.000 | H | -0.003 | -0.010 | 0.001 |
| H | -0.041 | 0.006 | 0.017 | H | 0.002 | -0.004 | -0.003 | H | -0.002 | -0.011 | 0.001 |
| H | -0.583 | 0.053 | 0.284 | H | 0.003 | 0.001 | 0.002 | H | 0.002 | 0.001 | 0.001 |
| H | -0.000 | 0.001 | -0.001 | H | -0.004 | 0.001 | -0.018 | H | 0.003 | 0.001 | -0.000 |
| H | -0.006 | 0.010 | -0.003 | H | 0.013 | -0.004 | 0.003 | H | 0.003 | 0.001 | 0.001 |
| H | -0.001 | 0.002 | -0.000 | H | -0.010 | -0.003 | 0.008 | H | -0.006 | -0.009 | 0.013 |
| H | -0.007 | -0.002 | -0.001 | H | 0.005 | -0.006 | 0.008 | H | -0.014 | 0.001 | 0.005 |
| H | 0.002 | -0.004 | 0.001 | H | 0.007 | -0.001 | 0.005 | H | -0.009 | 0.006 | 0.002 |
| H | 0.001 | -0.000 | 0.002 | H | 0.001 | -0.003 | 0.004 | H | -0.007 | 0.003 | 0.000 |
| H | -0.001 | -0.001 | 0.003 | H | -0.002 | -0.001 | -0.001 | H | -0.004 | -0.002 | 0.001 |
| H | -0.001 | 0.000 | -0.001 | H | -0.002 | 0.000 | 0.000 | H | 0.006 | -0.009 | 0.018 |
| H | -0.002 | -0.001 | -0.003 | H | -0.001 | -0.001 | -0.000 | H | 0.000 | -0.006 | 0.017 |
| H | -0.005 | 0.001 | -0.000 | H | 0.001 | -0.002 | -0.002 | H | 0.003 | -0.003 | 0.015 |
| H | 0.001 | 0.001 | 0.000 | H | -0.001 | 0.001 | -0.002 | H | -0.008 | -0.008 | 0.020 |
| H | 0.001 | 0.001 | 0.000 | H | -0.002 | -0.002 | -0.001 | H | -0.003 | -0.008 | 0.017 |
| H | 0.000 | 0.000 | -0.001 | H | 0.001 | 0.004 | 0.001 | H | -0.005 | -0.005 | 0.014 |
| H | 0.001 | -0.001 | -0.003 | H | 0.001 | -0.008 | -0.013 | C | -0.141 | -0.168 | 0.340 |
| H | -0.033 | 0.028 | 0.013 | H | -0.010 | 0.007 | 0.012 | O | -0.231 | 0.114 | 0.537 |
| H | 0.001 | 0.033 | 0.002 | H | -0.010 | 0.021 | -0.009 | O | 0.114 | -0.103 | -0.238 |
| H | -0.027 | 0.018 | 0.001 | H | -0.007 | 0.006 | 0.000 | H | -0.085 | -0.574 | 0.208 |
| H | 0.005 | -0.007 | -0.003 | H | -0.001 | 0.000 | 0.001 | | | | |
| H | 0.013 | -0.015 | -0.002 | H | -0.000 | 0.008 | 0.002 | | | | |
| H | 0.015 | -0.003 | 0.001 | H | -0.001 | -0.004 | 0.001 | | | | |
| H | -0.000 | -0.001 | 0.001 | H | -0.001 | -0.005 | 0.002 | | | | |
| H | -0.002 | 0.002 | 0.002 | H | -0.002 | -0.006 | 0.000 | | | | |
| H | 0.001 | 0.002 | 0.001 | H | -0.002 | -0.004 | -0.000 | | | | |
| H | 0.001 | 0.002 | 0.001 | H | -0.001 | -0.004 | -0.001 | | | | |
| H | 0.001 | 0.002 | 0.001 | H | -0.002 | -0.002 | -0.001 | | | | |
| H | 0.002 | 0.002 | 0.001 | C | 0.014 | -0.187 | -0.362 | | | | |
| C | -0.229 | 0.283 | 0.108 | O | 0.240 | 0.017 | -0.541 | | | | |
| O | 0.050 | 0.561 | 0.024 | O | -0.156 | -0.045 | 0.241 | | | | |
| O | -0.035 | -0.288 | 0.075 | H | -0.259 | -0.508 | -0.240 | | | | |
| 13-P TS | | | | 1 TS | | | | 2 TS | | | |
| | | -197.025 | | | | -183.939 | | | | -186.434 | |
| | ------------------------ | | | | ------------------------ | | | | ------------------------ | | |
| N | 0.018 | -0.021 | -0.015 | C | 0.034 | 0.033 | 0.151 | C | -0.004 | -0.007 | -0.009 |
| C | 0.035 | -0.018 | -0.008 | N | 0.020 | 0.016 | 0.027 | C | -0.001 | -0.017 | -0.013 |
| N | 0.018 | -0.013 | -0.009 | C | 0.026 | 0.007 | -0.006 | C | 0.007 | -0.011 | -0.008 |
| C | 0.003 | -0.017 | -0.017 | C | 0.024 | 0.011 | 0.014 | C | 0.008 | -0.004 | -0.004 |
| C | 0.001 | -0.021 | -0.022 | C | 0.032 | 0.019 | 0.058 | C | 0.002 | 0.000 | -0.001 |
| P | 0.037 | -0.047 | -0.021 | C | 0.010 | 0.009 | 0.013 | C | -0.003 | 0.001 | -0.004 |
| AU | -0.008 | 0.061 | 0.040 | C | 0.004 | 0.013 | 0.008 | N | 0.000 | -0.029 | -0.027 |
| C | 0.009 | -0.010 | 0.001 | C | -0.001 | 0.011 | 0.004 | C | -0.000 | -0.027 | 0.003 |
| C | 0.001 | -0.005 | 0.009 | C | -0.002 | 0.004 | 0.001 | C | 0.001 | -0.035 | -0.011 |
| C | 0.012 | -0.011 | -0.011 | C | -0.002 | 0.000 | 0.004 | C | -0.007 | -0.029 | -0.058 |
| C | 0.006 | -0.015 | -0.003 | C | 0.004 | 0.002 | 0.009 | C | -0.010 | -0.041 | -0.142 |
| C | -0.000 | -0.016 | -0.007 | C | 0.004 | 0.011 | 0.006 | C | -0.040 | 0.008 | -0.041 |
| C | 0.001 | -0.006 | -0.005 | C | 0.006 | 0.011 | 0.007 | C | -0.034 | 0.005 | -0.011 |
| C | -0.003 | -0.003 | -0.000 | C | 0.001 | -0.000 | 0.008 | C | -0.007 | -0.001 | 0.008 |
| C | -0.008 | -0.006 | 0.002 | C | -0.007 | 0.007 | -0.009 | C | 0.023 | -0.014 | -0.007 |
| C | -0.008 | -0.014 | 0.003 | C | 0.025 | -0.001 | -0.004 | C | 0.014 | -0.033 | -0.041 |
| C | -0.005 | -0.017 | 0.002 | C | 0.014 | 0.003 | -0.011 | C | -0.000 | -0.013 | 0.006 |
| C | 0.002 | -0.001 | -0.003 | C | 0.028 | 0.063 | 0.034 | C | -0.007 | -0.024 | 0.006 |
| C | 0.007 | -0.001 | -0.003 | C | 0.015 | -0.031 | 0.038 | AU | 0.015 | -0.000 | 0.078 |
| H | -0.005 | -0.016 | -0.018 | AU | 0.005 | -0.004 | -0.078 | C | 0.006 | -0.008 | -0.006 |
| H | -0.011 | -0.024 | -0.029 | C | -0.005 | -0.007 | 0.013 | C | 0.007 | -0.011 | -0.007 |
| H | -0.004 | -0.017 | 0.007 | C | 0.007 | 0.003 | 0.001 | C | -0.005 | -0.002 | -0.009 |
| H | -0.013 | -0.016 | 0.008 | H | 0.270 | 0.181 | 0.410 | C | 0.003 | 0.002 | 0.005 |
| H | -0.012 | -0.003 | 0.006 | H | -0.006 | -0.004 | 0.003 | C | -0.002 | -0.009 | -0.001 |
| H | -0.003 | 0.000 | 0.001 | H | -0.005 | 0.014 | 0.001 | C | -0.008 | 0.005 | -0.013 |
| H | 0.005 | -0.004 | -0.007 | H | -0.001 | 0.011 | 0.005 | H | 0.013 | 0.000 | -0.002 |
| H | 0.013 | -0.016 | 0.007 | H | 0.047 | 0.008 | 0.002 | H | -0.005 | 0.007 | -0.003 |
| H | -0.005 | -0.003 | 0.020 | H | 0.007 | 0.064 | 0.001 | H | -0.007 | -0.010 | -0.014 |
| H | -0.006 | 0.001 | 0.001 | H | 0.034 | 0.101 | 0.048 | H | 0.005 | -0.018 | 0.004 |
| H | 0.007 | -0.013 | 0.013 | H | 0.032 | 0.051 | 0.042 | H | -0.002 | -0.012 | 0.016 |
| H | -0.004 | -0.001 | 0.002 | H | 0.010 | -0.068 | 0.055 | H | -0.000 | -0.009 | 0.006 |
| H | 0.004 | -0.001 | -0.006 | H | 0.017 | -0.035 | -0.003 | H | 0.000 | -0.007 | 0.001 |
| H | 0.001 | 0.003 | -0.006 | H | 0.017 | -0.022 | 0.052 | H | -0.014 | -0.018 | 0.007 |
| H | 0.017 | -0.005 | -0.017 | H | 0.027 | -0.004 | 0.001 | H | -0.006 | -0.025 | 0.013 |
| H | 0.005 | 0.007 | 0.001 | H | 0.019 | -0.004 | -0.005 | H | -0.006 | -0.025 | -0.002 |
| H | 0.012 | 0.004 | -0.006 | H | 0.021 | -0.002 | -0.006 | H | -0.005 | -0.066 | -0.006 |
| H | -0.001 | -0.005 | 0.002 | H | 0.009 | 0.010 | -0.009 | H | 0.003 | 0.005 | 0.004 |
| H | 0.001 | -0.012 | -0.002 | H | 0.007 | -0.001 | -0.016 | H | -0.009 | 0.008 | -0.014 |
| H | 0.005 | -0.017 | -0.002 | H | 0.021 | -0.005 | -0.014 | H | -0.009 | 0.006 | -0.009 |
| H | 0.007 | -0.019 | -0.002 | H | -0.005 | 0.011 | 0.011 | H | -0.007 | 0.003 | -0.016 |
| C | 0.031 | -0.337 | -0.230 | H | -0.006 | 0.003 | -0.004 | H | 0.005 | 0.009 | 0.006 |

| | | | |
|---|---|---|---|
| O | -0.279 | -0.337 | -0.387 |
| O | 0.190 | 0.119 | 0.189 |
| H | 0.425 | -0.442 | -0.081 |

3 TS

-188.966

------------------------

| | | | |
|---|---|---|---|
| C | -0.000 | 0.006 | -0.008 |
| C | 0.005 | 0.014 | -0.012 |
| C | 0.010 | 0.008 | -0.007 |
| C | 0.010 | 0.003 | -0.004 |
| C | 0.004 | -0.000 | -0.001 |
| C | -0.000 | -0.001 | -0.004 |
| N | 0.008 | 0.026 | -0.024 |
| C | 0.005 | 0.026 | 0.005 |
| C | 0.008 | 0.031 | -0.011 |
| C | 0.003 | 0.027 | -0.047 |
| C | 0.011 | 0.036 | -0.135 |
| C | -0.029 | 0.004 | -0.034 |
| C | -0.025 | 0.003 | -0.013 |
| C | -0.005 | 0.001 | -0.003 |
| C | 0.016 | 0.005 | -0.011 |
| C | 0.015 | 0.017 | -0.023 |
| C | 0.014 | 0.004 | -0.017 |
| C | 0.007 | 0.004 | -0.016 |
| C | -0.032 | 0.002 | -0.009 |
| C | 0.003 | 0.013 | 0.007 |
| C | -0.003 | 0.023 | 0.005 |
| AU | 0.006 | 0.001 | 0.080 |
| C | 0.008 | 0.006 | -0.005 |
| C | 0.008 | 0.010 | -0.007 |
| C | -0.002 | 0.002 | -0.008 |
| C | 0.005 | -0.003 | 0.006 |
| C | -0.003 | 0.009 | 0.002 |
| C | -0.005 | -0.004 | -0.011 |
| C | 0.009 | -0.000 | -0.012 |
| H | 0.006 | 0.010 | 0.007 |
| H | 0.004 | 0.011 | 0.008 |
| H | 0.006 | 0.010 | 0.007 |
| H | 0.013 | 0.002 | -0.001 |
| H | 0.001 | -0.001 | -0.004 |
| H | 0.017 | 0.003 | -0.000 |
| H | -0.011 | 0.001 | -0.006 |
| H | -0.021 | 0.003 | -0.020 |
| H | 0.001 | 0.022 | -0.016 |
| H | 0.010 | 0.001 | 0.014 |
| H | -0.006 | -0.010 | 0.014 |
| H | -0.004 | -0.006 | 0.016 |
| H | -0.006 | -0.009 | 0.009 |
| C | -0.102 | -0.023 | 0.435 |
| O | 0.174 | 0.087 | -0.135 |
| O | -0.373 | -0.175 | 0.452 |

4 TS

-216.388

------------------------

| | | | |
|---|---|---|---|
| C | -0.012 | 0.012 | -0.016 |
| C | -0.002 | 0.012 | -0.025 |
| C | 0.007 | 0.000 | -0.017 |
| C | 0.004 | -0.005 | -0.007 |
| C | -0.003 | 0.001 | -0.002 |
| C | -0.011 | 0.008 | -0.007 |
| N | -0.002 | 0.020 | -0.053 |
| C | -0.002 | 0.034 | -0.174 |
| N | -0.001 | 0.026 | -0.051 |
| C | -0.000 | 0.018 | 0.034 |
| C | 0.002 | 0.023 | 0.036 |
| C | 0.001 | 0.014 | -0.024 |
| C | -0.008 | 0.001 | -0.017 |
| C | -0.005 | -0.005 | -0.008 |
| C | 0.003 | 0.001 | -0.002 |
| C | 0.011 | 0.009 | -0.007 |
| C | 0.011 | 0.013 | -0.016 |
| C | -0.012 | 0.000 | -0.019 |
| C | 0.017 | 0.006 | -0.018 |
| C | 0.003 | -0.000 | 0.007 |
| AU | 0.001 | 0.002 | 0.077 |
| C | 0.011 | 0.000 | -0.019 |
| C | -0.003 | -0.000 | 0.007 |
| C | -0.016 | 0.007 | -0.018 |
| H | -0.019 | 0.009 | -0.002 |
| H | 0.009 | -0.011 | -0.004 |
| H | -0.010 | -0.011 | -0.004 |
| H | 0.019 | 0.009 | -0.002 |
| H | -0.022 | -0.006 | -0.017 |
| H | 0.005 | -0.006 | 0.013 |
| H | 0.006 | 0.010 | 0.008 |
| H | -0.005 | -0.014 | 0.002 |
| H | -0.005 | -0.017 | 0.002 |
| H | -0.003 | -0.001 | 0.005 |
| H | 0.008 | -0.003 | -0.006 |
| H | 0.006 | -0.010 | -0.007 |
| H | 0.006 | -0.011 | -0.006 |
| H | 0.007 | -0.010 | -0.007 |
| H | -0.052 | 0.020 | -0.072 |
| H | -0.056 | 0.022 | -0.027 |
| H | -0.056 | 0.009 | 0.001 |
| H | -0.026 | 0.003 | -0.012 |
| H | -0.006 | 0.004 | 0.032 |
| H | -0.008 | 0.000 | 0.011 |
| H | 0.047 | -0.019 | -0.004 |
| H | 0.030 | -0.010 | 0.005 |
| H | -0.006 | -0.030 | -0.041 |
| H | 0.038 | -0.043 | -0.053 |
| C | -0.089 | 0.085 | -0.426 |
| O | -0.128 | 0.398 | -0.442 |
| O | 0.052 | -0.198 | 0.131 |
| H | 0.037 | -0.321 | -0.427 |

5 TS

-217.261

------------------------

| | | | |
|---|---|---|---|
| C | -0.004 | -0.012 | 0.003 |
| C | 0.003 | -0.018 | 0.014 |
| C | 0.011 | -0.012 | 0.010 |
| C | 0.012 | -0.006 | 0.005 |
| C | 0.004 | -0.001 | -0.000 |
| C | -0.002 | -0.006 | -0.003 |
| N | 0.002 | -0.039 | 0.024 |
| C | -0.005 | -0.157 | 0.056 |
| N | 0.002 | -0.029 | 0.032 |
| C | 0.004 | 0.030 | 0.011 |
| C | 0.001 | 0.042 | 0.018 |
| C | -0.002 | -0.016 | 0.011 |
| C | 0.004 | -0.010 | 0.002 |
| C | 0.002 | -0.007 | -0.004 |
| C | -0.004 | -0.003 | -0.000 |
| C | -0.010 | -0.006 | 0.005 |
| C | -0.010 | -0.011 | 0.008 |
| C | 0.003 | -0.009 | 0.002 |
| C | 0.003 | -0.005 | 0.003 |
| C | -0.009 | -0.010 | 0.004 |
| C | -0.010 | -0.010 | 0.006 |
| AU | 0.001 | 0.075 | -0.007 |
| C | 0.010 | -0.009 | 0.005 |
| C | 0.009 | -0.010 | 0.007 |
| C | -0.007 | -0.012 | 0.001 |
| C | -0.010 | -0.014 | -0.003 |
| C | 0.001 | -0.006 | 0.002 |
| C | 0.003 | -0.002 | -0.003 |
| C | 0.001 | -0.003 | 0.005 |
| H | 0.014 | -0.001 | -0.001 |
| H | -0.002 | -0.006 | -0.002 |
| H | -0.002 | 0.009 | -0.013 |
| H | 0.011 | 0.016 | -0.001 |
| H | -0.001 | 0.015 | 0.016 |
| H | 0.002 | 0.009 | 0.007 |
| H | 0.004 | 0.009 | 0.003 |
| H | -0.007 | 0.018 | 0.006 |
| H | -0.003 | 0.020 | 0.009 |
| H | -0.004 | 0.024 | -0.001 |
| H | 0.001 | 0.055 | -0.007 |
| H | 0.004 | -0.004 | 0.004 |
| H | -0.007 | -0.006 | -0.012 |
| H | -0.006 | -0.005 | -0.008 |
| H | -0.004 | -0.003 | -0.014 |
| H | 0.002 | -0.008 | 0.005 |
| H | 0.011 | 0.005 | 0.012 |
| H | 0.005 | -0.012 | 0.013 |
| H | -0.006 | 0.015 | 0.004 |
| H | -0.005 | 0.020 | 0.004 |
| H | -0.005 | 0.000 | 0.009 |
| H | 0.009 | 0.001 | -0.004 |
| H | 0.007 | 0.009 | -0.006 |
| H | 0.008 | 0.010 | -0.006 |
| H | 0.008 | 0.009 | -0.007 |
| H | -0.035 | -0.003 | -0.057 |
| H | -0.041 | -0.005 | -0.028 |
| H | -0.023 | 0.003 | -0.010 |
| H | -0.007 | -0.005 | 0.015 |
| H | -0.002 | 0.002 | -0.005 |
| H | 0.029 | 0.006 | -0.010 |
| H | 0.019 | 0.002 | -0.002 |
| H | 0.002 | 0.021 | -0.010 |
| H | 0.014 | 0.005 | -0.009 |
| H | 0.006 | -0.006 | -0.010 |
| H | 0.009 | -0.004 | -0.015 |
| H | 0.008 | 0.004 | -0.005 |
| H | 0.001 | 0.007 | -0.015 |
| H | 0.008 | 0.002 | -0.008 |
| H | 0.005 | 0.006 | -0.019 |
| H | -0.033 | -0.000 | -0.006 |
| H | -0.030 | 0.002 | -0.009 |
| H | -0.035 | 0.003 | -0.011 |
| C | -0.044 | -0.088 | -0.439 |
| O | -0.133 | -0.392 | -0.452 |
| O | 0.060 | 0.200 | 0.135 |
| H | 0.042 | 0.331 | -0.428 |

6 TS

-173.620

------------------------

| | | | |
|---|---|---|---|
| C | -0.005 | 0.003 | 0.020 |
| C | 0.002 | 0.017 | 0.029 |
| C | 0.013 | 0.019 | 0.021 |
| C | 0.014 | 0.015 | 0.010 |
| C | 0.006 | 0.005 | 0.003 |
| H | -0.021 | 0.010 | -0.031 |
| H | -0.009 | 0.008 | -0.008 |
| H | 0.020 | 0.021 | -0.018 |
| H | -0.007 | -0.005 | -0.007 |
| H | 0.023 | -0.014 | -0.034 |
| H | 0.001 | -0.002 | 0.008 |
| H | -0.002 | -0.000 | 0.009 |
| H | -0.006 | 0.002 | 0.008 |
| H | -0.019 | 0.015 | -0.016 |
| H | -0.021 | -0.009 | -0.030 |
| H | 0.002 | -0.005 | -0.012 |
| H | 0.024 | -0.010 | -0.020 |
| H | 0.008 | 0.006 | -0.006 |
| H | 0.021 | 0.012 | -0.034 |
| H | 0.007 | 0.001 | 0.009 |
| H | 0.001 | -0.001 | 0.008 |
| H | -0.001 | -0.002 | 0.009 |
| H | 0.005 | 0.059 | 0.062 |
| H | -0.005 | -0.032 | 0.051 |
| H | 0.001 | 0.072 | 0.058 |
| H | 0.003 | -0.021 | 0.062 |
| C | -0.017 | -0.067 | -0.425 |
| O | -0.015 | -0.387 | -0.433 |
| O | 0.017 | 0.231 | 0.148 |
| H | 0.025 | 0.370 | -0.416 |

7 TS

-197.491

------------------------

| | | | |
|---|---|---|---|
| C | -0.003 | -0.002 | -0.017 |
| C | 0.004 | -0.014 | -0.025 |
| C | 0.012 | -0.016 | -0.018 |
| C | 0.012 | -0.013 | -0.009 |
| C | 0.006 | -0.004 | -0.003 |
| C | 0.002 | -0.006 | 0.000 |
| H | 0.004 | -0.005 | -0.008 |
| H | -0.016 | -0.004 | 0.004 |
| H | 0.019 | -0.002 | 0.004 |
| H | -0.005 | -0.003 | -0.008 |
| H | 0.001 | -0.011 | 0.002 |
| H | -0.012 | -0.013 | -0.003 |
| H | 0.015 | -0.006 | 0.000 |
| H | -0.010 | -0.016 | 0.009 |
| H | 0.005 | 0.066 | -0.023 |
| H | -0.004 | 0.068 | 0.064 |
| H | 0.016 | 0.029 | -0.025 |
| H | -0.003 | 0.051 | 0.033 |
| H | 0.005 | 0.006 | -0.003 |
| H | -0.004 | 0.003 | -0.002 |
| H | -0.000 | -0.005 | -0.007 |
| H | 0.011 | 0.004 | 0.006 |
| H | 0.006 | 0.004 | -0.015 |
| H | -0.011 | -0.015 | -0.003 |
| H | -0.010 | -0.013 | -0.004 |
| H | -0.010 | -0.014 | -0.003 |
| H | -0.000 | -0.005 | 0.003 |
| H | -0.005 | -0.003 | 0.009 |
| H | -0.002 | -0.004 | -0.006 |
| H | 0.008 | -0.009 | 0.005 |
| H | 0.009 | -0.009 | 0.008 |
| H | 0.009 | -0.011 | 0.007 |
| H | 0.004 | -0.004 | 0.006 |
| H | 0.003 | -0.005 | 0.005 |
| H | 0.004 | -0.002 | -0.000 |
| H | 0.005 | -0.008 | -0.004 |
| H | 0.001 | -0.001 | -0.000 |
| H | 0.001 | -0.009 | 0.004 |
| H | 0.003 | -0.001 | 0.009 |
| H | 0.001 | 0.002 | -0.001 |
| H | 0.004 | 0.001 | 0.015 |
| H | -0.009 | -0.010 | 0.005 |
| H | -0.010 | -0.011 | 0.006 |
| H | -0.009 | -0.008 | 0.007 |
| C | 0.008 | -0.435 | -0.011 |
| O | -0.006 | -0.480 | -0.333 |
| O | -0.014 | 0.179 | 0.220 |
| H | -0.037 | -0.368 | 0.427 |

8 TS

-242.218

------------------------

| | | | |
|---|---|---|---|
| C | 0.013 | 0.002 | 0.004 |
| N | 0.015 | 0.007 | -0.080 |
| C | 0.021 | 0.010 | -0.185 |
| N | 0.019 | 0.004 | -0.081 |
| C | 0.016 | 0.003 | 0.004 |
| C | -0.002 | -0.003 | 0.010 |
| N | -0.001 | 0.031 | 0.048 |
| C | -0.003 | 0.049 | 0.170 |
| N | -0.003 | 0.029 | 0.050 |
| C | 0.002 | 0.023 | -0.004 |
| C | -0.000 | 0.023 | -0.005 |
| C | -0.003 | 0.026 | -0.005 |
| O | -0.000 | 0.018 | -0.022 |
| C | -0.006 | 0.015 | 0.030 |
| C | -0.014 | 0.020 | 0.022 |
| C | -0.015 | 0.017 | 0.011 |
| C | -0.007 | 0.006 | 0.004 |
| C | 0.000 | -0.002 | 0.010 |
| C | 0.003 | 0.002 | 0.021 |
| AU | 0.006 | 0.002 | -0.088 |
| O | -0.001 | 0.022 | -0.023 |
| C | -0.000 | 0.013 | -0.001 |
| C | 0.001 | 0.024 | 0.000 |
| C | -0.022 | 0.014 | 0.024 |
| C | -0.007 | 0.004 | -0.008 |
| C | 0.007 | -0.006 | 0.026 |
| C | 0.020 | 0.012 | 0.024 |
| C | 0.006 | 0.002 | -0.008 |
| C | -0.010 | -0.005 | 0.025 |
| H | -0.006 | -0.011 | 0.006 |
| H | 0.023 | 0.017 | 0.005 |
| H | -0.024 | 0.021 | 0.005 |
| H | 0.004 | -0.009 | 0.006 |
| H | -0.021 | 0.006 | 0.021 |
| H | -0.017 | -0.019 | 0.033 |
| H | 0.007 | -0.007 | 0.024 |
| H | 0.023 | 0.010 | 0.023 |
| H | 0.028 | 0.010 | 0.017 |
| H | 0.016 | 0.001 | 0.035 |
| H | 0.012 | 0.007 | -0.011 |
| H | 0.005 | 0.001 | -0.012 |
| H | 0.001 | -0.002 | -0.011 |
| H | -0.025 | 0.011 | 0.024 |
| H | -0.019 | 0.004 | 0.036 |
| H | -0.028 | 0.013 | 0.017 |
| H | 0.021 | 0.005 | 0.022 |
| H | -0.011 | -0.006 | 0.024 |
| H | 0.014 | -0.023 | 0.036 |
| H | -0.003 | -0.001 | -0.010 |
| H | -0.009 | 0.005 | -0.013 |
| H | -0.011 | 0.009 | -0.010 |
| H | 0.001 | 0.025 | -0.000 |
| H | 0.002 | 0.024 | 0.001 |
| H | 0.002 | 0.023 | 0.001 |
| H | 0.001 | 0.013 | 0.003 |
| H | -0.001 | 0.010 | -0.007 |
| H | 0.000 | 0.011 | 0.003 |
| C | -0.030 | -0.164 | 0.468 |
| O | -0.047 | -0.430 | 0.451 |
| O | 0.007 | 0.184 | -0.120 |
| C | -0.001 | 0.002 | -0.008 |
| N | 0.002 | -0.026 | -0.039 |
| C | -0.000 | -0.048 | -0.158 |
| N | 0.001 | -0.026 | -0.044 |
| C | 0.007 | -0.022 | -0.001 |
| C | 0.002 | -0.018 | 0.004 |
| C | 0.009 | -0.013 | 0.015 |
| O | 0.006 | -0.020 | 0.014 |
| C | -0.004 | -0.011 | -0.024 |
| C | -0.011 | -0.015 | -0.017 |
| C | -0.012 | -0.013 | -0.008 |
| C | -0.005 | -0.004 | -0.002 |
| C | 0.000 | 0.003 | -0.007 |
| C | 0.003 | -0.000 | -0.016 |
| AU | -0.002 | 0.003 | 0.082 |
| C | -0.007 | -0.014 | 0.004 |
| C | 0.001 | -0.017 | 0.001 |
| C | -0.015 | -0.012 | -0.018 |
| C | -0.005 | -0.003 | 0.007 |
| C | 0.007 | 0.004 | -0.018 |
| C | 0.016 | -0.010 | -0.019 |
| C | 0.006 | -0.002 | 0.006 |
| C | -0.009 | 0.001 | -0.020 |
| H | -0.004 | 0.007 | -0.005 |
| H | 0.020 | -0.016 | -0.005 |
| H | -0.019 | -0.016 | -0.004 |
| H | 0.004 | 0.008 | -0.004 |
| H | -0.019 | -0.010 | -0.016 |
| H | -0.015 | 0.012 | -0.031 |
| H | 0.005 | 0.003 | -0.016 |
| H | 0.021 | 0.003 | -0.021 |
| H | 0.012 | -0.008 | -0.008 |
| H | 0.017 | -0.011 | -0.035 |
| H | 0.010 | -0.006 | 0.008 |
| H | 0.006 | -0.001 | 0.009 |
| H | 0.003 | 0.001 | 0.007 |
| H | -0.017 | -0.008 | -0.018 |
| H | -0.013 | -0.007 | -0.025 |
| H | -0.017 | -0.011 | -0.013 |
| H | 0.016 | -0.004 | -0.016 |
| H | -0.006 | 0.004 | -0.016 |
| H | 0.011 | 0.014 | -0.026 |
| H | -0.002 | 0.001 | 0.008 |
| H | -0.005 | -0.004 | 0.011 |
| H | -0.008 | -0.007 | 0.008 |
| H | 0.002 | -0.018 | 0.001 |
| H | 0.001 | -0.017 | 0.001 |
| H | 0.001 | -0.017 | 0.001 |
| H | -0.010 | -0.010 | 0.006 |
| H | -0.004 | -0.014 | 0.004 |
| H | -0.011 | -0.017 | -0.000 |
| H | 0.003 | -0.031 | 0.046 |
| H | 0.036 | 0.018 | 0.019 |
| C | 0.005 | 0.081 | -0.451 |
| O | 0.014 | 0.389 | -0.468 |
| C | 0.007 | 0.007 | -0.038 |
| AU | 0.003 | -0.008 | 0.073 |
| C | 0.011 | 0.002 | -0.038 |
| C | 0.014 | 0.003 | 0.020 |
| C | 0.010 | -0.001 | 0.020 |
| C | -0.049 | 0.024 | -0.412 |
| O | 0.250 | -0.004 | 0.151 |
| O | -0.381 | 0.057 | -0.410 |
| H | 0.400 | -0.004 | -0.439 |
| H | 0.015 | 0.005 | 0.033 |
| H | 0.021 | -0.001 | 0.016 |
| H | 0.004 | 0.008 | 0.016 |
| H | 0.012 | -0.005 | 0.033 |
| H | -0.001 | -0.006 | 0.016 |
| H | 0.016 | 0.003 | 0.016 |
| H | 0.001 | 0.040 | -0.014 |
| H | 0.007 | 0.014 | -0.039 |
| H | 0.008 | -0.031 | -0.017 |
| H | 0.009 | 0.039 | -0.016 |
| H | 0.011 | -0.004 | -0.037 |
| H | 0.005 | -0.033 | -0.017 |

| | | | | | | | | | | | |
|---|---|---|---|---|---|---|---|---|---|---|---|
| H | 0.124 | 0.306 | 0.353 | O | -0.005 | -0.210 | 0.146 | | | | |
| | | | | H | -0.109 | -0.354 | -0.388 | | | | |
| 9 TS | | | | 10 TS | | | | 11 TS | | | |
| | | -229.952 | | | | -233.470 | | | | -236.085 | |
| | ------------------------ | | | | ------------------------ | | | | ------------------------ | | |
| C | 0.011 | 0.013 | -0.016 | C | 0.006 | -0.011 | -0.001 | C | 0.005 | -0.001 | 0.012 |
| C | 0.000 | 0.013 | -0.024 | C | -0.001 | -0.018 | -0.011 | C | -0.002 | 0.008 | 0.019 |
| C | -0.009 | 0.001 | -0.017 | C | -0.011 | -0.011 | -0.008 | C | -0.011 | 0.008 | 0.012 |
| C | -0.006 | -0.004 | -0.008 | C | -0.011 | -0.005 | -0.004 | C | -0.011 | 0.005 | 0.005 |
| C | 0.002 | 0.001 | -0.002 | C | -0.003 | -0.001 | 0.001 | C | -0.003 | -0.000 | 0.000 |
| C | 0.011 | 0.008 | -0.007 | C | 0.004 | -0.005 | 0.004 | C | 0.003 | -0.004 | 0.005 |
| N | 0.000 | 0.024 | -0.056 | N | -0.000 | -0.045 | -0.025 | N | -0.002 | 0.017 | 0.050 |
| C | -0.002 | 0.038 | -0.179 | C | 0.001 | -0.165 | -0.043 | C | -0.004 | 0.024 | 0.164 |
| N | -0.002 | 0.022 | -0.056 | N | 0.001 | -0.045 | -0.024 | N | 0.000 | 0.016 | 0.049 |
| C | 0.001 | 0.013 | 0.052 | C | -0.001 | 0.054 | -0.012 | C | -0.000 | 0.014 | -0.021 |
| C | -0.001 | 0.014 | 0.051 | C | 0.001 | 0.054 | -0.013 | C | 0.001 | 0.014 | -0.021 |
| C | -0.002 | 0.011 | -0.024 | C | 0.002 | -0.018 | -0.011 | C | -0.000 | 0.008 | 0.019 |
| C | 0.008 | 0.001 | -0.017 | C | -0.006 | -0.011 | -0.001 | C | 0.010 | 0.007 | 0.011 |
| C | 0.006 | -0.003 | -0.008 | C | -0.004 | -0.005 | 0.003 | C | 0.009 | 0.005 | 0.005 |
| C | -0.002 | 0.001 | -0.002 | C | 0.003 | -0.001 | 0.000 | C | 0.001 | 0.000 | 0.000 |
| C | -0.011 | 0.009 | -0.007 | C | 0.011 | -0.005 | -0.005 | C | -0.007 | -0.003 | 0.005 |
| C | -0.012 | 0.013 | -0.016 | C | 0.011 | -0.011 | -0.008 | C | -0.008 | 0.000 | 0.012 |
| C | 0.012 | 0.001 | -0.019 | C | -0.007 | -0.011 | 0.000 | C | 0.009 | 0.005 | 0.010 |
| C | -0.017 | 0.008 | -0.018 | C | -0.007 | -0.012 | 0.003 | C | -0.000 | 0.006 | 0.002 |
| C | -0.002 | -0.000 | 0.006 | C | 0.009 | -0.009 | -0.005 | C | -0.008 | -0.001 | 0.011 |
| AU | 0.002 | -0.001 | 0.075 | C | 0.008 | -0.010 | -0.006 | C | -0.002 | -0.002 | 0.004 |
| C | -0.013 | 0.002 | -0.018 | AU | -0.001 | 0.073 | 0.006 | AU | 0.004 | -0.000 | -0.075 |
| C | 0.002 | -0.001 | 0.007 | C | -0.009 | -0.009 | -0.005 | CL | 0.004 | 0.009 | -0.011 |
| C | 0.016 | 0.007 | -0.018 | C | -0.008 | -0.009 | -0.006 | CL | -0.003 | 0.010 | -0.011 |
| H | 0.001 | 0.011 | 0.120 | C | 0.007 | -0.011 | 0.000 | C | -0.009 | 0.005 | 0.009 |
| H | -0.000 | 0.008 | 0.120 | C | 0.007 | -0.012 | 0.003 | C | -0.002 | 0.004 | 0.005 |
| H | 0.019 | 0.008 | -0.002 | C | -0.001 | -0.005 | -0.003 | C | 0.007 | -0.002 | 0.012 |
| H | -0.012 | -0.009 | -0.005 | C | 0.000 | -0.003 | 0.002 | C | -0.000 | -0.004 | 0.003 |
| H | 0.011 | -0.009 | -0.005 | C | 0.001 | -0.005 | -0.004 | C | -0.007 | 0.006 | 0.010 |
| H | -0.019 | 0.010 | -0.002 | C | -0.000 | -0.003 | 0.002 | C | 0.008 | -0.005 | 0.012 |
| H | 0.021 | -0.003 | -0.018 | H | 0.001 | 0.121 | -0.006 | C | -0.007 | -0.002 | 0.011 |
| H | 0.018 | 0.008 | -0.031 | H | -0.001 | 0.121 | -0.007 | C | 0.009 | 0.007 | 0.010 |
| H | 0.014 | 0.006 | -0.010 | H | -0.007 | -0.003 | 0.007 | H | 0.007 | -0.008 | 0.001 |
| H | -0.016 | 0.009 | -0.016 | H | 0.018 | -0.002 | -0.003 | H | -0.018 | 0.004 | 0.001 |
| H | -0.007 | -0.001 | -0.014 | H | -0.018 | -0.002 | -0.003 | H | 0.016 | 0.004 | 0.001 |
| H | -0.016 | -0.000 | -0.025 | H | 0.007 | -0.003 | 0.008 | H | -0.011 | -0.006 | 0.002 |
| H | -0.002 | -0.002 | 0.008 | H | -0.009 | -0.012 | -0.004 | H | 0.009 | 0.001 | 0.016 |
| H | 0.001 | -0.001 | 0.008 | H | 0.013 | -0.008 | -0.000 | H | -0.013 | 0.002 | 0.007 |
| H | 0.006 | 0.001 | 0.009 | H | -0.013 | -0.008 | -0.000 | H | 0.012 | -0.000 | 0.010 |
| H | 0.018 | 0.013 | -0.016 | H | 0.009 | -0.013 | -0.004 | H | -0.010 | 0.003 | 0.011 |
| H | 0.018 | -0.005 | -0.029 | H | 0.004 | 0.006 | 0.002 | H | -0.003 | -0.001 | -0.008 |
| H | 0.002 | -0.002 | -0.012 | H | -0.002 | -0.003 | -0.006 | H | -0.000 | 0.006 | 0.004 |
| H | -0.021 | -0.001 | -0.017 | H | -0.001 | -0.001 | 0.003 | H | -0.001 | -0.003 | 0.002 |
| H | -0.015 | 0.007 | -0.011 | H | -0.004 | -0.003 | -0.011 | H | 0.003 | 0.009 | 0.004 |
| H | -0.018 | 0.008 | -0.031 | H | 0.007 | -0.009 | -0.004 | H | -0.006 | 0.005 | 0.009 |
| H | -0.006 | 0.002 | 0.009 | H | 0.008 | -0.011 | -0.005 | H | -0.008 | 0.004 | 0.010 |
| H | -0.001 | -0.001 | 0.008 | H | 0.008 | -0.007 | -0.007 | H | -0.006 | 0.006 | 0.009 |
| H | 0.000 | -0.002 | 0.008 | H | 0.001 | -0.003 | 0.006 | H | 0.002 | -0.007 | 0.005 |
| C | -0.026 | -0.041 | -0.419 | H | 0.003 | -0.002 | 0.008 | H | -0.004 | -0.013 | -0.004 |
| O | -0.028 | -0.365 | -0.437 | H | 0.001 | 0.001 | -0.002 | H | -0.007 | 0.005 | -0.001 |

| | | | |
|---|---|---|---|
| O | 0.029 | 0.231 | 0.160 |
| H | 0.021 | 0.399 | -0.400 |
| H | -0.008 | -0.012 | 0.003 |
| H | -0.006 | -0.011 | 0.003 |
| H | -0.008 | -0.010 | 0.002 |
| H | -0.004 | 0.006 | 0.003 |
| H | 0.002 | -0.003 | -0.006 |
| H | 0.004 | -0.003 | -0.010 |
| H | 0.000 | -0.002 | 0.003 |
| H | -0.007 | -0.008 | -0.003 |
| H | -0.008 | -0.007 | -0.007 |
| H | -0.008 | -0.011 | -0.005 |
| H | -0.001 | -0.003 | 0.006 |
| H | -0.001 | 0.001 | -0.002 |
| H | -0.003 | -0.002 | 0.008 |
| H | 0.008 | -0.012 | 0.003 |
| H | 0.008 | -0.010 | 0.002 |
| H | 0.006 | -0.011 | 0.003 |
| C | 0.004 | -0.423 | -0.001 |
| H | 0.011 | -0.369 | -0.442 |
| O | 0.002 | -0.467 | 0.318 |
| O | -0.000 | 0.184 | -0.229 |
| H | 0.009 | -0.004 | 0.012 |
| H | 0.007 | -0.005 | 0.011 |
| H | 0.009 | -0.003 | 0.011 |
| H | 0.001 | -0.001 | -0.008 |
| H | -0.003 | 0.011 | 0.000 |
| H | -0.005 | 0.016 | -0.001 |
| H | -0.002 | -0.001 | -0.004 |
| H | 0.008 | 0.006 | 0.010 |
| H | 0.008 | 0.008 | 0.009 |
| H | 0.009 | 0.005 | 0.011 |
| H | 0.000 | -0.005 | 0.003 |
| H | -0.003 | 0.001 | 0.001 |
| H | 0.000 | -0.005 | 0.005 |
| H | -0.007 | -0.002 | 0.011 |
| H | -0.008 | -0.001 | 0.010 |
| H | -0.006 | -0.002 | 0.011 |
| C | -0.040 | -0.052 | 0.432 |
| O | -0.039 | -0.375 | 0.437 |
| O | 0.038 | 0.241 | -0.155 |
| H | 0.003 | 0.413 | 0.415 |

12 TS

-229.052

------------------------

| | | | |
|---|---|---|---|
| C | 0.012 | -0.011 | -0.017 |
| C | 0.001 | -0.010 | -0.025 |
| C | -0.009 | -0.000 | -0.017 |
| C | -0.007 | 0.004 | -0.008 |
| C | 0.002 | -0.001 | -0.002 |
| C | 0.011 | -0.007 | -0.007 |
| N | 0.001 | -0.019 | -0.059 |
| C | -0.001 | -0.027 | -0.171 |
| N | -0.001 | -0.018 | -0.059 |
| C | 0.001 | -0.013 | 0.020 |
| C | -0.001 | -0.014 | 0.020 |
| C | -0.001 | -0.009 | -0.026 |
| C | 0.009 | 0.001 | -0.017 |
| C | 0.006 | 0.004 | -0.008 |
| C | -0.002 | -0.001 | -0.002 |
| C | -0.011 | -0.008 | -0.007 |
| C | -0.012 | -0.011 | -0.017 |
| C | 0.013 | 0.000 | -0.020 |
| C | -0.017 | -0.007 | -0.019 |
| C | -0.002 | 0.000 | 0.007 |
| AU | 0.001 | -0.002 | 0.077 |
| C | -0.002 | -0.012 | 0.026 |
| C | 0.002 | -0.010 | 0.026 |
| C | -0.013 | -0.000 | -0.020 |
| C | 0.001 | 0.000 | 0.007 |
| C | 0.016 | -0.006 | -0.019 |
| H | 0.019 | -0.008 | -0.002 |
| H | -0.013 | 0.009 | -0.004 |
| H | 0.012 | 0.009 | -0.004 |
| H | -0.019 | -0.009 | -0.002 |
| H | 0.020 | 0.002 | -0.017 |
| H | 0.017 | -0.005 | -0.030 |

13 TS

-240.120

------------------------

| | | | |
|---|---|---|---|
| C | 0.015 | 0.002 | 0.038 |
| N | 0.018 | 0.000 | -0.066 |
| C | 0.022 | 0.002 | -0.184 |
| N | 0.017 | 0.002 | -0.066 |
| C | 0.015 | 0.000 | 0.038 |
| C | 0.008 | -0.003 | -0.032 |
| C | 0.005 | 0.019 | -0.015 |
| AU | 0.003 | -0.000 | 0.072 |
| C | 0.008 | 0.004 | -0.032 |
| C | 0.001 | 0.025 | -0.020 |
| C | 0.006 | -0.018 | -0.015 |
| C | 0.004 | -0.024 | -0.019 |
| H | 0.015 | 0.001 | 0.110 |
| H | 0.015 | 0.001 | 0.109 |
| H | 0.008 | 0.007 | -0.031 |
| H | 0.009 | -0.020 | 0.001 |
| H | 0.006 | -0.011 | -0.013 |
| H | 0.006 | -0.034 | -0.023 |
| H | -0.006 | 0.028 | -0.009 |
| H | 0.001 | 0.037 | -0.029 |
| H | -0.000 | 0.019 | -0.015 |
| H | 0.009 | -0.006 | -0.031 |
| H | 0.007 | 0.021 | 0.001 |
| H | 0.003 | 0.035 | -0.023 |
| H | 0.006 | 0.012 | -0.013 |
| H | -0.003 | -0.028 | -0.009 |
| H | 0.002 | -0.019 | -0.014 |
| H | 0.005 | -0.037 | -0.029 |
| C | -0.051 | 0.017 | -0.408 |
| O | -0.375 | -0.007 | -0.409 |
| O | 0.248 | -0.009 | 0.153 |
| H | 0.396 | 0.032 | -0.428 |

8-PP TS

-208.196

------------------------

| | | | |
|---|---|---|---|
| C | -0.018 | -0.012 | -0.013 |
| C | -0.027 | -0.020 | -0.007 |
| C | -0.014 | -0.012 | -0.007 |
| C | -0.004 | -0.006 | -0.013 |
| C | -0.002 | -0.005 | -0.016 |
| C | -0.008 | -0.007 | -0.018 |
| P | -0.022 | -0.052 | -0.010 |
| AU | 0.023 | 0.059 | 0.013 |
| C | -0.150 | -0.339 | -0.046 |
| P | -0.009 | -0.007 | 0.001 |
| C | -0.007 | -0.004 | 0.005 |
| C | -0.008 | -0.003 | 0.004 |
| C | -0.008 | -0.001 | 0.001 |
| C | -0.004 | 0.000 | -0.000 |
| C | -0.002 | -0.001 | 0.001 |
| C | -0.003 | -0.004 | 0.002 |
| C | -0.007 | -0.006 | 0.004 |
| N | -0.003 | -0.005 | 0.001 |
| C | 0.000 | -0.003 | -0.000 |
| C | -0.000 | -0.001 | -0.000 |
| N | -0.005 | -0.003 | 0.002 |
| C | -0.003 | -0.005 | 0.001 |
| C | 0.000 | 0.007 | -0.008 |
| C | 0.004 | 0.004 | -0.001 |
| C | 0.002 | -0.002 | -0.002 |
| H | -0.006 | -0.005 | -0.022 |
| H | 0.010 | 0.013 | -0.002 |
| H | 0.005 | 0.011 | -0.013 |
| H | -0.001 | 0.001 | -0.000 |
| H | -0.007 | -0.006 | 0.006 |
| H | -0.002 | -0.010 | -0.003 |
| H | 0.007 | 0.004 | -0.001 |
| H | 0.015 | -0.005 | -0.013 |
| H | -0.017 | -0.006 | -0.017 |
| H | -0.010 | 0.002 | -0.017 |
| H | -0.016 | 0.001 | -0.026 |
| H | -0.002 | 0.003 | 0.008 |
| H | 0.002 | 0.000 | 0.009 |
| H | 0.005 | -0.002 | 0.008 |
| H | 0.017 | -0.006 | -0.017 |
| H | 0.016 | 0.002 | -0.026 |
| H | 0.009 | 0.002 | -0.017 |
| H | -0.021 | 0.000 | -0.017 |
| H | -0.015 | -0.006 | -0.013 |
| H | -0.018 | -0.006 | -0.030 |
| H | -0.006 | -0.002 | 0.009 |
| H | -0.000 | 0.001 | 0.008 |
| H | 0.001 | 0.002 | 0.008 |
| H | -0.004 | -0.013 | 0.033 |
| H | 0.005 | -0.013 | 0.021 |
| H | -0.008 | -0.008 | 0.021 |
| H | 0.005 | -0.011 | 0.033 |
| H | 0.009 | -0.006 | 0.020 |
| H | -0.004 | -0.011 | 0.021 |
| C | -0.020 | 0.053 | -0.420 |
| O | -0.023 | 0.388 | -0.428 |
| O | 0.028 | -0.243 | 0.149 |
| H | 0.022 | -0.383 | -0.434 |
| H | 0.006 | 0.004 | -0.002 |
| H | 0.000 | 0.012 | -0.002 |
| H | 0.002 | -0.001 | -0.001 |
| H | 0.002 | -0.003 | -0.003 |
| H | 0.003 | -0.001 | -0.002 |
| H | -0.017 | -0.011 | 0.000 |
| H | 0.000 | -0.002 | -0.016 |
| H | 0.003 | -0.003 | -0.019 |
| H | -0.005 | -0.007 | -0.024 |
| H | -0.021 | -0.015 | -0.010 |
| H | -0.001 | -0.006 | 0.001 |
| H | -0.001 | -0.001 | -0.001 |
| H | -0.004 | 0.002 | -0.002 |
| H | -0.010 | 0.001 | 0.000 |
| H | -0.009 | -0.003 | 0.005 |
| O | 0.182 | -0.531 | -0.024 |
| O | -0.163 | 0.261 | -0.032 |
| H | -0.576 | -0.315 | -0.057 |